\documentclass[12pt,a4paper,notitlepage]{report}%
\usepackage{times} 
\usepackage{bm}
\usepackage{amssymb}
\usepackage{amsmath}
\usepackage{amsthm}% for theorems and examples

\usepackage{amsfonts}
\usepackage{indentfirst}
\usepackage[none]{hyphenat}
\usepackage{afterpage}
\usepackage{longtable}
\usepackage{graphicx}
\usepackage[hang,small,bf]{caption}
\usepackage{setspace}
\usepackage[serbian]{babel}
\usepackage[T1]{fontenc}
\usepackage[cp1250]{inputenc}
\usepackage{epstopdf}%
\usepackage[hidelinks]{hyperref}
\usepackage{tabularx}
\usepackage{multirow}
\usepackage{multicol}
\usepackage{booktabs,makecell}
\usepackage{cite}
\usepackage{xcolor}
\usepackage{graphicx}
\usepackage{epstopdf}
\usepackage[pagestyles]{titlesec}%numeracija strana
\newpagestyle{main}{ \setfoot{}{}{\thepage}}%numeracija strana
\usepackage{titlesec}
\usepackage{fancyhdr}
\usepackage{lettrine}
\usepackage{lipsum}
\usepackage{algorithm}						% neophodno za algoritam
\usepackage{algorithmicx}				  % neophodno za algoritam
\usepackage{algpseudocode}			  % neophodno za algoritam
\usepackage{listings}						   % ovo je za kodove u nekom programskom jeziku, podešeno za MATLAB
\algnewcommand\algorithmicoutput{\textbf{Output:}} \algnewcommand\Output{\item[\algorithmicoutput]}
\algnewcommand\algorithmicinput{\textbf{Input:}} \algnewcommand\Input{\item[\algorithmicinput]}

\renewcommand\listfigurename{Lista slika}
\renewcommand\listtablename{Lista tabela}

\providecommand{\U}[1]{\protect\rule{.1in}{.1in}}
 \newenvironment{proof}[1][Proof]{\noindent\textbf{#1.} }{\
\makeatletter
 \long
\rule{0.5em}{0.5em}}
 \def\qquad{\hskip0.5cm\relax}

\theoremstyle{definition}
\newtheorem{primjer}{Primjer}[chapter]

\titleformat{\chapter}[display]
{\normalfont\LARGE\sffamily}{\chaptertitlename\ \thechapter}{0pt}  
{\vskip2pt \vskip20pt\Huge\bfseries\filleft}
\fancypagestyle{plain}{%
	\fancyhf{}
	\fancyfoot[R]{\bfseries\thepage}
	\renewcommand{\headrulewidth}{0pt}
}

\begin{document}

\begin{center}
	\centering
\normalfont\sffamily{ UNIVERZITET CRNE GORE }
 
\normalfont\sffamily{ ELEKTROTEHNIČKI FAKULTET}

\vspace{7cm} \bigskip

{\normalfont\sffamily\large Miloš Brajović }
\smallskip

\bigskip

%{\normalfont\LARGE\sffamily}
 \textbf{\normalfont\large\sffamily \textbf{ANALIZA ALGORITAMA ZA REKONSTRUKCIJU SIGNALA RIJETKIH U HERMITSKOM I FURIJEOVOM TRANSFORMACIONOM DOMENU}}
\smallskip

\bigskip
{\normalfont\sffamily\large -- Doktorska disertacija -- \ }

{\normalsize \vfill
}
  
{\normalfont\sffamily\normalsize \vspace{3cm} Podgorica, 2019. godine }
\end{center}

\thispagestyle{empty}

\clearpage

\begin{center}
\bigskip
	\text{ }
	\vspace{3cm} 
%		{\normalsize \vfill
%	}

	{\normalfont\sffamily\large\textbf{ON RECONSTRUCTION ALGORITHMS FOR SIGNALS SPARSE IN HERMITE AND FOURIER DOMAINS}}

		{\normalsize \vfill
	}
	
{\normalfont\sffamily\large by}
	
		{\normalsize \bigskip
	}
	
			{\normalfont\sffamily\large MSc Miloš Brajović }
	{\normalsize \vfill
	}
	{\normalfont\sffamily\large A thesis submitted for the degree of \\ \textit{Doctor of Philosophy}  \ }
		{\normalsize \vfill
	}
	
		\centering
	\normalfont\sffamily{Faculty of Electrical Engineering}
		
	\normalfont\sffamily{University of Montenegro}
	
	{\normalsize \vspace{3cm} \normalfont \sffamily{Podgorica 2019 }}
	{\normalsize \vfill
	}
\end{center}

\thispagestyle{empty}

\clearpage

%%%%%%%%%%%%%%%%%%%%%%%%%%%%%%%%%%%%%%%%%%%%%%%%%%%%%%%%%%%%%%%%%%%%%%%%%%%%%%%%%%%%%%%%%%%%%%%%%%%%%

\begin{center}
PODACI O DOKTORANDU, MENTORU I ČLANOVIMA KOMISIJE
\end{center}
\thispagestyle{empty}
\begin{flushleft}
	DOKTORAND
	
	\begin{minipage}[t]{6cm}
		\flushleft
		 Ime i prezime:\\
		 \vspace{5pt}
		 Datum i mjesto ro\dj enja: \\
		  \vspace{5pt}
		 Naziv zavr\v{s}enog postdiplomskog studijskog programa:\\
		  \vspace{3pt}
		 \text{ }\\
		 Godina zavr\v{s}etka:
	\end{minipage}
\null\hfill
	\begin{minipage}[t]{9.7cm}
	\flushleft
	 \textbf{Miloš Brajović} \\
	  \vspace{5pt}
	\textbf{24.05.1988. godine, Podgorica, Crna Gora}\\
	 \vspace{5pt}
	\textbf{Elektrotehnički fakultet, odsjek \textit{Elektronika, telekomunikacije, računari}, smjer \textit{Ra\v{c}unari} -- magistarske studije}\\
	 \vspace{3pt}
	 \textbf{2013}.
%	Ime i prezime:\\
%	Datum i mjesto ro\dj enja: \\
%	Naziv zavr\v{s}enog postdiplomskog studijskog programa:\\
%	Godina zavr\v{s}etka>
\end{minipage}

\vspace{0.6cm}

%\vspace{1cm}
%INFORMACIJE O DOKTORSKOJ DISERTACIJI \newline
%Naziv doktorskih studija: \textbf{Doktorske studije elektrotehnike} \\
%Naziv teze: \textbf{Analiza algoritama za rekonstrukciju signala rijetkih u Hermitskom i  Furijeovom transformacionom domenu}\\
%Fakultet na kojem je disertacija odbranjena: \textbf{Elektrotehni\v{c}ki fakultet, Podgorica}
	\begin{minipage}[t]{6cm}
		\flushleft
		MENTOR:\\
	\end{minipage}
	\null\hfill
	\begin{minipage}[t]{9.7cm}
		\flushleft
		\textbf{dr Miloš Daković}, redovni profesor, Univerzitet Crne Gore, Elektrotehnički fakultet
	\end{minipage}

\vspace{0.7cm}
	\begin{minipage}[t]{6cm}
	\flushleft
	KOMISIJA ZA OCJENU PODOBNOSTI TEZE I KANDIDATA: \\
	\vspace{3mm}
\end{minipage}
\null\hfill
\begin{minipage}[t]{9.7cm}
	\flushleft
\textbf{dr Srđan Stanković}, redovni profesor, Elektrotehnički fakultet, Univerzitet Crne Gore \\
\textbf{dr Miloš Daković}, redovni profesor, Elektrotehnički fakultet, Univerzitet Crne Gore\\
\textbf{dr Irena Orović}, vanredni profesor, Elektrotehnički fakultet, Univerzitet Crne Gore

\end{minipage}

	\begin{minipage}[t]{6cm}
	\flushleft
	 \text{ }\\
	 	\vspace{5mm}
KOMISIJA ZA OCJENU DOKTORSKE DISERTACIJE: \\
\text{ }\\
\end{minipage}
\null\hfill
\begin{minipage}[t]{9.7cm}
	\flushleft
	\text{ }\\
		\vspace{4mm}
	\textbf{dr Ljubiša Stanković}, redovni profesor, Elektrotehnički fakultet, Univerzitet Crne Gore \\
\textbf{dr Miloš Daković}, redovni profesor, Elektrotehnički fakultet, Univerzitet Crne Gore\\
\textbf{dr Danilo Mandić}, redovni profesor, Department of Electrical and Electronic Engineering, Faculty of Engineering, Imperial College London
\end{minipage}

	\begin{minipage}[t]{6cm}
	\flushleft

\text{ }\\
	\vspace{5mm}
KOMISIJA ZA ODBRANU DOKTORSKE DISERTACIJE: \\

\text{ }\\

\text{ }\\
\end{minipage}
\null\hfill
\begin{minipage}[t]{9.7cm}
	\flushleft
	\text{ }\\
		\vspace{4mm}
	\textbf{dr Ljubiša Stanković}, redovni profesor, Elektrotehnički fakultet, Univerzitet Crne Gore, predsjednik \\
\textbf{dr Miloš Daković}, redovni profesor, Elektrotehnički fakultet, Univerzitet Crne Gore, mentor\\
\textbf{dr Danilo Mandić}, redovni profesor, Department of Electrical and Electronic Engineering, Faculty of Engineering, Imperial College London, član\\
\textbf{dr Vesna Popović-Bugarin}, vanredni profesor, Elektrotehnički fakultet, Univerzitet Crne Gore, član\\
\textbf{dr Irena Orović}, vanredni profesor, Elektrotehnički fakultet, Univerzitet Crne Gore, član

\end{minipage}

\vspace{0.5cm}

	\begin{minipage}[t]{6cm}
	\flushleft
	DATUM ODBRANE:\\
\end{minipage}
\null\hfill
\begin{minipage}[t]{9.7cm}
	\flushleft
	\vspace{3mm}
\line(1,0){140} \\
	%	Ime i prezime:\\
	%	Datum i mjesto ro\dj enja: \\
	%	Naziv zavr\v{s}enog postdiplomskog studijskog programa:\\
	%	Godina zavr\v{s}etka>
\end{minipage}
%Datum prijave doktorske teze: \textbf{16.12.2015. godine} \\
%Datum sjednice Senata Univerziteta na kojoj je prihva\'{c}ena teza: \textbf{23.06.2016. godine}\\
%Komisija za ocjenu podobnosti teze i kandidata: \\
%\hspace{0.5cm} \textbf{dr Srđan Stanković}, redovni profesor, Univerzitet Crne Gore, Elektrotehnički fakultet \\
%\hspace{0.5cm} \textbf{dr Miloš Daković}, redovni profesor, Univerzitet Crne Gore, Elektrotehnički fakultet \\
%\hspace{0.5cm} \textbf{dr Irena Orović}, vanredni profesor, Univerzitet Crne Gore, Elektrotehnički fakultet \\
%\vspace{0.2cm}
%Mentor: \textbf{dr Miloš Daković}, redovni profesor, Univerzitet Crne Gore, Elektrotehnički fakultet\\
%\vspace{0.2cm}
%Komisija za ocjenu doktorske disertacije: \\
%\hspace{6cm} \line(1,0){240} \\
%\hspace{6cm} \line(1,0){240} \\
%\hspace{6cm} \line(1,0){240} \\
%\vspace{0.2cm}
%Komisija za odbranu doktorske disertacije: \\
%\hspace{6cm}  \line(1,0){240} \\
%\hspace{6cm}  \line(1,0){240} \\
%\hspace{6cm} \line(1,0){240} \\
%\vspace{0.2cm}
%Datum odbrane: \line(1,0){160} \\
%%Datum promocije: \line(1,0){150}
\end{flushleft}
\clearpage

\thispagestyle{empty}
\chapter*{Zahvalnica}
\fancyhf{}
%\fancyhead[ER]{\footnotesize\sffamily\leftmark}
 \fancyhead[EL,OL]{\nouppercase\leftmark}
\fancyheadoffset[R]{0.00\textwidth}
%\fancyhead[EL,OR]{\bfseries\thepage}
\rfoot{\thepage}

\renewcommand{\headrulewidth}{0.0pt}
\textit{Neizmjerno sam zahvalan svom mentoru, prof. dr Milošu Dakoviću, na svesrdno pruženoj pomoći, iskrenim savjetima, nemjerljivom stpljenju i dubokom razumijevanju, tokom svih godina mog angažovanja na Elektrotehničkom fakultetu u Podgorici. Zahvaljujući njemu, bavljenje naukom i nastavnim aktivnostima za mene nije predstavljalo samo profesiju, već sveprožimajući dio života i --  životni stil. Odmjerenim pristupom u svim aspektima svojih profesionalnih aktivnosti i uvijek duboko promišljenim sagledavanjem stvari, on je ovu doktorsku disertaciju pozicionirao ne kao krajnju, već kao početnu tačku jednog naučno-istraživačkog procesa.}

\textit{Posebnu zahvalnost dugujem prof. dr Srđanu Stankoviću i prof. dr Ireni Orović koji su, svojim predanim radom i međunarodno afirmisanim i visoko respektabilnim profesionalnim rezultatima, za mene predstavljali, ali i dalje predstavljaju, neiscrpan izvor inspiracije i motivacije za bavljenje naukom. Njihova nepokolebljiva vjera u mene i u moje sposobnosti bili su ključni izvor hrabrosti koja mi je bila potrebna da prve značajne istraživačke rezultate izložim otvorenoj naučnoj kritici kroz proces publikovanja radova u renomiranim naučnim časopisima. Iskustvo zajedničkog rada, kao i iskreni i dobronamjerni savjeti i kritike, duboko su oblikovali moj profesionalni identitet.}

\textit{Pravi naučno-istraživački rad širinom i univerzalnošću svojih dometa i kompleksnošću svoje suštine prevazilazi okvire uobičajenih ljudskih aktivnosti. Njegov tok podrazumijeva riječima teško uhvatljiva znanja i vještine, koji se prostiru daleko izvan granica jedne konkretne naučne oblasti, konkretne profesije ili  konkretnog konteksta. Iako odavno postoje pokušaji kategorizacije i materijalizacije ovakvih znanja i vještina kroz različite priručnike i metodologije naučno-istraživačkog rada, jedna direktna i kritički nastrojena komunikacija i iskustvo neposrednog rada sa renomiranim naučnikom nikada ne mogu biti kompenzovani pisanom riječju, niti iskazani suvoparnim skupom pravila. Na pružanju jednog takvog iskustva kroz zajednički rad, neizmjerno sam zahvalan prof. dr Ljubiši Stankoviću.
}

\textit{Tokom izrade ove disertacije, imao sam posebno zadovoljstvo da sarađujem sa prof. dr Vesnom Popović-Bugarin, prof. dr Igorem Đurovićem i prof. dr Slobodanom Đukanovićem. Kultura predanog rada i uvijek prisutno stremljenje najvišim standardima kvaliteta, koji i predstavljaju temelj institucionalnog renomea Elektrotehničkog fakulteta, kroz ovu saradnju postali su dio moje profesionalne agende. Na tome sam im posebno zahvalan.}

%\textit{Prof. dr Budimiru Lutovcu se zahvaljujem na lijepo provedenim trenucima tokom službenih boravaka u inostranstvu. Uvijek zanimljivi razgovori sa prof. Lutovcem i njegova široka poznanstva -- ovim putovanjima dali su pečat nezaboravnog iskustva i neponovljivih uspomena.}

\textit{Neizmjerno se zahvaljujem svojim kolegama i prijateljima dr Žarku Zečeviću, dr Anđeli Draganić i dr Branki Jokanović, sa kojima je svaki ciklus studija na Elektrotehničkom fakultetu u svakom trenutku imao dimenziju domaćeg i bliskog.  Zahvaljujem se mr Stefanu Vujoviću, mr Isidori Stanković, mr Nikoli Bulatoviću, dr Marku Simeunoviću i mr Predragu Rakoviću na zajedničkom radu, podršci i na uvijek prijatnoj atmosferi u Laboratoriji za obradu signala. }

	\textit{Zahvaljujem se svojim prijateljima iz djetinjstva, gimnazijskih dana i prijateljima koje sam stekao tokom boravka u ,,Prvoj fazi'' Novog studentskog doma u Podgorici, na beskrajnom razumijevanju, pružanju motivacije i na uvijek kvalitetno provedenom vremenu, iako ga u nekim periodima izrade ove teze možda nije bilo onoliko koliko sam želio da ga bude.}

\textit{Na kraju, najveću zahvalnost dugujem svojoj porodici na bezuslovnoj i neiscrpnoj podršci, koju su mi do sada pružali u svim aspektima života.}

\bigskip

\noindent \textit{Podgorica, oktobra 2018. godine}

\hspace*{\fill}  \textit{Miloš Brajović}
\clearpage
\thispagestyle{empty}
\begin{center}
	PODACI O DOKTORSKOJ DISERTACIJI
\end{center}
\begin{flushleft}
	\begin{minipage}[t]{5.5cm}
		\flushleft
		Naziv doktorskih studija:\\
		\vspace{6pt}
		Naslov doktorske disertacije: \\
			\text{ }\\
		\vspace{5pt}
		Ključne riječi:\\
		\vspace{5pt}
		\text{ }\\
		\text{ }\\
		\text{ }\\
		\text{ }\\
		\text{ }\\
	Datum prijave doktorske teze:  \\
		\vspace{4pt}
	Datum sjednice Senata UCG na kojoj je prihva\'{c}ena teza: \\
	\vspace{4pt}
	Naučna oblast:\\
	\vspace{4pt}
	Uža naučna oblast:
	\end{minipage}
	\null\hfill
	\begin{minipage}[t]{10cm}
		\flushleft
		\textbf{Doktorske studije elektrotehnike} \\
		\vspace{5pt}
		\textbf{Analiza algoritama za rekonstrukciju signala rijetkih u Hermitskom i Furijeovom transformacionom domenu}\\
		\vspace{5pt}
	{Dekompozicija multikomponentnih signala, digitalna obrada signala, diskretna kosinusna transformacija (DCT), Furijeova transformacija, Hermitska transformacija, kompresivno odabiranje, multivarijantni signali, nestacionarni signali, obrada rijetkih (\textit{sparse}) signala,  vremensko-frekvencijska analiza}\\
		\vspace{6pt}
		\textbf{16.12.2015. godine}\\
		\text{ }\\
		\vspace{5pt}
		\textbf{23.06.2016. godine}\\
			\vspace{4pt}
		\textbf{Elektrotehnika, Računarstvo}\\
			\vspace{4pt}
		\textbf{Digitalna obrada signala}
		%	Ime i prezime:\\
		%	Datum i mjesto ro\dj enja: \\
		%	Naziv zavr\v{s}enog postdiplomskog studijskog programa:\\
		%	Godina zavr\v{s}etka>
	\end{minipage}
\end{flushleft}
\vspace{0.2cm}
\noindent REZIME: 

\noindent Disertacija sadrži originalne naučne doprinose u oblasti digitalne obrade signala. Primarno se razmatra problematika  rekonstrukcije signala sa rijetkom (engl. \textit{sparse}), odnosno, visoko koncentrisanom reprezentacijom u različitim transformacionim domenima. Razmatraju se: Furijeova transformacija, diskretna Hermitska transformacija, varijante diskretne kosinusne transformacije, kao i nekoliko vremensko-frekvencijskih reprezentacija. Rekonstrukcija rijetkih signala se bazira na minimizaciji njihovih mjera rijetkosti (koncentracije) u razmatranim domenima. Teza sadrži  analizu i predlog novih rekonstrukcionih algoritama iz konteksta kompresivnog odabiranja, kao i opsežnu analizu uticaja nedostajućih odbiraka signala na odgovarajuće transformacione koeficijente. Predstavljeno je detaljno izvođenje relacija koje  izloženu analizu povezuju sa opštom teorijom kompresivnog odabiranja i koje egzaktno opisuju fenomene koji su posljedica nedostajućih mjerenja u signalima. Izvršena je detaljna analiza uticaja aditivnog šuma, zatim stepena rijetkosti, broja nedostajućih odbiraka i drugih faktora na performanse razmatranih algoritama; izvedeni su izrazi za greške u rekonstrukciji
i  vjerovatnoće grešaka.  U tezi su predstavljeni i originalni algoritmi za optimizaciju parametara  Hermitske transformacije;  napravljena je analiza uticaja aditivnog šuma i predložen postupak njegovog uklanjanja. Predstavljen je i algoritam za dekompoziciju multikomponentnih multivarijantnih signala u kontekstu vremensko-frekvencijske analize, kao i algoritam za estimaciju trenutne frekvencije na bazi Wigner-ove distribucije. Veći broj numeričkih eksperimenata i primjera sa realnim i sintetičkim signalima potvrđuju teorijske rezultate i zaključke, pritom čineći doprinose disertacije nerazdvojivim od konteksta praktičnih primjena.

\vspace{0.1cm}	

\noindent UDK:

\clearpage
\thispagestyle{empty}
\begin{center}
	INFORMATION ON DOCTORAL DISSERTATION
\end{center}
\begin{flushleft}
	\begin{minipage}[t]{5.5cm}
		\flushleft
		PhD study program:\\
		\vspace{6pt}
		Dissertation title: \\
		\text{ }\\
		\vspace{5pt}
		Keywords:\\
		\vspace{5pt}
		\text{ }\\
		\text{ }\\
		\text{ }\\
		\text{ }\\
		Thesis application date:  \\
		\vspace{5pt}
		Thesis acceptance date (UoM Senate Session): \\
		\vspace{5pt}
		Scientific area:\\
		\vspace{5pt}
		Specific scientific area:
	\end{minipage}
	\null\hfill
	\begin{minipage}[t]{10cm}
		\flushleft
		\textbf{PhD studies in Electrical Engineering} \\
		\vspace{5pt}
		\textbf{On Reconstruction Algorithms for Signals Sparse in Hermite and Fourier Domains}\\
		\vspace{5pt}
			{Multicomponent signal decomposition, digital signal processing, discrete cosine transform (DCT), Fourier transform, Hermite transform, compressive sensing, multivariate signals, sparse signal processing, non-stationary signals, time-frequency signal analysis}\\
		\vspace{6pt}
		\textbf{December 16, 2015}\\
		\text{ }\\
		\vspace{6pt}
		\textbf{June 23, 2016}\\
		\vspace{6pt}
		\textbf{Electrical Engineering, Computer Science}\\
		\vspace{5pt}
		\textbf{Digital Signal Processing}
		%	Ime i prezime:\\
		%	Datum i mjesto ro\dj enja: \\
		%	Naziv zavr\v{s}enog postdiplomskog studijskog programa:\\
		%	Godina zavr\v{s}etka>
	\end{minipage}
\end{flushleft}
\vspace{0.2cm}
\noindent ABSTRACT: 
\noindent This thesis consists of original contributions in the area of digital signal processing. The reconstruction of signals sparse (highly concentrated) in various transform domains is the primary problem analyzed in the thesis. The considered domains include Fourier, discrete Hermite, one-dimensional and two-dimensional discrete cosine transform, as well as various time-frequency representations. Sparse signals are reconstructed using sparsity measures, being, in fact, the measures of signal concentration in the considered domains. The thesis analyzes the compressive sensing reconstruction algorithms and introduces new approaches to the problem at hand. The missing samples influence on analyzed transform domains is studied in detail, establishing the relations with the general compressive sensing theory. This study provides new insights on phenomena arising due to the reduced number of signal samples. The theoretical contributions involve new exact mathematical expressions which describe performance and outcomes of reconstruction algorithms, also including the study of the influence of additive noise, sparsity level and the number of available measurements on the reconstruction performance,  exact expressions for reconstruction errors and error probabilities. Parameter optimization of the discrete Hermite transform is also studied, as well as the additive noise influence on Hermite coefficients, resulting in new parameter optimization and denoising algorithms. Additionally, an algorithm for the decomposition of multivariate multicomponent signals is introduced, as well as an instantaneous frequency estimation algorithm based on the Wigner distribution. Extensive numerical examples and experiments with real and synthetic data validate the presented theory and shed a new light on practical applications of the results.

\vspace{0.2cm}	

\noindent UDK:

\clearpage

\thispagestyle{empty}

\chapter*{Predgovor}
{\small
Ova doktorska disertacija je rezultat višegodišnjeg naučno-istraživačkog rada u Laboratoriji za obradu signala na Elektrotehničkom Fakultetu u Podgorici. Grupa za obradu signala (www.tfsa.ac.me) još od početka devedesetih godina XX vijeka prati naučne izazove nametnute ubrzanim razvojem tehnologije i svih aspekata  digitalizacije, kontinuirano produkujući značajne rezultate na polju fundamentalnih teorijskih istraživanja u oblasti digitalne obrade signala, kao i na polju njihovih praktičnih primjena, posebno u obradi: biomedicinskih, multimedijalnih, radarskih, sonarskih, telekomunikacionih i drugih vrsta signala. 

U tezi su predstavljeni originalni naučni doprinosi u oblasti kompresivnog odabiranja (engl. \textit{compressive sensing}) i obrade rijetkih (engl. \textit{sparse}) signala, kao i neki doprinosi u oblasti vremensko-frekvencijske analize. Kompresivno odabiranje nudi algoritme i tehnike za rekonstrukciju signala na osnovu redukovanog skupa mjerenja. Pomoću ovih tehnika je moguće rekonstruisati  signale na osnovu malog procenta dostupnih odbiraka: na primjer, digitalnu sliku na osnovu samo 10\% dostupnih piksela (tačaka) ili audio signal kod kojeg je 60\% sadržaja izgubljeno ili oštećenog tokom akvizicije, transmisije ili skladištenja. Pojava prvih algoritama kompresivnog odabiranja je u vezi sa biomedicinom, gdje je prilikom određenih dijagnostičkih procedura bilo veoma značajno skratiti vrijeme snimanja, smanjujući pritom izloženost pacijenata štetnim zračenjima ili štetnim hemikalijama, ali tako da kvalitet snimljene informacije bude potpuno očuvan ili poboljšan. Senzori razvijeni u kontekstu kompresivnog odabiranja mogu da rade sa značajno smanjenom potrošnjom energije, sa skraćenim vremenom akvizicije i uopšte -- sa smanjenim zahtjevima a očuvanim kvalitetom rezultata, što redukuje cijenu opreme i potrošnih materijala. Svi nabrojani izazovi su dobili posebnu važnost u današnjoj digitalnoj eri, gdje količina informacija, ali i broj uređaja za njihovu akviziciju, skladištenje i prenos bilježe eksponencijalni rast. Ostatak teze sadrži doprinose u oblasti vremensko-frekvencijske analize, koja nudi tehnike za reprezentaciju, analizu i obradu informacija iz vremenski promjenljivog spektralnog sadržaja signala. Posebno je značajna predložena tehnika za razdvajanje komponenti multivarijantnih signala.

Teza predstavlja rezultat višegodišnjeg učešća autora na većem broju naučno-istraživačkih i stručnih projekata, među kojima je posebno značajno istaći projekat ,,New ICT Compressive sensing-based trends applied to: multimedia, biomedicine and communications (CS-ICT)'', finansiran od strane Ministarstva nauke Crne Gore iz kredita Svjetske banke, realizovan pod rukovodstvom prof. dr Srđana Stankovića. Glavni doprinosi teze su verifikovani kroz publikovanje većeg broja radova u naučnim časopisima, od čega osam u renomiranim međunarodnim časopisima sa SCI/SCIE liste, kao i kroz prezentovanje rezultata na relevantnim međunarodnim naučnim konferencijama.
}
\clearpage

\thispagestyle{empty}

\chapter*{Izvod iz teze}
 
 U disertaciji se analizira rekonstrukcija signala zasnovana na mjerama rijetkosti (koncentracije) u specifičnim transformacionim domenima. Najveći dio sadržaja se odnosi na različite aspekte kompresivnog odabiranja (engl. \textit{compressive sensing}) i rekonstrukcije rijetkih (engl. \textit{sparse}) signala. Kao generalni koncept, mjere koncentracije su primjenjene u rekonstrukciji i poboljšanjima reprezentacija signala i mimo tog konteksta. U nastavku slijedi koncizan pregled sadržaja teze, uz posebno isticanje originalnih doprinosa.

U uvodu disertacije je uveden pojam kompresivnog odabiranja i koncept rijetkosti, odnosno, koncentrisanih reprezentacija signala u transformacionim domenima,  na kojima je zasnovano izlaganje u ostatku teze. Na ovom mjestu su precizirani motivacioni faktori koji stoje iza istraživanja prezentovanih u tezi. Napravljen je pregled doprinosa i strukture disertacije.

U prvoj glavi teze su predstavljeni osnovni teorijski principi koji stoje iza kompresivnog odabiranja i rekonstrukcije rijetkih signala. Definisani su formalni uslovi koji garantuju uspješnu rekonstrukciju i jedinstvenost rješenja koje se dobija primjenom rekonstrukcionih algoritama. Fundamentalni koncepti su uvedeni kroz razmatranje diskretne Furijeove tranformacije, kao reprezentativnog primjera domena rijetkosti signala. Predstavljeni su osnovni rekonstrukcioni algoritmi, koji su od posebnog značaja za dalja izlaganja u tezi.

Druga glava sadrži četiri odjeljka kroz koje su prezentovani glavni doprinosi u obradi signala koji imaju rijetku, odnosno, visoko koncentrisanu reprezentaciju u Hermitskom transformacionom domenu. Nakon formalnog uvođenja Hermitske transformacije, njenih diskretnih ekvivalenata i njihovih glavnih karakteristika, razmatra se optimizacija parametara (faktora skaliranja vremenske ose i pomjeraja po vremenskoj osi) jednog tipa diskretne Hermitske reprezentacije, sa primjenom u kompresiji QRS kompleksa -- posebno značajnih segmenata biomedicinskih EKG signala. Data je analiza uticaja aditivnog šuma na diskretnu Hermitsku transformaciju i prezentovan je jednostavan postupak za uklanje šuma u slučaju signala koji imaju koncentrisanu reprezentaciju u ovom domenu. Važan dio druge glave čini analiza uticaja nedostajućih odbiraka na koeficijente diskretne Hermitske transformacije, koja sadrži više novih teorijskih rezultata: matematičko modelovanje Hermitskih koeficijenata signala sa nedostajućim odbircima, probabilističku analizu uspješnosti rekonstrukcije koja je zasnovana na prezentovanoj teoriji, interpretaciju indeksa koherentnosti parcijalne mjerne matrice diskretne Hermitske transformacije u kontekstu predložene teorije, originalnu analizu uticaja aditivnog šuma na  performanse rekonstrukcije, kao i analizu rekonstrukcije signala koji nijesu rijetki, a koji su rekonstruisani uz pretpostavku o rijetkosti u ovom domenu. Predložena su dva nova algoritma za rekonstrukciju zasnovana na izloženoj teoriji i treći, iterativni gradijentni algoritam za rekonstrukciju, sa odgovarajućom interpretacijom u ovom domenu. Svi rezultati su potkrijepljeni numeričkim primjerima, uključujući i eksperimente sa realnim EKG i UWB signalima, kao i opsežne eksperimente u kojima je izvršena statistička evaluacija izloženih teorijskih rezultata.

Diskretna kosinusna transformacija (DCT) analizirana je u trećoj glavi disertacije. Razmatrana su dva oblika ove transformacije: jednodimenzioni i dvodimenzioni. Kod jednodimenzione transformacije, odrađena je detaljna analiza uticaja nedostajućih odbiraka na transformacione koeficijente, izvedeni su eksplicitni izrazi za vjerovatnoću uspješne rekonstrukcije u zavisnosti od broja dostupnih mjerenja, stepena rijetkosti i dužine signala. Uspostavljena je teorijska interpretacija indeksa koherentnosti parcijalne DCT matrice. U glavi je izvršena analiza uticaja aditivnog šuma na performanse rekonstrukcije, izveden je eksplicitini izraz za grešku u rekonstrukciji signala koji nijesu rijetki a rekonstruisani su uz pretpostavku o rijetkosti u ovom domenu i redefinisani su algoritmi za rekonstrukciju signala na osnovu predložene teorije. Bitan segment čini opsežna eksperimentalna analiza primjene u kontekstu uklanjanja impulsnih smetnji prisutnih u audio signalima. Glava sadrži brojne primjere i eksperimentalne rezultate sa sintetičkim i realnim audio signalima. U drugom dijelu ove glave razmatrana je rekonstrukcija signala rijetkih u dvodimenzionom DCT domenu, odnosno, rekonstrukcija digitalnih slika na osnovu redukovanog skupa dostupnih piksela. Izvedeni su izrazi koji modeluju statističke osobine 2D-DCT koeficijenata koji se, kao posljedica nedostajućih odbiraka signala, ponašaju kao slučajne varijable. I kod ove transformacije je izveden izraz za energiju greške u rekonstrukciji signala koji nijesu rijetki (tj. koji su aproksimativno rijetki), a koji su rekonstruisani uz pretpostavku o rijetkosti. Numerička evaluacija rezultata je izvršena u eksperimentima sa realnim signalima -- digitalnim slikama.

Četvrta glava disertacije se primarno bavi primjenama mjera koncentracije (rijetkosti) u problemima vremensko-frekvencijske analize. U glavi je prezentovan algoritam za rekonstrukciju krutog tijela (engl. \textit{rigid-body}) kod ISAR signala, na osnovu mjere koncentracije vremensko-frekvencijske reprezentacije i algoritma za kompresivno odabiranje. Glava sadrži i predlog novog algoritma za estimaciju trenutne frekvencije na bazi Wigner-ove distribucije, u uslovima jakog aditivnog šuma prisutnog u signalu. U drugom dijelu je prezentovan originalni pristup za dekompoziciju multivarijantnih multikomponentnih signala, koji se dobijaju mjerenjem fizičkih procesa pomoću više senzora. Predstavljeni dekompozicioni pristup, zasnovan na mjeri koncentracije iz konteksta kompresivnog odabiranja, omogućava rekonstrukciju pojedinačnih komponenti multikomponentnog signala, čak i u slučajevima kada se komponente preklapaju u vremensko-frekvencijskoj ravni.

U zaključnoj glavi su navedeni osnovni doprinosi disertacije i identifikovane su teme budućih istraživanja u ovoj oblasti.

\clearpage

\thispagestyle{empty}

\chapter*{Thesis overview}

The main topic of this thesis is the signal reconstruction based on concentration measures in specific transform domains. Mainly, the signal reconstruction in the thesis assumes the compressive sensing framework and sparse signal processing techniques. However, as a general concept, these measures are studied in the signal reconstruction framework even beyond this context. The rest of this overview briefly outlines each chapter, emphasizing original contributions.

The introductory chapter presents basic compressive sensing concepts, including the definition of sparsity and its relation with measures of signal concentration in relevant transform domains. This chapter also reveals the basic motivation behind the conducted research.

The first chapter presents fundamental compressive sensing principles and the basic theory behind the sparse signal reconstruction; it formulates the reconstruction problem and introduces formal conditions which guarantee the existence and the uniqueness of the solution. Fundamental concepts are mainly introduced by considering the Fourier transform as the representative example of a sparsity domain. To better support the further chapters, basic reconstruction algorithms are also presented. 

The second chapter consists of four sections, presenting the main contributions related to the processing of signals with a high concentration in the Hermite transform domain.  After the formal introduction of the Hermite transform and relevant discrete-time counterparts, along with their basic properties, we consider the parameter optimization of the transform. The presented optimization algorithm uses concentration measures of Hermite coefficients to automatically adjust the scaling factor of the time axis and the time-shift parameter for the considered signal. This optimization serves as the basis of a compression procedure developed for QRS complexes. Being important parts of ECG signals, QRS complexes are commonly used in medical diagnosis and treatment.
The chapter continues with the analysis of the additive noise influence on discrete Hermite transform coefficients. A simple denoising procedure is also introduced in this context.  Particularly important part analyses the Hermite coefficients calculated based on a reduced set of measurements. New theoretical contributions include mathematical models of Hermite coefficients, corresponding to under-sampled signals,  probabilistic analysis of the reconstruction process and interpretation of the coherence index which characterizes the Hermite sensing matrix. Additive noise influence on the reconstruction performance is also analyzed. The error in the reconstruction of non-sparse signals, performed under the sparsity assumption, is derived based on the presented theory. Two compressive sensing algorithms for signal recovery are proposed in light of the presented theoretical framework. Moreover, one additional algorithm based on the signal reconstruction in the measurements domain is proposed. A large number of numerical examples validate the presented theory. This includes examples with real ECG and UWB signals and a large number of comprehensive numerical experiments. 

The third chapter deals with the discrete cosine transform (DCT): one-dimensional in the first section and two-dimensional in the second section. Phenomena in the one-dimensional  DCT domain,  related to the unavailable measurements, are investigated in detail. Derivation of an explicit formula for the probability regarding the successful reconstruction additionally supports this analysis.
This probability depends on the number of available measurements, sparsity and signal length. The coherence index of the partial DCT matrix is interpreted in the context of the presented theory. This chapter also analyses the additive noise influence on the reconstruction performance and presents the derivation of the reconstruction error for signals reconstructed under the sparsity assumption, but which are not sparse. An important segment of this chapter is the development of two reconstruction algorithms, applied in the removal of impulsive disturbances in audio signal processing. Numerous examples and quite extensive numerical experiments, with synthetic and real audio data, support the presented theory. 
The second part of this chapter deals with the reconstruction from a reduced set of measurements, with the two-dimensional DCT acting as a domain of sparsity (high signal concentration). Digital images are the most important example of such signals. This part also investigates the statistical properties of transform coefficients acting as random variables due to the measurements unavailability. These properties serve as a starting point of the reconstruction error energy derivation. This error characterizes the performance of the reconstruction, in the case of approximately sparse signals being reconstructed under the sparsity assumption. Numerical examples with digital images validate the theory.

The last chapter primarily considers the application of concentration (sparsity) measures in the time-frequency analysis framework. The chapter presents an algorithm for the reconstruction of the rigid body in ISAR signals, which exploits the concentration measure of a time-frequency representation in conjunction with a compressive sensing reconstruction algorithm. In this chapter, an algorithm for the instantaneous frequency estimation, based on Wigner distribution, is also proposed. The second part of the chapter presents an original approach for the decomposition of multivariate multicomponent signals. Multivariate signals emerge in applications where multiple sensors are used to measure the physical processes. The presented decomposition approach, which exploits concentration measures from the compressive sensing framework, accurately extracts the signal components from multicomponent signals, even if they have overlapped supports in the time-frequency plane.

The thesis ends with concluding remarks, which also indicate topics of further research.

\clearpage

\chapter*{Popis akronima}
%\thispagestyle{empty}
%\addcontentsline{toc}{chapter}{Popis skraćenica}
\fancyhf{}
%\fancyhead[ER]{\footnotesize\sffamily\leftmark}
\fancyhead[EL,OL]{\nouppercase\leftmark}
\fancyheadoffset[R]{0.00\textwidth}
%\fancyhead[EL,OR]{\bfseries\thepage}
%\setcounter{page}{1}%
\rfoot{\thepage}
{\small
	\vspace{-3mm}
\begin{longtable}{p{3cm}p{0.5cm}p{11.7cm}}
ACO  & - & Ant Colony Optimization (Optimizacija mravlje kolonije);\\
AR  & - & Auto-regresivni;\\
2D-DCT & - & Dvodimenziona DCT;\\
CoSaMP & - & Compressive Sampling Matched Pursuit;\\
CS & - & Compressive Sensing (kompresivno odabiranje);\\
DCT & - & Discrete Cosine Transform (Diskretna kosinusna transformacija); \\
DFT & - & Diskretna Furijeova transformacija; \\
DHT & - & Diskretna Hermitska transformacija; \\
DHT1 & - & Diskretna Hermitska transformacija 1 (prvog tipa); \\
DHT2 & - & Diskretna Hermitska transformacija 2 (drugog tipa); \\
DWT & - & Diskretna \textit{wavelet} transformacija; \\
HT& - & Hermitska transformacija; \\
IF & - & Instantaneous Frequency (trenutna frekvencija); \\
LASSO-ISTA  & - &  Iterative Shrinkage Thresholding Algorithm for LASSO problem;\\
ISAR & - & Inverse Synthetic-Aperture Radar; \\
LSAR & - & Least-Squares AR Interpolator;\\
LFM & - & Linearno Frekvencijski Modulisan; \\
LPFT & - & Lokal-polinomijalna Furijeova transformacija; \\
FM & - & Frekvencijski modulisan; \\
FT & - & Furijeova transformacija; \\
m-D & - & Micro-Doppler; \\
MSE & - & Mean-Squared Error (srednja kvadratna gre{\v s}ka);\\
MP  & - & Matching Pursuit; \\
OMP & - & Orthogonal Matching Pursuit; \\
RB & - & Rigid Body (kruto tijelo); \\
RIP  & - & Restricted Isometry Property (Svojstvo ograničene izometrije);\\
RMSE & - & Root-Mean-Squared Error (Kvadratni korijen od MSE); \\
SASS & - & Sparsity-Assisted Signal Smoothing;\\
SNR & - & Signal To Noise Ratio (odnos signal-\v{s}um); \\
STFT & - & Short-Time Fourier Transform (Kratkotrajna FT); \\
UWB & - & Ultra-Wideband (ultra-širokopojasni); \\
WD & - & Wigner-ova distribucija. \\

\end{longtable}
}

\fancyhf{}
%\fancyhead[ER]{\footnotesize\sffamily\leftmark}
%\fancyhead[EL,OL]{\nouppercase\leftmark}
%\fancyheadoffset[R]{0.00\textwidth}
%\fancyhead[EL,OR]{\bfseries\thepage}
\rfoot{\thepage}

\renewcommand{\headrulewidth}{0.0pt}
\tableofcontents

\fancyhf{}
%\fancyhead[ER]{\footnotesize\sffamily\leftmark}
%\fancyhead[EL,OL]{\nouppercase\leftmark}
%\fancyheadoffset[R]{0.00\textwidth}
%\fancyhead[EL,OR]{\bfseries\thepage}
\rfoot{\thepage}

\renewcommand{\headrulewidth}{0.0pt}
\renewcommand{\listfigurename}{Popis slika}
\listoffigures
\fancyhf{}
%\fancyhead[ER]{\footnotesize\sffamily\leftmark}
%\fancyhead[EL,OL]{\nouppercase\leftmark}
\fancyheadoffset[R]{0.00\textwidth}
%\fancyhead[EL,OR]{\bfseries\thepage}
\rfoot{\thepage}

\renewcommand{\headrulewidth}{0.0pt}
\renewcommand{\listtablename}{Popis tabela}
\listoftables

\renewcommand{\headrulewidth}{0.0pt}

\renewcommand{\listalgorithmname}{Popis algoritama}
\listofalgorithms

\renewcommand{\headrulewidth}{1pt} %debljina
\chapter*{Uvod\markboth{Uvod}{}}
\fancyhf{}
%\fancyhead[ER]{\footnotesize\sffamily\leftmark}
\fancyhead[EL,OL]{\nouppercase\leftmark}
\fancyheadoffset[R]{0.00\textwidth}
%\fancyhead[EL,OR]{\bfseries\thepage}
\setcounter{page}{1}%
\rfoot{\thepage}

\addcontentsline{toc}{chapter}{Uvod}
Digitalna obrada signala je naučna oblast koja je svoje konstituisanje i ubrzan razvoj doživjela u drugoj polovini XX vijeka, zahvaljujući dostignućima u oblasti tehnologije integrisanih kola i digitalne računarske tehnike. Usljed razvoja digitalnog hardvera specijalizovane namjene i digitalnih računara, algoritmi i metode digitalne obrade signala postali su vrlo jednostavni za implementaciju, testiranje i modifikovanje, naročito imajući u vidu mogućnost realizovanja softverskim putem. Zahvaljujući navedenim činjenicama, digitalna obrada signala je vrlo brzo stekla širok dijapazon praktičnih primjena: telekomunikacije, biomedicina, radarska tehnika, obrada zvuka i govora, obrada slike, videa, odnosno generalno multimedijalnih podataka, zatim obrada seizmičkih i sonarskih signala -- samo su neki od reprezentativnih primjera praktičnih primjena. 

Klasične metode odabiranja signala podrazumijevaju ekspicitno ili implicitno zadovoljavanje uslova teoreme o odabiranju, po kojoj frekvencija odabiranja signala treba da bude najmanje jednaka dvostrukoj maksimalnoj frekvenciji prisutnoj u posmatranom signalu (u smislu spektralne komponente koja postoji na toj maksimalnoj frekvenciji). Na principima teoreme o odabiranju se zasnivaju gotovo sve klasične tehnike za akviziciju signala, bilo da se radi o audio signalima, digitalnim slikama, bio-medicinskim signalima ili, na primjer, radarskim signalima. Teorema o odabiranju takođe ima ključnu ulogu u tehnikama konverzije podataka, kao što je, na primjer, analogno-digitalna (A/D) konverzija signala. Može se smatrati da je teorema o odabiranju tokom prethodnih pola vijeka u velikoj mjeri determinisala ne samo utemeljenje, već i tok razvoja digitalne obrade signala.

Kompresivno odabiranje (engl. \textit{compressive sensing}, CS)  je nova i u naučnom smislu veoma atraktivna oblast u obradi signala, koja na izvjestan način prevazilazi spomenuti uslov teoreme o odabiranju, omogućavajući pritom potpunu rekonstrukciju određenih klasa signala na osnovu veoma malog broja dostupnih odbiraka/mjerenja, mnogo manjeg od onog koji se koristi u tradicionalnim tehnikama. Količina podataka čija se akvizicija vrši od ključne je važnosti  u praktično svim segmentima primjena informaciono-komunikacionih tehnologija. Sve obimnije količine podataka zahtijevaju veoma kompleksne kompresione algoritme u cilju uspješnog skladištenja, obrade i prenosa. Navedene činjenice postavile su oblast kompresivnog odabiranja u fokus intenzivnih naučnih istraživanja, i uslovile da se tokom posljednjih godina u okviru ove oblasti razviju veoma napredne tehnike i rekonstrukcioni algoritmi. Uređaji za akviziciju podataka, koji sadrže veliki broj senzora (na primjer, biomedicinski uređaji kao što je MRI, multimedijalni uređaji kao što su visokorezolucione kamere itd.) mogu biti značajno pojednostavljeni, a vrijeme i količina snimanja značajno redukovani (što je od naročito velike važnosti u biomedicinskim primjenama). Prilikom obrade različitih signala oštećenih jakim šumom, nekada se pristupa namjernom preskakanju/eliminisanju oštećenih odbiraka, kao na primjer, u slučaju primjene L-statistike i drugih robustnih tehnika. Sa druge strane, u nekim praktičnim okolnostima podaci mogu biti nedostupni zbog raznih fizičkih ograničenja: bilo instrumenata, bilo fizičkih procesa na kojima se bazira akvizicija. Bez obzira na uzrok njihove nedostupnosti, kompresivno odabiranje nudi mogućnost da se nedostajući podaci u potpunosti rekonstruišu, i to samo na osnovu dostupnih, tj. neoštećenih vrijednosti. Dakle, CS omogućava da se na osnovu malog broja slučajno snimljenih podataka obezbijedi isti kvalitet informacije koji bi postojao kada bi svi podaci bili dostupni. 

Uspješna rekonstrukcija signala sa nedostajućim odbircima je moguća ukoliko su ovi signali rijetki (engl. \textit{sparse}) u nekom određenom transformacionom domenu, odnosno, ukoliko mogu biti reprezentovani pomoću malog broja nenultih transformacionih koeficijenata. U praktičnim primjenama, od velikog značaja su i signali koji su aproksimativno rijetki, budući da se rekonstrukcioni koncepti razvijeni u kontekstu kompresivnog odabiranja (i obrade rijetkih signala, engl. \textit{sparse signal processing}) na njih mogu primjenjivati u gotovo neizmijenjenoj formi. Koncept rijetkosti se može  dovesti u vrlo usku vezu sa tzv. mjerama koncentracije. Naime, za rijetke signale se može reći da su visoko koncentrisani u odgovarajućem tranformacionom domenu. U tom smislu, u okviru ove doktorske disertacije je razmatrano  više različitih transformacionih domena: diskretna Furijeova transformacija (DFT), diskretna Hermitska transformacija (DHT), diskretna kosinusna transformacija (DCT), kao i vremensko frekvencijske reprezentacije - kratkotrajna Furijeova transformacija (STFT), Wigner-ova distribucija i S-metod. Zasebno razmatranje svakog od ovih transformacionih domena je važno iz dva razloga. Kao prvo -- rijetkost određenih klasa signala usko je vezana za transformacioni domen, i drugo -- uticaj nedostajućih (eliminisanih ili fizički nedostupnih) odbiraka na svaki transformacioni domen se manifestuje različito, što je posljedica specifičnih svojstava tih domena.
% Na ovom mjestu je ipak vrijedno spomenuti da će analiza prezentovana u ovoj disertaciji poslužiti kao osnova za dalja istraživanja u cilju formiranja opštije teorije, koja će prezentovane rezultate i izvedene zaključke na prikladan način objediniti.

Koncepti rijetkosti i mjera koncentracije signala u transformacionim domenima predstavljaju okosnicu istraživanja prezentovanih u ovoj doktorskoj tezi. U tom smislu, treba naglasiti da problematika rekonstrukcije nije sagledana isključivo kroz prizmu kompresivnog odabiranja. To ilustruje, na primjer, druga glava disertacije, gdje je koncept rijetkosti iskorišćen za optimizaciju parametara diskretne Hermitske transformacije, u cilju postizanja visoko koncentrisane reprezentacije signala. Takva reprezentacija je omogućila uklanjane šuma, kompresiju, ali i CS rekonstrukciju. Još jedan primjer je i dekompozicija multivarijantnih signala u kontekstu vremensko-frekvencijske analize, koja je prezentovana u sklopu četvrte glave disertacije. Iako se pri dekompoziciji multivarijantnih signala ne razmatra rekonstrukcija primjenom matematičke aparature kompresivnog odabiranja,  u pitanju je ipak rekonstrukcija komponenti signala koja je zasnovana na minimizaciji mjera koncentracije, što je jedna od fundamentalnih ideja u oblasti kompresivnog odabiranja.

Osnovni koncepti vezani za rekonstrukciju rijetkih signala prezentovani su u prvoj glavi. Ovdje je dat i kratak osvrt na Furijeov transformacioni domen, posebno u smislu matematičkog modelovanja uticaja nedostajućih odbiraka na tranformacione koeficijente. Navedena analiza je bila inspiracija za istraživanja u domenima diskretne Hermitske i diskretne kosinusne transformacije. Dodatno, ova glava sadrži kratak opis nekoliko osnovnih algoritama za rekonstrukciju, koji su detaljnije razrađeni u preostalim glavama teze, uzimajući u obzir specifičnosti razmatranih transformacionih domena. 

Hermitska transformacija i njene diskretne forme su sagledani iz više perspektiva  u drugoj glavi disertacije. Pored doprinosa koji se tiču analize uticaja nedostajućih odbiraka i razvoja novih rekonstrukcionih pristupa, prezentovani su i algoritmi za optimizaciju diskretne Hermitske transformacije, kao i analiza uticaja aditivnog šuma. Posebno je bitno naglasiti razmatrane aspekte praktičnih primjena: kompresiju i rekonstrukciju elektrokardiografskih (EKG) signala i rekonstrukciju ultra-širokopojasnih (engl. \textit{ultra-wideband}, UWB)  signala. Rezultati predstavljeni u ovoj glavi su verifikovani kroz publikovanje većeg broja konferencijskih radova, a glavni doprinosi su verifikovani kroz publikovanje četiri naučna rada u renomiranim naučnim časopisima. Jedan rad sadrži doprinose vezane za optimizaciju parametara diskretne Hermitske transformacije, drugi se bavi analizom uticaja šuma na Hermitske koeficijente, treći se bavi odgovarajućom analizom uticaja nedostajućih odbiraka  kao i analizom performansi procesa rekonstrukcije, dok se posljednji tiče gradijentnog algoritma za rekonstrukciju rijetkih signala.

Jednodimenziona i dvodimenziona diskretna kosinusna transformacija (DCT) predmet su proučavanja treće glave doktorske teze. Uticaj nedostajućih odbiraka na ove transformacione domene je detaljno teorijski razrađen, a analiza je obojena primjenama u obradi audio signala i digitalne slike. Važno je istaći da su svi teorijski rezultati evaluirani kroz brojne numeričke eksperimente sa realnim signalima. Predstavljeni doprinosi su verifikovani kroz publikovanje dva rada u renomiranim naučnim časopisima, od kojih se jedan odnosi na jednodimenzionu, a drugi na dvodimenzionu diskretnu kosinusnu transformaciju.

U posljednjoj glavi se razmatraju vremensko-frekvencijske reprezentacije -- kratkotrajna Furijeova transformacija, Wigner-ova distribucija i S-metod. Kroz rekonstrukciju djelova ISAR signala, STFT je povezana sa konceptima kompresivnog odabiranja. Kroz dekompoziciju multivarijantnih signala, koja je zasnovana na primjeni STFT, Wigner-ove distribucije i S-metoda, pokazano je da se koncepti mjera koncentracije mogu iskoristiti i za rekonstrukcije koje nijesu striktno vezane za kompresivno odabiranje. Rezultati iz ove glave su verifikovani kroz publikovanje većeg broja konferencijskih radova, kao i kroz publikovanje dva naučna rada u renomiranim međunarodnim časopisima, od kojih se jedan odnosi na algoritam za estimaciju trenutne frekvencije koji je baziran na Wigner-ovoj distribuciji, dok se drugi odnosi na dekompoziciju multikomponentnih multivarijantnih signala.

\chapter{Rekonstrukcija rijetkih signala sa osvrtom na Furijeov domen}

Brojni su inženjerski i matematički koncepti na kojima je utemeljana digitalna obrada signala. U naredom odjeljku će biti navedeno samo nekoliko osnovnih definicija, na koje će se nasloniti ostatak izlaganja u tezi.  Za dublje razumijevanje osnovnih teorijskih i praktičnih koncepata obrade signala, čitalac se upućuje na odgovarajuću stručnu literaturu, \cite{dos,multimedia,papulis_dsp,tfsa}.

 U ovoj glavi će nešto detaljnije će biti izloženi fundamentalni principi kompresivnog odabiranja i obrade rijetkih signala \cite{dos,tfsa, donoho,CandesRIP, baraniuk,candes,candes2, wohlberg, dftmiss, tutorial, enciklopedija, cs1, cs2,cs3,cs4, cosamp, RobSampling, erviniet,Ervin_CS,BJYDZ,li1,libi,irenacs,denoising,viktor,robustletters,BCS2,CSTF,More, automated,grad1, grad2,greedy, lasso,BCS1,impcs,DFT_nonsparse,csr1,csr4,sparkdef,convexity,binary,brajovic_mipro,brajovic_genetic_rec,brajovic_3dcs,sire0,sire1,sire3,sire4,sire5,sire6, jedinstvenost,ljubisa_isar,brajovic_cosera2}. Predstavljeni su osnovni rekonstrukcioni algoritmi, koji su od značaja za izlaganja u ostalim djelovima teze. Formalno se uvode pojmovi i koncepti kompresivnog odabiranja -- rijetkosti, mjerenja, indeksa koherentnosti i svojstva ograničene izometrije. Formalno se definišu i uslovi koji garantuju jedinstvenost rješenja problema rekonstrukcije. Rekonstrukcija se razmatra na primjerima relevantnih algoritama zasnovanih na minimizacijama $\ell_0$-norme i  $\ell_1$-norme transformacionih koeficijenata signala.

\section{Signali, Furijeova transformacija i odabiranje}

Pod pojmom signala se podrazumijevaju fizički procesi, matematičke funkcije, odnosno, bilo kakve fizičke ili simboličke reprezentacije informacija \cite{dos}. Informacija je skup podataka o određenoj pojavi ili procesu, na osnovu kojih se može doći do određenih zaključaka. U formi fizičkih procesa, signali se najčešće pojavljuju u vidu elektromagnetnih talasa, odnosno, karakterišu se elektromagnetnim veličinama koje su promjenljive u vremenu. Razvijeni su mnogi uređaji koji različite fizičke procese mogu konvertovati u elektromagnetnu formu. Tako se, na primjer, akustički talasi, zasnovani na mehaničkim pojavama, mogu lako konvertovati u vremenski promjenljiv napon ili struju.

Signali mogu biti kontinualni ili diskretni. Diskretni signali sa konačnim vrijednostima se nazivaju digitalnim. Pojmovi digitalnih i diskretnih signala se u savremenoj praksi često izjednačavaju, imajući u vidu dužine registara, memorijske kapacitete i ostale tehnološke karakteristike savremenih računara, odnosno specijalizovanih kola za obradu signala. 

Još jedna klasifikacija signala, koja uzima u obzir prirodu  njihovih promjena, jeste na determinističke i slučajne. Prve je moguće opisati eksplicitnim matematičkim izrazima, odnosno, funkcijama, dok druge nije moguće, pa se oni opisuju probabilističkim veličinama. Dok deterministički signali nijesu nosioci informacija, vrlo su važni u analizi i dizajnu sistema, bilo analognih, bilo diskretnih, a pomoću njih je moguće i modelovati realne signale. 

Formalno, kontinualni signali se predstavljaju funkcijom od jedne ili više kontinualnih varijabli. U tom smislu, možemo govoriti o jednodimenzionim ili višedimenzionim signalima. Jednodimenzioni signali najčešće predstavljaju funkciju vremenske varijable $t$, pa oni u opštem slučaju vrše preslikavanje skupa realnih brojeva  $\mathbb{R}$ (ili nekog njegovog povezanog podskupa) na
skup kompleksnih brojeva $\mathbb{C}$, odnosno, $x:t\rightarrow x(t)$. U slučaju diskretnih signala, nezavisno promjenljiva $n$, koja se najčešće interpretira kao indeks diskretnog vremena, vrši se preslikavanje skupa cijelih brojeva $\mathbb{Z}$ ili njegovog povezanog podskupa, na skup kompleksnih brojeva $\mathbb{C}$, odnosno,  $x:n\rightarrow x(n)$.

Kontinualni signal $x(t)$ je paran ako zadovoljava:
$x(-t)=x(t)$ za svako $t$, a neparan ako za svako $t$ važi: $x(-t)=-x(t)$. Ekvivalentne su i odgovarajuće definicije za slučaj diskretnih signala. Kontinualni signal je periodičan, sa periodom $T$, ukoliko za svako $t$ važi: $x(t)=x(t+T)$, gdje je
$T$ pozitivna realna konstanta. Ukoliko je uslov zadovoljen za $T=T_{0}$,
jasno je da je tada zadovoljen i za $T=kT_{0}$, za $k$ koje je cio broj. Veličina
$T_{0}$ se naziva osnovna perioda. Frekvencija
signala je jednaka recipročnoj vrijednosti periode, $f=\frac{1}{T}~$[Hz], dok je ugaona frekvencija po definiciji $\Omega=\frac{2\pi}{T}$ [rad/s].  Za diskretne signale važi da su periodični, ukoliko važi da je $x(n)=x(n+N),$ gdje je $N$ cio broj, za sve
vrijednosti promjenljive $n$. Ukoliko prethodno navedni uslovi nijesu ispunjeni, signale nazivamo aperiodičnim.

Kontinualni signali se često karakterišu totalnom energijom, veličinom definisanom izrazom:
$
E_x=
\int_{-\infty}^{\infty}
\left\vert x(t)\right\vert ^{2}dt,
$
odnosno prosječnom snagom,
$
P_x=\lim_{T\rightarrow\infty}\frac{1}{T}%
{\int_{-T/2}^{T/2}}
|x(t)|^{2}dt$. Potpuno ekvivalentno, energija diskretnog signala je definisana obrascem:
$
E_x=\sum_{n=-\infty}^{\infty}
|x(n)|^{2}, \label{energija}
$
a snaga:
$
P_x=\lim_{N\rightarrow\infty}\frac{1}{2N+1}
{\sum_{n=-N}^{N}}
|x(n)|^{2}
$. Signali sa konačnom energijom imaju snagu jednaku nuli, pa se često nazivaju energetskim signalima. Signali sa konačnom snagom imaju beskonačnu vrijednost energije, i poznati su kao signali snage,\cite{dos,papulis_dsp,tfsa}.
\subsection{Furijeova transformacija}
Furijeova transformacija (engl. \textit{Fourier transform}, FT) obezbjeđuje predstavljanje signala u frekvencijskom domenu. U tom smislu, signal se predstavlja preko svojih spektralnih komponenti, odnosno, preko skupa kompleksnih sinusoida odgovarajućih amplituda i faza. Vrijeme i frekvencija, koje ovdje imaju ulogu nezavisno promjenljivih, međusobno su isključivi.

Neka se posmatra kontinualni signal $x(t)$, koji zadovoljava Dirihleove uslove, odnosno, za koji važi: 1. da ima konačan broj diskontinuiteta prvog reda; 2. da ima konačan broj maksimuma i minimuma unutar bilo kojeg konačnog intervala; 3. da je apsolutno integrabilan, odnosno, zadovoljava
	$
	\int_{-\infty}^{\infty}\left\vert {x(t)}\right\vert <\infty
$. Za takav signal, Furijeova transformacija je definisana sljedećim izrazom \cite{dos,papulis_dsp,tfsa}:
\begin{equation} X(j\Omega)=\mathcal{F}\{x(t)\}=\int_{-\infty}^{\infty}{x(t){e^{-j\Omega t}}dt.}
\label{FTc}\end{equation}

U praktičnim okolnostima, Dirihleovi uslovi se ublažavaju nešto manje strogim uslovom da signal ima konačnu energiju, odnosno $
\int_{-\infty}^{\infty}\left\vert {x(t)}\right\vert ^{2}<\infty$.
Inverzna Furijeova transformacija data je u obliku:
\begin{equation}
x(t)=\mathcal{F}^{-1}\{X(j\Omega)\}=\frac{1}{{2\pi}}\int_{-\infty}^{\infty}{X(j\Omega){e^{j\Omega t}%
	}d\Omega.} \label{IFT}%
\end{equation}

Signal $x(t)$ i odgovarajuća Furijeova transformacija $X(j\Omega)$ čine Furijeov transformacioni par. 
Furijeova trasformacija $X(j\Omega)$ je kompleksna i kontinualna
funkcija. Moduo Furijeove transformacije, $\left\vert {X(j\Omega)}\right\vert
$, poznat je kao amplitudska karakteristika ili amplitudski spektar signala. U slučaju realnog signala $x(t)$, on je parno simetrična funkcija. Veličina $\arg\{X(j\Omega)\}=-\arctan\left(\frac{\Re
\{X(j\Omega)\}}{\Im\{X(j\Omega)\}}\right)$ se naziva fazna karakteristika. U slučaju realnog signala $x(t)$, ona je neparno simetrična funkcija od frekvencije. Furijeova transformacija posjeduje druga interesantna svojstva, \cite{dos}. Bitno je spomenuti da se Furijeova transformacija periodičnog signala $x(t)$ svodi na Furijeov red.

Neka se posmatra diskretni signal $x(n)$ koji je apsolutno sumabilan, tako da važi $\sum_{n=-\infty}^{\infty}|{x(n)|}<\infty$. Njegova Furijeova transformacija $X(e^{j\omega})$ je definisana izrazom \cite{dos,papulis_dsp,tfsa}:
\begin{equation}
X(e^{j\omega})=\mathcal{F}\{x(n)\}=\sum_{n=-\infty}^{\infty}
x(n)e^{-j\omega n}, \label{fds}%
\end{equation}
i predstavlja periodičnu funkciju po normalizovanoj frekvenciji $\omega$, sa periodom $2\pi$. U praktičnim okolnostima, smatra se da Furijeova transformacija diskretnog signala $x(n)$ postoji uz manje rigorozan uslov: $\sum_{n=-\infty}^{\infty}|{x(n)|}^{2}<\infty$. 
Furijeova transformacija diskretnog signala se može lako dovesti u vezu sa Furijeovom transformacijom analognog signala (\ref{FTc}). Naime, aproksimacija Furijeove transformacije kontinualnog signala, prema pravouganom pravilu numeričkog integraljenja, može biti zapisana u obliku:
\begin{equation}
X(j\Omega)\cong\sum_{n=-\infty}^{\infty}x(n\Delta t) e^{-j\Omega n\Delta t}\Delta t,
\end{equation}
što uz $x(n\Delta t)\Delta t \longrightarrow x(n)$ i $\Omega \Delta t\longrightarrow \omega$ dovodi do $\sum_{n=-\infty}^{\infty}x(n)e^{-j\omega n}=X(e^{j\omega})$, što jeste Furijeova transformacija diskretnog signala. Pod određenim uslovima, FT diskretnog signala $X(e^{j\omega})$  nije samo aproksimacija FT analognog signala $X(j\Omega)$, već važi $X(e^{j\omega})=X(j\Omega)$, uz $\Omega \Delta t = \omega$ gdje je $-\pi \leq \omega<\pi$, kao što je diskutovano u \cite{dos}. Inverzna Furijeova transformacija diskretnog signala $x(n)$ je definisana integralom:
\begin{equation}
x(n)=\mathcal{F}^{-1}\{X(e^{j\omega})\}=\frac{1}{{2\pi}}\int_{-\pi}^{\pi}{X({e^{j\omega}}){e^{j\omega n}%
	}d\omega,} \label{ifds}%
\end{equation}
gdje signal $x(n)$ i njegova Furijeova transformacija, $X(e^{j\omega})$ dati relacijama
(\ref{fds}) i (\ref{ifds}) predstavljaju Furijeov transformacioni par.  I pored toga što je Furijeova transformacija $X(e^{j\omega})$ definisana za diskretni signal $x(n)$, normalizovana frekvencija $\omega$ je kontinualna.

Pošto diskretni signali imaju Furijeovu transformaciju koja je funkcija od kontinualne frekvencije, u cilju numeričke analize i obrade ovih signala, javlja se potreba za uvođenjem diskretne forme Furijeove transformacije diskretnog signala. Diskretna Furijeova transformacija (DFT) je za diskretni signal $x(n)$ definisana izrazom \cite{dos,papulis_dsp,tfsa}:
\begin{equation}
X(k)=\left.  X(e^{j\omega})\right\vert _{\omega=2\pi k/N}=\sum_{n=0}%
^{N-1}{x(n){e^{-j\frac{{2\pi}}{N}kn},}}~k=0,1,\dots,N-1. \label{dft}%
\end{equation}
Veličina $k$ se naziva diskretni indeks frekvencije. Može se uočiti da je DFT periodična po promjenljivoj $k$ sa periodom $N$. Ova transformacija je, dakle, dobijena odabiranjem FT diskretnog signala, $X(e^{j\omega})$ sa korakom $\Delta \omega=2\pi/N$. Na intervalu $-\pi \leq \omega < \pi$, uzima se $N$ odbiraka Furijeove transformacije $X(e^{j\omega})$. Navedeno je moguće ukoliko je posmatrani signal $x(n)$ konačne dužine $N_0$, pri čemu je $N\geq N_0$. DFT je invertibilna, pri čemu je originalni signal moguće dobiti na osnovu DFT odbiraka $X(k)$ primjenom sume:
\begin{equation}
x(n)=\frac{1}{N}\sum_{k=0}^{N-1}X{(k){e^{j\frac{{2\pi}}{N}kn},}}~n=0,1,\dots,N-1.
\label{idft}%
\end{equation}

Ako se uvede vektor signala $\mathbf{x}=[x(0),x(1),\dots,x(N-1)]^T$ i vektor DFT transformacionih koeficijenata $\mathbf{X}=[X(0),X(1),\dots,X(N-1)]^T$, tada se DFT može napisati u formi:
\begin{equation}
	\mathbf{X=\Phi x},
\end{equation}
\begin{equation}
\left[
\begin{array}
[c]{c}%
X(0)\\
X(1)\\
\vdots  \\
X(N-1)
\end{array}
\right]=\left[
\begin{array}
[c]{cccc}%
1 & 1 &\dots   & 1\\
1 & e^{-j2\pi/N} &  &e^{-j2\pi (N-1)/N}\\
\vdots & \vdots &\ddots  &\vdots\\
1 & e^{-j2\pi(N-1)/N}&  & e^{-j2\pi (N-1)(N-1)/N})
\end{array}
\right] \left[
\begin{array}
[c]{c}%
x(0)\\
x(1)\\
\vdots  \\
x(N-1)
\end{array}
\right],
\end{equation}
gdje $\mathbf{\Phi}$ označava DFT transformacionu matricu. Ukoliko se uvede inverzna transformaciona matrica $\mathbf{\Psi=\Phi}^{-1}$, tada se inverzna DFT, data izrazom (\ref{idft}), može zapisati u obliku:
\begin{equation}
\mathbf{x=\Phi}^{-1}\mathbf{X}=\mathbf{\Psi}\mathbf{X}.
\end{equation}

Za inverznu i direktnu DFT transformacionu matricu važi $\mathbf{\Phi}^{-1}=\frac{1}{N}\mathbf{\Phi}^{*}$, gdje * označava operaciju konjugovanja. Vrijedi napomenuti da se konstanta $\frac{1}{N}$ kod inverzne DFT matrice (i transformacije) u nekim formulacijama zamjenjuje sa $\frac{1}{\sqrt N}$. U tom slučaju, i odgovarajuća direktna transformaciona matrica $\mathbf{\Phi}$ množi se sa konstantom $\frac{1}{N}$. Ovakvo skaliranje je posebno uobičajeno u kompresivnom odabiranju, kako bi se obezbijedila odgovarajuća normalizacija kolona relevantnih matrica. 
\subsection{Teorema o odabiranju kontinualnih signala}
Pod pojmom odabiranja (engl. \textit{sampling}) se podrazumijeva uzimanje diskretnih vrijednosti analognih signala, odnosno, sam proces konverzije kontinualnih signala u diskretne. Razmak između tačaka u kojima se vrši odabiranje je poznat pod nazivom korak odabiranja $\Delta t$. U ovom odjeljku će biti, bez dokaza, formulisana teorema o odabiranju kontinualnih signala, koja definiše uslove pod kojima je odabiranje moguće sprovesti tako da se originalni analogni signal može u potpunosti rekonstruisati na osnovu svojih odbiraka, \cite{dos,papulis_dsp,tfsa}. 

Kontinualni signal $x(t)$, čija je Furijeova transformacija $X(j\Omega)$ ograničena sa $\Omega_m=2\pi f_m$, odnosno, za koji važi $X(j\Omega)=0$, za $\Omega > \Omega_m$, može biti rekonstruisan, za bilo koje $t$, na osnovu odbiraka $x(n)=x(n\Delta t)\Delta t$ koji su uzeti sa korakom odabiranja $\Delta t$, koji zadovoljava uslov:
\begin{equation}
\Delta t<\frac{\pi}{\Omega_m}=\frac{1}{2f_{m}}. \label{to}
\end{equation}

Dokaz se može pronaći u literaturi, \cite{dos}. Rekonstrukciona formula je data u obliku:
\begin{equation}
	x(t)=\sum_{n=-\infty}^{\infty}x(n\Delta t)\frac{\sin\left(\frac{\pi}{\Delta t}(t-n\Delta t)\right)}{\frac{\pi}{\Delta t}(t-n\Delta t)},
\end{equation}
gdje je signal $x(t)$ za bilo koje $t$ predstavljen preko svojih odbiraka.
\section{Rijetkost signala i redukovani skup mjerenja}
Rijetki signali se mogu okarakterisati malim brojem nenultih koeficijenata u jednom od njihovih transformacionih domena \cite{dos,tfsa, donoho,CandesRIP, baraniuk,candes,candes2, wohlberg, dftmiss, tutorial, enciklopedija, cs1, cs2,cs3,cs4, cosamp, RobSampling, erviniet,Ervin_CS,BJYDZ,li1,libi,irenacs,denoising,viktor,robustletters,BCS2,CSTF,More, automated,grad1, grad2,greedy, lasso,BCS1,impcs,DFT_nonsparse,csr1,csr4,sparkdef,convexity,binary,brajovic_mipro,brajovic_genetic_rec,brajovic_3dcs}. Na primjer, radarski ISAR signali mogu imati rijetku reprezentaciju u domenu dvodimenzione Furijeove transformacije, neki specifični djelovi EKG signala rijetki su u domenu diskretne Hermitske transformacije, dok se digitalne slike mogu smatrati rijetkim ili aproksimativno rijetkim u domenu dvodimenzione diskretne kosinusne transformacije. 

Rijetki signali mogu biti rekonstruisani na osnovu redukovanog skupa dostupnih odbiraka -- mjerenja \cite{donoho,CandesRIP, baraniuk,candes,candes2, wohlberg,tutorial,enciklopedija}. Mjerenja su linearne kombinacije koeficijenata iz domena rijetkosti signala.  Odbirci signala se mogu smatrati mjerenjima u slučaju linearnih transformacija. U određenom broju primjena, redukovani broj odbiraka može biti posljedica njihove fizičke nedostupnosti, dok su u drugim primjenama oni  rezultat namjere da se redukuje broj mjerenja, a pri tome sačuva cijela informacija (kompresija signala), \cite{donoho,baraniuk}. Nedostupnost odbiraka signala može nastati i kao posljedica njihovog namjernog odbacivanja, na primjer u slučaju oštećenja jakim šumom \cite{dftmiss}. 

U obradi signala, najzastupljeniji je Furijeov transformacioni domen, \cite{dftmiss,dos}. Ulogu mjerenja u tom slučaju imaju odbirci signala. Odgovarajuće mjerne matrice su parcijalne matrice diskretne Furijeove transformacije (DFT), ali i parcijalne slučajne Furijeove matrice \cite{tutorial,enciklopedija}. Uticaj redukovanog skupa odbiraka na analizu i rekonstrukciju/sintezu signala rijetkih u DFT domenu su detaljno proučene u \cite{dftmiss}.

\subsection{Rijetkost i mjerenja}
Razmatra se skup od $N$ koeficijenata $X(k)$, za $k=0,1,\dots , N-1$, koji su elementi vektora koeficijenata $\mathbf{X}$. Po definiciji, ovaj vektor je rijedak (engl. \textit{sparse}) ukoliko je broj njegovih nenultih keficijenata, u oznaci $K$, mnogo manji od ukupnog broja koeficijenata $N$, $K\ll N$, odnosno, ako je zadovoljeno \cite{dos,candes, candes2, donoho, tutorial, enciklopedija}:
\begin{equation}
X(k)=0,\text{ za } k\notin\{k_{1},k_{2},\dots  ,k_{K}\}={\Pi}_K.
\end{equation}

Broj $K$ će dalje biti označen kao stepen rijetkosti. Mjerenje vektora $\mathbf{X}$ se može posmatrati kao linearna kombinacija njegovih elemenata (koeficijenata) $X(k)$. Neka je $i$-to mjerenje označeno sa $y(i)$. Ukoliko je dostupno samo $N_A$ mjerenja $y(0),y(1),\dots ,y(N_A-1)$, ona se mogu zapisati u vidu sistema od $N_A$ linearnih jednačina
\begin{align}
y(i)=\sum_{k=0}^{N-1}X(k)\varphi_{k}(i),~i=0,1,\dots ,N_A-1,~N_A<N,
\end{align}
gdje su $\varphi_{k}(i)$ težinski (ili mjerni) koeficijenti. U matričnoj formi, ovaj sistem je:
\begin{gather}
\left[
\begin{array}
[c]{c}%
y(0)\\
y(1)\\
\vdots \\
y(N_A-1)
\end{array}
\right]  =\left[
\begin{array}
[c]{cccc}%
\varphi_{0}(0) & \varphi_{1}(0) & \dots  & \varphi_{N-1}(0)\\
\varphi_{0}(1) & \varphi_{1}(1) & \dots  & \varphi_{N-1}(1)\\
\vdots & \vdots & \ddots & \vdots\\
\varphi_{0}(N_A-1) & \varphi_{1}(N_A-1) &  & \varphi_{N-1}(N_A-1)
\end{array}
\right]  \left[
\begin{array}
[c]{c}%
X(0)\\
X(1)\\
\vdots  \\
X(N-1)
\end{array}
\right],\notag \\
\mathbf{y=AX}.\notag
\end{gather}

Matrica $\mathbf{A}$, čiji su elementi težinski koeficijenti $\varphi_{k}(i)$, se naziva mjerna matrica.
\subsection{Interpretacija u obradi signala}
Koncepti rijetkih vektora i mjerenja će dalje biti razmatrani u kontekstu obrade signala. U tom cilju, posmatra se signal $x(n)$ i njegova linearna transformacija $X(k)$:
\begin{align}
x(n)=\sum_{k=0}^{N-1}X(k)\varphi_{k}(n),
\end{align}
odnosno, zapisano u matričnoj formi:
\begin{align}
\mathbf{x}\mathbf{=\Psi X,}%
\end{align}
pri čemu je inverzna transformaciona matrica označena sa $\mathbf{\Psi}$, njeni elementi su $\varphi_{k}%
(n)$, dok je $\mathbf{x}$ vektor-kolona koji sadrži odbirke signala, a  $\mathbf{X}$ je vektor-kolona sa odgovarajućim transformacionim koeficijentima. 

Pretpostavlja se da je signal $K$-rijedak u razmatranom transformacionom domenu, što znači da je 
$
X(k)\ne 0,\text{ za } k\in{\Pi}_K \text{ i }X(k)= 0,\text{ za } k\notin{\Pi}_K,
$
gdje je ${\Pi}_K=\{k_{1},k_{2},\dots  ,k_{K}\}$ skup pozicija nenultih transformacionih koeficijenata. Broj nenultih transformacionih koeficijenata je \cite{dos, tutorial, enciklopedija}:
\begin{align}
\left\Vert \mathbf{X}\right\Vert _{0}=\mathrm{card}\left\{  \mathbf{X}%
\right\}  =K,
\end{align}
pri čemu se veličina
$
\left\Vert \mathbf{X}\right\Vert _{0}=\sum_{k=0}^{N-1}\left\vert
X(k)\right\vert ^{0}%
$
naziva $\ell_{0}$-normom ili $\ell_{0}$-pseudo-normom. Ona ne posjeduje klasična svojstva normi, već predstavlja kardinalnost skupa, odnosno, broj nenultih transformacionih koeficijenata u vektoru $\mathbf{X}$. Po definiciji, važi $\left\vert X(k)\right\vert ^{0}=0$ za
$\left\vert X(k)\right\vert =0$ i $\left\vert X(k)\right\vert ^{0}=1$
za $\left\vert X(k)\right\vert \neq0$.%

Za signal $x(n)$ čiji transformacioni koeficijenti $X(k)=\mathcal{T}\{x(n)\}$ zadovoljavaju
$\mathrm{card}\left\{  \mathbf{X}\right\}  =K\ll N$,
kaže se da je rijedak u transformacionom domenu $\mathcal{T}$. Za linearne transformacije, $K$-rijetki signal može biti predstavljen kao linearna kombinacija $K$ koeficijenata $X(k)$, \cite{tutorial, enciklopedija}:
\begin{equation}
x(n)=\sum_{k\in{\Pi}_K}X(k)\varphi_{k}(n). \label{{N_A}easDFT}%
\end{equation}
Odbirak signala se može interpretirati kao mjerenje linearne kombinacije vrijednosti $X(k)$. 

Primarna problematika koja se razmatra u ovoj disertaciji jeste mogućnost rekonstrukcije signala sa stepenom rijetkosti $K$ na osnovu redukovanog seta od $N_A$ odbiraka -- mjerenja. 
Kompresivno odabiranje (odnosno, rekonstrukcija rijetkih signala) podrazumijeva da su odbirci signala $x(n)$, odnosno mjerenja, dostupni na slučajnim pozicijama
$
n_{i}\in{\mathbb{{N}}_A=}\{n_{1},n_{2},\dots ,n_{{N_A}}\}\subseteq\mathbb{N}%
=\{0,1,\dots ,N-1\}.
$
Sa $\mathbb{N}=\{0,1,\dots ,N-1\}$ je označen skup svih pozicija odbiraka signala $x(n)$ dok je sa ${\mathbb{{N}}_A=}\{n_{1},n_{2},\dots  ,n_{{N_A}}\}$ označen njegov podskup od 
${N_A}$ slučajno raspoređenih elemenata, gdje je ${N_A}\leq N$. U daljem izlaganju, vektor dostupnih mjerenja će biti označen sa
$
\mathbf{y=[}x(n_{1}),~x(n_{2}),\dots ,x(n_{{N_A}})]^{T},
$
gdje  važi $y(i)=x(n_{i+1})$, za $i=0,1,\dots ,N_A-1$.
U opštem slučaju, dostupni odbirci, odnosno mjerenja linearne kombinacije koeficijenata 
$X(k)$ definisana izrazom (\ref{{N_A}easDFT}) za $n_{i}\in{\mathbb{{N}}_A=}\{n_{1}%
,n_{2},\dots ,n_{{N_A}}\}$ mogu se zapisati kao sistem od ${N_A}$ linearnih jednačina:
\begin{gather}
\left[
\begin{array}
[c]{c}%
x(n_{1})\\
x(n_{2})\\
\vdots\\
x(n_{{N_A}})
\end{array}
\right]  =\left[
\begin{array}
[c]{cccc}%
\varphi_{0}(n_{1}) & \varphi_{1}(n_{1}) & \dots  & \varphi_{N-1}(n_{1})\\
\varphi_{0}(n_{2}) & \varphi_{1}(n_{2}) & \dots  & \varphi_{N-1}(n_{2})\\
\vdots & \vdots & \ddots & \vdots\\
\varphi_{0}(n_{{N_A}}) & \varphi_{1}(n_{{N_A}}) & \dots  & \varphi_{N-1}(n_{{N_A}})
\end{array}
\right]  \left[
\begin{array}
[c]{c}%
X(0)\\
X(1)\\
\vdots\\
X(N-1)
\end{array}
\right],\\
\mathbf{y=AX}%
\end{gather}
gdje je, kako je već istaknuto, $\mathbf{A}$ mjerna matrica, dimenzija ${N_A}\times N$.

U razmatranom kontekstu, nije poznato za koje pozicije $\Pi_K=\{k_{1}%
,k_{2},\dots ,k_{K}\}$ važi da je $X(k)=0,$ $k\notin\Pi_K$, odnosno, informacija o pozicijama nenultih koeficijenata nije ugrađena u mjernu matricu. Ukoliko bi nekim adekvatnim postupkom bilo moguće odrediti skup pozicija $\Pi_K$, sistem jednačina za mjerenja bi se mogao zapisati u sljedećoj formi \cite{tutorial, enciklopedija}:
\begin{gather}
\left[
\begin{array}
[c]{c}%
x(n_{1})\\
x(n_{2})\\
\vdots\\
x(n_{{N_A}})
\end{array}
\right]  =\left[
\begin{array}
[c]{cccc}%
\varphi_{k_{1}}(n_{1}) & \varphi_{k_{2}}(n_{1}) &\dots   & \varphi_{k_{K}}(n_{1})\\
\varphi_{k_{1}}(n_{2}) & \varphi_{k_{2}}(n_{2}) &  & \varphi_{k_{K}}(n_{2})\\
\vdots & \vdots &\ddots  &\vdots\\
\varphi_{k_{1}}(n_{{N_A}}) & \varphi_{k_{2}}(n_{{N_A}}) &  & \varphi_{k_{K}}(n_{{N_A}})
\end{array}
\right]  \left[
\begin{array}
[c]{c}%
X(k_{1})\\
X(k_{2})\\
\vdots \\
X(k_{K})
\end{array}
\right],\\
\mathbf{y=A}_{K}\mathbf{X}_{K}.\notag
\end{gather}
Dakle, matrica $\mathbf{A}_{K}$ bi se dobila na osnovu mjerne matrice $\mathbf{A}$, izostavljanjem onih kolona koje odgovaraju koeficijentima $X(k)=0$. Pod pretpostavkom da postoji $K$ nenultih koeficijenata $X(k)$ od ukupno $N$, ukupan broj mogućih različitih matrica 
$\mathbf{A}_{K}$ je jednak ukupnom broju mogućih kombinacija, $\binom{N}{K}$. Za jedan realni slučaj signala (odnosno transformacionog vektora) male dužine $N=256$ koji ima $K=6$ nenultih koeficijenata, broj ovih kombinacija je $\binom{N}{K}=\binom{256}{6}\approx 3.6853\cdot10^{11}$.

\subsection{Furijeova mjerna matrica}
U obradi signala, diskretna Furijeova transformacija predstavlja jednu od najznačajnijih transformacija. Za signal $x(t)$ ograničenog trajanja $T$ koji je odabran uniformno, u skladu sa teoremom o odabiranju, čiji su odbirci $x(n)=x(n\Delta t)$, gdje je $\Delta t$ korak odabiranja i $\Delta t=T/N$,
DFT $X(k)$  je data izrazom (\ref{dft}), dok je inverzna DFT definisana formulom (\ref{idft}).

Koeficijenti inverzne DFT matrice su $\varphi_k(n)=\frac{1}{N}e^{j2\pi nk/N}$. U slučaju redukovanog skupa odbiraka $\mathbf{y}=[x(n_1),x(n_2),\dots,x(n_{N_A})]^T$, koji su dostupni na pozicijama 
\begin{align}
n_{i}\in{\mathbb{{N}}_A=}\{n_{1},n_{2},\dots ,n_{{N_A}}\}\subseteq\mathbb{N}%
=\{0,1,2,\dots ,N-1\} 
\end{align}  uz $y(i)=x(n_{i+1})=\frac{1}{N}\sum_{k=1}^{N-1}X(k)e^{j2\pi n_i k/N}$, Furijeova mjerna matrica data je sljedećim izrazom \cite{multimedia,tutorial}:
\begin{equation}
\mathbf{A}=\frac{1}{N}\left[
\begin{array}
[c]{cccc}%
1 & e^{j2\pi n_1/N} &\dots   & e^{j2\pi n_1 (N-1)/N}\\
1 & e^{j2\pi n_2/N} &  &e^{j2\pi n_2 (N-1)/N}\\
\vdots & \vdots &\ddots  &\vdots\\
1 & e^{j2\pi n_{N_A}/N}&  & e^{j2\pi n_{N_A}(N-1)/N}
\end{array}
\right] .
\end{equation}

Navedena matrica je poznata pod nazivom parcijalna inverzna DFT matrica. U nekim slučajevima, kako bi se postigla normalizacija kolona tako da je njihova energija jednaka jedinici, faktor $\frac{1}{N}$ se može zamijeniti faktorom $\frac{1}{N_A}$.
\section{Uslovi za rekonstrukciju}
Kako je cilj rekonstruisati signale na osnovu redukovanog skupa mjerenja, postavlja se pitanje uslova pod kojima je to moguće uraditi. Čak i kada se dođe do rješenja problema rekonstrukcije, otvaraju se pitanja njegove egzaktnosti i jedinstvenosti. U ovom odjeljku se formalno tretiraju navedeni problemi.
\subsection{Direktna pretraga u prostoru koeficijenata}
Razmotrimo prvo simplifikovani problem rekonstrukcije signala čiji vektor koeficijenata $\mathbf{X}$ ima samo jedan koeficijent sa nenultom vrijednošću, odnosno, stepen rijetkosti $K=1$. Pozicija i amplituda nenultog koeficijenta $X(i)$ su nepoznati. Jedan pristup rješavanju problema jeste mjerenje svih $N$ koeficijenata $X(k)$, za $k=0,1,\dots ,N-1$. Međutim, kako je poznato da je samo jedan koeficijent različit od nule, problem određivanja nepoznatog koeficijenta se može riješiti sa značajno manjim brojem mjerenja.

Ako se uzme samo jedno dostupno mjerenje vektora $\mathbf{X}$:
\begin{equation}
y(0)=\sum_{k=0}^{N-1}X(k)\varphi_{k}(0). \label{Hip0}%
\end{equation}
sa težinskim koeficijentima $\varphi_{k}(0)=a_{k}$, $k=0,1,\dots ,N-1$, tada se ono u prostoru koeficijenata $X(0),X(1),\dots ,X(N-1)$ može interpretirati kao $N$-dimenziona hiperravan.  Budući da je poznato da tačno jedna promjenljiva $X(i)$ ima nenultu vrijednost, bilo koji presjek hiperravni sa bilo kojom koordinatnom osom predstavljao bi rješenje problema rekonstrukcije za $K=1$. Jedno mjerenje produkovalo bi $N$ mogućih pojedinačnih nenultih vrijednosti u formi \cite{dos}:
\begin{equation}
X(k)=y(0)/\varphi_{k}(0),\text{ }k=0,1,2,\dots ,N-1,
\end{equation}
pritom pretpostavljajući da za sve težinske koeficijente važi $\varphi_{k}(0)\neq 0$. Zaključuje se da jedno mjerenje nije dovoljno za rješavanje razmatranog problema. Potrebno je najmanje još jedno mjerenje.

Sa dva dostupna mjerenja, $y(0)$ i $y(1)$, u prostoru nepoznatih koeficijenata (odnosno promjenljivih) formiraju se dvije hiperravni, koje definišu dva seta mogućih rješenja. Ove hiperravni su date izrazima\cite{dos,tutorial}:
\begin{align}
X(k)&=y(0)/\varphi_{k}(0),\text{ }k=0,1,2,\dots ,N-1,\\
X(k)&=y(1)/\varphi_{k}(1),\text{ }k=0,1,2,\dots ,N-1.
\end{align}

Ukoliko ove dvije hiperravni imaju zajedničku tačku u $k=i$, tada je ta tačka, odnosno,
\begin{align}
X(i)=y(0)/\varphi_{i}(0)=y(1)/\varphi_{i}(1),
\end{align}
jedinstveno rješenje posmatranog problema. Može se jednostavno pokazati da je zajednička vrijednost dva mjerenja $X(i)=y(0)/\varphi_{i}(0)=y(1)/\varphi
_{i}(1)$ jedinstvena ako je zadovoljeno \cite{dos}:
\begin{align}
\det\left[
\begin{array}
[c]{cc}%
\varphi_{i}(0) & \varphi_{k}(0)\\
\varphi_{i}(1) & \varphi_{k}(1)
\end{array}
\right] =\varphi_{i}(0)\varphi_{k}(1)-\varphi_{i}(1)\varphi_{k}(0)\neq0.
\end{align}

U cilju generalizacije i izvođenja opštih zaključaka, definišimo rang i \textit{spark} matrice. Za matricu $\mathbf{A}$ koja ima ${N_A}$ vrsta i $N\geq {N_A}$ kolona, rang matrice $\mathbf{A}$ jednak je najvećem broju nezavisnih vrsta, odnosno kolona te matrice, gdje važi $
1\leq\mathrm{rank}\left\{  \mathbf{A}\right\}  \leq {N_A}.$ Za bilo koju matricu $\mathbf{A}_2$ dimenzija $2\times 2$, koja je podmatrica mjerne matrice, važi da je $\mathrm{rank}\{\mathbf{A}_2\}=2$. \textit{Spark} matrice je najmanji broj zavisnih kolona (vrsta) matrice \cite{tutorial,enciklopedija}. Po definiciji, ako kolona sadrži sve nule, tada važi $\mathrm{spark}\left\{  \mathbf{A}\right\}  =1$. Uočimo i da u opštem slučaju važi
$
2\leq\mathrm{spark}\left\{  \mathbf{A}\right\}  \leq {N_A}+1
$. Ukoliko ne postoji nijedna kolona sa svim nulama, a pri tome postoje dvije linearno zavisne kolone, tada je $\mathrm{spark}\left\{  \mathbf{A}\right\}  =2$. U razmatranom primjeru, rekonstrukcija je jedinstvena ako važi  $\mathrm{spark}\left\{  \mathbf{A}\right\}  > 2$, ili drugim riječima, ako u matrici $\mathbf{A}$ ne postoje dvije linearno zavisne kolone.

Razmotrimo sada opšti slučaj $K$-rijetkog vektora $\mathbf{X}$. Korišćenjem podskupa od $K$ mjerenja, u oznaci $\mathbf{y}^{(1)}_K$, može se dobiti skup rješenja koji predstavljaju odgovarajuće vektore nenultih koeficijenata. Za svaki skup pretpostavljenih pozicija nenultih koeficijenata iz $X(k)$, formira se odgovarajući vektor $\mathbf{X}_K$ koji ima $K$ nenultih elemenata na pozicijama $k\in \Pi_K={k_1,k_2,\dots,k_K}$. Sada imamo sistem $\mathbf{y}^{(1)}_K=\mathbf{A}_K\mathbf{X}_K$ koji se formira za svaki mogući skup pozicija $\{k_1,k_2,\dots,k_K\}$. Mogućih rješenja je $\binom{N}{K}$.

Ukoliko se uvede još jedan skup od $K$ mjerenja, u oznaci $\mathbf{y}^{(2)}_K$, ponovo rješavamo sistem oblika $\mathbf{y}^{(2)}_K=\mathbf{A}_K\mathbf{X}_K$, i to za svaki skup mogućih pozicija  $\{k_1,k_2,\dots,k_K\}$. Takvih kombinacija je ponovo  $\binom{N}{K}$.
Ukoliko dva dobijena skupa od po $\binom{N}{K}$ rješenja imaju jedan zajednički član $\mathbf{X}_K$, tada je pronađeno rješenje problema. 

U cilju ispitivanja jedinstvenosti, pretpostavimo da je moguće pronaći dva moguća rješenja $\mathbf{X}$ stepena rijetkosti $K$, kao i da je dostupno $N_A=2K$ mjerenja u vektoru $\mathbf{y}$. U tom slučaju, rješenje će biti jedinstveno ukoliko su determinante svih podmatrica $\mathbf{A}_{2K}$ matrice $\mathbf{A}$ različite od nule. To drugim riječima znači da sve podmatrice  $\mathbf{A}_{2K}$ moraju biti nesingularne. Ako  dva nenulta dijela rješenj\^{a} označimo sa $\mathbf{X}^{(1)}_K$ i  $\mathbf{X}^{(2)}_K$, tada će oba zadovoljavati jednačinu mjerenja, u smislu:
\begin{equation}
\left[
\begin{array}
[c]{cc}%
\mathbf{A}^{(1)}_K &
\mathbf{A}^{(2)}_K   
\end{array}
\right]
\left[
\begin{array}
[c]{cc}%
\mathbf{X}^{(1)}_K \\
\mathbf{0}_K   
\end{array}
\right]
=\mathbf{y} \text{ i }  
\left[
\begin{array}
[c]{cc}%
\mathbf{A}^{(1)}_K &
\mathbf{A}^{(2)}_K   
\end{array}
\right]
\left[
\begin{array}
[c]{cc}%
\mathbf{X}^{(2)}_K \\
\mathbf{0}_K   
\end{array}
\right]
=\mathbf{y}.
\end{equation}

Sa $\mathbf{A}^{(1)}_K$ i $\mathbf{A}^{(2)}_K$ su označene podmatrice matrice $\mathbf{A}$ koje odgovaraju elementima iz $\mathbf{X}^{(1)}_K $ i $\mathbf{X}^{(2)}_K $, respektivno.

Ukoliko je determinanta matrice $\mathbf{A}_{2K}=\left[
\begin{array}
[c]{cc}%
\mathbf{A}^{(1)}_K &
\mathbf{A}^{(2)}_K   
\end{array}
\right]$, dimenzija $2K\times 2K$, različita od nule, tada ne postoje nenulta rješenja $\mathbf{X}^{(1)}_K$ i  $\mathbf{X}^{(2)}_K$. U opštem slučaju, ako su sve moguće podmatrice  $\mathbf{A}_{2K}$ nesingularne, uključujući i sve podmatrice nižeg ranga, tada je nemoguće da postoje dva rješenja stepena rijetkosti $K$. Zato je rješenje u tom slučaju jedinstveno.

Iz prethodne diskusije, može se zaključiti da je rješenje razmatranog problema, sa signalom čiji je stepen rijetkosti $K$, jedinstveno, ukoliko je zadovoljen sljedeći uslov \cite{dos,tutorial,enciklopedija}:
\begin{equation}
\mathrm{spark}\left\{  \mathbf{A}\right\}  > 2K.
\end{equation}

U opštem slučaju, broj mjerenja je $N_A \geq 2K$. Za rješavanje problema ovim putem, potrebno je razmotriti sve moguće vektore $\mathbf{X	}_K$ koji imaju $K$ nenultih elemenata, gdje je $k\in \Pi_K$. Kako postoji $N_A \geq 2K$ jednačina a samo $K$ nepoznatih, potrebno je riješiti sistem u smislu najmanjih kvadrata \cite{dos,tutorial}:
\begin{align}
\mathbf{X=}\left(  \mathbf{A}^{H}_K\mathbf{A}_K\right)
^{-1}\mathbf{A}^{H}_K\mathbf{y},
\end{align}
i to za svaku moguću kombinaciju pozicija nenultih elemenata $\{k_1,k_2,\dots,k_K\}$, kojih je, kako je već istaknuto,  $\binom{N}{K}$, što je u praksi jako veliki broj. Traženo rješenje je ono koje minimizuje grešku $\left\Vert \mathbf{y}-\mathbf{A}_K\mathbf{X}_K \right\Vert_2^2$.  Ovakav problem je veoma teško riješiti u razumnom vremenu i spada u klasu teško izračunjivih, tzv. \textit{NP-hard} problema, budući da postoji nepolinomijalni broj kombinacija koje treba ispitati.
\subsection{Formalni uslovi}
U slučaju proizvoljnog rijetkog vektora $\mathbf{X}$ stepena rijetkosti $K$ i u slučaju najmanje ${N_A}\geq2K$ mjerenja -- dostupnih odbiraka, jedinstvenost rješenja se može garantovati ukoliko su mjerenja
$
\mathbf{y}=\mathbf{AX} \label{system_CS}%
$
nezavisna, u smislu da se na osnovu njih može  rekonstruisati proizvoljni $2K$-rijetki signal. To drugim riječima znači da sve podmatrice $\mathbf{A}_{2K}$ mjerne matrice $\mathbf{A}$ moraju biti nesingularne, odnosno, treba da važi da su sve odgovarajuće determinante reda $2K$ različite od nule,  \cite{dos, tutorial}:
\begin{equation}
\det(\mathbf{A}_{2K}\mathbf{)=}\det\left[
\begin{array}
[c]{cccc}%
\varphi_{k_{1}}(n_{1}) & \varphi_{k_{2}}(n_{1}) & \dots   & \varphi_{k_{2K}}(n_{1})\\
\varphi_{k_{1}}(n_{2}) & \varphi_{k_{2}}(n_{2}) & \dots   & \varphi_{k_{2K}}(n_{2})\\
\vdots  & \vdots   & \ddots   & \vdots \\
\varphi_{k_{1}}(n_{2K}) & \varphi_{k_{2}}(n_{2K}) & \dots   & \varphi_{k_{2K}}(n_{2K})
\end{array}
\right]  \neq0
\end{equation}
i to za bilo koju kombinaciju indeksa pozicija dostupnih odbiraka  $\{n_{1},n_{2}%
,\dots ,n_{2K}\}\subseteq\{n_{1},n_{2},\dots ,n_{{N_A}}\}$, za ${N_A}\geq2K$, i za bilo koju kombinaciju indeksa $\{k_{1},k_{2},\dots ,k_{2K}\}\subseteq\left\{
0,1,2,\dots ,N-1\right\}  $. Broj mogućih kombinacija klase $2K$ od $N$ elemenata je $\binom{N}{2K}$, što u realnim okolnostima predstavlja jako veliki broj.
Kao što je već diskutovano, uslov se može interpretirati i u sljedećem obliku:
\vspace{-3mm}
\begin{equation}
\mathrm{spark\{\mathbf{A}\}}>2K.
\end{equation}

Činjenica da je potrebno provjeriti da li postoji $2K$ nezavisnih mjerenja, znači da treba zapravo provjeriti da li za matricu $\mathbf{A}_{2K}$ važi da je
$\mathrm{rank}(\mathbf{A}_{2K})=2K$. Ukoliko je ${N_A}>2K$, nema potrebe ispitivati kombinacije po
$n_{i}$ u cilju formiranja kvadratne matrice dimenzija $2K\times2K$ iz matrice dimenzija ${N_A}\times2K$, imajući u vidu da se rang matrice  $\mathbf{A}_{2K}$  dimenzija ${N_A}\times2K$ može provjeriti testiranjem ranga matrice $\mathbf{A}_{2K}^{T}%
\mathbf{A}_{2K}$ dimenzija $2K\times2K$, korišćenjem poznate veze, \cite{dos}:
\begin{align}
\mathrm{rank}(\mathbf{A}_{2K})=\mathrm{rank}(\mathbf{A}_{2K}^{T}%
\mathbf{A}_{2K}).
\end{align}
Matrica $\mathbf{A}_{2K}^{T}\mathbf{A}_{2K}$ se naziva Gramovom matricom matrice 
$\mathbf{A}_{2K}$. Za matrice $\mathbf{A}_{2K}$ čiji su elementi kompleksni, koristi se ekvivalentna Hermitska matrica $\mathbf{A}_{2K}^{H}%
\mathbf{A}_{2K}$, pri čemu $(\cdot)^H$ označava Hermitsko transponovanje. Jedan način za određivanje da li je rang matrice $\mathbf{A}_{2K}^{T}%
\mathbf{A}_{2K}$ jednak $2K$ jeste korišćenje sljedećeg uslova \cite{tutorial}:
\begin{align}
\det(\mathbf{A}_{2K}^{T}\mathbf{A}_{2K})=d_{1}d_{2}\dots d_{2K}\neq0
\end{align}
gdje su $d_{1},d_{2},\dots, d_{2K}$ sopstvene vrijednosti kvadratne matrice $\mathbf{A}_{2K}^{T}%
\mathbf{A}_{2K}$. Važno je uočiti da su sve sopstvene vrijednosti $d_{i}$ simetrične matrice $\mathbf{A}
_{2K}^{T}\mathbf{A}_{2K}$, u oznaci
$
d_{i}=\mathrm{eig}(\mathbf{A}_{2K}^{T}\mathbf{A}_{2K}),
$
nenegativne. Rang matrice $\mathbf{A}_{2K}^{T}\mathbf{A}_{2K}$ jednak je $2K$ ukoliko minimalna sopstvena vrijednost matrice $\mathbf{A}_{2K}^{T}\mathbf{A}_{2K}$ zadovoljava \cite{dos,tutorial}:
\begin{align}
|d_{\min}|=\min \{|d_{1}|,|d_{2}|,\dots,|d_{2K}| \}>0,
\end{align}
za sve podmatrice $\mathbf{A}_{2K}$ matrice $\mathbf{A}$. 

U numeričkim i praktičnim realizacijama ne bi bilo povoljno da je $\det\{\mathbf{A}_{2K}\}$ ili $\det\{\mathbf{A}\}$ blisko nuli. Iako bi u tom slučaju bio zadovoljen uslov jedinstvenosti rješenja, došlo bi do visoke osjetljivosti na uticaj šuma u mjerenjima, bilo u analizi, bilo u inverziji koja je involvirana u procedurama za rekonstrukciju. Stoga, važno je istaći da postoji praktičan zahtjev da posmatrana determinanta bude ne samo različita od nule, već i da se u dovoljnoj mjeri razlikuje od nule, kako bi se obezbijedili stabilnost inverzije i robustnost na uticaj šuma.	

\subsubsection{Svojstvo ograničene izometrije}
U nastavku će biti podrazumijevana normalizacija energija kolona matrice $\mathbf{A}$. Energije kolona su u vezi sa elementima na dijagonali matrice $\mathbf{A}_{2K}^{T}%
\mathbf{A}_{2K}$. Norma matrice $\mathbf{A}_{2K}$ zadovoljava sljedeće svojstvo \cite{dos,CandesRIP}:
\begin{equation}
\lambda_{\min}  \leq \frac{\left\Vert \mathbf{A}_{2K} \mathbf{X}_{2K}\right\Vert _{2}^{2}}{\left\Vert \mathbf{X}_{2K}%
	\right\Vert _{2}^{2}}   \leq \lambda_{\max} \label{RIP_DEFb}
\end{equation}
gdje su $\lambda_{\min}$ i $\lambda_{\max}$ minimalna i maksimalna sopstvena vrijednost Gramove matrice $\mathbf{A}_{2K}^{T}\mathbf{A}_{2K}$, dok 
$
\left\Vert \mathbf{X}\right\Vert _{2}^{2}=\left\vert X(0)\right\vert
^{2}+\left\vert X(1)\right\vert ^{2}+\dots +\left\vert X(N-1)\right\vert ^{2}
$
predstavlja kvadriranu $\ell_2$-normu vektora $\mathbf{X}$. U izrazu (\ref{RIP_DEFb}), vektor $ \mathbf{X}_{2K}$ predstavlja proizvoljni $2K$-rijetki vektor. Linearna transformaciona matrica $\textbf{A}$ posjeduje svojstvo izometrije ukoliko očuvava intenzitet vektora u $N$-to dimenzionom prostoru, odnosno, ako važi \cite{dos,multimedia}:

\begin{equation}
\left\Vert \mathbf{A X}\right\Vert _{2}^{2}=\left\Vert
\mathbf{X}\right\Vert _{2}^{2} \text{ tj. } \frac{\left\Vert \mathbf{A X}\right\Vert _{2}^{2}}{\left\Vert
	\mathbf{X}\right\Vert _{2}^{2}}=1.\label{Izometrija}%
\end{equation}

Razvojem teorije kompresivnog odabiranja i obrade rijetkih signala, za mjerne matrice je bilo neophodno uvesti nešto blaži uslov izometrije. Svostvo ograničene izometrije (engl. \textit{Restricted Isometry Property - RIP}) matrice $\mathbf{A}_{2K}$ važi ako je ispunjeno \cite{dos,multimedia,CandesRIP}: 
\begin{align}
1-\delta_{2K}\leq\frac{\left\Vert \mathbf{A}_{2K} \mathbf{X}_{2K}\right\Vert _{2}^{2}}{\left\Vert \mathbf{X}_{2K}\right\Vert _{2}^{2}}%
\leq1+\delta_{2K},
\label{rip12}
\end{align}
i to za bilo koji $2K$-rijetki vektor $\mathbf{X}_{2K}$. Ovdje je uslov $\left\Vert \mathbf{A}_{2K} \mathbf{X}_{2K}\right\Vert _{2}^{2}=\left\Vert
\mathbf{X}_{2K}\right\Vert _{2}^{2} $ relaksiran u smislu da je relativna apsolutna vrijednost razlike $\left\Vert \mathbf{A}_{2K} \mathbf{X}_{2K}\right\Vert _{2}^{2}-\left\Vert
\mathbf{X}_{2K}\right\Vert _{2}^{2} $  dovoljno mala u poređenju sa energijom rijetkog signala, odnosno, da je u opsegu $0\leq\delta_{2K}<1$. Konstanta ograničene izometrije, $\delta_{2K}$, predstavlja mjeru koliko matrica $\mathbf{A}_{2K}$ odstupa od matrice koja zadovoljava svojstvo izometrije. Poređenjem izraza (\ref{RIP_DEFb}) i (\ref{rip12}) zaključuje se da važi:
\begin{equation}
\delta_{2K}= \max\left\{ 1-\lambda_{\min}, \lambda_{max}-1  \right\}  ,
\label{RIPCONSEIG}%
\end{equation}
pri čemu je izometrijska konstanta, u oznaci $\delta_{2K}$, uobičajeno definisana kao $\lambda_{max}-1=\max\left\{  \mathrm{eig}\left(  \mathbf{A}_{2K}^{T}\mathbf{A}_{2K}\mathbf{-I}\right)  \right\}$. Kod matrica sa kompleksnim vrijednostima, koristi se $\mathbf{A}_{2K}^{H}\mathbf{A}_{2K}$.

Za $K$-rijetki signal $\mathbf{X}_K$ i mjernu matricu $\mathbf{A}$, RIP uslov je zadovoljen ako (\ref{rip12}) važi za sve podmatrice $\mathbf{A}_K$, za $0\leq \delta_K <1$. Rješenje rekonstrukcije za $K$-rijetki signal će biti jedinstveno ukoliko mjerna matrica zadovoljava RIP uslov za $2K$-rijetki vektor $\mathbf{X}_{2K}$ i izometrijsku konstantu $0\leq \delta_{2K} <1$. Važno je primijetiti da je RIP uslov zadovoljen za $\lambda_{\min}>0$, što znači da su sve matrice $\mathbf{A}_{2K}^{T}\mathbf{A}_{2K}$ nesingularne. Drugim riječima, ako je $\mathbf{X}$
$K$-rijetki vektor dužine $N$, tada on može biti jedinstveno rekonstruisan iz redukovanog skupa od ${N_A}$ mjerenja/odbiraka $\mathbf{y}=\mathbf{AX}$, ukoliko je mjerna matrica $\mathbf{A}$ takva da njene podmatrice $\mathbf{A}_{2K}$
zadovoljavaju  $2K$ RIP sa izometrijskom konstantom $0\leq\delta_{2K}<1$, za sve kombinacije od $2K$ kolona, od ukupno $N$ kolona, \cite{dos,tutorial}.

Svojstvo ograničene izometrije je za malo $\delta_{2K}$ bliže svojstvu izometrije i poboljšava stabilnost rekonstrukcije. Invertibilnost i robustnost matrice je u vezi sa kondicionim brojem:
\begin{align}
\mathrm{cond}\left\{ \mathbf{A}_{2K}^{T}\mathbf{A}%
_{2K}\right\}  =\frac{\lambda_{\max}}{\lambda_{\min}}.%
\end{align}
Ako matrica $\mathbf{A}_{2K}$ zadovoljava RIP sa izometrijskom konstantom $\delta_{2K}$, tada važi \cite{dos}:
\begin{align}
\mathrm{cond}\left\{ \mathbf{A}_{2K}^{T}\mathbf{A}%
_{2K}\right\}  \leq\frac{1+\delta_{2K}}{1-\delta_{2K}},
\end{align}
pa manje $\delta_{2K}$ povlači da kondicioni broj bude bliži vrijednosti $1$, koja se, inače, vezuje za stabilnu invertibilnost matrice i slabiju osjetljivost na ulazni šum, što označava činjenicu da manje varijacije u mjerenjima neće uzrokovati veće varijacije u rezultatu rekonstrukcije.

\subsubsection{Uslov nekoherentnosti}
Indeks koherentnosti matrice $\mathbf{A}$ je veličina definisana izrazom \cite{dos,multimedia,donoho,CandesRIP}:
\begin{align}
\mu=\max\left\vert \mu(m,k)\right\vert \text{, za }m\neq k,
\end{align}
i predstavlja maksimalnu apsolutnu vrijednost normalizovanog skalarnog proizvoda dvije kolone posmatrane matrice: 
\begin{align}
\mu(m,k)=\frac{\sum_{i=1}^{N_A}\varphi_{m}(n_{i})\varphi_{k}^{\ast}(n_{i})}{\sum_{i=1}^{N_A}\left\vert \varphi_{k}(n_{i})\right\vert ^{2}%
},
\end{align}
pri čemu $\varphi_{k}(n_i)$ označava  $k$-tu kolonu matrice $\mathbf{A}$. Može se uočiti da su  $\mu(m,k)$ elementi mimo glavne dijagonale matrice $\mathbf{A}^{H}\mathbf{A}$ normalizovani odgovarajućim elementima na dijagonali. Indeks koherentnosti je veoma bitan u analizi mjernih matrica. Treba da bude što je moguće manji (odnosno, mjerna matrica treba da bude što je više nekoherentna). Sa manjim vrijednostima ovog indeksa, matrica $\mathbf{A}^{H}\mathbf{A}$ postaje bliža jediničnoj matrici.

Za matricu $\mathbf{A}$  dimenzija ${N_A}\times N$ indeks koherentnosti ne može biti proizvoljno mali. Pokazuje se da važi Velčova gornja granica (engl. \textit{Weltch upper bound}), \cite{tutorial}:
\begin{equation}
\mu\geq\sqrt{\frac{N-{N_A}}{{N_A}\left(  N-1\right)  }}. \label{WelchBound}%
\end{equation}

Rekonstrukcija $K$-rijetkog signala iz $N_A$ mjerenja je jedinstvena ako je zadovoljeno \cite{dos,tutorial}:
\begin{equation}
K<\frac{1}{2}\left(1+\frac{1}{\mu}\right).
\end{equation}

Indeks koherentnosti se može koristiti u određivanju donje granice za \textit{spark} matrice, \cite{tutorial}:
\begin{equation}
\mathrm{spark\{\mathbf{A}\}} \geq \left(1+\frac{1}{\mu}\right).
\end{equation}

Ako je $\mathbf{X}$ rješenje sistema jednačina $\mathbf{y=AX}$, i ako ono zadovoljava uslov:
\begin{equation}
\left\Vert\mathbf{X} \right\Vert_0=K<\frac{1}{2}\left(1+\frac{1}{\mu}\right)\leq\frac{1}{2}\mathrm{spark\{\mathbf{A}\}},
\end{equation}
tada je to rješenje $\mathbf{X}$ najmanjeg mogućeg stepena rijetkosti.

Indeks koherentnosti se može vezati i za izometrijsku konstantu na sljedeći način \cite{tutorial}:
\begin{equation}
\delta_K \leq (1-K)\mu.
\end{equation}

\section{Rekonstrukcija zasnovana na $\ell_0$-normi}
Formalno, problem rekonstrukcije rijetkih signala podrazumijeva da je signal moguće rekonstruisati iz redukovanog skupa mjerenja, koja formiraju vektor $\mathbf{y}$, nalaženjem vektora $\mathbf{X}$ sa najmanjim mogućim stepenom rijetkosti, a koji istovremeno odgovara mjerenjima $\mathbf{y}$. Ako su mjerenja $x(n_i)$ dostupna na slučajnim pozicijama $n_{i}\in{\mathbb{{N}}_A}=\{n_{1},n_{2},\dots ,n_{{N_A}}\}\subseteq\{0,1,2,\dots,{N-1}\}$ i ako nenultih koeficijenata ima $K=\left\Vert \mathbf{X}\right\Vert_0$, što je notacija zasnovana na $\ell_0$-normi, problem se formalno matematički definiše u obliku \cite{donoho,candes,candes2,CandesRIP, tutorial}:
\begin{align}
\min\left\Vert \mathbf{X}\right\Vert _{0}\text{ \ pod uslovom \ \ }%
\mathbf{y=AX}. \label{prviproblem}
\end{align}

Važno je istaći da se $\ell_0$-norma ne može direktno koristiti u minimizacionim procedurama, \cite{dos,tutorial}. Međutim, rješavanje problema (\ref{prviproblem}) moguće je obaviti implicitno, određivanjem pozicija $k\in\{k_{1},k_{2},\dots ,k_{K}\}=\Pi_K$ nenultih koeficijenata, gdje je $K\ll N$ i korišćenjem činjenice da je $X(k)=0$ za $k\notin\{k_{1},k_{2},\dots ,k_{K}\}=\Pi_K$.

\subsection{Rekonstrukcija u slučaju poznatih pozicija}

Posmatra se diskretni signal $x(n)$ koji je rijedak u transformacionom domenu definisanom baznim funkcijama $\varphi_{k}(n)$, $k=0,1,\dots ,N-1$. Broj nenultih koeficijenata  $K$ je mnogo manji od originalnog broja odbiraka signala,  $N$, odnosno, važi da je $X(k)=0$ za
$
k\notin\{k_{1},k_{2},\dots ,k_{K}\}=\Pi_K
$
i
$K\ll N$. Posmatrani signal
\begin{equation}
x(n)=\sum_{k\in\{k_{1},k_{2},\dots ,k_{K}\}}X(k)\varphi_{k}(n). \label{SIG_FS_SA{N_A}{N_A}}%
\end{equation}
stepena rijetkosti $K$ može biti rekonstruisan iz  ${N_A}$ odbiraka, gdje je  ${N_A}\leq N$. U tom slučaju, dakle, postoji $K$ nenultih vrijednosti $X(k_{1}),X(k_{2}),\dots,X(k_{K})$, a za sve ostale koeficijente $X(k)$, za $k\notin\{k_{1},k_{2},\dots ,k_{K}%
\}$ važi da su jednaki nuli. Pretpostavimo da su poznate pozicije  $\{k_{1}$, $k_{2}$, \dots , $k_{K}\}$. Tada se jednačina mjerenja može zapisati u obliku:
\begin{equation}
\sum_{k\in{\Pi}_K}X(k)\varphi_{k}(n_{i})=x(n_{i})\text{, za }i=1,2,\dots ,{N_A}\geq
K. \label{Sist_RJ}
\end{equation}
U matričnoj formi, ovaj sistem od $K$ nepoznatih i $N_A$ jednačina dat je u obliku:
\begin{equation}
\mathbf{A}_{K}\mathbf{X}_{K}\mathbf{=y} \label{Sus{N_A}at},
\end{equation}
gdje je $\mathbf{X}_{K}=[X(k_{1}),X(k_{2}),\dots,X(k_{K})]^{T}$ vektor nepoznatih nenultih koeficijenata (na poznatim pozicijama, po pretpostavci), dok je $\mathbf{y}=[x(n_{1}),x(n_{2}),\dots,x(n_{{N_A}})]^{T}$ vektor mjerenja, odnosno, dostupnih odbiraka signala. Matrica $\mathbf{A}_{K}$, dimenzija $N_A\times K$, je podmatrica mjerne matrice $\mathbf{A}$, dobijena odbacivanjem kolona koje odgovaraju pozicijama $k\notin
\{k_{1},k_{2}, \dots,k_{K}\}$ koeficijenata koji su jednaki nuli. Ona je sljedećeg oblika:
\begin{equation}
\mathbf{A}_{K}=\left[
\begin{tabular}
[c]{llll}%
$\varphi_{k_{1}}(n_{1})$ & $\varphi_{k_{2}}(n_{1})$ & \dots  & $\varphi_{k_{K} }(n_{1})$\\
$\varphi_{k_{1}}(n_{2})$ & $\varphi_{k_{2}}(n_{2})$ & \dots  & $\varphi_{k_{K}}(n_{2})$\\
$\vdots$  & $\vdots$  & $\ddots$  & $\vdots$ \\
$\varphi_{k_{1}}(n_{N_A})$ & $\varphi_{k_{2}}(n_{N_A})$ & \dots  & $\varphi_{k_{K}}(n_{N_A})$
\end{tabular}
\right]  . \label{{N_A}artr_Sampl}%
\end{equation}

Rješenje se dobija minimizacijom kvadrata razlike između dostupnih odbiraka i njihovih vrijednosti koje se dobijaju na osnovu rekonstruisanih koeficijenata, \cite{dos,tutorial}:
\begin{align}
e^{2}&=\sum_{n\in{\mathbb{{N}}_A}}\left\vert y(n)-\sum_{k\in{\Pi}_K}X(k)\varphi
_{k}(n)\right\vert ^{2}\nonumber\\
&=\left(  \mathbf{y-A}_{K}\mathbf{X}_{K}\right)  ^{H}\left(  \mathbf{y-A}%
_{K}\mathbf{X}_{K}\right)  =\left\Vert \mathbf{y}\right\Vert _{2}%
^{2}-2\mathbf{X}_{K}^{H}\mathbf{A}_{K}^{H}\mathbf{y+X}_{K}^{H}\mathbf{A}%
_{K}^{H}\mathbf{A}_{K}\mathbf{X}_{K} \label{e_vec_h}%
\end{align}

Uz simboličko diferenciranje po vektoru nepoznatih, dobija se minimum od $e^{2}$ u obliku:
\begin{align}
\frac{\partial e^{2}}{\partial\mathbf{X}_{K}^{H}}=-2\mathbf{A}_{K}%
^{H}\mathbf{y+2A}_{K}^{H}\mathbf{A}_{K}\mathbf{X}_{K}=0.
\end{align}

Rješenje je dato izrazom:
\begin{equation}
\mathbf{X}_{K}\mathbf{=}\left(  \mathbf{A}_{K}^{H}\mathbf{A}_{K}\right)
^{-1}\mathbf{A}_{K}^{H}\mathbf{y}=\mathrm{pinv}(\mathbf{A}_{K})\mathbf{y}, \label{Rjesenje}%
\end{equation}
a $\mathrm{pinv}(\mathbf{A}_{K})$ predstavlja pseudo-inverziju matrice $\mathbf{A}_{K}$.

\subsection{Estimacija nepoznatih pozicija -- OMP i CoSaMP algoritmi}
U opštem slučaju, pozicije nenultih koeficijenata u vektoru $\mathbf{X}$ nijesu poznate. One se mogu estimirati \textit{matching pursuit} (MP) pristupom za rekonstrukciju rijetkih signala. Napredna verzija ovog algoritma poznata je pod nazivom \textit{orthogonal matching pursuit} (OMP), \cite{tutorial,enciklopedija,cosamp}. Kod ovog pristupa, rekonstrukcija je zasnovana na iterativnom projektovanju vektora mjerenja $\mathbf{y}$ na kolone matrice $\mathbf{A}$ koje odgovaraju trenutno detektovanom setu pozicija nenultih koeficijenata, u oznaci $\hat{\mathbf{\Pi}}_K$.
Označimo inicijalnu estimaciju sa
\begin{equation}
\mathbf{X}_0=\mathbf{A}^{H}\mathbf{y}=\mathbf{A}^H\mathbf{A}\mathbf{X}.
\end{equation}

Mjerenja se dobijaju kao linearne kombinacije nenultih koeficijenata iz domena rijetkosti signala, gdje su vrste mjerne matrice težinski koeficijenti. To zapravo znači da se projekcijom mjerenja $\mathbf{y}$ na mjernu matricu $\mathbf{A}$ mogu estimirati pozicije nenultih koeficijenata. 

Pozicija prvog koeficijenta se dobija kao:
\begin{equation}
{{\hat{k}}_{1}}= \arg\max \left\{ \left| {{\mathbf{X}}_{0}} \right| \right\}.
\end{equation}

Zatim se rješava sistem (\ref{Sus{N_A}at}), u cilju nalaženja minimuma $\left\Vert \mathbf{y}-\mathbf{A}_{K}\mathbf{X}_{K} \right\Vert^2_2 $, gdje matrica $\mathbf{A}_K$ odgovara skupu $\mathbf{\hat{\Pi}}_K=\{{{\hat{k}}_{1}}\}$. U tu svrhu, određuju se odbirci signala ${{\mathbf{y}}_{1}}={{\mathbf{A}}_{1}}{{\mathbf{X}}_{1}}$, koji odgovaraju skupu $\mathbf{\hat{\Pi}}_K=\{{{\hat{k}}_{1}}\}$. Ukoliko važi $\mathbf{e}={{\mathbf{y}}_{\mathbf{1}}}$, tada je stepen rijetkosti traženog signala 1, i $\mathbf{X}_1$ predstavlja rješenje problema. Ako to nije slučaj, formira se signal ${{\mathbf{e}}_{1}}=\mathbf{y}-{{\mathbf{y}}_{1}}$.

Pozicija druge komponente se estimira korišćenjem signala $\mathbf{e}_1$ u obliku:
\begin{equation}
{{\hat{k}}_{2}}= \arg\max \left\{ \left| \mathbf{A}^H\mathbf{e}_1 \right| \right\},
\end{equation}
čime se formira skup $\mathbf{\hat{\Pi}}_K=\{{{\hat{k}}_{1}},\hat{k}_2\}$, i za njega se ponavlja prethodno opisani postupak, uz $K=2$. Sada se koriste koeficijenti $X(\hat{k}_1)$ i $X(\hat{k}_2)$, gdje se prvobitno dobijeni koeficijent $X(\hat{k}_1)$ reestimira. Postupkom se dobija novi vektor $\mathbf{X}_2$, estimirana mjerenja $\mathbf{y}_2$ i signal greške ${{\mathbf{e}}_{2}}=\mathbf{y}-{{\mathbf{y}}_{2}}$. Ako je vektor ${{\mathbf{e}}_{2}}$ vektor nula, ili su njegovi elementi sa zadovoljavajuće malim vrijednostima, postupak se završava i rješenje je nađeno u obliku $\mathbf{X}_2$. Ako to nije slučaj, procedura se nastavlja, estimira se pozicija treće komponente. Procedura se završava kada greška bude jednaka nula, ili je u nekim prihvatljivim granicama, definisanim preciznošću $\varepsilon$. Rezime pristupa dat je u Algoritmu \ref{Norm0Alg}.

\begin{algorithm}[!b]
	\floatname{algorithm}{Algoritam}
	\caption{OMP rekonstrukcioni algoritam}
	\label{Norm0Alg}
	\begin{algorithmic}[1]
		\Input
		\Statex
		\begin{itemize}
			\item Vektor mjerenja $\mathbf{y}$
			\item Mjerna matrica $\mathbf{A}$
			\item Broj odabranih koeficijenata u svakoj iteraciji $r$
			\item Zahtijevana tačnost $\varepsilon$
		\end{itemize}
		\Statex
		\State $\hat{\mathbf{\Pi}}_K \gets \emptyset$
		\State $\mathbf{e} \gets \mathbf{y} $
		\While{$\left\| \mathbf{e} \right\|_2 > \varepsilon$} \Comment Mogu se dodati i ograničenja u pogledu maksimalnog broja iteracija
		\State $(\hat{k}_1,\hat{k}_2,\ldots,\hat{k}_r) \gets $ \text{pozicije $r$ najvećih vrijednosti u vektoru $\mathbf{A}^H\mathbf{e}$}
		\smallskip
		\State $\hat{\mathbf{\Pi}}_K \gets \hat{\mathbf{\Pi}}_K \cup \{\hat{k}_1,\hat{k}_2,\ldots,\hat{k}_r\} $
		\State $\mathbf{A}_K \gets \mathbf{A}(:,\hat{\mathbf{\Pi}}_K)$ \Comment kolone matrice $\mathbf{A}$ odabrane na osnovu $\hat{\mathbf{\Pi}}_K $
		\State $\mathbf{X}_K \gets \operatorname{pinv}(\mathbf{A}_K)\mathbf{y}$
		\State $\mathbf{y}_K \gets \mathbf{A}_K\mathbf{X}_K$
		\State $\mathbf{e} \gets \mathbf{y} -  \mathbf{y}_K$
		\EndWhile
		\smallskip
		\State $\displaystyle \mathbf{X} \gets
		\begin{cases} \mathbf{0}, & \text{za pozicije koje nijesu u }\hat{\mathbf{\Pi}}_K \\
		\mathbf{X}_K, & \text{za pozicije iz }\hat{\mathbf{\Pi}}_K
		\end{cases}$
		\Statex
		\Output
		\Statex
		\begin{itemize}
			\item Rekonstruisani koeficijenti $\mathbf{X}$
		\end{itemize}
	\end{algorithmic}
\end{algorithm}

Nešto izmijenjena forma OMP algoritma poznata je pod nazivom \textit{Compressive Sampling Matched Pursuit} (CoSaMP), \cite{tutorial,cosamp}. U slučaju ovog pristupa, signal sa željenim stepenom rijetkosti $K$ se dobija iterativnim putem. Ovdje se vrši projekcija vektora mjerenja $\mathbf{y}$ na kolone mjerne matrice $\mathbf{A}$, i selekcija $2K$ pozicija sa najvećim magnitudama projekcije.  Skup odabranih pozicija se proširuje pozicijama nenultih elemenata u trenutnoj estimaciji rijetkog vektora $\mathbf{X}$. Zatim se nalazi rješenje u smislu najmanjih kvadrata, i $K$ koeficijenata sa najvećim vrijednostima se proglašavaju za rekonstruisani vektor $\mathbf{X}$. Mjerni vektor se ažurira, oduzimanjem trenutnog rješenja, a zatim se procedura iterativno ponavlja. CoSaMP rekonstrukciona procedura je predstavljena u Algoritmu \ref{CoSaMPAlg}. Procedura se ili ponavlja do predefinisanog broja iteracija, ili dok je zadovoljen kriterijum definisan normom vektora greške.

\begin{algorithm}[htb]
	\floatname{algorithm}{Algoritam}
	\caption{CoSaMP rekonstrukcioni algoritam}
	\label{CoSaMPAlg}
	\begin{algorithmic}[1]
		\Input
		\Statex
		\begin{itemize}
			\item Mjerni vektor $\mathbf{y}$
			\item Mjerna matrica $\mathbf{A}$
			\item Željeni stepen rijetkosti $K$
		\end{itemize}
		\Statex
		\State $\mathbf{X} \gets \mathbf{0}_{N\times 1}$
		\State $\mathbf{e} \gets \mathbf{y}$		
		\Repeat
		\State $\mathbb{T}_{1} \gets $ {pozicije $2K$ najvećih vrijednosti u vektoru $\mathbf{A}^H\mathbf{e}$}
		\State $\mathbb{T}_{2} \gets $ {pozicije nenultih koeficijenata u vektoru $\mathbf{X}$}
		\State $\mathbb{T} \gets \mathbb{T}_{1} \cup \mathbb{T}_{2} $
		\State $\mathbf{A}_T \gets $ kolone iz matrice $\mathbf{A}$ definisane skupom $\mathbb{T}$
		\State $\mathbf{B} \gets \operatorname{pinv}(\mathbf{A}_T) \mathbf{y}$
		\vspace{0.5ex}		 
		\State \parbox[t]{\dimexpr\linewidth-\algorithmicindent}{Postaviti $K$ koeficijenata sa najvećim vrijednostima iz $\mathbf{B}$ na odgovarajuće pozicije u $\mathbf{X}$, a ostale koeficijente postaviti na nulu.}
		\vspace{0.1ex}
		\State $\mathbf{e} \gets \mathbf{y} -  \mathbf{A} \mathbf{X} $
		\Until{zadovoljen kriterijum zaustavljanja}
		\Statex
		\Output
		\Statex
		\begin{itemize}
			\item Rekonstruisani $K$-rijetki vektor  $\mathbf{X}$
		\end{itemize}
	\end{algorithmic}
\end{algorithm}

\subsection{Šum u DFT domenu koji je uzrokovan nedostajućim odbircima}
Razmatra se signal oblika
\begin{align}
x(n)=\sum_{p=1}^{K}A_{p}e^{j2\pi nk_{p}/N},
\end{align}
sa $N_A\leq N$ dostupnih odbiraka na pozicijama  $n\in{\mathbb{{N}}_A}=\{n_{1},n_{2}%
,\dots ,n_{{N_A}}\}$
koji je $K$-rijedak u DFT domenu. Inicijalna DFT estimacija dobija se računanjem transformacije na bazi dostupnih mjerenja, pretpostavljajući nule na pozicijama nedostupnih odbiraka, na sljedeći način:
\begin{equation}
X(k)=\sum_{n\in{\mathbb{{N}}_A}}x(n)e^{-j2\pi nk/N}=\sum_{n\in{\mathbb{{N}}_A}}\sum
_{p=1}^{K}A_{p}e^{-j2\pi n(k-k_{p})/N}.\label{{N_A}S_SU{N_A}{N_A}{N_A}}%
\end{equation}

Nedostajući odbirci se manifestuju kao šum u DFT domenu. DFT koeficijenti su stoga slučajne varijable. Mogu se razlikovati dvije grupe koeficijenatata, sa međusobno različitim statističkim karakteristikama. U nastavku slijedi  izvođenje tih karakteristika, dato u \cite{dos,dftmiss}.

\subsubsection{Srednja vrijednost}

\paragraph{Slučaj 1.} Za $k=k_{i}\in\{k_{1},k_{2},\dots ,k_{K}\}$ i uz ${N_A}=\mathrm{card}
({\mathbb{{N}}_A})$, imaćemo:
\begin{align}
X(k_{i})=A_{i}{N_A}+\sum_{n\in{\mathbb{{N}}_A}}\sum_{\substack{p=1\\p\neq i}}^{K}%
A_{p}e^{-j2\pi n(k_{i}-k_{p})/N}.
\end{align}

Drugi član prethodne sume,
$
\chi=\sum_{n\in{\mathbb{{N}}_A}}\sum_{\substack{p=1,p\neq i}}^{K}A_{p}e^{-j2\pi
	n(k_{i}-k_{p})/N}\label{noise_EPS}%
$,
za skup slučajnih pozicija ${\mathbb{{N}}_A}=\{n_{1},n_{2},\dots ,n_{{N_A}}\}$, može biti interpretiran kao slučajna varijabla. Njena srednja vrijednost, po različitim realizacijama signala sa slučajno pozicioniranim dostupnim odbircima, jednaka je nuli, odnosno, važi
$E\{\chi\}=0$. Stoga je srednja vrijednost koeficijenta $X(k_{i})$:
\begin{align}
{E}\{X(k_{i})\}=A_{i}{N_A}.
\end{align}
\paragraph{Slučaj 2.} Za $k\notin\{k_{1},k_{2},\dots ,k_{K}\}$ razmatrana slučajna varijabla je:
\begin{align}
X(k)=\sum_{n\in{\mathbb{{N}}_A}}\sum_{p=1}^{K}A_{p}e^{-j2\pi n(k-k_{p})/N}.
\end{align}
Njena srednja vrijednost jednaka je nuli,
$
{E}\{X(k)\}=0.
$
Dakle, za DFT koeficijente inicijalne estimacije, date izrazom (\ref{{N_A}S_SU{N_A}{N_A}{N_A}}), srednja vrijednost je definisana na sljedeći način \cite{dftmiss}:
\begin{align}
\mu_{X(k)}={E}\{X(k)\}={N_A}\sum_{p=1}^{K}A_{p}\delta(k-k_{p}),
\end{align}
gdje je $\delta(n)=1$ za $n=0$ i $\delta(n)= 0$ za $n\neq 0$. 

\subsubsection{Varijansa}
Varijansa DFT koeficijenata inicijalne estimacije (\ref{{N_A}S_SU{N_A}{N_A}{N_A}}) definisana je sljedećim izrazom \cite{dftmiss}:
\begin{equation}
\sigma_{X(k)}^{2}=\mathrm{var}\{X(k)\}=\sum_{p=1}^{K}|A_{p}|^{2}{N_A}\frac{N-{N_A}}%
{N-1}\left[  1-\delta(k-k_{p})\right]  \text{.}\label{sign}
\end{equation}
U nastavku će biti izveden ovaj izraz. Za $K=1$ i za $k\neq k_{1}$, varijansa je, po definiciji
\begin{align}
\mathrm{var}\{X(k)\}&={E}\left\{  \sum_{n\in{\mathbb{{N}}_A}}\sum
_{m\in{\mathbb{{N}}_A}}\left\vert A_{1}\right\vert ^{2}e^{-j2\pi m(k-k_{1}%
	)/N}e^{j2\pi n(k-k_{1})/N}\right\}.
\end{align}

Prethodni izraz se dalje može zapisati u obliku:
\begin{align}
\mathrm{var}\{X(k)\}=\sum_{m\in{\mathbb{{N}}_A}}{E}\left\{  \left\vert A_{1}\right\vert ^{2}%
+\sum_{n\in{\mathbb{{N}}_A,}n\neq m}\left\vert A_{1}\right\vert ^{2}e^{-j2\pi
	m(k-k_{1})/N}e^{j2\pi n(k-k_{1})/N}\right\}  \label{var1}
\end{align}
Jasno je da važi
$
{E}\left\{  \sum_{n\in{\mathbb{{N}}_A}}\left\vert A_{1}\right\vert
^{2}\right\}  =\vert A_{1}\vert^2{N_A}.
$
Lako se pokazuje i da je:
\begin{align}
e^{-j2\pi m(k-k_{1})/N}\sum_{n=0}^{N-1}e^{j2\pi n(k-k_{1})/N}=0,
\end{align}
budući da je suma po svim diskretnim indeksima vremena deterministička, dok je $X(k)=0$ za $k\neq
k_{1}$. Očekivanje ovog izraza je:
\begin{equation}
\sum_{n=0}^{N-1}{E}\{e^{-j2\pi m(k-k_{1})/N}e^{j2\pi n(k-k_{1}%
	)/N}\}=0\label{veza1}%
\end{equation}
Pošto su, za slučajno $n$, sve vrijednosti $e^{j2\pi n(k-k_{1})/N}$ jednako distribuirane, za očekivanje po velikom broju realizacija važi \cite{dos,dftmiss}:
\begin{equation}
{E}\{e^{-j2\pi m(k-k_{1})/N}e^{j2\pi n(k-k_{1})/N}\}=
\left\{
\begin{array}
{ll}%
B,& n\neq m,\\
1, & n=m,%
\end{array}
\right.\label{veza2}
\end{equation} 
Iz (\ref{veza1}) i (\ref{veza2}) dalje slijedi
$
(N-1)B+1=0.
$

Članovi $\Xi={E}\left\{  \sum_{n\in{\mathbb{{N}}_A,}n\neq m}\left\vert A_{1}\right\vert
^{2}e^{-j2\pi m(k-k_{1})/N}e^{j2\pi n(k-k_{1})/N}\right\}$ iz (\ref{var1}) sada se mogu izračunati u obliku:
\begin{align*}
\Xi=\left\vert A_{1}\right\vert ^{2}({N_A}-1)B=\left\vert A_{1}\right\vert
^{2}({N_A}-1)\left(  -\frac{1}{N-1}\right)  .
\end{align*}
Varijansa slučajne varijable $X(k)$, za $k\neq k_{1}$, konačno se može zapisati sljedećim izrazom:
\begin{align}
\sigma_{X(k)}^{2}(k)=\mathrm{var}(X(k))=\left\vert A_{1}\right\vert
^{2}{N_A}\frac{N-{N_A}}{N-1}.
\end{align}
Za $k=k_{1}$ lako se dobija da važi: $\sigma_{X(k)}^{2}(k_{1})=0$, pošto se svi članovi
$X(k)$ sabiraju u fazi. 

U slučaju multikomponentnih (višekomponentnih) signala, tj. za $K>1$, varijansa je jednaka sumi varijansi pojedinačnih komponenti na svim frekvencijama $k$:
\begin{equation}
\sigma^2_N=\sigma_{X(k)}^{2}=\sum_{p=1}^{K}|A_{p}|^{2}{N_A}\frac{N-{N_A}}%
{N-1},~k\neq k_i,~i=1,2,\dots,K.
\end{equation}

Na frekvencijama $k_{i}%
\in\{k_{1},k_{2},\dots ,k_{K}\}$,
vrijednosti varijanse su od $\sigma^2_N$ manje za $\left\vert
A_{i}\right\vert ^{2}{N_A}\frac{N-{N_A}}{N-1}$, s obzirom na to da se sve vrijednosti $i$-te komponente sabiraju u fazi, bez slučajnih varijacija na poziciji $p=i$, kao što je diskutovano u \cite{dos}. Stoga, za pozicije $p=i$ varijansa postaje:
\begin{equation}
\sigma_{X(k)}^{2}=\mathrm{var}\{X(k)\}=\sum_{\substack{p=1\\p\neq i}}^{K}|A_{p}|^{2}{N_A}\frac{N-{N_A}}%
{N-1},~k= k_i,~i=1,2,\dots,K.
\end{equation}
Prema centralnoj graničnoj teoremi, za $1\ll {N_A}\ll N$ realni i imaginarni djelovi DFT koeficijenata na pozicijama šuma, tj. za $k\notin
\{k_{1},k_{2},\dots ,k_{K}\}$ mogu biti opisani Gausovom distribucijom,
$\mathcal{N}(0,\sigma_{N}^{2}/2)$ sa srednjom vrijednošću nula i varijansom $\sigma_{N}
^{2}=\sigma_{X(k)}^{2}(k\neq k_i)$ gdje je $i=1,2,\dots, K$. 

Za DFT koeficijent na poziciji
$p$-te komponente signala, tj. za $k_{p}\in\{k_{1},k_{2},\dots ,k_{K}\}$, realni i imaginarni djelovi mogu biti opisani sljedećim Gausovim distribucijama:
$
 \mathcal{N}({N_A}\Re\{A_{p}\},\sigma_{X(p)}^{2}/2)\text{i
}
  \mathcal{N}({N_A}\Im\{A_{p}\},\sigma_{X(p)}^{2}/2),
$
respektivno, gdje su $\sigma_{X(p)}^{2}=\sigma_{N}^{2}-A_{p}^{2}{N_A}\frac
{N-{N_A}}{N-1}$, prema dobijenom izrazu (\ref{sign}).

Dublja interpretacija dobijenih izraza, i njihova generalizacija na slučaj diskretne Hermitske i diskretne kosinusne transformacije, biće prezentovani u drugoj i trećoj glavi ove disertacije. Na ovom mjestu, zadržaćemo se isključivo na primjeni prezentovanih rezultata u definisanju novog postupka za rekonstrukciju rijetkih signala, prezentovanog u Algoritmu  \ref{sira}.

Vjerovatnoća da se $N-K$ koeficijenata koji odgovaraju šumu u DFT domenu, uzrokovanog nedostajućim odbircima, nalazi ispod praga $T$, može se definisati sljedećim izrazom:
\begin{equation}
P(T)=\left(1-\exp\left( -\frac{T^2}{\sigma^2_N}\right)\right)^{N-K}.
\end{equation}

Kako bi odredili pozicije komponenti signala, fiksiraćemo vjerovatnoću na  $P(T)=P$.

\begin{algorithm}[bh]
	\floatname{algorithm}{Algoritam}
	\caption{Jednoiterativna rekonstrukcija signala rijetkih u DFT domenu}
	\label{sira}
	\begin{algorithmic}[1]
		\Input
		\Statex
		\begin{itemize}
			\item Mjerni vektor $\mathbf{y}$
			\item Mjerna matrica $\mathbf{A}$
			\item Vjerovatnoća za detekciju komponenti $P$
		\end{itemize}
		\Statex
		\State $\sigma^2_N \gets\frac{1}{N_A} \sum_{n \in \mathbb{N}_A}|x(n)|^2{N_A}\frac{N-{N_A}}%
		{N-1}$
		\smallskip
		\State $T\gets\sqrt{-\sigma_N^2\log(1-P^{\frac{1}{N}})}$
		\State $\mathbf{X}_0\gets \mathbf{A}^{H}\mathbf{y} $
		\smallskip
		\State $\hat{\mathbf{\Pi}}_K \gets \{ k:|X_0(k)|>T\}$
		\State $\mathbf{A}_K \gets \mathbf{A}(:,\hat{\mathbf{\Pi}}_K)$ \Comment kolone matrice $\mathbf{A}$ odabrane na osnovu $\hat{\mathbf{\Pi}}_K $
		\State $\mathbf{X}_K \gets \operatorname{pinv}(\mathbf{A}_K)\mathbf{y}$
		
		\smallskip
		\State $\displaystyle \mathbf{X} \gets
		\begin{cases} \mathbf{0} & \text{za pozicije koje nijesu u }\hat{\mathbf{\Pi}}_K \\
		\mathbf{X}_K & \text{za pozicije iz }\hat{\mathbf{\Pi}}_K
		\end{cases}$
		
		\Statex
		\Output
		\Statex
		\begin{itemize}
			\item Rekonstruisani koeficijenti $\mathbf{X}$
		\end{itemize}
		
	\end{algorithmic}
\end{algorithm}

Prag se može izračunati izrazom:
\begin{equation}
T=\sqrt{-\sigma_N^2\log(1-P^{\frac{1}{N-K}})}\approx\sqrt{-\sigma_N^2\log(1-P^{\frac{1}{N}})}.
\end{equation}
Na bazi ovog praga, može se definisati jednostavan jednoiterativni postupak za rekonstrukciju rijetkih signala, prezentovan u Algoritmu  \ref{sira}.
Uočimo da važi sljedeća aproksimacija: $\sigma^2_N=\sum_{p=1}^{K}A_{p}^{2}{N_A}\frac{N-{N_A}}%
{N-1}\approx \frac{1}{N_A} \sum_{n \in \mathbb{N}_A}|x(n)|^2{N_A}\frac{N-{N_A}}%
{N-1},~k\neq k_i,~i=1,2,\dots,K.$

\section{Rekonstrukcija zasnovana na $\ell_1$-normi}

Razmatra se $N$-dimenzioni vektor $\mathbf{X}$, sa stepenom rijetkosti  $K$ i ukupno
${N_A}$ mjerenja $\mathbf{y=AX}$, gdje je mjerna matrica $\mathbf{A}$
dimenzija ${N_A}\times N$, uz $K<{N_A}\leq N$.  Vektor
$\mathbf{X}$ se može rekonstruisati na osnovu posmatranog redukovanog skupa mjerenja, datog vektorom $\mathbf{y}$, korišćenjem minimizacije mjere rijetkosti vektora $\mathbf{X}$. Takva mjera je, pored ranije razmatrane $\ell_0$-norme i dobro poznata $\ell_1$-norma.

Rješenje zasnovano na minimizaciji $\ell_{1}$-norme:
\begin{equation}
\min\left\Vert \mathbf{X}\right\Vert _{1}\text{ \ \ \ pod uslovom
	\ \ }\mathbf{y=AX}, \label{minL1}%
\end{equation}
gdje je $\left\Vert \mathbf{X}\right\Vert _{1}=\sum_{k=0}^{N-1}X(k)$, identično je rješenju minimizacije zasnovane na $\ell_{0}$-normi:
\begin{align}
\min\left\Vert \mathbf{X}\right\Vert _{0}\text{ \ \ \ pod uslovom
	\ \ }\mathbf{y=AX},%
\end{align}
ako mjerna matrica $\mathbf{A}$ zadovoljava svojstvo ograničene izometrije za $2K$-rijetki vektor:%
\begin{align}
1-\delta_{2K}\leq\frac{\left\Vert \mathbf{A}_{2K}%
	\mathbf{X}_{2K}\right\Vert _{2}^{2}}{\left\Vert \mathbf{X}_{2K}\right\Vert
	_{2}^{2}}\leq1+\delta_{2K}%
\end{align}
uz $0\leq\delta_{2K}<\sqrt{2}-1$ i za sve podmatrice reda $2K$ mjerne matrice $\mathbf{A}$. Važno je uočiti da je izometrijska konstanta u slučaju $\ell_0$-norme zadovoljavala $0\leq\delta_{2K}<1$.

\subsection{Gradijentni algoritam za rekonstrukciju signala}

Kao reprezentativni primjer rekonstrukcije zasnovane na $\ell_1$-normi, ovdje će biti razmotren gradijenti rekonstrukcioni algoritam \cite{grad1,grad2}, u kojem se nedostajući odbirci tretiraju kao minimizacione varijable. Njihove vrijednosti se variraju sve dok se ne dostigne minimum mjere rijetkosti, odnosno, $\ell_1$-norme vektora transformacionih koeficijenata $\left\Vert \mathbf{X}\right\Vert _{1}$. U cilju pronalaženja rješenja koje odgovara minimumu mjere rijetkosti, koristi se gradijent mjere. Suštinski, procedura odgovara poznatom \textit{metodu najbržeg spuštanja}. Bitno je istaći da je mjera koncentracije  $\left\Vert \mathbf{X}\right\Vert _{1}$ konveksna funkcija u prostoru koeficijenata. Za slučaj parcijalne DFT matrice, konveksnost je dokazana u radu \cite{convexity}.

\begin{algorithm}[!b]
	\floatname{algorithm}{Algoritam}
	\caption{Gradijentni algoritam za rekonstrukciju rijetkih signala}
	\label{gradijentni}
	\begin{algorithmic}[1]
		\Input
		\Statex
		\begin{itemize}
			\item Skup pozicija dostupnih odbiraka $\mathbb{N}_{A}$
			\item Skup pozicija nedostajućih odbiraka $\mathbb{N}_{Q}$
			\item Dostupni odbirci (mjerenja) $y(n)$
			\item Transformaciona matrica $\mathbf{\Phi}$
			\item Korak $\mu$
		\end{itemize}
		\Statex
		\State  $m \gets 0$  
		\State Inicijalizovati vektor $\mathbf{x}^{(0)}$ sa:  
		$$ \displaystyle 
		x^{(0)}(n) \gets
		\left\{
		\begin{array}{ll}
		y(n) & \text{ for } n\in \mathbb{N}_{A} \\
		0 & \text{ for } n\in \mathbb{N}_{Q}
		\end{array}
		\right.
		$$
		\State 
		$\displaystyle \Delta \gets \max_n{|x^{(0)}(n)|}$
		\Repeat 
		\Repeat 
		\State $\mathbf{x}^{(m+1)} \gets \mathbf{x}^{(m)}$
		\For{ $n_i\in \mathbb{N}_{Q}$}
		\State $\mathbf{z}_1 \gets  \mathbf{x}^{(m)}$
		\State $z_1(n_i) \gets z_1(n_i)+\Delta$
		\State $\mathbf{z}_2 \gets  \mathbf{x}^{(m)}$
		\State $z_2(n_i) \gets z_2(n_i)-\Delta$
		
		\State $\mathbf{Z}_1 \gets\mathbf{\Phi} \mathbf{z_1}$
		\State $\mathbf{Z}_2 \gets\mathbf{\Phi} \mathbf{z_2}$
		\State $\displaystyle 
		g(n_i) \gets \left\|\mathbf{Z}_1\right\|_1- \left\|\mathbf{Z}_2\right\|_1$
		\smallskip
		\State $x^{(m+1)}(n_i) \gets x^{(m)}(n_i)- \mu\, g(n_i)$
		\EndFor 
		\State $m \gets m+1$
		\Until{ zadovoljen kriterijum zaustavljanja} 
		\State $\Delta \gets \Delta/3$ 
		\Until{ postignuta zadata tačnost} 
		\State  $\mathbf{x} \gets \mathbf{x}^{(m)}$
		
		\Statex
		\Output
		\Statex
		\begin{itemize}
			\item Vektor rekonstruisanog signala $\mathbf{x}$
		\end{itemize}
	\end{algorithmic}
\end{algorithm}

Algoritam počinje od inicijalnog signala $x^{(0)}(n)$ koji se formira tako da sadrži mjerenja na pozicijama dostupnih odbiraka, i nule na pozicijama nedostupnih odbiraka, odnosno,
\begin{equation}
x^{(0)}(n) =
\left\{
\begin{array}{ll}
y(n), & \text{ za } n\in \mathbb{N}_{A} \\
0, & \text{ za } n\in \mathbb{N}_{Q},
\end{array}
\right.
\end{equation}
gdje $
{\mathbb{{N}}_A=}\{n_{1},n_{2},\dots ,n_{{N_A}}\}\subseteq\mathbb{N}%
=\{0,1,\dots ,N-1\}
$ označava skup pozicija dostupnih odbiraka,  a $\mathbb{{N}}_Q=\mathbb{N}\setminus\mathbb{{N}}_A$ predstavlja skup pozicija distupnih odbiraka.
Zatim se, sve dok ne bude zadovoljen neki predefinisani uslov (kao, na primjer, dok varijacije u uzastopnim iteracijama nijesu dovoljno male) ponavlja sljedeća iterativna procedura:
\vspace{2mm}

\noindent\textbf{Korak 1:} Za svaki nedostajući odbirak na poziciji $n\in\mathbb{N}_Q$ formiraju se po dva signala, $z_{1}(n)$ i $z_{2}(n)$, prema sljedećem pravilu:
\begin{equation}
z_{1}^{(m)}(n)=\left\{
\begin{array}
[c]{ll}%
y^{(m)}(n)+\Delta, & \quad\text{za $n \in \mathbb{N}_{Q}$}\\
y^{(m)}(n), & \quad\text{za $n \in \mathbb{N}_{A}$}%
\end{array}
\right.
\end{equation}
\begin{equation}
z_{2}^{(m)}(n)=\left\{
\begin{array}
[c]{ll}%
y^{(m)}(n)-\Delta, & \quad\text{za $n \in \mathbb{N}_{Q}$}\\
y^{(m)}(n), & \quad\text{za $n \in \mathbb{N}_{A}$}%
\end{array}
\right.
\end{equation}
gdje \textit{m} označava redni broj iteracije. Konstanta $\Delta$ se koristi za određivanje da li vrijednost razmatranog nedostajućeg odbirka treba uvećati ili umanjiti, u odnosu na postojeću vrijednost.

\noindent\textbf{Korak 2:} Estimirati razliku mjera koncentracije
\begin{equation}
g(n)=\left\|\mathbf{Z}_1\right\|_1- \left\|\mathbf{Z}_2\right\|_1,~\text{za } n\in\mathbb{N}_{Q}, 
\label{eq:mjera1}%
\end{equation}
gdje su $\mathbf{Z}_1 =\mathbf{\Phi} \mathbf{z_1}$ i $\mathbf{Z}_2 =\mathbf{\Phi} \mathbf{z_2}$.

\noindent\textbf{Korak 3:} Formirati vektor gradijenta $\mathbf{G}^{(k)}$:
\begin{equation}
G^{(k)}(n)=\left\{
\begin{array}
[c]{ll}%
g(n), & \quad\text{za $n \in \mathbb{N}_{Q}$}\\
0, & \quad\text{za $n\in \mathbb{N}_{A}.$}%
\end{array}
\right.
\end{equation}
Drugim riječima, biće ažurirani samo nedostajući odbirci, dok dostupni odbirci ostaju neizmijenjeni, diktirajući pritom uslove za minimizaciju.

\noindent\textbf{Korak 4:} Korigovati vrijednosti $x(n)$ na sljedeći način:
\begin{equation}
x^{(m+1)}(n)=x^{(m)}(n)-\mu  G^{(m)}(n).
\end{equation}

Ponavljanjem prezentovane procedure, nedostajući odbirci će konvergirati pravim vrijednostima signala, produkujući pritom minimalnu mjeru koncentracije koeficijenata u transformacionom domenu. Budući da se za estimaciju gradijenta koristi konačna razlika mjera koncentracije, kada se algoritam približi optimalnoj tački, gradijent $\ell_1$-norme će biti konstantan, i neće moći da se približi rješenju sa proizvoljnom tačnošću. Stoga će, umjesto približavanja rješenju, doći do oscilacija, što znači da će gradijentni vektor mijenjati pravac u susjednim iteracijama. Taj problem se može riješiti uvođenjem varijabilnog koraka $\Delta$, kao što je prezentovano u \cite{dos,grad1,grad2}. Kada se detektuju oscilacije, vrijednost parametra $\Delta$ se smanjuje, a postepeno se ovim postupkom postiže željena tačnost rezultata. Rezime cijelog postupka dat je u Algoritmu \ref{gradijentni}, a dodatna razmatranja, koja se, između ostalog, tiču kriterijuma zaustavljanja i kriterijuma za mjerenje postignute tačnosti, biće prezentovana na kraju sljedeće glave, u kontekstu diskretne Hermitske transformacije.

\begin{primjer}
	Posmatra se signal $x(n)=4\sin(4\pi n/128)+3\cos(42\pi n/128+\pi/8)+5.7\sin(240\pi n/128)$ dužine $N=128$, od kojih je dostupno samo $N_A=28$ slučajno pozicioniranih odbiraka. Signal i dostupni odbirci su prikazani na slici \ref{primjer00}, prvi i drugi red. Ovaj signal je rijedak u DFT domenu, što ilustruje slika \ref{primjer00}, treći red (lijevo). Nedostajući odbirci dovode do pojave šuma u DFT domenu, slika \ref{primjer00}, treći red (desno). Gradijentni algoritam uspješno rekonstruiše signal, slika \ref{primjer00}, četvrti red. Mjera koncentracije za slučaj signala  $x(n)=\cos(32\pi n/128)+2\cos(24\pi n/128+\pi/8)-4\sin(240\pi n/128)$, dužine $N=128$ sa dva nedostajuća odbirka je prikazana na slici \ref{primjer01}. Slika \ref{primjer01} ilustruje kako varijacije vrijednosti nedostajućih odbiraka utiču na mjeru koncentracije (drugi, treći i četvrti red).

	\begin{figure}[tb]%
		\centering
		\includegraphics[
		]%
		{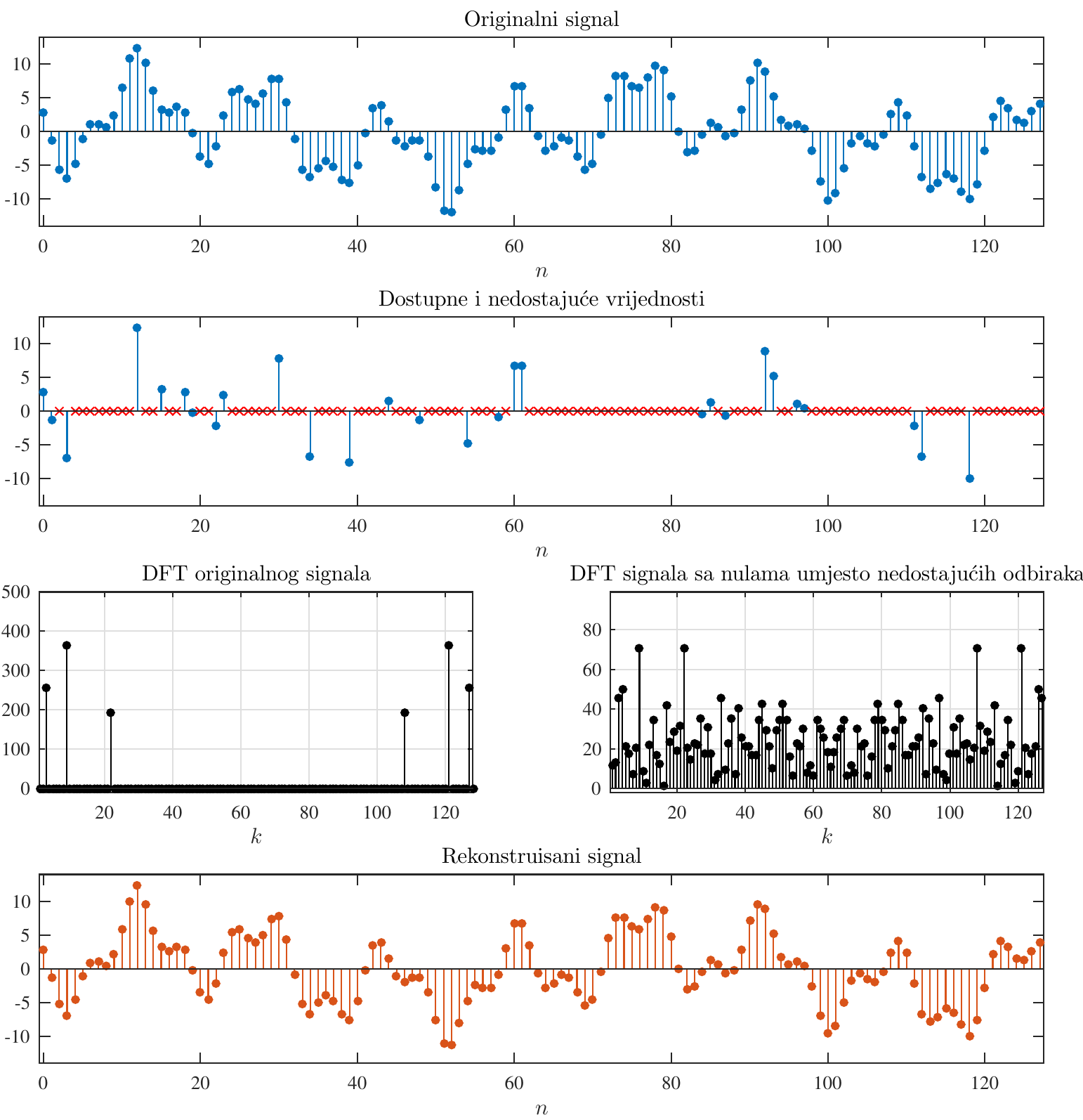}%
		\caption[Primjer rekonstrukcije primjenom gradijentnog algoritmom.]{Primjer rekonstrukcije primjenom gradijentnog algoritma: prvi red -- originalni signal, drugi red -- dostupni odbirci (krstići označavaju nedostupne odbirke), treći red -- DFT originalnog signala (lijevo) i DFT signala sa nulama na pozicijama nedostajućih odbiraka (desno), četvrti red -- rekonstruisani signal.}%
		\label{primjer00}%
	\end{figure}

\begin{figure}[tb]%
	\centering
	\includegraphics[
	]%
	{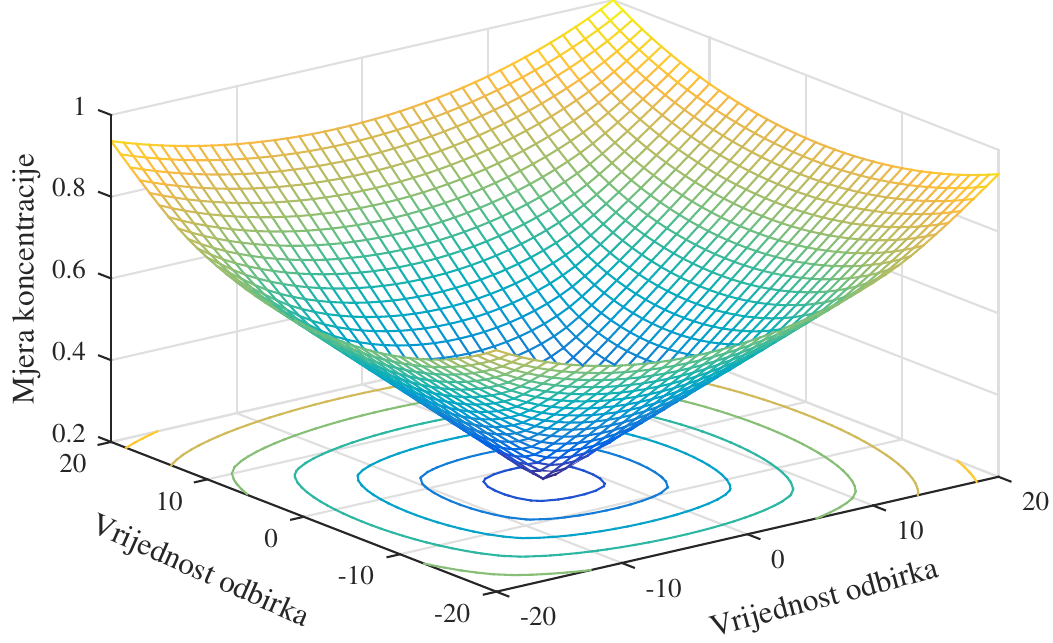}
	
	\vspace{10mm}
	\includegraphics[
	]%
	{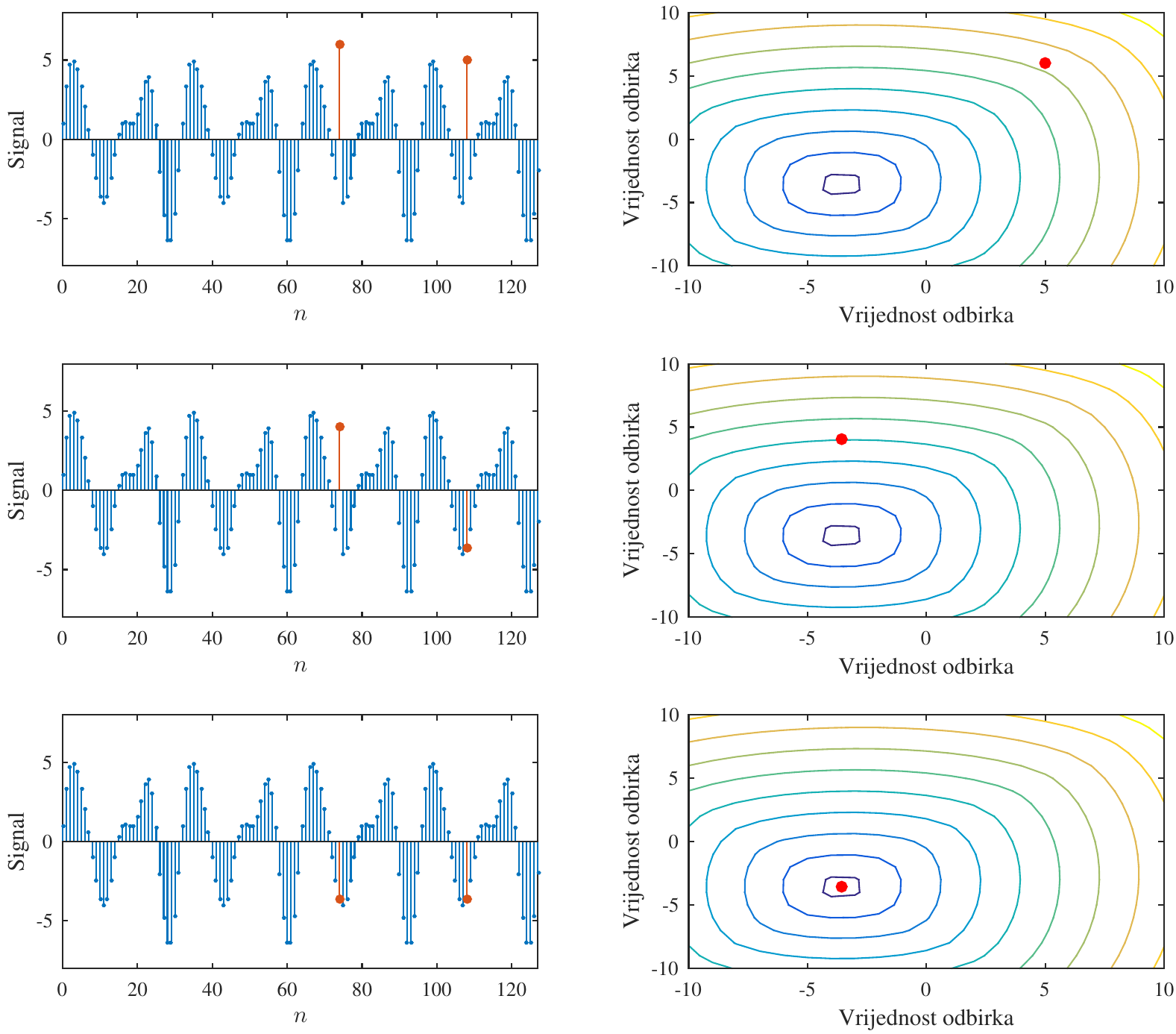}%%konv2d
	\caption[Mjera koncentracije u prostoru koeficijenata.]{Mjera koncentracije u prostoru koeficijenata u slučaju dva nedostajuća odbirka signala: prvi red -- normalizovana mjera u prostoru dva koeficijenta, drugi, treći i četvrti red lijevo -- dostupni odbirci (plavo) i varijacije dva nedostajuća odbirka (crveno), drugi, treći i četvrti red desno - pozicija trenutnog rješenja (crvena tačka) na konturnom grafiku mjere koncentracije. Tačno rješenje (četvrta vrsta) odgovara minimumu mjere koncentracije.}%
	\label{primjer01}%
\end{figure}
\end{primjer}

\chapter{Diskretna Hermitska transformacija kao domen rijetkosti signala}
Koncepti kompresivnog odabiranja i rekonstrukcije rijetkih signala u ovoj glavi su primijenjeni na Hermitsku transformaciju (HT). Ovdje su izučavani i drugi aspekti primjene činjenice da se neki signali mogu visoko koncentrisano reprezentovati u ovom domenu. Glava sadrži više originalnih doprinosa. Biće prezentovan novi algoritam za optimizaciju parametara Hermitske transformacije, i to faktora skaliranja vremenske ose i vremenskog pomjeraja signala, za slučaj jedne forme ove transformacije. Algoritam je zasnovan na optimizaciji mjere koncentracije Hermitskih koeficijenata signala. Nakon toga, biće izvedene statističke karakteristike Hermitskih koeficijenata zašumljenih signala. U ovoj glavi je predstavljena i originalna analiza uticaja nedostajućih odbiraka u signalima koji su rijetki (odnosno, dobro koncentrisani) u Hermitskom domenu. Analiza sadrži interpretaciju indeksa koherentnosti Hermitske mjerne matrice, izvođenje eksplicitnog izraza grešku u rekonstrukciji signala koji nijesu rijetski a rekonstruisani su uz pretpostavku da jesu rijetki, zatim pristupe rješavanju problema rekonstrukcije, uključujući i posebnu interpretaciju gradijentnog algoritma koji je skiciran u prvoj glavi disertacije.

\section{Hermitske funkcije i Hermitska transformacija signala}
 Hermitska transformacija je već decenijama predmet brojnih naučnih istraživanja, posebno kao alternativa Furijeovoj transformaciji \cite{ht1,ht2,ht3,ht3appis,brajovic_eurocon,ht4,ht5,ht6,ht7,ht9,ht10,zoja,zoja2,ht10-noises,ht10-telforj,brajovic_fhss2,brajovic_perspectives,brajovic_uniform_ht,brajovic_2dht,ht11,ht12,ht15,ht16,ht17,ht24,ht25,brajovic_ht1,htn1,hta16,hta17,ht_ecg1,ht_ecg2,ht_ecg4,hta1,hta3,hta4,hta5}. Zbog velikog broja interesantnih osobina, Hermitska transformacija je našla primjenu u mnogim naučnim oblastima, ali je najviše povezivana sa mogućnošću koncizne reprezentacije QRS kompleksa, koji predstavljaju specifične djelove EKG signala,\cite{ht1,ht2,ht3appis,ht10}. Primjene obuhvataju i obradu cijelih EKG signala \cite{htn1,ht_ecg1,ht_ecg2,ht_ecg4}. Mogućnost koncizne, odnosno, visoko koncentrisane reprezentacije, može se iskoristiti u kompresiji, segmentaciji, uklanjanju šuma itd. Druge primjene Hermitske transformacije uključuju \cite{ht4}: molekularnu biologiju \cite{ht15}, digitalnu obradu slike i računarsku tomografiju \cite{ht17}, obradu radarskih signala, biomedicinu \cite{ht24}, fizičku optiku \cite{ht7}, i druge.

\subsection{Kontinualne Hermitske funkcije i Hermitski razvoj}
Hermitski polinom reda $p$ je definisan izrazom \cite{ht21,ht22,ht8,ht1,ht2,ht3}:
\begin{equation}
{{H}_{p}}(t)={{(-1)}^{p}}{{e}^{{{t}^{2}}}}\frac{{{d}^{p}}}{d{{t}^{p}}}e^{-t^2},	 
\end{equation}
i pored Lagerovog, Jakobijevog, i njihovih specijalnih slučajeva - Gegenbauerovog, Čebiševljevog i Ležandrovog polinoma, pripada familiji najčešće korišćenih ortogonalnih polinoma. Prvih deset Hermitskih polinoma dati su sljedećim izrazima:
\begin{align*}
H_0(t)&=1\\
H_1(t)&=2t\\
H_2(t)&=4t^2-2\\
H_3(t)&=8t^3-12t\\
H_4(t)&=16t^4-48t^2+12\\
H_5(t)&=32t^5-160t^3+120t\\
H_6(t)&=64t^6-480t^4+720t^2-120\\
H_7(t)&=128t^7-1344t^5+3360t^3-1680t\\
H_8(t)&=256t^8-3584t^6+13440t^4-13440t^2+1680\\
H_9(t)&=512t^9-9216t^7+48384t^5-80640t^3+30240t.
\end{align*}

Hermitski polinomi zadovoljavaju niz interesantnih svojstava. Među njima, važno je spomenuti rekurzivnu relaciju svojstvenu ortogonalnim polinomimima\cite{ht1,ht22,ht8}:
 \begin{align}
& H_0(t)=1,~ H_1(t)=1,\notag
 \\ 
& H_p(t)=2tH_{p-1}(t)-2(p-1)H_{p-2}(t),~p\geq 2 .
\label{hprec}
\end{align} 
Ovi polinomi su ortogonalni u odnosu na težinsku funkciju (mjeru) $w(t)=e^{-t^2}$:
\begin{equation}
	\int_{-\infty}^{\infty}{H_p(t)H_q(t)w(t)}dt=0, \text{ za } p\neq q,
\end{equation}
odnosno, u opštem slučaju, lako se dokazuje da važi sljedeće svojstvo:
\begin{equation}
\int_{-\infty}^{\infty}{H_p(t)H_q(t)e^{-t^2}}dt=\sqrt{\pi}2^pp!\delta_{pq},\label{hport}
\end{equation}
gdje je $\delta_{pq}$ Kronekerova delta, definisana izrazom:
\begin{equation}
\delta_{pq}=\left\{
\begin{array}
{ll}%
1,&p=q,\\
0,&p\neq q.
\end{array}
\right.
\label{kron}
\end{equation}

Hermitski polinomi formiraju ortogonalnu bazu u Hilbertovom prostoru i ona predstavlja kompletan ortogonalni sistem \cite{ht1}. Naime, iz (\ref{hport}) direktno slijedi ortonormalnost u odnosu na standardni unutrašnji proizvod
\begin{equation}
\langle{{\psi }_{p}}(t,\sigma ),{{\psi }_{q}}(t,\sigma )\rangle=	\int_{-\infty}^{\infty}{{\psi }_{p}}(t,\sigma ){{\psi }_{q}}(t,\sigma )dt=\delta_{pq}
\end{equation}
funkcija ${{\psi }_{p}}(t,\sigma )$ definisanih izrazom:
\begin{align}
{{\psi }_{p}}(t,\sigma )&= \frac{1}{\sqrt{{  \sigma {{2}^{p}}p!\sqrt{\pi }  }}}{{e}^{\frac{-{{t}^{2}}}{2{{\sigma }^{2}}}}}{{H}_{p}}\left(\frac{t}{\sigma}\right ) = \frac{1}{\sqrt{{  \sigma {{2}^{p}}p!\sqrt{\pi }  }}}{{(-1)}^{p}}{{e}^{\frac{{{t}^{2}}}{{{2\sigma }^{2}}}}}\frac{{{d}^{p}}}{d{{t}^{p}}}{{e}^{-\frac{{{t}^{2}}}{{{\sigma }^{2}}}}} .
\end{align} 
Set funkcija $\big\{{{\psi }_{p}}\big\}_{p\geq 0}$, koje su poznate pod nazivom kontinualne Hermitske funkcije, ortonormalan je u Hilbertovom prostoru kontinualnih funkcija definisanih na posmatranom intervalu u skupu $\mathbb{R}$. Stoga se realna funkcija (signal) $x(t)$ iz ovog prostora može predstaviti Hermitskim razvojem \cite{ht1,ht2, ht4,ht16}:
\begin{equation}
x(t)=\sum\limits_{p=0}^{\infty}{{{C(p)}}{{\psi }_{p}}(t,\sigma )},\label{he}
\end{equation}	  
gdje je sa $C(p)$ označen Hermitski koeficijent reda $p$, definisan sljedećom relacijom:
\begin{equation}
C(p)=\langle{x(t),{\psi }_{p}}(t,\sigma )\rangle=\int_{-\infty }^{\infty }{x(t){{\psi }_{p}}(t,\sigma)dt},\,p=0,1,2,\dots\label{coef}
\end{equation}

Na osnovu rekurzivnog svojstva Hermitskih polinoma (\ref{hprec}),  slijedi da Hermitske funkcije zadovoljavaju rekurzivnu relaciju \cite{ht1,multimedia}:
 \begin{align}
& {{\psi }_{0}}(t,\sigma)=\frac{1}{\sqrt{\sigma\sqrt{\pi }}}{{e}^{-\frac{{t}^{2}}{2\sigma}}},~{{\psi }_{1}}(t,\sigma)=\frac{\sqrt{2}t}{\sigma^2\sqrt{\sigma\sqrt{\pi} }}{{e}^{-\frac{{t}^{2}}{2\sigma}}},\notag \\ 
& {{\psi }_{p}}(t,\sigma)=t\sqrt{\frac{2}{p}}\,\,{{\psi }_{p-1}}(t,\sigma)-\sqrt{\frac{p-1}{p}}\,\,{{\psi }_{p-2}}(t,\sigma).\label{hfrec}
\end{align} 

Bitno je istaći da sa rastom $|t|$, sve funkcije $\psi _p(t,\sigma)$ brzo teže nuli -- naime, kako je $H_p\left(\frac{t}{\sigma}\right)$ polinom reda $p$, važi:
\begin{equation}
\lim\limits_{|t|\to\infty}H_p\left(\frac{t}{\sigma}\right){{e}^{-\frac{{t}^{2}}{2\sigma}}}=0.
\end{equation}	

Ovo svojstvo će biti i eksperimentalno potvrđeno prilikom razmatranja diskretne Hermitske transformacije. Parametar $\sigma$ predstavlja skalirajući faktor vremenske ose, koji se koristi za ,,širenje'' i ,,skupljanje'' Hermitskih funkcija. Promjena ovog parametra ne remeti ortogonalnost funkcija \cite{ht1,ht2}. Budući da se u praktičnim aplikacijama uvijek koristi konačan skup od $P$ Hermitskih funkcija, može se smatrati da sve funkcije ${{\psi }_{0}}(t,\sigma),~{{\psi }_{1}}(t,\sigma),\dots {{\psi }_{P-1}}(t,\sigma)$ imaju nenulte (značajne) vrijednosti na konačnom intervalu $t\in[-T_{\sigma},T_{\sigma}]$. Naime, u slučaju mnogih važnih klasa realnih signala \cite{ht1}, aproksimaciju visoke tačnosti je moguće postići sa konačnim brojem funkcija $P$. Bitno je uočiti da $T_{\sigma}$ zavisi od izbora broja funkcija $P$ i faktora skaliranja $\sigma$. Stoga, za konačan broj Hermitskih funkcija važi:
\begin{equation}
	{{\psi }_{p}}(t,\sigma)=0,~t\notin [-T_{\sigma},T_{\sigma}],
\end{equation}
za $0 \leq p \leq P$. U slučaju da je razmatrani signal (funkcija) $x(t)$ takođe konačnog trajanja, odnosno, ako ima nenulte (značajne) vrijednosti samo na intervalu $[-T_{\sigma},T_{\sigma}]$, tada se koeficijenti Hermitskog razvoja mogu računati na sljedeći način \cite{ht1}:
\begin{equation}
{{C}({p})}=\int_{-\infty }^{\infty }{x(t){{\psi }_{p}}(t,\sigma)dt}=\int_{-T_{\sigma} }^{T_{\sigma} }{x(t){{\psi }_{p}}(t,\sigma)dt},\,p=0,1,2,\dots\label{htc}
\end{equation}

 Nameće se zaključak da se izborom faktora skaliranja $\sigma$, bazne funkcije Hermitskog razvoja mogu prilagoditi boljoj reprezentaciji posmatranog signala $x(t)$, posebno u smislu poboljšanja koncentracije signala u Hermitskom domenu i mogućnosti dobijanja njegove rijetke reprezentacije \cite{ht2}.
 \subsection{Diskretna Hermitska transformacija}
 Kao što je već naglašeno, problem diskretizacije Hermitskih funkcija je široko proučavan u literaturi \cite{hta1,hta3,hta4,hta5}. Hermitske funkcije dobijene direktnim uniformnim odabiranjem odgovarajućih kontinualnih funkcija nijesu ortogonalne. Proučavanje u ovoj glavi je ograničeno na dva pristupa definisanju diskretnih Hermitskih funkcija i diskretne Hermitske transformacije, iako se zaključci, izvođenja i algoritmi mogu generalizovati i na druge forme ove transformacije.
 \subsubsection{Diskretna Hermitska transformacija zasnovana na Gaus-Hermitskoj kvadraturi}
 Inverzna diskretna Hermitska transformacija se može posmatrati kao diskretna verzija kontinualnog Hermitskog razvoja (\ref{he}), dok se odgovarajuća diskretizovana forma integrala (\ref{htc}) iz definicije koeficijenata ovog razvoja može smatrati diskretnom Hermitskom transformacijom \cite{ht3,ht5}. U tom slučaju, diskretni signal dužine $N$ može biti reprezentovan kompletnim setom od $N$ diskretnih baznih funkcija. Integral (\ref{htc}) može biti veoma tačno aproksimiran Gaus-Hermitskom kvadraturnom relacijom \cite{ht3,ht5,ht6,ht7,ht9}:
\begin{equation}
	C(p)=\frac{1}{N}\sum\limits_{n=1}^{N}{\frac{{{\psi }_{p}}({{t}_{n}},\sigma)}{{{\left[ {{\psi }_{N-1}}({{t}_{n}},\sigma) \right]}^{2}}}x({{t}_{n}})},
	\label{dht1}
\end{equation}
 gdje su sa ${{t}_{n}},\,\,1\le n\le N$ označene tačke odabiranja koje se poklapaju sa nulama Hermitskog polinoma reda $N$. Funkcije $ {{\psi }_{p}}({{t}_{n}},\sigma),~1\le n\le N,0\le p\le N-1$ koje se dobijaju odabiranjem kontinualnih Hermitskih funkcija u tačkama $t_n$ su ortogonalne, za razliku od funkcija koje bi se dobile direktnim uniformnim odabiranjem. Kada su tačke odabiranja proporcionalne tačkama $t_n$, transformacija (\ref{dht1}) je kompletna reprezentacija. Ovu formu Hermitske transformacije ćemo u daljem izlaganju označavati skraćenicom DHT1. Podrazumijevana vrijednost parametra $\sigma$ je $\sigma=1$. U daljem izlaganju, ukoliko se pretpostavlja ova podrazumijevana vrijednost, parametar $\sigma$ će biti izostavljen iz notacije baznih funkcija, ${{\psi }_{p}}({{t}_{n}})$, bez gubljenja opštosti izlaganja.
 
 Može se uočiti da pravilno računanje izraza (\ref{dht1}) zahtijeva specifičnu formu odabiranja signala $x(t)$. Međutim, uz određene pretpostavke koje su realne za klase signala koje su od interesa za proučavanje u kontekstu ove transformacije, uniformno odabrani signali takođe mogu biti uspješno reprezentovani, što će biti pokazano u narednim sekcijama. Inverzna Hermitska transformacija \cite{multimedia} se direktno dobija iz Hermitskog razvoja (\ref{he}):
 \begin{equation}
 x(t_n)=\sum\limits_{p=0}^{N}{{{C(p)}}{{\psi }_{p}}(t_n,\sigma )}.\label{idht}
 \end{equation}	
 
Direktna i inverzna DHT1, date relacijama (\ref{dht1}) i (\ref{idht}) mogu biti predstavljene u matričnoj formi. Transformaciona matrica DHT1 dimenzija $N\times N$ je definisana na sljedeći način: 
\begin{equation}
{{\mathbf{T}}_{H}}=\frac{1}{M}\left[ \begin{matrix}
 \frac{{{\psi }_{0}}({{t}_{1}},\sigma)}{{{({{\psi }_{N-1}}({{t}_{1}},\sigma))}^{2}}} & \frac{{{\psi }_{0}}({{t}_{2}},\sigma)}{{{({{\psi }_{N-1}}({{t}_{2}},\sigma))}^{2}}} & \ldots  & \frac{{{\psi }_{0}}({{t}_{N}},\sigma)}{{{({{\psi }_{N-1}}({{t}_{N}},\sigma))}^{2}}}  \\
 \frac{{{\psi }_{1}}({{t}_{1}},\sigma)}{{{({{\psi }_{N-1}}({{t}_{1}},\sigma))}^{2}}} & \frac{{{\psi }_{1}}({{t}_{2}},\sigma)}{{{({{\psi }_{N-1}}({{t}_{2}},\sigma))}^{2}}} & \cdots  & \frac{{{\psi }_{1}}({{t}_{N}},\sigma)}{{{({{\psi }_{N-1}}({{t}_{N}},\sigma))}^{2}}}  \\
 \vdots  & \vdots  & \ddots  & \vdots   \\
 \frac{{{\psi }_{N-1}}({{t}_{1}},\sigma)}{{{({{\psi }_{N-1}}({{t}_{1}},\sigma))}^{2}}} & \frac{{{\psi }_{N-1}}({{t}_{2}},\sigma)}{{{({{\psi }_{N-1}}({{t}_{2}},\sigma))}^{2}}} & \cdots  & \frac{{{\psi }_{N-1}}({{t}_{N}},\sigma)}{{{({{\psi }_{N-1}}({{t}_{N}},\sigma))}^{2}}}  \\
 \end{matrix} \right],
 \label{htm1}
 \end{equation}
 dok je odgovarajuća inverzna transformaciona matrica  $\mathbf{T}_{H}^{-1}$ (takođe dimenzija $N\times N$) data izrazom:
 \begin{equation}\mathbf{T}_{H}^{-1}=\left[ \begin{matrix}
 {{\psi }_{0}}({{t}_{1}},\sigma) & {{\psi }_{1}}({{t}_{1}},\sigma) & \ldots  & {{\psi }_{N-1}}({{t}_{1}},\sigma)  \\
 {{\psi }_{0}}({{t}_{2}},\sigma) & {{\psi }_{1}}({{t}_{2}},\sigma) & \cdots  & {{\psi }_{N-1}}({{t}_{2}},\sigma)  \\
 \vdots  & \vdots  & \ddots  & \vdots   \\
 {{\psi }_{0}}({{t}_{N}},\sigma) & {{\psi }_{1}}({{t}_{N}},\sigma) & \cdots  & {{\psi }_{N-1}}({{t}_{N}},\sigma)  \\
 \end{matrix} \right].\label{ihtm1}
 \end{equation}

Ako je sa $\mathbf{C}={{[C(0),\,C(1),\dots,C(N-1)]}^{T}}$ označen vektor Hermitskih koeficijenata dok vektor $\mathbf{x}={{[x({{t}_{1}}),\,x({{t}_{2}}),\dots,x({{t}_{N}})]}^{T}}$ sadrži $N$ odbiraka posmatranog signala, tada zapis:
 \begin{equation}
 \mathbf{C}={{\mathbf{T}}_{H}}\mathbf{x}
 \end{equation}
 predstavlja matričnu formu DHT1. Inverzna DHT1 je u matričnoj formi data sljedećim izrazom:
  \begin{equation}
\mathbf{x}=\mathbf{T}_{H}^{-1}\mathbf{C}.
 \end{equation}
 Matrica inverzne DHT1 se  može zapisati u obliku proizvoda:
\begin{equation}
\mathbf{T}_{H}^{-1}=\mathbf{T}_{H}^{T}\mathbf{D},
\label{l14}
\end{equation}
gdje je $\mathbf{D}$ dijagonalna matrica čija je analitička forma predstavljena u \cite{htn1}, što potvrđuje činjenicu da transformaciona matrica DHT1 nije ortogonalna. Standardna QR dekompozicija transformacione matrice $\mathbf{T}_H$ daje proizvod $\mathbf{T}_H=\mathbf{ QR}$, gdje je $\mathbf{Q}$ ortogonalna matrica, odnosno, $\mathbf{Q}\mathbf{Q}^T=\mathbf{ I}$, pri čemu je $\mathbf{I}$ jedinična matrica, dok je matrica  $\mathbf{R}$ dijagonalna, sa elementima:
\begin{equation}
{{r}_{n}}={{\left( -1 \right)}^{n-1}}{{\left[ \sqrt{N}{{\psi }_{N-1}}\left( {{t}_{n}} \right) \right]}^{-1}},~n=1,2\dots,N.\label{l15}
\end{equation}
 
Može se pokazati da se matrica $\mathbf{Q}$  može zapisati u sljedećem obliku:
 \begin{equation}
 \mathbf{Q}=\frac{1}{\sqrt{M}}\left[ \begin{matrix}
 \frac{{{\psi }_{0}}({{t}_{1}},\sigma)}{{{\psi }_{N-1}}({{t}_{1}},\sigma)} & \frac{{{\psi }_{0}}({{t}_{2}},\sigma)}{{{\psi }_{N-1}}({{t}_{2}},\sigma)} & \ldots  & \frac{{{\psi }_{0}}({{t}_{N}},\sigma)}{{{\psi }_{N-1}}({{t}_{N}},\sigma)}  \\
 \frac{{{\psi }_{1}}({{t}_{1}},\sigma)}{{{\psi }_{N-1}}({{t}_{1}},\sigma)} & \frac{{{\psi }_{1}}({{t}_{2}},\sigma)}{{{\psi }_{N-1}}({{t}_{2}},\sigma)} & \cdots  & \frac{{{\psi }_{1}}({{t}_{N}},\sigma)}{{{\psi }_{N-1}}({{t}_{N}},\sigma)}  \\
 \vdots  & \vdots  & \ddots  & \vdots   \\
 \frac{{{\psi }_{N-1}}({{t}_{1}},\sigma)}{{{\psi }_{N-1}}({{t}_{1}},\sigma)} & \frac{{{\psi }_{N-1}}({{t}_{2}},\sigma)}{{{\psi }_{N-1}}({{t}_{2}},\sigma)} & \cdots  & \frac{{{\psi }_{N-1}}({{t}_{N}},\sigma)}{{{\psi }_{N-1}}({{t}_{N}},\sigma)}  \\
 \end{matrix} \right].
 \end{equation}
 
 Usljed toga što važe (\ref{l14}) i (\ref{l15}), a pri tome $r_n = 1$ ne važi za svako $n = 1, 2,\dots, N,$ može se očekivati da standardni aditivni bijeli Gausov šum utiče na ovu transformaciju drugačije, nego što je to slučaj sa ortogonalnim transformacijama, kao što je, na primjer, DFT.
 \subsubsection{Diskretna Hermitska transformacija zasnovana na simetričnoj tridijagonalnoj matrici koja je povezana sa centriranom Furijeovom matricom}
Prethodna forma diskretne Hermitske transformacije je razmatrana u širokom kontekstu primjena \cite{ht3,ht6,ht9}. Između ostalog, primjenjivana je u kompresiji QRS kompleksa EKG signala \cite{ht1,ht2,ht10}, kao i rekonstrukciji kompresivno odabranih UWB signala \cite{brajovic_ht1}. Uspješnost u ovim primjenama je direktno vezana za veliku sličnost talasnih oblika baznih funkcija transformacije sa razmatranim signalima. Međutim, iako su diskretne Hermitske funkcije dobijene odabiranjem u tačkama koje su direktno proporcionalne nulama Hermitskog polinoma reda $N$ ortogonalne, sama transformaciona matrica (\ref{htm1}) nije ortogonalna. Budući da ni uniformno odabiranje kontinualnih Hermitskih funkcija ne vodi do kompatibilne diskretne baze, alternativni oblici definisanja diskretnih Hermitskih funkcija su široko proučavani u literaturi, na primjer, \cite{hta1,hta16,hta17,hta3,hta4,hta5}. 

Između ostalog, pokazano je da se diskretne Hermitske funkcije mogu dobiti kao sopstveni vektori centrirane ili pomjerene Furijeove matrice \cite{htn1}. U cilju konzistentnosti izlaganja, alternativne diskretne Hermitske funkcije, dobijene ovim pristupom, biće označene sa ${{\tilde{\psi }}_{p}}(n,\sigma ),$ za $\sigma \ge 1$, odnosno sa ${{\tilde{\psi }}_{p}}(n)$, u slučaju faktora skaliranja  $\sigma =1$. Potrebno je naglasiti da je sa $n$ označen diskretni vremenski indeks dobijen uniformnim odabiranjem u skladu sa teoremom o odabiranju, odnosno, indeks se odnosi na \textit{uniformni vremenski grid}. Postoje i drugi alternativni pristupi definisanju diskretnih Hermitskih funkcija.

Neka se posmatra diskretni signal $x(n)$ od $N$ odbiraka, koji mogu biti dobijeni odabiranjem analognog signala $x(t)$ u skladu sa teoremom o odabiranju, gdje je $0\le n\le N-1.$  Za takav signal postoji set baznih funkcija ${{\tilde{\psi }}_{p}}(n,\sigma ),~p=0,1,\dots,N-1,$ koje su suštinski vezane za kontinualne Hermitske funkcije ${{\psi }_{p}}(t,\sigma)$. Nedavno je pokazano   da se navedene funkcije mogu dobiti kao sopstveni vektori centrirane Furijeove matrice, u oznaci $\mathbf{F}_C$, gdje je zadovoljeno \cite{htn1}:
\begin{equation}
{{\mathbf{F}}_{C}}{{\tilde{\psi }}_{p}}(n,\sigma )={{j}^{n}}{{\tilde{\psi }}_{p}}(n,\sigma ),
\end{equation}
gdje je $j=\sqrt{-1}$. Ovaj set funkcija se dobija dekompozicijom na sopstvene vrijednosti posmatrane Furijeove matrice:
\begin{equation}
 {{\mathbf{F}}_{C}}=\mathbf{Q\Lambda }{{\mathbf{Q}}^{T}} 	\label{dek1}
\end{equation}
kao kolone matrice $\mathbf{Q}=\left[  
 {{{\boldsymbol{\tilde{\psi }}}}_{0}}, ~ {{{\boldsymbol{\tilde{\psi }}}}_{1}}, ~ \ldots , ~{{{\boldsymbol{\tilde{\psi }}}}_{N}}  \\
  \right]$. Vektori kolone ${{\boldsymbol{\tilde{\psi }}}_{p}},~p=0,1,\dots,N-1$ sadrže vrijednosti diskretnih Hermitskih funkcija ${{\tilde{\psi }}_{p}}(n,\sigma ),~n=0,1,\dots,N-1$ i one su sopstveni vektori matrice $\mathbf{F}_C$, dok je $\mathbf{\Lambda }$ dijagonalna matrica njenih sopstvenih vrijednosti. U literaturi je pokazano da se ovaj oblik diskretnih Hermitskih funkcija može numerički efikasno generisati kao set sopstvenih vektora simetrične tridijagonalne matrice, definisane na sljedeći način \cite{htn1}:
 \begin{equation}
 \mathbf{T}=\left[ \begin{matrix}
 {{\varphi }_{0}}(0) & {{\varphi }_{1}}(1) & 0 & \cdots  & 0  \\
 {{\varphi }_{1}}(1) & {{\varphi }_{0}}(1) & {{\varphi }_{1}}(2) & \cdots  & 0  \\
 0 & {{\varphi }_{1}}(2) & {{\varphi }_{0}}(2) & \ddots  & 0  \\
 \vdots  & \vdots  & \ddots  & \ddots  & {{\varphi }_{1}}(N-1)  \\
 0 & 0 & 0 & {{\varphi }_{1}}(N-1) & {{\varphi }_{0}}(N-1)  \\
 \end{matrix} \right], \label{tridiag}
 \end{equation}
 gdje su funkcije  ${{\varphi }_{0}}(n) $ i ${{\varphi }_{1}}(n)$ date izrazima:
 \begin{align}
   {{\varphi }_{0}}(n)&=-2\cos \big( \tfrac{\pi }{{{\sigma }^{2}}} \big)\sin \big( \tfrac{\pi n}{N{{\sigma }^{2}}} \big)\sin \big( \tfrac{\pi }{M{{\sigma }^{2}}}(N-1-n) \big),\\
    {{\varphi }_{1}}(n)&=\sin \left( \tfrac{\pi n}{N{{\sigma }^{2}}} \right)\sin \left( \tfrac{\pi }{M{{\sigma }^{2}}}(N-n) \right),
 \end{align}
za $~0\le n\le N-1$. Dekompozicija na sopstvene vrijednosti,
$
 \mathbf{T}=\mathbf{Q\Lambda }{{\mathbf{Q}}^{T}},
$
  dovodi do iste matrice sopstvenih vektora kao (\ref{dek1}), ali na numerički efikasniji način. Funkcije $\{{{\tilde{\psi }}_{p}}(n,\sigma )\}_{0 \leq p \leq N-1}$ formiraju $N$-dimenzionu ortogonalnu bazu alternativne diskretne Hermitske transformacije, koja će u nastavku izlaganja biti označena sa DHT2. Dobijene na ovaj način, diskretne Hermitske funkcije imaju vremenski oblik koji je veoma sličan odgovarajućim kontinualnim funkcijama. Slično kao u kontinualnom slučaju, ove funkcije imaju nenulte vrijednosti u ograničenom vremenskom intervalu, zatim, parne su ili neparne u zavisnosti od vrijednosti indeksa $p$ (koji ujedno predstavlja broj presjeka funkcije sa apscisnom osom). 
  
\begin{figure}[tb]%
  	\centering
  	\includegraphics[
  	]%
  	{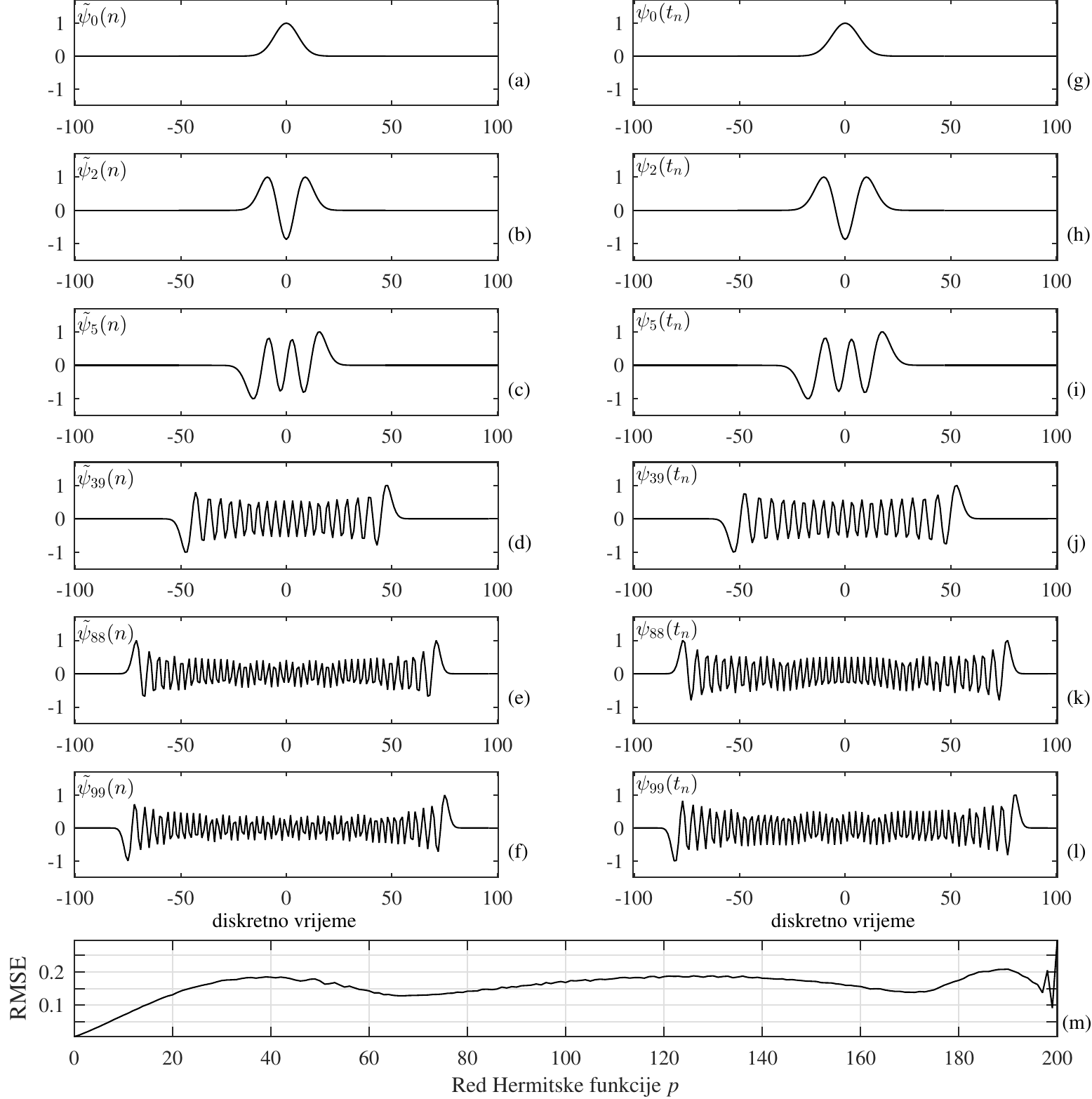}%
  	\caption[Primjeri baznih funkcija dviju formi diskretne Hermitske transformacije]{Primjeri baznih funkcija dviju formi diskretne Hermitske transformacije: (a) -- (f) DHT2 i (g) -- (l) DHT1 za $N = 201$; (m) RMSE između funkcija}%
  	\label{fight1}%
  \end{figure}

  Razlika između kontinualnih i na ovaj način definisanih odgovarajućih diskretnih Hermitskih funkcija raste sa proporcionalno sa rastom indeksa $p$. Navedena činjenica je ilustrovana na slici \ref{fight1}, gdje je prikazan određeni broj Hermitskih baznih funkcija koje odgovaraju DHT2 (lijeva kolona), odnosno DHT1 (desna kolona) - koje su dobijene odabiranjem kontinualnih Hermitskih funkcija u tačkama koje se poklapaju sa nulama Hermitskog polinoma reda $N=201$. U cilju boljeg poređenja, funkcije su u oba razmatrana slučaja normalizovane odgovarajućim maksimalnim amplitudama. Za funkcije reda $p=0,~p=2~\text{i}~p=5$ evidentan je visok stepen sličnosti između ${{\tilde{\psi }}_{p}}(n)$ and ${{\psi }_{p}}(t_n)$, slika \ref{fight1}, prva, druga i treća vrsta, gdje je korišćeno $\sigma=1$. Za $p=39$, razlika između njih postaje očiglednija (četvrta vrsta na istoj slici) i ona raste sa povećanjem reda $p$, što potvrđuje izgled funkcija iz pete i šeste vrste na slici \ref{fight1}, prikazanih za  $p=88$ i $p=99$. 
    
  Slika \ref{fight1} naglašava relevantnost forme DHT1 u primjenama koje zahtijevaju visok nivo sličnosti diskretnih Hermitskih funkcija sa odgovarajućim kontinualnim parovima. Korijen srednje kvadratne greške između svih baznih funkcija DHT1 i DHT2, računat po formuli
  \begin{equation}
  	RMSE(p)=\sqrt{\frac{1}{N}\sum_{n=0}^{N-1}\left|{{\psi }_{p}}(t_{n+1})-\tilde{\psi }_{p}(n)\right|^2}
  \end{equation}
  je prikazan na slici \ref{fight1} (m). Bitno je naglasiti da se ova greška može smanjiti stavljanjem adekvatnog faktora skaliranja $\sigma$ za jedan od setova baznih funkcija. Bitno je naglasiti da je računanje DHT2 numerički efikasnije od odgovarajućeg računanja DHT1, za istu dužinu signala, odnosno isti broj baznih funkcija $N$. Tako na primjer, za $N=201$, prosječno vrijeme potrebno za generisanje baznih funkcija DHT2 iznosilo je 0.0025 sekundi, u poređenju sa 0.0060 sekundi, koliko je bilo potrebno za generisanje baznih funkcija DHT1 korišćenjem rekurzivne relacije (\ref{hfrec}). Testiranje je izvršeno na istom računaru, sa Intel(R) Core(TM) i7-6700HQ CPU @ 2.60 GHz i 8 GB RAM-a. Za $N = 500$, vrijeme potrebno za generisanje baznih funkcija DHT2 iznosilo je 0.012 sekundi, dok je u slučaju baznih funkcija DHT1 bilo potrebno 0.016 sekundi. Međutim, važno je napomenuti i da je DHT1 dodatno numerički složenija i zbog potrebe računanja Gaus-Hermitske kvadrature u (\ref{dht1}), kao i zbog procesa reodabiranja, ukoliko ga je potrebno primijeniti u slučaju uniformno odabranog polaznog signala. Bolje poklapanje baznih funkcija DHT2 može biti postignuto generisanjem većeg broja baznih funkcija (dodavanjem nula u polaznom diskretnom signalu), kao i adekvatnim podešavanjem faktora skaliranja vremenske ose $\sigma$.

 Transformaciona matrica DHT2 je sljedećeg oblika
 \begin{equation}
 	 {{\mathbf{\tilde{T}}}_{H}}=\left[ \begin{matrix}
 	{{{\tilde{\psi }}}_{0}}(0,\sigma) & {{{\tilde{\psi }}}_{0}}(1,\sigma) & \ldots  & {{{\tilde{\psi }}}_{0}}(N-1,\sigma)  \\
 	{{{\tilde{\psi }}}_{1}}(0,\sigma) & {{{\tilde{\psi }}}_{1}}(1,\sigma) & \cdots  & {{{\tilde{\psi }}}_{1}}(N-1,\sigma)  \\
 	\vdots  & \vdots  & \ddots  & \vdots   \\
 	{{{\tilde{\psi }}}_{N-1}}(0,\sigma) & {{{\tilde{\psi }}}_{N-1}}(1,\sigma) & \cdots  & {{{\tilde{\psi }}}_{N-1}}(N-1,\sigma)  \\
 	\end{matrix} \right] 
 	\label{DHT2m}
 \end{equation}
 
Označimo sa $\mathbf{\tilde{C}}={{[\tilde{C}(0),\tilde{C}(1),\dots,\tilde{C}(N-1)]}^{T}}$ vektor DHT2 koeficijenata, i sa $\mathbf{x}={{[x(0),x(1),\dots,x(N-1)]}^{T}}$ vektor od $N$ odbiraka signala na uniformnom vremenskom gridu. Tada se DHT2 može zapisati u matričnoj formi:
 \begin{equation}
 \mathbf{\tilde{C}}={{\mathbf{\tilde{T}}}_{H}}\mathbf{x}.
 \end{equation} 
Pošto je formirana na osnovu sopstvenih vektora tridijagonalne matrice (\ref{tridiag}) matrica ${{\mathbf{\tilde{T}}}_{H}}$ je ortogonalna, odnosno važi $\mathbf{\tilde{T}}_{H}^{-1}=\mathbf{\tilde{T}}_{H}^{T}$, gdje je $\mathbf{\tilde{T}}_{H}^{{}}\mathbf{\tilde{T}}_{H}^{T}=\mathbf{I}.$ Stoga se inverzna DHT2 u matričnoj formi definiše na sljedeći način:
\begin{equation}
\mathbf{x}=\mathbf{\tilde{T}}_{H}^{T}\mathbf{\tilde{C}}.
\end{equation}
\section{Optimizacija parametara Hermitske transformacije}
Koncentracija i stepen rijetkosti signala u domenu diskretne Hermitske transformacije u velikoj mjeri može zavisiti od izbora faktora skaliranja vremenske ose $\sigma$, \cite{ht1,ht17}. Dodatno, koncentracija signala može zavisiti i od pozicije signala na vremenskoj osi u odnosu na pozicije baznih funkcija, koju ćemo kvantifikovati parametrom pomjeraja po vremenskoj osi \cite{ht17}. U ovoj sekciji će biti razmatrana problematika optimizacije ovih parametara i odgovarajuća primjena u kontekstu kompresije elektrokardiografskih (EKG) signala. Specijalno, biće prezentovan algoritam za automatsko određivanje ovih parametara, zasnovan na $\ell_1$-normi Hermitskih koeficijenata. Rezultati su publikovani u naučnom radu \cite{ht10}.

QRS kompleksi, kao najprepoznatljiviji talasi u EKG signalima, važni su u medicinskoj dijagnostici i liječenju, \cite{ht1,ht2,ht17}. Generalno govoreći, u obradi i kompresiji EKG signala i QRS kompleksa, mnogi autori su primjenjivali različite vrste \textit{wavelet}-a i sličnih transformacija \cite{ecg_compr1,ecg_compr2}. Nedavno je, međutim, pokazano da Hermitska transformacija u tom kontekstu može pružiti značajno bolje performanse, u slučaju kada se primijeni adekvatna optimizacija \cite{ht10}. Ova transformacija je pogodna za predstavljanje QRS kompleksa usljed njihove sličnosti sa talasnim oblicima njenih baznih funkcija \cite{ht1,ht2,ht17}. To znači da navedeni signali mogu biti reprezentovani korišćenjem malog broja Hermitskih koeficijenata sa značajnim vrijednostima. To je poslužilo za razvijanje više algoritama za kompresiju QRS kompleksa \cite{ht1,ht2}. Jedan dio njih je razvijen na osnovu kontinualnih funkcija \cite{ht2}, a drugi na osnovu diskretnih \cite{ht1,ht10}. U radu \cite{ht1} je prezentovan algoritam koji pokazuje bolje performanse u kompresiji, od odgovarajućih ekvivalenata razvijenih na osnovu DFT, DCT i diskretne \textit{wavelet} transformacije. U tom algoritmu korišćena je eksperimentalno dobijena vrijednost faktora skaliranja vremenske ose, koji omogućava ,,skupljanje'' i ,,širenje'' QRS kompleksa, kako bi se u što je moguće većoj mjeri poklopili sa baznim funkcijama Hermitske transformacije.

 Biće pokazano da se inkorporiranjem  predloženog algoritma za optimizaciju parametara Hermitske transformacije u postojeću proceduru za kompresiju QRS kompleksa postižu bolji rezultati u odnosu na aktuelne pristupe u oblasti, što će biti verifikovano obimnijim eksperimentom zasnovanim na opsežnoj bazi realnih signala.

\subsection{Optimizacija faktora skaliranja vremenske ose}
U definiciji Hermitskih funkcija
$
{{\psi }_{p}}(t_n,\sigma )$, faktor skaliranja $\sigma$ vremenske ose direktno utiče na širinu opsega u kojima  funkcije imaju nenulte vrijednosti. Drugim riječima, podešavanjem ovog parametra, bazne funkcije diskretne Hermitske transformacije se mogu ,,skupljati'' i ,,širiti''. Umjesto podešavanja faktora skaliranja, alternativno, njegova vrijednost se može fiksirati na $\sigma=1$ i uvesti ekvivalentni parametar $\lambda$ koji daje mogućnost ,,skupljanja'' i ,,širenja'' razmatranog signala $x(\lambda t_n)$ na vremenskoj osi, \cite{ht1}. 

\subsubsection{DHT1 uniformno odabranih signala}
Uvedimo realnu pretpostavku da je kontinualni signal $x(t)$ takav da se njegove nenulte (značajne) vrijednosti pojavljuju u nekom konačnom vremenskom opsegu širine $2T$, tako da važi $x(t)=0\text{, za }t\notin \left[ -T,~T \right]$. Neka je ovaj signal odabran uniformno sa korakom  $\Delta t$, u skladu sa teoremom o odabiranju (\ref{to}), tako da je dobijen njegov diskretni ekvivalent $x(m)$ koji je konačnog trajanja. Pretpostavimo da je diskretni signal neparne dužine $ N = 2D+1$. Po teoremi o odabiranju, originalni signal može biti rekonstruisan, odnosno reodabran u željenim tačkama $\lambda{t_1},\lambda t_2,\dots, \lambda t_N$  pomoću sljedeće interpolacione formule:	
\begin{equation}x(\lambda {{t}_{n}})\approx \sum\limits_{m=-D}^{D}{x(m\Delta t)}\frac{\sin \left( \pi (\lambda {{t}_{n}}-m\Delta t)/\Delta t \right)}{\pi (\lambda {{t}_{n}}-m\Delta t)/\Delta t},\label{sincint}
\end{equation}
gdje je $n = 1,2,\dots, N, ~m = -D,-D+1,\dots,D-1, D$. Budući da je signal konačnog trajanja, doći će do određene greške u odsječenoj \textit{sinc} interpolaciji, \cite{papulis_dsp,errors_dsp1,errors_dsp3}. Međutim, biće ilustrovano da je ova greška veoma mala u slučaju razmatranih klasa signala. U matričnoj formi, prethodna interpolacija postaje:
\begin{equation}
\mathbf{\hat{s}}\approx {{\mathbf{A}}_{\lambda }}\mathbf{x}_{u}
\end{equation}
gdje je $\mathbf{\hat{s}}$ vektor koji sadrži vrijednosti signala odabranog u željenim tačkama $t=\lambda {{t}_{n}},$ koje su proporcionalne nulama Hermitskog polinoma reda $N$, dok je vektor  sastavljen od odbiraka uniformno odabranog signala $x(m)$ označen sa $\mathbf{x}_{u}=[x(-D),~x(-D+1),\dots,x(D)]^T$. U slučaju signala parne dužine $N = 2D$, u ($\ref{sincint}$) treba koristiti indekse $m= -D,\dots, K - 1$. Zapišimo interpolacionu formulu   (\ref{sincint}) u matričnoj formi:
\begin{equation}
\left[ \begin{matrix}
\hat{s}(\lambda {{t}_{1}})  \\
\hat{s}(\lambda {{t}_{2}})  \\
\vdots   \\
\hat{s}(\lambda {{t}_{N}})  \\
\end{matrix} \right]\approx \left[ \begin{matrix}
{{a}_{11}} & {{a}_{12}} & \cdots  & {{a}_{1N}}  \\
{{a}_{21}} & {{a}_{22}} & \cdots  & {{a}_{2N}}  \\
\vdots  & \vdots  & \ddots  & \vdots   \\
{{a}_{M1}} & {{a}_{M2}} & \cdots  & {{a}_{NN}}  \\
\end{matrix} \right]\left[ \begin{matrix}
x(-D)  \\
x(-D+1)  \\
\vdots   \\
x(D)  \\
\end{matrix} \right]
\end{equation}
uz $N = 2D + 1$ i sa elementima  $a_{ij}$ definisanim sljedećim izrazom:
\begin{equation}
{{a}_{ij}}=\frac{\sin\left[   \pi (\lambda {{t}_{i}}-(j-D-1)\Delta t)/\Delta t\right]}{ \pi (\lambda {{t}_{i}}-(j-D-1)\Delta t)/\Delta t },
\label{htinterp}
\end{equation}
pri čemu je $i,j\in \{1,~2,\dots,M\}$. Kao što je nedano prezentovano, greška odsijecanja \textit{sinc} interpolatora je najveća za indekse vremena koji su bliski ivicama diskretnog grida. Međutim, u slučaju signala od interesa, čiji talasni oblici liče Hermitskim funkcijama ili njihovim linearnim kombinacijama, male vrijednosti na krajevima intervala garantuju i malu grešku usljed odsijecanja interpolacionog jezgra. Ova činjenica će biti i numerički evaluirana. 

Problem interpolacije signala konačnog trajanja je takođe razmatran sa stanovišta \textit{sinc} interpolacije zasnovane na FIR filtrima (engl. \textit{finite impulse response}), odnosno filtrima sa konačnom dužinom impulsnog odziva \cite{ht10,errors_dsp1,errors_dsp3}. Navedena greška se može značano smanjiti množenjem interpolacionog jezgra sa prozorskom funkcijom.

Sada uniformno odabrane signale i odgovarajuću DHT1 možemo povezati sljedećom matričnom relacijom:
\begin{equation}
\mathbf{C}={{\mathbf{W}}_{H}}\mathbf{\hat{s}}\approx {{\mathbf{T}}_{H}}{{\mathbf{A}}_{\lambda }}\mathbf{x}_u.
\label{ht_resamp}
\end{equation}

\subsubsection{Algoritam za optimizaciju faktora skaliranja $\lambda$}

Optimizacija faktora skaliranja $\lambda$ vremenske ose može biti urađena korišćenjem mjere koncentracije vektora Hermitskih koeficijenata $\mathbf{C}$. Cilj je pronaći vrijednost parametra $\lambda$ za koji je vektor koeficijenata $\mathbf{C}$ najviše koncentrisan (odnosno, za koji ima najmanji mogući stepen rijetkosti $K$). Mjere koncentracije, kao što je $\ell_1$-norma vektora transformacionih koeficijenata, mogu biti korišćene u ovu svrhu. Za vektor Hermitskih koeficijenata, $\ell_1$-norma se može izračunati po sljedećoj formuli:
\begin{equation}
 \mathcal{M}={{\left\| \mathbf{C} \right\|}_{1}}=\sum\limits_{p=0}^{N-1}{\left| {{c}_{p}} \right|}. 
\end{equation}

Optimalna vrijednost parametra $\lambda$ se može dobiti rješavanjem optimizacionog problema:
\begin{equation}
\lambda =\underset{\lambda }{\mathop{\arg \min }}\,{{\left\| {{\mathbf{T}}_{H}}{{\mathbf{A}}_{\lambda }}\mathbf{x}_u \right\|}_{1}}. 
\label{lambda_opt}
\end{equation}

U pitanju je problem jednodimenzionog pretraživanja u prostoru mogućih vrijednosti parametra $\lambda$, a cilj je naći onu vrijednost parametra koja minimizuje mjeru koncentracije DHT1 razmatranog signala. Opseg pretrage je moguće odrediti, između ostalog, postavljanjem zahtjeva da se nule Hermitskog polinoma reda $N$ pozicioniraju između odgovarajućih tačaka na uniformnom gridu odabranog signala, kao što je diskutovano u \cite{ht1}. Glavna ideja algoritma optimizacije jeste da se omogući iterativna i automatska pretraga parametra, počevši od neke inicijalne vrijednosti $\lambda^{(0)}$. U svakoj iteraciji $k$, mala vrijednost $\Delta$ je dodata i oduzeta od postojećeg $\lambda^{(k)}$, kako bi se odredilo na koji način ove promjene utiču na mjeru koncentracije. Na osnovu mjera izračunatih u oba slučaja, aproksimira se gradijent mjere, a zatim se $\lambda^{(k)}$ ažurira na način koji je ekvivalentan \textit{metodu najbržeg spuštanja}. Optimizacija je detaljno predstavljena Algoritmom \ref{SigmaOpt}.

\begin{algorithm}[hbt]
	\floatname{algorithm}{Algoritam}
	\caption{Optimizacija faktora skaliranja vremenske ose za DHT1}
	\label{SigmaOpt}
	\begin{algorithmic}[1]
		\Input
		
		\Statex
		\begin{itemize}
			\item 	Vektor signala $\mathbf{x}_u$  dužine $N= 2D + 1$
			\item Korak $\mu$
			\item 	Transformaciona matrica DHT1 $\mathbf{W}_H$, izračunata po formuli (\ref{htm1})
		\end{itemize}	
		\Statex
		\State  ${{\lambda }^{(0)}}\leftarrow N\Delta t/\left[ 2\left( \sqrt{\pi (N-1)/1.7}+1.8 \right) \right]$ \Comment Inicijalna vrijednost faktora skaliranja
		\State $\Delta \leftarrow 2/{{t}_{N}}$ \Comment Parametar koji određuje brzinu promjena
		\State $\varepsilon \leftarrow {{10}^{-10}}$ \Comment Zadata tačnost
		
		\While{$\Delta >\varepsilon $} \Comment Alternativno/dodatno: ukoliko je broj iteracija manji od predefinisanog
		
		\smallskip
		\State  
		$\mathbf{A}_{\lambda }^{+}\leftarrow \left[ \begin{matrix}
		a_{11}^{+} & a_{12}^{+} & \cdots  & a_{1N}^{+}  \\
		a_{12}^{+} & a_{22}^{+} & \cdots  & a_{2N}^{+}  \\
		\vdots  & \vdots  & \ddots  & \vdots   \\
		a_{N1}^{+} & a_{N2}^{+} & \cdots  & a_{NN}^{+}  
		\end{matrix} \right],$
		\Comment $a_{ij}^{+ }=\frac{\sin \big( \frac{\pi \left( (\lambda +\Delta ){{t}_{i}}-(j-D-1)\Delta t \right)}{\Delta t} \big)}{\frac{\pi \left( (\lambda + \Delta ){{t}_{i}}-(j-D-1)\Delta t \right)}{\Delta t}},~i,j\in \{1,\dots,N\}$
		\smallskip
		\State $\mathbf{A}_{\lambda }^{-}\leftarrow \left[ \begin{matrix}
		a_{11}^{-} & a_{12}^{-} & \cdots  & a_{1N}^{-}  \\
		a_{12}^{-} & a_{22}^{-} & \cdots  & a_{2N}^{-}  \\
		\vdots  & \vdots  & \ddots  & \vdots   \\
		a_{N1}^{-} & a_{N2}^{-} & \cdots  & a_{NN}^{-}  \\
		\end{matrix} \right],$
		\Comment $a_{ij}^{- }=\frac{\sin \big( \frac{\pi \left( (\lambda -\Delta ){{t}_{i}}-(j-D-1)\Delta t \right)}{\Delta t} \big)}{\frac{\pi \left( (\lambda - \Delta ){{t}_{i}}-(j-D-1)\Delta t \right)}{\Delta t}},~i,j\in \{1,\dots,N\}$
		\smallskip
		
		\State ${{\mathcal{M}}^{+}}\leftarrow {{\left\| {{\mathbf{C}}^{+}} \right\|}_{1}}=\sum\limits_{p=0}^{N-1}{\left| {{\mathbf{T}}_{H}}\mathbf{A}_{\lambda }^{+}\mathbf{x}_u \right|}$ \Comment Mjera koncentracije u slučaju malog povećanja $\lambda$
		
		\State ${{\mathcal{M}}^{-}}\leftarrow {{\left\| {{\mathbf{C}}^{-}}  \right\|}_{1}}=\sum\limits_{p=0}^{N-1}{\left| {{\mathbf{T}}_{H}}\mathbf{A}_{\lambda }^{-}\mathbf{x}_u \right|}$ \Comment Mjera koncentracije u slučaju malog smanjenja $\lambda$
		\State ${{\nabla }^{(k)}}\leftarrow \left( {{\mathcal{M}}^{+}}-{{\mathcal{M}}^{-}} \right)/N$\Comment Aproksimacija gradijenta
		\State ${{\lambda }^{(k+1)}}\leftarrow {{\lambda }^{(k)}}-\mu {{\nabla }^{(k)}}$ \Comment Ažuriranje faktora skaliranja
		\State $\beta \leftarrow \operatorname{sign}\left( {{\nabla }^{(k)}}{{\nabla }^{(k-1)}} \right)$ 
		\If{$\beta <0$}\Comment U slučaju promjene znaka gradijenta u uzastopnim iteracijama,
		\State $\Delta \leftarrow \Delta /2$ \Comment smanjiti $\Delta$
		
		\EndIf
		
		\EndWhile
		
		\Statex
		\Output
		\Statex
		\begin{itemize}
			\item Optimalni faktor skaliranja ${{\lambda }^{(k)}}$
		\end{itemize}
	\end{algorithmic}
\end{algorithm}

U algoritmu je kao polazna vrijednost faktora skaliranja uzeto ${{\lambda }^{(0)}}=N\Delta t/\left[ 2(\sqrt{\pi (N-1)/1.7}+1.8) \right]$, što je u skladu sa donjom granicom  koja obezbjeđuje konvergenciju algoritma, \cite{ht10}. Vrijednosti $\mu$ i $\Delta$ su odabrane empirijski i korišćene su za dobijanje rezultata prezentovanih u ovoj disertaciji i radu \cite{ht10}. Premala vrijednost koraka $\mu$ će dovesti do previše spore konvergencije, dok veći korak vodi bržoj konvergenciji. Sa druge strane, $\mu$ treba biti dovoljno malo, tako da može garantovati stabilnost algoritma (drugim riječima, vrijednost koraka treba da obezbijedi da je u svakoj iteraciji zadovoljen uslov $\lambda <\left[ \sqrt{\pi N/1.7}+1.8 \right]/\left[ 2\pi W \right]$, gdje je $W$ frekvencijski opseg razmatranog signala). Dakle, izbor koraka $\mu$ je stvar pravljenja kompromisa između navedena dva zahtjeva. Važno je napomenuti da se u predstavljenom algoritmu u praktičnim primjenama stavlja i dodatni uslov koji će ograničiti maksimalan broj iteracija, kako ne bi došlo do stvaranja beskonačnih petlji u slučajevima kada nije moguće dostići zadatu tačnost.

Maksimalni broj iteracija odgovara dužini razmatranog signala (u mnogim numeričkim eksperimentima, konvergencija algoritma je postignita i za polovinu od navedenog broja iteracija). Numerička složenost algoritma može biti aproksimirana na sljedeći način. U jednoj iteraciji je potrebno: a) generisati argumente \textit{sinc} funkcija u koracima 5 i 6, za što je potrebno  $2N^2+2$ sabiranja (odnosno oduzimanja) i  $6N^2$ množenja sa konstantama; b) za interpolaziju je potrebno $N^2$ množenja i $N(N-1)$ sabiranja; c) za proračun dvije DHT1, numerička složenost je $2N^2$ sabiranja i $2N^2$ množenja; d) za proračun mjera koncentracija potrebno je $2N-2$ sabiranja. Stoga, prezentovani algoritam zahtijeva ukupno $5N^2+N$ sabiranja i $9M^2$ množenja. 
Dvodimenziona HT može biti realizovana računanjem odgovarajućih jednodimenzionih transformacija za svaku dimenziju posebno. Stoga je prezentovani algoritam moguće primijeniti i na dvodimenzionu formu ove transformacije.

\begin{primjer}
Razmatra se signal zadat izrazom:
\begin{equation}
x(t)=3\sin (5\pi t)\exp \left( -\frac{{N^{2}t^{2}}}{2\sigma _{0}^{2}} \right)
\label{sigop}
\end{equation}
gdje je $N=77$, ${{\sigma }_{0}}=2.1$, $-\frac{1}{2}<t<\frac{1}{2}$. Signal je odabran sa korakom $\Delta t=\frac{1}{N}$, tako da su dobijene uniformno raspoređene diskretne vrijednosti sa indeksima $m\in \{-\frac{N-1}{2},\dots,~\frac{N-1}{2}\}$. Odbirci originalnog signala su prikazani na slici \ref{Case3sin} (a), dok su DFT koeficijenti i koeficijenti standardne DHT1 sa $\sigma =1$  prikazani na slici \ref{Case3sin} (b) i (c), respektivno. Može se uočiti da signal ima kompaktnu reprezentaciju u vremenskom domenu i talasni oblik koji je sličan Hermitskim baznim funkcijama. Ovakva vrsta signala (uprozorene ili filtrirane sinusoide, QRS segmenti, kratkotrajni signali kao što su npr. FHSS ili UWB signali) su pogodni za reprezentaciju u Hermiskom transformacionom domenu.

  \begin{figure}[!h]%
	\centering
	\includegraphics[
	]%
	{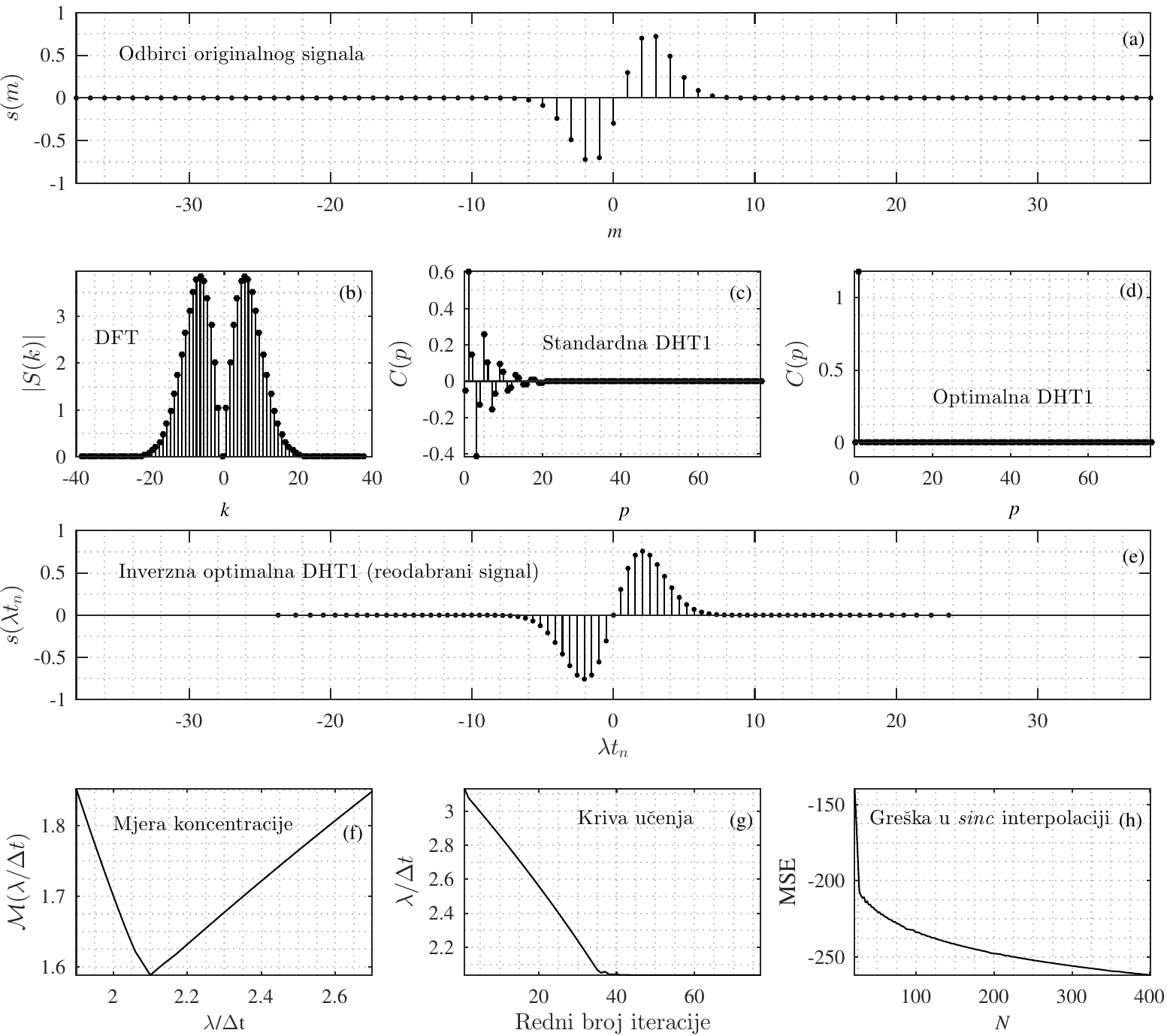}%
	\caption[Optimizacija faktora skaliranja DHT1 na primjeru sinusoide ograničenog trajanja]{Optimizacija faktora skaliranja DHT1 na primjeru sinusoide ograničenog trajanja: (a) originalni uniformno odabrani signal, (b) DFT polaznog signala, (c) standardna DHT1 originalnog signala, (d) DHT1 nakon primjene algoritma optimizacije, (e) odbirci reodabranog signala, (f) mjera koncentracije u funkciji od faktora skaliranja, (g) promjene faktora skaliranja tokom iteracija, (h) srednja kvadratna greška u \textit{sinc} interpolaciji signala (\ref{sigop}) u zavisnosti od zadate dužine signala $N$.}%
	\label{Case3sin}%
\end{figure}

U cilju određivanja DHT1 reodabranog signala sa najboljom koncentracijom, primijenjena je predložena procedura za optimizaciju faktora skaliranja. Dobijeni rezultat je $\lambda ={{\sigma }}\Delta t=\frac{2.0368}{N}$. Signal reodabran u tačkama koje su proporcionalne nulama Hermitskog polinoma $N$-tog reda sa prethodno određenim faktorom skaliranja predstavljen je na slici \ref{Case3sin} (e). Odgovarajuća DHT1 je prikazana na slici \ref{Case3sin} (d), gdje se može uočiti da je dobijeni faktor skaliranja omogućio reprezentaciju istog signala sa samo jednim Hermitskim koeficijentom koji ima značajnu vrijednost, sa indeksom $p=1$, dok su svi ostali koeficijenti ili jednaki nuli ili su sa zanemarljivo malim vrijednostima. 

Kako bi potvrdili da predloženi algoritam zaista nalazi optimalnu vrijednost, izračunata je mjera koncentracije za različite vrijednosti faktora skaliranja $\lambda $: $1/\Delta t\le \lambda /\Delta t\le 2/\Delta t$, koji je variran sa korakom $0.01/\Delta t$. Rezultati su prikazani na slici \ref{Case3sin} (f), gdje se lako uočava globalni mininum za ${{\lambda }_{\min }}=2.0368/\Delta t$. Ovdje je pretpostavljeno da su zadovoljene donja i gornja granica iz \cite{ht10}. Donja granica se može kontrolisati adekvatnom inicijalizacijom algoritma, dok pogodno odabrani korak $\mu$ osigurava da gornja teorijska granica nikad ne bude dostignuta. U intervalu između donje i gornje granice očekuje se da postoji globalni minimum koji korespondira najboljoj mogućoj koncentraciji DHT1. Promjene parametra $\lambda$ tokom iteracija predstavljene su na slici \ref{Case3sin} (g). Sa ove slike evidentna je stabilizacija algoritma sa dostizanjem minimuma mjere koncentracije.  

U ovom primjeru je takođe ispitan i uticaj \textit{sinc} interpolacionog jezgra konačne dužine. Naime, u ovu svrhu je izračunata srednja kvadratna greška (MSE) između interpoliranog signala  ${{x}_{\operatorname{int}}}(\lambda {{t}_{m}})$ (interpolacija je sprovena iz uniformnih odibaraka $x(m)$ korišćenjem izraza (\ref{htinterp})) i originalnog (analitičkog) signala (\ref{sigop}), posmatranog u tačkama $t=\lambda {{t}_{n}}$: 
\begin{equation}
 MSE=\frac{1}{N}\sum\nolimits_{n=1}^{N}{{{\left| {{s}_{\operatorname{int}}}(\lambda {{t}_{n}})-x(\lambda {{t}_{n}}) \right|}^{2}}}.
\end{equation}
Dužina signala varirana je u opsegu od 21 do 401, sa korakom 2. Rezultati su prikazani na slici \ref{Case3sin} (h), ilustrujići činjenicu da je za razmatranu klasu signala konačnog vremenskog trajanja ova greška jako mala.
\end{primjer}

\begin{primjer}
	\label{uwbden}
Jedna vrsta signala čiji su talasni oblici slični baznim funkcijama Hermitske transformacije, pa time imaju potencijal za reprezentaciju sa malim brojem transformacionih koeficijenata, jesu i ultra-širokopojasni ili UWB signali ({engl.} \textit{ultra-wideband}). Jedna klasa karakterističnih talasnih oblika ovih signala poznata je pod nazivom Gausovi \textit{doublet}-i \cite{brajovic_ht1,ht_uwb1,ht_uwb2}. Promjena talasnog oblika (filtriranje) Gausovih impulsa na odgovarajućim prijemnim antenama tipično se modeluje operacijom diferenciranja tih impulsa, koja rezultuje sljedećom formom Gausovih doublet-a:
\begin{equation}
	x(t)=\left[ 1-4\pi {{\left( t/{{\tau }_{m}} \right)}^{2}} \right]{{e}^{-2\pi {{\left( t/{{\tau }_{m}} \right)}^{2}}}}.
\end{equation}
U ovom primjeru se razmatra diskretna verzija ovih signala, odabranih na frekvenciji od 2 GHz, trajanja 100 ns, sa ${{\tau }_{m}}=22.2$ ns. Signal je prikazan na slici \ref{ExUWB} (a), dok je standardna DHT1 sa faktorom skaliranja $\sigma=1$ prikazana na slici \ref{ExUWB} (b). Nakon primjene algoritma za optimizaciju faktora skaliranja, signal reodabran u tačkama $\lambda t_n$, prikazan na slici \ref{ExUWB} (c), moguće je predstaviti sa samo dva koeficijenta sa značajnim vrijednostima, slika \ref{ExUWB} (d). Potencijalna primjena ovog pristupa leži u mogućnosti dizajna tehnologije prijemnika UWB signala sa pojednostavljenom procedurom za njihovu detekciju.
 
  \begin{figure}[ptb]%
	\centering
	\includegraphics[
	]%
	{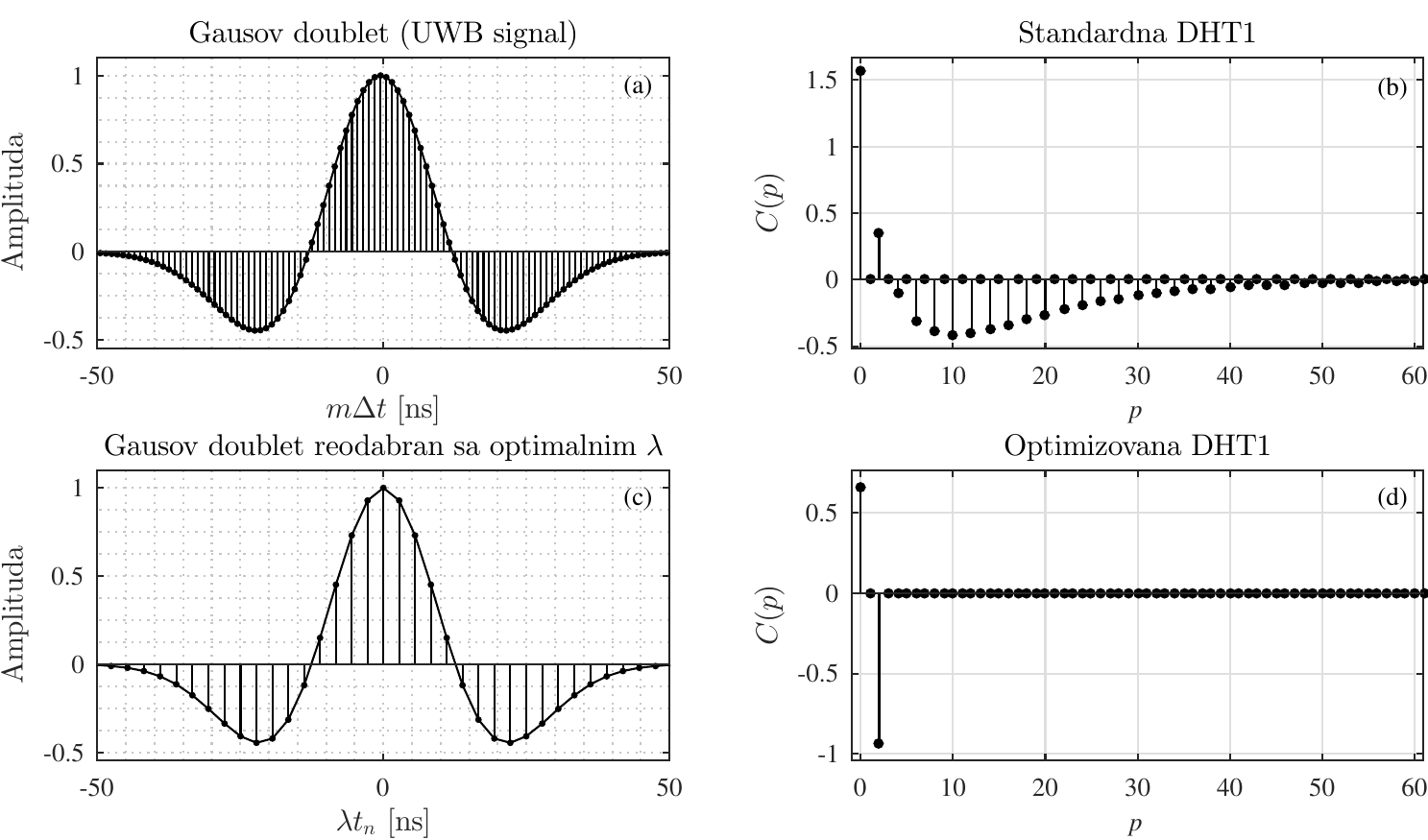}%
	\caption[Optimizacija faktora skaliranja u slučaju UWB signala.]{Optimizacija faktora skaliranja u slučaju UWB signala: (a) signal u vremenskom domenu, (b) DHT1 sa faktorom skaliranja vremenske ose $\sigma=1$, (c) isti signal reodabran u tačkama  $\lambda t_n$, sa optimalnim $\lambda$ i (d) DHT1 reodabranog signala.}%
	\label{ExUWB}%
\end{figure}
\end{primjer}

\subsection{Optimizacija parametra pomjeraja po vremenskoj osi}
\label{pomjeranje}
Bazne funkcije diskretnih Hermitskih transformacija moguće je pomjerati po vremenskoj osi. Umjesto centriranja signala oko koordinatnog početka na vremenskoj osi, vršiće se pomjeranje centralne tačke signala na lijevo i na desno od koordinatnog početka, prije računanja Hermitskih koeficijenata. Navedeni pristup će biti korišćen u kombinaciji sa optimizacijom faktora skaliranja, koji je razmatran u prethodnoj podsekciji. Drugim riječima, umjesto $x(m\Delta t)$, u analizi, odnosno obradi,  koristiće se signal: ${{s}_{l}}(m\Delta t)=x((m\pm l)\Delta t)$, sa $l=\left[ -{{l}_{\max }},{{l}_{\max }} \right]$.
Za svaku diskretnu vrijednost $l$, rješavanjem problema (\ref{lambda_opt}), nalazi se optimalna vrijednost $\lambda$. Za svaku razmatranu vrijednost $l$ se formira vektor mjere $\mathbf{L}$, koji sadrži minimalnu vrijednost mjere (\ref{lambda_opt}), koja se dobija za optimalno $\lambda$. Optimalna vrijednost parametra pomjeraja $l$ se dobija  rješavanjem problema:
\begin{equation}
l=\underset{l}{\mathop{\arg \min }}\,\mathbf{L}.
\label{lopt}
\end{equation}

Potrebno je naglasiti da je granična vrijednost $l_{max}$ u praktičnim aplikacijama relativno mala, npr. $l_{max} = 3$ u slučaju kompresije djelova EKG signala, koja će biti razmatrana u ovoj tezi. Zato je prilikom rješavanja optimizacionog problema (\ref{lopt}) moguće koristiti direktno pretraživanje. Takođe, u razmatranoj primjeni, biće korišćene samo cjelobrojne vrijednosti ovog parametra. 

\subsection{Primjena u kompresiji QRS kompleksa}
\begin{figure}[ptb]%
	\centering
	\includegraphics[
	]%
	{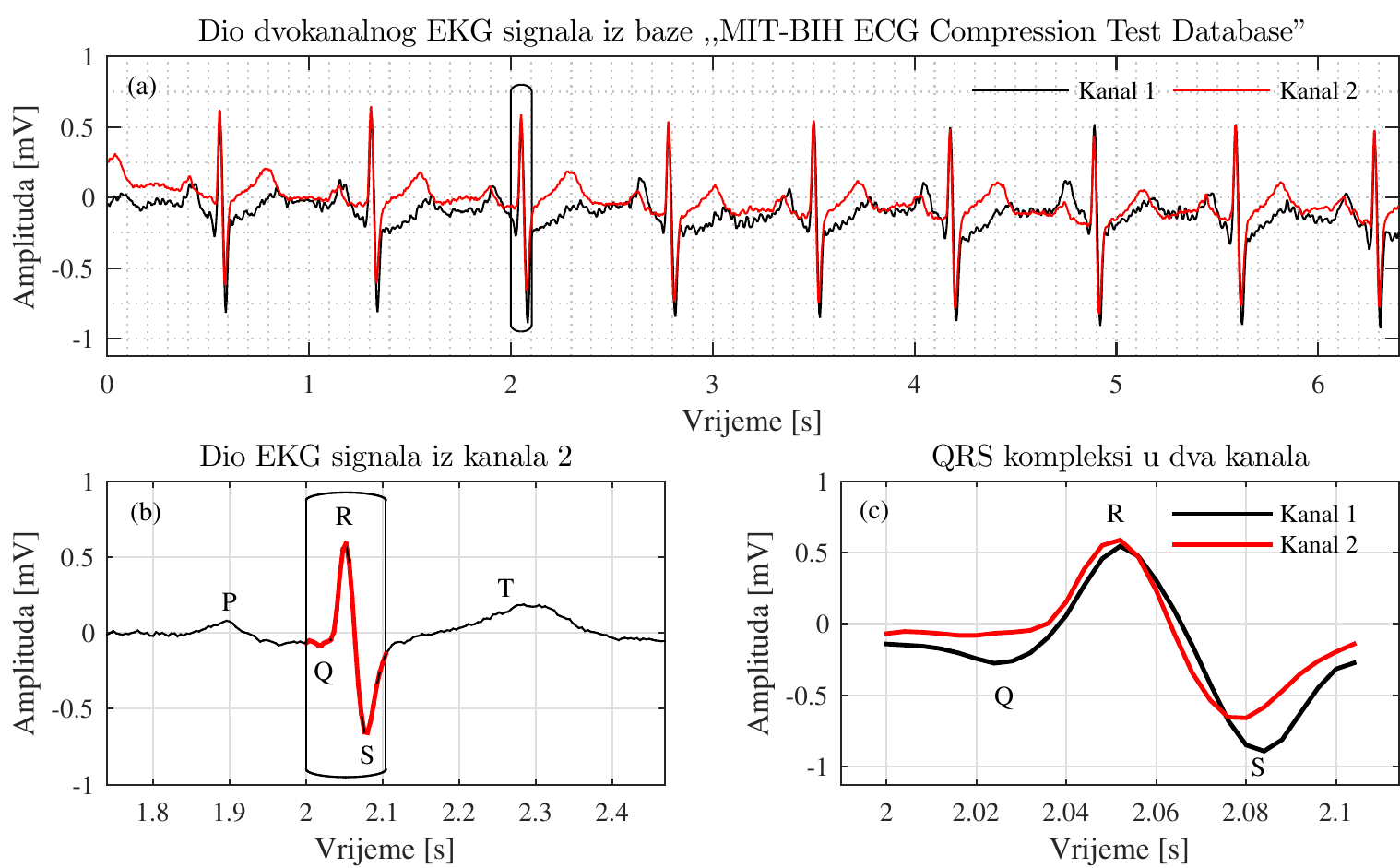}%
	\caption[Dio dvokanalnog EKG signala i QRS kompleksi.]{Dio dvokanalnog EKG signala i QRS kompleksi: (a) dio dvokanalnog EKG signala sa označenim parom QRS kompleksa, (b) okolina jednog QRS kompleksa iz drugog kanala, (c) selektovani QRS kompleksi iz oba kanala.}%
	\label{ECG_example}%
\end{figure}

Elektrokardiografski signali (EKG) su značajna grupa biomedicinskih signala, koji su od velike važnosti u medicinskoj dijagnostici i liječenju. Uklanjanje šuma, detekcija karakterističnih djelova, segmentacija, filtriranje, karakterizacija, klasifikacija, kompresija kao i drugi oblici obrade ovih signala su velikim intenzitetom razvijani i proučavani u proteklim decenijama \cite{ht1,ht2,ht17,ht_ecg1,ht_ecg2,ht_ecg4,ecg_compr1,ecg_compr2,ecgden1,ecgden2,ecgden3,htn3}. EKG signali se tipično snimaju kao višekanalni, pomoću odgovarajućih elektroda pozicioniranih na karakterističnim djelovima tijela pacijenata i oni reprezentuju elektrofiziološke obrasce rada srčanog mišića koji se manifestuju u vidu sitnih bioelektričnih promjena na koži. Primjer dvokanalnog EKG signala signala predstavljen je na slici \ref{ECG_example} (a). Signal je preuzet iz poznate baze biomedicinskih signala ,,MIT-BIH ECG Compression Test Database'', snimljenih u okviru djelatnosti laboratorije ,,Laboratory for Computational Physiology'' pri programu ,,Harvard-MIT Health Sciences and Technology'', kao dio šire baze koja je u cjelosti dostupna \textit{online}, \cite{ecg_baza}.

Mnogi segmenti EKG signala igraju odgovarajuće važne uloge u dijagonostičkim procedurama. Ovi signali su, kao što je već istaknuto, decenijama dovođeni u vezu sa Hermitskom transformacijom, najčešće u kontekstu uklanjanja šuma, određivanja karakteristika signala i kompresije. Jedan od segmenata EKG signala koji je važan u medicinskoj dijagnostici jeste i QRS kompleks \cite{ht1}. Sastavljen od tri talasa, u oznaci Q, R i S, vizuelno je vjerovatno najprepoznatljiviji dio EKG signala. Po jedan od QRS kompleksa iz oba kanala posmatranog signala je uokviren na slici \ref{ECG_example} (a). Njihov uvećan prikaz može se vidjeti na slici \ref{ECG_example} (c), sa označenim karakterističnim djelovima. Na slici \ref{ECG_example} (b) prikazana je i neposredna okolina selektovanog QRS kompleksa u okviru posmatranog EKG signala iz drugog kanala, gdje se prvenstveno mogu uočiti tzv. P i T talasi. 

Imajući u vidu  veliku dijagnostičku važnost QRS kompleksa, ali i činjenicu da su već razvijeni brojni efikasni algoritmi za njihovu detekciju, godinama su proučavane i odgovarajuće kompresione procedure \cite{ht1,ht2}. Značaj ovakvih algoritama ogleda se, između ostalog, u činjenici da u komprimovanoj formi može biti sačuvan veći broj QRS kompleksa u sklopu odgovarajućih baza signala pacijenata, koji se dalje mogu koristiti u praćenju zdravstvenog stanja, istorije bolesti pacijenta itd. Značaj se ogleda i u mogućnosti njihovog efikasnog transfera između medicinskih institucija, u slučaju potrebe za dodatnim ekspertizama. Dodatno, čuvanje velikog broja komprimovanih QRS kompleksa otvara mogućnost i za njihovo proučavanje ne samo u kontekstu medicine, već i u kontekstu obrade signala i multidisciplinarne oblasti biomedicine, u cilju razvoja novih automatizovanih dijagnostičkih procedura primjenom vještačke inteligencije, odnosno, algoritama za klasifikaciju, detekciju itd.

U okviru razmatranog kompresionog problema, ključno je reprezentovati QRS komplekse sa što je moguće manjim brojem transformacionih koeficijenata i pri tome praviti grešku koja je medicinski prihvatljiva. Cilj je iskoristiti predstavljene algoritme za optimizaciju parametara Hermitske transformacije u cilju postizanja najbolje moguće koncentracije QRS kompleksa u domenu DHT1,  a zatim ove optimizovane reprezentacije inkorporirati u  kompresionu proceduru koja je predstavljena u \cite{ht1} i \cite{ht2}. 

Kompresioni algoritam iz  \cite{ht1} i \cite{ht2} funkcioniše na sljedeći način. Pretpostavlja se da je razmatrani QRS kompleks $x(t)$ odabran u tačkama $\lambda t_1,\lambda t_2,\dots, \lambda t_N$, tako da se dobija vektor  $\mathbf{\hat{x}}$, i za njega se računaju Hermitski koeficijenti $\mathbf{C}$ korišćenjem obrasca (\ref{ht_resamp}). Nakon toga, na osnovu $L$ koeficijenata vektora $\mathbf{C}$ sa najvećim vrijednostima, formira se vektor $\mathbf{\tilde{C}}$, u kojem su svi preostali koeficijenti, kojih ima $N-L$,
postavljeni na nulu. Aproksimacija razmatranog signala se može izračunati inverznom formulom
\begin{equation}
\mathbf{\tilde{x}}=\mathbf{T}_{H}^{-1}\mathbf{\tilde{C}},
\end{equation}
gdje je $\mathbf{T}_{H}^{-1}$ inverzna transformaciona matrica (\ref{ihtm1}).
Algoritam prezentovan u \cite{ht2} može biti unaprijeđen korišćenjem optimizacije parametara Hermitske transformacije (DHT1). Naime, u ovom algoritmu, kontinualni signal $x(t)$ je odabran u tačkama $\lambda t_1, \lambda t_2,\dots,  \lambda t_M$, gdje je faktor skaliranja $\lambda$ odabran tako da se dobije najmanji mogući broj koeficijenata u vektoru $\mathbf{\tilde{C}}$, pod uslovom da je greška u rekonstrukciji 
\begin{equation}
E=\frac{{{\left\| \mathbf{\tilde{x}}-\mathbf{x} \right\|}_{2}}}{{{\left\| \mathbf{x} \right\|}_{2}}}
\label{qrsmse}
\end{equation}
manja od 10\%, što je medicinski prihvatljivo \cite{ht1}. U ovakvom pristupu postoji nekoliko problema. Da bi se odredila optimalna vrijednost faktora skaliranja $\lambda$, polazeći od kontinualnih ECG signala (QRS kompleksa), proces odabiranja mora da se ponovi za svaku  vrijednost parametra $\lambda$ iz nekog mogućeg opsega vrijednosti, što može predstavljati problem za uređaje kojima se odabiranje sprovodi. Zatim, Hermitska transformacija se računa za svako moguće $\lambda$, a (\ref{qrsmse}) se koristi za određivanje onog $\lambda$ za koje je $E \leq 10\%$. Druga mogućnost je da se koristi fiksna vrijednost $\lambda$ za sve komplekse. Međutim, lako se pokazuje da neadekvatno odabran parametar $\lambda$  vodi do većeg broja Hermitskih koeficijenata sa značajnim vrijednostima u vektoru $\mathbf{\tilde{C}}$. Štaviše, naši eksperimentalni rezultati, ali i rezultati u \cite{ht1} i \cite{ht2} pokazuju da svaki QRS kompleks ima drugačiju optimalnu vrijednost faktora skaliranja $\lambda$, što znači da bi uređaj za odabiranje morao da se ručno podešava za svaki kompleks, što je, naravno, praktično neprihvatljivo. Sa druge strane, QRS kompleksi razmatrani u \cite{ht1} su originalno  uniformno odabrani, i reodabiraju se korišćenjem interpolacije  (\ref{sincint}), dok se optimalno $\lambda$  traži mjerenjem kompresionog odnosa (engl. \textit{compression ratio}) koji treba da bude maksimizovan uz uslov da je greška $E \leq 10\%$. Međutim, ovakav pristup je numerički zahtjevan, budući da i inverzna i direktna Hermitska transformacija moraju biti računate za svaki posmatrani broj koeficijenata sa najvećim vrijednostima i za svaki posmatrani faktor skaliranja $\lambda$.

Navedene probleme moguće je riješiti korišćenjem predloženih optimizacionih procedura za faktor skaliranja vremenske ose  $\lambda$ i parametra pomjeraja $l$ koje su zasnovane na mjerama koncentracije, prije nego što se primijeni kompresioni algoritam. Sama kompresija QRS kompleksa  se i dalje obavlja korišćenjem algoritma iz \cite{ht1}.

U svrhu testiranja razmatra se 168 EKG signala iz \textit{online} baze ,,MIT-BIH Compression Test Database'' \cite{ecg_baza}. U svim slučajevima se posmatra samo prvi kanal. Automatizovanim algoritmom, koji je dostupan \textit{online} u sklopu pratećih programa rada \cite{ht1}, izdvojeno je ukupno $Q=1486$ kompleksa, koji se pojavljuju u prvih 10 s u svakom od razmatranih signala. Signali su uniformno odabrani, sa korakom $\Delta t=1/250$ [s]. U proceduri testiranja razmatrane su tri moguće dužine: $2D+1\in \left\{ 27,29, 31 \right\}$. Rezultati kompresije su prikazani u tabeli \ref{table_qrs}. Kao kriterijumi poređenja korišćeni su: broj transformacionih koeficijenata koji izazivaju grešku u aproksimaciji $E\leq 10\%$, kao i prosječni kompresioni odnos koji se računa po formuli:
\begin{equation}
\operatorname{ACR}=\frac{\sum\nolimits_{i=1}^{Q}{(2{{D}_{i}}+1)}}{\sum\nolimits_{i=1}^{Q}{{{L}_{i}}}},
\end{equation}
gdje je $L_i$ broj nenultih transformacionih koeficijenata koji izazivaju grešku manju od $10\%$, a $2Di+1$ je dužina $i$-tog QRS kompleksa. Druga, treća i četvrta kolona sadrže odgovarajuće rezultate publikovane u \cite{ht1}, koji uključuju kompresiju zasnovanu na Hermitskoj transformaciji računatoj algoritmom iz \cite{ht1}, kao i kompresione pristupe zasnovane na DFT i DCT.
	\begin{table}[!tbp]
	\small
	%% increase table row spacing, adjust to taste
	\renewcommand{\arraystretch}{1.3}
	% if using array.sty, it might be a good idea to tweak the value of
	% \extrarowheight as needed to properly center the text within the cells
	\caption{Prosječni broj transformacionih koeficijenata i prosječni kompresioni odnos prilikom kompresije testiranih 1486 QRS kompleksa. Dobijene vrijednosti garantuju da je greška u aproksimaciji manja od 10\%.}
	\label{table_qrs}
	\centering
	%% Some packages, such as MDW tools, offer better commands for making tables
	%% than the plain LaTeX2e tabular which is used here.
	
	\begin{tabular}{lcccc} 
		\toprule  
		{Kriterijum  poređenja} & \parbox[c]{3cm} {Predloženi   pristup} & \parbox[c]{3cm} {HT kompresioni \\ algoritam iz \cite{ht1}} & \parbox[c]{1.5cm} {DFT kompresija} & \parbox[c]{1.5cm} {DCT kompresija} \\
		\midrule 
		Prosječni broj koeficijenata & 5.0 &5.8 &8.3 &7.3\\
		Prosječni kompresioni odnos & 6.2 &5.3 &3.7  &4.3\\
		
		\bottomrule
	\end{tabular}
\end{table}
U originalnom pristupu \cite{ht1} koji koristi numerički zahtjevnu proceduru pretraživanja mogućih faktora skaliranja $\lambda$, u prosjeku je potrebno 5.8 Hermitskih koeficijenata za adekvatnu rekonstrukciju komprimovanih QRS kompleksa sa greškom $E\leq 10\%$. U predloženom metodu (prva kolona u tabeli \ref{table_qrs}) postoji poboljšanje ovog rezultata, ukoliko se iskoristi dodatna optimizacija parametra vremenskog pomjeranja sa  ${l}_{\max }=3$, $l\in \{-3,-2,-1,~0,~1,~2,~3\}$. Ukoliko se iskoristi i optimizacija faktora skaliranja i optimizacija parametra vremenskog pomjeraja korišćenjem pristupa predloženih u ovoj sekciji, u prosjeku je potrebno samo 5 Hermitskih koeficijenata, što predstavlja poboljšanje rezultata za 13.8\% (ne računajući poboljšanje u brzini izvršavanja algoritma). Prosječna vrijednost faktora skaliranja na nivou svih $Q = 1486$ QRS kompleksa iznosi $\lambda /\Delta t=\text{ 4}\text{.2495}$ (u sekundama $\lambda =\text{ 4}\text{.2495}/250=0.017$), što je vrijednost koja je eksperimentalno dobijena u \cite{ht2}, što dodatno potvrđuje tačnost predloženog pristupa. U eksperimentu je korišćena vrijednost koraka $\mu =0.05$.

Razmotrimo i prosječan broj bita po odbirku: a) u vremenskom domenu ovaj broj je 9 bps, b) u slučaju optimizovane DHT1 postoji 5 najznačajnijih koeficijenata koji su čisto realni, od ukupno 31 vrijednosti unutar QRS kompleksa, (jedna nenulta vrijednost se može pojaviti između 5 nenultih koeficijenata, pa se smatra da je potrebno kodirati 6 koeficijenata), što daje rezultat od 2 bps; c) u slučaju DFT, u prosjeku postoji 8 najznačajnijih koeficijenata (koji su kompleksni, odnosno, imaju i realne i imaginarne djelove) koje je potrebno kodirati, od ukupno 31, što je aproksimativno 7 bps u prosjeku.

  \begin{figure}[ptb]%
	\centering
	\includegraphics[
	]%
	{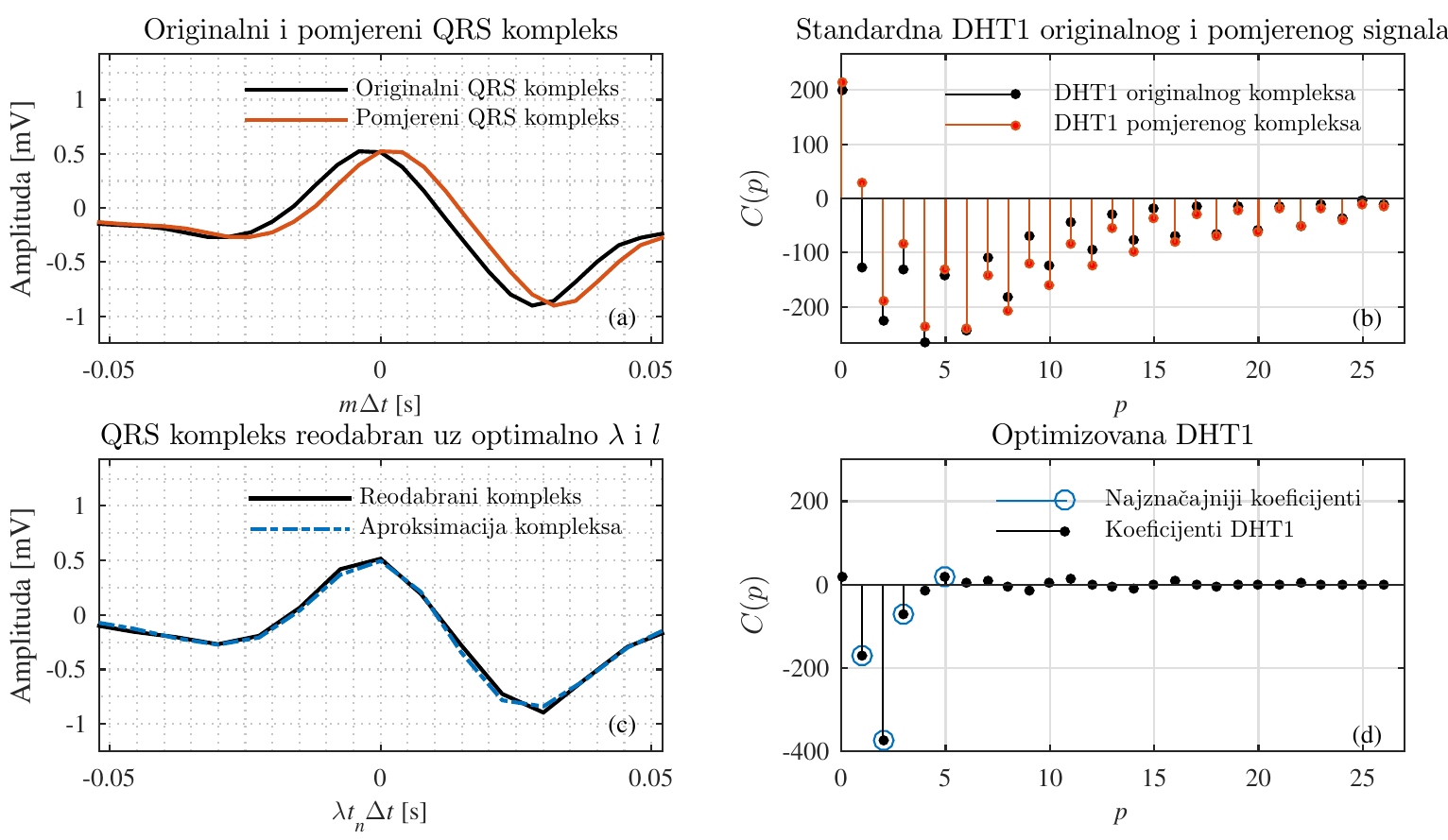}%
	\caption[Optimalno skaliranje, pomjeranje i reodabiranje QRS kompleksa.]{Optimalno skaliranje, pomjeranje i reodabiranje QRS kompleksa: (a) originalni  i optimalno pomjereni signal, (b) Hermitski koeficijenti originalnog i pomjerenog signala, (c) pomjereni reodabrani signal sa optimalnim faktorom skaliranja i signal rekonstruisan na osnovu samo 4 najznačajnija Hermitska koeficijenta (relativna greška manja od 10\%), (d) optimizovana Hermitska transformacija reodabranog pomjerenog signala i 4 najznačajnija koeficijenta.}%
	\label{mart14_ex2}%
\end{figure}
Primjer QRS kompleksa iz baze \cite{ecg_baza} je prikazan na slici \ref{mart14_ex2}. Originalni signal je prikazan na slici \ref{mart14_ex2} (a), dok su odgovarajući Hermitski koeficijenti prikazani na slici \ref{mart14_ex2} (b). Mjera koncentracije je najmanja za pomjeraj $l=1$ (signal pomjeren za ovu vrijednost je prikazan na slici \ref{mart14_ex2} (a) crvenom bojom), sa odgovarajućim faktorom skaliranja  $\lambda /\Delta t=0.4352$ (u sekundama $\lambda =0.0176$). Optimalno pomjereni signal, koji je reodabran u tačkama proporcionalnim nulama Hermitskog polinoma (reda $N=27$, i sa konstantom proporcionalnosti $\lambda /\Delta t=0.4352$) prikazan je na slici \ref{mart14_ex2} (c), dok su odgovarajući Hermitski koeficijenti prikazani na slici \ref{mart14_ex2} (d). Greška $E\leq 10\%$ se postiže  sa samo četiri značajna koeficijenta, koji su na slici označeni kružićima. Signal koji je rekonstruisan pomoću samo ova četiri koeficijenta predstavljen je crta-tačka-crta linijom na slici \ref{mart14_ex2} (c). Evidentan je visok nivo poklapanja sa originalnim reodabranim signalom.

\begin{figure}[ptb]%
	\centering
	\includegraphics[
	]%
	{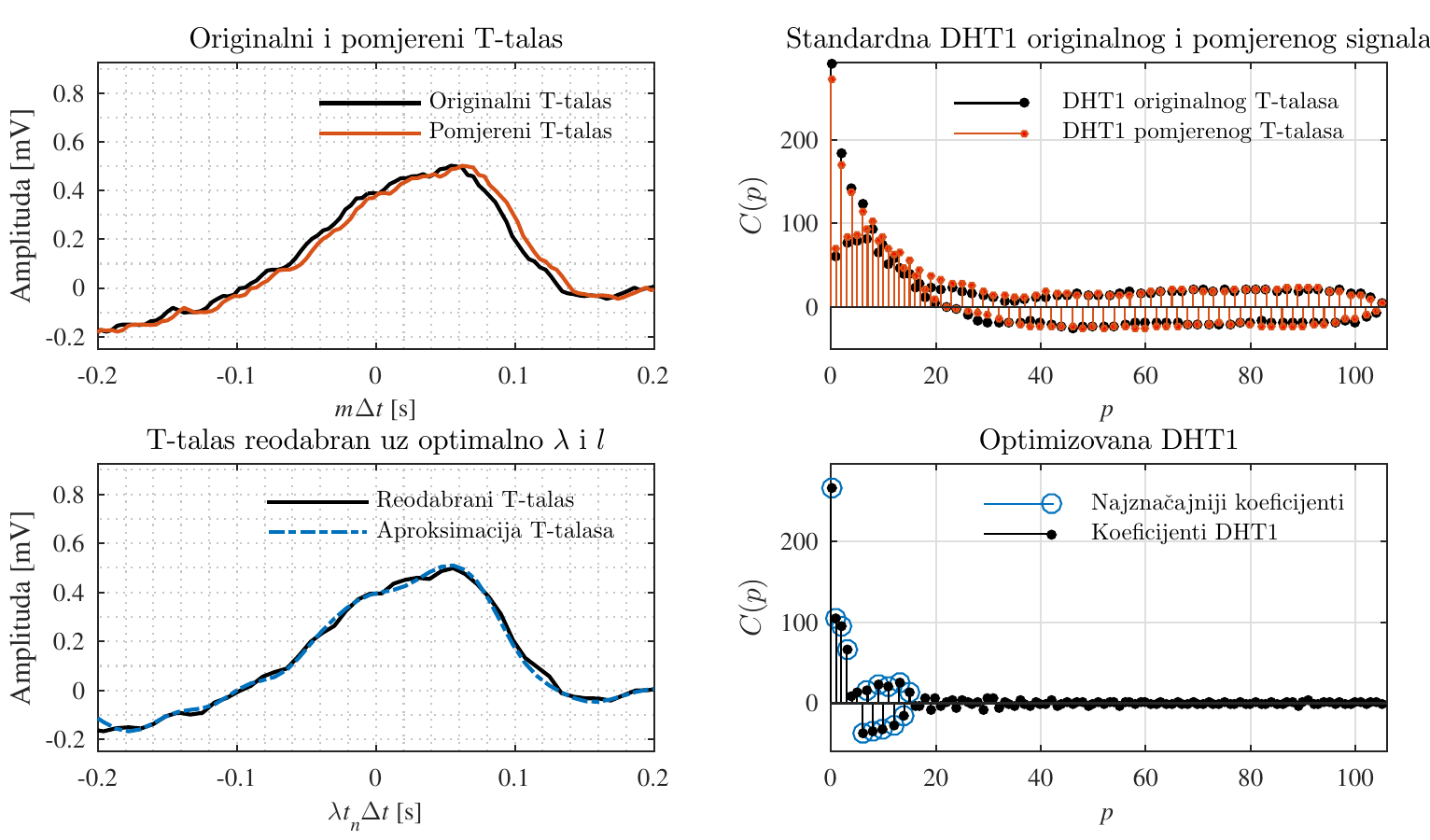}%
	\caption[Optimalno skaliranje, pomjeranje i reodabiranje T-talasa EKG signala.]{Optimalno skaliranje, pomjeranje i reodabiranje T-talasa EKG signala: (a) originalni  i optimalno pomjereni signal, (b) Hermitski koeficijenti originalnog i pomjerenog signala, (c) pomjereni reodabrani signal sa optimalnim faktorom skaliranja i signal rekonstruisan na osnovu 14 najznačajnijih Hermitskih koeficijenata (relativna greška manja od 10\%), (d) optimizovana Hermitska transformacija reodabranog pomjerenog signala i 14 najznačajnijih koeficijenata.}%
	\label{Twave}%
\end{figure}
I drugi djelovi EKG signala mogu biti reprezentovani sa malim brojem Hermitskih koeficijenata, iako ne jednako efikasno kao u slučaju QRS kompleksa. Ilustrujmo ovo na primjeru T-talasa, prikazanom na slici \ref{Twave} (a), koji je dio jednog od EKG signala iz baze \cite{ecg_baza}. Nakon primjene prezentovanog algoritma, T-talas je reprezentovan sa samo 14 koeficijenata sa značajnim vrijednostima (od ukupno 106), što je 13\% dužine signala, koji garantuju  relativnu rekonstrukcionu grešku koja je manja od 10\%. Na nivou cijele razmatrane baze ,,MIT-BIH Compression Test Database'' (prvi kanal, prvih 10 sekundi od svih 168 EKG signala) \cite{ecg_baza} greška $E\leq 10\%$ se postiže sa oko 23\% značajnih koeficijenata (u odnosu na dužinu signala). Hermitske transformacije originalnog i optimalno pomjerenog signala prikazane su na slici \ref{Twave} (b), signal reodabran sa optimalnim faktorom skaliranja prikazan je na slici \ref{Twave} (c), dok je odgovarajuća optimizovana DHT1 prikazana na slici \ref{Twave} (d). Na istoj slici su najznačajniji koeficijenti označeni kružićima, dok je aproksimacija zasnovana na tim koeficijentima upoređena sa reodabranim signalom na slici \ref{Twave} (c).
 
\section{Uticaj aditivnog šuma na Hermitske koeficijente}
U ovoj sekciji će biti izvedeni izrazi za varijansu i srednju vrijednost koeficijenata diskretne Hermitske transformacije, u slučaju prisustva aditivnog bijelog Gausovog šuma. Izraz za varijansu će poslužiti za definisanje jednostavnog pristupa za uklanjanje šuma. Prezentovani rezultati će u narednoj sekciji biti inkorporirani i u algoritme za rekonstrukciju signala u kontekstu kompresivnog odabiranja. Rezultati su publikovani u radu \cite{ht10-noises}.
 \subsection{Hermitski transformacioni koeficijenti zašumljenog signala}
Posmatra se vektor aditivnog bijelog Gausovog šuma srednje vrijednosti nula $\mathbf{\varepsilon }.$  Bez gubljenja opštosti izlaganja, podrazumijevano je da su odbirci šuma dostupni u diskretnim tačkama ${{t}_{n}},~n=1,2,\dots,N$ koje odgovaraju nulama Hermitskog polinoma $N$-tog reda, kao i da je faktor skaliranja vremenske ose $\sigma=1$.
 DHT1 ovog vektora ima sljedeći oblik:
 \begin{equation}
\mathbf{\Xi }={{\mathbf{T}}_{H}}\mathbf{\varepsilon }=\mathbf{QR\varepsilon }=\mathbf{Q}\,\left[ \begin{matrix}
{{d}_{1}}\varepsilon ({{t}_{1}}) & 0 & \cdots  & 0  \\
0 & {{d}_{2}}\varepsilon ({{t}_{2}}) & \cdots  & 0  \\
\vdots  & \vdots  & \ddots  & 0  \\
0 & 0 & \cdots  & {{d}_{N}}\varepsilon ({{t}_{M}})  \\
\end{matrix} \right].
 \end{equation}
  
Zaključuje se da se skalirani odbirci šuma  ${{d}_{n}}\varepsilon ({{t}_{n}}),~n=1,2\dots,N$ razvijaju u ortogonalnom vektorskom prostoru sastavljenom od vrsta matrice $\mathbf{Q}$. U slučaju ortogonalnih transformacionih matrica, kao što je DFT matrica, odbirci šuma se pojavljuju neskalirani u kvadratnoj matrici $\mathbf{ R}$, što znači da će aditivni bijeli Gausov šum na DHT1 uticati drugačije nego na DFT. Stoga je potrebno posebno izvesti statističke osobine Hermitskih koeficijenata koji odgovaraju signalu $x(t_n)$ koji je zahvaćen ovom aditivnim bijelim Gausovim šumom $\varepsilon(t_n)$: 
\begin{equation}
x({{t}_{n}})=s({{t}_{n}})+\varepsilon({{t}_{n}}).
\label{htnoisesig}
\end{equation}

Pošto je DHT1 linearna transformacija, važiće
$X(p)=S(p)+\Xi (p),
$
gdje je $S(p)$ Hermitska tranformacija signala $s(t_n)$. Pošto je šum $\varepsilon$ slučajni Gausov proces, kao posljedica teoreme o odabiranju i definicije DHT1 (\ref{dht1}), zaključuje se da su $\Xi(p)$ i $X(p)$ takođe slučajne varijable sa Gausovskom prirodom. Dalje ćemo analizirati statističke osobine slučajne promjenljive $X(p)$. Srednja vrijednost se može izraziti u sljedećem obliku:
\begin{equation}
 {{\mu }_{X}}(p)=E\left\{ X(p) \right\}=S(p)+E\left\{ \Xi (p) \right\}.
\end{equation}
 
Kako za aditivni šum sa srednjom vrijednošću nula važi $E\{\varepsilon(n)\}=0$, dalje se može pisati:
\begin{equation}
  {{\mu }_{X}}(p)=S(p)+\frac{1}{N}\sum\limits_{n=1}^{N}{\frac{{{\psi }_{p}}({{t}_{n}})}{{{\left[ {{\psi }_{N-1}}({{t}_{n}}) \right]}^{2}}}E\left\{ \varepsilon ({{t}_{n}}) \right\}}=S(p). 
\end{equation}

Uz pretpostavku da je srednja vrijednost šuma $\varepsilon(t_n)$, polazeći od definicije varijanse slučajne promjenljive $X(p)$ slijedi:
  \begin{align}
  \sigma _{X}^{2}(p)&=E\left\{ {{\left| X(p)-{{\mu }_{X}}(p) \right|}^{2}} \right\}=E\left\{ {{\left| S(p)+\Xi (p) \right|}^{2}} \right\}-\mu _{X}^{2}(p)\notag \\ 
 & =\frac{1}{{{N}^{2}}}\sum\limits_{{{n}_{1}}=1}^{M}{\sum\limits_{{{n}_{2}}=1}^{N}{\frac{{{\psi }_{p}}({{t}_{{{n}_{1}}}})}{{{\left[ {{\psi }_{N-1}}({{t}_{{{n}_{1}}}}) \right]}^{2}}}\frac{{{\psi }_{p}}({{t}_{{{n}_{2}}}})}{{{\left[ {{\psi }_{N-1}}({{t}_{{{n}_{2}}}}) \right]}^{2}}}}E\left\{ \varepsilon ({{t}_{{{n}_{1}}}})\varepsilon ({{t}_{{{n}_{2}}}}) \right\}}. 
 \label{varadit} 
 \end{align} 
 
U slučaju aditivnog bijelog šuma sa varijansom $\sigma_{\varepsilon}^2$, autokorelaciona funkcija je
\begin{equation}
 {{r}_{\varepsilon \varepsilon }}({{t}_{{{n}_{1}}}},{{t}_{{{n}_{2}}}})=E\{\varepsilon ({{t}_{{{n}_{1}}}})\varepsilon ({{t}_{{{n}_{2}}}})\}=\sigma _{\varepsilon }^{2}\delta({{t}_{{{n}_{1}}}}-{{t}_{{{n}_{2}}}}). \label{autokor}
\end{equation}
Inkorporiranjem (\ref{autokor}) u izraz za varijansu (\ref{varadit}), dobija se
\begin{equation}
\sigma _{X}^{2}(p)=\frac{\sigma _{\varepsilon }^{2}}{{{N}^{2}}}\sum\limits_{n=1}^{N}{\frac{\psi _{p}^{2}({{t}_{n}})}{{{\left[ {{\psi }_{N-1}}({{t}_{n}}) \right]}^{4}}}}=\gamma \left( p,N\right)\sigma _{\varepsilon }^{2}.
\label{varhtawgn} 
\end{equation}
  
 Rezultat (\ref{varhtawgn}) ukazuje na to da je varijansa DHT1 koeficijenata zavisna od indeksa $p$. Usrednjavanje izraza (\ref{varhtawgn}) dalje daje
 \begin{align}
  \bar{\sigma }_{X}^{2}&=\frac{1}{N}\sum\limits_{p=0}^{N-1}{\sigma _{X}^{2}(p)}\frac{\sigma _{\varepsilon }^{2}}{{{N}^{2}}}\sum\limits_{n=1}^{N}{\frac{1}{{{\left[ {{\psi }_{N-1}}({{t}_{n}}) \right]}^{2}}}\frac{1}{N}\sum\limits_{p=0}^{N-1}{\frac{\psi _{p}^{{}}({{t}_{n}})\psi _{p}^{{}}({{t}_{n}})}{{{\left[ {{\psi }_{N-1}}({{t}_{n}}) \right]}^{2}}}}}\notag \\ 
 & =\frac{\sigma _{\varepsilon }^{2}}{{{N}^{2}}}\sum\limits_{n=1}^{N}{{{\left[ {{\psi }_{N-1}}({{t}_{n}}) \right]}^{-2}}}=\xi (N)\sigma _{\varepsilon }^{2}.  
 \label{varhtawgn_mean}
 \end{align} 
    \begin{figure}[ptb]%
 	\centering
 	\includegraphics[
 	]%
 	{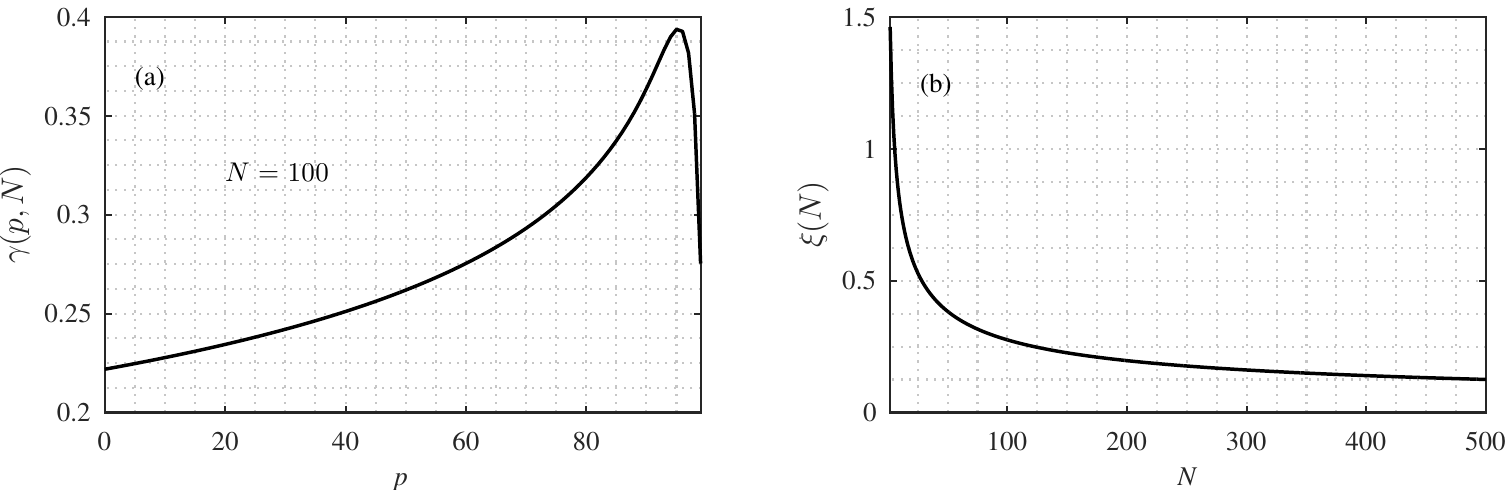}%
 	\caption[Funkcije koje se pojavljuju u izrazima za varijansu Hermitskih koeficijenata zašumljenog signala.]{Funkcije koje se pojavljuju u izrazima za varijansu Hermitskih koeficijenata zašumljenog signala: (a) funkcija $\gamma(p,N)$ i (b) funkcija $\xi(N)$.}%
 	\label{slika0}%
 \end{figure}
 Funkcija $\gamma(p,N)$ za $N=100$ je prikazana na slici \ref{slika0} (a), dok je funkcija $\xi(N)$ prikazana na istoj slici pod (b). Srednja vrijednost prve funkcije je 0.2753. To znači da prvih 60 koeficijenata imaju varijansu (skaliranu sa $\gamma(p,N)$) koja je manja od ove srednje vrijednosti, dok su preostali koeficijenti skalirani sa većim vrijednostima, pa su zbog toga osjetljiviji na uticaj šuma. Bitno je uočiti da su signali koji su dobro koncentrisani u DHT1 domenu (QRS kompleksi, UWB signali itd.) takvi da su transformacioni koeficijenti sa najznačajnijim vrijednostima upravo u dijelu gdje je varijansa manja. Ovo takođe znači da je moguće uvesti adekvatan prag za razdvajanje koeficijenata koji odgovaraju signalu od koeficijenata koji predstavljaju šum. Slika \ref{slika0} (b) pokazuje da se srednja vrijednost varijanse smanjuje sa povećavanjem dužine signala $N$.
 
 U slučaju DHT2, budući da je transformaciona matrica (\ref{DHT2m}) ortogonalna i da su vrste matrice normalizovane (jer su sopstveni vektori matrice $\mathbf{T}$), zaključuje se da za varijansu DHT2 koeficijenata signala  (\ref{htnoisesig}) važi:
 \begin{equation}
   \tilde{\sigma }_{X}^{2}(p)=\sigma _{\varepsilon }^{2}, 
 \end{equation}
vodeći računa da je u ovom slučaju signal uniformno odabran.

 \subsection{Uklanjanje šuma postavljanjem praga u DHT1 domenu}

 Prezentovana diskusija se može direktno iskoristiti u proceduri za uklanjanje šuma, Dodatno, rezultati će u sljedećoj sekciji biti inkorporirani u kontekst kompresivnog odabiranja. Posmatra se signal (\ref{htnoisesig}) zahvaćen aditivnim bijelim Gausovim šumom čija je varijansa  $\sigma_{\varepsilon}^2$. Iz dosadašnjeg izlaganja zaključili smo da je DHT1 koeficijent $X(p)$ predstavlja slučajnu promjenljivu opisanu Gausovom distribucijom $\mathcal{N}(0,\sigma_{X}^2 (p))$. Uklanjanje šuma može biti obavljeno direktnim postavljanjem praga za odvajanje koeficijenata koji odgovaraju komponentama signala od  koeficijenata koji odgovaraju šumu (engl. \textit{hard-thresholding}):
 \begin{align}
 {{X}_{den}}(p)=\left\{ 
 \begin{matrix}
 X(p), & \left| X(p) \right|>T(p)  \\
 0, & \left| X(p) \right|\le T(p).  \\
 \end{matrix} \right.
 \label{htht}
 \end{align}	 
 Treba uočiti da je prag $T(p)$ funkcija od indeksa $p$, usljed matematičke forme varijanse (\ref{varhtawgn}). Naime, koristeći izraz za varijansu, odnosno standardnu devijaciju Hermitskih koeficijenata $X(p)$, prag se može definisati u sljedećoj formi: 
 \begin{equation}
 T(p)=\alpha\sigma _{X}^{{}}(p)=\alpha\sqrt{\gamma (p,M)}\sigma _{\varepsilon },
 \label{htpragden} 
 \end{equation}
gdje konstanta $\alpha$ obezbjeđuje da koeficijenti koji odgovaraju šumu budu ispod nivoa praga. Za mnoge realne signale, kao što su UWB signali i QRS kompleksi, koeficijenti sa malim vrijednostima indeksa $p$ su najznačajniji. Nelinearna forma praga (\ref{htpragden}) povećava vjerovatnoću razdvajanja koeficijenata koji korespondiraju komponentama signala od koeficijenata koji predstavljaju čisti šum, slika \ref{slika0} (a). Za $\alpha=3$ , prema poznatom $3\sigma$ empirijskom pravilu, koeficijenti šuma će biti ispod praga sa vjerovatnoćom od 99.73\%. U slučaju DHT2, za \textit{hard-thresholding} proceduru (\ref{htht}) treba koristiti standardni prag:
\begin{equation}
T=\alpha\sigma _{\varepsilon }^{{}}.
\end{equation}

Kako prezentovani metod za uklanjanje šuma može poslužiti kao alternativa uklanjanju šuma zasnovanom na DFT, razmotrimo i povećanje numeričke složenosti pristupa u odnosu na DFT ekvivalent. Kako DHT1 (odnosno ekvivalentne forme koje podrazumijevaju upotrebu numeričkih kvadratura) može biti izračunata brzim algoritmima \cite{ht1}, \cite{ht2}, aproksimativna numerička složenost je $O(N\log _{2}^{2}N)$ operacija sa realnim vrijednostima, u poređenju sa $O(N\log _{2}^{{}}N)$ operacija sa kompleksnim vrijednostima potrebnih u slučaju DFT. Povećanje složenosti je, dakle, malo i za veliku dužinu signala $N$. Za računanje praga 
(\ref{htpragden}) potrebno je dodatnih $O(N)$ operacija.
 
 \subsection{Numerički rezultati i primjene}
\begin{primjer}
	   \begin{figure}[ptb]%
	\centering
	\includegraphics[
	]%
	{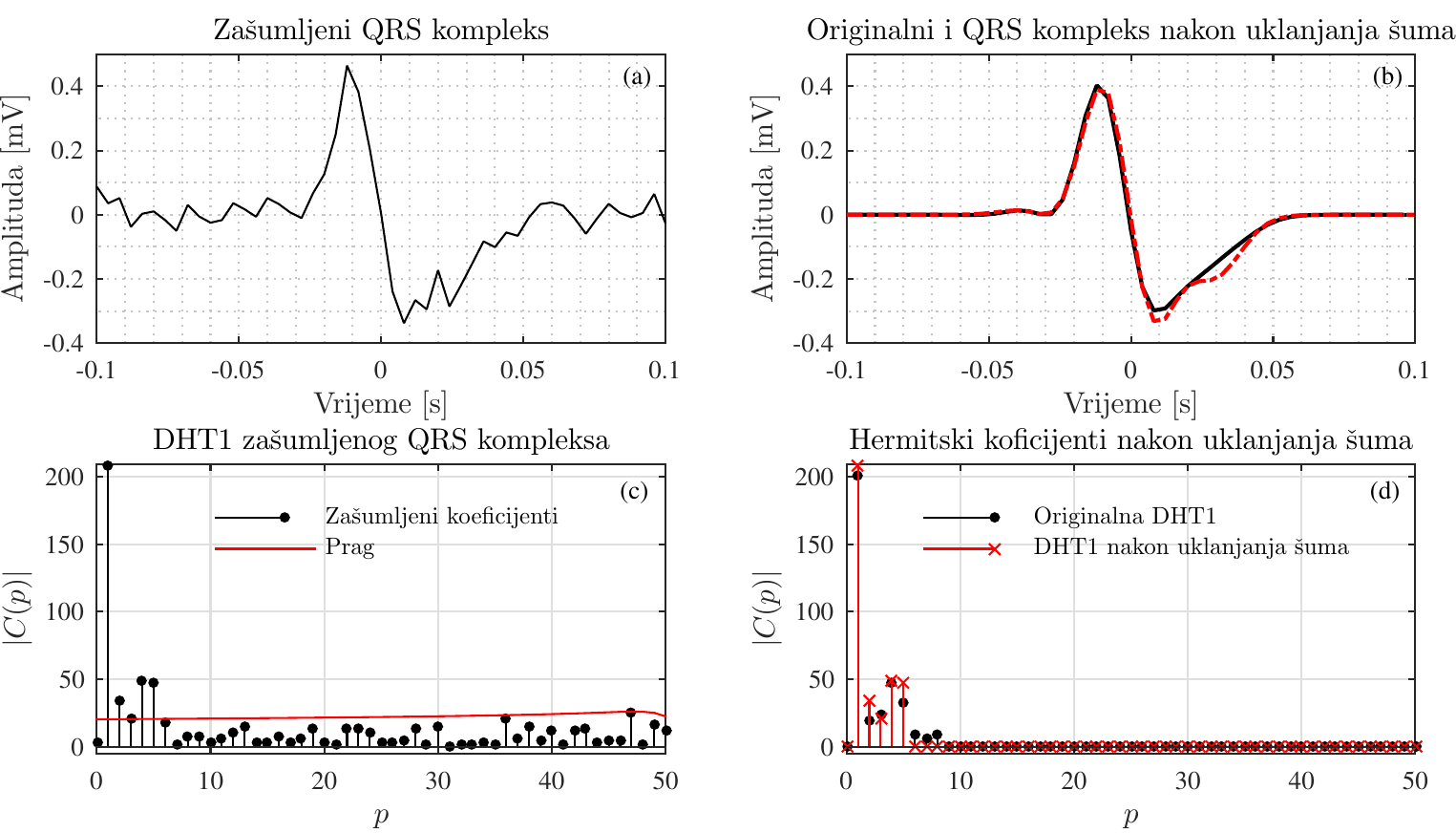}%
	\caption[Uklanjanje aditivnog bijelog Gausovog šuma iz QRS kompleksa.]{Uklanjanje aditivnog bijelog Gausovog šuma iz QRS kompleksa: (a) zašumljeni QRS kompleks, (b) originalni signal (puna linija) i signal iz kojeg je uklonjen šum (crta-tačka-crta linija), (c) DHT1 zašumljenog kompleksa i prag, (d) DHT1 koeficijenti originalnog signala i signala iz kojeg je uklonjen šum.}%
	\label{NoisyQRS}%
\end{figure}
Razmatra se QRS kompleks $s(\lambda t_n)$, odabran u tačkama proporcionalnim nulama Hermitskog polinoma, sa faktorom skaliranja koji je određen primjenom Algoritma \ref{SigmaOpt}, koji je opisan u prethodnoj sekciji, kako bi bila postignuta visoka koncentracija DHT1 koeficijenata. QRS kompleks dužine $N=51$ je preuzet iz jednog od signala dostupnih u bazi ,,MIT-BIH ECG Compression Test Database'' \cite{ecg_baza}, i originalno je bio odabran u skladu sa teoremom o odabiranju, sa periodom od $\Delta t= 1/250$ [s]. Signal je zatim oštećen vještačkim aditivnim bijelim Gausovim šumom srednje vrijednosti nula, tako da je odnos signal-šum (SNR) jednak $6$ dB. Rezultati uklanjanja šuma za posmatrani signal prikazani su na slici \ref{NoisyQRS} (a) -- (d). Na slici \ref{NoisyQRS} (b) uočena je značajna redukcija šuma. U \textit{hard-thresholding} proceduri korišćen je parametar $\alpha=3$ za prag (\ref{htpragden}). Zanimljivo je uočiti da je koeficijent koji predstavlja komponentu signala na poziciji $p=3$, iako po vrijednosti manji od koeficijenta šuma na poziciji $p=47$, pravilno selektovan pomoću praga (ima veću vrijednost od praga), usljed njegove nelinearne forme. Samo nekoliko koeficijenata signala koji su mnogo slabiji od šuma ostali su ispod praga, i to na pozicijama $p = 6$, $p = 7$ i $p = 8$, slika \ref{NoisyQRS} (d). Srednja kvadratna greška (MSE) između originalnog i zašumljenog signala od $-28.71$ dB je smanjena na $-35.83$ dB, nakon primjene procedure za uklanjanje šuma, što je poboljšanje od oko $7.1$ dB.
\label{ex_qrs_ht}
\end{primjer}
\begin{primjer}
	\label{ex_uwb_ht}
	Razmatra se realni UWB signal na $1.3$ GHz, koji je poslat između dvije UWB antene na udaljenosti od 1 m, u sobnom okruženju. Signal je dobijen u eksperimentu koji je opisan u \cite{ht_uwb2}. Posmatra se prvih $N = 165$ odbiraka signala \texttt{ACW7FD45.dat} iz \textit{online} baze \cite{ht_uwb3}.
	Signalu je dodat vještački aditivni bijeli Gausov šum srednje vrijednosti nula, tako da je $SNR=3$ dB, a zatim je izvršeno njegovo uklanjanje opisanim \textit{hard-thresholding} pristupom u DHT1 domenu. Zašumljeni signal je prikazan na slici \ref{UWB_threshodling_noise_new} (a), dok su odgovarajući DHT1 koeficijenti prikazani na slici \ref{UWB_threshodling_noise_new} (c), zajedno sa pragom (\ref{htpragden}). Originalni nezašumljeni signal i signal nakon uklanjanja šuma su upoređeni na slici \ref{UWB_threshodling_noise_new} (b), dok su njihove DHT1 upoređene na slici \ref{UWB_threshodling_noise_new} (d). Prilikom uklanjanja, korišćen je parametar $\alpha = 4$. Može se uočiti da je šum značajno redukovan. Pretpostavljeno je da je varijansa šuma $\sigma_{\varepsilon}^2$ unaprijed poznata, budući da se može estimirati postupkom koji je opisan u \cite{ecgden1}.
  \begin{figure}[!h]%
	\centering
	\includegraphics[
	]%
	{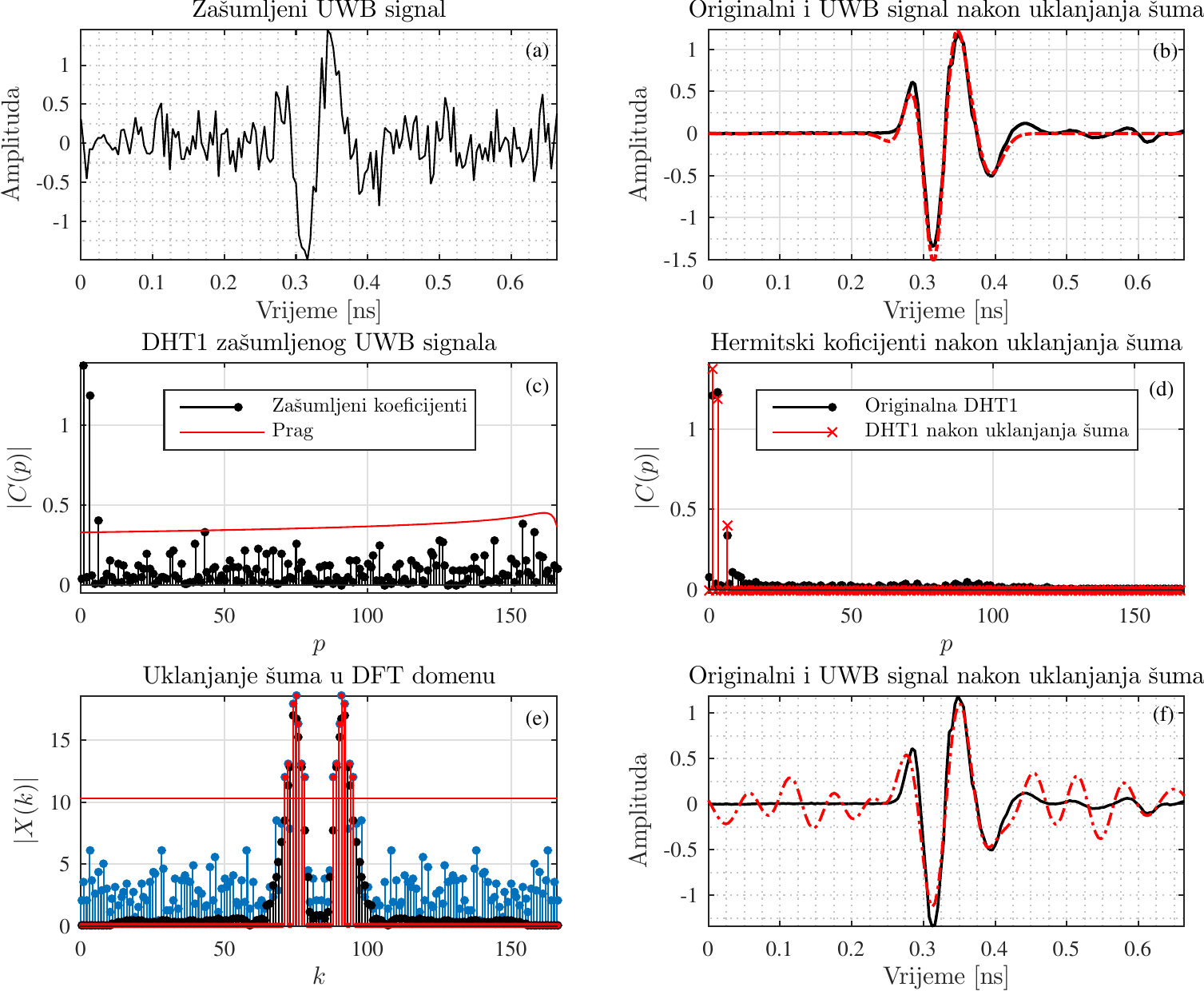}%
	\caption[Uklanjanje aditivnog bijelog Gausovog šuma iz realnog UWB signala.]{Uklanjanje aditivnog bijelog Gausovog šuma iz realnog UWB signala: (a) zašumljeni UWB signal, (b) nezašumljeni originalni signal (puna linija) i signal iz kojeg je uklonjen šum (crta-tačka-crta linija), (c) Hermitski koeficijenti zašumljenog signala i prag, (d) DHT1 zašumljenog optimalno pomjerenog kompleksa i prag, (e) DFT zašumljenog signala (plave tačke), prag (horizontalna linija), odabrani koeficijenti (crvene tačke) i originalni koeficijenti (crne tačke), (f) poređenje originalnog nezašumljenog signala i signala dobijenog uklanjanjem šuma u DFT domenu.}%
	\label{UWB_threshodling_noise_new}%
\end{figure}
 Nelinearni prag uspješno selektuje koeficijent sa indeksom $p=6$ koji predstavlja komponentu signala, iako on ima gotovo istu vrijednost kao koeficijent na poziciji šuma $p=154$ (koji ostaje ispod praga). Samo koeficijenti signala sa najmanjim vrijednostima ostaju ispod praga. U cilju poređenja, uklanjanje šuma je obavljeno i u DFT domenu, ekvivalentnim \textit{hard-thresholding} pristupom, gdje je korišćen odgovarajući DFT prag  $T_{DFT} = \alpha N\sigma_{\varepsilon}^2$, pri čemu je $\alpha = 4$, kao za slučaj DHT1. Rezultati uklanjanja šuma su prikazani na slici  \ref{UWB_threshodling_noise_new} (e) -- (f), gdje su lošije performanse u odnosu na DHT1 pristup evidentne. U slučaju DHT1 pristupa, rezultujući MSE je iznosio $-24.23$ dB, dok je u slučaju DFT pristupa MSE jednak $-16.07$ dB. Redukcija MSE u odnosu na polazni zašumljeni signal je iznosila 11.8 dB u DHT1 pristupu, i $3.63$ dB u DFT slučaju.
\end{primjer}
\begin{primjer}
U cilju što je moguće bolje numeričke validacije rezultata, prethodno opisani eksperimenti su sprovedeni za opseg SNR vrijednosti: od $-10$ dB do $15$ dB, sa korakom $1$ dB. Za svaku SNR vrijednost izračunata je MSE između originalnog signala i signala iz kojeg je uklonjen šum, na bazi 500 nezavisnih slučajnih realizacija vještačkog aditivnog bijelog Gausovog šuma koji je dodat signalu. Rezultati su prikazani na slici \ref{denstatistika2}.

Rezultati uklanjanja šuma u slučaju UWB signala iz primjera \ref{ex_uwb_ht} su prikazani na slici \ref{denstatistika2} (a). Rezultati potvrđuju da primjenom DHT1 metoda sa nelinearnim pragom dolazi do značajnog smanjenja srednje kvadratne greške. Štaviše, eksperiment je ponovljen i za slučaj uklanjanja šuma \textit{hard-thresholding} pristupom u kombinaciji sa diskretnom \textit{wavelet} transformacijom (DWT). Razmatrano je uklanjanje šuma zasnovano na \textit{Symlet 8} (,,sym8'') i \textit{Daubechies 8} (,,db8'') \textit{wavelet}-ima. U oba slučaja je korišćena dekompozicija nivoa 5. Ove specifične vrste \textit{wavelet}-a su odabrane usljed velike sličnosti talasnih oblika njihovih baznih funkcija sa razmatranim signalom. Uklanjanje šuma je izvršeno pomoću dva različita pravila za odabir praga za \textit{wavelet} koeficijente: Štajnov ,,Unbiased Risk Estimate'' u slučaju \textit{Daubechies 8} \textit{wavelet}-a, i ,,Universal threshold with level-dependent estimation of the noise'' predložen od strane Donohoa i Džonstona za \textit{Symlet 8} \textit{wavelet}  \cite{ht10-noises,wavelets_den2}. U svim slučajevima primjenjena je metoda \textit{hard-thresholding}-a. U ovom numeričkom eksperimentu je korišćena implementacija funkcije \texttt{wden} dostupna u okviru \textit{Wavelet Toolbox}-a softverskog paketa  MATLAB\textsuperscript{\textregistered}. Skaliranje praga je obavljeno korišćenjem jedne estimacije nivoa šuma zasnovane na koeficijentima prvog nivoa. Na slici  \ref{denstatistika2} (a) je uočljivo da oba \textit{wavelet} pristupa daju slične rezultate. Eksperiment potvrđuje da mogućnost visoko koncentrisane reprezentacije signala u DHT1 domenu može biti ključni faktor u uklanjanju šuma, čak i kada je uklanjanje zasnovano na veoma jednostavnim metodama, kao što je \textit{hard-thresholding}.

Dodatno, performanse uklanjanja šuma su testirane i u slučaju pristupa zasnovanom na postavljanju praga u DCT domenu. U ovom slučaju, prag (\ref{htpragden}) je postavljen na nivo $T_{DCT} = \alpha\sigma_{\varepsilon}^2$, imajući u vidu ortogonalnost DCT transformacione matrice. I u ovom slučaju je korišćen parametar $\alpha=4$. Uklanjanje šuma zasnovano na DHT1 metodu je i u ovom slučaju dalo bolje rezultate, \ref{denstatistika2} (a).

Konačno, izvršeno je i poređenje sa metodom uklanjanja šuma bazirano na \textit{hard-thresholding} pristupu i DHT2, sa pragom $T_{DHT2}(p)=\alpha\sigma _{\varepsilon }^{{}}$, uz istu vrijednost parametra $\alpha=4$. U slučaju ove verzije diskretne Hermitske transformacije korišćena je ista vrijednost faktora skaliranja vremenske ose $\sigma=1$, kao u slučaju DHT1. Sa slike \ref{denstatistika2} (a) uočljivo je da je uklanjanje šuma u DHT2 domenu dalo nešto lošije rezultate od uklanjanja u DHT1 domenu. Zanimljivo je uočiti da postoji visoka vizuelna sličnost između MSE krivih, što je u skladu sa prethodno diskutovanom sličnošću između baznih funkcija ovih transformacija.
   \begin{figure}[!htb]%
 	\centering
 	\includegraphics[
 	]%
 	{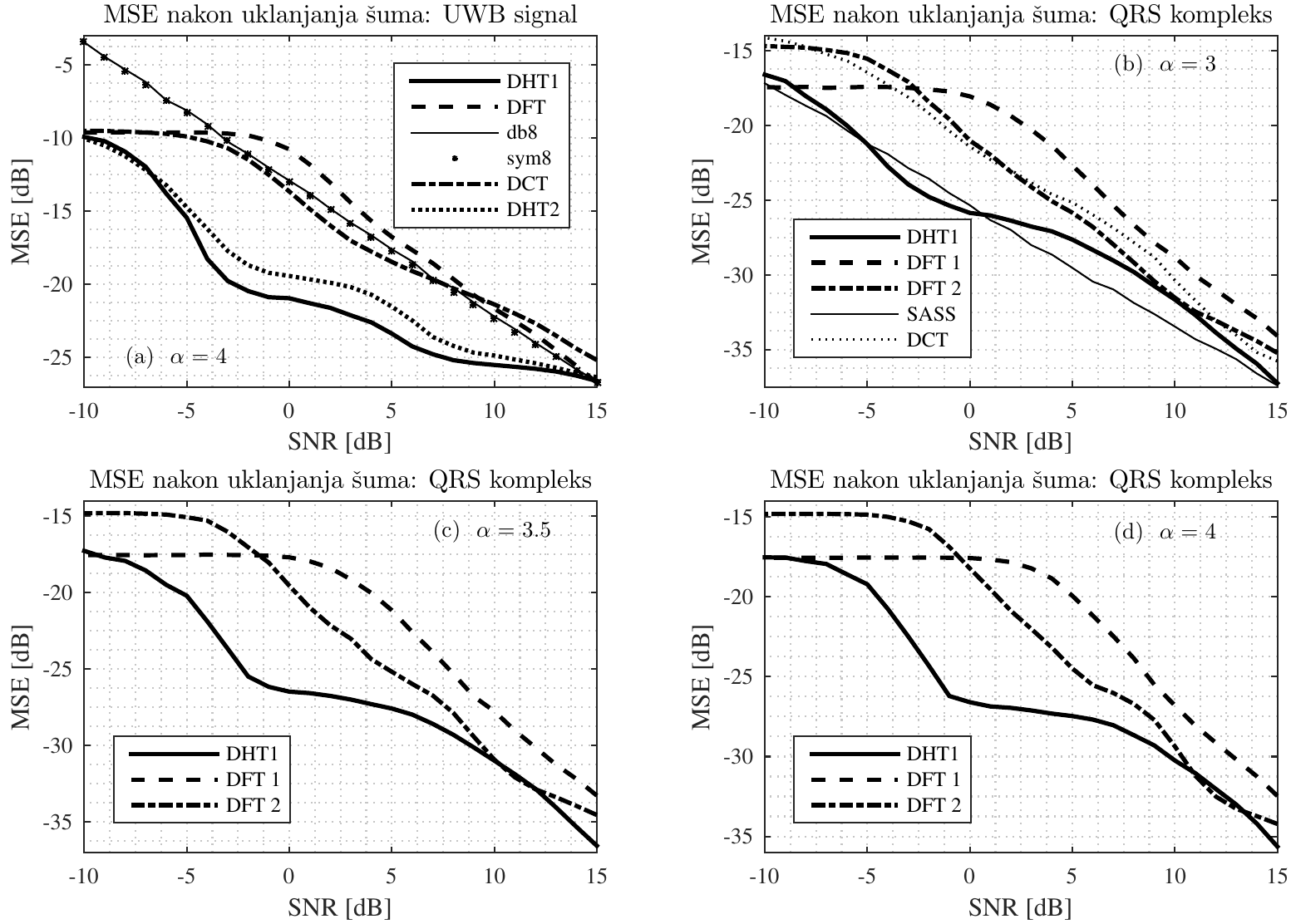}%
 	\caption[Srednja kvadratna greška (MSE) između originalnog (nezašumljenog) signala i signala iz kojeg je uklonjen šum, prikazana za različite odnose signal-šum (SNR).]{Srednja kvadratna greška (MSE) između originalnog (nezašumljenog) signala i signala iz kojeg je uklonjen šum, prikazana za različite odnose signal-šum (SNR): (a) u slučaju UWB signala, (b) -- (d) u slučaju QRS kompleksa za različite nivoe parametra $\alpha$.}%
 	\label{denstatistika2}%
 \end{figure}
   
Performanse DHT2 mogu biti poboljšane ukoliko se izvrši optimizacija faktora skaliranja vremenske ose, kao što je to urađeno za DHT1. Označimo sa ${{\mathbf{\tilde{T}}}_{H}}(\sigma )$ DHT2 transformacionu matricu sa baznim funkcijama ${{\tilde{\psi }}_{p}}(n,\sigma )$, gdje je $n=0,1,\dots,N-1,~p=0,1,\dots,N-1$. Važno je istaći da promjena parametra $\sigma$ ne narušava svojstvo ortogonalnosti [5]. Poboljšana koncentracija DHT2 se može postići rješavanjem optimizacionog problema:
\begin{equation}
{{\sigma }_{opt}}=\underset{\sigma }{\mathop{\min }}\,{{\left\| {{{\mathbf{\tilde{T}}}}_{H}}(\sigma )\mathbf{x} \right\|}_{1}}=\underset{\sigma }{\mathop{\min }}\,\sum\limits_{p=0}^{N-1}{\left| \sum\limits_{n=0}^{N-1}{s(n){{{\tilde{\psi }}}_{p}}(n,\sigma )} \right|},\label{dht2opt}
\end{equation}
 koji predstavlja minimizaciju $\ell_1$-norme transformacionih koeficijenata, koja se koristi u kontekstu obrade rijetkih signala i kompresivnog odabiranja. Rješenje problema (\ref{dht2opt}) je faktor skaliranja koji daje najbolju moguću koncentraciju transformacionih koeficijenata. Rješenje se može pronaći direktnim pretraživanjem u zadatom prostoru mogućih vrijednosti parametra $\sigma$. Dalje istraživanje ovog koncepta i komparativna analiza prevazilazi okvire ove disertacije i predmet je budućih istraživanja.
 
 Eksperiment sa opsegom SNR vrijednosti je ponovljen i za QRS komplekse iz primjera \ref{ex_uwb_ht}. Na slici \ref{denstatistika2} rezultati su prikazani za nekoliko vrijednosti parametra $\alpha$: slika \ref{denstatistika2} (b) $\alpha=3$, slika \ref{denstatistika2} (c) $\alpha=3.5$, slika \ref{denstatistika2} (c) $\alpha=4$. U cilju poređenja pod jednakim uslovima, uklanjanje šuma postavljanjem praga u DFT domenu je urađeno i za uniformno odabrani signal iz baze \cite{ecg_baza} (slučaj je označen sa ,,DFT1'' na slikama \ref{denstatistika2} (b) -- (d)), ali i za slučaj neuniformno odabranog signala korišćenog za DHT1 (u oznaci .,,DFT2'' na slikama \ref{denstatistika2} (b) -- (d)). U oba slučaja, isti šum je dodat odbircima signala. Rezultati pokazuju da uklanjanje šuma u DHT1 domenu daje dominantno niže MSE vrijednosti u svim slučajevima.
 
 Za $\alpha=3$, izvršeno je poređenje i sa procedurom za uklanjanje šuma postavljanjem praga u DCT domenu. U ovom slučaju prag je psotavljen u skladu sa varijansom DCT koeficijenata  $T_{DCT} = l\sigma_{\varepsilon}^2$. Na slici \ref{denstatistika2} (b) uočljive su niže MSE vrijednosti dobijene DHT1 metodom za uklanjanje šuma. Rezultati su dodatno upoređeni i sa veoma naprednim modernim algoritmom za uklanjanje šuma iz EKG signala -- \textit{Sparsity-Assisted Signal Smoothing} (SASS),\cite{ecgden3,ecgden2,htn3}. U eksperimentu je korišćena implementacija dostupna \textit{online}, sa originalnim setom parametara,\cite{ecgden3}. Rezultati su prezentovani na slici \ref{denstatistika2} (b). MSE kriva prikazana na slici je dobijena na način čiji opis slijedi. SASS algoritam je napravljen tako da radi sa čitavim EKG signalima, umjesto da se procesiranje vrši samo sa selektovanim QRS kompleksom. Stoga se algoritmu kao ulaz prosljeđuje cio EKG signal iz baze \cite{ecg_baza}, čiji je dio i razmatrani QRS kompleks. Ovaj signal je oštećen aditivnim bijelim Gausovim šumom sa istim varijansama koje su korišćene pri evaluaciji ostalih razmatranih tehnika. Uklanjanje šuma je obavljeno SASS algoritmom, sa podrazumijevanim setom parametara iz dostupne implementacije. Nakon toga, QRS kompleks od interesa je izdvojen iz signala iz kojeg je šum otklonjen, a zatim je taj rezultat upoređen sa originalnim QRS kompleksom. Kao i u ostalim slučajevima, eksperiment je ponovljen po 500 puta (sa slučajnim realizacijama šuma) za svaku posmatranu SNR vrijednost, na bazi kojih je izračunata MSE koja je prikazana na slici \ref{denstatistika2} (b). Na ovaj način, SASS algoritam je testiran bez ikakvih promjena parametara. Sa rezultata na slici \ref{denstatistika2} (b) može se uočiti da ovaj metod daje umjereno bolje rezultate od metoda uklanjanja šuma primjenom DHT1 sa pragom, i to u SNR opsegu od $0$ dB do $10$ dB. Međutim, cijena ovog poboljšanja je značajno povećana numerička složenost u slučaju SASS algoritma. Jednostavna (primitivna) procedura za uklanjanje šuma primjenom DHT1 sa pragom je dala rezultate koji su uporedivi sa rezultatima napredne tehnike za uklanjanje šuma (u razmatranom kontekstu QRS kompleksa).
\end{primjer}  
 
 \subsubsection{Numerička validacija izraza za varijansu}

\begin{figure}[htb]%
 	\centering
 	\includegraphics[
 	]%
 	{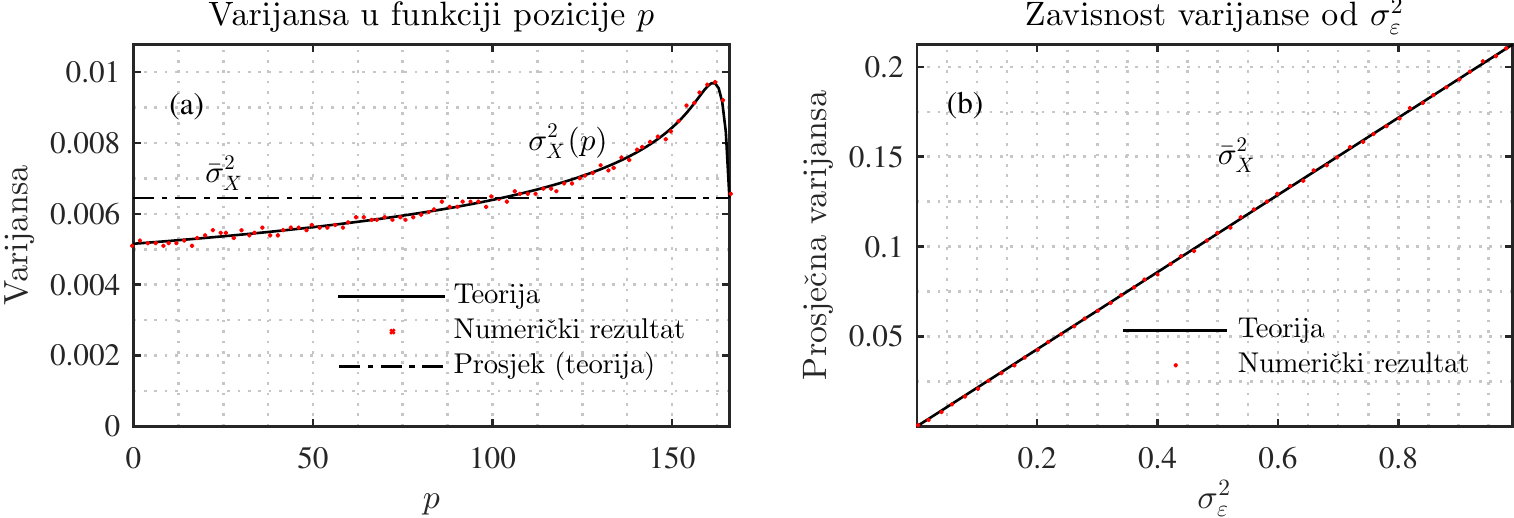}%
 	\caption[Varijansa DHT1 koeficijenata za UWB signal oštećen aditivnim bijelim Gausovim šumom.]{Varijansa DHT1 koeficijenata za UWB signal oštećen aditivnim bijelim Gausovim šumom: (a) varijansa u funkciji od indeksa $p$ transformacionog koeficijenta, (b) prosječna varijansa transformacionih koeficijenata u funkciji od varijanse ulaznog šuma.}%
 	\label{var_den}%
 \end{figure}

U cilju numeričke potvrde validnosti izvedenog izraza za variansu (\ref{varhtawgn}) i njene srednje vrijednosti (\ref{varhtawgn_mean}), sprovedena su dva eksperimenta sa zašumljenim UWB signalom iz primjera \ref{ex_uwb_ht}. U prvom testu, signal je zašumljen vještačkim aditivnim bijelim Gausovim šumom sa varijansom  $\sigma_{\varepsilon}^2= 0.030$ , koja daje SNR nivo od $5$ dB. Varijansa DHT1 koeficijenata je numerički izračunata na osnovu 10000 nezavisnih realizacija šuma. Rezultat prikazan na slici \ref{var_den} (a) se potpuno poklapa sa teorijskim izrazom (\ref{varhtawgn}). Na istoj slici je srednja vrijednost (\ref{varhtawgn_mean}) varijanse prikazana horizontalnom linijom. U drugom eksperimentu, varijansa ulaznog šuma $\sigma_{\varepsilon}^2$ je varirana u opsegu od $10^{-6}$ do $1$, sa korakom $10^{-2}$. Za svaku posmatranu SNR vrijednost, izračunata je srednja varijansa DHT1 koeficijenata, koja se u velikoj mjeri poklapa sa teorijskim rezultatom (\ref{varhtawgn_mean}), što je prikazano na slici (\ref{var_den}) (b).
Na obije posmatrane slike je, zbog bolje vizuelne prezentacije, prikazana svaka peta vrijednost numeričkog rezultata.

 \section {Rekonstrukcija signala sa rijetkom reprezentacijom u Hermitskom transformacionom domenu}

  Motivacija za proučavanje diskretne Hermitske transformacije u kontekstu kompresivnog odabiranja ogleda se u činjenici da određene vrste realnih signala, kao što su QRS kompleksi i UWB signali, mogu biti visoko koncentrisano (rijetko) reprezentovane u ovom transformacionom domenu. U okviru ove sekcije, biće prezentovana analiza uticaja nedostajućih odbiraka na koeficijente diskretne Hermitske transformacije. Oni će biti interpretirani kao slučajne varijable, čije su statističke karakteristike predmet prezentovane analize. Pored toga, biće uspostavljena veza sa indeksom koherentnosti parcijalne matrice diskretne Hermitske transformacije, zatim će biti analiziran uticaj aditivnog Gausovog šuma na performanse CS rekonstrukcije. Dodatno, biće izveden eksplicitni izraz za grešku u rekonstrukciji signala koji nijesu rijetki, ali koji su rekonstruisani pod pretpostavkom da jesu rijetki u DHT1 domenu. Analiza će biti potkrijepljena većim brojem numeričkih i statističkih eksperimenata. Na kraju, biće prezentovana originalna interpretacija gradijentnog algoritma za rekonstrukciju signala rijetkih u Hermitskom transformacionom domenu. Glavni rezultati ove sekcije su publikovani u radovima \cite{ht3} i \cite{brajovic_ht1}.

 \subsection{Uticaj nedostajućih odbiraka}
Razmatra se DHT1 signala $x(t_n)$, gdje tačke odabiranja $t_n~,n=1,2,\dots,N$, odgovaraju nulama Hermitskog polinoma reda $N$. Tada se Hermitski koeficijenti računaju primjenom Gaus-Hermitske kvadrature (\ref{dht1}), odnosno,
\begin{align}
{C(p)}=\frac{1}{N}\sum\limits_{n=1}^{N}{\frac{{{\psi }_{p}}(t_n)}{{{({{\psi }_{N-1}}(t_n))}^{2}}}x(t_n)},
\end{align}
pri čemu  je uzeto da $\sigma=1$ bez gubljenja opštosti izlaganja. Budući da je pokazano da parametar $\sigma$ može biti inkorporiran u razmatrani signal, ali i zbog činjenice da je irelevantnan sa stanovišta  ortogonalnosti Hermitskih funkcija, on će biti izostavljen iz notacije u ovoj podsekciji, uz pretpostavljenu vrijednost $\sigma=1$.

Pretpostavlja se da je posmatrani signal $x(t_n)$ rijedak u Hermitskom transformacionom domenu (DHT1), što podrazumijeva da se on može zapisati na sljedeći način:
\begin{align}
x(t_n)=\sum\limits_{l=1}^{K}{{{A}_{l}}{{\psi }_{{{p}_{l}}}}(t_n)},
\label{htsparse}
\end{align}
gdje je $K\ll N$ broj nenultih komponenti signala, odnosno stepen rijetkosti, dok  su $A_l$ amplitude komponenti signala i $p_l$ indeksi njihovih pozicija u Hermitskom transformacionom domenu. Skup pozicija nenultih komponenti biće označen sa ${\Pi}_K=\{{{p}_{1}},\,{{p}_{2}},\dots,{{p}_{K}}\}$. Za multikomponentni signal (\ref{htsparse}), Hermitski koeficijenti zadati su sljedećim izrazom:
\begin{align}
{C(p)}=\sum\limits_{n=1}^{N}{\sum\limits_{l=1}^{K}{\frac{{{A}_{l}}}{N}\frac{{{\psi }_{p}}(t_n){{\psi }_{{{p}_{l}}}}(t_n)}{{{({{\psi }_{N-1}}(t_n))}^{2}}}}},\,p=0,1,\dots,N-1.
\end{align}
Normalizovane komponente signala se množe sa baznim funkcijama ${{\psi }_{p}}(t_n)/{{({{\psi }_{N-1}}(t_n))}^{2}}$, dajući signal $z(p_l,p,t_{n_i})$ definisan izrazom:
\begin{align}
z(p_l,p,t_{n})=\frac{{{A}_{l}}}{N}\frac{{{\psi }_{p}}(t_n){{\psi }_{{{p}_{l}}}}(t_n)}{{{({{\psi }_{N-1}}(t_n))}^{2}}}.
\end{align}

Vrijednosti signala $z(p_l,p,t_{n})$ pripadaju skupu
\begin{align}\mathbf{\Omega }=\left\{ \frac{{{A}_{l}}}{N}\frac{{{\psi }_{p}}(t_n){{\psi }_{{{p}_{i}}}}(t_n)}{{{\left( {{\psi }_{N-1}}(t_n) \right)}^{2}}},~n=1,2,\dots,N,~l=1,2,\dots,K,~p=0,1,\dots,N-1 \right\}.
\end{align}

Imajući u vidu svojstvo ortogonalnosti
\begin{equation}
\frac{A_l}{N}\sum\limits_{n=1}^{N}{\frac{{{\psi }_{p}}({{t}_{n}})}{{{({{\psi }_{N-1}}({{t}_{n}}))}^{2}}}{{\psi }_{p_l}}({{t}_{n}})}=A_l\delta (p-p_l),
\label{htort}
\end{equation} 
slijedi da članovi skupa $\mathbf{\Omega }$ zadovoljavaju relaciju:
\begin{align}
\sum\limits_{n=1}^{N}z(p_l,p,t_{n})=z(p_l,p,t_{1})+z(p_l,p,t_{2})+\dots+z(p_l,p,t_{N})=0
\label{hto1}
\end{align}	
za  $p\ne {{p}_{l}},~l=1,2,\dots,K$, odnosno,
\begin{align}
\sum\limits_{n=1}^{N}z(p_l,p,t_{n})=z(p_l,p,t_{1})+z(p_l,p,t_{2})+\dots+z(p_l,p,t_{N})=A_l
\label{hto2}
\end{align}	
za  $p={{p}_{l}},~l=1,2,\dots,K$.

Neka je dostupno $N_A<N$ odbiraka signala (nedostupno $N_Q=N-N_A$), tako da formiraju vektor:
\begin{gather}
\mathbf{y}=\left\{  x(t_{n_{1}}),x(t_{n_{2}}),\ldots,x(t_{n_{N_A}})\right\}  \subseteq
\mathbf{x}=\left\{x(t_1),x(t_2),\dots,x(t_N)\right\}%
\end{gather}
gdje je
$
x(t_{n_i})=\sum_{p=0}^{N}{{{C(p)}}{{\psi }_{p}}(t_{n_i} )}$, uz $i=1,2,\dots,N_A,
$
na pozicijama  $t_n\in {{\mathbb{N}}_{{A}}}=\{t_{n_1},t_{n_2},\dots,t_{n_{N_A}}\}\subseteq \mathbb{N}=\{t_1,t_2,\dots,t_N\}$. Skup pozicija nedostupnih odbiraka je podskup skupa pozicija svih dostupnih odbiraka $\mathbb{N}_Q=\mathbb{N\setminus}\mathbb{N}_A$ i prećutno je podrazumijevana uniformna funkcija gustine raspodjele vjerovatnoća pozicija odbiraka  $n_i\in \mathbb{N}_A$. U matričnom zapisu, dostupni odbirci se mogu zapisati u formi
\begin{equation}\mathbf{y}=\mathbf{AC},\end{equation}
gdje $\mathbf{A} $ predstavlja $N_A\times N$ matricu mjerenja, dok je $\mathbf{C}=[C(0),C(1),\dots,C(N-1)]^T$ vektor Hermitskih koeficijenata signala koji odgovara vektoru signala $\mathbf{x}$. Mjerna matrica je parcijalna inverzna DHT1 matrica, čije su vrste jednake onim vrstama matrice $\mathbf{T}_H^{-1}$, koje odgovaraju pozicijama dostupnih odbiraka $t_{n_i},~i=1,2,\dots,N_A$:
\begin{equation}\mathbf{A}=\left[ \begin{matrix}
{{\psi }_{0}}({{t}_{n_1}}) & {{\psi }_{1}}({{t}_{n_1}}) & \ldots  & {{\psi }_{N-1}}({{t}_{n_1}})  \\
{{\psi }_{0}}({{t}_{n_2}}) & {{\psi }_{1}}({{t}_{n_2}}) & \cdots  & {{\psi }_{N-1}}({{t}_{n_2}})  \\
\vdots  & \vdots  & \ddots  & \vdots   \\
{{\psi }_{0}}({{t}_{n_{N_A}}}) & {{\psi }_{1}}({{t}_{n_{N_A}} }) & \cdots  & {{\psi }_{N-1}}({{t}_{n_{N_A}}})  \\
\end{matrix} \right].
\end{equation}

Inicijalna DHT1 zasnovana na $\ell_2$-normi ima sljedeću formu:
\begin{align}
{C_0(p)}= \sum_{t_n \in \mathbb{N}_A}\sum\limits_{l=1}^{K} z(p_l,p,t_n),
\end{align}
ili u matričnom zapisu:
$
\mathbf{C}_{0}=\mathbf  {A}^{T}\mathbf{y}.%
$
U slučaju kompresivnog odabiranja, razmatra se, dakle, podskup skupa $\mathbf{\Omega}$ od  ${{N}_{A}}\le N$ slučajno pozicioniranih odbiraka:
\begin{align}
\mathbf{\Theta }=\{ z(p_l,p,t_{n_1}),z(p_l,p,t_{n_2}),\dots,z(p_l,p,t_{n_{N_A}})\}\subseteq \mathbf{\Omega }.
\end{align}
 Budući da je DHT1 linearna transformacija u kojoj se vrši unutrašnje množenje  (engl. \textit{inner product}) signala sa baznim funkcijama, nedostupni odbirci u signalu daju isti rezultat kao kada su ti odbirci jednaki nuli. Kao posljedica toga, redukovani broj obiraka (mjerenja) se može posmatrati kao skup svih odbiraka, u kojem su neki od njih oštećeni aditivnim šumom:
\begin{align}
\eta (p_l,p,t_n)=\left\{ \begin{matrix}
-z(p_l,p,t_n),&t_n\in \mathbb{N}_Q  \\
0,&~~t_n\in {{\mathbb{N}}_{{A}}}.  \\
\end{matrix} \right.
\end{align}

Inicijalna DHT1 (zasnovana na $\ell_1$ normi) se na bazi dostupnih odbiraka može izraziti kao: 
\begin{align}
{C_0(p)}&= \sum_{t_n \in \mathbb{N}_A}\sum\limits_{l=1}^{K} z(p_l,p,t_n)=\sum\limits_{n=1}^{N}\sum\limits_{l=1}^{K}{[z(p_l,p,t_n)+\eta (p_l,p,t_n)]},
\end{align}
za $p=0,1,\dots N-1$. Ovo je slučajna promjenljiva.

\subsubsection{Statističke karakteristike inicijalne DHT1 estimacije}

\paragraph{Monokomponentni signali.} Posmatra se slučaj jednokomponentnog signala sa $K=1$, ${{A}_{l}}=1$ i ${{p}_{l}}={{p}_{1}}$. Inicijalna DHT1  se na bazi dostupnih odbiraka može izračunati na sljedeći način: 
\begin{align}
{C_0(p)}= \sum_{t_n \in \mathbb{N}_A} z(p_1,p,t_n)= \sum_{i=1}^{N_A} z(p_1,p,t_{n_i}),
\end{align}
za $p=0,1,\dots N-1$. Ona je zasnovana na $\ell_2$ normi. Ovo je slučajna varijabla, formirana kao suma $N_A$ slučajno pozicioniranih odbiraka. Po centralnoj graničnoj teoremi, varijabla ${C_0(p)}$ ima normalnu (Gausovu) distribuciju. Određivanje srednje vrijednosti i varijanse ove slučajne promjenljive je predmet daljeg izlaganja. Navedena varijabla ima drugačije statističke osobine na pozicijama $p=p_1$ od statističkih osobina na pozicijama $p\neq p_1$. Pozicije $p=p_1$ će u daljem izlaganju biti označene kao pozicije komponenti signala, dok će pozicije $p\neq {p}_{1}$ biti označene kao pozicije (CS) šuma, jer one ne odgovaraju komponentama signala.

\vspace{2mm}
\noindent\emph{Slučaj kada je $p\neq {p}_{1}$.} Za Hermitske koeficijente koji ne odgovaraju komponentama signala, može se smatrati da korespondiraju aditivnom šumu u transformacionom domenu. Imajući u vidu da su posmatrani odbirci $z(p_1,p,t_{n_i}),~i=1,2,\dots,N_A$ slučajno pozicionirani (sa uniformnom raspodjelom), na osnovu (\ref{hto1}) se može zaključiti da je $E\left\{ z(p_1,p,t_{n_i})\right\}=0$, gdje se operator matematičkog očekivanja primjenjuje po vremenskoj varijabli $t_{n_i}$. Stoga je  srednja vrijednost posmatrane slučajne varijable jednaka nuli:
\begin{align}
{{\mu }_{C_0(p)}}&=E\left\{C_0(p)\right\}=E\left\{ \sum\limits_{i=1}^{{{N}_{A}}}z(p_1,p,t_{n_i}) \right\}=0,~p\neq p_1.
\end{align}	

Varijansa slučajne promjenljive $C_0(p),~p\neq p_l$ je definisana  sljedećim izrazom:
\begin{align}
\sigma_{C_0(p)}^{2}&=E\left\{ |C_0(p){{|}^{2}} \right\}=E\left\{ \sum\limits_{i=1}^{{{N}_{A}}}{\sum\limits_{j=1}^{{{N}_{A}}}{z(p_1,p,t_{n_i})z(p_1,p,t_{n_j})}} \right\} \notag\\ 
& =\underbrace{E\left\{ \sum\limits_{\begin{smallmatrix} 
		i=1 \\ 		
		\end{smallmatrix}}^{{{N}_{A}}}z^2(p_1,p,t_{n_i}) \right\}}_{S1}+\underbrace{E\left\{ \sum\limits_{i=1}^{{{N}_{A}}}{\sum\limits_{\underset{i\ne j}{\mathop{j=1}}\,}^{{{N}_{A}}}{z(p_1,p,t_{n_i})z(p_1,p,t_{n_j})}} \right\}}_{S2}.  
\label{ht_cs_n_s1}
\end{align}

Množeći lijevu i desnu stranu izraza (\ref{hto1}) sa $z(p_l,p,t_{n_i}),~i=1,2,\dots,N_A$ i primjenom operatora matematičkog očekivanja na lijevu i desnu stranu ovog izraza, dobija se: 
\begin{gather}
E\left\{z(p_l,p,t_{n_i})[ z(p_l,p,t_{1})+z(p_l,p,t_{2})+\dots+z(p_l,p,t_{N})]\right\}=0,\notag
\\
E\left\{z(p_l,p,t_{n_i})z(p_l,p,t_{1})\right\}+\dots+E\left\{z(p_l,p,t_{n_i})z(p_l,p,t_{N})\right\} =0,
\label{htn_iz}
\end{gather}
za $i=1,2,\dots,N_A$. Vrijednosti $z(p_l,p,t_{n_i})$ su jednako distribuirane, pa su očekivanja $E\left\{z(p_l,p,t_{n_i})z(p_l,p,t_{j})\right\},~ i\neq j,~t_{n_i}\in\mathbb{N}_A,~t_{j}\in\mathbb{N}$ ista i jednaka konstanti $B$. Za $n_i=j$ i $p\neq p_1$ očekivanje postaje:
\begin{align}
E\{z^2(p_l,p,t_{n_i})\}&=E\left\{ \frac{1}{{{N}^{2}}}\frac{{{\psi }_{p}}({t_{n_i}}){{\psi }_{p}}({t_{n_i}})}{{{({{\psi }_{N-1}}({t_{n_i}}))}^{2}}}\frac{{{\psi }_{{{p}_{1}}}}({t_{n_i}}){{\psi }_{{{p}_{1}}}}({t_{n_i}})}{{{({{\psi }_{N-1}}({t_{n_i}}))}^{2}}} \right\} \notag \\
&=E\left\{ \frac{1}{N}\frac{{{\psi }_{p}}({t_{n_i}}){{\psi }_{p}}({t_{n_i}})}{{{({{\psi }_{N-1}}({t_{n_i}}))}^{2}}} \right\}E\left\{ \frac{1}{N}\frac{{{\psi }_{{{p}_{1}}}}({t_{n_i}}){{\psi }_{{{p}_{1}}}}({t_{n_i}})}{{{({{\psi }_{N-1}}({t_{n_i}}))}^{2}}} \right\},
\end{align}
gdje je iskorišćena statistička nezavisnost vrijednosti Hermitskih funkcija reda $p_1$ i $p$ po $t_{n_i}$, pa su matematička očekivanja razdvojena. Dalje, usljed ortogonalnosti (\ref{hto2}) slijedi:
\begin{align}
E\left\{ \frac{1}{N}\frac{{{\psi }_{p}}({t_{n_i}}){{\psi }_{p}}({t_{n_i}})}{{{({{\psi }_{N-1}}({t_{n_i}}))}^{2}}} \right\}=E\left\{ \frac{1}{N}\frac{{{\psi }_{{{p}_{1}}}}({t_{n_i}}){{\psi }_{{{p}_{1}}}}({t_{n_i}})}{{{({{\psi }_{N-1}}({t_{n_i}}))}^{2}}} \right\}=\frac{1}{N}.
\end{align}

Dakle, zaključuje se da važi
\begin{align}
E\{z^2(p_l,p,t_{n_i})\}=E\{z(p,p,t_{n_i})\}E\{z(p_l,p_l,t_{n_i})\}=\frac{1}{{N}^{2}}. \label{ht_izvodjenje3}
\end{align}

Pošto u izrazu (\ref{htn_iz}) postoji $N-1$ očekivanja jednakih $B$ za  $n_i\neq j$ i jedno očekivanje dato izrazom (\ref{ht_izvodjenje3}) za $n_i=j$, izraz (\ref{htn_iz}) dalje postaje:
\begin{align} \frac{1}{N^2}+(N-1)B=0.\end{align}
Odavde je dalje moguće izraziti nepoznatu vrijednost $B$ u obliku
$ E\left\{z(p_l,p,t_{n_i})z(p_l,p,t_{j})\right\}=B=\frac{-1}{{{N}^{2}}(N-1)},~n_i\ne j$. Budući da u članu S1 izraza (\ref{ht_cs_n_s1}) postoji $N_A$ sumanada, dok u članu S2 postoji $N_A(N_A-1)$ sumanada, traženi izraz za varijansu Hermitskih koeficijenata u slučaju signala sa nedostajućim odbircima, na pozicijama koeficijenata $p\neq p_1$ koje ne odgovaraju komponentama signala postaje
$ \sigma_{C_0(p)}^{2}=\frac{{{N}_{A}}}{{{N}^{2}}}+{{N}_{A}}({{N}_{A}}-1)B,$
odnosno, nakon preuređivanja i uvođenja oznake $\sigma_{csN}^{2}=\sigma_{C_0(p)}^{2}$ za $p\ne p_1$:
\begin{equation}
\sigma_{csN}^{2}=\frac{{{N}_{A}}(N-N_{A})}{{{N}^{2}}(N-1)}.
\label{ht_noisecs}
\end{equation}

Može se zaključiti da varijansa Hermitskog transformacionog šuma u koeficijentima na pozicijama $p\ne {{p}_{1}}$ koje ne odgovaraju komponentama signala   zavisi samo od broja dostupnih odbiraka $N_A$ i dužine orignalnog signala $ N$. Prema centralnoj graničnoj teoremi, slučajna varijabla $C_0(p)$ za $p\neq p_1$ ima normalnu (Gausovu) raspodjelu. 

\vspace{2mm}

\noindent \emph{Slučaj kada je $p=p_1$}. Statističke osobine Hermitskih koeficijenata se u slučaju pozicija $p = {p}_{1}$, koje odgovaraju komponentama signala, značajno razlikuju od osobina koeficijenata na pozicijama $p\neq p_1$. Budući da proizvod ${{\psi }_{{{p}_{1}}}}({t_{n_i}}){{\psi }_{{{p}_{1}}}}({t_{n_i}})$ u razmatranom slučaju zavisi od vrijednosti specifičnih Hermitskih funkcija  ${{\psi }_{{{p}_{1}}}}({t_{n_i}})$, čiji su odbirci dostupni na slučajnim pozicijama, može se zaključiti da su koeficijenti $C_0(p),~p\neq p_1$ takođe slučajne varijable sa normalnom distribucijom, što je i u ovom slučaju posljedica centralne granične teoreme.

Na osnovu definicije signala $z(p_1,p_1,t_{n})$ iz skupa $\mathbf{\Theta }$ sa $N_A$ slučajno pozicioniranih dostupnih odbiraka, srednja vrijednost slučajne varijable $C_0(p)~p=p_1$ (definisana kao suma tih signala) direktno slijedi kao posljedica principa ortogonalnosti (\ref{hto2}):
\begin{align}{{\mu }_{C_0(p)}}=\frac{1}{N}E\left\{ \sum\limits_{i=1}^{{{N}_{A}}}{\frac{{{\psi }_{{{p}_{1}}}}({t_{n_i}}){{\psi }_{{{p}_{1}}}}({t_{n_i}})}{{{({{\psi }_{N-1}}({t_{n_i}}))}^{2}}}} \right\}=\frac{{{N}_{A}}}{N},~p=p_1,
\label{htmono_srvr}
\end{align}	
budući da su vrijednosti jednako distribuirane. 

Za slučajnu promjenljivu sa nenultom srednjom vrijednošću i normalnom raspodjelom, varijansa se definiše u obliku:
\begin{align}\sigma_{C_0(p)}^{2}=E\left\{ \sum\limits_{i=1}^{{{N}_{A}}}z^2(p_1,p,t_{n_i}) \right\}+E\left\{ \sum\limits_{i=1}^{{{N}_{A}}}{\sum\limits_{\underset{i\ne j}{\mathop{j=1}}\,}^{{{N}_{A}}}z(p_1,p,t_{n_i})z(p_1,p,t_{n_j})} \right\}-\left|{{\mu }_{C_0(p)}}\right|^{2}\label{pomhtht},\end{align}
gdje je $p=p_1$. Računanje pojedinačnih djelova u (\ref{pomhtht}) će se razlikovati od prethodno razmatranog slučaja za ($p\ne {{p}_{1}}$). Polazeći od svojstva ortogonalnosti (\ref{hto2}) uz $A_1=1$, množenjem lijeve i desne strane izraza sa  $z(p_1,p,t_{n_i}),~t_{n_i}\in \mathbb{N}_A$ i primjenom operatora matematičkog očekivanja dobija se: 
\begin{gather}
E\{z(p_1,p,t_{n_i})z(p_l,p,t_{1})\}+\dots+E\{z(p_1,p,t_{n_i})z(p_l,p,t_{N})\}=E\{z(p_1,p,t_{n_i})\},
\notag
\\
E\{z(p_1,p,t_{n_i}) z(p_l,p,t_{1})\}+\dots+E\{z(p_1,p,t_{n_i}) z(p_l,p,t_{N})\}=\frac{1}{N}, \label{pomhtht2}
\end{gather}

Za $n_i\ne j,~t_{n_i}\in\mathbb{N}_A,~t_{j}\in\mathbb{N}$, važiće:
\begin{gather}
E\{z(p_1,p,t_{n_i})\}E\{z(p_l,p,t_{1})\}=\dots=E\{z(p_1,p,t_{n_i})\}E\{z(p_l,p,t_{N})\}\notag \\
=E\{z(p_1,p,t_{n_i})\}E\{z(p_l,p,t_{j})\}=D.
\end{gather}

Za $n_i=j$, matematičko očekivanje $E\{z^2(p_1,p,t_{n_i})\}$ ne može biti estimirano u obliku proizvoda $E\{z(p_1,p,t_{n_i})\}E\{z(p_l,p,t_{j})\}$, budući da članovi nijesu statistički nezavisni. Stoga, u cilju određivanja  $E\{z^2(p_1,p,t_{n_i})\}$ posmatraće se sljedeća suma:
\begin{align}
\sum\limits_{i=1}^{N}{z^2(p_1,p,t_{n_i})}  &=\frac{1}{{{N}^{2}}}{{\sum\limits_{i=1}^{N}{\left( \frac{\psi _{{{p}_{1}}}^{2}({t_{n_i}})}{{{({{\psi }_{N-1}}({t_{n_i}}))}^{2}}} \right)^{2}}}} =\frac{1}{{{N}^{2}}}\sum\limits_{i=1}^{N}{a({t_{n_i}},{{p}_{1}},N)}=\frac{1}{{{N}^{2}}}{{P}_{{{p}_{1}}}},
\end{align}
što odgovara energiji monokomponentnog signala definisanog Hermitskom funkcijom reda ${p}_{1}$. Treba uočiti da je uvedena notacija
${{\left( \frac{\psi _{{{p}_{1}}}^{2}({t_{n_i}})}{{{({{\psi }_{N-1}}({t_{n_i}}))}^{2}}} \right)}^{2}}=z^2(p_1,p,t_{n_i})=a({t_{n_i}},{{p}_{1}},N)$,
gdje su ${t_{n_i}}\in \mathbb{N}_A$ slučajne pozicije $N_A$ dostupnih odbiraka. Pokazuje se da važi
$E\left\{ a({t_{n_i}},{{p}_{1}},N) \right\}= a({{p}_{1}},N) =\frac{{P}_{{{p}_{1}}}}{{N}^{3}}$.
Sada (\ref{pomhtht2}) postaje:
$a({{p}_{1}},N)+(N-1)D=\frac{1}{N}, $
dok se nepoznato $D$ može izraziti u obliku
$D=\frac{1-Na({{p}_{1}},N)}{N(N-1)}$.
Varijansa (\ref{pomhtht}) razmatrane slučajne promjenljive $C_0(p), ~p\ne p_1$ se sada može zapisati u obliku:
\begin{align}
\sigma_{C_0(p)}^{2} &={{N}_{A}}a({{p}_{1}},N)+{{N}_{A}}\left( {{N}_{A}}-1 \right)D-{{\left( \frac{{{N}_{A}}}{N} \right)}^{2}} \notag \\ 
& ={{N}_{A}}a({{p}_{1}},N)+{{N}_{A}}\left( {{N}_{A}}-1 \right)\frac{1-Na\left( {{p}_{1}},N \right)}{N(N-1)}-{{\left( \frac{{{N}_{A}}}{N} \right)}^{2}}. 
\end{align}

Nakon jednostavnog sređivanja prethodnog izraza, varijansa se može predstaviti tako da se dovede u vezu sa prethodno određenim izrazom (\ref{ht_noisecs}) za $\sigma^2_{csN}$, na sljedeći način:
\begin{align}
\sigma_{C_0(p_1)}^{2}&=\frac{N{{N}_{A}}-N_{A}^{2}}{{{N}^{2}}(N-1)}\left( {{N}^{2}}a({{p}_{1}},N)-1 \right) \notag \\ 
& =\sigma _{csN}^{2}\left( \frac{1}{N}{{\sum\limits_{{t_{i}}=1}^{N}{\left( \frac{\psi _{{{p}_{1}}}^{2}({t_{n_i}})}{{{({{\psi }_{N-1}}({t_{n_i}}))}^{2}}} \right)}}^{2}}-1 \right)=\sigma _{csN}^{2}\left( \frac{{{P}_{{{p}_{1}}}}}{N}-1 \right). \label{varht_com}
\end{align}
\begin{primjer}
Relacija (\ref{varht_com}), koja opisuje na koji način varijansa zavisi od pozicije $p_1$ komponente signala u Hermitskom transformacionom domenu, verifikovana je i numerički. Rezultati su prikazani na slici \ref{Fig_vars_p0} za slučaj signala dužine  $N = 200$, sa $N_A = 120$ dostupnih odbiraka. Numerički proračun varijanse je dobijen na bazi 5000 nezavisnih realizacija signala sa slučajno pozicioniranim nedostajućim odbircima.

   \begin{figure}[!tb]%
	\centering
	\includegraphics[
	]%
	{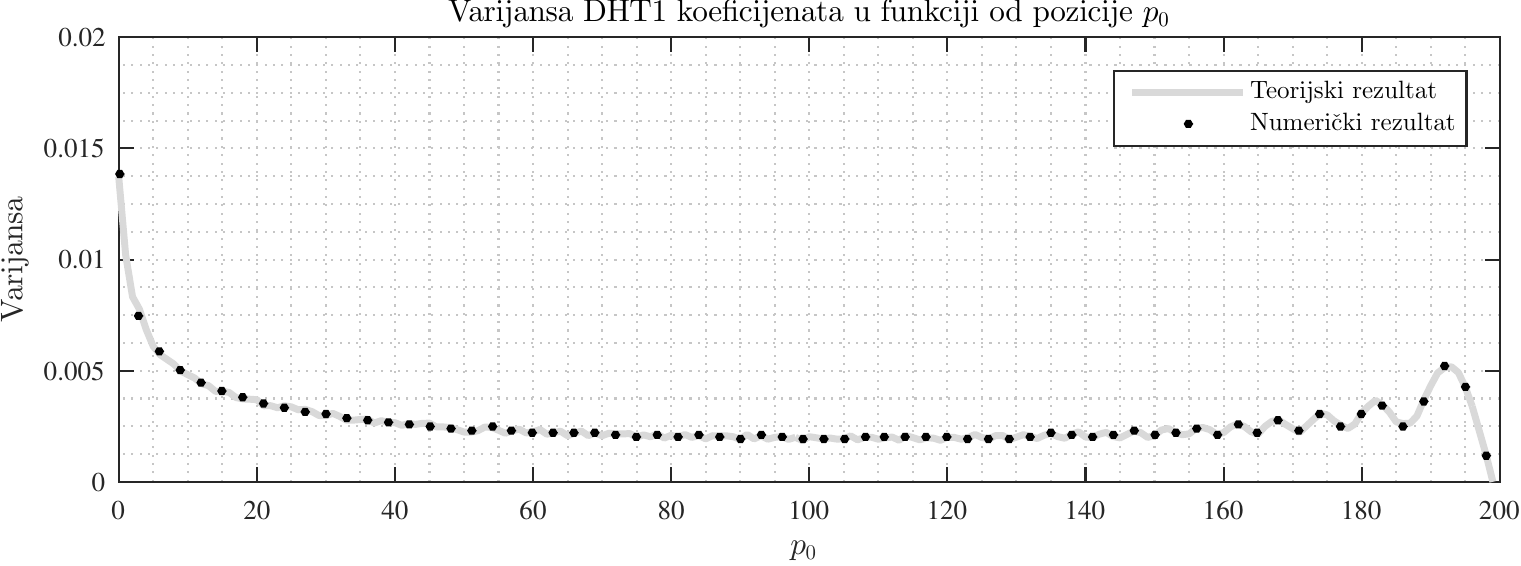}%
	\caption[Varijansa na poziciji komponente signala, kao funkcija od indeksa ${p}_{1}.$]{Varijansa na poziciji komponente signala, kao funkcija od indeksa ${p}_{1}$.}%
	\label{Fig_vars_p0}%
\end{figure}
\end{primjer}
Kada je dostupno samo $N_A$ od ukupno  $N$ odbiraka, očekivani \textit{bias} u amplitudi se može kompenzovati sa $\frac{N}{N_A}$, dok član ${{P}_{{{p}_{1}}}} $ može biti estimiran na osnovu dostupnih odbiraka u obliku:
\begin{align}{{\tilde{P}}_{{{p}_{1}}}}={{\sum\limits_{i=1}^{{{N}_{A}}}{\left( \frac{\psi _{{{p}_{1}}}^{2}({t_{n_i}})}{{{({{\psi }_{N-1}}({t_{n_i}}))}^{2}}} \right)}}^{2}}.
\end{align}
Korišćenjem ove estimacije, varijansa Hermitskog koeficijenta $C_0(p)$ na poziciji komponente signala $p={p}_{1}$ u slučaju nedostajućih odbiraka može se aproksimirati na sljedeći način:
\begin{align}\tilde{\sigma }_{C_0(p_1)}^{2}=\frac{N}{{{N}_{A}}}\sigma _{csN}^{2}\left( \frac{{{{\tilde{P}}}_{{{p}_{1}}}}}{N}-1 \right).
\label{var_monoaprox}
\end{align}

Za amplitude $A_1\ne 1$, dobijene izraze za srednje vrijednosti je potrebno pomnožiti sa $A_1$, a dobijene izraze za varijanse je potrebno pomnožiti sa $A_1^2$.

\paragraph{Analiza greške u detekciji komponente signala.} Na osnovu predstavljene teorije o stohastičkim karakteristikama Hermitskih koeficijenata monokomponentnih signala rijetkih u DHT1 domenu, moguće je izvršiti probabilističku analizu greške u detekciji Hermitskog koeficijenta koji predstavlja komponentu signala, odnosno njegovog razdvajanja od ostalih koeficijenata. Prema centralnoj graničnoj teoremi, uspostavljeno je da je slučajna varijabla $C_0(p)$ sa Gausovom raspodjelom. Srednje vrijednosti i varijanse su različite za indekse $p=p_1$ i $p\ne p_1$. Za $p\ne p_1$ raspodjela je: $ \mathcal{N}(0,\sigma_{csN}^{2})$, dok za $p=p_1$ važi raspodjela $
\mathcal{N}(\frac{N_A}{N}A_{1},\sigma^{2}_{{C_0}(p_1)})$. Izvedene statističke karakteristike koeficijenata će, između ostalog, 
 poslužiti u definisanju metoda za razdvajanje komponenti signala od šuma u transformacionom domenu indukovanog nedostajućim odbircima u signalu. Navedeni postupak se može posmatrati kao detekcija komponente signala. U tu svrhu, posmatraće se apsolutne vrijednosti slučajnih varijabli $C_0(p)$ za $p=p_1$ i $p\ne p_1$.

\begin{figure}[ptb]%
	\centering
	\includegraphics[
	]%
	{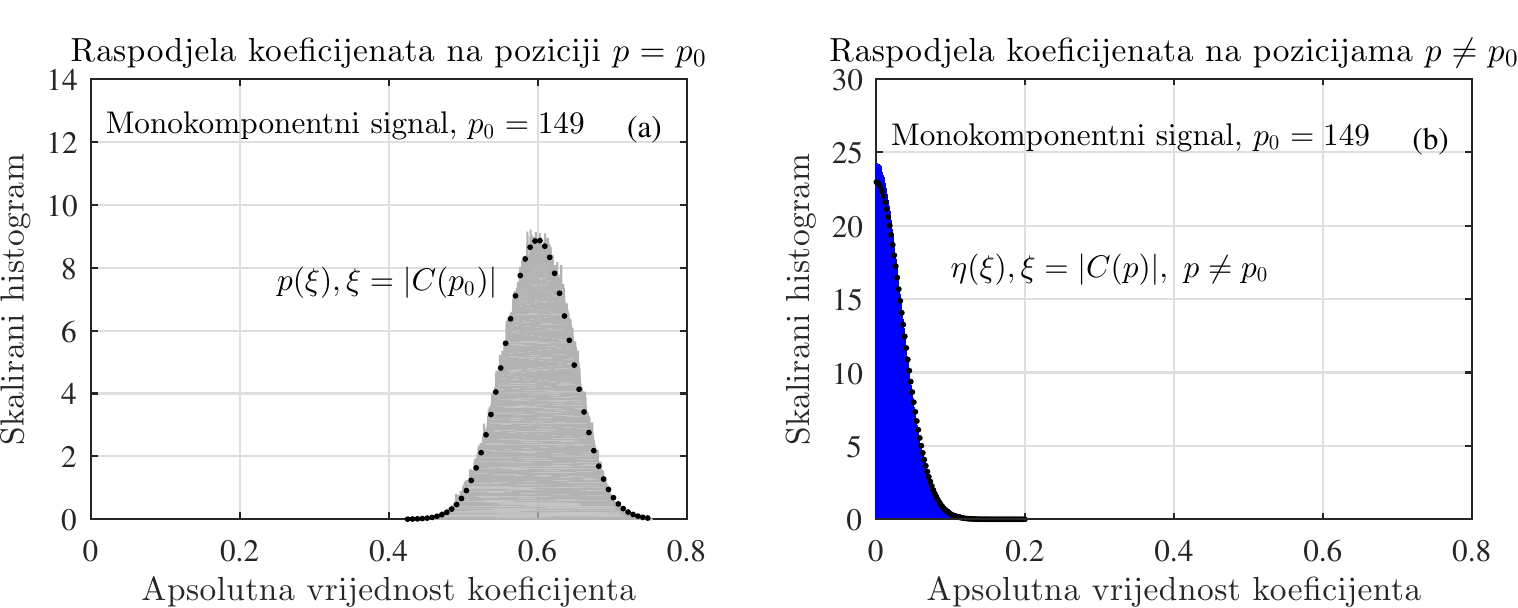}%
	\caption[Histogrami i funkcije gustine raspodjele za Hermitske koeficijente monokomponentnog signala]{Histogrami i funkcije gustine raspodjele za Hermitske koeficijente monokomponentnog signala na: (a) pozicijama koje odgovaraju komponentama signala i (b) na ostalim pozicijama. Histogrami ((a) siva površina za koeficijent na poziciji komponente signala (b) plava površina za koeficijent mimo pozicije signala) su simulirani sa $N_A=120$ od $N=200$ dostupnih odbiraka, amplitudom $A_0=1$, na bazi 20000 nezavisnih realizacija signala sa slučajno pozicioniranim nedostajućim odbircima. Teorijski rezultati su dobijeni korišćenjem iskrivljene normalne distribucije (\ref{folded1koef}) računate sa varijansom (\ref{varht_com}) i polunormalne distribucije (\ref{halfnorht})  sa varijansom (\ref{ht_noisecs}).}%
	\label{Fig_histogramsMONO_HT}%
\end{figure}

Apsolutna vrijednost slučajne varijable $C(p),~p=p_1$ sa raspodjelom $
\mathcal{N}(\frac{N_A}{N}A_{1},\sigma^{2}_{{C_0}(p_1)})$ sa srednjom vrijednošću (\ref{htmono_srvr}) i varijansom (\ref{varht_com})
ima ,,savijenu'' normalnu distribuciju (engl. \textit{folded normal distribution}):
\begin{align}p(\xi)=\frac{1}{\sigma_{{C_0}(p_1)}\sqrt{2\pi }}\left( \exp \left( -\frac{{{(\xi -\frac{N_A}{N}A_{1})}^{2}}}{2\sigma^{2}_{{C_0}(p_1)}} \right)+\exp \left( -\frac{{{(\xi +\frac{N_A}{N}A_{1})}^{2}}}{2\sigma^{2}_{{C_0}(p_1)}} \right) \right),
\label{folded1koef}
\end{align}	
gdje je slučajna promjenljiva $\xi=|C(p)|$. Raspodjela je prikazana na slici \ref{Fig_histogramsMONO_HT} (a). Na istoj slici prikazan je i numerički dobijeni histogram, baziran na 20000 nezavisnih realizacija jednokomponentnog signala rijetkog u Hermitskom domenu, sa slučajno pozicioniranim nedostajućim odbricima.

Slučajna varijabla koja odgovara šumu u Hermitskom transformacionom domenu, $C(p),~p \ne p_1$, ima takođe normalnu raspodjelu $ \mathcal{N}(0,\sigma_{csN}^{2})$. Njena aposolutna vrijednost, $\xi =\left| C_0(p), \right|~p\ne p_1$  podliježe polunormalnoj distribuciji (engl. \textit{half-normal distribution}):
\begin{align}\eta (\xi )=\frac{\sqrt{2}}{\sigma_{csN}\sqrt{\pi }}\exp \left( -\frac{{{\xi }^{2}}}{2\sigma_{csN}^{2}} \right),
\label{halfnorht}
\end{align}
gdje je varijansa $\sigma_{csN}^{2}$ data izrazom (\ref{ht_noisecs}). Navedena distribucija, sa eksperimentalno dobijenim histogramom (u istom eksperimentu sa 20000 ponovljenih realizacija signala sa slučajno pozicioniranim nedostajućim odbircima), prikazana je na slici \ref{Fig_histogramsMONO_HT} (b).

Vjerovatnoća da je slučajna promjenljiva $\xi =\left| C(p) \right|,~p\ne p_1$ manja od vrijednosti $\Xi $ je data izrazom:
\begin{align}{{P}_{N}}(\Xi )=\int_{0}^{\Xi }{\frac{\sqrt{2}}{{\sigma_{csN}}\sqrt{\pi }}\exp \left( -\frac{{{\xi }^{2}}}{2{\sigma^{2}_{csN}}} \right)d\xi =}\operatorname{erf}\left( \frac{\Xi }{\sqrt{2}{\sigma_{csN}}} \right).
\end{align}	

Ukupan broj Hermitskih koeficijenata koji ne odgovaraju komponentama signala je $N-1$. Vjerovatnoća da su $N-1$ nezavisnih koeficijenata šuma manji od praga $\Xi $ je:
\begin{align}{{P}_{NN}}(\Xi )=\operatorname{erf}{{\left( \frac{\Xi }{\sqrt{2}{\sigma_{csN}}} \right)}^{N-1}}.
\end{align}
Vjerovatnoća da je najmanje jedan šumni koeficijent veći od $\Xi $ iznosi ${{P}_{NL}}(\Xi )=1-{{P}_{NN}}(\Xi )$. U terminologiji standarde teorije detekcije, nulta hipoteza ${{\mathcal{H}}_{0}}$ može biti formulisana kao: ${C(p)}$ je Hermitski koeficijent koji odgovara čistom šumu, odnosno, ${{\mathcal{H}}_{0}}:{C(p)},~p\ne p_1$. Druga hipoteza može biti formulisana u obliku: ${C(p)}$ je Hermitski koeficijent na poziciji signala, odnosno, ${{\mathcal{H}}_{1}}:{C(p)},~p=p_1$. 

Ako je vrijednost koeficijenta signala u opsegu od $\xi $ do $\xi +d\xi $ sa vjerovatnoćom $x(\xi )d\xi$, on neće biti detektovan ukoliko je najmanje jedan Hermitski koeficijent šuma iznad vrijednosti $\xi $. Navedeno će se desiti sa vjerovatnoćom
${{P}_{NL}}(\xi )x(\xi )d\xi =(1-{{P}_{NN}}(\xi ))x(\xi )d\xi.$ Razmatranjem svih mogućih vrijednosti $\xi $, greška u detekciji će nastupiti sa vjerovatnoćom:
\begin{align}
{{P}^{(1)}_{E}}&=\int_{0}^{\infty }{(1-{{P}_{NN}}(\xi ))x(\xi )d\xi }\notag\\
&=\frac{1}{{\sigma_{C_0(p_1)}}\sqrt{2\pi }}\int\limits_{0}^{\infty }{\left( 1-\operatorname{erf}{{\left( \frac{\xi }{\sqrt{2}{\sigma_{csN}}} \right)}^{N-1}} \right)} \notag \\
&~~~~~~~~~~~~~~~~~~~~~~~~~~\times \left[ \exp \left( -\frac{{{(\xi -{\frac{N_A}{N}A_{1}})}^{2}}}{2\sigma^{2}_{C_0(p_1)}} \right)+\exp \left( -\frac{{{(\xi +{\frac{N_A}{N}A_{1}})}^{2}}}{2\sigma^{2}_{C_0(p_1)}} \right) \right]d\xi.  
\end{align}

Prethodna relacija je vjerovatnoća greške u detekciji Hermitskog koeficijenta koji predstavlja komponentu signala, u jednokomponentnom signalu rijetkom u Hermitskom domenu sa slučajno pozicioniranim nedostajućim odbircima. Budući da je za jednokomponentne signale komponenta (odnosno Hermitski koeficijent na poziciji $p=p_1$) deterministička i jednaka vrijednosti ${\frac{N_A}{N}A_{1}}$, može se izvesti jednostavna apoksimacija prethodnog izraza za grešku:
\begin{align}{{P}^{(1)}_{E}}\approx 1-\operatorname{erf}{{\left( \frac{{\frac{N_A}{N}A_{1}}}{\sqrt{2}{\sigma_{csN}}} \right)}^{N-1}}.
\end{align}	
Ovaj aproksimativni izraz može biti korigovan za 1.5 standardnu devijaciju komponente signala, uzimajući u obzir činjenicu da komponente signala koje su manje od srednje vrijednosti više doprinose veličini greške u detekciji od komponenti koje bi bile veće od srednje vrijednosti:
\begin{align}{{P}^{(1)}_{E}}\approx 1-\operatorname{erf}{{\left( \frac{{\frac{N_A}{N}A_{1}}-1.5{\sigma_{C_0(p_1)}}}{\sqrt{2}{\sigma_{csN}}} \right)}^{N-1}}.
\end{align}

Vjerovatnoća lažnog alarma (engl. \textit{false alarm probability}, odnosno, vjerovatnoća da je koeficijent šuma iznad praga $\Xi$) data je izrazom:
\begin{align}{{P}_{FA}}(\Xi )=1-{{P}_{NN}}(\Xi )=1-\operatorname{erf}{{\left( \frac{\Xi }{\sqrt{2}{\sigma_{csN}}} \right)}^{N-1}},\end{align}
dok vjerovatnoća prave detekcije (engl. \textit{true detection probability}, odnosno, vjerovatnoća da je koeficijent komponente signala iznad praga $\Xi $) može biti izračunata na sljedeći način:
\begin{align}{{P}_{TD}}(\Xi )=\int\limits_{\Xi }^{\infty }{\left( \exp \left( -\frac{{{(\xi -{\frac{N_A}{N}A_{1}})}^{2}}}{2\sigma^{2}_{C_0(p_1)}} \right)+\exp \left( -\frac{{{(\xi +{\frac{N_A}{N}A_{1}})}^{2}}}{2\sigma^{2}_{C_0(p_1)}} \right) \right)d\xi },\end{align}
za zadati prag $\Xi$. Za datu dužinu signala $N$, ove vjerovatnoće su funkcije od broja dostupnih odbiraka $N_A$, budući da se on pojavljuje u izrazima za varijanse i srednje vrijednosti.

\begin{primjer}
Na osnovu izraza ${{P}_{FA}}(\Xi )$ i ${{P}_{TD}}(\Xi )$ može se izračunati tzv. ROC kriva (engl. \textit{ receiver operating characteristic}). Neka se posmatra jednokomponentni signal dužine $N = 200$ sa jediničnom amplitudom $A_1=1$. Variranjem praga $0\le \Xi \le 1$ i računanjem ROC krivih (prikaz ${{P}_{TD}}(\Xi )$ u funkciji od ${{P}_{FA}}(\Xi )$) za različite brojeve dostupnih odbiraka ${{N}_{A}}\in \left\{ 10,20,40,60 \right\}$, dobijeni su rezultati prikazani na slici \ref{roc_curve_single_component}. Šum usljed kompresivnog odabiranja nastaje kao posljedica nedostajućih odbiraka u samo jednoj komponenti signala, pa je stoga očekivano da se dekcija komponente u Hermitskom domenu može uspješno obaviti na osnovu malog broja dostupnih odbiraka, što je i potvrđeno prikazanim rezultatima. Ako se izračuna of ${{P}^{(1)}_{E}}$ za razmatrane brojeve dostupnih odbiraka, dobija se sljedeći niz vjerovatnoća: ${{P}^{(1)}_{E}}({{N}_{A}}=10)=0.44$, ${{P}^{(1)}_{E}}({{N}_{A}}=20)=0.16$, ${{P}^{(1)}_{E}}({{N}_{A}}=40)=0.01$ and ${{P}^{(1)}_{E}}({{N}_{A}}=60)=0.0$. Navedene vrijednosti su saglasne sa rezultatima sa slike \ref{roc_curve_single_component}.

   \begin{figure}[ptb]%
	\centering
	\includegraphics[
	]%
	{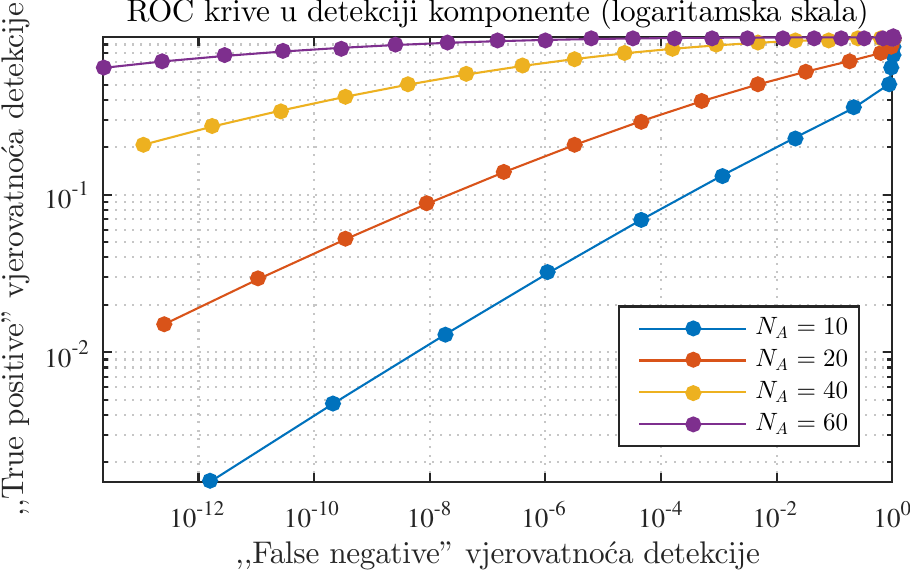}%
	\caption[ROC krive u detekciji komponente signala, prikazane za nekoliko zadatih brojeva dostupnih odbiraka.]{ROC krive u detekciji komponente signala, prikazane za nekoliko zadatih brojeva dostupnih odbiraka.}%
	\label{roc_curve_single_component}%
\end{figure}
\end{primjer}
\paragraph{Multikomponentni signali.}
Prethodnu analizu proširićemo na slučaj multikomponentnih signala koji su rijetki u DHT1 domenu. Razmatrana slučajna promjenljiva, za signal $z(p_l,p,t_n)\in \mathbf{\Omega }$ postaje
\begin{align}
{C_0(p)}&= \sum_{t_n \in \mathbb{N}_A}\sum\limits_{l=1}^{K} z(p_l,p,t_n)=\sum_{t_n \in \mathbb{N}_A}\sum\limits_{l=1}^{K} {\frac{{{A}_{l}}}{N}\frac{{{\psi }_{p}}({{t}_{n}}){{\psi }_{{{p}_{l}}}}({{t}_{n}})}{{{({{\psi }_{N-1}}({{t}_{n}}))}^{2}}}},
\end{align}
i sada ima $K$ komponenti, gdje je $p_l\in{\Pi}_K$. Hermitski koeficijenti $C_0(p)$ imaju karakteristike Gausove slučajne varijable, sa srednjom vrijednošću:
\begin{align}{{\mu }_{C_0(p)}}=\sum\limits_{l=1}^{K}{{{A}_{l}}\frac{{{N}_{A}}}{N}\delta (p-{{p}_{l}})},\end{align}
gdje je $p=p_l\in{\Pi}_K$, budući da šum uzrokovan nedostajućim odbricima u svakoj komponenti ima srednju vrijednost nula, što je pokazano u analizi monokomponentnih signala. Varijansa Hermitskih koeficijenata na pozicijama mimo komponenti signala je data izrazom:
\begin{align}\sigma^2_{{C_0}(p)}=\sigma _{csN}^{2}=\frac{{{N}_{A}}N-N_{A}^{2}}{{{N}^{2}}(N-1)}\sum\limits_{l=1}^{K}{A_{l}^{2}},~p\neq p_l, ~p_l\in{\Pi}_K, 
\label{htsigman}
\end{align}
budući da, u slučaju koeficijenata sa indeksima $p\ne {{p}_{l}}$, nedostajući odbirci iz svih komponenti izazivaju šum koji se na ovim pozicijama sabira. Komponente šuma koje potiču iz različitih komponenti signala su nekorelisane i imaju srednju vrijednost nula. Budući da se sabiraju slučajne varijable sa normalnim raspodjelama koje su nezavisne, njihove varijanse se sabiraju.

Sa druge strane, na osnovu prezentovane analize za monokomponentne signale, komponenta signala sa indeksom $q=1,2,\dots, K$, na poziciji $p={{p}_{q}}\in{\Pi}_K$ ima varijansu
\begin{align}\tilde{\sigma }_{C_0(p_q)}^{2}\approx \frac{{{N}_{A}}N-N_{A}^{2}}{{{N}^{2}}(N-1)}\left[ \frac{N}{N_{A}^{{}}}A_{q}^{2}\left( {{\sum\limits_{i=1}^{{{N}_{A}}}{\left( \frac{\psi _{{{p}_{i}}}^{2}({t_{n_i}})}{{{({{\psi }_{N-1}}({t_{n_i}}))}^{2}}} \right)^{2}}}}-1 \right) \right].\end{align}

Dodatno, šum izazvan nedostajućim odbircima u preostalih $K-1$ komponenti je takođe prisutan na poziciji $q$-te komponente. Navedeno znači da, pored šuma izazvanog nedostajućim odbircima u posmatranoj komponenti signala sa indeksom $p=p_q$,  čija je varijansa $\tilde{\sigma }_{C_0(p_q)}^{2}$, na poziciji $p_q$ se sa tim šumom sabiraju šumovi uzrokovani nedostajućim odbricima iz preostalih komponenti, čiji su indeksi $p=p_l\in{\Pi}_K=\{{{p}_{1}},{{p}_{2}},\dots,{{p}_{K}}\},~p\ne p_q$. Ove slučajne promjenljive su sa normalnom raspodjelom, sa srednjim vrijednostima nula i sa varijansama: $
\frac{{{N}_{A}}(N-N_{A})}{{{N}^{2}}(N-1)}A_l^2,~l\ne q,~l=1,2,\dots, K.
$

Zaključujemo da je srednja vrijednost komponente signala na poziciji $p_q\in {\Pi}_K$ jednaka ${{A}_{q}}\frac{{N}_{A}}{N},~q=1,2,\dots, K$, dok je njena varijansa definisana izrazom:
\begin{align} {\sigma }_{C_0(p_q)}^{2}=\frac{N-N_{A}^{{}}}{N\left( N-1 \right)}\left( A_{q}^{2}\left( \frac{{{{\tilde{P}}}_{{{p}_{q}}}}}{N}-1 \right)+\sum\limits_{\begin{smallmatrix} 
	l=1 \\ 
	l\ne q
	\end{smallmatrix}}^{K}{A_{l}^{2}} \right),
\label{htcvar}
\end{align}
gdje je
\begin{align}
{{\tilde{P}}_{{{p}_{q}}}}={{\sum\limits_{i=1}^{{{N}_{A}}}{\left( \frac{\psi _{{{p}_{q}}}^{2}({t_{n_i}})}{{{({{\psi }_{N-1}}({t_{n_i}}))}^{2}}} \right)}}^{2}}.
\end{align}

\begin{figure}[ptb]%
	\centering
	\includegraphics[
	]%
	{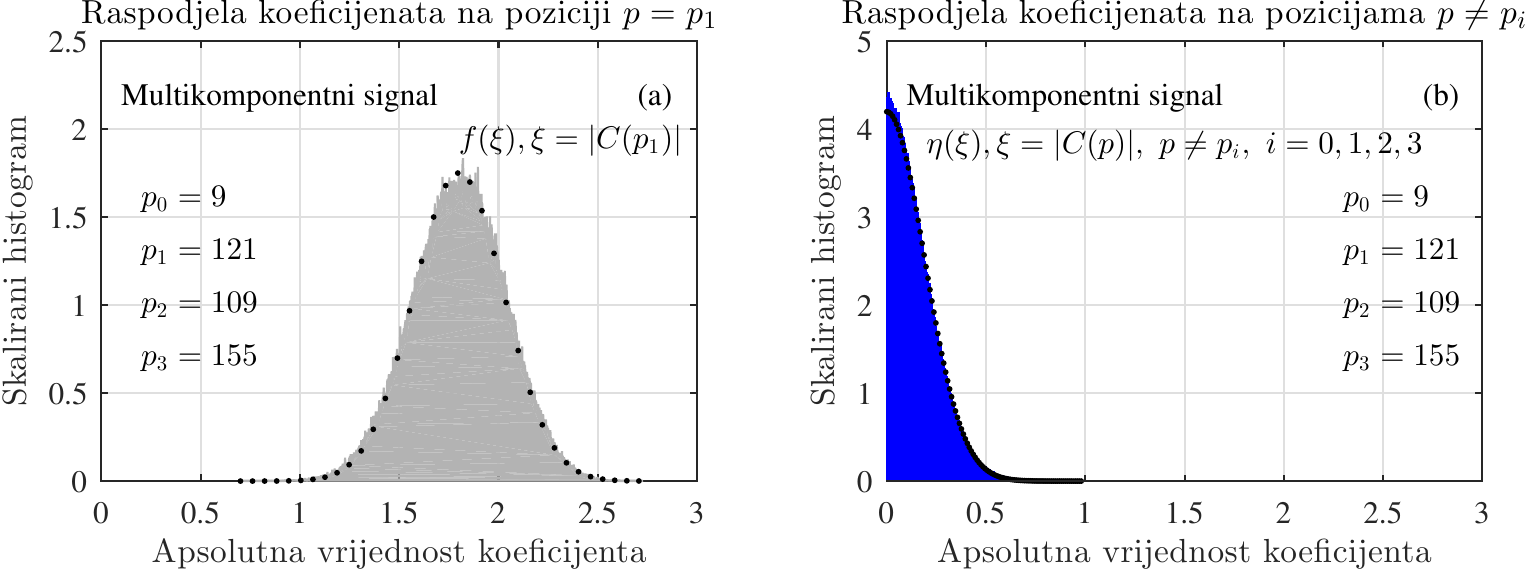}%
	\caption[Histogrami i funkcije raspodjele za apsolutne vrijednosti Hermitskih koeficijenata multikomponentnog signala.]{Histogrami i funkcije raspodjele za apsolutne vrijednosti Hermitskih koeficijenata multikomponentnog signala: (a) na poziciji jedne od komponenti signala (b) na pozicijama šuma. Histogrami ((a) siva površina za koeficijent na poziciji signala i (b) plava površina za koeficijent na poziciji šuma)  su simulirani za multikomponentni signal sa $ N_A = 120$ dostupnih od ukupno $N=200$  odbiraka, gdje su amplitude komponenti  $A_1 = 1$, $A_2 = 3$, $A_3 = 4$ i $A_4 = 2$, na bazi 20000 nezavisnih realizacija signala sa slučajno pozicioniranim dostupnim odbircima. Teorijski rezultati su dobijeni korišćenjem iskrivljene Gausove raspodjele sa estimiranom varijansom (\ref{htcvar}), odnosno polunormalne raspodjele sa varijansom (\ref{htsigman}). }%
	\label{Fig_histogramsMULTI_HT}%
\end{figure}

Može se zaključiti da koeficijent DHT1 na poziciji signala $p=p_{p_q}\in {\Pi}_K$ predstavlja slučajnu promjenljivu sa Gausovom raspodjelom, modelovanu kao
$
\mathcal{N}\left({A}_{q} \frac{{{N}_{A}}}{N},{\sigma }_{C_0(p_q)}^{2} \right).
$ Koeficijenti koji odgovaraju šumu se modeluju slučajnom varijablom sa normalnom raspodjelom $\mathcal{N}(0,\sigma _{csN}^{2})$, gdje je varijansa $\sigma _{csN}^{2}$ data izrazom (\ref{htsigman}).
\paragraph{Analiza greške u detekciji komponenti signala.}
Ekvivalentno prethodno analiziranom slučaju monokomponentnog signala, moguće je izvesti vjerovatnoću greške u detekciji Hermitskih koeficijenata koji odgovaraju komponentama mulikomponentnog signala, u uslovima kada postoje nedostajući odbirci u vremenskom domenu. Pogrešna detekcija komponente signala se dešava kada je barem jedan Hermitski koeficijent šuma, sa pozicija $p\ne {{p}_{q}},~q\in \left\{ 1,2,\dots,K \right\}$, iznad koeficijenta na poziciji ${{p}_{q}}$. Apsolutne vrijednosti razmatranih slučajnih varijabli  $C(p_q),~p_q\in {\Pi}_K$ i $C(p),p\ne p_l$ podliježu polunormalnoj i iskrivljenoj normalnoj distribuciji, respektivno, kao što je i prikazano na slici \ref{Fig_histogramsMULTI_HT}.
Slučajna varijabla $\xi =\left|C(p_q) \right|$ ima funkciju raspodjele data izrazom (\ref{folded1koef}), srednjom vrijednošću ${{A}_{q}}\frac{{N}_{A}}{N},~q=1,2,\dots, K$ i varijansom (\ref{htcvar}). Slučajna varijabla koja reprezentuje koeficijente šuma $\xi =\left| C(p) \right|,~p\neq p_q$, ima srednju vrijednost nula i varijansu (\ref{htsigman}), dok je njena funkcija raspodjele (\ref{halfnorht}). Vjerovatnoća da $N-K$ nezavisnih koeficijenata šuma budu manji od $\Xi $ data je sljedećim izrazom:
\begin{align}{{P}_{NN}}(\Xi )=\operatorname{erf}{{\left( \frac{\Xi }{\sqrt{2}{\sigma_{csN}}} \right)}^{N-K}}.
\label{ht_prob}
\end{align}

Slično kao u slučaju monokomponentnog signala, lako se pokazuje da je vjerovatnoća greške u detekciji $q$-te komponente signala u uslovima nedostajućih odbiraka u multikomponentnom signalu data izrazom:
\begin{align}
{{P}^{(q)}_{E}}&\frac{1}{{\sigma_{C_0(p_q)}}\sqrt{2\pi }}\int\limits_{0}^{\infty }{\left( 1-\operatorname{erf}{{\left( \frac{\xi }{\sqrt{2}{\sigma_{csN}}} \right)}^{N-K}} \right)} \notag \\
&~~~~~~~~~~~~~~~~~~~~~~~~~~\times \left[ \exp \left( -\frac{{{(\xi -{\frac{N_A}{N}A_{q}})}^{2}}}{2\sigma^{2}_{C_0(p_q)}} \right)+\exp \left( -\frac{{{(\xi +{\frac{N_A}{N}A_{q}})}^{2}}}{2\sigma^{2}_{C_0(p_q)}} \right) \right]d\xi.  
\label{htprob_det}
\end{align}

Uz iste pretpostavke kao u slučaju monokomponentnog signala, ova greška može biti aproksimirana izrazom:
\begin{align}{{P}^{(q)}_{E}}\approx 1-\operatorname{erf}{{\left( \frac{{\frac{N_A}{N}A_{q}}-1.5{\sigma_{C_0(p_q)}}}{\sqrt{2}{\sigma_{csN}}} \right)}^{N-K}}.
\label{htprobapr}
\end{align}

\subsubsection{Numerička provjera izvedenih izraza}
Predstavljena teorija će biti evaluirana kroz nekoliko numeričkih eksperimenata. Sproveden je veći broj eksperimenata koji potvrđuju izvedene izraze za srednje vrijednosti i varijanse DHT1 koeficijenata kao slučajnih varijabli u kontekstu kompresivnog odabiranja, kao i izraza za vjerovatnoću greške u detekciji komponenti. 
\begin{primjer}
Na slikama \ref{Fig_histogramsMONO_HT} i \ref{Fig_histogramsMULTI_HT} su prikazani histogrami Hermitskih koeficijenata $C(p)$, na pozicijama komponenti signala i na pozicijama šuma, za slučaj monokomponentnog i multikomponentnog signala, respektivno. Histogrami su dobijeni na osnovu 20000 realizacija signala sa slučajno pozicioniranim nedostajućim odbircima. Rezultati na slikama \ref{Fig_histogramsMONO_HT} i \ref{Fig_histogramsMULTI_HT} numerički potvrđuju kako izraze za varijanse i srednje vrijednosti u razmatranim slučajevima, tako i prirodu odgovarajućih raspodjela Hermitskih koeficijenata kao slučajnih varijabli. Detalji eksperimenata predstavljeni su u sklopu opisa slika.
\end{primjer}
\begin{primjer}
Razmatra se monokomponentni signal rijedak u DHT1 domenu sa jediničnom amplitudom, definisan izrazom:
\begin{align}x(t_n)={{\psi }_{{{p}_{1}}}}(t_n).
\label{modelhtn}
\end{align}	
U signalu je dostupno $N_A$ od ukupno $N$ odbiraka. Indeks ${p}_{1}$ komponente signala u DHT1 domenu je variran između:
\begin{itemize}
\item 0 i 199 za signal dužine $N=200$;
\item 0 i 399 za signal dužine $N=400$.
\end{itemize}

Za svaku zadatu vrijednost ${p}_{1}$, izvedeno je 7000 nezavisnih realizacija signala sa $N_A$ slučajno pozicioniranih odbiraka u svakoj realizaciji, i na bazi njih je izračunata eksperimentalna vrijednost varijanse $\bar\sigma_{C_0(p_1)}^{2}$ koeficijenta na poziciji ${p}_{1}$. 

   \begin{figure}[ptb]%
	\centering
	\includegraphics[
	]%
	{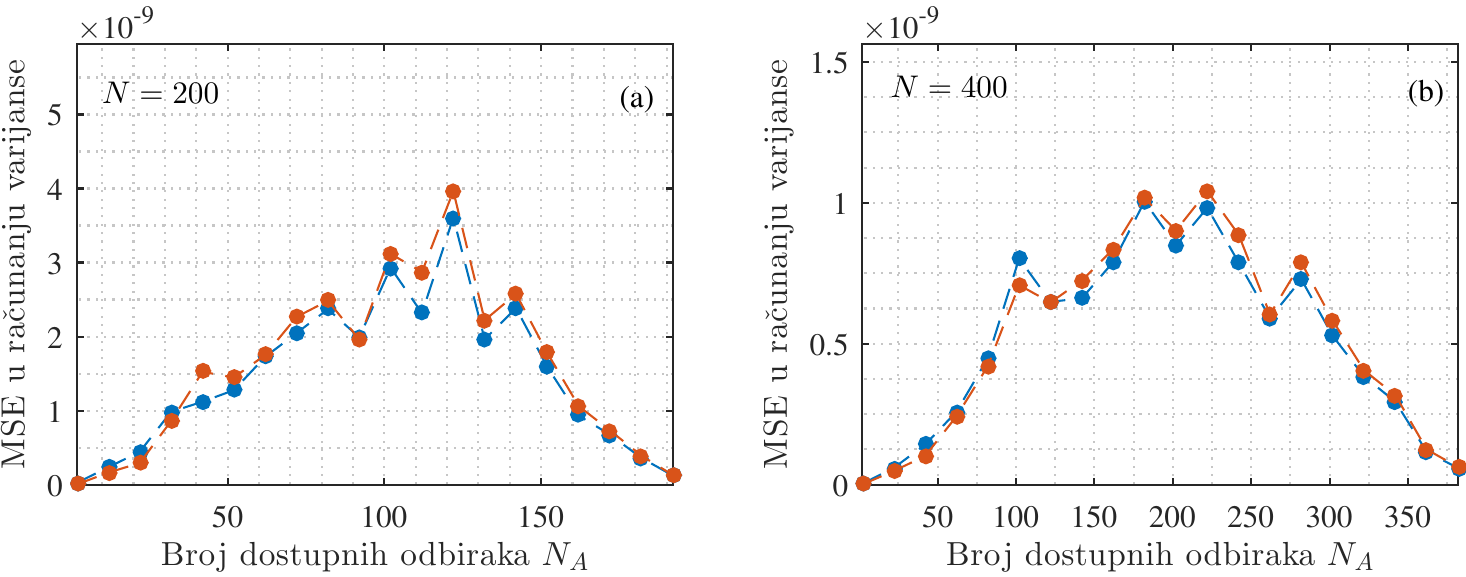}%
	\caption[Srednja kvadratna greška u proračunu varijanse korišćenjem izraza (\ref{varht_com}) i (\ref{var_monoaprox}), za Hermitske koeficijente na poziciji komponente signala ${p}_{1}$, za različite brojeve dostupnih odbiraka $N_A$.]{ Srednja kvadratna greška u proračunu varijanse korišćenjem izraza (\ref{varht_com}) i (\ref{var_monoaprox}), za Hermitske koeficijente na poziciji komponente signala ${p}_{1}$, za različite brojeve dostupnih odbiraka $N_A$. Za posmatrano $N_A$, MSE je reačunata za sve moguće pozicije ${p}_{1}$. Razmatrane su dužine signala: (a) $N=200$ i (b) $N=400$. Za svaku poziciju ${p}_{1}$, numeričke vrijednosti varijanse su sračunate na bazi 7000 nezavisnih realizacija signala sa slučajno pozicioniranim odbircima.}%
	\label{Fig_errorsHT}%
\end{figure}

Numerički dobijena varijansa je upoređena sa teorijskom, $\sigma_{C_0(p_1)}^{2}$, koja je data izrazom (\ref{varht_com}) i sa aproksimativnim izrazom (\ref{var_monoaprox}). MSE je sračunata usrednjavanjem rezultata dobijenih za 7000 posmatranih realizacija. Rezultati su prikazani na slici \ref{Fig_errorsHT} (a) i (b), za dvije razmatrane dužine signala, respektivno. Poređenje je sprovedeno za različite vrijednosti brojeve dostupnih odbiraka: (a) između 2 i 200 sa korakom 10, i (b) između 4 i 400, sa korakom 20. Plava isprekidana linija sa tačkama predstavlja srednju kvadratnu grešku između eksperimentalnog rezultata i teorijskog modela (\ref{varht_com}), dok odgovarajuća crvena linija odgovara srednjoj kvadratnoj grešci između eksperimentalnih rezultata i aproksimativnog modela (\ref{var_monoaprox}), uz pretpostavku da ${p}_{1}$ poznato. Može se uočiti da su MSE vrijednosti reda $10^{-9}$, što potvrđuje tačnost izvedenih teorijskih izraza.
\end{primjer}
\begin{primjer}
Razmatra se monokomponentni signal (\ref{modelhtn}), za tri moguće pozicije komponente signala (a) ${p}_{1} = 1$, (b) ${p}_{1} = 266$ i (c) ${p}_{1} = 390$. Dužina signala je $N = 400$. Broj dostupnih odibraka  $N_A$ se varira između $1$ i $N$. Za svaku zadatu vrijednost $N_A$, varijansa  $\sigma_{C_0(p)}^{2}$ slučajne varijable $C_0(p)$ je izračunata numerički, na bazi 5000 nezavisnih realizacija signala, sa slučajno raspoređenim pozicijama nedostajućih odbiraka. Varijansa  $\sigma_{C_0(p_1)}^{2}$  koja odgovara Hermitskim koeficijentima na poziciji $p=p_1$ komponente signala je računata teorijski, po aproksimativnom izrazu (\ref{var_monoaprox}), za svaku realizaciju signala. Rezultati su usrednjeni na bazi 5000 realizacija, za svako zadato $N_A$. Za varijanse  $\sigma_{C_0(p_1)}^{2}$, rezultati su prikazani na slici \ref{Fig_parabole} u slučaju: (a) ${p}_{1} = 1$, (b) ${p}_{1} = 266$ i (c) ${p}_{1} = 390$. Značajno poklapanje teorijskog i numeričkih rezultata evidentno je za sve razmatrane pozicije ${p}_{1}$. Dodatno, varijansa $\sigma_{csN}^2$ na pozicijama $p\ne p_1$ šuma je evaluirana eksperimentalno za sve tri razmatrane pozicije Hermitskog koeficijenta komponente $p_1$. Za Hermitske koeficijente na pozicijama šuma, rezultati numeričke validacije izraza (\ref{ht_noisecs}) prikazani su na slici \ref{Fig_parabole} (d), potvrđujući da je ova varijansa nezavisna od razmatrane pozicije komponente signala ${p}_{1}$.

   \begin{figure}[ptb]%
	\centering
	\includegraphics[
	]%
	{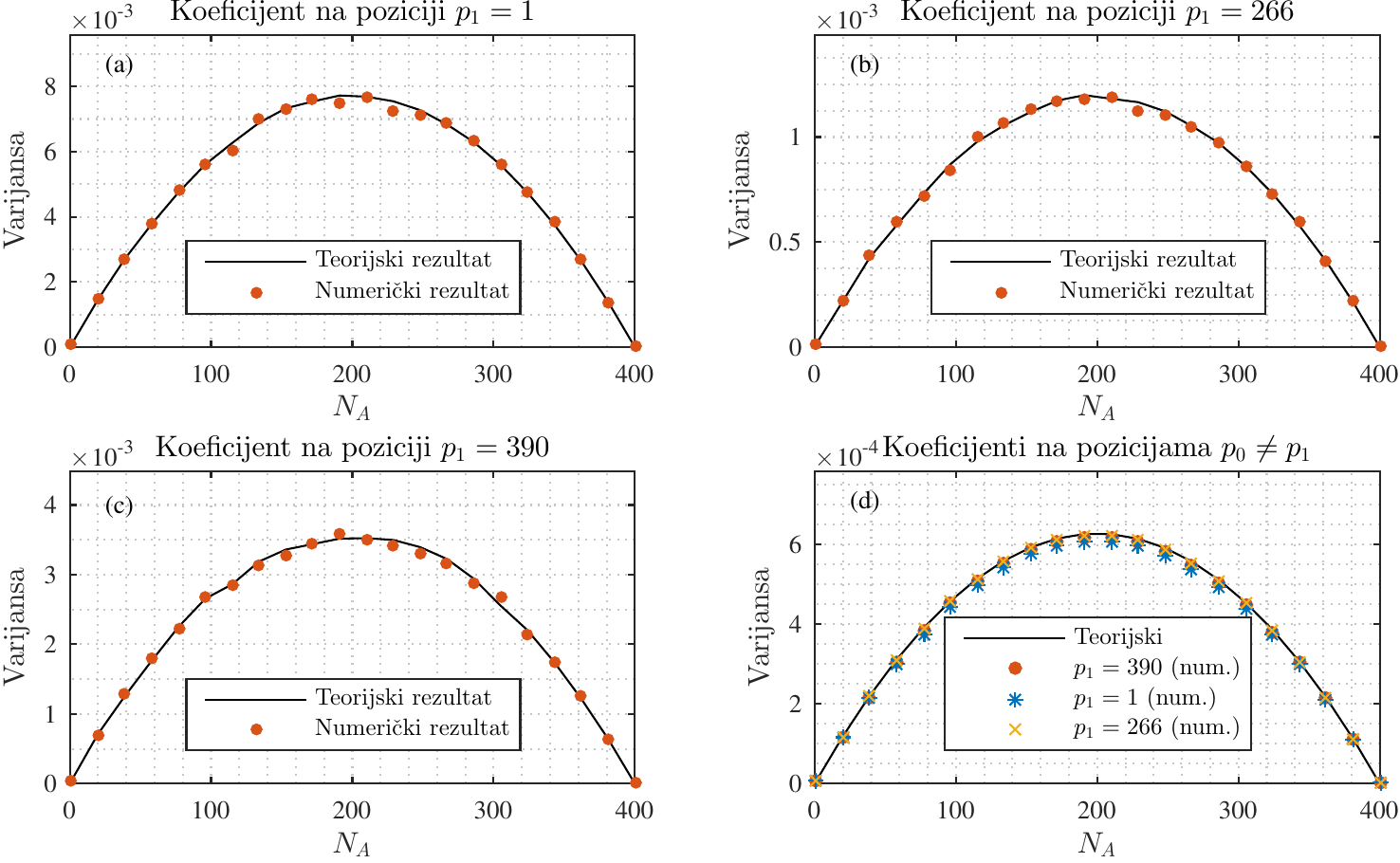}%
	\caption[Statistička evaluacija izraza za varijansu Hermitskih koeficijenata jednokomponentnog signala u funkciji od broja dostupnih odbiraka $N_A$.]{Statistička evaluacija izraza za varijansu Hermitskih koeficijenata jednokomponentnog signala u funkciji od broja dostupnih odbiraka $N_A$: (a) -- (c) varijansa koeficijenata na pozicijama komponenti signala, (d) prosječna varijansa Hermitskih koeficijenata jednokomponentnog signala na pozicijama šuma, $p\ne p_1$, prikazana u slučajevima tri različite pozicije komponenti $p_1$.}%
	\label{Fig_parabole}%
\end{figure}
\end{primjer}
\begin{primjer}
	\label{pozex}
U ovom primjeru se vrši evaluacija izraza (\ref{htprob_det}) za vjerovatnoću greške u detekciji Hermitskih koeficijenata koji odgovaraju komponentama signala, kao i odgovarajućeg izraza za aproksimaciju ove vjerovatnoće (\ref{htprobapr}). Posmatra se signal dužine $N = 200$ sa $K=5$ komponenti, zadat izrazom:
 \begin{align}s(t_n)=\sum\limits_{l=1}^{K}{{{A}_{l}}{{\psi }_{{{p}_{l}}}}}(t_n)\end{align}
 uz $A_l=\{1, 0.7, 0.5, 0.3, 0.2\}$ i $p_l = \{20, 54, 94, 162, 192\}$ za $l = 1,2,\dots,K$. Na slici \ref{Fig_probabilities} (a) prikazana je vjerovatnoća greške u detekciji za svaku od komponenti signala posebno, računata korišćenjem izraza (\ref{htprob_det}). Na istoj slici prikazane su i krive odgovarajućeg aproksimativnog izraza (\ref{htprobapr}). Broj distupnih odbiraka $N_A$ je variran između 1 i 200. Horizontalna isprekidana linija označava vjerovatnoću greške $P_E=10^{-2}$. Može se zaključiti da se tačni i aproksimativni izraz u velikoj mjeri poklapaju za vjerovatnoću greške od $10^{-2}$. Treba uočiti da slika \ref{Fig_probabilities} (a) specificira koliki je broj dostupnih odbiraka $N_A$ neophodan za uspješnu detekciju posmatrane komponente signala, sa zadatom vjerovatnoćom. Na primjer, $N_A=80$ dostupnih odbiraka je dovoljno za detekciju komponente sa amplitudom $A_1 = 0.1$, sa vjerovatnoćom greške u detekciji koja je bliska nuli. Isti broj odbiraka je dovoljan za detekciju komponente sa amplitudom $ A_2 = 0.7$, sa vjerovatnoćom greške u detekciji koja je jednaka $10^{-2}$. Takođe, može se zaključiti da je za detekciju svih komponenti signala odjednom sa vjerovatnoćom $10^{-2}$ potrebno oko $N_A=176$ dostupnih odbiraka.
 
Posmatrane vjerovatnoće su takođe eksperimentalno verifikovane. Broj dostupnih odibraka $N_A$ je variran između 1 i 200. Za svaku posmatranu vrijednost $N_A$, vjerovatnoće greške u detekciji su računate na bazi 3000 realizacija signala sa slučajno pozicioniranim dostupnim odbircima. U svakoj realizaciji i za svaku komponentu signala, vršeno je prebrojavanje pogrešnih detekcija. Pogrešna detekcija $l$-te komponente signala se dešava ako je najmanje jedan Hermitski koeficijent šuma (na poziciji mimo pozicija komponenti signala, $p\ne p_l,~l=1,2,\dots,K$) ima amplitudu koja je jednaka ili veća od amplitude $l$-te komponente signala, odnosno koeficijenta na poziciji $p\ne p_l,~l=1,2,\dots,K$. Broj pogrešnih detekcija je zatim podijeljen sa brojem realizacija signala. Postupak je ponovljen za svako posmatrano $N_A$. Rezultati su prikazani na slici \ref{Fig_probabilities} (b). Može se uočiti velika vizuelna sličnost između teorijski dobijenih krivih vjerovatnoća (a) sa linijama koje su dobijene numeričkim putem (b).

    \begin{figure}[ptb]%
 	\centering
 	\includegraphics[
 	]%
 	{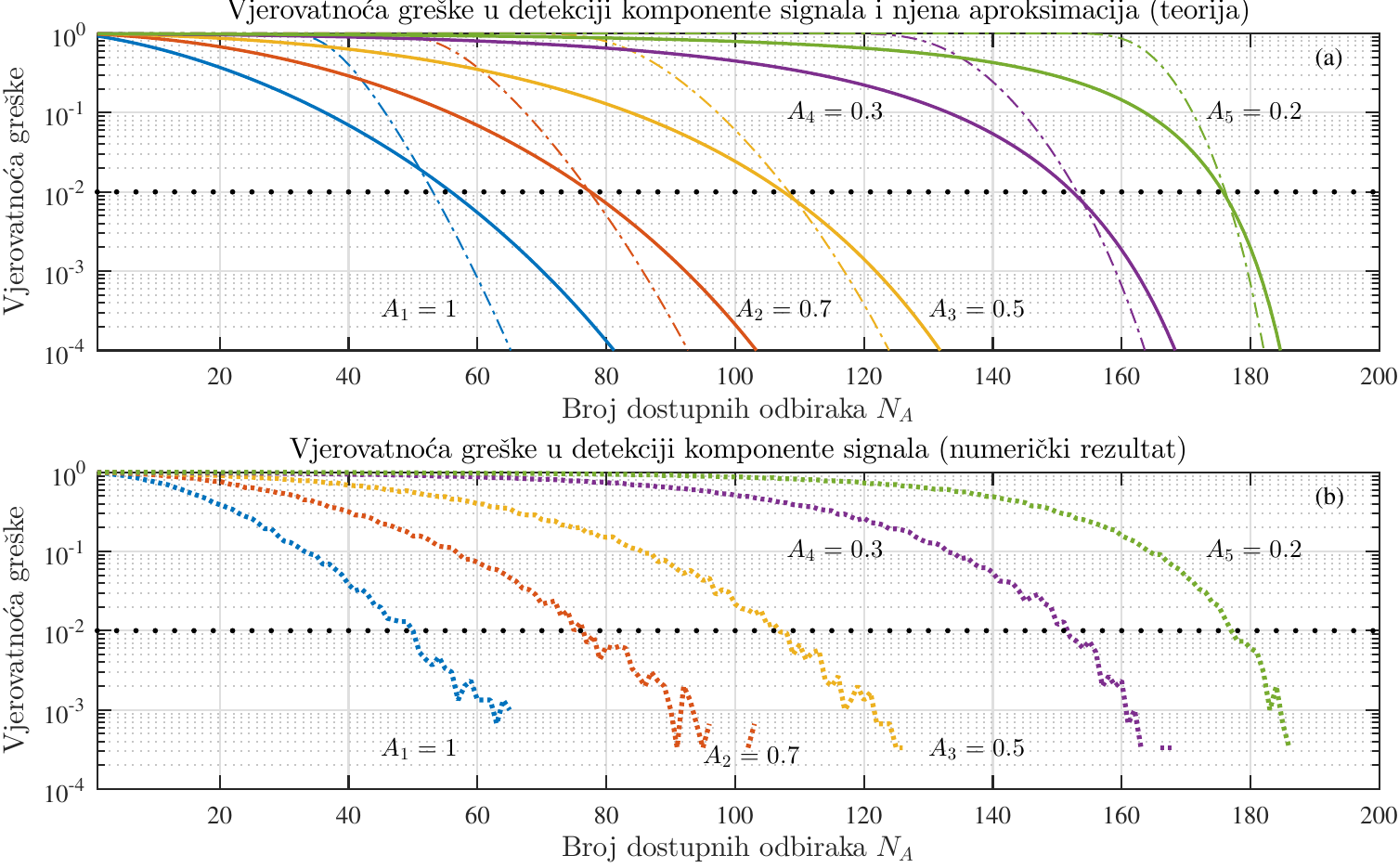}%
 	\caption[Vjerovatnoća pogrešne detekcije komponente signala, prezentovana u vidu funkcije od broja dostupnih odbiraka signala $N_A$.]{Vjerovatnoća pogrešne detekcije komponente signala, prezentovana u vidu funkcije od broja dostupnih odbiraka signala $N_A$: (a) vjerovatnoća dobijena primjenom teorijskog izraza (\ref{htprob_det}) i njegove aproksimacije (\ref{htprobapr}), (b) eksperimentalni rezultati. }%
 	\label{Fig_probabilities}%
 \end{figure} 
\end{primjer}
\begin{primjer}
Razmatra se signal sa nedostajućim odbircima iz primjera \ref{pozex}. Posmatraju se različiti brojevi dostupnih odbiraka $N_A$, koji se koriste za izračunavanje očekivanih vjerovatnoća greške u detekciji, prikazanih na slici \ref{Fig_probabilities}:
\begin{itemize}
\item  $N_A = 56$, koji obezbjeđuje detekciju komponenti signala sa sljedećim vjerovatnoćama greške:  $P_E^{(1)} = 0$, $P_E^{(2)} =  0.0086$, $P_E^{(3)} = 0.8679$, $P_E^{(4)} = 1$, $P_E^{(5)} = 1$. Navedeno znači da će prva i druga komponenta biti detektovane sa vjerovatnoćom većom od 0.99, treća komponenta će biti detektovana sa vjerovatnoćom od 0.13, dok četvrta i peta komponenta gotovo sigurno neće biti detektovane.
\item  $N_A = 108$, pri čemu su vjerovatnoće greške u detekciji različitih komponenti: $P_E^{(1)} = 0$, $P_E^{(2)} = 0$, $P_E^{(3)} = 0.0109$, $P_E^{(4)} = 1$ i $P_E^{(5)} = 1$; 
\item $N_A = 154$, sa vjerovatnoćama greške u detekciji: $P_E^{(1)} = 0$, $P_E^{(2)} = 0$, $P_E^{(3)} = 0$, $P_E^{(4)} = 0.0073$ i $P_E^{(5)} = 0.9944$; 
\item $N_A = 176$, sa odgovarajućim vjerovatnoćama greške u detekciji: $P_E^{(1)} = 0$, $P_E^{(2)} = 0$, $P_E^{(3)} = 0$, $P_E^{(4)} = 0$ i $P_E^{(5)} =0.0106$. U ovom slučaju, u oko $99\%$ realizacija signala sve komponente signala će biti uspješno detektovane, odnosno, biće iznad praga za detekciju $T=\Xi$ iz (\ref{ht_prob}). 
\end{itemize} 

DHT1 koeficijenti i probabilistički prag (razmatran detaljno u narednom odjeljku) prikazani su na slici \ref{Hermitetreshold} (a) -- (d), za jednu realizaciju signala sa različitim brojevima dostupnih odbiraka $N_A$.
    \begin{figure}[ptb]%
	\centering
	\includegraphics[
	]%
	{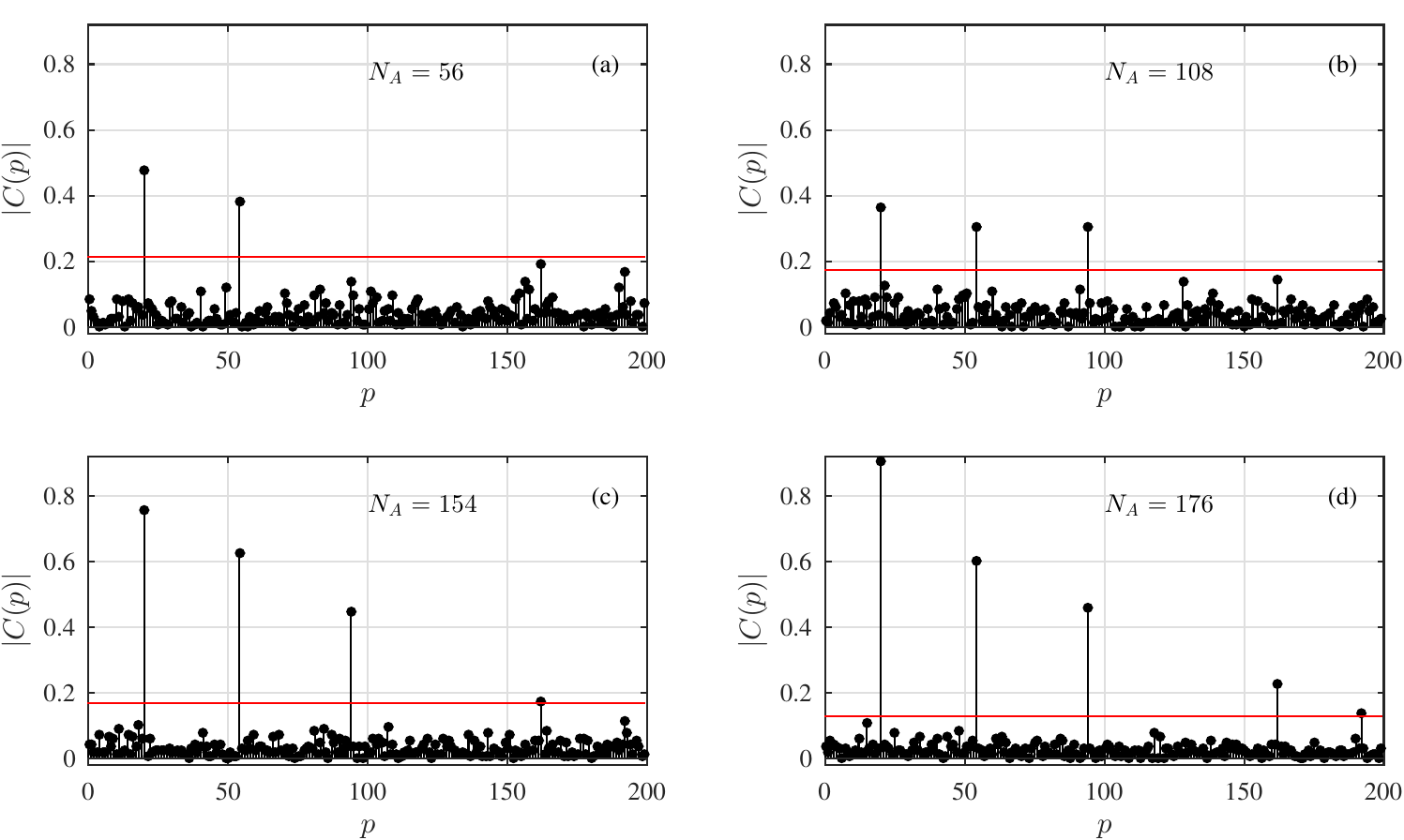}%
	\caption[Ilustracija automatizovanog praga za detekciju komponenti čiji je nivo diktiran brojem dostupnih odbiraka $N_A$.]{Ilustracija automatizovanog praga za detekciju komponenti čiji je nivo diktiran brojem dostupnih odbiraka $N_A$.}%
	\label{Hermitetreshold}%
\end{figure}
\end{primjer}

\subsection{Rekonstrukcija zasnovana na analizi uticaja nedostajućih odbiraka}
\subsubsection{Detekcija komponenti signala}
Uzimajući u obzir njenu važnost, razmotrimo detaljno vjerovatnoću  (\ref{ht_prob}) da je $N-K$ koeficijenata CS šuma u DHT1 domenu manje od neke zadate vrijednosti $\Xi$. Ova relacija može pomoći u definisanju praga sa slike \ref{Hermitetreshold} za razdvajanje komponenti signala (Hermitskih koeficijenata koji odgovaraju komponentama signala) od šuma. Na osnovu izraza (\ref{ht_prob}), za $\Xi=T$ ovaj prag se izvodi u obliku:
\begin{align}T=\sqrt{2}{\sigma_{csN}}{{\operatorname{erf}}^{-1}}\left( {{\left( P_{NN}^{{}}(T) \right)}^{\frac{1}{N-K}}} \right)\approx \sqrt{2}{\sigma_{csN}}{{\operatorname{erf}}^{-1}}\left( {{\left( P_{NN}^{{}}(T) \right)}^{\frac{1}{N}}} \right),
\label{htprag111}
\end{align}	
gdje je $P_{NN}^{{}}(T)$ zadata vjerovatnoća. Uočava se da se nivo rijetkosti $K$ može zanemariti u prethodnom izrazu, imajući u vidu da je on, po definiciji, u kontekstu kompresivnog odabiranja mnogo manji od dužine signala $N$ ($K\ll N$). Prag se može izračunati za bilo koju zadatu (željenu) vjerovatnoću $P_{NN}(T)$, korišćenjem varijanse šuma definisane izrazom (\ref{htsigman}) Dodatno, funkcija $\operatorname{erf}(x)$ može biti aproksimirana na sljedeći način \cite{ht13,ht14}:
\begin{align}
\operatorname{erf}(x)\approx \operatorname{sgn} (x)\sqrt{1-\exp \left( -{{x}^{2}}\frac{4/\pi +a{{x}^{2}}}{1+a{{x}^{2}}} \right)},
\label{htppp}
\end{align}
sa $a\approx 0.147$, i $x=\frac{T}{\sqrt{2}{\sigma_{csN}}}$. Pošto važi $T\ge 0$ i $\sigma_{csN}>0$ a posljedično i $x > 0$, zaključuje se da je uvijek zadovoljeno  $\operatorname{sgn}(x)=1$. Tada se iz (\ref{htppp}) dobija:
\begin{align}{{\left( {{P}_{NN}}(T) \right)}^{\frac{1}{N}}}=\sqrt{1-\exp \left( -{{x}^{2}}\frac{{4}/{\pi} +a{{x}^{2}}}{1+a{{x}^{2}}} \right)}.\end{align}
Primjenom operacije $\log(\cdot)$ na obije strane prethodne jednačine, nakon preuređivanja se dobija
$a{{x}^{4}}+( \tfrac{4}{\pi }+a\log ( 1-{{\left( {{P}_{NN}}(T) \right)}^{\tfrac{2}{N}}})){{x}^{2}}+\log( 1-{{\left( {{P}_{NN}}(T) \right)}^{\tfrac{2}{N}}})=0.$
Ova jednačina se rješava uvođenjem smjene $t=x^2$, i postoji samo jedno pozitivno rješenje (od ukupno četiri rješenja), koje izraz za prag:
\begin{align}T={\sigma_{csN}}\sqrt{\frac{1}{a}\left( -\frac{4}{\pi} -aL+\sqrt{{{\left(\frac{4}{\pi} +aL\right)}^{2}}-4aL} \right)},
\label{praghttt}
\end{align}
odnosno aproksimaciju izraza (\ref{htprag111}) pogodnu za hardverske realizacije, sa $L=\log \left( 1-{{\left( {{P}_{NN}}(T) \right)}^{\frac{2}{N}}} \right)$ i $a\approx 0.147$.

\subsubsection{Algoritmi za rekonstrukciju zasnovani na izvedenom pragu}
Dosadašnja analiza može poslužiti kao osnov za definisanje algoritama za rekonstrukciju signala rijetkih u DHT1 domenu, koji imaju nedostajuće odbirke. Navedeni algoritmi su insipirisani OMP odnosno MP pristupima rekonstrukciji i imaju za cilj povećavanje efikasnosti procesa rekonstrukcije. Prag $T$ se koristi za određivanje pozicija
${\Pi}_K=\{{{p}_{1}},\,{{p}_{2}},\dots,{{p}_{K}}\}$ komponenti signala u DHT1 domenu. 
Problem pronalaženja rješenja koje predstavlja DHT1 reprezentaciju signala koja ima najmanji mogući stepen rijetkosti (i koji odgovara vrijednostima nedostajućih odbiraka, pod uslovom da je signal rijedak, $K \ll N$) formalno se definiše u obliku:
\begin{align}\min {{\left\| \mathbf{C} \right\|}_{0}}~~\text{pod uslovom~~ }{{\mathbf{y}}}\mathbf{=}{{\mathbf{A}}}\mathbf{C}.
\end{align}

Kao što je već naglašeno, budući da ,,${{\ell}_{0}}$-norma'' ne može biti korišćena u direktnoj minimizaciji, navedeni optimizacioni problem se, u kontekstu brojnih procedura za rekonstrukciju, preformuliše korišćenjem  ${{\ell }_{1}}$-norme, otvarajući mogućnosti za primjenu linearnog programiranja i drugih tehnika. Sa druge strane, ukoliko je poznat set pozicija transformacionih koeficijenata koji reprezentuju komponente signala, odnosno ukoliko je na adekvatan način on estimiran u okviru skupa  $\mathbf{\hat{\Pi}}_K$ od $K\le \hat{K}\le N$ elemenata, tako da je zadovoljeno ${\Pi}_K\subseteq \mathbf{\hat{\Pi}}_K$, $\operatorname{card}\left\{ {\mathbf{\hat{\Pi}}_K} \right\}\le {{N}_{A}}$ uz $K\ll N$, tada pseudoinverzija (\ref{Rjesenje}):
\begin{align}{{\mathbf{C}}_{K}}={{\left( \mathbf{A}_{K}^{T}{{\mathbf{A}}_{K}} \right)}^{-1}}\mathbf{A}_{K}^{T}{{\mathbf{y}}},\end{align}
predstavlja rješenje problema rekonstrukcije. Ovo je jedan od ključnih momenata u mnogim rekonstrukcionim pristupima (predstavlja optimalno rješenje u srednjem kvadratnom smislu). Matrica $\mathbf{A}_{K}$ predstavlja podmatricu matrice $\mathbf{A}$, sa izostavljenim kolonama koje odgovaraju pozicijama nedostajućih odbiraka $p\notin \mathbf{\hat{\Pi}}_K$. Procedure za rekonstrukciju zasnovane na pragu (\ref{praghttt}) su predstavljene Algoritmom \ref{htrec1} (postupak za rekonstrukciju zanovan na jednoj iteraciji) i Algoritmom \ref{htrec2} (iterativni postupak za rekonstrukciju). Rekonstrukcija u jednoj iteraciji, sa adekvatno postavljenim pragom koji je zasnovan na prethodnoj teoriji, može biti primijenjen u slučaju kada su amplitude komponenti signala $A_l,~l=1,2,\dots,K$ relativno bliskih vrijednosti. Iterativni algoritam se može smatrati njegovom generalizacijom.

\begin{algorithm}[ptb]
			\floatname{algorithm}{Algoritam}
	\caption{Rekonstrukcija postavljanjem praga u DHT1 domenu (jednoiterativni postupak)}
	\label{htrec1}
	\begin{algorithmic}[1]
		\Input
		\Statex
		\begin{itemize}
			\item Vektor mjerenja $\mathbf{y}$
			\item Mjerna matrica $\mathbf{A}$
		\end{itemize}
		\Statex
		\State $a\leftarrow 0.147$
		\State ${{P}_{NN}}(T)\leftarrow 0.99$ \Comment Zadata vjerovatnoća da DHT1 koeficijenti šuma budu ispod praga
		\State $L\leftarrow\log \left( 1-{{\left( {{P}_{NN}}(T) \right)}^{\frac{2}{N}}} \right)$ 
		\State ${{\mathbf{C}}_{0}}\leftarrow \mathbf{A}^{-1}{{\mathbf{y}}}$ \Comment Računa se inicijalna DHT1 estimacija
		\State $\sigma _{csN}^{{}}\leftarrow \sqrt{\frac{{{N}_{A}}N-N_{A}^{2}}{{{N}^{2}}(N-1)}\sum\limits_{p=0}^{N-1}{\frac{N}{{{N}_{A}}}{{\left| {C_0(p)} \right|}^{2}}}}$ \Comment Izraz $\frac{N}{{{N}_{A}}}\sum_{p=0}^{N-1}{{{\left| {C_0(p)} \right|}^{2}}}$ aproksimira $\sum_{l=1}^{K}{A_{l}^{2}}$
		\State $T\leftarrow{\sigma_{csN}}\sqrt{\frac{1}{a}\left( -\frac{4}{\pi} -aL+\sqrt{{{\left(\frac{4}{\pi} +aL\right)}^{2}}-4aL} \right)}$ \Comment Prag se računa na osnovu $\sigma_{csN}$
		\State $\mathbf{\hat{\Pi}}_K\leftarrow \arg \left\{ \left| {{\mathbf{C}}_{0}} \right|>T \right\}$ \Comment Pragom se biraju koeficijenti komponenti signala
		\smallskip   
		\State ${{\mathbf{A}}_{K}}\leftarrow {{\mathbf{A}}}{{(:,\mathbf{\hat{\Pi}}_K)}}$ \Comment Matrica $\mathbf{A}_K$ sadrži kolone matrice $\mathbf{A}$  sa indeksima $\mathbf{\hat{\Pi}}_K$
		\State${{\mathbf{C}}_{K}}\leftarrow {{\left( \mathbf{A}_K^{T}{{\mathbf{A}_K}} \right)}^{-1}}\mathbf{A}_{K}^{T}{{\mathbf{y}}}$ \Comment Izraz ${{\left( \mathbf{A}_K^{T}{{\mathbf{A}_K}} \right)}^{-1}}\mathbf{A}_{K}^{T}$ predstavlja pseudoinverziju matrice $\mathbf{A}_K$
		\State $
		C_{T_z}(k)\leftarrow\left\{
		\begin{array}
		{ll}%
		C_K(k),&k\in \mathbf{\hat{\Pi}}_K,\\
		0,&k \notin \mathbf{\hat{\Pi}}_K%
		\end{array}
		\right.
		$
		\Statex
		\Output
		\Statex
		\begin{itemize}
			\item	Vektor rekonstruisanih koeficijenata $\mathbf{C}_{T_z}$
			
		\end{itemize}
	\end{algorithmic}
\end{algorithm}

\begin{algorithm}[tb]
			\floatname{algorithm}{Algoritam}
	\caption{Rekonstrukcija postavljanjem praga u DHT1 domenu (iterativni postupak)}
	\label{htrec2}
	\begin{algorithmic}[1]
		\Input
		\Statex
		\begin{itemize}
			\item Vektor mjerenja $\mathbf{y}$
			\item Mjerna matrica $\mathbf{A}$
%			\item Pozicije dostupnih odbiraka ${{\mathbb{N}}_{{A}}}=\{{{t}_{n_1}},\,{{t}_{n_2}},\dots,{{t}_{{n_{N_A}}}}\}$
			\item Zahtijevana tačnost $\delta$
		\end{itemize}
		\Statex		
		\State	$\mathbf{\hat{\Pi}}_K\leftarrow \varnothing $  \Comment Skup estimiranih pozicija komponenti; na početku je prazan
		\State	$\mathbf{e}\leftarrow {{\mathbf{y}}}$ \Comment Vektor greške na početku je jednak vektoru dostupnih odbiraka
		\State	$a\leftarrow 0.147$
		\State ${{P}_{NN}}(T)\leftarrow 0.99$ \Comment Zadata vjerovatnoća da DHT1 koeficijenti šuma budu ispod praga
		\State	 $L\leftarrow\log \left( 1-{{\left( {{P}_{NN}}(T) \right)}^{\frac{2}{N}}} \right)$
		\While{ $\left\| \mathbf{e} \right\|_{2}^{2}>\delta $ }
		\State	${{\mathbf{C}}_{0}}\leftarrow \mathbf{A}^{-1}\mathbf{e}$	
		\State $\sigma _{csN}^{{}}\leftarrow \sqrt{\frac{{{N}_{A}}N-N_{A}^{2}}{{{N}^{2}}(N-1)}\sum\limits_{p=0}^{N-1}{\frac{N}{{{N}_{A}}}{{\left| {C_0(p)} \right|}^{2}}}}$ \Comment Izraz $\frac{N}{{{N}_{A}}}\sum_{p=0}^{N-1}{{{\left| {C_0(p)} \right|}^{2}}}$ aproksimira $\sum_{l=1}^{K}{A_{l}^{2}}$
		\State $T\leftarrow{\sigma_{csN}}\sqrt{\frac{1}{a}\left( -\frac{4}{\pi} -aL+\sqrt{{{\left(\frac{4}{\pi} +aL\right)}^{2}}-4aL} \right)}$ \Comment Prag se računa na osnovu $\sigma_{csN}$

		\State 	$\mathbf{\hat{\Pi}}_K\leftarrow \mathbf{\hat{\Pi}}_K\cup \arg \left\{ \left| {{\mathbf{C}}_{0}} \right|>T \right\}$ 
		
		\State ${{\mathbf{A}}_{K}}\leftarrow {{\mathbf{A}}}{{(:,\mathbf{\hat{\Pi}}_K)}}$ \Comment Matrica $\mathbf{A}_K$ sadrži kolone matrice $\mathbf{A}$  sa indeksima $\mathbf{\hat{\Pi}}_K$
		\State${{\mathbf{C}}_{K}}\leftarrow {{\left( \mathbf{A}_K^{T}{{\mathbf{A}_K}} \right)}^{-1}}\mathbf{A}_{K}^{T}{{\mathbf{y}}}$ \Comment Izraz ${{\left( \mathbf{A}_K^{T}{{\mathbf{A}_K}} \right)}^{-1}}\mathbf{A}_{K}^{T}$ je pseudoinverzija matrice $\mathbf{A}_K$
		\State	${{\mathbf{y}}_K}\leftarrow {{\mathbf{A}}_{K}}{{\mathbf{C}}_{K}}$
		\State	$\mathbf{e}\leftarrow {{\mathbf{y}}}-{{\mathbf{y}}_K}$
		\EndWhile
		\State $
		C_{T_z}(k)\leftarrow\left\{
		\begin{array}
		{ll}%
		C_K(k),&k\in \mathbf{\hat{\Pi}}_K,\\
		0,& k \notin \mathbf{\hat{\Pi}}_K%
		\end{array}
		\right.
		$

		\Statex
		\Output
		\Statex
		\begin{itemize}
			\item	Vektor rekonstruisanih koeficijenata $\mathbf{C}_{T_z}$
			
		\end{itemize}
	\end{algorithmic}
\end{algorithm}

Princip rada predstavljenih algoritama je baziran na MP (engl. \textit{matching pursuit}) pristupima rekonstrukciji, kao što je, na primjer, OMP, dat pseudokodom u Algoritmu \ref{Norm0Alg}. Iterativna forma algoritma se zapravo može smatrati OMP generalizacijom, koja smanjuje broj iteracija ovog algoritma bazirajući detekciju komponenti signala adekvatnim izborom praga. Stoga, na ovom mjestu će biti napravljena paralela između ova dva pristupa. Neka je na početku algoritma uveden signal greške, inicijalizovan elementima vektora dostupnih odbiraka, $\mathbf{e=y}$. U OMP algoritmu, pozicija prvog elementa u skupu $\mathbf{\hat{\Pi}}_K$ se estimira kao pozicija maksimuma u vektoru ${{\mathbf{C}}_{0}}=\mathbf{A}^{-1}\mathbf{e}$, koji zapravo predstavlja DHT1 signala čije su nedostajuće vrijednosti postavljene na nulu:
${{\hat{p}}_{1}}\leftarrow \arg\max \left\{ \left| {{\mathbf{C}}_{0}} \right| \right\}$. Ovaj indeks se dodaje u inicijalno prazni skup $\mathbf{\hat{\Pi}}_K=\{{{\hat{p}}_{1}}\}$. Zatim se formira pacijalna matrica odabiranja ${{\mathbf{A}}_{1}}={{\mathbf{A}}}{{(:,{{\hat{p}}_{1}})}}$ uzimanjem samo kolone matrice $\mathbf{A}$ koja ima indeks $\hat{p}_1$. Prva komponenta se dobija rješavanjem linearnog sistema mjernih jednačina, korišćenjem pseudoinverzije:
${{\mathbf{C}}_{1}}={{\left( \mathbf{A}_{1}^{T}{{\mathbf{A}}_{1}} \right)}^{-1}}\mathbf{A}_{1}^{T}{{\mathbf{y}}}$.

U sljedećem koraku, određuju se odbirci signala ${{\mathbf{y}}_{1}}={{\mathbf{A}}_{1}}{{\mathbf{C}}_{1}}$. Ukoliko važi $\mathbf{e}={{\mathbf{y}}_{\mathbf{1}}}$, tada je stepen rijetkosti traženog signala 1, i $\mathbf{C}_1$ predstavlja rješenje problema. Ukoliko navedeni uslov nije ispunjen, tada se estimirana komponenta oduzima od vektora $\mathbf{e}$, formirajući signal ${{\mathbf{e}}_{1}}=\mathbf{e}-{{\mathbf{y}}_{1}}$. Nakon ovog koraka, vrši se estimacija pozicije sljedeće komponente. Prvo se izračunava novi vektor ${{\mathbf{C}}_{0}}=\mathbf{A}^{-1}{{\mathbf{e}}_{1}}$ i određuje pozicija druge komponente
${{\hat{p}}_{2}}\leftarrow \max \left\{ \left| {{\mathbf{C}}_{0}} \right| \right\}$. Nakon toga se formira skup pozicija $\mathbf{\hat{\Pi}}_K=\{{{\hat{p}}_{1}},~{{\hat{p}}_{2}}\}$. Pseudoinverzija
${{\mathbf{C}}_{2}}={{\left( \mathbf{A}_{2}^{T}{{\mathbf{A}}_{2}} \right)}^{-1}}\mathbf{A}_{2}^{T}{{\mathbf{y}}}$
se računa za drugu parcijalnu matricu ${{\mathbf{A}}_{1}}={{\mathbf{A}}}{{(:,\mathbf{\hat{\Pi}}_K)}}$ sa dvije kolone iz matrice $\mathbf{A}$ koje odgovaraju pozicijama detektovanih komponenti. Nakon određivanja ${{\mathbf{y}}_{2}}={{\mathbf{A}}_{2}}{{\mathbf{C}}_{2}}$, izračunava se vektor ${{\mathbf{e}}_{2}}=\mathbf{e}-{{\mathbf{y}}_{2}}$. Ukoliko je to vektor-nula, tada je pronađeno rješenje, i algoritam se zaustavlja. Ukoliko nije, proces se iterativno nastavlja, sve dok se ne dobije $\mathbf{y=0}$, ili neki prihvatljivi nivo greške, određen zadatom tačnošću $\delta$. 

Prethodno opisani princip rada OMP algoritma (Algoritam \ref{Norm0Alg}) se koristi i u Algoritmu \ref{htrec1} i Algoritmu \ref{htrec2}. Teorijska analiza prezentovana u ovoj sekciji omogućila je promjenu kriterijuma za detekciju komponenti. Naime, detektovanjem seta komponenti, broj iteracija OMP algoritma je, u opštem slučaju, značajno smanjen. Kao što je već naglašeno, u slučaju kada komponente imaju relativno bliske vrijednosti amplituda, Algoritam \ref{htrec1} obezbjeđuje rekonstrukciju u samo jednoj iteraciji, budući da se predloženi prag $T$ koristi u kriterijumu \begin{align}
	\mathbf{\hat{\Pi}}_K\leftarrow \arg \left\{ \left| {{\mathbf{C}}_{0}} \right|>T \right\}
\end{align}
koji će rezultirati skupom $\mathbf{\hat{\Pi}}_K$ koji sadrži pozicije svih komponenti. Ukoliko ovakav oblik detekcije nije moguć, kao, na primjer, u slučaju kada su neke amplitude komponenti značajno manje od ostalih, može se koristiti Algoritam \ref{htrec2}. On obezbjeđuje detekciju pozicija komponenti u blokovima (sve komponente iznad trenutne vrijednosti praga će biti detektovane istovremeno). Stoga, može se smatrati varijacijom CoSaMP algoritma. Glavna prednost predstavljenih algoritama u odnosu na klasične rekonstrukcione algoritme iz konteksta kompresivnog odabiranja jeste redukovanje broja iteracija. Na ovom mjestu je važno napomenti da se, u slučaju kada je nemoguće postići zadati nivo tačnosti, može dodati i uslov koji će ograničiti mogući broj iteracija algoritama, kako ne bi došlo do stvaranja beskonačnih petlji.
\subsubsection{Rekonstrukcija UWB signala}
Uzimajući u obzir njihovu relevantnost u komunikacionim, radarskim i drugim primjenama, razmotrićemo rekonstukciju UWB signala u kontekstu kompresivnog odabiranja. Model UWB signala je uveden  u primjeru \ref{uwbden}, gdje je navedeno da usljed činjenice da je talasni oblik ovih signala  usko povezan sa Gausovim funkcijama i njihovim izvodima, ovi signali imaju potencijal za rijetku, odnosno visoko koncentrisanu reprezentaciju u Hermitskom transformacionom domenu, za razliku od DFT domena u kojem imaju širok frekvencijski opseg. Izučavanje ove vrste signala u kontekstu obrade rijetkih signala prezentovano je u \cite{ht_uwb5,ht_uwb6}. \textit{Sparsifikacija}, odnosno, povećanje stepena rijetkosti, bila je predmet proučavanja u \cite{ht_uwb5,sparse_rad1}. Primjena kompresivnog odabiranja u raznim aspektima radarskih signala i sistema: estimacija kanala, dizajn talasnih oblika, obrada radarskih slika (engl. \textit{radar imaging}) predmet su skorašnjih istraživanja \cite{sparse_rad1}. %%[10]-[16] druge [17],[18] zatim [18], [19]-[20],[22],
\begin{primjer}
	\label{uwbrekonstrukcija}
Efikasnost predloženih algoritama za rekonstrukciju je testirana  na realnom UWB signalu iz baze razmatrane u primjeru \ref{uwbden}. Razmatrani signal je dobijen pri testiranju UWB radara na 1.3 GHz, u eksperimentu koji je detaljno opisan u \cite{ht_uwb2} i dostupan je \textit{online} \cite{ht_uwb3}. Posmatra se prvih $N = 165$ odbiraka iz fajla pod nazivom \texttt{ACW7FD45.dat} koji je dio razmatrane baze signala. Signal je reodabran, tako da su odbirci dostupni u tačkama koje odgovaraju nulama Hermitskog polinoma reda $N$, primjenom $\operatorname{sinc}$ interpolatora i adekvatnog faktora skaliranja, korišćenjem relacije (\ref{sincint}). Nakon toga, razmatran je scenario sa samo $N_A = 55$ ($33.33 \%$) dostupnih, slučajno pozicioniranih odbiraka. Imajući u vidu izraz (\ref{htprob_det}) za vjerovatnoću greške u detekciji komponenti, određeno je da je sa zadatim brojem dostupnih odbiraka moguće rekonstruisati sve komponente signala, tako da vjerovatnoća greške u detekciji komponenti bude manja od $10^{-2}$ za sve komponente. Rezultati rekonstrukcije su prikazani na slici \ref{uwb_ht_rec}.

Rekonstrukcija je sprovedena korišćenjem Algoritma \ref{htrec2}. Originalni signal i mjerenja prikazani su na slici \ref{uwb_ht_rec} (a). Rekonstrukcija je sprovedena u samo četiri iteracije posmatranog algoritma. DHT1 koeficijenti tokom ovih iteracija, kao i teorijski prag inkorporiran u sklopu algoritma su prikazani na slici \ref{uwb_ht_rec} (b) -- (e). Rezultati rekonstrukcije u vremenskom domenu su prikazani na slici \ref{uwb_ht_rec} (f), gdje je rekonstruisani signal upoređen sa originalnim (sa svim dostupnim odbircima), dok se njihovo poređenje u Hermitskom transformacionom domenu može vidjeti na slici \ref{uwb_ht_rec} (g). Mali broj iteracija postignut je zahvaljujući blokovskoj detekciji pozicija komponenti. Rezultujuća srednja kvadratna greška (MSE) između originalnog i rekonstruisanog signala je $2.1451\cdot {10}^{-27}$, odnosno, $-266.66$ dB. Vrijeme izvršavanja je testirano na laptopu sa Intel(R) Core(TN) i7-6700HQ CPU @2.60 GHz procesorom i 8 GB RAM-a. Algoritam za rekonstrukciju je izvršen u softverskom paketu MATLAB\textsuperscript{\textregistered} R2015a. Vrijeme izvršavanja je 0.0156 sekundi.

    \begin{figure}[ptb]%
	\centering
	\includegraphics[
	]%
	{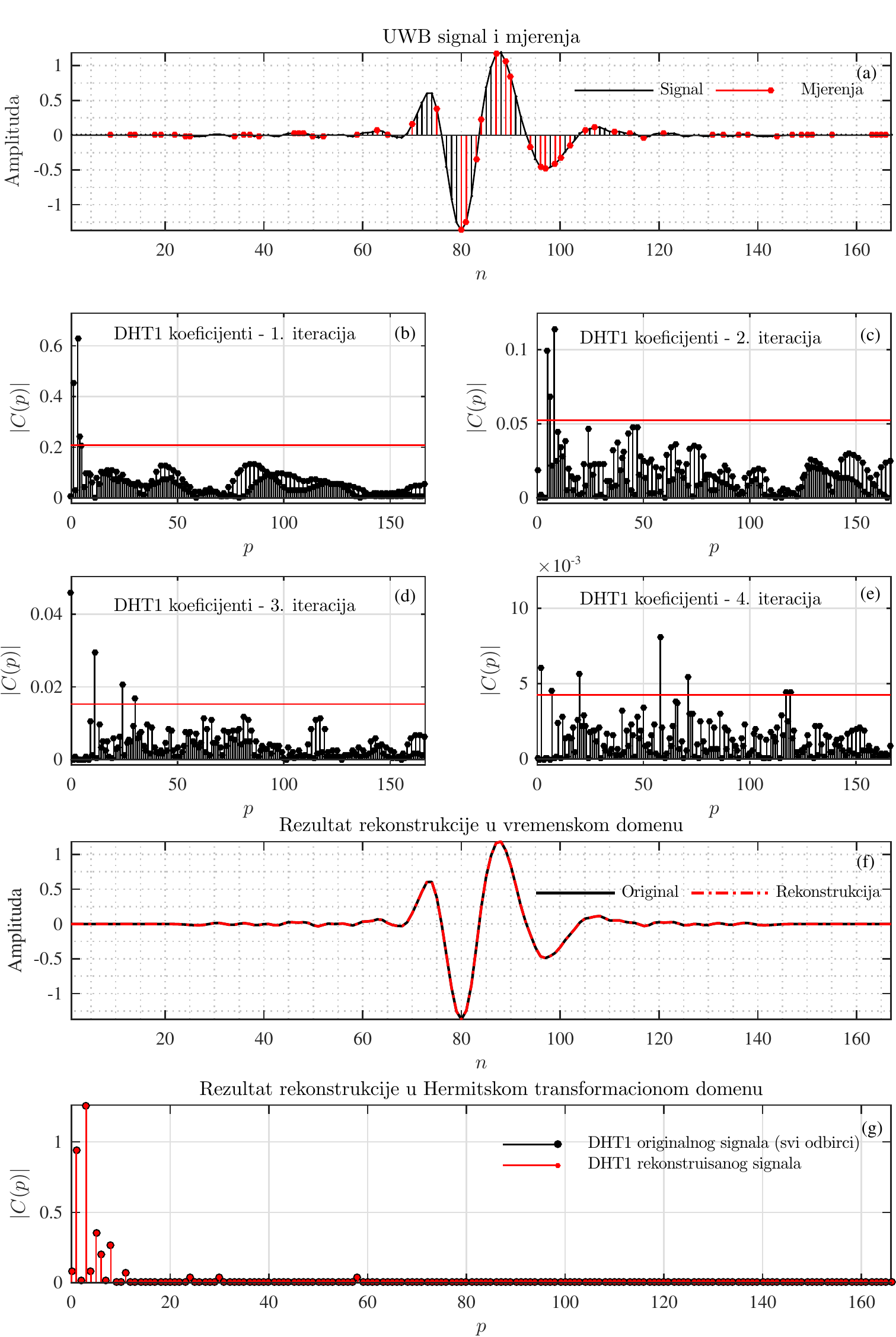}%
	\caption[Rekonstrukcija UWB signala.]{Rekonstrukcija UWB signala: (a) originalni signal i dostupna mjerenja (crvene tačke), (b) -- (e) proces detekcije komponenti tokom 4 iteracije, sa adaptivnim pragom (horizontalna crvena linija), (f) originalni i rekonstruisani signal, (g) poređenje Hermitskih koeficijenata originalnog signala (sa svim dostupnim odbircima) i rekonstruisanog signala.}%
	\label{uwb_ht_rec}%
\end{figure}
\end{primjer}
 \begin{primjer}
Razmatra se rekonstrukcija eksperimentalnog UWB signala iz primjera \ref{uwbrekonstrukcija}. Navedeni signal je rijedak u DHT1 domenu, međutim, postoji određeni broj nenultih odbiraka sa jako malim vrijednostima. Od ukupno $N = 165$, dostupno je samo $N_A$ slučajno pozicioniranih odbiraka, gdje je ${{N}_{A}}=\{30,~50,~70,~90,~110,~130\}$. Rekonstrukcija je izvršena korišćenjem predloženog Algoritma \ref{htrec2}, kao i klasičnog OMP algoritma. U ovom primjeru je izvršeno poređenje njihovih performansi. Kao kriterijum zaustavljanja u oba algoritma, koriščen je stepen tačnosti $\delta ={{10}^{-13}}$. Maksimalni broj iteracija u oba algoritma je limitiran na 50, u slučaju kada zahtijevanu tačnost nije moguće postići. Poređenje sa stanovišta rekonstrukcione srednje kvadratne greške, vremena izvršavanja i prosječnog broja iteracija je  predstavljeno u tabeli \ref{tabuwb}. Rezultati su dobijeni na osnovu usrednjavanja na osnovu 500 nezavisnih rekonstrukcija signala sa slučajno pozicioniranim dostupnim odbircima. Iz tabele je jasno uočljivo da smanjenje broja iteracija i kraće vrijeme izvršavanja, uz zadovoljavajući nivo tačnosti, kod predloženog Algoritma \ref{htrec2} u odnosu na OMP, postoji dok god su ispunjeni uslovi za rekonstrukciju, odnosno za ${{N}_{A}}\ge 50$ u ovom primjeru. Navedene prednosti postaju manje evidentne kada se broj dostupnih odbiraka $N_A$ približava dužini signala $N$, zbog toga što su slabije komponente postaju manje izložene šumu nastalom usljed nedostajućih odbiraka.

\begin{table}
	\small
	\centering
	\caption{Poređenje procesa i rezultata rekonstrukcije realnog UWB signala primjenom Algoritma \ref{htrec2} i OMP algoritma, za različite vrijednosti $N_A$, sa stanovišta: srednje kvadratne greške (MSE), vremena izvršavanja algoritama i prosječnog broja iteracija.}
	\label{tabuwb}
	\begin{tabular}{lrrrrrr}
		\toprule
		& \multicolumn{2}{c}{MSE} & \multicolumn{2}{c}{Vrijeme izvršavanja}  & \multicolumn{2}{c}{Prosječni broj iteracija} \\
		\cmidrule(lr){2-3}\cmidrule(lr){4-5}\cmidrule(lr){6-7}
	  $N_A$&	\textbf{Algoritam \ref{htrec2}}    & \textbf{OMP} & \textbf{Algoritam \ref{htrec2}} & \textbf{OMP}  & \textbf{Algoritam \ref{htrec2}} & \textbf{OMP}  \\ \midrule
	 {30}    	& 10.02 dB               & 9.05 dB  	&	0.007 s 		& 0.006  s	&	48.49	&	29.83 \\
	 {50}  & 3.63 dB    	  & 3.38 dB 	&	0.008 s 		&0.015  s 	&	27.10	&	40.91 \\
	 {70}    & -5.18 dB 	 &-5.35 dB 	&	0.003 s 		&0.008 s	 &	9.31	&	22.77\\
	 {90}   &-21.08 dB 	 &-12.81 dB	&	0.001 s 		& 0.003 s	&	4.70	&	14.61\\
 {110}   &-268.80 dB 	&-31.65 dB	&	0.0016  s  	 & 0.0018  s	&	4.08	&	12.15\\
 {130}   & -270.09 dB 	&-270.83 dB     &	0.0017 s	 & 0.0019 s 	&	3.68	&	12.00\\

 \bottomrule
	\end{tabular}
\end{table}

 \end{primjer}
\subsection{Veza sa indeksom koherentnosti}
Pretpostavimo da se rekonstrukcija vrši MP klasom algoritama, na primjer, OMP Algoritmom \ref{Norm0Alg}. U najgorem mogućem slučaju, uticaj drugih komponenti na detekciju najjače komponente će biti najveći kada svih $K$ komponenti imaju jednake amplitude (bez gubljenja opštosti -- amplitude jednake jedinici). Pretpostavlja se da je dostupno samo $N_A$ od ukupno $N$ odbiraka signala. Srednja vrijednost posmatrane komponente je $\frac{N_A}{N}$. Indeks koherentnosti za matricu $\mathbf{A}^{T}{{\mathbf{A}}}$ definisan je sljedećim izrazom:
\begin{align}\mu=\max \left| \frac{N}{{{N}_{A}}}\sum\limits_{i=1,~p\ne {{p}_{l}}}^{{{N}_{A}}}{\frac{{{A}_{l}}}{N}\frac{{{\psi }_{p}}(t_{n_i}){{\psi }_{{{p}_{l}}}}(t_{n_i})}{{{({{\psi }_{N-1}}(t_{n_i}))}^{2}}}} \right|,~l=1,2,\dots, K.\end{align}
Sa druge strane, komponenta na poziciji $p$ koja odgovara šumu u DHT1 domenu usljed nedostajućih odbiraka, odnosno, za čiju poziciju važi $p \notin {\Pi}_K$, može biti zapisana u sljedećoj formi:
\begin{align}{{Q}(p,p_l)}=\sum\limits_{i=1,~p\ne {{p}_{l}}}^{{{N}_{A}}}{\frac{{{A}_{l}}}{N}\frac{{{\psi }_{p}}(t_{n_i}){{\psi }_{{{p}_{l}}}}(t_{n_i})}{{{({{\psi }_{N-1}}(t_{n_i}))}^{2}}}}.\end{align}
Očigledno je da se može uspostaviti relacija sa indeksom koherentnosti. Ukoliko se šum koji potiče od nedostajućih odbiraka u svim komponentama sabira na poziciji koja ne odgovara komponentama, odnosno, $p\ne p_l$, tada koeficijent šuma na posmatranoj poziciji ima najveću moguću vrijednost: 
\begin{align}
K\max \left| Q(p,p_l) \right|=K\frac{{{N}_{A}}}{N}\mu.
\end{align}
Posmatrajmo sada poziciju komponente signala $p={{p}_{l}}\in {\Pi}_K.$ Posmatrana komponenta će biti oštećena šumom koji potiče od nedostajućih odbiraka iz preostalih $K-1$ komponenti. U najgorem mogućem slučaju, sve ove komponente šuma se sabiraju, i to u suprotnom smjeru od srednje vrijednosti signala (u posmatranom slučaju $\frac{N_A}{N}$, u opštem slučaju $A_l\frac{N_A}{N}$). Pretpostavljajući najveću moguću vrijednost šuma $K\max \left| Q(p,p_l) \right|$, rezultujuća, najgora moguća vrijednost koeficijenta signala je
\begin{align}
\min \left\{ {{c}_{{{p}_{l}}}} \right\}=\frac{{{N}_{A}}}{N}-(K-1)\max \left| {Q(p,p_l)} \right|=\frac{{{N}_{A}}}{N}-(K-1)\frac{{{N}_{A}}}{N}\mu.
\end{align} 

Posmatrana komponenta signala može biti detektovana ako je veća od najvećeg koeficijenta šuma:
\begin{align}
\frac{{{N}_{A}}}{N}-(K-1)\frac{{{N}_{A}}}{N}\mu>K\frac{{{N}_{A}}}{N}\mu.
\end{align} 
Ovaj izraz se može preurediti u poznati uslov za rekonstrukciju, zasnovan na sparku posmatrane matrice \cite{sparkdef}:
\begin{align}
K<\frac{1}{2}\left( 1+\frac{1}{\mu} \right),
\end{align} 
koji predstavlja dobro poznati rezultat u oblasti teorije kompresivnog odabiranja.
\subsection{Uticaj aditivnog šuma i rekonstrukcija zašumljenih signala}

Dosadašnja analiza u ovoj sekciji podrazumijevala je da mjerenja nijesu zahvaćena eksternim aditivnim šumom. Budući da je šum uobičajen u praktičnim aplikacijama, u ovom odjeljku će prethodna analiza biti proširena na signale koji su zahvaćeni aditivnim šumom. Pretpostavimo da su mjerenja zahvaćena bijelim Gausovim šumom varijanse $\sigma _{\varepsilon }^{2}$. U takvom slučaju, ukupna smetnja u transformacionom domenu uzrokovana je i eksternim šumom i šumom usljed nedostajućih odbiraka. U prethodnoj sekciji je pokazano da je srednja vrijednost varijanse aditivnog šuma u DHT1 domenu zadata izrazom (\ref{varhtawgn_mean}):
\begin{align}\bar{\sigma} _{e}^{2}=\frac{\sigma _{\varepsilon }^{2}}{{{N}^{2}}}\sum\limits_{n=1}^{N}{\frac{1}{\psi _{N-1}^{2}({{t}_{n}})}}=\xi (N)\sigma _{\varepsilon }^{2}\label{ovdje1},\end{align}
dok je tačna varijansa definisana kao (\ref{varhtawgn}):
\begin{equation}
\sigma _{e}^{2}(p)=\frac{\sigma _{\varepsilon }^{2}}{{{N}^{2}}}\sum\limits_{n=1}^{N}{\frac{\psi _{p}^{2}({{t}_{n}})}{{{\left[ {{\psi }_{N-1}}({{t}_{n}}) \right]}^{4}}}}=\gamma \left( p,N\right)\sigma _{\varepsilon }^{2},
\label{ovdje2}
\end{equation}
koja je, u cilju boljeg naglašavanja njenog porijekla, označena sa $\sigma_{e}^{2}$ u okviru ove sekcije. Budući da su posmatrane slučajne varijable Gausovske prirode i međusobno nekorelisane, koeficijenti šuma u DHT1 domenu u uslovima nedostajućih odbiraka i aditivnog eksternog šuma imaju ukupnu varijansu:
\begin{align}\sigma _{T}^{2}(p)=\sigma _{csN}^{2}+\sigma_{e}^{2}(p)=\frac{{{N}_{A}}N-N_{A}^{2}}{{{N}^{2}}(N-1)}\sum\limits_{l=1}^{K}{A_{l}^{2}}+\sigma_{e}^{2}(p),\end{align}
pri čemu $\xi (N)=\frac{1}{N^2}\sum\nolimits_{n=1}^{N}{\psi _{N-1}^{-2}({{t}_{n}})}$ ne zavisi od pozicije koeficijenata, dok $\gamma \left( p,N\right)=\frac{{1}}{{{N}^{2}}}\sum\nolimits_{n=1}^{N}{\frac{\psi _{p}^{2}({{t}_{n}})}{{{\left[ {{\psi }_{N-1}}({{t}_{n}}) \right]}^{4}}}}$ zavisi.
Kako bi se zadržala ista vjerovatnoća $P_{NN}^{{}}(T)$ da $N-K$ nezavisnih koeficijenata šuma bude manje od praga $\Xi =T$, koji je dat izrazom (\ref{htprag111}), sljedeći uslov treba da bude zadovoljen:
\begin{align}\frac{\frac{{{N}_{A}}(N-N_{A}^{{}})}{{{N}^{2}}(N-1)}\sum_{l=1}^{K}{A_{l}^{2}}}{\frac{{{N}_{i}}(N-N_{i}^{{}})}{{{N}^{2}}(N-1)}\sum_{l=1}^{K}{A_{l}^{2}}+\sigma_{e}^{2}(p)}=1.\end{align} 

Aproksimacija izraza $\sum\nolimits_{l=1}^{K}{A_{l}^{2}}$ se može jednostavno dobiti na osnovu dostupnih odbiraka, korišćenjem $\sum\nolimits_{l=1}^{K}{A_{l}^{2}}\approx \tfrac{N}{{{N}_{A}}}\sum\nolimits_{p=0}^{N-1}{c_{p}^{2}}$, pa se stoga taj izraz može smatrati poznatim. Za zadatu dužinu signala $N$, dostupni broj odbiraka $N_A$ i varijansu eksternog šuma $\sigma _{\varepsilon }^{2}$, može se odrediti $N_i$ koji reprezentuje povećanje broja mjerenja (dostupnih odbiraka), neophodno da se u prisustvu šuma zadrži ista vjerovatnoća $P_{NN}(T)$, koja je važila u slučaju kada eksternog aditivnog šuma nije bilo.

U cilju postizanja optimalnih rezultata prilikom rekonstrukcije signala zašumljenog aditivnim bijelim šumom varijanse  $\sigma _{\varepsilon }^{2}$, prag (\ref{htprag111}) može biti zamijenjen nelinearnim pragom
\begin{align}T (p)=\sqrt{2}\sigma _{T}^{{}}(p){{\operatorname{erf}}^{-1}}\left( {{\left( P_{NN}^{{}}(T ) \right)}^{\frac{1}{M}}} \right),\label{nonlintr}\end{align}
ili se u okviru $\sigma _{T}^{{}}(p)$ može koristiti prosječna varijansa $\bar{\sigma} _{e}^{2}$ koja je nezavisna od pozicija $p$ u DHT1 domenu.

Ulazni odnos signal-šum (SNR), za signal sa aditivnim šumom i u slučaju kada su svi odbrici prisutni je dat izrazom: \begin{align}SNR=10\log \frac{\sum\nolimits_{n=1}^{N}{{{\left| x({{t}_{n}}) \right|}^{2}}}}{\sum\nolimits_{n=1}^{N}{{{\left| \varepsilon ({{t}_{n}}) \right|}^{2}}}}=10\log \frac{{{E}_{x}}}{{{E}_{\varepsilon }}},\end{align} 
gdje je ${{E}_{\varepsilon }}=N\sigma _{\varepsilon }^{2}$  i ${{E}_{f}}=\sum\nolimits_{n=1}^{N}{{{\left| x({{t}_{n}}) \right|}^{2}}}$. Komponente signala u DHT1 domenu imaju srednju vrijednost $\tfrac{{{N}_{A}}}{N}{{A}_{l}},~l\in {\Pi}_K.$ U procesu rekonstrukcije, koeficijenti koji odgovaraju komponentama signala su pojačani za faktor $\tfrac{N}{{{N}_{A}}}.$ Pretpostavljajući uspješnu CS rekonstrukciju, koeficijent komponente signala postaje jednak originalnoj vrijednosti, odnosno onoj vrijednosti koju bi imao kada su svi odbirci signala dostupni. Ako u $N_A$ mjerenja postoji mali aditivni šum varijanse $\sigma _{\varepsilon }^{2}$, tada u jednom koeficijentu inicijalne DHT1 (računate uz pretpostavku da nedostajući odbirci imaju vrijednosti jednake nuli) varijansa šuma se množi sa $\tfrac{{{N}_{A}}}{N}.$ Energija aditivnog šuma će biti uvećana za ${{\left( \tfrac{N}{{{N}_{A}}} \right)}^{2}}$ u svakom nenultom koeficientu rekonstruisanog signala. Budući da uspješna rekonstrukcija daje tačno $K$ nenultih DHT1 koeficijenata koji odgovaraju komponentama signala, pri čemu je preostalih $N - K$ koeficijenata jednako nuli, energija šuma ostaje prisutna samo u $K$ nenultih koeficijenata. Odnos signal-šum u rekonstruisanom signalu je stoga jednak: 
\begin{align}
SN{{R}_{r}}=10\log \frac{\sum\nolimits_{n=1}^{N}{{{\left| x({{t}_{n}}) \right|}^{2}}}}{K\tfrac{{{N}_{A}}}{N}{{\left( \tfrac{N}{{{N}_{A}}} \right)}^{2}}\sigma _{\varepsilon }^{2}}=10\log \frac{{{E}_{f}}}{\tfrac{K}{{{N}_{A}}}{{E}_{\varepsilon }}}.
\end{align}
To znači da je inicijalni SNR povećan za $-10\log \left( K/{{N}_{A}} \right)$:
\begin{align}
SN{{R}_{r}}=SNR-10\log \left( \frac{K}{{{N}_{A}}} \right),
\label{htmanjisum}
\end{align} 
budući da je $K<{{N}_{A}}$. 

Važno je naglasiti da navedeni rezultat važi u prisustvu slabog aditivnog šuma u mjerenjima, odnosno kada je moguća rekonstrukcija $K$ nenultih komponenti signala. To takođe znači da predstavljeni algoritmi mogu biti primijenjeni u svrhu redukcije aditivnog šuma namjernim smanjivanjem broja dostupnih odbiraka signala. Naime, ukoliko se pri rekonstrukciji koristi najmanje moguće $K$, zaostali aditivni šum će biti nakon rekonstrukcije prisutan u samo $K$ koeficijenata selektovanih algoritmom. Navedeni zaključci će biti numerički verifikovani u okviru ove sekcije.
\subsubsection{Numerički rezultati}
\begin{primjer}
Razmatra se zašumljeni signal definisan sljedećim izazom:
\begin{align}
x(t_n)=\sum\limits_{l=1}^{K}{{{A}_{l}}{{\psi }_{{{p}_{l}}}}}(t_n)+\varepsilon (t_n),\label{htnoiserecmod}
\end{align}
koji je rijedak u DHT1 domenu i koji ima $K=6$ komponenti sa amplitudama ${{A}_{1}}=2,$ ${{A }_{2}}=-3,$ ${{\alpha }_{3}}=2.7,$ ${{A}_{4}}=2.1,$ ${{A}_{5}}=2.1,$ ${{A}_{6}}=1.4,$ na pozicijama ${{p}_{1}}=0,$ ${{p}_{2}}=1,$ ${{p}_{3}}=3,$ ${{p}_{4}}=6,$ ${{p}_{5}}=7$ i ${{p}_{6}}=18$. Signal je zašumljen aditivnim bijelim Gausovim šumom, tako da je rezultujući odnos signal-šum $SNR=7$ dB. Iz signala je odabrano $N_A = 54$ slučajno pozicioniranih odbiraka, od ukupno $N = 128$ (42.19\%). Rekonstrukcija je obavljena u samo dvije iteracije Algoritma \ref{htrec1}, sa modifikovanim nelinearnim pragom (\ref{nonlintr}). Rezultati rekonstrukcije su prikazani na slici \ref{sinteticki_denoising_CS}. Inicijalna MSE između originalnog (nezašumljenog) i zašumljenog signala od $-11.67$ dB primijenjenom rekonstrukcijom je smanjena za oko $12$ dB. Rezultujuća MSE između originalnog (nezašumljenog signala) i rekonstruisanog signala iznosi $-23.66$ dB. Ovaj primjer ilustruje činjenicu da je moguće postići smanjivanje nivoa šuma redukovanjem broja mjerenja, što je u skladu sa relacijom (\ref{htmanjisum}). 

Zašumljeni signal i dostupni odbirci (mjerenja) prikazani su na slici \ref{sinteticki_denoising_CS} (a), proces biranja komponenti primjenom nelinearnog praga u DHT1 domenu prikazan je na slikama \ref{sinteticki_denoising_CS} (b) i (c), rezultat rekonstrukcije je upoređen sa originalnim nezašumljenim signalom na slici \ref{sinteticki_denoising_CS} (d), dok je poređenje ovih rezultata u DHT1 domenu prikazano na slici \ref{sinteticki_denoising_CS} (e).
    \begin{figure}[htb]%
	\centering
	\includegraphics[
	]%
	{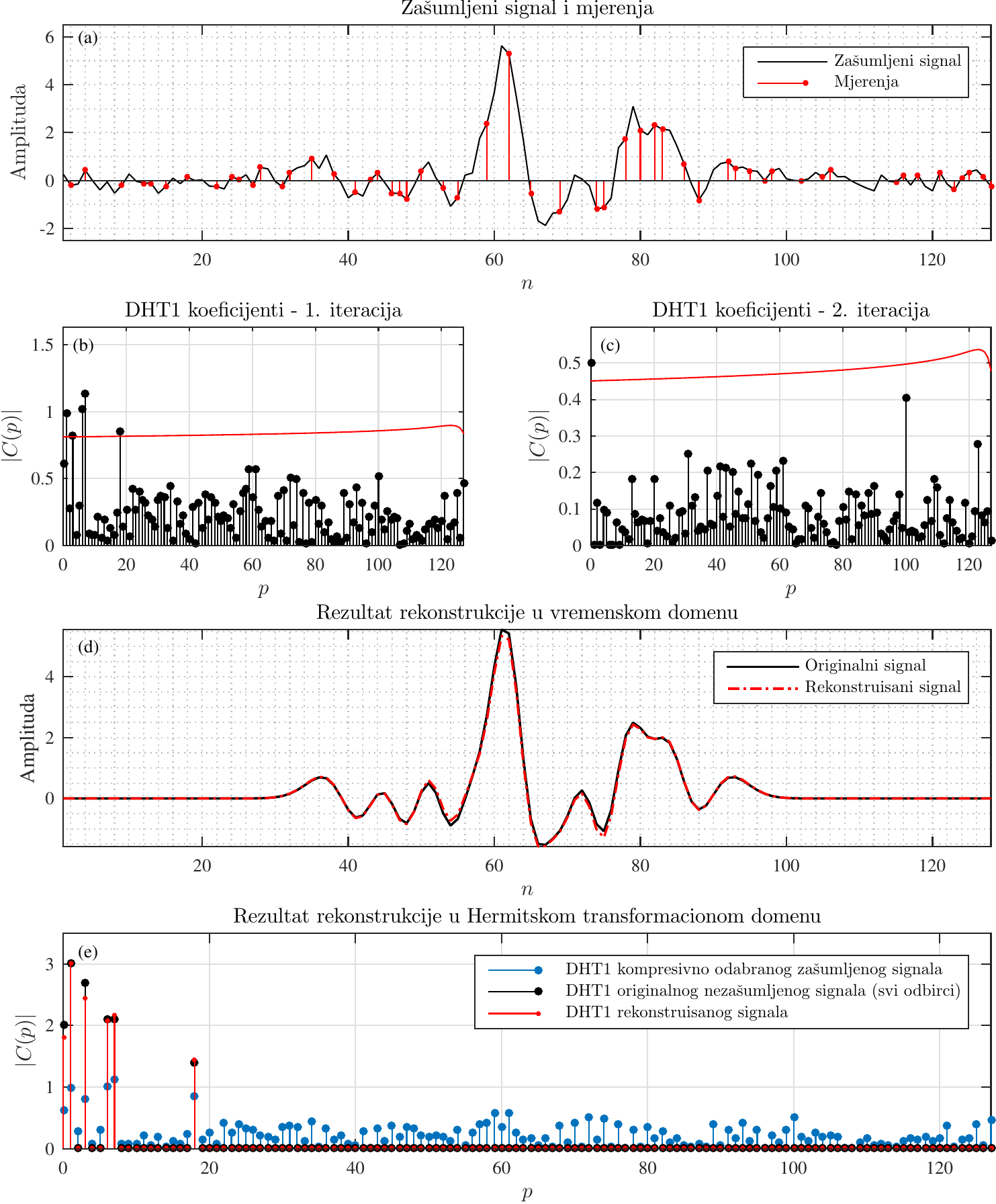}%
	\caption[Rekonstrukcija zašumljenog signala koji je rijedak u DHT1 domenu.]{Rekonstrukcija zašumljenog signala koji je rijedak u DHT1 domenu sa $N_A=54$ od $N=128$ dostupnih odbiraka: (a) originalni signal i dostupna mjerenja (crvene tačke), (b), (c) proces detekcije komponenti tokom dvije iteracije, sa adaptivnim nelinearnim pragom (horizontalna linija), (f) originalni i rekonstruisani signal, (g) poređenje Hermitskih koeficijenata originalnog signala (sa svim dostupnim odbircima), zašumljenog signala i rekonstruisanog signala.}%
	\label{sinteticki_denoising_CS}%
\end{figure}
\end{primjer}
\begin{primjer}
Razmatra se signal (\ref{htnoiserecmod}) sa $K = 3$ komponente, čije su amplitude ${{A}_{1}}=1$ ${{A}_{2}}=0.9$ i ${{A}_{3}}=0.6$. Pozicije komponenti $p_l,~l=1,2,3$ biraju se slučajno sa uniformnom raspodjelom iz skupa $p_l\in \{1,2,\dots,N-1\}$. Varijansa aditivnog bijelog Gausovog šuma $\varepsilon(t_n)$ je $\sigma _{\varepsilon }^{2}=0.1$, dok je (ulazni) SNR za sve dostupne odbirke $SNR_i=\text{7.67}$ dB. Izlazni prosječni SNR, odnosno odnos signal-šum nakon rekonstrukcije, dat izrazom (\ref{htmanjisum}) je evaluiran eksperimentalno. Numerički rezultat za rezultujući odnos signal-šum dobijen je na bazi 500 usrednjavanja rezultata rekonstrukcije (dobijenih primjenom OMP algoritma) za signale sa slučajnim pozicijama dostupnih odbiraka, slučajnim šumom $\varepsilon (t_n)$ i slučajnim pozicijama komponenti $p_l$. Odnos signal-šum rekonstrukcije je u svakoj realizaciji računat na bazi $N$ rezultujućih odbiraka. Rezultati su prikazani u tabeli \ref{tabnoise}, za nekoliko različitih brojeva dostupnih odbiraka (500 eksperimenata je ponovljeno za svaku zadatu vrijednost $N_A$). Može se uočiti da postoji veliko poklapanje teorijskih i eksperimentalnih rezultata, što potvrđuje validnost relacije (\ref{htmanjisum}).
	\begin{table}[!t]
	\small
	%% increase table row spacing, adjust to taste
	\renewcommand{\arraystretch}{1.3}
	% if using array.sty, it might be a good idea to tweak the value of
	% \extrarowheight as needed to properly center the text within the cells
	\caption{Odnos signal šum polaznog zašumljenog signala (zašumljenih dostupnih odbiraka) i rekonstruisanog signala (teorijski i eksperimentalni rezultat), prikazan za različite brojeve dostupnih odbiraka $N_A$.}
	\label{tabnoise}
	\centering
	%% Some packages, such as MDW tools, offer better commands for making tables
	%% than the plain LaTeX2e tabular which is used here.
	\begin{tabular*}{\textwidth}{l@{\extracolsep{\fill}}rrrr}
		\toprule
		Broj mjerenja $N_A$  & 	\textbf{60} &  	\textbf{120} & \!	\textbf{180} & \!	\textbf{240} \!\\
		\midrule
		Ulazni SNR  &7.82 dB &	7.45 dB &	7.48 dB &	7.27 dB\\
		SNR rekonstrukcije (teorija)	& 20.83 dB &	23.47 dB &	25.26 dB &	26.31 dB\\
		SNR rekonstrukcije (numerički) &	20.13 dB &	23.28 dB &	25.12 dB &	26.59 dB \\
		\bottomrule
	\end{tabular*}
\end{table}
\end{primjer}
\subsection{Analiza rekonstrukcije signala koji nijesu rijetki}
Ukoliko se izvrši rekonstrukcija signala koji nije rijedak u DHT1 domenu pod pretpostavkom da je rijedak, i to sa stepenom rijetkosti $K$, doći će do greške u rekonstrukciji, čija će energija biti proporcionalna energiji nerekonstruisanih $N-K$ komponenti signala. Budući da realni signali najčešće nijesu potpuno rijetki, već su rijetki samo aproksimativno, analiza i zaključci u vezi rekonstrukcije takvih signala uz pretpostavku o zadatom stepenu rijetkosti ima veliki značaj. 
\subsubsection{Teorema o grešci u rekonstrukciji signala koji nijesu rijetki}
Posmatra se signal koji nije čisto rijedak,  čije su najveće amplitude $A_{l}$,
$l=1,2,\dots,K$, i koji ima $N_A$ od ukupno $N$ slučajno pozicioniranih dostupnih odbiraka, gdje je $1\ll N_A\ll N$. Pretpostavimo da je taj signal rekonstruisan kao da je rijedak, sa stepenom rijetkosti $K$. Energija greške u $K$ rekonstruisanih koeficijenta $\left\Vert \mathbf{C}_{K}%
-\mathbf{C}_{T}\right\Vert _{2}^{2}$ je direktno povezana sa energijom nerekonstruisanih komponenti $\left\Vert \mathbf{C}_{T_z}-\mathbf{C}\right\Vert
_{2}^{2}$ sljedećom relacijom
\begin{equation}
\left\Vert \mathbf{C}_{K}-\mathbf{C}_{T}\right\Vert _{2}^{2} =\frac
{K(N-N_A)}{N_A(N-1)}\left\Vert \mathbf{C}_{T_z}-\mathbf{C}\right\Vert
_{2}^{2},
\end{equation}
gdje je $\mathbf{C}_{K}$  vektor rekonstruisanih komponenti dimenzija $K \times 1$, $\mathbf{C}_{T}$  je vektor dimenzija $K \times 1$  koji zadrži prave vrijednosti koeficijenata na pozicijama rekonstruisanih koeficijenata, $\mathbf{C}$ je vektor dimenzija $N \times 1$ i sadrži DHT1 koeficijente originalnog signala (kada su svi odbirci dostupni), dok je
$\mathbf{C}_{T_z}$ vektor dimenzija $N \times 1$ koji sadrži $K$ originalnih koeficijenata na rekonstruisanim (detektovanim) pozicijama i nule na preostalih $N-K$ pozicija.

\subsubsection{Dokaz teoreme} U signalu se nerekonstruisana $l$-ta komponenta manifestuje kao ulazni Gausov šum, čija je varijansa
\begin{equation} 
\sigma_{csN}^{2}=A_{l}^{2}\frac{N_A(N-N_A)}{N^{2}(N-1)}. \label{htvarNON}%
\end{equation}
Sve nerekonstruisane komponente će u transformacionom domenu predstavljati Gausov šum, čija je ukupna varijansa jednaka sumi:
\begin{equation}
\sigma_{T}^{2}=\sum_{l=K+1}^{N}A_{l}^{2}\frac{N_A(N-N_A)}{N^{2}(N-1)}.
\label{var_noise-f1_ht}%
\end{equation}

Nakon završene rekonstrukcije, ukupna energija šuma koji potiče od nerekonstruisanih komponenti (koji je prisutan u $K$ rekonstruisanih komponenti) biće jednaka:
\begin{equation}
\left\Vert \mathbf{C}_{K}-\mathbf{C}_{T}\right\Vert _{2}^{2}=K\frac{N^2}{N_A^2} \sigma^{2}_{T} =\frac{K(N-N_A)}{N_A(N-1)} \sum
_{l=K+1}^{N}A_{l}^{2}\label{vvvar_noise_ht}.
\end{equation}

Šum koji potiče od nerekonstruisanih komponenti se može direktno povezati sa energijom nerekonstruisanih komponenti:
\begin{equation}
\left\Vert \mathbf{C}_{T_z}-\mathbf{X}^{C}\right\Vert _{2}^{2}=\sum_{l=K+1}^{N}
A_{l} ^{2}.%
\end{equation}

Dakle, ukupna energija nerekonstruisanih komponenti data je sljedećim izrazom:
\begin{equation}
\left\Vert \mathbf{C}_{K}-\mathbf{C}_{T}\right\Vert _{2}^{2} =\frac
{K(N-N_A)}{N_A(N-1)}\left\Vert \mathbf{C}_{T_z}-\mathbf{C}\right\Vert,
\end{equation}
čime je dokaz završen.

U slučaju kada razmatrani signal sadrži eksterni aditivni šum, sa vrijednostima ispod nivoa rekonstruisanih komponenti u DHT1 domenu, posmatrana energija greške postaje:
\begin{equation}
\left\Vert \mathbf{C}_{K}-\mathbf{X}^{C}_{T}\right\Vert _{2}^{2} =\frac
{K(N-N_A)}{N_A(N-1)}\left\Vert \mathbf{C}_{T_z}-\mathbf{C}\right\Vert
_{2}^{2}+\frac{K}{N_A} \sigma^{2}_{\varepsilon}N,
\label{nonsparseht}
\end{equation}
imajući u vidu da je aditivni eksterni šum nezavisna slučajna varijabla sa Gausovom raspodjelom, pa se odgovarajuće varijanse sabiraju.
\begin{primjer}
	    \begin{figure}[ptb]%
		\centering
		\includegraphics[
		]%
		{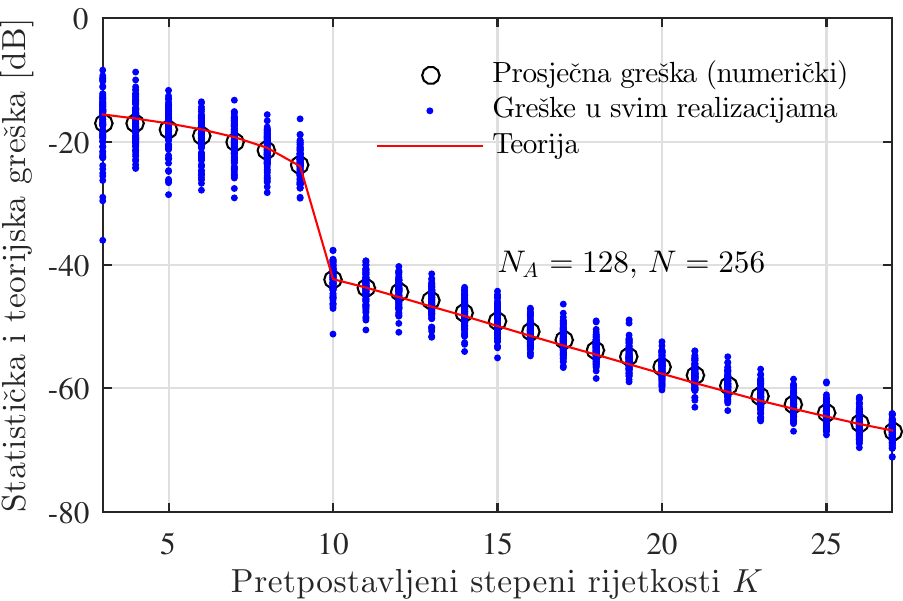}%
		\caption[Numerička evaluacija izraza za grešku u rekonstrukciji signala koji nije rijedak u DHT1 domenu, ali je rekonstruisan uz pretpostavku o rijetkosti.]{Numerička evaluacija izraza za grešku u rekonstrukciji signala koji nije rijedak u DHT1 domenu, ali je rekonstruisan uz pretpostavku o rijetkosti. Rekonstrukcija je obavljena OMP algoritmom. Greška je prikazana za različite pretpostavljene stepene rijetkosti.}%
		\label{h5t_nonsparse}%
	\end{figure}
Razmatra se signal koji nije rijedak u DHT1 domenu:
\begin{align}s(t_n)=\sum\limits_{l=1}^{N}{{{A}_{l}}{{\psi }_{{{p}_{l}}}}}(t_n)+\varepsilon (t_n),\end{align}
sa amplitudama ${{A}_{l}}=1$ za $l<S$ i ${{A}_{l}}=0.5{{e}^{-2l/(S+1)}}$ za $S+1\le l\le N$, uz $S = 10$. Pozicije komponenti $p_l\in {\Pi}_K$ su selektovane slučajno. Ovaj signal je, dakle, aproksimativno $S$-rijedak u DHT1 domenu. Dostupno je samo ${{N}_{A}}=128$ od ukupno $N=256$ slučajno pozicioniranih odbiraka. Signal je oštećen slabim bijelim Gausovim šumom srednje vrijednosti nula i standardne devijacije ${{\sigma }_{\varepsilon }}=0.1/N$. Rekonstrukcija nedostajućih odbiraka je obavljena OMP algoritmom, pretpostavljajući različite stepene rijetkosti: $3\le K\le 27$. Srednje kvadratne greške u rekonstrukciji su izračunate na bazi 100 nezavisnih realizacija zašumljenog signala sa slučajno pozicioniranim dostupnim odbircima i slučajnim realizacijama šuma. Dobijeni rezultati su upoređeni sa izvedenim izrazom za grešku. Obavljena je odgovarajuća normalizacija sa pretpostavljenim stepenom rijetkosti, tako da se razmatraju sljedeći izrazi: 
\begin{equation}
{{E}_{N}}=10\log \left( \tfrac{1}{K}\left\| {{\mathbf{C}}_{K}}-{{\mathbf{C}}_{R}} \right\|_{2}^{2} \right),
\end{equation}
za numerički dobijenu grešku i
\begin{equation}
{{E}_{T}}=10\log \left( \frac{(N-{{N}_{A}})}{{{N}_{A}}(N-1)}\left\| \mathbf{C}-{{\mathbf{C}}_{K}} \right\|_{2}^{2}+\frac{N}{{{N}_{A}}}\sigma _{\varepsilon }^{2} \right),
\end{equation}
za teorijsku grešku. Rezultati su prikazani na slici \ref{h5t_nonsparse}. Može se uočiti da se numerički i teorijski rezultati poklapaju u velikoj mjeri, što predstavlja eksperimentalnu potvrdu teorijski izvedenih izraza za grešku u rekonstrukciji signala koji nijesu rijetki, ali i za uticaj aditivnog šuma.
\end{primjer}

\subsection{Gradijentni rekonstrukcioni algoritam}
Do sada prezentovane tehnike za rekonstrukciju signala rijetkih u DHT1 domenu podrazumijevaju indirektnu minimizaciju u cilju rješavanja optimizacionog problema
\begin{align}\min {{\left\| \mathbf{C} \right\|}_{0}}~~\text{pod uslovom~~ }{{\mathbf{y}}}\mathbf{=}{{\mathbf{A}}}\mathbf{C},
\end{align}
gdje se, na bazi analize uticaja nedostajućih odbiraka na vektor DHT1 koeficijenata estimiraju pozicije komponenti signala, a zatim rješenje traži u vidu pseudoinverzije ${{\mathbf{C}}_{K}}={{\left( \mathbf{A}_{K}^{T}{{\mathbf{A}}_{K}} \right)}^{-1}}\mathbf{A}_{K}^{T}{{\mathbf{y}}}$, gdje je matrica $\mathbf{A}_{K}$ sastavljena od kolona mjerne matrice $\mathbf{A}$ koje odgovaraju estimiranim pozicijama komponenti. Pomoću ove pseudoinverzije određuju se vrijednosti koeficijenata. Kao što je već istaknuto, drugačiji pristup rješavanju ovog problema podrazumijeva njegovu reformulaciju u obliku:
\begin{align}\min {{\left\| \mathbf{C} \right\|}_{1}}~~\text{pod uslovom~~ }{{\mathbf{y}}}\mathbf{=}{{\mathbf{A}}}\mathbf{C},
\end{align}
gdje se vrši minimizacija $\ell_1$ norme ${{\left\| \cdot \right\|}_{1}}$, koja  se može obaviti, na primjer, pomoću algoritama iz oblasti linearnog programiranja. Pod određenim, strogo definisanim uslovima, navedeni minimizacioni problemi daju ista rješenja \cite{CandesRIP,dos}. Sa druge strane, minimizaciju je moguće obaviti i korišćenjem varijante gradijentnog algoritma, Algoritam \ref{gradijentni}. 

Na ovom mjestu algoritam će biti interpretiran sa stanovišta DHT1, uz notaciju koja će bolje naglasiti mehanizam rekonstrukcije u pojedinim koracima, Algoritam \ref{gradht}. Ideja je da se u vremenskom domenu vrši varijacija vrijednosti odbiraka signala, malom konstantom $\pm \Delta$ i mjeri uticaj tih varijacija na mjeru koncentracije. Na početku se inicijalizuje vektor:
\begin{align}
 \displaystyle 
x_r^{(0)}(t_n) \gets
\left\{
\begin{array}{ll}
y(t_n) & \text{ za } t_n\in \mathbb{N}_{Q} \\
0 & \text{ za } t_n\in \mathbb{N}_{A}
\end{array}
\right.
\end{align}

Formirajmo $N\times N$ matricu sastavljenu od mjernih vektora, sa nulama na nedostajućim pozicijama. U $m$-toj iteraciji ova matrica je:
\begin{align}
{{\mathbf{X}}^{(m)}}={{\mathbf{x}}_r^{(m)}}\,{{\mathbf{1}}_{1\times N}}=\left[ {{\begin{matrix}
		{{\mathbf{x}}_r^{(m)}} & {{\mathbf{x}}_r^{(m)}} & \ldots  & \mathbf{x} _r^{(m)} \\
		\end{matrix}}} \right].
\end{align}
čime se zadržavaju dostupne vrijednosti, dok se nedostajući odbirci posmatraju kao varijable minimizacije.
Odgovarajuća matrica $\mathbf{X}_{\mathbf{\Delta }+}^{(m)}$ kao kolone sadrži nasložene vektore ${{\mathbf{x}_r}^{(m)}}$, kojima je na pozicijama $t_n\in \mathbb{N}_{Q}=\mathbb{N}\backslash \mathbb{N}_A$ dodato $\Delta$, dok ekvivalentna matrica $\mathbf{X}_{\mathbf{\Delta }-}^{(m)}$ sadrži ove vektore sa oduzetim vrijednostima $\Delta$ na pozicijama nedostajućih odbiraka. Svaka kolona odgovara jednoj poziciji $t_n$. Ukoliko je $t_n\in \mathbb{N}_{A}$, u koloni će biti samo prepisan vektor  ${{\mathbf{x}_r}^{(m)}}$. Formirane matrice su (koraci 9. i 10. u Algoritmu \ref{gradht}):
	\begin{align}
\mathbf{X}_{\mathbf{\Delta }+}^{(m)}	= \left[ \begin{matrix}
\begin{matrix}
{{x}_r^{(m)}}(t_1)  \\
{{x}_r^{(m)}}(t_2)  \\
\vdots   \\
{{x}_r^{(m)}}(t_N)  \\
\end{matrix} & \ldots  & \begin{matrix}
{{x}_r^{(m)}}(t_1)  \\
{{x}_r^{(m)}}(t_2)  \\
\vdots   \\
{{x}_r^{(m)}}(t_N)  \\
\end{matrix}  \\
\end{matrix} \right]+\Delta {{\left[ \begin{matrix}
		\delta (t_1-{t_{{n}_{1}}})   & \cdots  & 0  \\
		0 &  \cdots  & 0  \\
		\vdots  & \ddots  & \vdots   \\
		0  & \cdots  & \delta (t_N-t_{{n}_{N}})  \\
		\end{matrix} \right]}_{N\times N}} \end{align} odnosno
\begin{align}
	\mathbf{X}_{\mathbf{\Delta }-}^{(m)}	= \left[ \begin{matrix}
\begin{matrix}
{{x}_r^{(m)}}(t_1)  \\
{{x}_r^{(m)}}(t_2)  \\
\vdots   \\
{{x}_r^{(m)}}(t_N)  \\
\end{matrix} & \ldots  & \begin{matrix}
{{x}_r^{(m)}}(t_1)  \\
{{x}_r^{(m)}}(t_2)  \\
\vdots   \\
{{x}_r^{(m)}}(t_N)  \\
\end{matrix}  \\
\end{matrix} \right]-\Delta {{\left[ \begin{matrix}
		\delta (t_1-{t_{{n}_{1}}})   & \cdots  & 0  \\
		0 &  \cdots  & 0  \\
		\vdots  & \ddots  & \vdots   \\
		0  & \cdots  & \delta (t_N-t_{{n}_{N}})  \\
		\end{matrix} \right]}_{N\times N}} \end{align}
gdje $\delta(\cdot)$ označava diskretnu, odnosno, Kronekerovu delta funkciju. Dalje, za obije matrice je potrebno, po kolonama, izračunati odgovarajuće DHT1. U tu svrhu uvodi se operator $ {{\mathcal{H}}^{(c)}}\left\{ \cdot \right\} $ koji, za svaku kolonu matrice računa proizvod $\mathbf{T}_H\mathbf{x}_{i\mathbf{\Delta }\pm},i=1,2,\dots,N$. Uvedimo i simboličku oznaku za $\ell_1$-normu matrice $\left\|\cdot\right\|_1^{(c)}$ koji ovu normu računa za svaku kolonu matrice zasebno.

 Cilj je aproksimirati gradijent mjere koncentracije, što se može postići oduzimanjem odgovarajućih mjera dobijenih za $ \left\| {{\mathcal{H}}^{(c)}}\left\{ \mathbf{X}_{\mathbf{\Delta }+}^{(m)} \right\} \right\|_{{{1}}}^{(c)}$ i $ \left\| {{\mathcal{H}}^{(c)}}\left\{ \mathbf{X}_{\mathbf{\Delta }-}^{(m)} \right\} \right\|_{{{1}}}^{(c)}$, odnosno, na bazi konačnih razlika. Računanjem ove razlike, formira se vektor gradijenta mjere koncentracije, ${{\mathbf{g}}^{(m)}}$, čijim se oduzimanjem od aktuelnog rješenja $x_r^{(m)}$ klasičnim metodom najbržeg spuštanja, korak 12 u Algoritmu \ref{gradht}, ide ka rješenju koje minimizuje mjeru koncentracije ($\ell_1$-normu DHT1 koeficijenata). Cilj je naći minimum mjere koncetracije, koji odgovara pravim vrijednostima nedostajućih odbiraka.

Korišćenjem uvedenih simboličkih oznaka, gradijent se računa na sljedeći način, korak 11 u Algoritmu \ref{gradht}:
\begin{align}
{{\mathbf{g}}^{(m)}}&=\frac{1}{2\Delta }\left[ \left\| {{\mathcal{H}}^{(c)}}\left\{ \mathbf{X}_{\mathbf{\Delta }+}^{(m)} \right\} \right\|_{{{1}}}^{(c)}-\left\| {{\mathcal{H}}^{(c)}}\left\{ \mathbf{X}_{\mathbf{\Delta }-}^{(m)} \right\} \right\|_{{{1 }}}^{(c)} \right]\notag\\&=\frac{1}{2\Delta }\left\{ \left[ \begin{matrix}
{{\left\| {{\mathbf{T}}_{H}}\mathbf{x}_{1\mathbf{\Delta }+}^{(m)} \right\|}_{{{1}}}} &\!\!\! \cdots  & {{\left\| {{\mathbf{T}}_{H}}\mathbf{x}_{N\mathbf{\Delta }+}^{(m)} \right\|}_{{{1}}}}  \\
\end{matrix} \right]\right.\left.-\left[ \begin{matrix}
{{\left\| {{\mathbf{T}}_{H}}\mathbf{x}_{1\mathbf{\Delta }-}^{(m)} \right\|}_{{{1}}}} &\!\!\! \cdots  & {{\left\| {{\mathbf{T}}_{H}}\mathbf{x}_{N\mathbf{\Delta }-}^{(m)} \right\|}_{1}}  \\
\end{matrix} \right] \right\}\notag.
\end{align}
U našoj analizi i numeričkim rezultatima, koristi se korak $\mu=1$ u liniji 12 posmatranog algoritma. U blizini minimuma mjere koncentracije, algoritam će dostići stacionarno stanje sa stanovišta srednje kvadratne greške, i počeće proces oscilovanja oko rješenja. Kako bi se povećala tačnost rješenja, oscilovanje je moguće detektovati mjerenjem ugla:
\begin{align}
\beta_m=\arccos\frac{\sum_{n=1}^{N}{g^{(m-1)}(t_n)g^{(m)}(t_n)}}{\sqrt{\sum_{n=1}^{N}{(g^{(m-1)}(t_n)})^2}\sqrt{\sum_{n=1}^{N}{(g^{(m)}(t_n)})^2}}
\end{align}
između dva gradijenta iz dvije susjedne iteracije. Ako je ovaj ugao, na primjer, veći od $170^{\circ}$, može se zaključiti da su vrijednosti nedostajućih odbiraka započele oscilacije oko pozicije koja odgovara minimumu mjere koncentracije. Kada se ovaj događaj detektuje, neophodno je smanjiti vrijednost parametra $\Delta$, na primjer, $\Delta\gets\Delta/3$ u koracima 14 i 15 Algoritma \ref{gradht}.  
\begin{algorithm}[!htb]
		\floatname{algorithm}{Algoritam}
	\caption{Gradijentni algoritam - interpretacija u DHT1 domenu}
	\label{gradht}
	\begin{algorithmic}[1]
		\Input
		\Statex
		\begin{itemize}
			\item Skup pozicija dostupnih odbiraka $\mathbb{N}_{A}$
			\item Dostupni odbirci (mjerenja) $y(n)$
			\item Korak $\mu$
			\item Zahtijevana tačnost $T_{max}$
		\end{itemize}
		\Statex
		\State  $m \gets 0$ \Comment Inicijalizacija indeksa iteracije
		\State Inicijalizovati vektor $\mathbf{x}_r^{(0)}$ sa vrijednostima: \Comment Vektor inicijalne estimacije signala
		$$ \displaystyle 
		x_r^{(0)}(t_n) \gets
		\left\{
		\begin{array}{ll}
		y(t_n) & \text{ za } t_n\in \mathbb{N}_{Q} \\
		0 & \text{ za } t_n\in \mathbb{N}_{A}
		\end{array}
		\right.
		$$
		\State 
		$\displaystyle \Delta \gets \max{|x^{(0)}(n)|}$ \Comment Inicijalizacija koraka
		\Repeat 
	\State	$x^{p}_r(t_n)\gets x_r^{(m)}(t_n)$
		\Repeat 
		\State $\mathbf{x}_r^{(m+1)} \gets \mathbf{x}_r^{(m)}$
		\State ${{\mathbf{X}}^{(m)}}\gets {{\mathbf{x}}_r^{(m)}}\,{{\mathbf{1}}_{1\times N}}=\left[ {{\begin{matrix}
				{{\mathbf{x}}_r^{(m)}} & {{\mathbf{x}}_r^{(m)}} & \ldots  & \mathbf{x}_r ^{(m)} \\
				\end{matrix}}} \right]$	\Comment ${{\mathbf{1}}_{1\times N}}$ je vektor jedinica

		\State 	
		$\mathbf{X}_{\mathbf{\Delta }+}^{(m)}	\gets\left[ \begin{matrix}
		\mathbf{x}_{1\mathbf{\Delta }+}^{(m)} & \mathbf{x}_{2\mathbf{\Delta }+}^{(m)} & \cdots  & \mathbf{x}_{N\mathbf{\Delta }+}^{(m)}  \\
		\end{matrix} \right]$

				\State 	
			$\mathbf{X}_{\mathbf{\Delta }-}^{(m)}	\gets\left[ \begin{matrix}
			\mathbf{x}_{1\mathbf{\Delta }-}^{(m)} & \mathbf{x}_{2\mathbf{\Delta }-}^{(m)} & \cdots  & \mathbf{x}_{N\mathbf{\Delta }-}^{(m)}  \\
			\end{matrix} \right]$
			
			\State  ${{\mathbf{g}}^{(m)}}\gets \frac{1}{2\Delta }\left[ \left\| {{\mathcal{H}}^{(c)}}\left\{ \mathbf{X}_{\mathbf{\Delta }+}^{(m)} \right\} \right\|_{{{1}}}^{(c)}-\left\| {{\mathcal{H}}^{(c)}}\left\{ \mathbf{X}_{\mathbf{\Delta }-}^{(m)} \right\} \right\|_{{{1 }}}^{(c)} \right]$ \Comment Gradijent mjere

		\State ${{\mathbf{x}}^{(m+1)}}\gets{{\mathbf{x}}^{(m)}}-\mu{{\mathbf{g}}^{(m)}}$ 
		%\Comment ${\mathcal{H}}^{(c)}\{\cdot\}$ računa DHT1 po kolonama matrice

		\State $m \gets m+1$
		\Until $\arccos\frac{\sum_{n=1}^{N}{g^{(m-1)}(t_n)g^{(m)}(t_n)}}{\sqrt{\sum_{n=1}^{N}{(g^{(m-1)}(t_n)})^2}\sqrt{\sum_{n=1}^{N}{(g^{(m)}(t_n)})^2}}<170^{\circ}$ \Comment Adaptacija koraka
		\State $\Delta \gets \Delta/3$  
		\Until 
		$\frac{\sum_{t_n\in\mathbb{N}_Q}\left|x_r^{p}(t_n)-x_r^{(m)}(t_n)\right|^2}{\sum_{t_n\in\mathbb{N}_Q}\left|x_r^{(m)}(t_n)\right|^2}<T_{max}$  \Comment Kriterijum zaustavljanja
		\State  $\mathbf{x}_r \gets \mathbf{x}_r^{(m)}$
		
		\Statex
		\Output
		\Statex
		\begin{itemize}
			\item Rekonstruisani vektor signala $\mathbf{x}_r$
		\end{itemize}
	\end{algorithmic}
\end{algorithm}

Konačno, kriterijum zaustavljanja se može definisati na osnovu promjena vrijednosti nedostajućih odbiraka tokom zadnjih itereracija:
\begin{align}
T_r=\frac{\sum_{t_n\in\mathbb{N}_Q}\left|x_r^{p}(t_n)-x_r^{(m)}(t_n)\right|^2}{\sum_{t_n\in\mathbb{N}_Q}\left|x_r^{(m)}(t_n)\right|^2},
\end{align}
koji predstavlja grubu estimaciju odnosa greške u rekonstrukciji i signala. U navedenoj relaciji, $x_r^{p}(t_n)$ označava signal koji je rekonstruisan prije redukcije koraka $\Delta$, dok je $x_r^{(m)}(t_n)$ rekonstruisan nakon iteracija sa redukovanim korakom. Ako je vrijednost $T_r$ veća od zahtijevane tačnosti, na primjer, $T_r>-100$ dB, tada se rekonstrukcija treba nastaviti sa redukovanim vrijednostima $\Delta$.
\begin{primjer}
	 \begin{figure}[ptb]%
		\centering
		\includegraphics[
		]%
		{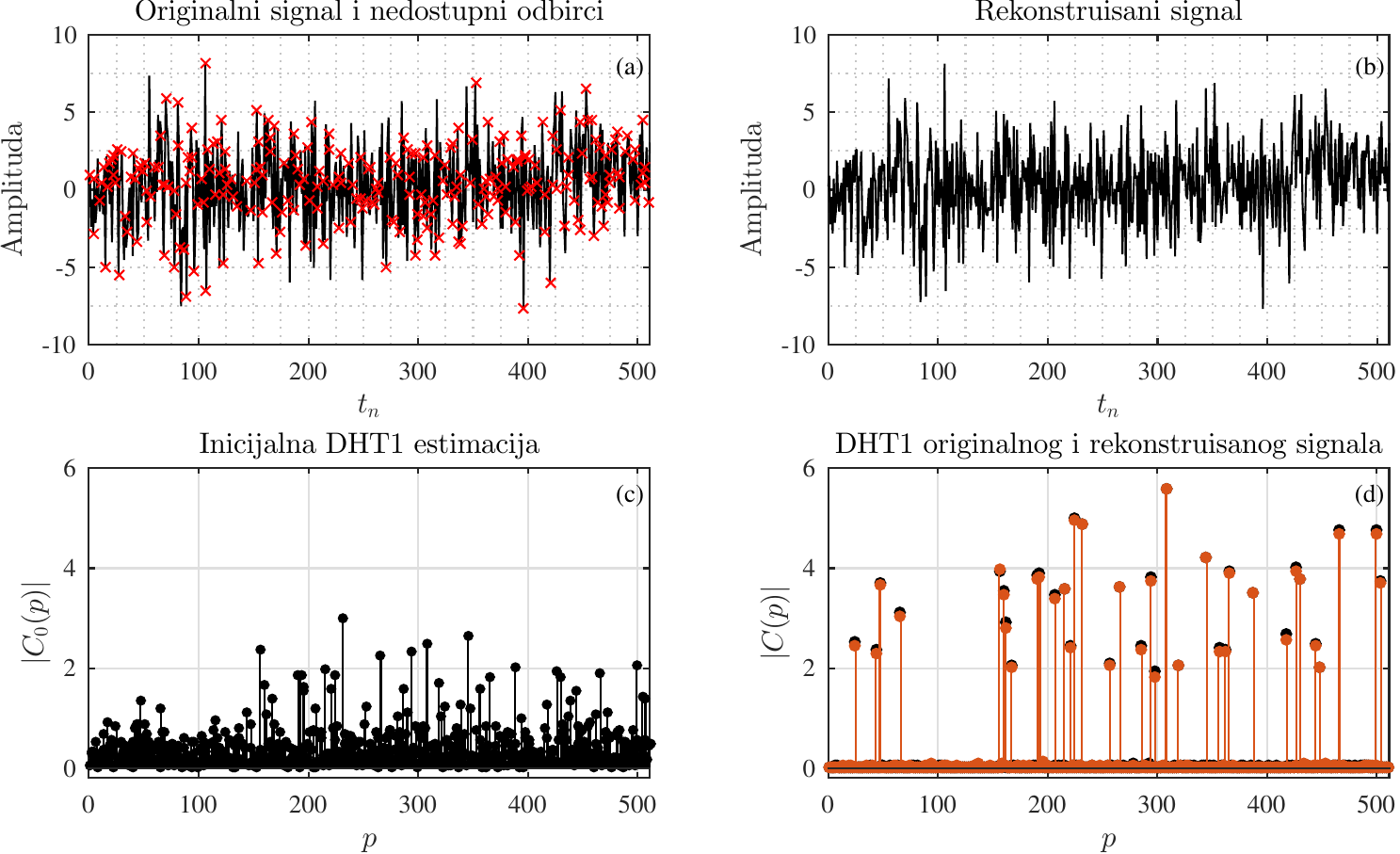}%
		\caption[Rekonstrukcija zašumljenog sintetičkog signala, rijetkog u DHT1 domenu primjenom gradijentnog algoritma.]{Rekonstrukcija sintetičkog signala rijetkog u DHT1 domenu primjenom gradijentnog algoritma: (a) originalni signal i nedostajući odbirci (crveni krstići), (b) rekonstruisani signal, (c) DHT1 signala sa nulama na mjestima nedostajućih odbiraka, (d) poređenje originalnog i rekonstruisanog signala u DHT1 domenu.}%
		\label{primjer_electronic_letters_sinteticiki2}%
	\end{figure}
	Razmatra se signal
	\begin{align}s(t_n)=\sum\limits_{l=1}^{N}{{{A}_{l}}{{\psi }_{{{p}_{l}}}}}(t_n)+\varepsilon (t_n),\end{align}
	dužine $N=512$ odbiraka, sa $K=35$ komponenti. Signal je kontaminiran aditivnim bijelim Gausovim šumom, gdje je $SNR=30$ dB i dostupno je samo 50\% odbiraka na slučajnim pozicijama. Amplitude $A_l$ i pozicije $p_l$ su slučajno odabrane. Rekonstrukcija signala je obavljena Algoritmom \ref{gradht}, i rezultati su prikazani na slici \ref{primjer_electronic_letters_sinteticiki2} (a) -- (d). Iako signal nije potpuno rijedak u DHT1 domenu, i pored činjenice što je zašumljen, postignuta MSE u rekonstrukciji je reda $10^{-3}$. U poređenju sa ,,$\ell_1$-magic'' algoritmom koji pripada grupi algoritama za konveksnu optimizaciju, vrijeme izvršavanja na istom računaru i u istom MATLAB\textsuperscript{\textregistered} okruženju je između 2 i 4 puta manje (zavisno od slučajnih pozicija mjerenja, pozicija koeficijenata signala i vrijednosti amplituda). Povećanjem broja dostupnih mjerenja, brzina gradijentnog algoritma se proporcionalno povećava.   

\end{primjer}
\begin{primjer}
	Algoritam je testiran i u kontekstu rekonstrukcije QRS kompleksa realnog EKG signala iz baze \cite{ecg_baza}. Rezultati su prikazani na slici \ref{primjer_electronic_letters_realni}. Budući da su odbirci signala bili dostupni u uniformnim tačkama (kao u dosadašnjim primjerima), on je reodabran u tačkama proporcionalnim nulama Hermitskog polinoma i u ovom slučaju koncentrisan korišćenjem optimizacije (\ref{lambda_opt}). Relativna greška usljed \textit{sparsifikacije} (anuliranja koeficijenata sa malim vrijednostima, u cilju postizanja bolje koncentracije) iznosi 5.23\%, što je medicinski prihvatljivo (manja je od 10 \%). Rekonstrukciona MSE od $1.262\cdot 10^{-7}$  je ostvarena nakon 18 iteracija gradijentnog algoritma. \begin{figure}[ptb]%
		\centering
		\includegraphics[
		]%
		{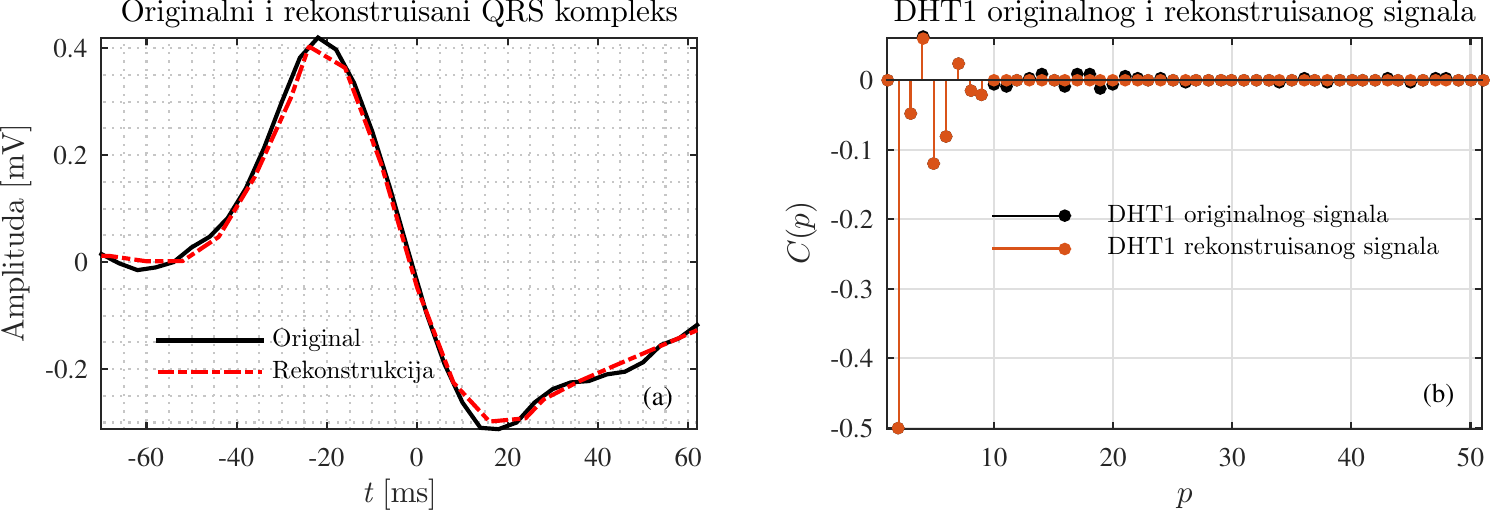}%
		\caption[Rekonstrukcija QRS kompleksa realnog EKG signala primjenog gradijentnog algoritma.]{Rekonstrukcija QRS kompleksa realnog EKG signala primjenog gradijentnog algoritma: (a) Originalni i rekonstruisani signal, (b) poređenje originalnog  i rekonstruisanog signala u DHT1 domenu.}%
		\label{primjer_electronic_letters_realni}%
	\end{figure}
\end{primjer}
%\subsection{Numerička analiza procesa rekonstrukcije i neki aspekti primjene}

\chapter{Diskretna kosinusna transformacija kao domen rijetkosti signala}
Diskretna kosinusna transformacija (DCT) je važan i često korišćen alat u analizi digitalnih slika i digitalnih audio signala. Jedna je od fundamentalnih transformacija u oblasti digitalne obrade signala. Određene osobine i specifičnosti DCT čine pogodnom reprezentacijom za analizu i obradu pomenutih vrsta signala, slično kao što je to bio slučaj sa Hermitskom transformacijom i QRS kompleksima.  Doprinosi prezentovani u ovoj glavi se odnose na kompresivno odabiranje i obradu signala koji imaju rijetku reprezentaciju u jednodimenzionom (1D) DCT i dvodimenzionom (2D) DCT domenu (u oznaci 2D-DCT). Za oba slučaja, biće izvršena detaljna analiza uticaja nedostajućih odbiraka na koeficijente u odgovarajućim transformacionim domenima, koja će dalje poslužiti kao osnova za analizu performansi rekonstrukcionih algoritama, unapređenje rekonstrukcionih pristupa, kao i za rasvjetljavanje fenomena vezanih za rekonstrukciju realnih signala u kontekstu kompresivnog odabiranja.  U okviru ove glave, teorijski rezultati će biti primjenjeni na realne audio signale za slučaj 1D DCT, odnosno, na digitalne slike za slučaj 2D verzije ove transformacije. Razvijena teorija i prezentovani rezultati su opšti i mogu biti primijenjeni i mimo razmatranih praktičnih konteksta.  Glava je podijeljena na dvije  sekcije, u okviru kojih se izučavaju: 1. oblast jednodimenzione DCT sa primjenom na audio signale, 2. oblast dvodimenzione DCT sa primjenom na digitalnu sliku.

\section{Jednodimenziona diskretna kosinusna transformacija kao domen rijetkosti signala}
Glavni doprinosi prezentovani u ovoj sekciji su publikovani u radu \cite{dctaudio}.
 Pored opštih teorijskih rezultata, biće prezentovana njihova primjena u kontekstu obrade audio signala \cite{dos,dctaudio,app_audio1,app_audioCS,app_audioCS2}. Izvršena je detaljna analiza uticaja nedostajućih odbiraka na DCT transformacione koeficijente. Na osnovu toga, uspostavljena je veza sa indeksom koherentnosti odnosno, sa opštom CS teorijom, u smislu uslova za rekonstrukciju. Izvedena je probabilistička analiza uspješnosti rekonstrukcije, eksplicitni izraz za energiju greške pri rekonstrukcji signala koji nijesu rijetki a rekonstruisani su pod pretpostavkom da jesu rijetki, kao i analiza uticaja aditivnog šuma na performanse rekonstrukcije. Rezultati su primijenjeni na rekonstrukciju realnih audio signala sa impulsnim smetnjama koje se pojavljuju pojedinačno i u blokovima. Teorija je potkrijepljena opsežnom eksperimentalnom analizom.
\subsection{Osnovne definicije}
Jednodimenziona diskretna kosinusna transformacija drugog tipa (poznata i kao DCT-II) se definiše na sljedeći način \cite{dos,dctknjiga}:
\begin{equation}
X^{C}(k)=\sum\limits_{n=0}^{N-1}a_{k}^{}x(n)\cos\left(  \frac{\pi(2n+1)}%
{2N}k\right),
\end{equation}
gdje $x(n)$ predstavlja diskretni signal dužine $N$, dok  $k=0,\dots,N-1,$ označava odgovarajuće frekvencijske indekse. Uzimajući u obzir činjenicu da je ovo najzastupljenija forma navedene transformacije, ona će u nastavku izlaganja biti implicitno podrazumijevana, i indeks II će biti izostavljen iz odgovarajućih zapisa i skraćenica. Diskretna kosinusna transformacija je invertibilna, i relacija 
\begin{equation}
x(n)=\sum\limits_{k=0}^{N-1}a_{k}^{}X^{C}(k)\cos\left(  \frac{\pi(2n+1)}%
{2N}k\right),
\end{equation}
omogućava računanje  odbiraka signala $x(n)$ na osnovu DCT koeficijenata $X^{C}(k)$, pri čemu indeksi diskretnog vremena uzimaju vrijednosti $n=0,1,\dots,N-1$. Skalirajući koeficijenti $a_k$ uzimaju vrijednosti $a_{k}=\sqrt{1/N}$ za $k=0$ i $a_{k}=\sqrt{2/N}$ za $k\neq0$. U matričnoj formi, transformacija ima sljedeći oblik
\begin{equation}
\mathbf{X}^{C}=\mathbf{\Phi} \mathbf{x},
\end{equation}
gdje $\mathbf{X}^{C}$, $\mathbf{\Phi}$, i
$\mathbf{x}$ predstavljaju: vektor DCT koeficijenata, DCT transformacionu matricu i vektor signala, respektivno. Za odgovarajuću inverznu transformaciju, matrična forma postaje: 
\begin{equation}
\mathbf{x}=\mathbf{\Phi}^{-1} \mathbf{X}^C.%
\end{equation}
Imajući u vidu ortogonalnost razmatrane transformacije, bitno je istaći  da važi $\mathbf{\Phi}^{-1}=\mathbf{\Phi}^{T}$, \cite{dctknjiga}. 
Za signal oblika%
\begin{equation}
x(n)=\sum\limits_{l=1}^{K}a_{k_l}A_{l}\cos\left(  \frac{\pi(2n+1)}{2N}%
k_{l}\right)   \label{modelprep}%
\end{equation}
može se reći da je rijedak u DCT domenu ukoliko je odgovarajući broj komponenti (odnosno DCT koeficijenata sa nenultim vrijednostima) $K$ mnogo manji od broja odbiraka signala $N$, tj. $K\ll N$. Amplitude komponenti označene su sa $A_l,~l=1,2,\dots,K$. U nastavku izlaganja, pozicije $k_1,k_2,\ldots,k_K$ će biti označene kao pozicije komponenti signala.

Računata po definiciji, DCT razmatranog modela signala (\ref{modelprep}) uzima oblik
\begin{equation}
\!X^{C}(k)\!=\!\sum\limits_{n=0}^{N-1}\sum\limits_{l=1}^{K}\!A_{l}a_{k}a_{k_l}\!\cos\left(\!
\tfrac{\!\pi(2n+1)}{2N}k_{l}\!\right)\!  \cos\left(\!  \tfrac{\pi(2n+1)}{2N}k\!\right) \label{DCTSS}
\end{equation}
gdje $k=0,\dots,N-1$. Komponente signala oblika $A_{l}\cos\left(\pi
k_{l}(2n+1)/(2N)\right)  $ se množe sa DCT baznim funkcijama, produkujući u (\ref{DCTSS}) članove
\begin{equation}
z(k_{l},k,n)=A_la_{k}a_{k_l}\cos\left(  \tfrac{\pi(2n+1)}{2N}k_{l}\right)
\cos\left(  \tfrac{\pi(2n+1)}{2N}k\right)\label{defxcc}.
\end{equation}
U slučaju kada su svi odbirci dostupni, odgovarajuća DCT je  $X^{C}(k)=\sum_{l=1}^{K}A_{l}\delta(k-k_l). $

\subsection{Uticaj nedostajućih odbiraka signala}
\label{dct_teorema}
Pretpostavimo da je dostupno samo ${N_A}\le N$ slučajno raspoređenih odbiraka signala
na pozicijama $n_i\in \mathbb{N}_A\mathbf{=}\left\{  n_{1},n_{2},\dots,n_{N_A}\right\}
\subseteq\mathbb{N}=\{0, 1, \dots, N-1\}$, iz skupa
\begin{gather*}
\mathbf{y}=\left\{  x(n_{1}),x(n_{2}),\ldots,x(n_{N_A})\right\}  \subseteq
\mathbf{x}%
\end{gather*}
gdje je
$
x(n_{i})=\sum_{k=0}^{N-1}a_{k}^{}X^{C}(k)\cos\left(  \frac{\pi(2n_{i}%
	+1)}{2N}k\right),~ i=1,2,\dots,N_A.%
$
Ne rizikujući gubitak opštosti izlaganja i zaključaka, prećutno je podrazumijevana uniformna funkcija gustine raspodjele pozicija odbiraka  $n_i\in \mathbb{N}_A$. U matričnom zapisu, dostupni odbirci se mogu zapisati u formi
$$\mathbf{y}=\mathbf{A} \mathbf{X}^{C},$$
gdje $\mathbf{A} $ predstavlja mjernu matricu dimenzija $N_A\times N$. Ona je definisana kao parcijalna inverzna DCT matrica, čije su vrste jednake onim vrstama matrice $(\mathbf{C}_N^{-1})$, koje odgovaraju pozicijama dostupnih odbiraka $n_i$:
\[
\setlength{\arraycolsep}{2.4pt}
\mathbf  {A}\!\! = \! {\tfrac{\sqrt{2}}{\sqrt{N}}} \!
\begin{bmatrix}
\frac{\sqrt 2}{2} &\operatorname{cos}\left(\frac{\pi(2n_1+1)}{2N}\right) & \!\!\cdots \!\! & \operatorname{cos}\left(\frac{\pi(2n_1+1)(N-1)}{2N}\right)  \\
\frac{\sqrt 2}{2} &  \operatorname{cos}\left(\frac{\pi(2n_2+1)}{2N}\right) & \!\!\cdots\!\! & \operatorname{cos}\left(\frac{\pi(2n_2+1)(N-1)}{2N}\right)  \\
\vdots & \vdots & \!\!\ddots\!\! & \vdots  \\
\frac{\sqrt 2}{2} & \operatorname{cos}\left(\frac{\pi(2n_{{N_A}}+1)}{2N}\right) & \!\!\cdots\!\! & \operatorname{cos}\left(\frac{\pi(2n_{{N_A}}+1)(N-1)}{2N}\right) \end{bmatrix}\!\!.
\]

U teoriji kompresivnog odabiranja (očitavanja), uobičajena je normalizacija srednjih vrijednosti energija kolona  (tj. elemenata na glavnoj dijagonali matrice $\mathbf  {A}^T\mathbf  {A}$). 
U tom slučaju, koristio bi se faktor $\sqrt{N_A/2}$, umjesto faktora $\sqrt{N/2}$.

Inicijalna DCT estimacija, koja je zasnovana na $\ell_2$-normi, definiše se kao DCT koja je računata samo na bazi dostupnih odbiraka:
\begin{align}
X^C_{0}(k)&=\sum\limits_{n \in\mathbb{N}_A}a_{k}^{}x(n)\cos\left(  \frac
{\pi(2n+1)}{2N}k\right)=\sum\limits_{i=1}^{N_A}\sum\limits_{l=1}^{K}z(k_l,k,n_i),
\end{align}
gdje je $k=0, 1,\dots,N-1$. Identičan rezultat bi se dobio ukoliko bi nedostajući odbirci imali vrijednosti jednake nuli \cite{dftmiss}. U matričnoj formi, prethodni rezultat postaje:
\begin{gather}
\mathbf{X}^{C}_{0}=\mathbf  {A}^{T}\mathbf{y}.%
\end{gather}
Članovi $z(k_l,k,n_i)$ pripadaju skupu
\begin{gather}
\mathbf{\Theta}=\left\{  z(k_{l},k,n_{1}),~z(k_{l},k,n_{2}),\dots,~z(k_{l}%
,k,n_{N_A})\right\},%
\end{gather}
koji je podskup kompletnog skupa odbiraka
\begin{gather*}
\Big\{  A_{l}a_{k}a_{k_l}\cos\left(  \frac{\pi(2n+1)}{2N}%
k_{l}\right)  \cos\left(  \frac{\pi(2n+1)}{2N}k\right)
,~n,~k=0,\dots,N-1,\ l=1,\dots,K\Big\}  .
\end{gather*}

{  
	Zato je skup pozicija nedostajućih odbiraka  $\mathbb{N}_Q$ zapravo podskup skupa pozicija svih odbiraka, $\mathbb{N}_Q=\mathbb{N\backslash N}_A$.}
Kako je ranije razmatrano \cite{dftmiss}, može se smatrati da su originalni odbirci signala na pozicijama definisanim skupom  $\mathbb{N}_Q$ zapravo zahvaćeni šumom:%
\begin{equation}
\eta(k_l,k,n)=\left\{
\begin{array}
{ll}%
-z(k_{l},k,n),&n\in\mathbb{N}_Q\\
0, &n\in\mathbb{N}_A,%
\end{array}
\right.
\end{equation}
gdje je $k=0,\dots,N-1,~l=1,\dots,K$. DCT koeficijenti%
\begin{align}
X^C_0(k)&=\sum\limits_{i=1}^{N_A}\sum_{l=1}^{K}z(k_{l},k,n_{i})=\sum\limits_{n=0}^{N-1}\sum_{l=1}^{K}\left[
z(k_{l},k,n)+\eta(k,k_l,n)\right]
\end{align}
sa dostupnim odbircima koji su slučajno pozicionirani i koji se za $1 \ll N_A \ll N$ mogu posmatrati kao slučajne promjenljive. U nastavku će biti sprovedena analiza statističkih osobina koeficijenata $X^C_0(k)$. U tom smislu, biće izvedeni srednja vrijednost i varijansa ovih koeficijenata, ali i utvrđena funkcija raspodjele kojoj oni podliježu.

\subsubsection{Statističke karakteristike DCT koeficijenata u slučaju signala sa nedostajućim odbircima} Pretpostavimo da se posmatra rijetki signal sa $K$ koeficijenata koji imaju nenulte vrijednosti u DCT domenu, na slučajnim pozicijama $k_{l}$ i sa amplitudama $A_{l}$, $l=1,2,\dots,K$.
Pretpostavimo da je od ukupno $N$ odbiraka signala dostupno samo njih $N_A$, pri čemu je $1\ll
N_A \ll N$. 
DCT koeficijenti  $X^C_0(k)$ računati na osnovu dostupnih odbiraka su slučajne promjenljive sa aproksimativno Gausovom raspodjelom. Njihove srednje vrijednosti i varijanse su date  sljedećim relacijama:
\begin{gather}
\mu_{{X^C_0}(k)}=\frac{N_A}{N}\sum_{l=1}^{K}A_{l}\delta(k-k_{l})\text{, i}
\\
\sigma^{2}_{{X^C_0}(k)}=\frac{N_A(N-N_A)}{N^{2}(N-1)}   {\sum_{l=1}^{K}}
A_{l}^{2}[ 1-\frac{1}{2}\delta(k-(N-k_l))
-\frac{1}{2}[1 +\delta(k_{l})]\delta(k-k_{l})]  , \label{varsm}%
\end{gather}
respektivno, gdje su $a_{k}$ ranije definisane DCT konstante. Dokaz će biti predstavljen u narednoj podsekciji.

\subsubsection{Dokaz}
U cilju očuvanja konciznosti i preglednosti izlaganja, dokazivanje ove teoreme ćemo razdvojiti u dva segmenta. Prvo ćemo posmatrati specijalni slučaj monokomponentnih (jednokomponentnih), a zatim i opšti slučaj multikomponentnih (višekomponentnih) signala.

\paragraph{Monokomponentni signali.} Posmatrajmo specijalni slučaj monokomponentnog signala, gdje je $K=1$, $k_{l}%
=k_{1}$. Na početku, smatraće se da je amplituda $A_1=1$. Polazeći od pretpostavki teoreme, inicjalna DCT signala sa $N_A$ dostupnih odbiraka se može računati kao:
\begin{equation}
X^C_0(k)=\sum\limits_{i=1}^{N_A}z(k_{1},k,n_{i}) \text{,} \label{dctk}%
\end{equation}
gdje su $z(k_{1},k,n_{i})$ definisani relacijom (\ref{defxcc}) sa $A_1=1$. Budući da su signali i bazne funkcije ortogonalni, važiće sljedeće svojstvo:%
\begin{equation}\label{orth}
\sum\limits_{n=0}^{N-1}z(k_{1},k,n)=\delta(k-k_{1}).
\end{equation}

Drugim riječima, za kompletan skup odbiraka se može pisati:%
\begin{align}
z(k_{1},k,0)+z(k_{1},k,1)+\dots+z(k_{1},k,N-1)&=1\text{, za }k=k_{1}\notag\\
z(k_{1},k,0)+z(k_{1},k,1)+\dots+z(k_{1},k,N-1)&=0\text{, za }k\neq k_{1}\text{.}
\label{orth0}%
\end{align}

\noindent\emph{Slučaj kada je $k=k_{1}.$} Za $k=k_{1}$, odgovarajući DCT koeficijent \ $X^C_0(k_{1})\,$ je slučajna promjenljiva. Imajući u vidu relaciju (\ref{orth}), kao i činjenicu da su sve vrijednosti $z(k_{1},k_1,n_{i})$ jednako distribuirane sa očekivanom vrijednošću $1/N$, lako se dolazi do zaključka da je srednja vrijednost promjenljive $X^C_0(k_{1})$ jednaka:%
\begin{align}
\mu_{{X^C_0}(k_1)}&=E\left\{  X^C_0(k_{1})\right\}  =E\left\{  z(k_{1},k_1,n_{1}%
)+\dots+z(k_{1},k_1,n_{N_A})\right\}  =\frac{N_A}{N}\text{.}
\label{srvr_sig}%
\end{align}
U slučaju kada je $A_1\neq1$, srednju vrijednost (\ref{srvr_sig}) treba pomnožiti sa $A_1$. Po definiciji, varijansa posmatrane slučajne promjenljive je data relacijom
\begin{equation}
\sigma^{2}_{{X^C_0}(k_1)}=E\left\{  \left\vert X^C_0(k)-\mu_{{X^C_0}(k_1)}\right\vert ^{2}
\right\}  =E\left\{  [X^C_0(k)]^{2}\right\}  -\mu_{{X^C_0}(k_1)}^2.%
\label{c0}
\end{equation}
Razvijanjem prvog člana prethodnog izraza koristeći definiciju (\ref{dctk}), dalje se dobija
\begin{align}
\sigma^{2}_{{X^C_0}(k_1)}&=E\Bigg\{
\left(  \sum\limits_{i=1}^{N_A}z(k_{1},k_1,n_{i})\right)  ^{2}\Bigg\}
-\mu^{2}_{C_0}(k_1) \notag
\\&
=E\Bigg\{  \sum\limits_{i=1}^{N_A}z^{2}(k_{1},k_1,n_{i}) +\sum\limits_{i=1}%
^{N_A}\sum\limits_{\substack{j=1\\i\neq j}}^{N_A}z(k_{1},k_1,n_{i}%
)z(k_{1},k_1,n_{j})\Bigg\}  -\mu^{2}_{{X^C_0}(k_1)}. \label{vars}%
\end{align}

Polazeći od (\ref{orth}) za $k=k_1$, nakon množenja lijeve i desne strane relacije sa
$z(k_{1},k_1,n)$, operator matematičkog očekivanja primijenjen na lijevu i desnu stranu daje:%
\begin{gather}
E\left\{  z(k_{1},k_1,0)z(n,k_1,k_1)+\dots+z(k_{1},k_1,N\!-\!1)z(k_{1}%
,k_1,n)\right\} \notag   =E\left\{  z(k_{1},k_1,n)\right\} =\frac{1}{N}.\label{kk1}
\end{gather}
Vrijednosti $z(k_{1},k_1,n)$ su jednako distribuirane \cite{dftmiss}. Kao posljedica ove činjenice, očekivanja $E\{ z(k_{1},k_1,m)z(k_{1},k_1,q)\}  $  za
$m \neq q,~m,q\in \mathbb{N}$ su ista i jednaka konstanti $B$. Stoga dalje možemo pisati:%
\begin{gather}
\left(  N-1\right)  B+E\left\{  z^{2}(k_{1},k_1,m)\right\}  =\frac{1}%
{N}\text{, ili}\nonumber\\
B=\frac{\frac{1}{N}-E\left\{  z^{2}(k_{1},k_1,m)\right\}  }{N-1}, m\in \mathbb{N}. \label{Aa}%
\end{gather}

Varijansa (\ref{vars}) sada dobija oblik:%
\begin{equation}
\sigma^{2}_{{X^C_0}(k_1)}\!\!=\!N_AE\left\{  z^{2}(k_{1},k_1,m)\right\} \! +\!N_A\left(
N_A-1\right)  B- \frac{N_A^2}{N^2},~ m\in\mathbb{N}_A\subseteq\mathbb{N}.\label{var_signala}%
\end{equation}

Posmatrajmo sljedeće očekivanje:
\begin{align}
E\left\{  z^{2}(k_{1},k_1,m)\right\}   &  =E\left\{  \left[  a_{k_1}^{2}%
\cos^{2}\left(  \tfrac{\pi(2m+1)}{2N}k_{1}\right)  \right]  ^{2}\right\}=E\left\{  a_{k_1}^{4}\left[  \frac{1}{2}+\frac{1}{2}\cos\left(  2\tfrac
{\pi(2m+1)}{2N}k_{1}\right)  \right]  ^{2}\right\} \nonumber\\
&  =\tfrac{a_{k_1}^{4}}{4}\left[  1+E\left\{  2\cos\left(  2\tfrac{\pi(2m%
	+1)}{2N}k_{1}\right)  \right\}  +E\left\{  \cos^{2}\left(  2\tfrac{\pi
	(2m+1)}{2N}k_{1}\right)  \right\}  \right] \nonumber\\
&  =\frac{a_{k_1}^{4}}{4}+\frac{a_{k_1}^{2}}{4N} ~\text{za}~ k_{1} \ne0.
\label{raz}%
\end{align}

Za $k_{1}=0$ očekivanje postaje $a_{k_1}^{4}$.

U prethodnom izvođenju, srednji član je kosinus čija je srednja vrijednost 0 za slučajno $m\in\mathbb{N}_A$, dok je posljednji član
DCT računat za $z(k_{1},k,n)$, $k=2k_{1}$. Koristeći
(\ref{srvr_sig}) i periodičnost DCT, možemo pisati:%
\begin{equation}
E\left\{  a_{k_1}^{2}\cos^{2}\left(  \frac{\pi(2m+1)}{2N}2k_{1}\right)
\right\}  =\frac{1}{N}\text{.}%
\end{equation}

Inkorporiranjem (\ref{raz}) u (\ref{var_signala}) i korišćenjem
(\ref{Aa}), nakon jednostavnog sređivanja se dobija:%
\begin{align}
\sigma^{2}_{{X^C_0}(k_1)}&=\frac{N_A(N-N_A)}{N^{2}(N-1)}\left[ N^{2}E\left\{
z^{2}(k_{1},k,m)\right\}  -1\right]\notag \\ &=\frac{N_A(N-N_A)}{N^{2}%
	(N-1)}\left[  N^{2}\left(  \frac{a_{k_1}^{4}}{4}+\frac{a_{k_1}^{2}}{4N}\right)
-1\right], ~\text{za } k_1\neq 0,
\label{sv111}
\end{align}
odnosno,
\begin{equation}
\sigma^{2}_{{X^C_0}(k_1)}=\frac{N_A(N-N_A)}{N^{2}%
	(N-1)}\left[  N^{2}a^4_{k_1}
-1\right]=0,~ \text{za} ~k_1=0.
\label{sv11}
\end{equation}
Za $A_1\neq1$, dobijene izraze treba pomnožiti sa $A_1^2$. Zamjenom odgovarajućih vrijednosti za $a_{k_1}$, dobija se rezultat iz teoreme:

%Matematičko očekivanje koje se pojavljuje u prvom članu relacije (\ref{var_signala}) je za $k_{1} \ne0$ jednako $
%E\left\{  z^{2}(k_{1},k_1,m)\right\}   
%=\frac{a_{k_1}^{4}}{4}+\frac{a_{k_1}^{2}}{4N} .
%\label{kvadratni}%
%$
%U slučaju kada je $k_{1}=0$, ovo očekivanje postaje $a_{k_1}^{4}$. Inkorporiranjem ovih rezultata u (\ref{var_signala}), sa $B$ expressed from (\ref{A}), then multiplying the variance expression with $A_1^2$ and replacing the values of $a_{k_1}$ we get the result as in Theorem 1:
\begin{equation}
\sigma^{2}_{{X^C_0}(k_1)}=\frac{N_A(N-N_A)}{N^{2}%
	(N-1)}\left[1-\frac{1}{2}(1+\delta(k_1))\right]A^2_1.
\label{vars_un}
\end{equation}

\noindent\emph{Slučaj kada je $k\neq k_{1}$}. DCT koeficijent na pozicijama koje ne odgovaraju komponentama signala, $X^C_0(k), k\neq k_{1}$ predstavlja slučajnu varijablu čije su statističke karakteristike drugačije od onih iz prethodno razmatranog slučaja. Naime, usljed svojstva ortogonalnosti (\ref{orth}) i činjenice da su sve vrijednosti 
$z(k_{1},k,n_{i})$ jednako distribuirane, lako se zaključuje da je srednja vrijednost koeficijenta sada jednaka nuli, odnosno:%
\begin{equation}
\mu_{{X^C_0}(k)}=E\left\{  X^C_0(k)\right\}  =0,~ k\neq k_{1}.
\end{equation}

Imajući u vidu da je srednja vrijednost nula, varijansa po definiciji postaje:%
\begin{align}
\sigma^{2}_{{X^C_0}(k)}&=E\left\{  \left\vert C_0(k)\right\vert^2 \right\}=E\left\{  \left(  \sum\limits_{i=1}^{N_A}z(k_{1},k,n_{i})\right)
^{2}\right\} \notag\\ 
&=E\Bigg\{  \sum\limits_{i=1}^{N_A}z^{2}(k_{1},k,n_{i}%
)+\sum\limits_{i=1}^{N_A}\sum\limits_{\substack{j=1\\i\neq j}}^{N_A%
}z(k_{1},k,n_{i})z(k_{1},k,n_{j})\Bigg\},~ k\neq k_{1}.  \label{sn_razvoj}%
\end{align}

Polazeći od (\ref{orth}), množenjem lijeve i desne strane izraza sa
$z(k_{1},k,n)$ i primjenjivanjem operatora matematičkog očekivanja na lijevu i desnu stranu izraza, dobija se
\begin{equation}
E\left\{  z(k_{1},k,0)z(k_{1},k,n)\right\}  +\dots+E\{z(k_{1},k,N-1)z(k_{1},k,n)\}=0. \label{noise_ort}%
\end{equation}

Analogno prethodno razmatranom slučaju, može se pretpostaviti da su vrijednosti $z(k_{1},k,n)$ jednako distribuirane, i da su očekivanja $E\left\{  z(k_{1}%
,k,m)z(k_{1},k,q)\right\}  $ za $m\neq q,~\text{uz}~ m= {0, 1, \dots, N-1},~q = {0, 1, \dots, N-1}$, ista i jednaka konstanti $D$. Ovih članova je ukupno $N-1$, dok postoji jedan član za koji je $m=n$. Navedene činjenice vode do
\begin{gather}
\left(N-1\right) D+E\left\{  z^{2}(k_{1},k,m)\right\} =0. \label{aconst}%
\end{gather}

Kako je $k\neq k_{1}$, uz pretpostavku da je $k\neq N-k_{1}$, nepoznato očekivanje $E\left\{  z^{2}(k_{1},k,m)\right\}
$ se može izraziti u formi
\begin{align}
E\left\{  z^{2}(k_{1},k,m)\right\}   &  =E\left\{  a_{k_1}^{2}\cos
^{2}\left(  \frac{\pi(2m+1)}{2N_A}k_{1}\right)  a_{k_1}^{2}\cos^{2}\left(
\frac{\pi(2m+1)}{2N_A}k\right)  \right\}  =\nonumber\\
&  E\left\{
z(k_{1},k_{1},m)\right\}  E\left\{  z(k,k,m)\right\}  =\frac{1}{N^{2}%
},\label{kvadrat}%
\end{align}
$m=0,1,\dots,N-1$. Sve vrijednosti $z(k_{1},k_{1},m)$, odnosno $z(k,k,m)$, jednako su distribuirane. Lako se pokazuje da važi $E\left\{  z(k_{1},k_{1},m)\right\}  =E\left\{
z(k,k,m)\right\}  =1/N$. 

U specijalnom slučaju kada je $k = N-k_{1}$, nepoznato očekivanje postaje
\begin{align}
E\left\{  z^{2}(k_{1},N-k_{1},m)\right\}   &  =E\left\{  a_{k_1k}^{2}\cos
^{2}\left(  \frac{\pi(2m+1)}{2N}k_{1}\right)  a_{k_1}^{2}\cos^{2}\left(
\frac{\pi(2m+1)}{2N}(N-k_{1})\right)  \right\}  \notag\\
&  =E\left\{  a_{k_1}^{2}\cos^{2}\left(  \frac{\pi(2m+1)}{2N}k_{1}\right)
a_{k_1}^{2}\cos^{2}\left(  \frac{\pi(2m+1)}{2N}k_{1}\right)  \right\}  \notag\\
&  =E\left\{  z^{2}(k_{1},k_1,m)\right\} ,~m=0,1,\dots,N-1,
\notag
\end{align}
\begin{equation}
E\left\{ z^{2}(k_{1},k_1,m)\right\}=\left\{
\begin{array}
{ll}%
\frac{a_{k_1}^{4}}{4}+\frac
{a_{k_1}^{2}}{4N},&k=N-k_1 \text{ i } k_1\neq0.\\
a_{k_1}^{4},&k=N-k_1 \text{ i } k_1=0.%
\end{array}
\right.
\label{izraz}
\end{equation}

Može se, dakle, zaključiti da je za koeficijent na poziciji $k=N-k_1$ varijansa definisana izrazom (\ref{vars_un}). Ovdje valja napomenuti da je iskorišćeno da je $\sin\left(  \frac{\pi(2m+1)}{2N}N\right)  =0$ i $\cos\left(
\frac{\pi(2m+1)}{2N}N\right)  =-1$, što se pojavljuje u izrazu
\begin{gather}
\cos\left(  \frac{\pi(2m+1)}{2N}(N-k_{1})\right)  =\cos\left(  \frac
{\pi(2m+1)}{2N}N\right)  \cos\left(  \frac{\pi(2m+1)}{2N}%
k_{1}\right) \notag\\ +\sin\left(  \frac{\pi(2m+1)}{2N}N\right)  \sin\left(
\frac{\pi(2m+1)}{2N}k_{1}\right).\notag
\end{gather}

Konačno, polazeći od definicije varijanse (\ref{sn_razvoj}), odnosno izraza $
\sigma^{2}_{{X^C_0}(k)}=N_AE\left\{  z^{2}(k,k_1,n_{i})\right\}  +N_A\left(
N_A-1\right)  D,~k\neq k_1
$, prateći prethodno izvedene izraze i inkorporirajući u rezultat nenultu amplitudu $A_1\neq1$, dobija se
\begin{align}
\sigma^{2}_{{X^C_0}(k)}&=\frac{N_A(N-N_A)}{N^{2}(N-1)}A_1^2\left[1-\frac{1}{2}\delta(k-(N-k_1))\right],~k \neq k_1,\label{var_noise2}%
\end{align}
što vodi do rezultata predstavljenog u teoremi.

\paragraph{Gausova distribucija.} Posmatrajmo distribuciju koeficijenata $X^C_0(k_1)=\sum\limits_{i=1}^{N_A}z(k_{1},k,n_{i})$ za veliko $N_A$ i $k_1 \ne k$. Funkcija gustine raspodjele vjerovatnoća normalizovane slučajne promjenljive sa srednjom vrijednošću koja je jednaka nuli, $c=X^C_0(k_1)/\sigma_{{X^C_0}(k_1)}$, prema Edžvortovom izrazu \cite{prob} je  
\begin{gather}
f(c)=\phi (c)+\frac{1}{4!N_A}\Big[\frac{\kappa_4}{\sigma^4}\phi ^{(4)}(c)+\frac{\kappa^2_3}{3\sigma^6}\phi ^{(6)}(c)\Big]+O(\frac{1}{N_A^2})(c).
\end{gather}
Prvi član je Gausova distribucija $\phi(c)=e^{-c^2/2}/\sqrt{2\pi}$, dok preostali članovi predstavljaju devijaciju od ove distribucije. Varijansa, zatim treći i četvrti momenti od $z(k_{1},k,n_{i})$ su označeni sa $\sigma^2$, $\kappa_3$, i $\kappa_4$, respektivno. U našem slučaju, za veliko $N_A$, imamo $\sigma^2 \to 1/N^2$, $\kappa_3 \to 0$, i $\kappa_4 \to 9/(4N^4)$. Zaključujemo da važi $\kappa_4/(4!N_A\sigma^4) \to3/(32N_A) \to 0$ i $f(c) \to \phi(c)$.   

\paragraph{Slučaj multikomponentnih signala.} Prethodno predstavljenu analizu za slučaj monokomponentnih signala sada možemo generalizovati na slučaj multikomponentnih signala. Posmatrana slučajna promjenljiva (\ref{dctk})  postaje:%
\begin{gather}
X^C_0(k)=%
%TCIMACRO{\dsum \limits_{l=1}^{K}}%
\sum_{i=1}^{N_A}\sum_{l=1}^{K}a_{k}^{2}A_{l}\cos\left(  \frac{\pi(2n_i+1)}{2N}k_{l}\right)\cos\left(
\frac{\pi(2n_i+1)}{2N}k\right).
\end{gather}

Prethodno sprovedena analiza ukazuje na to da i slučaju multikomponentnih signala, DCT koeficijenti na
$l$-toj poziciji u inicijalnoj estimaciji $X^C_0(k),~k=k_{l}$ predstavljaju Gausove slučajne promjenljive, sa srednjim vrijednostima različitim od nule $
\mu_{{X^C_0}(k)}=A_{l}\frac{N_A}{N},~l=1, 2, \dots, K,
$
dok  DCT koeficijenti na ostalim pozicijama, $X^C_0(k),~k\neq k_{l}$ takođe imaju karakteristike Gausove slučajne promjenljive, u ovom slučaju sa srednjim vrijednostima jednakim 0, budući da na tim pozicijama postoji samo šum koji je posljedica nedostajućih odbiraka u komponentama signala, čija je srednja vrijednost nula. Ovi zaključci slijede iz centralne granične teoreme \cite{clt}, i činjenice da sumiranje Gausovih slučajnih varijabli produkuje nove Gausove slučajne varijable. Srednja  vrijednost DCT koeficijenata za slučaj multikomponentnog signala može se predstaviti sljedećim unificiranim rezultatom:
\begin{align}
\mu_{{X^C_0}(k)}=\frac{N_A}{N}\sum_{l=1}^{K}A_{l}\delta(k-k_{l}).
\end{align}

Za DCT koeficijente $X^C_0(k)$ na pozicijama koje ne odgovaraju komponentama signala, tj. $k\neq k_{l}$
varijansa je:%
\begin{align}
\sigma^{2}_{{X^C_0}(k)}\!=\!\frac{N_A(N-N_A)}%
{N^{2}(N-1)}%
{\sum \limits_{l=1}^{K}}%
A_{l}^{2}\left[1-\frac{1}{2}\delta(k-(N-k_l))\right]\!\!. \label{varnm}%
\end{align}

Ovaj izraz se dobija jednostavno, imajući u vidu da na pozicijama $k\neq k_{l}$ nedostajući odbirci iz svih komponenti signala doprinose šumu, a šumovi koji potiču od svake pojedinačne komponente su Gausovi, međusobno nekorelisani slučajni procesi, sa srednjim vrijednostima nula. Varijansa šuma koji potiče od $l$-te komponente signala je $A_{l}^{2}\frac{N_A(N-N_A)}{N^{2}(N-1)}\left[1-\frac{1}{2}\delta(k-(N-k_l))\right],~l=1,\dots,K%
$.

Na ovom mjestu treba istaći da rezultat (\ref{varnm}) važi u smislu srednje varijanse DCT koeficijenata na pozicijama šuma (koje ne korespondiraju komponentama signala), budući da je u njenom izvođenju pretpostavljena statistička nezavisnost slučajnih promjenljivih. Međutim, strogo gledano, komponente signala pomnožene sa baznim funkcijama mogu izazvati tzv. efekat spajanja (engl. \textit{coupling effect}) ukoliko su pozicionirane tako da zadovoljavaju određene uslove. Tako na primjer, u dvokomponentnom rijetkom signalu, sa komponentama na pozicijama $k_1$ i $k_2$, ako je zadovoljen uslov $k_1+k_2=2k$, efekat spajanja (uparivanja) izazvaće povećanje varijansi na pozicijama $k_{c1}=(k_1+k_2)/2$ i $k_{c2}=(k_2-k_1)/2$. Međutim, na pozicijama $N-k_{c1}$ i $N-k_{c2}$ varijansa će biti u tom slučaju smanjena za iste vrijednosti. Kao posljedica navedenog, prosječna varijansa DCT koeficijenata na pozicijama koje ne odgovaraju komponentama signala $k\neq k_l$  ostaje ista i data je formulom  (\ref{varnm}), bez obzira na opisani efekat.

Prema analizi koja je predstavljena za monokomponentne signale, $k$-ta komponenta signala na poziciji $k=k_{p},~p\in\{1,2,\dots,K\}$ ima varijansu $A_{p}^{2}\frac{N_A(N-N_A)}{N^{2}%
	(N-1)}\left[1-\frac{1}{2}(1+\delta(k_p))\right]  
$
i srednju vrijednost $\mu_{{X^C_0}(k_p)}=A_{p}^{2}N_A/N$. Dodatno, na poziciji
$k=k_{p}$ postoji takođe šum izazvan nedostajućim odbircima u preostalih $K-1$ komponenti. Ovo znači da da je zbir slučajnih promjenljivih koje potiču od drugih komponenti signala sa pozicija $k_l, l=\left\{  1,2,\dots,K\right\},~l\neq p  $ dodat postojećoj slučajnoj promjenljivoj na poziciji $k_p$. Ovih $K-1$ slučajnih promjenljivih predstavljaju Gausove procese sa srednjom vrijednošću nula, i imaju varijanse $A_l^2\frac{N_A(N-N_A)}{N^{2}(N-1)}\left[1-\frac{1}{2}\delta(k-(N-k_l))\right]$ gdje je $l \neq p$ i $l=1,2,\dots,K,~p=1,2,\dots,K$. Rezultujuća slučajna promjenljiva je takođe Gausova, sa srednjom vrijednošću $\mu_{{X^C_0}(k_p)}=A_{p}^{2}N_A/N$ i varijansom%
\begin{gather}
\sigma^{2}_{{X^C_0}(k_p)}\!\!=\frac{N_A(N-N_A)}{N^{2}%
	(N-1)}\Big\{  A_{p}^{2}\left[1\!-\frac{1}{2}(1+\delta(k_p))\right] \!+\!
%BeginExpansion
{\displaystyle\sum\limits_{\substack{l=1\\l\neq p}}^{K}}
%EndExpansion
A_{l}^{2}\left[1-\frac{1}{2}\delta(k-(N-k_l))\right]\Big\}  . \label{varsm2}%
\end{gather}

Unifikacijom izraza (\ref{varnm}) i (\ref{varsm2}) se dolazi do izraza za varijansu (\ref{varsm}) koji je predstavljen u formulaciji teoreme.

\begin{figure*}[!h]%
	\centering
	\includegraphics
	{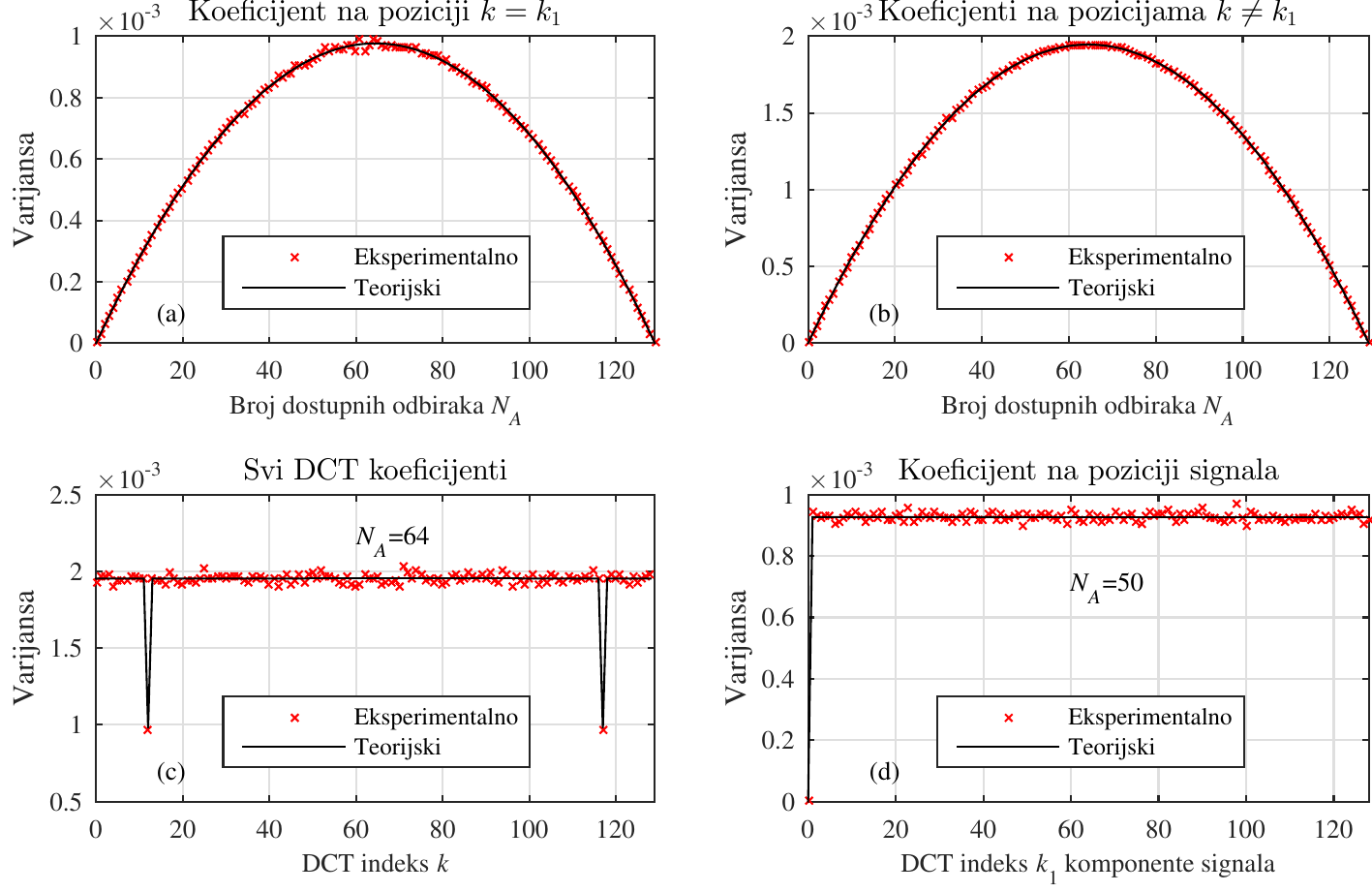}%
	\caption[Numerička evaluacija varijanse DCT koeficijenata u slučaju signala sa nedostajućim odbircima]{Numerička evaluacija varijanse DCT koeficijenata u slučaju signala sa nedostajućim odbircima:
		(a) Varijansa DCT koeficijenta na poziciji signala $k=k_{1}$, predstavljena kao funkcija od broja dostupnih odbiraka,
		(b) prosječna varijansa DCT koeficijenata na pozicijama koje ne odgovaraju komponentama signala $k\neq
		k_{1}$ kao funkcija od broja dostupnih odbiraka, (c) varijansa svih DCT koeficijenata monokomponentnog rijetkog signala sa $k_1=12$ i $N_A=64$ dostupnih odbiraka, (d) varijansa DCT koeficijenata $X^C_0(k=k_{1})$ u funkciji od
		$k_{1}$, računata za signal sa $N_A=50$ dostupnih odbiraka.}%
	\label{fig_stat}%
\end{figure*}
\subsubsection{Numerička provjera izraza za varijanse koeficijenata}
\begin{primjer}
	Posmatrajmo jednokomponentni signal koji je rijedak u DCT domenu i zadat relacijom
	(\ref{modelprep}), sa $K=1$, $k_{1}=12$, i $N=129$. Od ukupno $N$, dostupno je samo $N_A$ odbiraka signala na slučajnim pozicijama. 
	
	U prvom dijelu eksperimenta, broj dostupnih odbiraka $N_A$ variramo od 0 do
	$N-1$. Za svaki posmatrani broj dostupnih odbiraka $N_A$, varijansa  DCT koeficijenta $X^C_0(k_{1})$ na poziciji  komponente signala $k_1$ je izračunata numerički, usrednjavanjem rezultata dobijenih na bazi  30000 nezavisnih realizacija posmatranog signala sa slučajno raspoređenim nedostajućim (odnosno dostupnim) odbircima. Rezultat poređenja numerički dobijene varijanse sa teorijskim modelom (\ref{varsm}) predstavljen je na slici \ref{fig_stat} (a).
	
	Zatim je eksperiment sproveden i za DCT koeficijente na pozicijama koje ne odgovaraju komponentama signala, $k\neq k_1$ (kao i $k\neq N-k_1$). Pošto se posmatra $N-2=127$ koeficijenata, broj slučajnih realizacija je sada smanjen na 200 (ukupan broj posmatranih koeficijenata u svim realizacijama je 25400). Prosječna numerički dobijena varijansa i varijansa zadata teorijskim izrazom (\ref{varsm}%
	) su prikazane na slici. \ref{fig_stat} (b). 
	
	Kako bi istakli razliku između varijansi koeficijenata na pozicijama koje odgovaraju komponentama signala i varijansi na pozicijama koje ne odgovaraju komponentama signala, varijanse svih DCT koeficijenata signala sa  $N_A=64$ dostupnih odbiraka su izračunate numerički na osnovu  10000 nezavisnih realizacija signala i prikazane su na slici \ref{fig_stat} (c), gdje su upoređene sa teorijskim izrazom (\ref{varsm}). 
	
	Konačno, ispitana je i zavisnost varijanse od pozicije komponente signala $k_1$. U tom cilju, pozicija komponente $k_{1}$ je varirana u opsegu od 0 do $N-1$, pri čemu je broj dostupnih odbiraka $N_A=50$. Na bazi 10000 nezavisnih realizacija signala sa slučajno raspoređenim nedostajućim odbircima, dobijen je numerički rezultat za varijansu koeficijenata $X^C_0(k_{1})$, i on je upoređen sa teorijskim rezultatom (\ref{varsm}). Kao što je očekivano, varijansa nije zavisna od posmatrane pozicije $k_1$, osim za $k_{1}=0$, kada je jednaka nuli (što je predviđeno teorijskim izrazom). Rezultati za ovaj eksperiment su prikazani na slici \ref{fig_stat} (d). 
	
	U svim razmatranim slučajevima, dobijeno je veliko poklapanje teorijskog i numeričkih rezultata.
	\end{primjer}

\begin{figure}[ptb]%
	\centering
	\includegraphics[
	]%
	{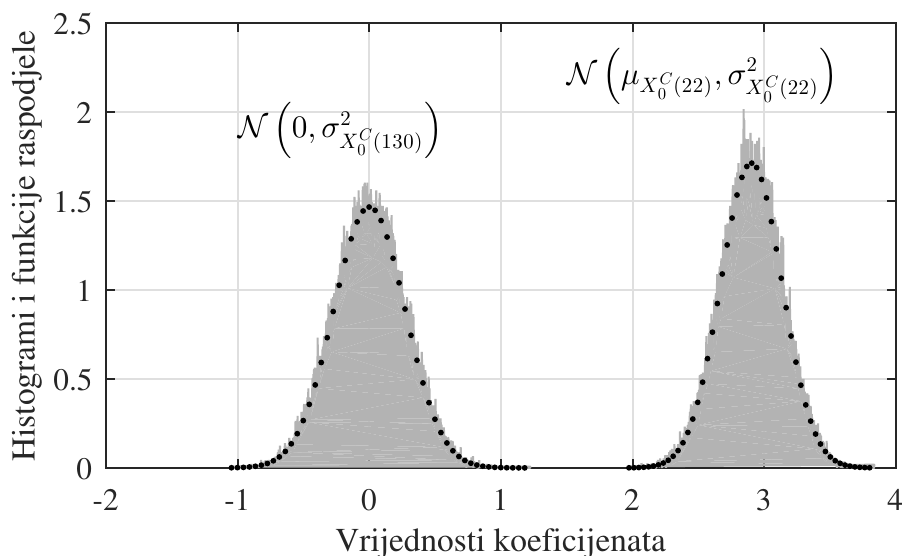}%
	\caption[Skalirani histogrami DCT koeficijenata i odgovarajuće teorijske funkcije gustine raspodjele vjerovatnoća]{Skalirani histogrami DCT koeficijenata i odgovarajuće teorijske funkcije gustine raspodjele vjerovatnoća: za koeficijent koji ne odgovara komponenti signala (lijevo) i za koeficijent koji odgovara poziciji signala (desno). Teorijski rezultat je označen tačkama.}%
	\label{histograms}%
\end{figure}

\begin{primjer} Posmatra se multikomponentni signal definisan izrazom (\ref{modelprep}), sa $K=5$. Dužina kompletnog signala je $N=155$. Pozicije komponenti signala su zadate indeksima
$k_{l}=\left\{  22,49,47,89,100\right\},  $  dok su odgovarajuće amplitude komponenti
$A_{l}=\left\{  5,3.5,1.5,2.5,1\right\} $. U cilju numeričke provjere raspodjele slučajne promjenljive $X^C_0(k),$  posmatra se 60000 nezavisnih realizacija signala sa slučajno raspoređenih $N_A=90$ odbiraka. Histogrami koeficijenta koji odgovara jednoj od komponenti, konkretno, $X^C_0(22)$, i koeficijenta čiji indeks ne odgovara nijednoj komponenti signala, konkretno, $X^C_0(130)$ su prikazani na slici \ref{histograms}. Histogrami su skalirani sa brojem realizacija. Histogram koeficijenta $X^C_0(22)$
je upoređen sa Gausovom distribucijom (predstavljenom tačkama) koja ima srednju vrijednost
$\mu_{{X^C_0}(22)}=A_{1}N_A/N\approx2.90$ i varijansu $\sigma^{2}_{{X^C_0}(22)}=0.0542,$
izračunatu po formuli (\ref{varsm}). Rezultat je prezentovan na slici \ref{histograms} (desno). Za koeficijent $X^C_0(130)$
skalirani histogram je upoređen sa Gausovom distribucijom čija je srednja vrijednost nula, a varijansa $\sigma^{2}_{{X^C_0}(130)}=0.0739$, izračunata po formuli  (\ref{varsm}). Rezultat je predstavljen na slici \ref{histograms} (lijevo).
\end{primjer}

\subsection{Rekonstrukcija zasnovana na analizi nedostajućih odbiraka} Koristeći rezultat (\ref{varsm}), varijansa DCT koeficijenata koji ne odgovaraju pozicijama komponenti se može izraziti u obliku:
\begin{align}
\sigma^{2}_{{X^C_0}(k)}=\frac{N_A(N-N_A)}%
{N^{2}(N-1)}%
{\sum \limits_{l=1}^{K}}%
A_{l}^{2}\left[1-\frac{1}{2}\delta(k-(N-k_l))\right],
\end{align}
gdje je $k\neq k_{l}$. Koeficijentima na pozicijama $k=N-k_l, ~l=1,2,\dots,K$ varijansa je smanjena za $A^2_l/2$ u poređenju sa ostalim koeficijentima na pozicijama koje ne odgovaraju indeksima komponentama signala. Može se pretpostaviti da je varijansa koeficijenata na svim pozicijama koje ne odgovaraju komponentama signala nezavisna od samih pozicija:
\begin{align}
\sigma_{csN}^2=\frac{N_A(N-N_A)}%
{N^{2}(N-1)}%
{\sum \limits_{l=1}^{K}}%
A_{l}^{2}.\label{var_s_n}
\end{align}
Na pozicijama $k=N-k_l$ njena vrijednost je precijenjena. Međutim, dalje izlaganje će pokazati da se to neće negativno odraziti na analizu performansi rekonstrukcije.
Varijansa (\ref{var_s_n}) zavisi od ukupne snage signala, koja se na osnovu dostupnih odbiraka može estimirati u obliku:
\begin{align}
\sum_{l=1}^{K} A_{l}^{2}=E_{s} \cong\frac{N}{N_A} \sum_{n \in \mathbb{N}_A}%
s^{2}(n).
\end{align}

Ukoliko bi postaivili prag na vrijednosti $4\sigma_{csN}$, tada (prema poznatom $4\sigma$ empirijskom pravilu) znamo da će manje od 1 u 15000 koeficijenata (koji ne odgovaraju pozicijama signala) biti iznad ovog praga. Koeficijenti koji su iznad praga se mogu smatrati koeficijentima koji odgovaraju komponentama signala, i oni mogu biti rekonstruisani nakon što su njihove pozicije detektovane. U Algoritmu \ref{dctrec1} i Algoritmu \ref{dctrec2} predstavljene su procedure za rekonstrukciju signala, koje su zasnovane na predstavljenoj analizi.

Ukoliko bi ovakav pristup detekciji pogrešno uključio koeficijent šuma (koji ne odgovara nijednoj komponenti signala), vrijednost tog koeficijenta bi tokom rekonstrukcije bila postavljena na nulu. U slučaju da postoje koeficijenti komponenti sa malim vrijednostima, odnosno vrijednostima koje su ispod nivoa šuma, procedura za rekonstrukciju se može ponoviti nakon što su najjače komponente detektovane, rekonstruisane i njihov doprinos oduzet od vrijednosti dostupnih odbiraka.

\begin{figure}[ptb]%
	\centering
	\includegraphics[
	]%
	{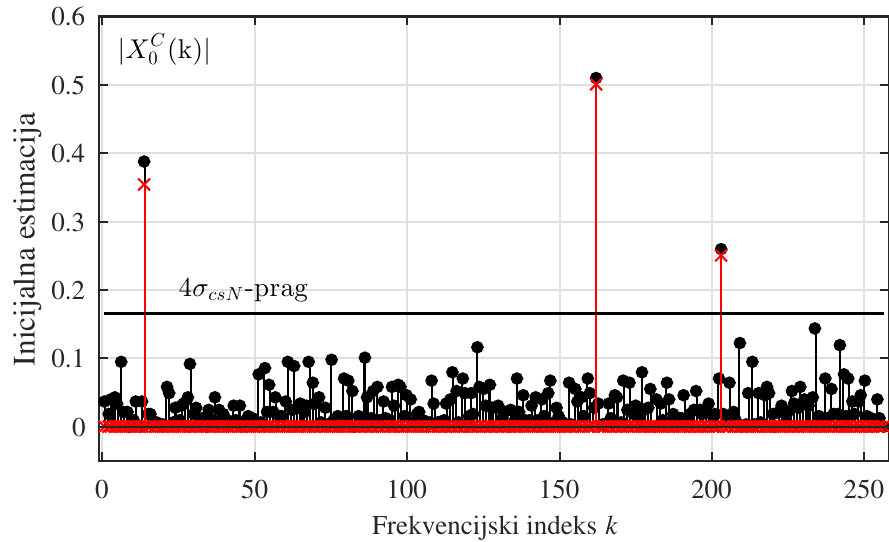}%
	\caption[Inicijalni DCT trokomponentnog rijetkog signala i odgovarajući $4\sigma$ empirijski prag] {Inicijalni DCT trokomponentnog rijetkog signala i odgovarajući $4\sigma$ empirijski prag (dat horizontalnom linijom): crveni krstići -- DCT koeficijenti signala sa svim dostupnim odbircima, crne tačke -- DCT koeficijenti signala sa pola slučajno raspoređenih nedostajućih odbiraka.} %
	\label{four-sigma}%
\end{figure}

\begin{primjer} Posmatra se trokomponentni signal dužine $N=256$. Pozicije komponenti i odgovarajuće amplitude su $k_1=14$, $k_2=162$, $k_3=203$ i $A_{1}=1$, $A_{2}=1/\sqrt{2}$, $A_{3}=1/2$, respektivno. Dostupno je samo $N_A=128$ slučajno pozicioniranih odbiraka signala. U cilju rekonstrukcije signala, računamo inicijalnu estimaciju $X^C_{0}(k)$. Zatim definišemo prag na osnovu varijanse koeficijenata izazvane nedostajućim odbircima, čija je vrijednost $\sigma_{csN}^{2}=\frac{128(256-128)}{256^{2}(256-1)} (1+1/2+1/4)=0.0017.$
Na slici \ref{four-sigma} je prikazana inicijalna estimacija, i prag na $4\times\sqrt
{0.0017}=0.1657$. Može se uočiti da je srednja vrijednost najslabije komponente
$1/4$ daleko iznad nivoa praga. 
\end{primjer}
Sprovedena analiza se može generalizovati i na $K$-komponentne signale. U najgorem slučaju, detektabilna je barem najjača komponenta (detekcija ostalih komponenti bi se sprovela nakon rekonstrukcije i oduzimanja doprinosa najjače komponente iz dostupnih odbiraka signala). Najgori mogući slučaj za detekciju najjače komponente je slučaj kada su sve ostale komponente iste amplitude (i bez gubljenja opštosti, jednake jedinici), odnosno, ${A_{1}=A_{2}=\dots= A_{K}=1}$. U najgorem slučaju $K$-rijetkog signala sa jednakim komponentama, odgovarajući $4\sigma$ prag bi bio
\begin{equation}
	T=4\sqrt{\frac{N_A(N-N_A)}{N^{2}(N-1)}K}.
\end{equation}

Možemo zaključiti da je srednja vrijednost koeficijenata koji odgovaraju komponentama signala iznad praga (sa vjerovatnoćom koju diktira posmatrano empirijsko pravilo) ako je ispunjeno:
\begin{equation}
\frac{N_A}{N}>4\sqrt{\frac{N_A(N-N_A)}{N^{2}(N-1)}K}.
\end{equation}

Navedeni uslov diktira gornju granicu za stepen rijetkosti $K$:
\begin{equation}
K<\frac{N_A(N-1)}{16(N-N_A)}.\label{MKM}
\end{equation}
Tako za $N=256$ i $N_A=128$ dobijamo $K<15.94$ ($K \le15$). Ako bi umjesto $4\sigma$ posmatrali $3\sigma$ empirijsko pravilo, dobili bi uslov $K \le28$. Potrebno je istaći da u slučaju kada je nekoliko komponenti šuma pogrešno detektovano (kao da su komponente signala), to neće uticati na uspješnost rekonstrukcije, dok god su odgovarajući uslovi zadovoljeni. Rekonstrukcioni algoritam će ove pogrešno detektovane koponente jednostavno postaviti na nule.
\begin{algorithm}[!htb]
	\floatname{algorithm}{Algoritam}
	\caption{Rekonstrukcija postavljanjem praga u DCT domenu (jednoiterativni postupak)}
	\label{dctrec1}
	\begin{algorithmic}[1]
		\Input
		\Statex
		\begin{itemize}
			\item Vektor mjerenja $\mathbf{y}$
			\item Mjerna matrica $\mathbf{A}$
			\item Inverzna DCT matrica $\mathbf{\Phi}^{-1}$
		\end{itemize}
		\Statex
		\State $a\leftarrow 0.147$
		\State ${{P}_{NN}}(T)\leftarrow 0.99$ \Comment Zadata vjerovatnoća da DCT koeficijenti šuma budu ispod praga
		\State $L\leftarrow\log \left( 1-{{\left( {{P}_{NN}}(T) \right)}^{\frac{2}{N}}} \right)$ 
		\State $\mathbf{X}^{C}_{0} \gets (\mathbf  {A}^{T}\mathbf  {A})^{-1}\mathbf  {A}^{T}\mathbf{y}$ \Comment Računa se inicijalna DCT estimacija
		\State $\sigma _{csN}^{{}}\leftarrow \sqrt{\frac{(N_A(N-N_A)}{N^{2}(N-1)}
			\frac{\left\Vert \mathbf{y} \right\Vert ^{2}_{2}}{N_A}}$
		\Comment Izraz $\frac{\left\Vert \mathbf{y} \right\Vert ^{2}_{2}}{N_A}$ aproksimira $\sum_{l=1}^{K}{A_{l}^{2}}$
		\State $T\leftarrow{\sigma_{csN}}\sqrt{\frac{1}{a}\left( -\frac{4}{\pi} -aL+\sqrt{{{\left(\frac{4}{\pi} +aL\right)}^{2}}-4aL} \right)}$ \Comment Alternativno,  $T\leftarrow 4\sigma_{csN}$
		\State $\mathbf{\hat{\Pi}}_K\leftarrow \arg \left\{ \left| {\mathbf{X}^{C}_{0}} \right|>T \right\}$ \Comment Pragom se biraju koeficijenti komponenti signala
		\smallskip   
		\State ${{\mathbf{A}}_{K}}\leftarrow {{\mathbf{A}}}{{(:,\mathbf{\hat{\Pi}}_K)}}$ \Comment Matrica $\mathbf{A}_K$ sadrži kolone matrice $\mathbf{A}$  sa indeksima $\mathbf{\hat{\Pi}}_K$
		\State${\mathbf{X}^{C}_{K}}\leftarrow {{\left({\mathbf{X}^{C}_{K}} \mathbf{A}_K^{T}{{\mathbf{A}_K}} \right)}^{-1}}\mathbf{A}_{K}^{T}{{\mathbf{y}}}$ 
		\State $
		X^C_{Kz}(k)\leftarrow\left\{
		\begin{array}
		{ll}%
		X^C_K(k),&k\in \mathbf{\hat{\Pi}}_K,\\
		0,&k \notin \mathbf{\hat{\Pi}}_K%
		\end{array}
		\right.
		$
		\Statex
		\Output
		\Statex
		\begin{itemize}
			\item	Vektor rekonstruisanih koeficijenata $\mathbf{X}^{C}_{Kz}$
			\item Rekonstruisani signal $\mathbf{x}=\mathbf{\Phi}^{-1} \mathbf{X}^{C}_{Kz}$
			
		\end{itemize}
	\end{algorithmic}
\end{algorithm}

\begin{algorithm}[!htb]
	\floatname{algorithm}{Algoritam}
	\caption{Rekonstrukcija postavljanjem praga u DCT domenu (iterativni postupak)}
	\label{dctrec2}
	\begin{algorithmic}[1]
		\Input
		\Statex
		\begin{itemize}
			\item Vektor mjerenja $\mathbf{y}$
			\item Mjerna matrica $\mathbf{A}$
			\item Zahtijevana tačnost $\delta$
			\item Inverzna DCT matrica $\mathbf{\Phi}^{-1}$
		\end{itemize}
		\Statex		
		\State	$\mathbf{\hat{\Pi}}_K\leftarrow \varnothing $  \Comment Skup estimiranih pozicija komponenti; na početku je prazan
		\State	$\mathbf{e}\leftarrow {{\mathbf{y}}}$ \Comment Vektor greške na početku je jednak vektoru dostupnih odbiraka
		\State	$a\leftarrow 0.147$
		\State ${{P}_{NN}}(T)\leftarrow 0.99$ \Comment Zadata vjerovatnoća da DCT koeficijenti šuma budu ispod praga
		\State	 $L\leftarrow\log \left( 1-{{\left( {{P}_{NN}}(T) \right)}^{\frac{2}{N}}} \right)$
		\While{ $\left\| \mathbf{e} \right\|_{2}^{2}>\delta $ }
		\State	$\mathbf{X}^{C}_{0}\leftarrow \mathbf{A}^{-1}\mathbf{e}$	
		\State $\sigma _{csN}^{{}}\leftarrow \sqrt{\frac{(N_A(N-N_A)}{N^{2}(N-1)}
			\frac{\left\Vert \mathbf{y} \right\Vert ^{2}_{2}}{N_A}}$
		\Comment Izraz $\frac{\left\Vert \mathbf{y} \right\Vert ^{2}_{2}}{N_A}$ aproksimira $\sum_{l=1}^{K}{A_{l}^{2}}$
		\State $T\leftarrow{\sigma_{csN}}\sqrt{\frac{1}{a}\left( -\frac{4}{\pi} -aL+\sqrt{{{\left(\frac{4}{\pi} +aL\right)}^{2}}-4aL} \right)}$ \Comment Prag se računa na osnovu $\sigma_{csN}$

		\State 	$\mathbf{\hat{\Pi}}_K\leftarrow \mathbf{\hat{\Pi}}_K\cup \arg \left\{ \left| \mathbf{X}^{C}_{0} \right|>T \right\}$ 
		
		\State ${{\mathbf{A}}_{K}}\leftarrow {{\mathbf{A}}}{{(:,\mathbf{\hat{\Pi}}_K)}}$ \Comment Matrica $\mathbf{A}_K$ sadrži kolone matrice $\mathbf{A}$  sa indeksima $\mathbf{\hat{\Pi}}_K$
		\State${\mathbf{X}^{C}_{K}}\leftarrow {{\left( \mathbf{A}_K^{T}{{\mathbf{A}_K}} \right)}^{-1}}\mathbf{A}_{K}^{T}{{\mathbf{y}}}$ \Comment Izraz ${{\left( \mathbf{A}_K^{T}{{\mathbf{A}_K}} \right)}^{-1}}\mathbf{A}_{K}^{T}$ je pseudoinverzija matrice $\mathbf{A}_K$
		\State	${{\mathbf{y}}_K}\leftarrow {{\mathbf{A}}_{K}}{\mathbf{X}^{C}_{K}}$
		\State	$\mathbf{e}\leftarrow {{\mathbf{y}}}-{{\mathbf{y}}_K}$
		\EndWhile
		\State $
		X^C_{Kz}(k)\leftarrow\left\{
		\begin{array}
		{ll}%
		X^C_K(k),&k\in \mathbf{\hat{\Pi}}_K,\\
		0,& k \notin \mathbf{\hat{\Pi}}_K%
		\end{array}
		\right.
		$

		\Statex
		\Output
		\Statex
		\begin{itemize}
			\item	Vektor rekonstruisanih koeficijenata $\mathbf{X}^{C}_{Kz}$
			\item Rekonstruisani signal $\mathbf{x}=\mathbf{\Phi}^{-1} \mathbf{X}^{C}_{Kz}$
			
		\end{itemize}
	\end{algorithmic}
\end{algorithm}
Nakon detekcije pozicija (frekvencijskih indeksa) komponenti, sistem mjernih jednačina postaje: 
\begin{equation}
\mathbf{y}=\mathbf{A}_{K}\mathbf{X}^{C}_K,
\end{equation}
gdje $\mathbf{X}^{C}_K$ predstavlja vektor sastavljen od $K$ nepoznatih koeficijenata na detektovanim pozicijama. Mjerna matrica je sada redukovana na dimenzije $N_A\times K$, odbacivanjem kolona koje odgovaraju koeficijentima čije su vrijednosti nula. Budući da je $K<N_A$, posmatrana jednačina može biti riješena u srednjem kvadratnom smislu. Rezultat je vektor od $K$ elemenata:
\begin{equation}
\mathbf{X}^{C}_K=(\mathbf{A}_{K}^{T}\mathbf{A}_{K})^{-1}\mathbf{A}_{K}^{T}\mathbf{y}. 
\end{equation}
Ukoliko su dobijeni koeficijenti takvi da je greška $\mathbf{e}=\mathbf{y}-\mathbf{A}_{K}\mathbf{X}^{C}_K$ jednaka nuli (ili ima vrijednost koja je u nekim prihvatljivim granicama), tada je nađeno rješenje problema. Ukoliko ova greška nije mala, tada neki nenulti koeficijenti nijesu detektovani i uključeni u $\mathbf{X}^{C}_K$. Stoga računanje treba ponoviti, sa greškom $\mathbf{e}$ sada u ulozi signala. Kandidati za pozicije nenultih koeficijenata se sada detektuju na osnovu inicijalne DCT estimacije ovog novog signala, i dodaju se u prethodni skup pozicija nenultih koeficijenata $\mathbf{\hat{\Pi}}_K$. Računanje $\mathbf{X}^{C}_K$ treba ponoviti sa ovim novim, tj. ažuriranim setom pozicija, sve dok se ne dostigne zadovoljavajući nivo greške (ili greška ne padne na nulu). 

U nastavku će biti predstavljena probabilistička analiza greške u detekciji koeficijenata koji korespondiraju komponentama signala.

Neka se posmatra $K$-komponentni rijetki signal definisan modelom (\ref{modelprep}).  DCT koeficijenti koji ne odgovaraju komponentama signala predstavljaju slučajne promjenljive opisane aproksimativno Gausovom funkcijom raspodjele, $ \mathcal{N}(0,\sigma_{csN}^{2})$ sa srednjom vrijednošću nula i varijansom $\sigma_{csN}^{2}$ definisanom sa (\ref{var_s_n}). DCT koeficijent koji odgovara
$l$-toj komponenti signala takođe predstavlja slučajnu promjenljivu sa aproksimativno Gausovom distribucijom,  $
\mathcal{N}(\frac{N_A}{N}A_{l},\sigma^{2}_{{X^C_0}(k_l)}), ~l=1, 2, \dots, K,%
$ sa srednjom vrijednošću $\mu_{{X^C_0}(k_l)}=\frac{N_A}{N}A_{l}$ i varijansom $\sigma^{2}_{{X^C_0}(k_l)}$, definisanom relacijom (\ref{varsm}). Apsolutna vrijednost DCT koeficijenta na poziciji $l$-te komponente signala imaće ,,savijenu'' normalnu distribuciju (engl. \textit{folded normal distribution})
\begin{gather}
p(\xi)=\frac{1}{{{\sigma_{X^C_0}(k_l)}\sqrt {2\pi } }} \bigg[ \exp \left( { - \frac{{{{(\xi  - {\frac{N_A}{N}A_{l}})}^2}}}{{2{\sigma^2_{{X^C_0}(k_l)}}}}} \right) +\exp \left( { - \frac{{{{(\xi  + {\frac{N_A}{N}A_{l}})}^2}}}{{2{\sigma^2_{{X^C_0}(k_l)}}}}} \right)\bigg].
\label{x}%
\end{gather}

Apsolutne vrijednosti DCT koeficijenata šuma (koji ne odgovaraju komponentama signala) podliježu tzv. polunormalnoj funkciji gustine raspodjele (engl. \textit{half-normal distribution}) 
\begin{equation}
q(\xi)= \frac{{\sqrt 2 }}{{{\sigma _N}\sqrt \pi  }}\exp \left( { - \frac{{{\xi ^2}}}{{2{\sigma^2 _{csN}}}}} \right).
\end{equation}

DCT koeficijent na poziciji šuma imaće vrijednost manju od $\Xi$ sa vjerovatnoćom
\begin{equation}
Q(\Xi)\!= \!\!\int_0^\Xi \! {\frac{{\sqrt 2 }}{{{\sigma _N}\sqrt \pi  }}\exp \left( \!{ - \frac{{{\xi ^2}}}{{2{\sigma^2 _{csN}}}}} \!\right)\!d\xi  = } \operatorname{erf} \left( {\frac{\Xi }{{\sqrt 2 {\sigma _{csN}}}}}\right)\!\!.
\end{equation}

Ukupan broj DCT koeficijenata na pozicijama šuma je $N-K$. Vjerovatnoća da je $N-K$ nezavisnih DCT koeficijenata koji ne odgovaraju komponentama signala manji od $\Xi$ je $ {Q(\Xi)}  ^{N-K}$.
Vjerovatnoća da je barem jedan od $N-{K}$ DCT koeficijenata šuma veći od
$\Xi$ iznosi
$
G(\Xi)=1-Q(\Xi)  ^{N-K}. \label{x2}%
$
Greška u detekciji komponente signala nastaje kada DCT koeficijent koji ne odgovara poziciji signala svojom vrijednošću nadmaši DCT koeficijent koji je na poziciji komponente signala. U cilju izračunavanja vjerovatnoće ove greške, razmatrajmo apsolutnu vrijednost DCT koeficijenta na poziciji komponente signala na i oko vrijednosti $\xi$. DCT koeficijent na poziciji signala ima vrijednost u opsegu od $\xi$ do $\xi+d\xi$ sa vjerovatnoćom
$p(\xi)d\xi$, gdje je $p(\xi)$ definisano relacijom (\ref{x}). Vjerovatnoća da najmanje jedan od $N-{K}$ DCT koeficijenata na poziciji šuma ima vrijednost veću od $\xi$ iznosi
$G(\xi)=1-Q(\xi) ^{N-K}$. Može se zaključiti da vjerovatnoća da je apsolutna vrijednost DCT koeficijenta na poziciji signala u opsegu od $\xi$ do $\xi+d\xi$ i da je apsolutna vrijednost najmanje jednog DCT koeficijenta na poziciji šuma veća od vrijednosti DCT koeficijenta komponente signala zadata izrazom $G(\xi)p(\xi)d\xi$. Uzimajući u obzir sve moguće vrijednosti $\xi$, zaključuje se da je vjerovatnoća pogrešne detekcije $l$-te komponente signala definisana relacijom:
\begin{align}
P_{E}^{(l)}=\frac{1}{{{\sigma_{{X^C_0}(k_l)}}\sqrt {2\pi } }}&\int\limits_0^\infty  {\left( {1 - \operatorname{erf} {{\left( {\frac{\xi }{{\sqrt 2 {\sigma _{csN}}}}} \right)}^{N- K}}} \right)} \notag\\&\times\left[ {\exp \left( { - \frac{{{{(\xi  - {\frac{N_A}{N}A_{l}})}^2}}}{{2{\sigma^2_{{X^C_0}(k_l)}}}}} \right) + \exp \left( { - \frac{{{{(\xi  + {\frac{N_A}{N}A_{l}})}^2}}}{{2{\sigma^2_{{X^C_0}(k_l)}}}}} \right)} \right]d\xi. \label{errT}%
\end{align}
Navedeni izraz se može pojednostaviti korišćenjem aproksimacije predstavljene u \cite{dftmiss}. 

\subsection{Veza sa indeksom koherentnosti} U cilju analize najgoreg mogućeg scenarija, pretpostavimo maksimalni mogući uticaj DCT koeficijenata šuma na detekciju  koeficijenta najjače komponente signala. Maksimalan međusobni uticaj posmatranih $K$ komponenti postoji kada sve one imaju jednake (bez gubljenja opštosti - jedinične) amplitude.

U slučaju kada je dostupno $N_A$ odbiraka signala, srednja vrijednost komponente (DCT koeficijenta na poziciji komponente signala) je $N_A/N$. Šumna komponenta (DCT koeficijent koji ne odgovara nijednoj komponenti signala) na frekvencijskom indeksu $k$, koja potiče od komponente signala na poziciji $k_l$ je jednaka:
\begin{gather}
Q(k,k_{l})=\sum\limits_{n \in\mathbb{N}_A, k\ne k_{l}}a_{k}a_{k_{l}}%
\cos\left(  \frac{\pi(2n+1)}{2N}k_{l}\right)
\cos\left(  \frac{\pi
	(2n+1)}{2N}k\right)  .
\end{gather}
Ona se može povezati sa indeksom koherentnosti matrice $\mathbf  {A}^{T}\mathbf  {A}$, koji je definisan na sljedeći način:
\begin{gather}
\mu=\max\Big|\frac{N}{N_A}\sum\limits_{n \in\mathbb{N}_A, k\ne k_{l}}%
a_{k}a_{k_{l}}\cos\left(  \frac{\pi(2n+1)}{2N}k_{l}\right)  \cos\left(
\frac{\pi(2n+1)}{2N}k\right)\Big|.
\end{gather}

Ukoliko je šum koji potiče od svih komponenti signala takav da se sumira po fazi na poziciji $k$ koja ne odgovara pozicijama komponenti signala, tada, pretpostavljajući maksimalnu moguću vrijednost za $\mu$, maksimalna moguća vrijednost DCT koeficijenta šuma je
\begin{equation}
K\max{|Q(k,k_{l})|}=K\mu\frac{N_A}{N}.
\end{equation}

Ukoliko se istovremeno na poziciji komponente signala svi šumovi koji potiču od preostalih $K-1$ komponenti sabiraju u fazi, i to tako da su suprotnog znaka od znaka srednje vrijednosti posmatrane komponente signala, opet pretpostavljajući njihove maksimalne moguće apsolutne vrijednosti $\mu$, tada će rezultujuća najgora moguća amplituda komponente signala biti
\begin{equation}
\min\{X^C_0(k_{l})\}=\frac{N_A}{N}-(K-1)\max{|Q(k,k_{l})|}.
\end{equation}

Detekcija komponente signala će biti još uvijek moguća ukoliko je zadovoljen uslov $\min\{X^C_0(k_{l})\}>K\max{|Q(k,k_{l})|}$, odnosno
$
\frac{N_A}{N}-(K-1)\frac{N_A}{N}\mu>K\mu\frac{N_A}{N}.
$

Uslov $1-(K-1)\mu>K\mu~$ je zapravo ekvivalentan dobro poznatom uslovu za rekonstrukciju signala zasnovanom na sparku posmatrane matrice  \cite{sparkdef}:
\begin{equation}
K<\frac{1}{2}\left(1+\frac{1}{\mu}\right).\label{dct_uslov}
\end{equation}

Prethodna diskusija odlično ilustruje činjenicu da je, u literaturi inače široko prihvaćena, relacija između stepena rijetkosti $K$ i indeksa koherentnosti $\mu$ (\ref{dct_uslov}) zapravo ekstremno pesimističan uslov za rekonstrukciju nedostajućih odbiraka signala. Kada bi bili u poziciji da pravimo strategiju odabiranja na osnovu izbora pozicija, tada bi to trebalo učiniti tako da se minimizuje vrijednost indeksa koherentnosti $\mu$. Minimalna vrijednost je definisana Velčovom granicom (engl. \textit{Welch bound}) $\mu \ge \sqrt\frac{(N-N_A)}{N_A(N-1)}$, \cite{welch}. Jednakost važi u slučaju specifičnih formi transformacija i mjernih matrica koje se označavaju skraćenicom ETF (engl. \textit{equiangular tight
frames}). Važno je istaći da DCT ne zadovoljava svojstva ETF. Međutim, čak i kada bi DCT predstavljala ETF, tada za $N=256$ i $N_A=128$ prema Velčovoj granici važi uslov $\mu \ge \sqrt{(256-128)/128/255}=0.0626$. Minimalna moguća vrijednost $\mu$ garantuje rekonstrukciju za $K<8.5$. Navedeni uslov je ekstremno pesimističan za realne slučajeve. Na osnovu naše numeričke analize, za $N=256$ i $N_A=128$ smo zaključili da je moguće  rekonstruisati signale sa mnogo većim brojem komponenti $K$. Na primjer, u najgorem slučaju jednakih amplituda i sa pesimističnim $3\sigma$ empirijskim pravilom, potpuna rekonstrukcija je moguća za $K$ koje ide do vrijednosti
$28$.

\subsection{Uticaj aditivnog šuma}

Budući da DCT koeficijenti koji odgovaraju komponentama signala imaju srednju vrijednost $\mu_{{X^C_0}(k_l)}=A_{l}\frac{N_A}{N},~l=1,2\dots,K$, tokom procesa rekonstrukcije oni se pojačavaju sa faktorom $N/N_A$, kako bi se produkovale tačne vrijednosti amplituda $A_{l}$. Ukoliko u signalu postoji slabi aditivni šum varijanse $\sigma^{2}_{\varepsilon}$, tada će, u odnosu na inicijalnu DCT estimaciju šuma
\begin{gather*}
\sigma_{X^C_{0}(k)}^2=\sum\limits_{n \in\mathbb{N}_{A}}\sum\limits_{m \in
	\mathbb{N}_{A}}a_{k}^{2} E \bigg\{ \varepsilon(n) \varepsilon(m) \cos\left(
\frac{\pi(2n+1)}{2N}k\right)  \cos\left(  \frac{\pi(2m+1)}{2N}k\right) \bigg\} 
=\frac{N_A}{N} \sigma^{2}_{\varepsilon},%
\end{gather*}
ova varijansa će biti pojačana $(N/N_A)^{2}$ puta. Stoga je varijansa u jednom estimiranom koeficijentu jednaka:
\begin{equation}
\sigma_{X^{C}_0(k)}^2
=\frac{N}{N_A} \sigma^{2}_{\varepsilon}. \label{varcoefff}
\end{equation}

Energija šuma u rekonstruisanom signalu sa $K$ komponenti (odnosno energija šuma prisutna u $K$ rekonstruisanih DCT koeficijenata) će biti $K$ puta veća:
\begin{equation}
E_{\varepsilon R}=\frac{N_A}{N} \sigma^{2}_{\varepsilon}K \left(\frac{N}{N_A%
}\right)^{2}=\frac{K}{N_A} \sigma^{2}_{\varepsilon}N.
\end{equation}

Sada se može definisati odnos signal-šum u rekonstruisanom signalu kao:
\begin{align}
SNR&=10\log\left(\frac{E_{s}}{E_{\varepsilon R}}\right) =10\log\left(\frac{E_{s}}{\frac
	{K}{N_A} \sigma^{2}_{\varepsilon}N}\right)
\notag \\
&=10\log\left(\frac{E_{s}}{\frac{K}{N_A} E_{\varepsilon}}\right)=SNR_{i}-10\log\left( \frac{K}{N_A}\right),
\end{align}
pri čemu je $SNR_{i}=10\log(E_{s}/E_{\varepsilon})$ ulazni odnos signal-šum u svim odbircima signala. 

Ovaj rezultat, dobijen jednostavnim i intuitivno jasnim izvođenjem, može se uporediti sa rezultatom dobijenim u kontekstu tzv. Bajesovog kompresivnog odabiranja \cite{BCS2}. Naime, matrica kovarijanse estimiranih koeficijenata je, u ovom kontekstu rekonstrukcije, definisana kao $\Sigma=(\mathbf  {A}^T\mathbf  {A}/\sigma_{\varepsilon}^2+\mathbf{D)}^{-1}$, gdje je $\mathbf{D}$ dijagonalna matrica tzv. hiperparametara. Nakon nalaženja pozicija nenultih koeficijenata, korišćenjem iterativne procedure u Bajesovom pristupu, koeficijenti sa velikim hiperparametrima su isključeni zajedno sa odgovarajućim elementima matrice $\mathbf{D}$ i kolonama $\mathbf  {A}$. Za našu mjernu matricu, varijansa estimiranih koeficijenata je jednaka dijagonalnim elementima matrice $\mathbf  {A}^T\mathbf  {A}$, pošto su hiperparametri jednaki nuli za nenulte koeficijente. Srednja vrijednost dijagonalnih elemenata matrice $\mathbf  {A}^T\mathbf  {A}$ je $N_A/N$. Ona se ne mijenja izostavljanjem kolona u mjernoj matrici. Stoga su dijagonalni elementi matrice kovarijanse, u finalnoj iteraciji rekonstrukcije zasnovane na Bajesovom metodu, jednaki $\sigma_{\varepsilon}^2 N/N_A$, vodeći do (\ref{varcoefff}).

Prezentovani rezultati će  za uticaj aditivnog šuma biti korišćeni i numerički provjereni u kontekstu analize rekonstrukcije signala koji nijesu čisto rijetki.

\subsection{Analiza rekonstrukcije signala koji nijesu rijetki}
Signali koji nijesu čisto rijetki, a pri tome posjeduju dobru koncentraciju u određenom transformacionom domenu, često se susrijeću u praktičnim primjenama. U slučaju DCT domena, ovaj fenomen odlično ilustruju audio signali. Zbog njihove dobre koncentracije, ovakve signale vrlo često ima smisla rekonstruisati algoritmima iz konteksta kompesivnog odabiranja, uz pretpostavku da su $K$-rijetki, odnosno, uz pretpostavku da se preostalih $N-K$ odbiraka signala mogu zanemariti. Greška koja se pravi takvom pretpostavkom je neizbježna, a u realnim scenarijima - njen nivo je često više nego prihvatljiv u kontekstu primjene. Na primjer, ukoliko se radi rekonstrukcija audio signala nakon odbacivanja odbiraka teško oštećenih impulsnim šumom, distorzijama, ili jednostavno rekonstrukcija odbiraka koji nedostaju zbog prirode procesa akvizicije signala, gubitaka tokom prenosa, oštećenja medijuma za skladištenje i slično, greška u rekonstrukciji će biti zanemarljivo mala. 

U sekciji \ref{audioapp} će biti pokazano da objektivne perceptualne mjere kvaliteta u eksperimentima sa većim brojem realnih signala pokazuju potpunu perceptualnu prihvatljivost rekonstruisanih signala. Drugim riječima, naši numerički testovi, ali i testovi slušanja, pokazuju da ljudsko uho ne može detektovati greške i napraviti razliku između originalnih i rekonstruisanih signala. U ovom odjeljku, navedena greška će biti izvedena u vidu eksplicitnog izraza. Pored njene analitičke forme, numeričkim eksperimentima sa generisanim i realnim signalima biće potvrđena validnost ovog bitnog teorijskog rezultata. Izvedena greška kvantitativno određuje performanse rekonstrukcionog procesa, i za posmatrane klase signala može predvidjeti očekivane ishode rekonstrucije.
\subsubsection{Teorema o grešci u rekonstrukciji signala koji nijesu rijetki}
 Posmatra se signal koji nije čisto rijedak, i čije su najveće amplitude $A_{l}$,
$l=1,2,\dots,K$. Pretpostavimo da je od ukupnog broja $N$ dostupno samo $N_A$ odbiraka posmatranog signala, uz
 $1\ll N_A\ll N$. Pretpostavimo i da je rekonstrukcija signala sprovedena kao da je on bio $K$-rijedak. Energija greške u $K$ rekonstruisanih koeficijenata $\left\Vert \mathbf{X}^{C}_{K}%
-\mathbf{X}^C_{T}\right\Vert _{2}^{2}$ je direktno povezana sa energijom nerekonstruisanih komponenti $\left\Vert \mathbf{X}^{C}_{T_z}-\mathbf{X}^C\right\Vert
_{2}^{2}$ sljedećom relacijom
\begin{equation}
\left\Vert \mathbf{X}^{C}_{K}-\mathbf{X}^C_{T}\right\Vert _{2}^{2} =\frac
{K(N-N_A)}{N_A(N-1)}\left\Vert \mathbf{X}^{C}_{T_z}-\mathbf{X}^C\right\Vert
_{2}^{2},
\end{equation}
gdje je $\mathbf{X}^{C}_{K}$ vektor dimenzija $K \times 1$ koji sadrži rekonstruisane komponente, $\mathbf{X}^{C}_{T}$ je vektor dimenzija $K \times 1$ koji zadrži prave vrijednosti koeficijenata na pozicijama rekonstruisanih koeficijenata, $\mathbf{X}^{C}$ je vektor dimenzija $N \times 1$ i on sadrži DCT koeficijente originalnog signala (sa svim dostupnim odbircima), dok je
	$\mathbf{X}^{C}_{T_z}$ vektor dimenzija $N \times 1$ koji sadrži $K$ originalnih koeficijenata na rekonstruisanim (detektovanim) pozicijama, i nule na preostalih $N-K$ pozicija.

\subsubsection{Dokaz teoreme} Nerekonstruisana $l$-ta komponenta u signalu se manifestuje kao ulazni Gausov šum varijanse
\begin{equation} 
\sigma_{csN}^{2}=A_{l}^{2}\frac{N_A(N-N_A)}{N^{2}(N-1)}. \label{var_noise-f}%
\end{equation}
Sve nerekonstruisane komponente će predstavljati Gausov šum ukupne varijanse
\begin{equation}
\sigma_{T}^{2}=\sum_{l=K+1}^{N}A_{l}^{2}\frac{N_A(N-N_A)}{N^{2}(N-1)}.
\label{var_noise-f1}%
\end{equation}

Nakon rekonstrukcije, ukupna energija šuma koji potiče od nerekonstruisanih komponenti (koji je prisutan u $K$ rekonstruisanih komponenti) biće jednaka:
\begin{equation}
\left\Vert \mathbf{X}^{C}_{K}-\mathbf{X}^{C}_{T}\right\Vert _{2}^{2} =E_{\varepsilon
	R}=K\frac{N^2}{N_A^2} \sigma^{2}_{T} =\frac{K(N-N_A)}{N_A(N-1)} \sum
_{l=K+1}^{N}A_{l}^{2}\label{vvvar_noise}.
\end{equation}

Šum koji potiče od nerekonstruisanih komponenti se može direktno povezati sa energijom nerekonstruisanih komponenti
\begin{equation}
\left\Vert \mathbf{X}^{C}_{T_z}-\mathbf{X}^{C}\right\Vert _{2}^{2}=\sum_{l=K+1}^{N}
A_{l} ^{2}.%
\end{equation}

Ovo znači da je ukupna energija nerekonstruisanih komponenti jednaka
\begin{equation}
\left\Vert \mathbf{X}^{C}_{K}-\mathbf{X}^C_{T}\right\Vert _{2}^{2} =\frac
{K(N-N_A)}{N_A(N-1)}\left\Vert \mathbf{X}^{C}_{T_z}-\mathbf{X}^C\right\Vert,
\end{equation}
čime smo završili dokaz.

Rezultat prethodne teoreme može se lako generalizovati na slučaj zašumljenih signala. Ukoliko posmatrani signal sadrži ulazni aditivni šum varijanse $\sigma_{\varepsilon}^2$, čije su vrijednosti ispod nivoa rekonstruisanih komponenti u transformacionom domenu, tada je posmatrana greška definisana relacijom:
\begin{equation}
\left\Vert \mathbf{X}^{C}_{K}-\mathbf{X}^{C}_{T}\right\Vert _{2}^{2} =\frac
{K(N-N_A)}{N_A(N-1)}\left\Vert \mathbf{X}^{C}_{T_z}-\mathbf{X}^C\right\Vert
_{2}^{2}+\frac{K}{N_A} \sigma^{2}_{\varepsilon}N.
\label{nonsparse}
\end{equation}

\begin{primjer}
	Posmatrajmo signal koji nije potpuno rijedak:
\begin{equation}
	x(n)=\sum\limits_{l=1}^{N}a_{k_l}A_{l}\cos\left(  \frac{\pi(2n+1)}{2N}%
k_{l}\right) +\varepsilon(n) ,
\end{equation}
pri čemu je $A_l=1$ za $l\leq S$ i $A_l=0.5e^{-2l/(S+1)}$ ta $S+1 \leq l \leq N$, gdje su odgovarajući DCT indeksi $0\leq k_l<N$ slučajno raspoređeni. Neka je samo $N_A=192$ od ukupno $N=256$ odbiraka signala dostupno na slučajnim pozicijama. Odgovarajuće normalizacione DCT konstante su označene sa $a_{k_l}$. Ovaj signal je aproksimativno $S$-rijedak, pri čemu je $S=10$. Signal je kontaminiran bijelim Gausovim šumom, srednje vrijednosti nula i standardne devijacije $\sigma_{\varepsilon}=0.11/N$. Signal je rekonstruisan korišćenjem  OMP algoritma (Algoritam \ref{Norm0Alg}), sa različitim pretpostavljenim vrijednostima $K$ koji definišu koliko je signal rijedak, $3 \leq K \leq 32$. Na bazi 200 realizacija signala sa slučajnim DCT indeksima, slučajnim pozicijama dostupnih odbiraka i slučajnim realizacijama šuma, numerički je izračunata srednja kvadratna greška (MSE) i upoređena je sa teorijskim izrazom. Greška (\ref{nonsparse}) je izračunata sa odgovarajućom normalizacijom  pretpostavljenom vrijednošću $K$. Greške su računate po formulama:
\begin{equation}
	E_{num}=10\operatorname{log}\left(
\frac{1}{K}\Vert \mathbf{X}^{C}_{K}-\mathbf{X}^{C}_{T}\Vert _{2}^{2}
\right)
\end{equation}
za slučaj numeričke greške, i
	\begin{equation}
	\label{greska_dct}
	E_{teor}=10\operatorname{log}\left(
\frac
{N-N_A}{N_A(N-1)}\left\Vert \mathbf{X}^{C}_{T_z}-\mathbf{X}^{C}\right\Vert
_{2}^{2}+\frac{1}{N_A} \sigma^{2}_{\varepsilon}N
\right) .
	\end{equation}
za slučaj teorijske greške.	
	Rezultati su predstavljeni na slici \ref{dctnonsparse}.
	Crvena linija predstavlja teorijski MSE, plave tačke označavaju dobijene numeričke podatke, čije usrednjavanje produkuje vrijednosti označene crnim kružićima. Numerički i teorijski rezultati pokazuju vrlo visok nivo poklapanja.
	
	\begin{figure}[ptb]%
		\centering
		\includegraphics[
		]%
		{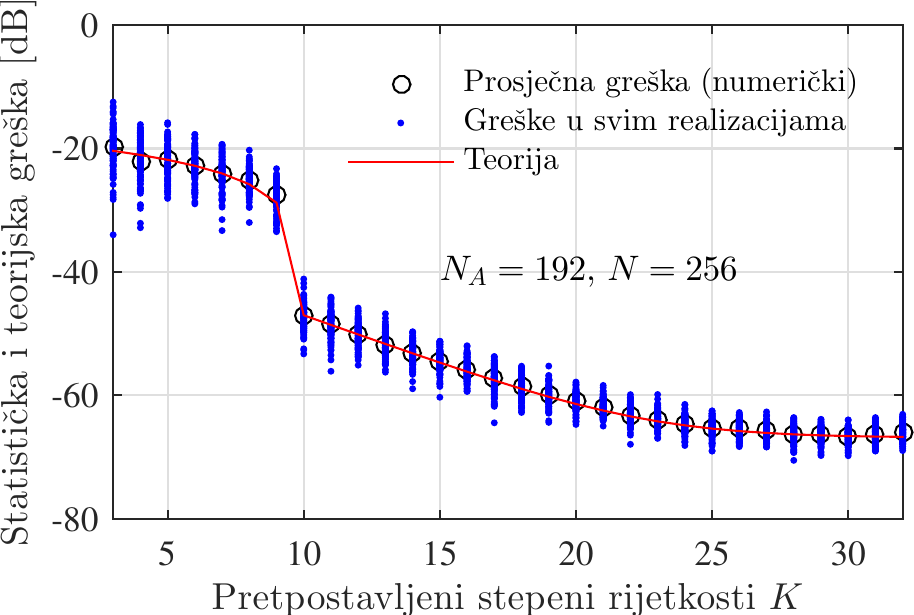}%
		\caption[Energija greške u rekonstrukciji zašumljenog signala koji nije rijedak u DCT domenu]{Energija greške u rekonstrukciji zašumljenog signala koji nije rijedak u DCT domenu: računata numerički, i na osnovu prezentovane teorije. Greška je prikazana za različite pretpostavljene stepene rijetkosti $K$.}%
		\label{dctnonsparse}%
	\end{figure}
\end{primjer} 

\subsection{Primjena u obradi audio signala}\label{audioapp}

U kontekstu primjena, DCT je zastupljena u obradi različitih vrsta signala: radarskih \cite{app_rad}, biomedicinskih \cite{app_ecg}, zatim audio signala i digitalnih slika \cite{dos,app_audio1,app_audioCS,app_audioCS2}. Važno je istaći da se ova transformacija nalazi u jezgru čitave klase  algoritama za kompresiju EKG signala \cite{ht1} i digitalnih slika \cite{dos}, ali i algoritama za različite vrste obrade audio signala (kompresiju, poboljšanje govora (engl. \textit{speech enhancement}) \cite{aud1}, uklanjanje šuma, zatim \textit{inpainting} - popunjavanje nedostajućih ili nedostupnih segmenata signala \cite{audio_cs3,app_audioCS2,za_rec,inpainting}). 
Kompresivno odabiranje u primijenjeno u obradi audio signala je relativno nova tema \cite{aud1,aud2,audio_cs3}.

Navedene činjenice ukazuju na potencijal DCT domena za predstavljanje audio signala pomoću malog broja transformacionih koeficijenata (odnosno, na  potencijal DCT domena za njihovu visokokoncentrisanu reprezentaciju). Imajući u vidu fundamentalne teorijske zahtjeve kompresivnog odabiranja, ovo ukazuje na mogućnost primjene razvijenih rekonstrukcionih algoritama na audio signale sa nedostajućim ili nedostupnim odbircima, što potrvrđuju i  noviji naučni radovi u ovoj oblasti, kao na primjer \cite{inpainting,aud1}. 

  Audio signali su često podložni oštećenjima, pod uticajem impulsnih šumova i drugih jakih smetnji \cite{inpainting,book_audio ,ar,godsill_r5,godsill_r8,godsill_r10,godsill_r12,godsill_r13,gosill_r100,godsill_r1, janssen, savremeni_por, savremeni_konf_por, denoising_arrec, denoising_viterbi, bayesnoise}. Ona nastaju i usljed odsijecanja \cite{inpainting}, gubitka paketa tokom prenosa, \cite{godsill_r4,ofir1,ofir2,packetloss,packetloss2,packetloss3,packetloss4,packetloss5,packetloss6}, fizičkih oštećenja medijuma za skladištenje podataka \cite{book_audio,inpainting}. Mogućnosti korigovanja nabrojanih oštećenja predmet su mnogih aktuelnih istraživanja, \cite{book_audio,godsill_r1,savremeni_por,savremeni_konf_por,denoising_arrec,denoising_viterbi,bayesnoise,ref3_1,audio_cs3,r3_2,r3_3,clicks1,godsill_r3}.

Audio signali su nestacionarni, odnosno, imaju vremenski promjenljiv spektralni sadržaj. Generalno govoreći, iako audio signali nijesu rijetki u DCT domenu \cite{app_audioCS}, njihov stepen rijetkosti se može značajno poboljšati ukoliko se posmatraju  odgovarajući lokalizovani segmenti \cite{aud1, app_audioCS,app_audioCS2}. Ova vrsta signala se u tom slučaju može smatrati aproksimativno rijetkom u DCT domenu. U cilju poboljšanja stepena rijetkosti, odnosno koncentracije reprezentacije, u praksi se koriste prozorske forme (engl. \textit{windowed forms}) DCT, kao na primjer tzv. MDCT \cite{mdct}. Ova forma transformacije je široko zastupljena u procedurama za kompresiju inkorporiranim u modernim audio formatima \cite{app_audio1}. Audio signali $x(n)$ dužeg trajanja se analiziraju (odnosno obrađuju) pomoću DCT primjenjenog na uzastopne blokove uprozorenih segmenata
\begin{equation}
x_i(n)=w(n)x(n+i N/2),
\end{equation}
gdje je $w(n)$ prozorska funkcija unutar $0\le n \le N-1$. 
Susjedni blokovi su preklopljeni tako da je druga polovina jednog bloka preklopljena sa prvom polovinom sljedećeg bloka. Važno je napomenuti da je ovakav blokovski pristup analizi i obradi audio signala od velike važnosti i za analizirane algoritme za rekonstrukciju rijetkih signala, budući da se segmentiranjem signala velikog trajanja smanjuju dimenzije parcijalne DCT matrice, čija se inverzija mora vršiti tokom procesa rekonstrukcije. Blokovski pristup nije stran u primjenama -- on je široko zastupljen i u analizi slika zasnovanoj na DCT. Ukoliko je forma prozorske funkcije takva da je zadovoljen uslov $w(n)+w(n+N/2)=1$ unutar intervala preklapanja $N/2\le n \le N-1$, tada se rekonstrukcija čitavog signala jednostavno sprovodi na osnovu rekonstruisanih uprozorenih segmenata $\hat x_i(n)$, u formi
\begin{equation}
\hat x(n)=\sum_i \hat x_i(n-iN/2).
\end{equation}

Veliki broj široko zastupljenih prozorskih funkcija zadovoljava spomenuti uslov \cite{dos}. Takav je i često korišćeni Hanov prozor $w(n)=0.5(1+\cos(\frac{2\pi}{N}(n+\frac{N}{2})))
=\sin^2(\frac{\pi}{N} n)$. 
%Condition that $w(i)+w(i+N/2)=1$
%within the overlapping interval, corresponds to well known Princen-Bradley condition $w^2(i)+w^2(i+N/2)=2$
%within the overlapping interval, if the same windowing function is used on both, the analysis and reconstruction side.
%Then the windowing function $w(i)=\sin(\frac{\pi}{2N}(N+\frac{1}{2}))$ would be used. 
\begin{figure}[!ht]%
	\centering
	\includegraphics[
	]%
	{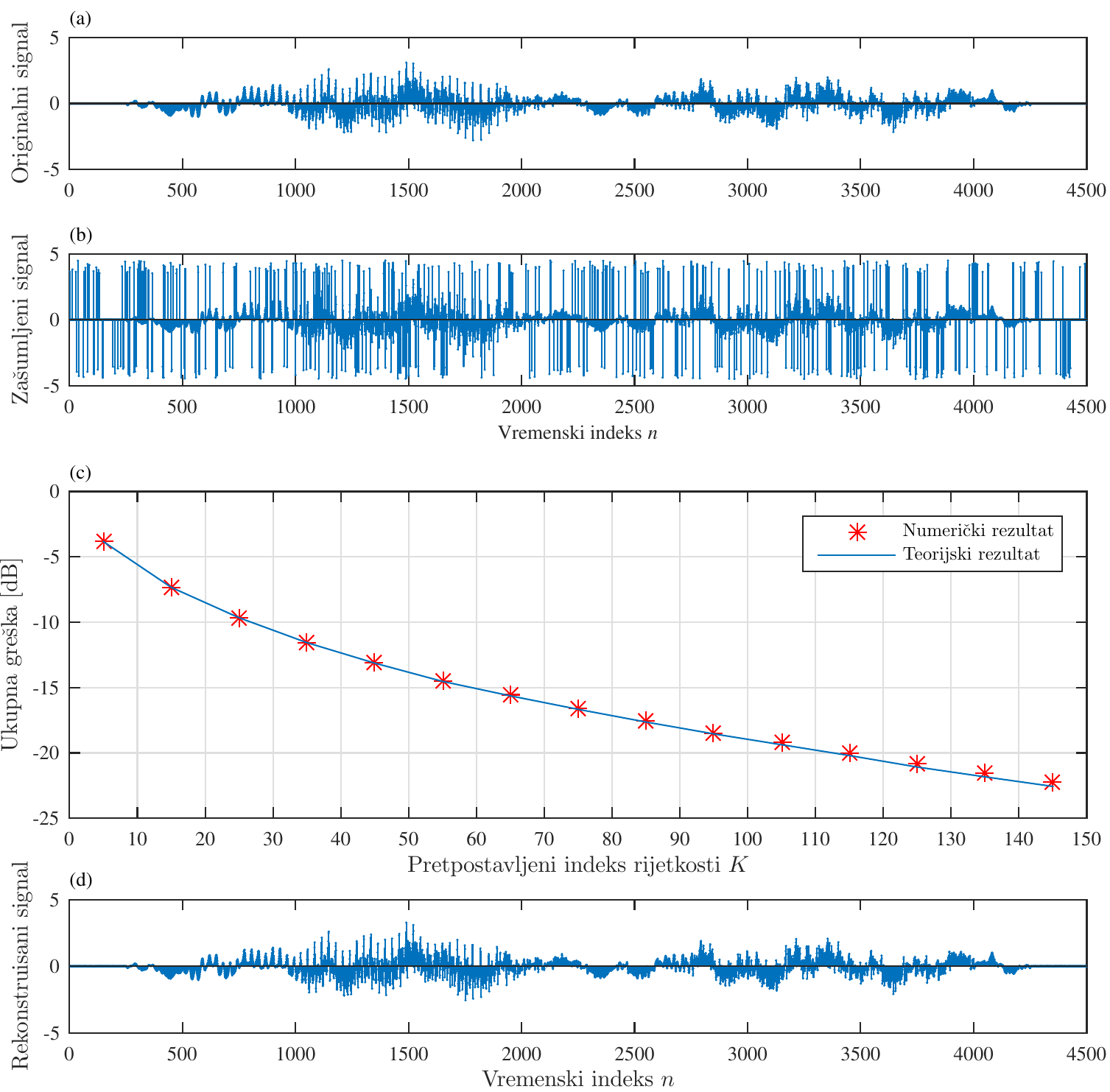}%
	\caption[Rekonstrukcija audio signala \texttt{mtlb.mat} nakon uklanjanja impulsnog šuma]{Rekonstrukcija audio signala \texttt{mtlb.mat} nakon uklanjanja impulsnog šuma:
		(a) originalni signal,
		(b) signal oštećen impulsnim smetnjama, (c) ukupna greška nakon rekonstrukcije sa različitim pretpostavljenim stepenima rijetkosti, (d) rekonstruisani signal.}%
	\label{app1}%
\end{figure}

U daljem razmatranju, biće pretpostavljeno da je u svakom bloku dostupan redukovani skup nezašumljenih obiraka signala, na pozicijama $n\in \mathbb{N}_A\mathbf{=}\left\{  n_{1},n_{2},\dots,n_{N_A}\right\}$. Različite okolnosti mogu biti uzrok nedostupnosti odbiraka audio signala.  Ilustrativan primjer su smetnje u obliku klikova (engl. \textit{clicks and pops}) koje su prisutne u starim audio snimcima, karakteristične po tome da teško oštećuju određeni procenat odbiraka signala na pozicijama $n \in \mathbb{N}_Q$ \cite{book_audio,app_audioCS,app_audioCS2,inpainting,bayesnoise}. U našem razmatranju, skup $\mathbb{N}_Q$ ćemo smatrati poznatim, budući da se za detekciju impulsnih smetnji može koristiti neki od široko prihvaćenih i efikasnih algoritama za ovu namjenu. Nakon uklanjanja impulsnih smetnji, odbirci na ovim, slučajno raspoređenim pozicijama $n \in \mathbb{N}_Q$ se mogu posmatrati kao nedostupni. Oni će biti rekonstruisani CS Algoritmom \ref{dctrec2}, odnosno, OMP pristupom (Algoritam \ref{Norm0Alg}). Ovakva forma primjene će biti ilustrovana primjerima \ref{aud1} i \ref{aud2}. U kontekstu kompresivnog odabiranja, redukovani skup slučajno pozicioniranih odbiraka može da bude inicijalni rezultat akvizicije. Takva vrsta signala ilustrovana je primjerom \ref{aud3}. Bez obzira na to da li su teško oštećeni odbirci na pozicijama $n \in \mathbb{N}_Q$ namjerno odbačeni, ili je akvizicija signala vršena na slučajnim pozicijama $n \in \mathbb{N}_A$, pristupima kompresivnog odabiranja oni se procesiraju na potpuno isti način.

\begin{primjer} \label{aud1}
Razmatra se testni signal iz fajla \texttt{mtlb.mat}, koji je standardno dostupan u softverskom paketu MATLAB\textsuperscript{\textregistered}. U pitanju je audio snimak niskog kvaliteta, u kojem ženski glas na engleskom jeziku izgovara riječ ,,Matlab'', pri čemu je odgovarajuća frekvencija odabiranja 7418 Hz. Vremenska forma signala je prikazana na slici \ref{app1} (a). Ukupno 15\% slučajno raspoređenih odbiraka signala je oštećeno impulsnim šumom,  slika \ref{app1} (b). Pozicije impulsa šuma mogu biti jednostavno detektovane korišćenjem limitera. U slučaju složenijeg oblika impulsnog šuma, u kojem se pojavljuju impulsi u opsegu vrijednosti odbiraka signala, mogu se koristiti napredniji metodi detekcije, na primjer, algoritam predstavljen u \cite{impcs}). Odbirci signala na pozicijama impulsa su proglašeni za nedostupne, i rekonstrukcija se sprovodi na osnovu preostalih odbiraka unutar bloka. Za svaki blok, korišćen je Hanov prozor dužine $N=500,$ sa preklapanjem na polovini dužine prozorske funkcije. Rekonstrukcija je vršena korišćenjem MP pristupa (Algoritam \ref{Norm0Alg}, Algoritam \ref{dctrec2}), sa različitim pretpostavljenim stepenima rijetkosti $K$. Za rekonstruisani signal je numerički estimirana srednja kvadratna greška, i izračunata odgovarajuća teorijska greška (\ref{greska_dct}). Na slici \ref{app1} (c) estimirana greška je predstavljena simbolom ,,*'', dok je teorijska greška (\ref{greska_dct}) predstavljena punom linijom. Može se uočiti veliko poklapanje numeričkog i teorijskog rezultata. Krajnji signal formiran na osnovu rekonstruisanih segmenata je predstavljen na slici \ref{app1} (d), za slučaj kada je pretpostavljeni stepen rijetkosti $K=150$. Kvadratni korijen srednje kvadratne greške, u literaturi poznat i kao RMSE (engl. \textit{root-mean-squared error}), računat za signale predstavljene na slikama \ref{app1} (a) i \ref{app1} (d) iznosi 0.0738.

\end{primjer}
 
\begin{figure}
	\centering
	%[scale=0.9]
	\includegraphics%
	{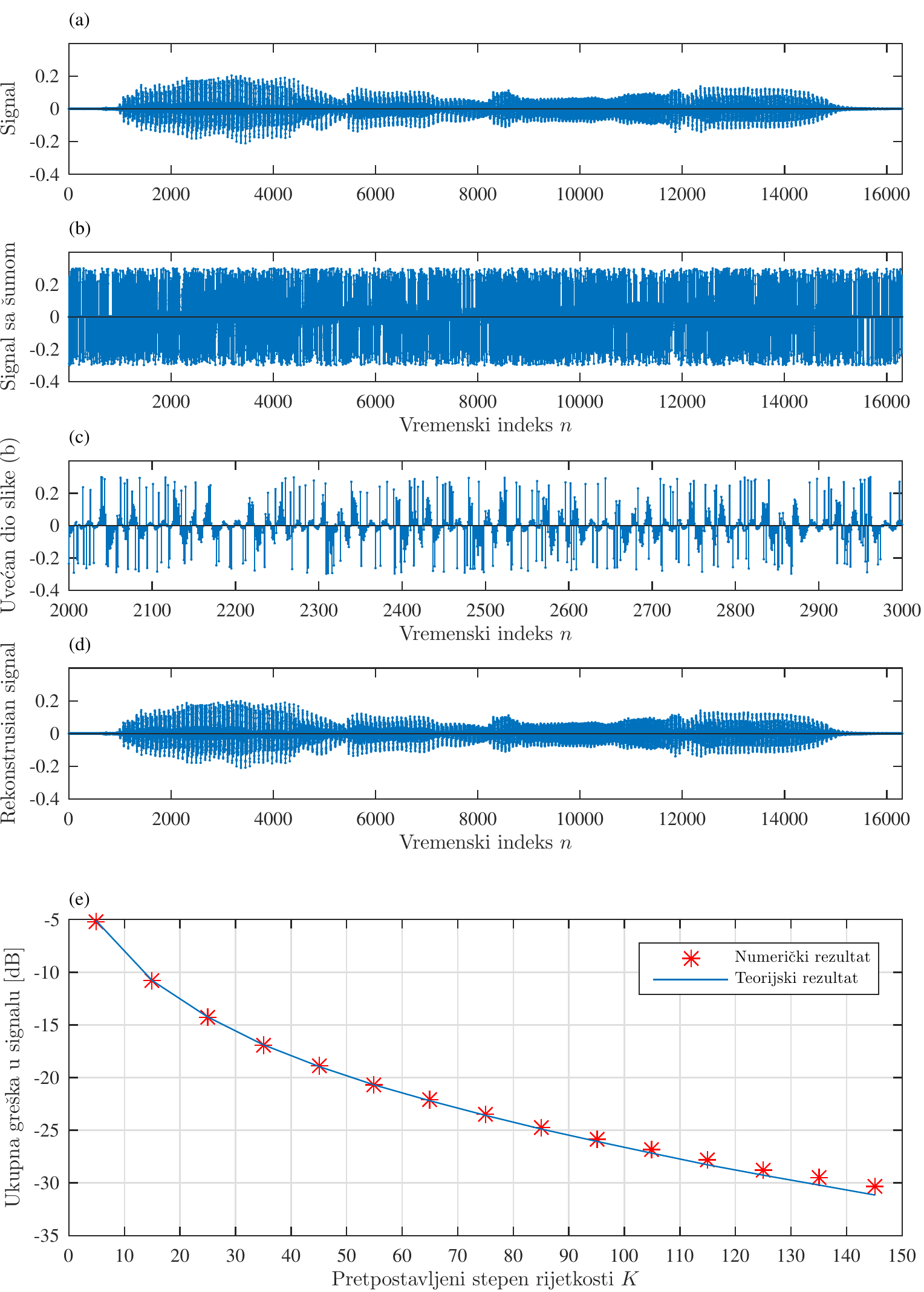}%
	\caption[Rekonstrukcija snimljenog audio signala nakon uklanjanja impulsnog šuma]{Rekonstrukcija snimljenog audio signala nakon uklanjanja impulsnog šuma: (a) originalni signal, (b) signal oštećen impulsnim smetnjama, (c) uvećano prikazani dio oštećenog signala od 1000 odbiraka,
		(d) rekonstruisani signal, (e) ukupna greška nakon rekonstrukcije sa različitim pretpostavljenim stepenima rijetkosti. }
	\label{app2}%
\end{figure}

\begin{primjer}\label{aud2}
	 U ovom primjeru se razmatra snimljeni audio signal u kojem je riječ ,,Aleluja'' izgovorena na našem jeziku. Signal je snimljen korišćenjem \textit{MacBook} računara sa softverskim paketom MATLAB\textsuperscript{\textregistered}.  Frekvencija odabiranja je 11025 Hz. U ovom slučaju je 20\% of slučajno pozicioniranih odbiraka signala oštećeno jakim impulsnim šumom. Ovi odbirci su izostavljeni iz signala, i rekonstrukcija je sprovedena samo na osnovu preostalih odbiraka. Rezultati rekonstrukcije su prezentovani na slici \ref{app2}, gdje je, pored elemenata iz prethodnog primjera, dodatno uvećano prikazan i segment polaznog signala, u cilju bolje vizuelizacije rezultata.
\end{primjer} 

\begin{figure}
	\centering
	\includegraphics%
	{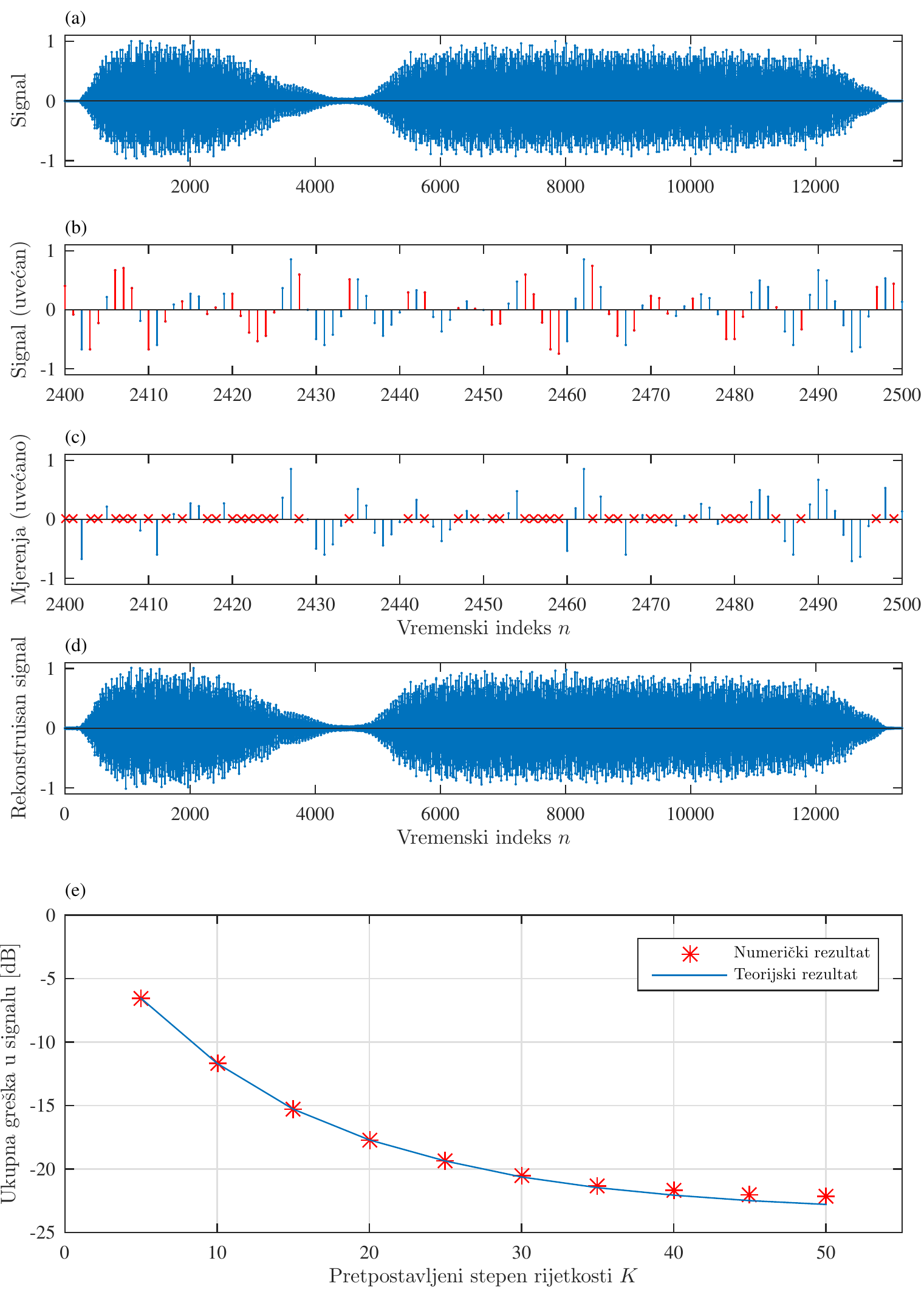}%
	\caption[Rekonstrukcija kompresivno odabranog audio signala sa 50\% dostupnih odbiraka na slučajnim pozicijama]{Rekonstrukcija kompresivno odabranog audio signala sa 50\% dostupnih odbiraka na slučajnim pozicijama:
		(a) originalni signal, (b) uvećani dio signala: crveno -- nedostajući odbirci, plavo -- dostupni odbirci, (c) uvećani dio signala sa krstićima na pozicijama nedostajućih odbiraka, (d) rekonstruisani signal,
		(e) ukupna energija nakon rekonstrukcije računata za različite stepene rijetkosti. }
	\label{app3}%
\end{figure}

\begin{primjer}\label{aud3}
	Razmatra se ugrađeni MATLAB\textsuperscript{\textregistered} signal, \texttt{train.mat}, prikazan na slici \ref{app3} (a). U pitanju je audio zapis zvižduka voza, odabran sa frekvencijom 8192 Hz. Pretpostavljeno je da je signal kompresivno odabran i da je samo 50\% odbiraka dostupno na slučajnim pozicijama, što je i ilustrovano na uvećanim (zumiranim) slikama \ref{app3} (b) i \ref{app3} (c). I u ovom slučaju je signal rekonstruisan uz različite stepene rijetkosti, dok su odgovarajuće greške u rekonstrukciji prikazane na slici \ref{app3} (e). Rekonstruisani signal za $K=50$ je prikazan na slici \ref{app3} (d).
\end{primjer}

	\subsection{Eksperimentalna evaluacija u kontekstu obrade audio signala}
	Teoriju prezentovanu u ovoj sekciji ćemo eksperimentalno evaluirati na primjeru tri seta realnih audio signala iz \cite{inpainting}. Svaki set se sastoji od 10 signala pojedinačnog trajanja 5 s. Signali su dobijeni tokom kampanje ,,2008 Signal Separation Evaluation Campaign'' i dostupni su \textit{online} \cite{campaign,database}. Bazu čine sljedeća tri skupa podataka:
	\begin{itemize}
		\item Muzika @ 16 kHz: set od 10 muzičkih signala odabranih na 16 kHz,
		\item Govor @ 16 kHz: set od 10 muških i ženskih govornih signala odabranih na 16 kHz,
		\item Govor @ 8 kHz: set od 10 govornih signala odabranih na 8 kHz, koji reprezentuju govor sa telefonskim kvalitetom. Ovi signali su dobijeni pododabiranjem signala iz drugog seta.
	\end{itemize}
	Signali su pažljivo odabrani tako da sadrže veliki broj raznovrsnih zvučnih izvora i zvučnih kombinacija: muški i ženski govor različitih govornika, glas koji pjeva, zvuk različitih muzičkih instrumenata itd. \cite{inpainting}.
	
U našoj eksperimentalnoj evaluaciji, smetnje su simulirane kroz dva scenarija. U prvom slučaju, pretpostavljeno je da su audio signali oštećeni na slučajnim pozicijama. U drugom slučaju, oštećeni odbirci su grupisani u slučajno pozicionirane blokove različitih dužina, simulirajući na taj način oštećenja kao što su klikovi (\textit{clicks}), gubici paketa tokom prenosa putem komunikacionog kanala, zatim posljedice oštećenja prenosnih medijuma itd. \cite{bayesnoise}.
	  
	  	U oba razmatrana slučaja, signali su rekonstruisani i tačnost predloženog teorijskog izraza (\ref{greska_dct}) za MSE je numerički testirana.
		
		% The perceptual quality improvement of the reconstructed speech signals is evaluated with respect to the PESQ metrics \cite{pesq}.
		
		Prezentovani CS algoritam je upoređen sa drugim tehnikama za rekonstrukciju i restauraciju audio signala (engl. \textit{audio restoration techniques}): median i niskopropusnim filtriranjem, zatim sa dvije često korišćena algoritma zasnovana na modelovanju audio signala, kao i sa algoritmom iz CS konteksta zasnovanom na optimizaciji $\ell_1$-norme. U eksperimentu su razmatrani median filtri dužine 3 i 5, i niskopropusni filtri Batervortovog (engl. \textit{Butterworth}) tipa sa dvije presječne frekvencije, postavljene u skladu sa spektralnim karakteristikama razmatranih signala. 
		
		Što se tiče tehnika za rekonstrukciju i restauraciju audio signala, razmatrane su dvije reprezentativne interpolacione procedure  zasnovane na autoregresivnom (AR) modelovanju \cite{book_audio},\cite{janssen}, \cite{lsarsin}. Prvi razmatrani metod je interpolator zasnovan na AR modelovanju u smislu najmanjih kvadrata (LSAR, engl. \textit{least-squares AR interpolator}), originalno uveden za popravljanje ozbiljnih grešaka koje nastaju u CD (engl. \textit{compact disk}) sistemima \cite{book_audio},\cite{janssen}. Korišćena je javno dostupna implementacija u sklopu \textit{online} dostupnog softvera \textit{Audio Inpainting Toolbox} \cite{inpainting}. U razmatranim eksperimentima, dužina AR modela je postavljena na 30, dok se interpolacija sprovodi u blokovima dužine 500 odbiraka, kako je predloženo u originalnom pristupu. Drugi razmatrani algoritam je iz klase AR, kombinovan sa  reprezentacijom bazne funkcije \cite{book_audio}, \cite{lsarsin}. U našoj komparativnoj analizi korišćena je implementacija sa sinusoidama u ulozi baznih funkcija (LSAR+SIN), \cite{lsarsin}, koja je nedavno testirana od strane eminentnih istraživača iz oblasti \cite{denoising_arrec,savremeni_por}. Algoritam je korišćen sa podrazumijevanim podešavanjima, a koja uključuju: AR model dužine $P=31$, broj baznih funkcija $Q=31$ i dužina blokova $1024$ odbiraka, što je u skladu sa eksperimentalnim rezultatima prezentovanim u \cite{savremeni_por} i \cite{lsarsin}. 
		
		U svim razmatranim tehnikama, performanse rekonstrukcije jako zavise od uspješnosti detekcije pozicija oštećenih odbiraka. Kako bi obezbijedili poređenje pod istim uslovima, u svim algoritmima su umjesto rezultata procedura za detekciju impulsnih smetnji prosleđivane tačne pozicije impulsa. Na ovaj način, obezbijeđeno je isključivo poređenje performansi rekonstrukcije, bez uticaja detekcije na dobijene rezultate.
		
		 Ulogu reprezentativne tehnike iz skupa CS algoritama zasnovanih na minimizaciji $\ell_1$-norme, imao je LASSO-ISTA (engl. \textit{Iterative Shrinkage Thresholding Algorithm for LASSO problem}), \cite{dos}, \cite{lasso}. Korišćena je regularizaciona konstanta $\lambda=0.01$.
		
		U drugom scenariju, pored prethodno pobrojanih standardih tehnika, u komparativnu analizu su uključena i dva nedavno publikovana pristupa visoko adaptirana za uklanjanje i rekonstrukciju klikova u audio signalima \cite{savremeni_por,savremeni_konf_por}. 		
		U ovim algoritmima, rekonstrukcija se obavlja nakon detekcije koja je sastavni i nerazdvojni dio pristupa. Za veliki broj impulsa/blokova u oba razmatrana scenarija, zadovoljavajuću detekciju nije bilo moguće postići. Implementacije algoritama su korišćene sa njihovim podrazumijevanim podešavanjima.
		
		\subsubsection{Slučajno pozicionirane impulsne smetnje}
		U prvom eksperimentu, signali iz sva tri seta signala su oštećeni jakim impulsnim šumom u $p\%$ slučajno pozicioniranih odbiraka, kao što je ilustrovano u primjeru \ref{aud1} i na slici \ref{app1}. Posmatrani signali, u blokovima od $N=500$ odbiraka i pomnoženi Hanovom prozorskom funkcijom  su dobro koncentrisani u DCT domenu. Međutim, oni su samo aproksimativno rijetki u razmatranom domenu, pa će postupak CS rekonstrukcije neminovno uvesti grešku, čija je energija u ovoj sekciji eksplicitno izvedena. 		
		Pozicije jakih impulsa se mogu jednostavno detektovati korišćenjem limitera. Naprednije tehnike detekcije, čije izučavanje prevazilazi okvire ove teze, opisani su u {\cite{impcs,book_audio,savremeni_konf_por,savremeni_por,bayesnoise}}. Bilo koja od ovih tehnika može biti korišćena u postupku detekcije koji prethodi rekonstrukciji. 
		
		U kontekstu kompresivnog odabiranja, detektovani oštećeni odbirci na pozicijama $n\in\mathbb{N}_Q$ se posmatraju kao nedostupna mjerenja. Rekonstrukcija se sprovodi na bazi preostalih odbiraka u blokovima, koji se posmatraju kao dostupna CS mjerenja na pozicijama $n \in \mathbb{N}_A$. Blokovi su preklopljeni na polovinama dužina prozorskih funkcija. Rekonstrukcija se sprovodi prezentovanim CS Algoritmom \ref{dctrec2}, sa različitim pretpostavljenim stepenima rijetkosti $K$.

\begin{figure*}[tb]%
			\centering
			\includegraphics
			{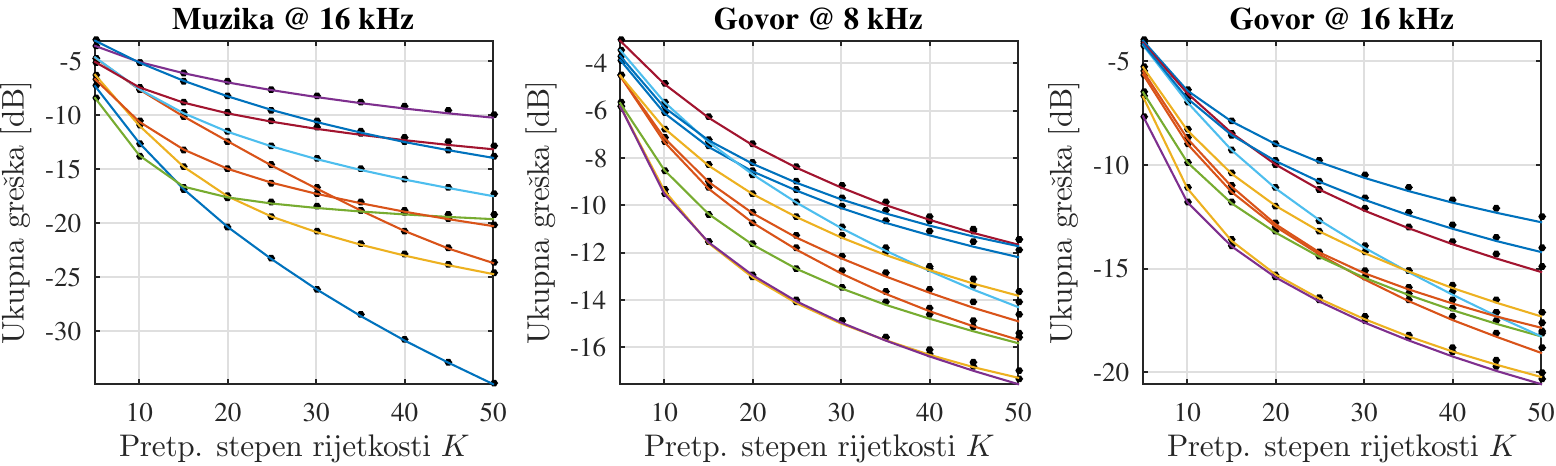}%
			\caption[Energije grešaka u rekonstrukciji zašumljenih audio signala iz setova ,,Muzika @ 16 kHz'', ,,Govor @ 8 kHz'' ,,Govor @ 16 kHz'',  u eksperimentalnom scenariju sa slučajno pozicioniranim impulsnim smetnjama]{Energije grešaka u rekonstrukciji zašumljenih audio signala koji nijesu rijetki -
				računate numerički (crne tačke) i u skladu sa prezentovanim teorijskim izrazom (pune linije), za slučajno pozicionirane impulsne smetnje. Greške su prikazane sa različitim pretpostavljenim stepenima rijetkosti $K$. Prva slika prikazuje rezultat za 10 muzičkih signala odabranih na 16 kHz, druga slika prikazuje greške za 10 govornih signala odabranih na 8 kHz, dok treća slika prikazuje odgovarajuće greške za 10 različitih govornih signala sa frekvencijom odabiranja od 16 kHz.}%
			\label{e1_error_stat}%
		\end{figure*}

		\paragraph{Teorijska greška.} Na početku, biće izvršena evaluacija predloženog izraza za MSE. Razmatra se slučaj sa $p=30\%$ oštećenih odbiraka. Za $i$-ti blok, numerička greška se računa po formuli:
		
		\begin{equation}
		E_{num}^{(i)}=10\operatorname{log}\left(
		\frac{1}{K}\Vert \mathbf{X}^{C}_{K}-\mathbf{X}^{C}_{T}\Vert _{2}^{2}
		\right),
		\label{enumer}
		\end{equation}
		dok je teorijska greška definisana izrazom:
		\begin{equation}
		E_{teor}^{(i)}\!=\!10\operatorname{log}\left(\!
		\tfrac
		{N-N_A}{N_A(N-1)}\left\Vert \mathbf{X}^{C}_{T_z}-\mathbf{X}^{C}\right\Vert
		_{2}^{2}\right) .
		\label{eteor}
		\end{equation}
		Kvadratne greške su usrednjene po blokovima i upoređene na slici \ref{e1_error_stat}, kao funkcije od pretpostavljenog stepena rijetkosti $K$, za svaki signal iz svakog razmatranog seta. Punim linijama su prezentovane teorijske MSE vrijednosti, dok tačke označavaju rezultate dobijene numeričkim putem. Može se uočiti da postoji visok stepen njihovog poklapanja, potvrđujući validnost predloženog izraza za MSE.
		
		%The reconstruction results are evaluated from the perceptual quality perspective, for two selected sparsity levels. The  PESQ \cite{pesq} is used as a quality measure. It is commonly used in the evaluation of speech quality in the CS-based speech enhancement in various noisy environments \cite{aud1}, \cite{aud2} with the  DCT of windowed audio signal frames as the sparsity domain \cite{aud1}. Results for the PESQ-based perceptual evaluation obtained for datasets Speech @ 8kHz and Speech @ 16kHz are shown in Fig. \ref{e1_preceptual}. The PESQ score is calculated for corrupted signals and for signals reconstructed with two assumed sparsity levels: $K=40$ and $K=50$. A significant PESQ score improvement, after the CS-based reconstruction, is obvious. An increase of the PESQ score with an increase of the assumed sparsity $K$ corresponds to the MSE decrease observed in Fig. \ref{e1_error_stat} (the second and the third subplot).
		\paragraph{Poređenje sa stanovišta srednje kvadratne greške (MSE).}
		\label{comparison1}
		Rezultate rekonstrukcije korišćenjem razmatranog CS algoritma sa pretpostavljenim stepenom rijetkosti $K=80$ ćemo uporediti sa rezultatima dobijenim standardnim tehnikama zasnovanim na filtiranju, algoritmima za rekonstrukciju i restauraciju audio signala zasnovanim na modelovanju, kao i sa CS algoritmom zasnovanim na minimizaciji $\ell_1$-norme \cite{grad1,grad2,lasso}. U ovom dijelu eksperimentalne analize, $p=40\%$ obiraka signala je zahvaćeno impulsnim smetnjama. 		
		Rezultati poređenja sa stanovišta srednje kvadratne greške (MSE) su predstavljeni u tabeli \ref{table_example1}. 
		
		Prezentovani CS algoritam za rekonstrukciju je prvo upoređen sa klasičnim metodom za uklanjanje impulsnih smetnji zasnovanim na median filtru. Razmatrani su filtri dužine 3 i 5. Za sva tri seta podataka, median filtar dužine 3 je smanjio MSE za $17.26$ dB u prosjeku, a slični rezultati su dobijeni i korišćenjem median filtra dužine 5 (vrste označene sa ,,Med3'' i ,,Med5'' u tabeli \ref{table_example1}).	
				Razmatrani algoritam je zatim upoređen i sa niskopropusnim filtrom Batervortovog tipa, kao reprezentativnom klasičnom tehnikom filtriranja. Za setove signala ,,Muzika @ 16 kHz'' i ,,Govor @ 16 kHz'', razmatrani su filtri sa dvije presječne učestanosti, koje su određene na osnovu analize spektralnog sadržaja ovih signala. Prvi filtar (vrsta označena sa ,,LPF1'') je dizajniran sa normalizovanom presječnom frekvencijom od $0.375$, dok drugi filtrar (vrsta ,,LPF2'') ima normalizovanu presječnu frekvenciju od $0.625$. Za treći set signala, ,,Govor  @ 8 kHz'' odgovarajuće normalizovane frekvencije su $0.5,$ za prvi i $0.7$ za drugi dizajnirani filtar, takođe dobijene empirijski. Oba filtra su pokazala slične performanse, sa MSE padom od $16.25$ dB u prosjeku (u poređenju sa MSE  zašumljenog signala). 
			
		Rezultati dobijeni LSAR i LSAR+SIN algoritmima  su  bili značajno bolji u poređenju sa rezultatima dobijenim klasičnim tehnikama zasnovanim na filtriranju. Prosječno MSE poboljšanje je iznosilo $25.42$ dB za LSAR tehniku, odnosno $26.55$ dB za tehniku LSAR+SIN. Bitno je napomenuti da je u razmatranom scenariju implementacija LSAR algoritma \cite{inpainting} imala određene ispade koji su izazvali pojavu impulsa velikih vrijednosti, za koje se smatra da su posljedica nesposobnosti algoritma da podesi parametre AR modela usljed velikog broja nedostajućih (oštećenih) odbiraka. Ova hipoteza je potvrđena činjenicom da do navedenih ispada algoritma nije dolazilo u testovima sa manjim brojem oštećenih/nedostajućih odbiraka, kada su MSE rezultati bili približno jednaki kao u slučaju LSAR+SIN algoritma. 
		
	Rezultati dobijeni predloženom CS rekonstrukcionom tehnikom su bili u prosjeku za $4$ dB bolji od LSAR+SIN rezultata i za $5$ dB bolji od rezultata LSAR pristupa. Zanimljivo je uočiti da je poboljšanje bilo najmanje u slučaju skupa pododabranih signala ,,Govor @ 8 kHz'', i ono je iznosilo $1.69$ dB. Razlog bi mogao biti redukovani stepen rijetkosti ovih signala u DCT domenu, nastao kao posljedica njihovog pododabiranja.
			CS algoritam za rekonstrukciju primjenom minimizacije $\ell_1$-norme (LASSO-ISTA) pokazao je gore rezultate od onih dobijenih predloženim CS algoritmom. Oni su u prosjeku bili za $0.5$ dB gori od odgovarajućih LSAR rezultata. 
				
		\begin{table}[!t]
			\small
			%% increase table row spacing, adjust to taste
			\renewcommand{\arraystretch}{1.3}
			% if using array.sty, it might be a good idea to tweak the value of
			% \extrarowheight as needed to properly center the text within the cells
			\caption{Srednja kvadratna greška (MSE) dobijena usrednjavanjem rezultata za sve signale iz razmatranih setova, u eksperimentalnom scenariju sa slučajno pozicioniranim impulsnim smetnjama.}
			\label{table_example1}
			\centering
			%% Some packages, such as MDW tools, offer better commands for making tables
			%% than the plain LaTeX2e tabular which is used here.
			\begin{tabular}{lccc}
				\toprule & \!\textbf{Muzika @ 16 kHz} & \! \textbf{Govor @ 8 kHz}\!\!& \textbf{Govor @ 16 kHz}\!\\
				\midrule
				Zašumljeni signali & -12.40dB & -25.85dB & -25.60dB \\
			 
				Med3 & -27.42dB &  -44.04dB &  -44.16dB \\
			 
				Med5 & -27.41dB & -43.52dB &  -44.20dB \\
				 
				LPF1 &   -26.92dB & -42.42dB & -43.24dB \\
			 
				LPF2 & -26.89dB & -42.55B & -43.37dB \\
			 
				LSAR &  -36.22dB &  -52.58dB & -51.31dB \\
			 
				LSAR+SIN & -37.48dB & -53.59dB & -52.44dB \\
				 
				LASSO & -40.69dB &  -49.80dB &  -51.15dB \\
				MP & -41.33dB & -55.28dB &  -59.02dB \\
				\bottomrule
			\end{tabular}
		\end{table}
		
\begin{figure*}[tb]%
			\centering
			\includegraphics%[scale=0.9]
			{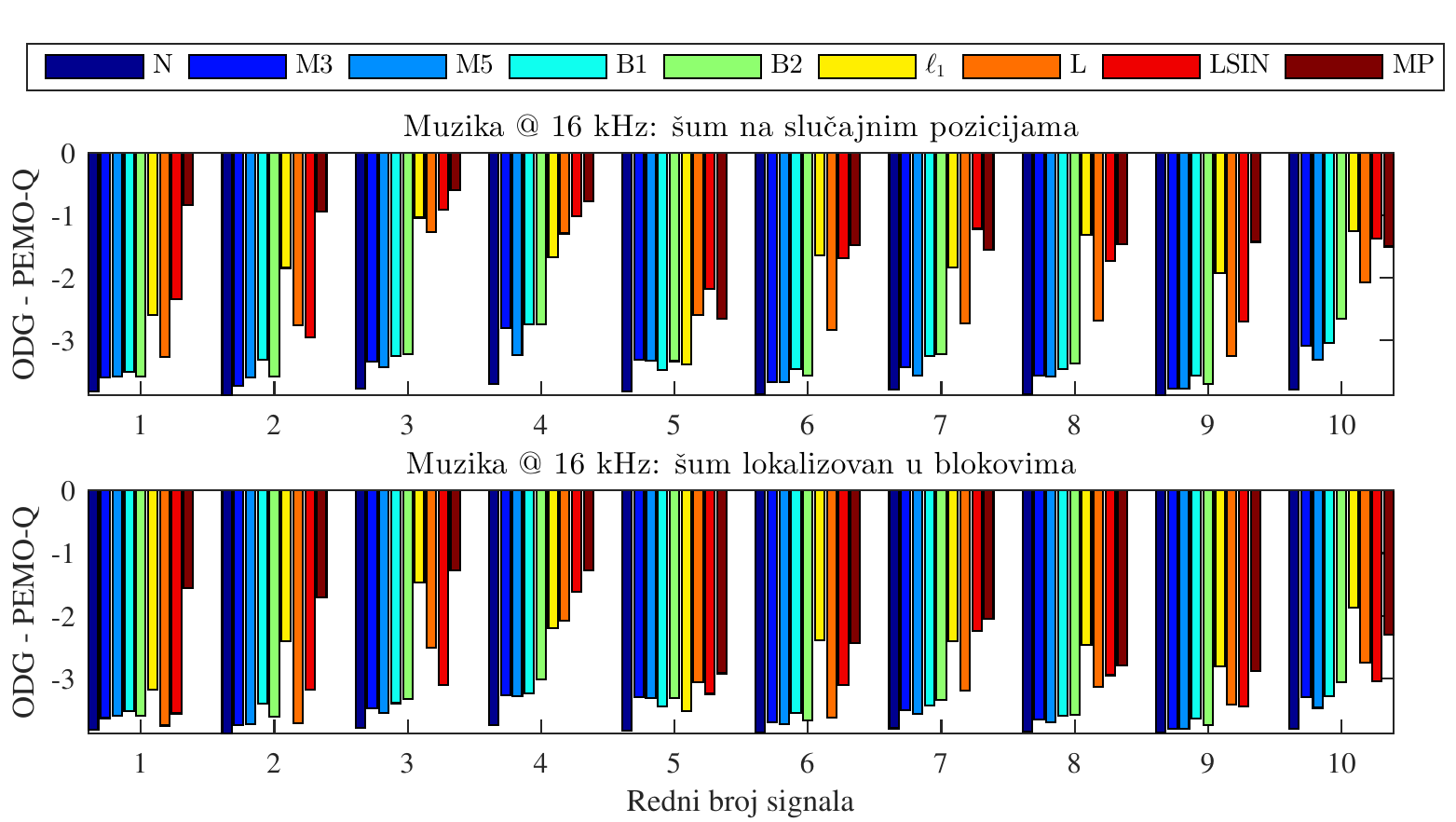}%
			\caption[Perceptualna evaluacija signala iz seta ,,Muzika @ 16 kHz'' korišćenjem PEMO-Q ODG metrike]{Perceptualna evaluacija signala iz seta ,,Muzika @ 16 kHz'' korišćenjem PEMO-Q ODG metrike. Gornja slika prikazuje PEMO-Q ODG ocjene za zašumljene i rekonstruisane signale u scenariju sa slučajno pozicioniranim impulsnim smetnjama, koji odgovaraju MSE rezultatima prezentovanim u tabeli \ref{table_example1}. Donja slika prikazuje rezultate u slučaju impulsnih smetnji lokalizovanih u vremenskim blokovima, koji odgovaraju MSE vrijednostima iz tabele \ref{table_example2}. Oznake za primijenjene tehnike: N -- zašumljeni signali, M3 i M5  --  median filtri dužine 3 i 5, B1 i B2  --  razmatrani Batervortovi niskopropusni filtri, $\ell_1$  --  LASSO-ISTA, L  --   LSAR, LSIN  --  LSAR+SIN, MP  --  predloženi MP pristup.}
			\label{pemoq}%
		\end{figure*}

		\paragraph{Poređenje sa stanovišta objektivnih mjera perceptualnog kvaliteta.}
		\label{comparison2} Audio signali dobijeni prethodno opisanim tehnikama, čije su MSE vrijednosti prezentovane u tabeli \ref{table_example1}, evaluirani su i sa stanovišta mjera perceptualnog kvaliteta. U cilju predikcije kvaliteta percepcije (od strane ljudskih slušalaca) audio signala iz seta ,,Muzika @ 16 kHz'', koriščen je široko prihvaćeni PEMO-Q method \cite{pemoq1}. U  eksperimentima je korišćena implementacija ovog algoritma koja je dostupna \textit{online}, u sklopu softverskih dodataka radova \cite{pemoq2imp,pemoq3imp}. Mapiranje izlaza algoritma, odnosno odgovarajuće mjere perceptualnog kvaliteta (PSM$_t$, engl. \textit{Perceptual Similarity Measure -- PSM}), na standardno prihvaćenu skalu poznatu kao ODG (engl. \textit{objective difference grade}) izvršeno je pomoću procedure koja je opisana u \cite{pemoq1}. Navedena skala obuhvata opseg ocjena od $-4$ (veoma iritirajuće pogoršanje) do $0$ (pogoršanje koje nije moguće percipirati). Odgovarajući PEMO-Q ODG su izračunati za oštećene signale i za sve rekonstruisane signale koji su razmatrani u prethodnoj sekciji  za set podataka ,,Muzika @ 16 kHz''. Kao i u slučaju srednje kvadratne greške, poređenje se u svim slučajevima vrši sa originalnim, nezašumljenim signalima. Rezultati poređenja, prikazani na slici \ref{pemoq} ({prvi grafik}), ukazuju na to da je prezentovana CS tehnika bila dominantna u odnosu na ostale metode, imajući prosječnu ocjenu od $-1.32$, u poređenju sa LSAR+SIN ocjenom od $-1.81$, i LASSO-ISTA ocjenom od $-1.85$. Međutim, potrebno je i istaći činjenicu da u slučaju signala br. 5 i signala br. 7, prezentovani metod je imao nešto gore rezultate od LSAR-SIN tehnike, a u slučaju signala br. 8 i signala br. 10, rezultati su malo gori od onih koje je dao LASSO-ISTA algoritam.
		
\begin{figure*}[!h]%
			\centering
			\includegraphics
			{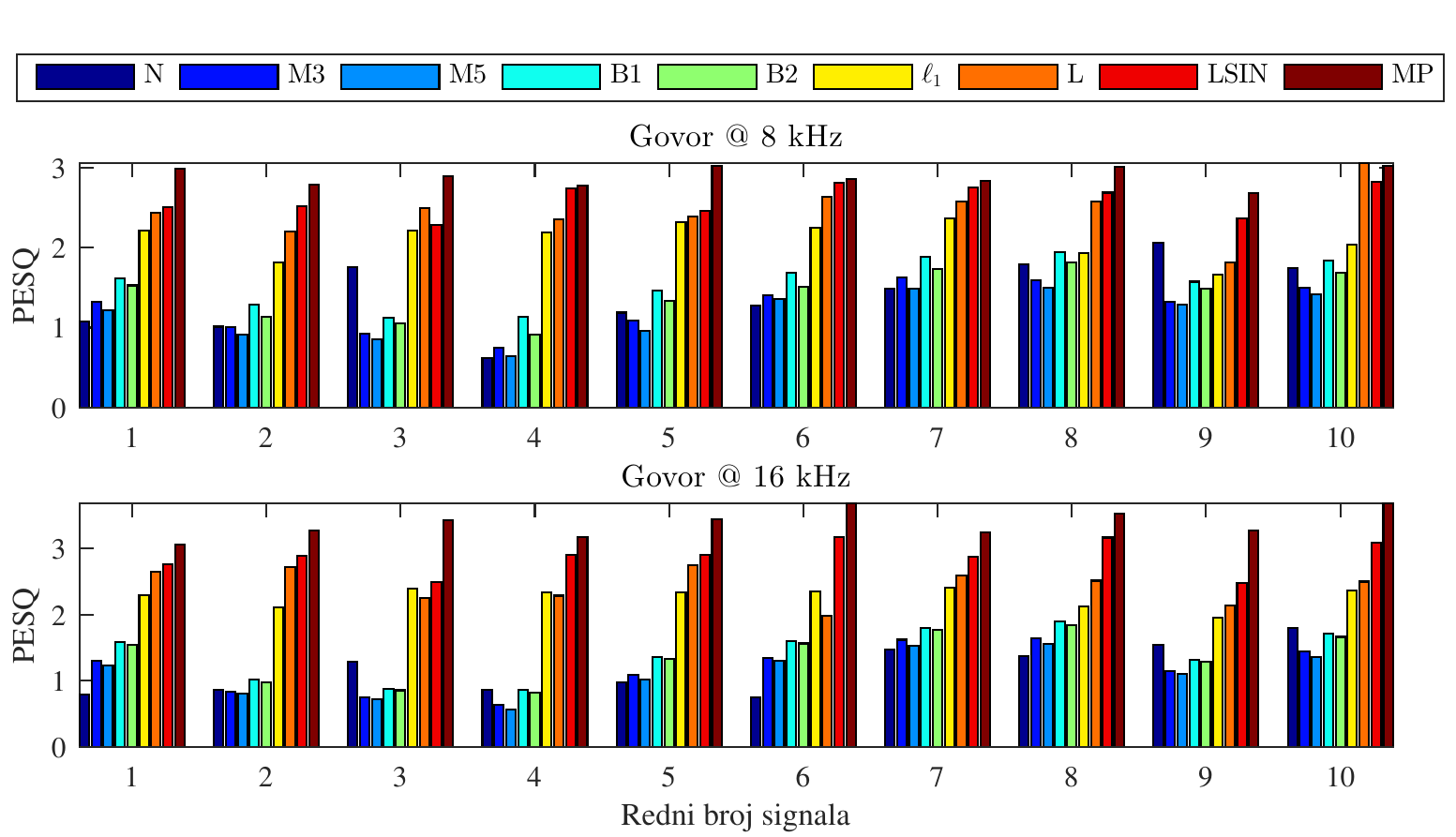}%
			\caption[Perceptualna evaluacija govornih signala korišćenjem PESQ metrike u scenariju sa slučajno pozicioniranim impulsnim smetnjama]{
				Perceptualna evaluacija govornih signala korišćenjem PESQ metrike u scenariju sa slučajno pozicioniranim impulsnim smetnjama. Rezultati su u vezi sa MSE vrijednostima iz tabele \ref{table_example1}. Oznake za primijenjene tehnike: N -- zašumljeni signali, M3 i M5 -- median filtri dužine 3 i 5, B1 i B2 -- razmatrani Batervortovi niskopropusni filtri, $\ell_1$ -- LASSO-ISTA, L --  LSAR, LSIN -- LSAR+SIN, MP -- predloženi MP pristup.}%
			\label{e1_preceptual}%
		\end{figure*}

		Za dva seta govornih signala, poređenje sa stanovišta perceptualnog kvaliteta je obavljeno korišćenjem PESQ metrike (engl. \textit{Perceptual Evaluation of Speech Quality}), \cite{pesq}. Kao jedna od standarnih mjera kvaliteta, PESQ se koristio u evaluaciji kvaliteta govornih signala nakon primjene CS algoritama u kontekstu poboljšanja govora (\textit{speech enhancement}), \cite{aud1,aud2}, gdje je ulogu domena rijetkosti takođe igrala DCT primijenjena na odgovarajuće uprozorene blokove (frejmove) audio signala \cite{aud1}. Na odgovarajućoj PESQ-MOS (engl. \textit{Mean opinion score} -- MOS) skali ocjene idu od $1$ do $5$, gdje se sa $5$ ocjenjuje najbolji rezultat. Rezultati evaluacije zasnovane na PESQ algoritmu, dobijeni za setove audio signala ,,Govor @ 8 kHz'' and ,,Govor @ 16 kHz'' prikazani su na slici \ref{e1_preceptual}. Odgovarajuće PESQ ocjene su izračunate kako za oštećene signale, tako i za signale rekonstruisane svim razmatranim metodama. 
		
		Za slučaj signala iz seta ,,Govor @ 8 kHz'', prosječna PESQ ocjena za signale rekonstruisane prezentovanim CS metodom iznosi $2.89$. U prosjeku, ona je veća od LSAR+SIN prosječne ocjene od $2.59$, LSAR ocjene od $2.45$, kao i od prosječne ocjene signala rekonstruisanih korišćenjem LASSO-ISTA algoritma, koja iznosi $2.1$. Svi ovi algoritmi doveli su do značajnog poboljšanja perceptualnog kvaliteta u poređenju sa oštećenim (zašumljenim) signalima. U slučaju seta signala ,,Govor @ 16 kHz'', poboljšanje u rekonstrukciji je još očiglednije. U ovom slučaju, prosječna ocjena prezentovanog CS algoritma je bila $3.37$. Ostali razmatrani algoritmi pokazali su sljedeće rezultate: LSAR+SIN je imao ocjenu 2.87, zatim slijedi LSAR sa ocjenom $2.44$, dok je rezultat LASSO-ISTA rekonstrukcije ocijenjen sa $2.27$. Ocjene preostalih razmatranih metoda su niže. U razmatranom slučaju, poboljšanje PESQ ocjene u odnosu na zašumljene signale je veće, nego u slučaju seta signala ,,Govor @ 8 kHz''.


		\subsubsection{Smetnje lokalizovane u blokovima u vremenskom domenu}
\begin{figure*}[tb]%
			\centering
			\includegraphics
			{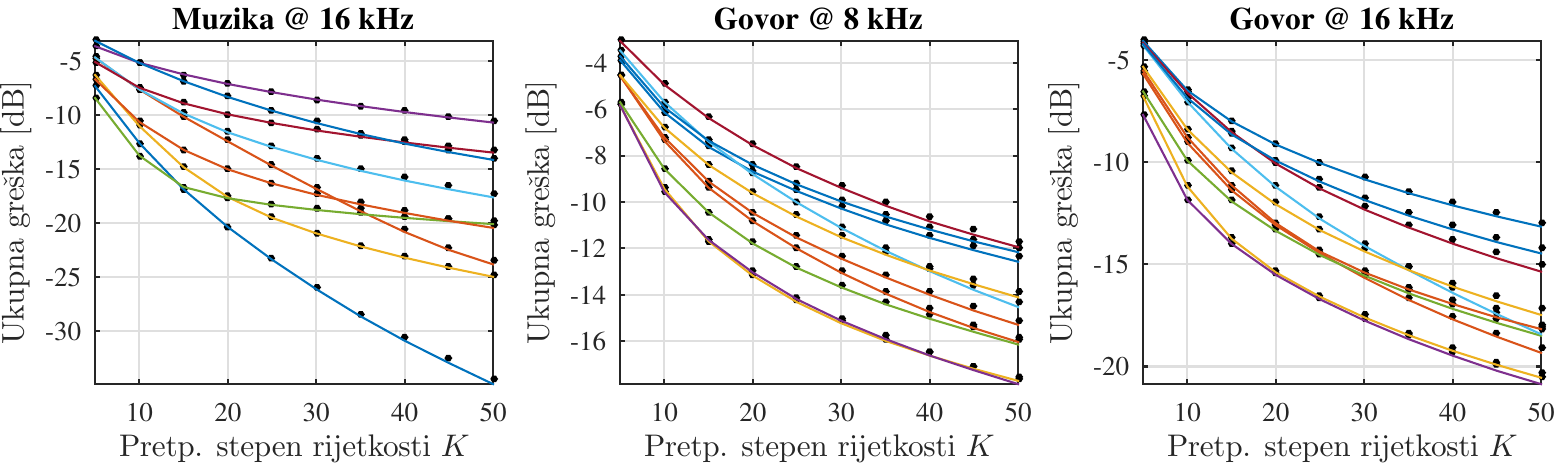}%
			\caption[Energija greške u rekonstrukciji audio signala iz setova ,,Muzika @ 16 kHz'', ,,Govor @ 8 kHz'' ,,Govor @ 16 kHz'', u slučaju kada se impulsni šum pojavljuje u vremenskim blokovima varijabilne dužine]{Energija greške u rekonstrukciji audio signala iz setova ,,Muzika @ 16 kHz', ,,Govor @ 8 kHz'' ,,Govor @ 16 kHz'': računata numerički (crne tačke) i u skladu za prezentovanom teorijom (pune linije), u slučaju kada se impulsni šum pojavljuje u vremenskim blokovima varijabilne dužine. Greške su prikazane za različite pretpostavljene stepene rijetkosti.}%
			\label{e1_error_stat2}%
		\end{figure*}

		U ovom eksperimentu, razmatraju se impulsne smetnje locirane u blokovima uzastopnih odbiraka. Svi signali iz tri razmatrana seta oštećeni su šumom lociranim u slučajno pozicioniranim blokovima, varijabilne dužine između 1 i 5, tako da je, u prosjeku, $p\%$  odbiraka zahvaćeno šumom. Ovakva vrsta impulsnih smetnji se razmatra u cilju ispitivanja performansi procesa rekonstrukcije, kao i tačnosti teorijski izvedene greške (\ref{nonsparse}) u slučaju postojanja vremenski lokalizovanih distorzija u signalu (klikovi u zvučnim zapisima, ogrebotine na kompakt diskovima, zatim odsijecanje (engl. \textit{clipping}) itd.), \cite{inpainting}, \cite{savremeni_konf_por,bayesnoise}. Rekonstrukcija se i u ovom slučaju sprovodi korišćenjem polupreklopljenih frejmova signala dužine $N=500$, obuhvaćenih Hanovom prozorskom funkcijom. Oštećeni odbrici su detektovani limiterom i tretirani kao nedostupni.
		\paragraph{Teorijska greška.}
	Numerički dobijene srednje kvadratne greške - MSE (\ref{enumer}) se i u ovom slučaju u velikoj mjeri poklapaju sa teorijskim rezultatom (\ref{eteor}), kao što je pokazano na slici \ref{e1_error_stat2}. Rezultati su predstavljeni za sve signale iz sva tri razmatrana seta, za slučaj kada je, u prosjeku, procenat oštećenih odbiraka $p=10\%$. Pune linije predstavljaju krive teorijske greške, dok su crnim tačkama predstavljeni numerički rezultati. Tačnost predloženog teorijskog izraza za MSE je očekivana dok god su ispunjeni uslovi za CS rekontrukciju, čak i u slučajevima kada  se oštećenja pojavljuju u blokovima sukcesivnih odbiraka (koji se posmatraju kao nedostupni).

		\paragraph{Poređenje sa stanovišta srednje kvadratne greške (MSE).}
		Rezultati rekonstrukcije ostvareni primjenom prezentovanog CS metoda upoređeni su
		sa rezultatima dobijenim primjenom median filtara, niskopropusnih filtara Batervortovog tipa (sa istim parametrima kao u prvom eksperimentalnom scenariju), zatim sa LSAR i LSAR+SIN rekonstrukcionim algoritmima, kao i sa rezultatima  LASSO-ISTA pristupa. Poređenje je obavljeno sa stanovišta srednje kvadratne greške (MSE). Rezultati su prezentovani u tabeli \ref{table_example2}, za slučaj kada je, u prosjeku, procenat oštećenih odbiraka $p=50\%$. Pored toga, u ovom eksperimentalnom scenariju poređenje je obavljeno i sa nedavno publikovanim algoritmom za uklanjanje impulsnih smetnji i klikova (u pitanju je rekonstrukcija zasnovana na AR modelovanju), \cite{savremeni_por,savremeni_konf_por}. Detekcija i rekonstrukcija zasnovana na AR modelu su sprovedeni korišćenjem algoritama autora navedenih radova, sa podrazumijevanim podešavanjima i parametrima u programima (tzv. \textit{semi-causal with decision-feedback scheme}), \cite{savremeni_konf_por}. Vrsta ,,FTR'' u tabeli \ref{table_example2} sadrži rezultate za tzv. \textit{forward-time} pristup, dok vrsta ,,BDR'' sadrži rezultate sa tzv. bidirekcionim pristupom, originalno predstavljenim u \cite{savremeni_por}. Ovi algoritmi su visoko adaptirani za primjenu u uklanjanju klikova u audio signalima. Detekcija oštećenih odbiraka je obavljena primjenom ugrađenih detekcionih procedura. Veliki broj oštećenih odbiraka u razmatranom eksperimentu je značajno redukovao performanse rekonstrukcije navedenim pristupima. 
			\begin{table}[!t]
			\small
			%% increase table row spacing, adjust to taste
			\renewcommand{\arraystretch}{1.3}
		
			\caption{Srednja kvadratna greška (MSE) dobijena usrednjavanjem rezultata za sve signale iz razmatranih setova, za slučaj impulsnih smetnji lokalizovanih u blokovima u vremenskom domenu.}
			\label{table_example2}
			\centering
			%% Some packages, such as MDW tools, offer better commands for making tables
			%% than the plain LaTeX2e tabular which is used here.
			\begin{tabular}{lccc}
				\toprule  & \!\textbf{Muzika @ 16 kHz} & \! \textbf{Govor @ 8 kHz}\!\!& \textbf{Govor @ 16 kHz}\!\\
				\midrule
				Zašumljeni signali &  -11.33dB &   -23.55dB &  -24.53dB \\
				
				Med3 & -26.15dB & -42.70dB & -42.83dB \\
				
				Med5 &  -25.96dB & -42.25dB & -42.68dB \\
				
				LPF1 &  -25.72dB &  -41.62dB & -42.21dB \\
				
				LPF2 & -25.77dB &  -41.72dB & -42.31dB \\
				
				FTR & -26.24dB &-42.92dB & -42.91dB\\
				
				BDR & -26.66dB & -42.93dB & -42.95dB\\
				
				LSAR & -32.52dB & -45.48dB &  -46.58dB \\
				
				LSAR+SIN & -34.17dB & -48.61dB & -50.20dB \\
				
				LASSO & -37.64dB &  -48.41dB & -49.65dB \\
				
				MP& -37.80dB & -50.58dB &  -53.17dB \\
				\bottomrule
			\end{tabular}
		\end{table}

		Kao i u prethodnom eksperimentu, rekonstrukcione tehnike LSAR i LSAR+SIN su pokazale u prosjeku bolje rezultate od ostalih klasičnih pristupa. Prosječno poboljšanje sa stanovišta MSE, korišćenjem median filtra, iznosi $17.5$ dB. Sličan rezultat je dobijen za obije razmatrane dužine filtra. Niskopropusno filtriranje dovodi do prosječnog poboljšanja od $16.7$ dB, za oba razmatrana niskopropusna filtra. Poboljšanje u slučaju FTR i BDR algoritama je bilo značajno gore u razmatranom eksperimentalnom kontekstu sa $50\%$ nedostajućih odbiraka, nego što je to slučaj kada je ovaj procenat niži (e.g. na primjer, za $p=10\% \text{ ili } 15\%$). Poboljšanje je iznosilo samo $17.6$ dB. CS rekonstrukciona tehnika zasnovana na minimizaciji $\ell_1$ norme (LASSO-ISTA) dovela je do MSE poboljšanja od $25.43$ dB u prosjeku, je MSE poboljšanje primjenom LSAR i LSAR-SIN tehnika iznosilo $21.72$ dB i $24.52$ dB, respektivno. Zanimljivo je uočiti da je LASSO-ISTA algoritam u prosjeku pokazao bolji rezultat u slučaju seta signala ,,Muzika @ 16 kHz'' u odnosu na LSAR-SIN pristup. U prosjeku, prezentovani CS algoritam je u prosjeku produkovao rezultat za $1.95$ dB bolji od rezultata LASSO-ISTA rekonstrukcije. Najznačajnija razlika u rezultatima je uočljiva za set signala ,,Govor @ 16 kHz''.
		
\begin{figure*}[tb]%
			\centering
			\includegraphics
			{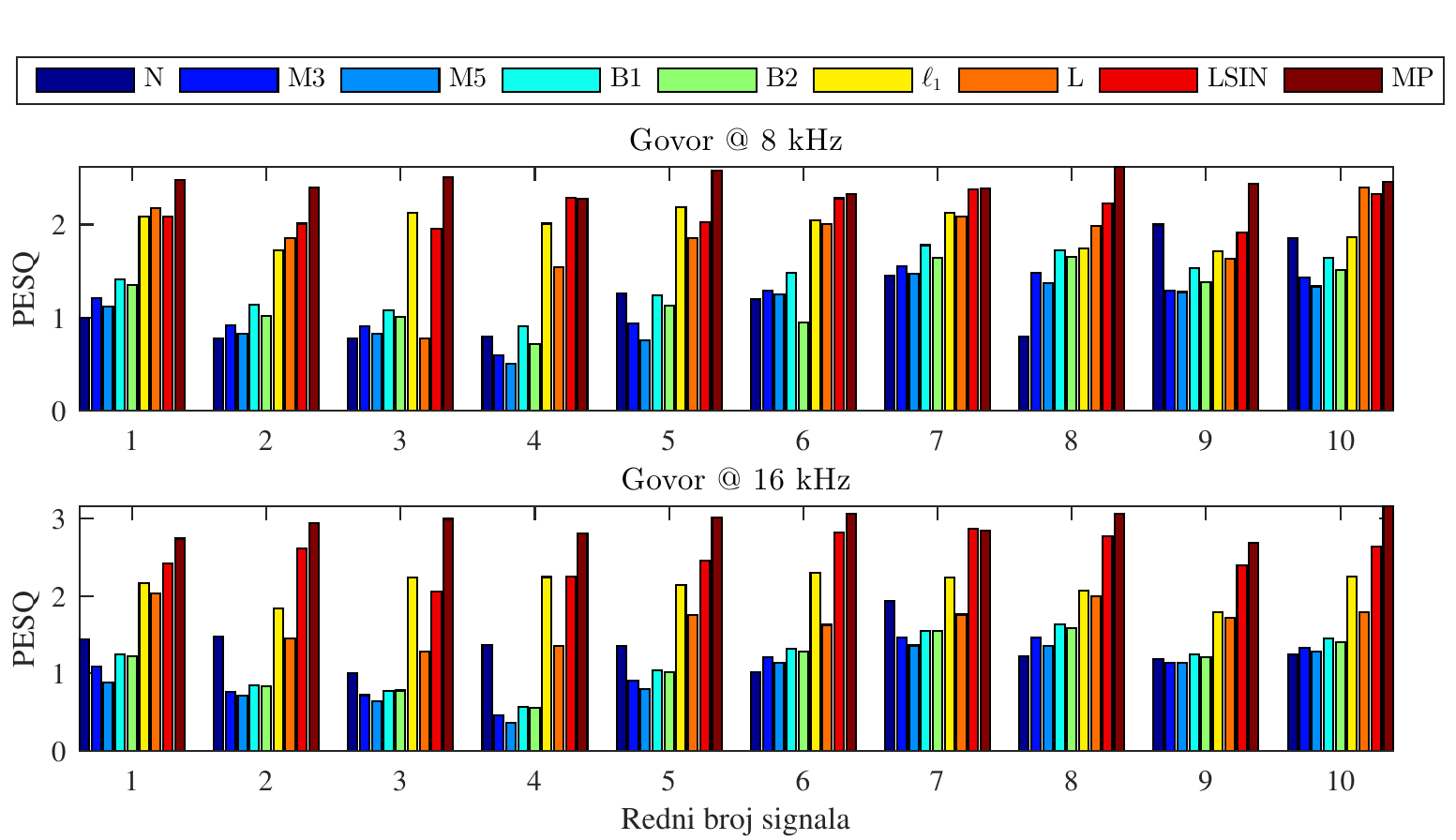}%
			\caption[Perceptualna evaluacija govornih signala korišćenjem PESQ metrike u eksperimentalnom scenariju sa impulsnim smetnjama lokalizovanim u vremenskim blokovima]{Perceptualna evaluacija govornih signala korišćenjem PESQ metrike u eksperimentalnom scenariju sa impulsnim smetnjama lokalizovanim u vremenskim blokovima. Rezultati su u vezi sa MSE vrijednostima iz tabele \ref{table_example2}. Oznake za primijenjene tehnike: N -- zašumljeni signali, M3 i M5 -- median filtri dužine 3 i 5, B1 i B2 -- razmatrani Batervortovi niskopropusni filtri, $\ell_1$ -- LASSO-ISTA, L --  LSAR, LSIN -- LSAR+SIN, MP -- predloženi MP pristup.}%
			\label{e2_preceptual}%
		\end{figure*}

		\paragraph{Poređenje sa stanovišta objektivnih mjera perceptualnog kvaliteta.} Za evaluaciju perceptualnog kvaliteta, i u ovom eksperimentalnom kontekstu su korišćenje PEMO-Q i PESQ metrike. Rezultati za set signala ,,Muzika @ 16'' kHz su prikazani na slici \ref{pemoq} (druga podslika). Rezultati evaluacije perceptivnog kvaliteta za setove signala ,,Govor @ 8 kHz'' i ,,Govor @ 16 kHz'' su prezentovani na slici \ref{e2_preceptual}. 
		
		Na osnovu PEMO-Q ocjena na slici \ref{pemoq} (druga podslika) može se zaključiti da je prezentovana CS rekonstrukciona tehnika, sa prosječnom ocjenom $-2.12$, bolja od LASSO-ISTA algoritma ($-2.48$),  kao i od LSAR+SIN ($-2.95$) i LSAR ($-3.13$) algoritama. Međutim, u slučaju malog broja signala, LASSO-ISTA algoritam je pokazao bolji rezultat od prezentovane CS tehnike. Kao što je to već nagoviješteno MSE rezultatima u tabeli \ref{table_example2}, LASSO-ISTA pristup je bio bolji od LSAR i LSAR+SIN pristupa u ovom eksperimentu.
		
		 PESQ ocjene, prikazane na slici \ref{e2_preceptual} odgovaraju rezultatima prezentovanim u tabeli \ref{table_example2}. U slučaju seta signala ,,Govor @ 8 kHz'', prosječna PESQ ocjena kvaliteta signala dobijenih prezentovanim CS pristupom je $2.44$ i ona je bolja od odgovarajućih ocjena dobijenih u slučaju drugih algoritama: LSAR+SIN ($2.15$), LASSO-ISTA ($1.96$) i LSAR ($1.83$). Za set signala ,,Govor @ 16 kHz'' dobijene prosječne ocjene su: prezentovani CS pristup ($2.93$), LSAR+SIN ($2.53$), LASSO-ISTA ($2.13$) i LSAR ($1.68$). Kao što je primijećeno u prethodnom scenariju, poboljšanje perceptualnog kvaliteta je veće za slučaj ,,Govor @ 16 kHz''.
		
%\noindent1. Iterative algorithm:
%\small
%\begin{align*} 
%&\noindent\mathbf{K}=\emptyset, ~~~ \mathbf{y}_{r}=\mathbf{y}\\
%&\textbf{for}~~i=1:K\\
%&~~~{\mathbf{X}^{C}_{0}=(\mathbf  {A}^{T}\mathbf  {A})^{-1}\mathbf  {A}^{T}\mathbf{y}_{r} }\\
%&~~~k=\text{arg}\{\max_{k}|X^C_{0}(k)|\}\\
%&~~~\mathbf{K}=\{\mathbf{K},k\}\\
%&~~~{\mathbf{A}_{K}=\mathbf  {A}(K,:)}\\
%&~~~{\mathbf{X}^{C}_{K}=(\mathbf{A}_{K}^{T}\mathbf{A}_{K})^{-1}\mathbf{A}_{K}^{T}\mathbf{y}}\\
%&~~~X^C_{Kz}(k)=X^C_{K}(k), k \in\mathbf{K},~~\text{and} ~~~X^C_{Kz}(k)=0, k \not \in
%\mathbf{K},\\
%&~~~\mathbf{x}_{r}=\mathbf{\Phi}^{-1} \mathbf{X}^{C}_{Kz}\\
%&~~~\mathbf{y}_{r}=\mathbf{y}-\mathbf{x}_{r},~~~ \text{for}~~~ n \in\mathbf{ M}\\
%&\textbf{end}\\
%&\mathbf{x}=\mathbf{x}_{r}%
%\end{align*}
%\normalsize
%\noindent 2. One step algorithm:
%\small
%\begin{align*}
%&\noindent{\mathbf{X}^{C}_{0}=(\mathbf  {A}^{T}\mathbf  {A})^{-1}\mathbf  {A}^{T}\mathbf{y}}\\
%&\mathbf{K}=\underset{k}{\text{arg}}\{(|X^C_{0}(k)|-4\sqrt{\frac{(M(N-M)}{N^{2}(N-1)}
%	\frac{\left\Vert \mathbf{y} \right\Vert ^{2}_{2}}{M}} )>0\}\\
%&\mathbf{A}_{K}=\mathbf  {A}(K,:)\\
%&\mathbf{X}^{C}_{K}=(\mathbf{A}_{K}^{T}\mathbf{A}_{K})^{-1}\mathbf{A}_{K}^{T}\mathbf{y}\\
%&X^C_{Kz}(k)=X^C_{K}(k), k \in\mathbf{K}, ~~\text{and} ~~~X^C_{Kz}(k)=0, k \not \in
%\mathbf{K},\\
%&\mathbf{x}=\mathbf{\Phi}^{-1} \mathbf{X}^{C}_{Kz}%
%\end{align*}
%
%\normalsize
%The algorithms can be combined.

\section{Dvodimenziona diskretna kosinusna transformacija kao domen rijetkosti signala}
\label{sekcija32}
Već je ustanovljena činjenica da je DCT važna u brojnim oblastima primjene \cite{libi,irenacs,denoising,dos,dctaudio,app_audioCS,bi1,app_audio1,brajovic_dct2}. Zahvaljujući svojstvu da digitalne slike visoko koncentriše u manjem broju transformacionih koeficijenata, dvodimenziona DCT (2D-DCT) je jedna od najkorišćenijih transformacija u kompresionim algoritmima za digitalne slike \cite{dos}. Štaviše, pokazalo se da je ova transformacija pogodna i za rekonstrukciju digitalnih slika sa nedostajućim piskelima i/ili pikselima koji su oštećeni šumom. 
Pretpostavka o rijetkosti reprezentacije slike u domenu dvodimenzione DCT pokazala se opravdanom, \cite{libi,irenacs,denoising}.  Mjerenje koncentracije 2D-DCT koeficijenata pomoću  $\ell_1$-norme i variranje vrijednosti nedostajućih odbiraka u cilju dobijanja rješenja sa najmanjim stepenom rijetkosti dovelo je do razvoja veoma naprednih tehnika za kompresivno odabiranje \cite{impcs,denoising}. U slučaju OMP pristupa, jasno je da se rekonstrukcija se može postići ukoliko su poznate pozicije komponenti signala, odnosno slike \cite{cosamp,More,greedy,lasso}. U tom slučaju, prave vrijednosti koeficijenata se mogu  izračunati ukoliko se identifikovane pozicije komponenti kombinuju sa 2D-DCT mjernom matricom \cite{automated,dos}. 

Važno je, međutim, naglasiti da su u suštini realne digitalne slike najčešće samo aproksimativno rijetke u 2D-DCT domenu \cite{libi,irenacs,denoising,dos}.  To znači da, pored koeficijenata sa značajnim vrijednostima koji nose najveći dio energije signala, mogu postojati i koeficijenti sa malim vrijednostima, umjesto koeficijenata koji bi bili jednaki nuli. Budući da algoritmi za rekonstrukciju rijetkih signala podrazumijevaju određeni stepen rijetkosti, navedene komponente će ostati nerekonstruisane \cite{DFT_nonsparse,dos}. To neizbježno dovodi do  grešaka u rekonstrukciji. Već je ustanovljeno da se redukcija broja dostupnih odbiraka manifestuje kao šum u transformacionom domenu \cite{dftmiss,automated}. Tokom rekonstrukcije, taj šum se u potpunosti redukuje u slučaju kada je pretpostavka o rijetkosti striktno zadovoljena. Međutim, koeficijenti sa malim vrijednostima u signalima koji nijesu potpuno rijetki ostaju nerekonstruisani, pa u rekonstruisanom signalu/slici, njihov doprinos u vidu šuma i dalje ostaje. Ako se signal koji nije rijedak rekonstuiše iz redukovanog skupa mjerenja, tada će se šum usljed nedostajućih odbiraka u nerekonstruisanim koeficijentima razmatrati kao aditivni šum u rekonstruisanom signalu \cite{DFT_nonsparse}. 

Široko prihvaćena literatura u oblasti kompresivnog odabiranja, kako je već naglašeno, daje samo opšte granice za grešku u rekonstrukciji signala koji nijesu rijetki, a koji su rekonstruisani uz pretpostavku o rijetkosti \cite{donoho,CandesRIP,candes,candes2,More}. 
%Granice za grešku u slučaju DCT i DFT razmatrane su u kontekstu jedinstvenosti rješenja u radu \cite{wohlberg}. 

U ovoj sekciji, biće prezentovan egzaktan izraz za očekivanu kvadratnu grešku u rekonstrukciji aproksimativno rijetkih signala, odnosno signala koji nijesu rijetki u 2D-DCT domenu. Pretpostavlja se da su ti signali rekonstruisani iz redukovanog skupa mjerenja pod pretpostavkom rijetkosti u ovom domenu. Uticaj nedostajućih odbiraka će i u ovom slučaju biti modelovan aditivnim šumom \cite{dftmiss,dctaudio}. Šum koji potiče od nedostajućih odbiraka u svakoj komponenti signala se modeluje kao Gausov slučajni proces, pa će stoga biti izvedene srednje vrijednosti i varijanse odgovarajućih slučajnih promjenljivih. Upravo ti rezultati će biti uključeni u izvođenje egzaktnog izraza za grešku u rekonstrukciji signala koji nijesu rijetki, a koja se sprovodi uz pretpostavku o rijetkosti. Prezentovana teorija će biti verifikovana numeričkim eksperimentima sa realnim slikama. Rezultati su predstavljeni u radu \cite{brajovic_dct2}.

\subsection{Osnovne definicije}
\label{BacSec}
Posmatrajmo 2D diskretni signal (digitalnu sliku) dimenzija $M\times N$, označen sa $x(m,n)$. 2D-DCT ovog signala je definisana izrazom \cite{dos}:
\begin{align}
X{^C}(p,q)=&\sum\limits_{m=0}^{M-1}\sum\limits_{n=0}^{N-1}x(m,n)\varphi_M(m,p)\varphi_N(n,q)
,\end{align}
gdje su $p=0,\dots,M-1$ i $q=0,\dots,N-1$ indeksi transformacionih koeficijenata, dok 
\begin{gather}
\nonumber
\varphi_M(m,p)=\sqrt{\frac{2}{M}}\cos\left(  \frac{\pi(2m+1)p}{2M}\right), p \ne 0\\ 
\varphi_N(n,q)=\sqrt{\frac{2}{N}}\cos\left(\frac{\pi(2n+1)q}{2N}\right), q \ne 0
\label{base_f}
\end{gather}
predstavljaju normalizovane bazne funkcije ove transformacije. Za $p=0$ ili $q=0$, ove funkcije su definisane izrazima $\varphi_M(m,0)=\sqrt{1/M}$ i $\varphi_N(n,0)=\sqrt{1/N}$, respektivno. Odgovarajuća inverzna transformacija ima sljedeću formu:
\begin{align}
s(m,n)=&\sum\limits_{p=0}^{M-1}\sum\limits_{q=0}^{N-1}X{^C}(p,q)\varphi_M(m,p)\varphi_N(n,q)
\end{align}
gdje je $m=0,1,\dots,M-1, n=0,1,\dots,N-1$. 2D-DCT transformacija se može zapisati i u matričnom obliku \footnote{U okviru sekcije \ref{sekcija32} će u matričnom zapisu 2D-DCT i vezanih relacija biti implicitno i po potrebi podrazumijevane operacije pretvaranja matrica signala (slika) i transformacionih matrica u odgovarajuće vektore-kolone, kao i odgovarajuće inverzne operacije. Adekvatan oblik direktnih i inverznih transformacionih matrica je takođe podrazumijevan.} \cite{dos,dctaudio}:
\begin{equation}
\mathbf{X}^C=\mathbf{\Phi x},
\end{equation}
gdje je $\mathbf{X}^C$ matrica 2D-DCT koeficijenata razmatranog signala, $\mathbf{\Phi}$ je 2D-DCT transformaciona matrica, dok matrica $\mathbf{x}$ sadrži odbirke posmatranog signala (odnosno, piksele digitalne slike). Za inverznu 2D-DCT važi relacija:
$$
\mathbf{x=\Psi }\mathbf{X}^C,
$$
pri čemu je inverzna 2D-DCT označena sa $\mathbf{\Psi=\Phi}^{-1}$.
Dvodimenzioni signal ili digitalna slika oblika:
\begin{align}
x(m,n)=&\sum\limits_{l=1}^{K}A_{l}\varphi_M(m,{p_l})\varphi_N(n,{q_l})
\label{modeld2}
\end{align}
je rijedak u 2D-DCT domenu ukoliko je broj nenultih 2D-DCT koeficijenata $K$ mnogo manji od ukupnog broja piksela (odbiraka), odnosno, ako važi $K\ll MN$. Komponente su locirane na DCT indeksima $(p_l,q_l)$ i imaju amplitude $A_l,~l=1,2,\dots,K$.

%\subsection{Rekonstrukcija digitalnih slika i digitalnog videa}

\subsection{Uticaj nedostajućih odbiraka (piksela)}
\label{RecProcSec}
U kontekstu kompresivnog odabiranja, pretpostavlja se da je samo $N_A\le M N$ slučajno pozicioniranih piksela 
sa indeksima $(m_{i}, n_{i}) \in \{(m_1,n_1), (m_2,n_2),\ldots,(m_{N_A},n_{N_A})\}= \mathbb{N}_A
\subseteq\mathbb{N}=\{(0, 0), (0, 1),\ldots,(M-1, N-1)\}$
dostupno. Ukoliko se od dostupnih odbiraka formira vektor $\mathbf{y}$ sa elementima
\begin{align}
y(i)=x(m_{i},n_i)=&\sum\limits_{p=0}^{M-1}\sum\limits_{q=0}^{N-1}X{^C}(p,q)\varphi_M(m_i,p)\varphi_N(n_i,q),
\end{align}
gdje je $i=0,1,\dots,N_A$, tada dolazimo do matrične forme
$
\mathbf{y=A}\mathbf{X}^C,
$
 pri čemu je $\mathbf{A}$ mjerna matrica dimenzija $N_A\times M N$. Po definiciji, ona je parcijalna inverzna 2D-DCT matrica, koja sadrži vrste matrice $\mathbf{\Psi}$ koje korespondiraju pozicijama dostupnih odbiraka (piksela).

Digitalna slika $x(m,n)$ je rijetka u 2D-DCT domenu, sa stepenom rijetkosti $K$, ukoliko je samo $K$ 2D-DCT koeficijenata tog signala različito od nule. Nenulti koeficijenati pozicionirani na indeksima $(p,q) \in \Pi_K=\{(p_1,q_1),(p_2,q_2),\dots,(p_K,q_K)\}$ formiraju vektor $\mathbf{X}^C_{K}$. 
Vektor 2D-DCT koeficijenata dobijen rekonstrukcijom slike, uz pretpostavku da je ona rijetka sa stepenom $K$, će biti označen sa $\mathbf{X}^C_{R}$. Ovo je vektor od $K$ rekonstruisanih nenultih koeficijenata sa pozicija $(p,q) \in \Pi_K$. Digitalna slika je aproksimativno rijetka, ukoliko su koeficijenti $X{^C}(p,q)$, $(p,k) \notin \Pi_K$ malih vrijednosti, a ako su reda veličine koeficijenata $X{^C}(p,q)$, $(p,k) \in \Pi_K$, smatraće se da slika nije rijetka u 2D-DCT domenu. Tada vektor $\mathbf{X}^C_{K}$ sadrži $K$ najvećih vrijednosti iz matrice $\mathbf{X}^C$. Vektor $\mathbf{X}^C_{K}$ produžen nulama do veličine originalne matrice $\mathbf{X}^C$, i zapisan u obliku matrice $\mathbf{X}^C$ odgovarajućom preraspodjelom indeksa, biće označen sa $\mathbf{X}^C_{K0}$.

Inicijalna estimacija 2D-DCT koeficijenata, zasnovana na $\ell_2$-normi, računa se samo na osnovu dostupnih odbiraka (piksela):
\begin{align}\label{ff}
X^C_0(p,q)=&\sum\limits_{(m,n) \in\mathbb{N}_A}x(m,n)\varphi_M(m,p)\varphi_N(n,q),
\end{align}
gdje je $p=0, 1,\dots,M-1,~ q=0, 1, \dots, N-1$. Kao i u slučaju jednodimenzione DCT, jasno je da se isti rezultat dobija ukoliko su svi nedostupni pikseli jednaki nuli, \cite{dftmiss}. U matričnoj formi možemo pisati
\begin{gather}
\mathbf{X}^C_0=\mathbf{A}^{T}\mathbf{y}.
\end{gather}

Koeficijenti u (\ref{ff}) predstavljaju slučajne promjenljive. Kao u ranije razmatranim transformacijama, pokazaćemo da su njihove statističke karakteristike različite na pozicijama komponenti signala (slike), $(p,q)=(p_l,q_l)$, od osobina koje imaju na pozicijama koje ne odgovaraju komponentama slike, odnosno, $(p,q)\neq(p_l, q_l)$. 

\subsubsection{Statističke karakteristike 2D-DCT koeficijenata u slučaju nedostajućih piksela} 
Prvo će biti razmatran slučaj monokomponentnog signala, a zatim i opšti slučaj multikomponentnih signala. Dobijeni rezultati će biti osnov za izvođenje izraza za grešku u rekonstrukciji signala koji nijesu rijetki, ali koji su rekonstruisani pod pretpostavkom da su rijetki.

\paragraph{Monokomponentni signali.}

Prvo će biti razmatran slučaj monokomponentnog signala za $(p,q)\neq
(p_1,q_1)$, $K=1$, a zatim će rezultati  biti generalizovani i za  multikomponentne signale. Bez gubljenja opštosti, pretpostavićemo jediničnu amplitudu komponente signala, $A_1=1$. Iz (\ref{modeld2}) i (\ref{ff}) se dobija:
\begin{align}
X{^C}(p,q)=&\!\sum\limits_{(m,n) \in\mathbb{N}_A}\varphi_M(m,{p_1})\! \varphi_N(n,q_1)\varphi_M(m,p)\varphi_N(n,q).
\end{align}
Varijabla definisana izrazom
\begin{gather}
z_{p_1q_1}(m,n,p,q)=\varphi_M(m,{p_1}) \varphi_N(n,q_1)\varphi_M(m,p)\varphi_N(n,q) \label{daefx_1} 
\end{gather}
je slučajna za slučajne vrijednosti $(m,n)$. Prvo će biti razmatrane njene statističke karakteristike za $(p,q)\neq (p_1,q_1)$.  Inicijalna 2D-DCT estimacija se može zapisati na sljedeći način:
\begin{align}
X{^C}(p,q)=\sum\limits_{(m,n) \in\mathbb{N}_A}z_{p_1q_1}(m,n,p,q).
\end{align}

Kada važi $(p,q)\neq (p_1,q_1)$, 2D-DCT koeficijenti ne odgovaraju komponentama signala (odgovaraju CS šumu), dok se $X^C_0(p,q)$ ponaša kao slučajna Gausova promjenljiva \cite{dftmiss,dctaudio}. Imajući u vidu da su bazne funkcije ortogonalne,
\begin{equation}
\sum_{m=0}^{M-1}\sum_{n=0}^{N-1}z_{p_1q_1}(m,n,p,q)=\delta(p-p_{1},q-q_1)
\end{equation}
i da su sve vrijednosti $z_{p_1q_1}(m,n,p,q)$ jednako distribuirane, može se zaključiti da je srednja vrijednost slučajne promjenljive $X^C_0(p,q)$ jednaka nuli, odnosno:
\begin{equation}
\mu_{{X{^C_0}(p,q)}}=E\left\{  X{^C_0}(p,q)\right\}  =0,~ (p,q)\neq (p_1,q_1).
\end{equation}

U slučaju kada koeficijenti odgovaraju komponentama signala (slike), korišćenjem svojstva ortogonalnosti i pretpostavke o jednakoj distribuciji vrijednosti $z_{p_1q_1}(m,n,p,q)$, slijedi
\begin{equation}
\mu_{{X{^C_0}(p,q)}}=E\left\{  X^C_0(p,q)\right\}  =\frac{N_A}{MN},~ (p,q)=(p_1,q_1).
\end{equation}
Za slučajnu varijablu srednje vrijednosti nula, varijansa je definisana sljedećim izrazom:
\begin{align}
\sigma_{{X{^C_0}(p,q)}}^{2}=E\Big\{&\sum\limits_{(m,n) \in\mathbb{N}_A}  z_{p_1q_1}^{2}(m,n,p,q
)+\sum\limits_{(m,n) \in\mathbb{N}_A}\sum\limits_{\substack{(i,j) \in\mathbb{N}_A\\(i,j)\neq(m,n)}}z_{p_1q_1}(m,n,p,q)z_{p_1q_1}(i,j,p,q)\Big\}.\notag
\end{align}

Pošto se razmatra slučaj kada je $(p,q)\neq (p_1,q_1)$, lako se zaključuje da važi:
\begin{align}
\sum_{m=0}^{M-1}\sum_{n=0}^{N-1}z_{p_1q_1}(m,n,p,q)=0.\label{ort}
\end{align} 

Množenjem lijeve i desne strane izraza (\ref{ort}) sa
$z_{p_1q_1}(i,j,p,q)$ i primjenom operatora matematičkog očekivanja na obije strane jednakosti, slijedi:
\begin{align}
E\Big\{\sum_{m=0}^{M-1}\sum_{n=0}^{N-1}z_{p_1q_1}(m,n,p,q)z_{p_1q_1}(i,j,p,q)\Big\}=0,\label{ort1}
\end{align} 
gdje je $(i,j)\in \mathbb{N}$. Vrijednosti $z_{p_1q_1}(m,n,p,q)$ su jednako distribuirane. Stoga su članovi $E\{z_{p_1q_1}(m,n,p,q)z_{p_1q_1}(i,j,p,q)\}$  za
$(m,n)\neq (i,j)$ međusobno jednaki i jednaki konstanti $D$. Ukupan broj ovih članova je $MN-1$. Dalje, na osnovu (\ref{ort1}) dobijamo
\begin{gather}
\left(  MN-1\right)D+E\left\{  z_{p_1q_1}^{2}(m,n,p,q)\right\}  =0\label{D}.
\end{gather}

Polazni definicioni obrazac varijanse sada možemo zapisati u obliku
\begin{align}
\sigma_{{X{^C_0}(p,q)}}^{2}&= {N_A}E\{  z_{p_1q_1}^{2}(m,n,p,q)\}+(N_A^2-N_A)D,\label{A}
\end{align}
budući da je tačno ${N_A}$ matematičkih očekivanja sa kvadratnim članovima u prvoj sumi, dok je $N_A(N_A-1)$ članova u drugoj sumi jednako $D$. U cilju određivanja nepoznatog člana $E\left\{  z^{2}_{p_1q_1}(m,n,p,q)\right\}$, potrebno je razmotriti više slučajeva, u zavisnosti od uslova koje zadovoljavaju indeksi komponente signala.
	
\noindent\textbf{Podrazumijevani slučaj:} Razmatrajmo prvo najčešće zadovoljeni uslov, gdje za indekse važi: $p\neq p_1,~p\neq M-p_1,~q\neq q_1,~q\neq N-q_1$. Tada se može pisati da
$
E\{  z_{p_1q_1}^{2}(m,n, p ,q)\}=
E\{\varphi_M^2(m,{p_1})\varphi^2_N(n,{q_1})\}E\{\varphi^2_M(m,p)\varphi^2_N(n,q)\}=\tfrac{1}{M^2N^2}.
$
Inkorporirajući ovaj međurezultat u (\ref{D}), dobija se
$ D=-\frac{1}{M^2N^2}\frac{1}{MN-1}.$
Dalje, na osnovu izraza (\ref{A}), varijansa može biti zapisana u obliku:
\begin{align}
\sigma_{{X{^C_0}(p,q)}}^{2}=\frac{N_A(MN-N_A)}{M^2N^2(MN-1)}.\label{v1}
\end{align}
Ovaj rezultat važi i za $(p_1,q_1)=(0,0)$. Za $A_1\neq1$, rezultat je neophodno pomnožiti sa $A^2_1$.

\bigskip

\noindent \textbf{Slučaj 1:} Kada koeficijenti nijesu na pozicijama signala, a zadovoljavaju $p=p_1,~q\neq
q_1,~q\neq N-q_1$ dobija se
\begin{align}
E \Big\{  z_{p_1q_1}^{2}(m,n,p_1,&q)\Big\}=
E\Big\{\varphi^4_M(m,{p_1})\Big\}E\Big\{\varphi^2_N(n,{q_1})\Big\}
E\Big\{\varphi^2_N(n,q)\Big\}.
\end{align}

Korišćenjem osobine da je $E\{\varphi^2_N(n,{q_1})\}
E\Big\{\varphi^2_N(n,q)\Big\}=1/N^2$, dalje se može pisati:
\begin{align*}
E \Big\{  z_{p_1q_1}^{2}(m,n,p_1,q)\Big\}=\frac{1}{N^2} \bigg[\frac{1}{M^2}
+\frac{1}{2M}E\Big\{ \varphi^2_M(m,{2p_1})\Big\}  \bigg].
\end{align*}
Navedeno važi za ${p_1}\neq0$. U prethodnom izrazu je korišćena činjenica da funkcija $\varphi_M(m,{2p_1})$ ima srednju vrijednost jednaku nuli za slučajno $m$. Budući da je kosinus periodična funkcija, može se pisati: 
$
E\{ \varphi^2_M(m,{2p})\}=\frac{1}{M}\text{.}%
$
Konačno, dobija se: 
$E\left\{  z^{2}_{p_1q_1}(m,n,p_1,q)\right\}=\frac{3}{2M^2N^2}.$

Inkorporiranjem ovog rezultata u (\ref{D}) i (\ref{A}) dolazimo do:
\begin{align}
\sigma_{X^C(p,q)}^{2}=\frac{3N_A(MN-N_A)}{2M^2N^2(MN-1)},\label{v11}
\end{align}
za $p_1\neq0$. Ukoliko, dodatno, važi $p=p_1=0$, za nepoznato očekivanje se dobija:
\begin{equation}
E \left\{
z^{2}_{p_1q_1}(m,n,p,q)\right\}=\frac{1}{M^2N^2},
\end{equation} 
što dalje daje rezultat (\ref{v1}).

\bigskip

\noindent\textbf{Slučaj 2:} Korišćenjem istih izvođenja kao za Slučaj 1, lako se pokazuje da se, u slučaju kada pozicija razmatranog koeficijenta zadovoljava $p\neq p_1,~p\neq M-p_1,~q= q_1\neq0$, dobija rezultat (\ref{v11}). Dodatno, uslov $q= q_1=0$ vodi rezultatu (\ref{v1}).

\bigskip

\noindent\textbf{Slučaj 3:} Posmatrajmo slučaj kada je zadovoljeno $p=M-p_1,~q\neq q_1,~q\neq N-q_1$.
Dodatno, neka je zadovoljeno i $p_1\neq 0$. Nepoznati kvadratni član tada postaje:
\begin{align}\setstretch{2}
E \Big\{\!  z^{2}_{p_1q_1}(m,n,M-p_1,q)\Big\}&=\!
E\Big\{\varphi^2_M(m,{p_1}) \varphi^2_M(m,{M-p_1}) \Big\} E\Big\{\varphi^2_N(m,q)\varphi^2_N(m,{q_1})\Big\}\notag
\\
&=\frac{1}{N^2}E\Big\{\varphi^2_M(m,{p_1})\psi^2_M(m,{p_1})\Big\},
\end{align}
gdje je $\psi_M(m,{p_1})=\sqrt{\frac{2}{M}}\sin\big(  \frac{\pi(2m+1)p_1}{2M}\big)$ za $p_1 \ne 0$ i $\psi_M(m,0)=\sqrt{\frac{1}{M}}$.

Ovdje su korišćeni dobro poznati identiteti $\psi^2_{M}(m,M)  ={\frac{2}{M}}$ i $
\varphi_{M}(m,M)  =0$ koji se pojavljuju u formuli
$
\varphi^2_M(m,{M-p_1})$, kada je izražena u obliku $\varphi_{M}(m,M)\varphi_M(m,{p_1})
)+\psi_{M}(m,M) \psi_M(m,{p_1}).\notag
$

Korišćenjem identiteta za sinus dvostrukog ugla i očekivanje
$E\{\psi^2_{M}(m,2p_1)\}=\frac{1}{M}$, analogno sa slučajem očekivanja kvadrata kosinusa, dobija se:
\begin{align}
E& \left\{  x^{2}_{p_1q_1}(m,n,M-p_1,q)\right\}=
\frac{1}{2M^2N^2}.
\end{align}

Uvrštavanjem ovog međurezultata u (\ref{A}), dobija se sljedeći izraz za varijansu:
\begin{align}
\sigma_{X^C(p,q)}^{2}=\frac{N_A(MN-N_A)}{2M^2N^2(MN-1)},\label{v10}
\end{align}
što važi uz $p_1\neq 0$. Kada je $p_1=0$, lako se pokazuje da važi izraz (\ref{v1}). 

\bigskip

\noindent\textbf{Slučaj 4:} U ekvivalentnom slučaju kada je $p\neq p_1,~p\neq M-p_1,~q=
N-q_1$, rezultati su isti kao u Slučaju 3. Za $q_1=0$, važi rezultat iz (\ref{v1}).

\bigskip

\noindent\textbf{Slučaj 5:} Posmatra se uslov $p=p_1,~q = N-q_1$. Kombinujući izvođenja predstavljena za Slučaj 1 i Slučaj 3, lako se pokazuje da varijansa postaje:
\begin{align}
\sigma_{X^C(p,q)}^{2}=\frac{3N_A(MN-N_A)}{4M^2N^2(MN-1)}.\label{v12}
\end{align}
Za $p_1\neq0$ ili $q_1\neq0$ važi izraz (\ref{v1}). Navedeni izraz važi i za uslov $(p_1,q_1)=(0,0)$.

\bigskip

\noindent\textbf{Slučaj 6:} U analognom slučaju, kada je zadovoljeno, važi rezultat
result (\ref{v12}), a uz dodatni uslov $p_1\neq0$
ili $q_1\neq0$ ili  $(p_1,q_1)=(0,0)$, varijansa postaje (\ref{v1}).

\bigskip

\noindent \textbf{Slučaj 7:} Kada je $p = M-p_{1}$ i $q=N-q_1$, nepoznato očekivanje dobija oblik:
\begin{align}
E\left\{  z^{2}_{p_1q_1}(m,n,M-p_1,N-q_1)\right\}=\frac{1}{4 M^2 N^2} \notag
\end{align}
za $(p_1,q_1)\neq(0,0)$, što vodi do:
\begin{align}
\sigma_{X^C(p,q)}^{2}=\frac{N_A(MN-N_A)}{4M^2N^2(MN-1)}.\label{v13}
\end{align}

Za $(p_1,q_1)=(0,0)$ važiće $E\{  x^{2}_{00}(m,n,M,N)\}=\frac{1}{M^2N^2}$.
Stoga, u slučaju koeficijenta koji ne odgovara komponenti signala, na poziciji $(M-p_1, N-q_1)$, varijansa dobija oblik zadat izrazom ({\ref{v1}}). Lako se pokazuje da je za $p_1=0$ ili $q_1=0$
razmatrana varijansa definisana izrazom (\ref{v10}).

U svim razmatranim slučajevima, u slučaju nejediničnih amplituda $A_1\neq1$, dobijene izraze za varijansu treba pomnožiti sa
$A^2_1$, što je očigledno iz definicije varijanse.

Konačno, izvedeni izrazi mogu biti objedinjeni sljedećom relacijom:
\begin{align}
\sigma_{{X{^C_0}(p,q)}}^{2}&=A^2_1\frac{N_A(MN-N_A)}{M^2N^2(MN-1)}\notag
\\ & \times\bigg[1+\Big(1-\delta(p_1,q_1)\Big)\Big(\frac{1}{2}\sum_{i=0}^{N-1}(1-\delta(p_1,0))\delta(p-p_1,q-i)\notag\\
& +\frac{1}{2}\sum_{i=0}^{M-1}(1-\delta(0,q_1))\delta(p-i,q-q_1)\notag\\
&-\frac{1}{2}\sum_{i=0}^{N-1}(1-\delta(M-p_1,0))\delta(p-(M-p_1),q-i)\notag\\
& -\frac{1}{2}\sum_{i=0}^{M-1}(1-\delta(0,N-q_1))\delta(p-i,q-(N-q_1))\notag\\
&-\frac{7}{4}(1-\delta(p_1,0)-\delta(0,q_1))\delta(p-(M-p_1),q-(N-q_1))\notag\\
&-\frac{5}{4}(1-\delta(p_1,0)-\delta(0,q_1))\delta(p-p_1,q-(N-q_1))\notag\\
&-\frac{5}{4}(1-\delta(p_1,0)-\delta(0,q_1))\delta(p-(M-p_1),q_1)
\Big)\bigg].\label{vg}
\end{align}
%-7/4 i -5/4 zato sto iz gornjih suma imamo sigurno sabrane dvije 1/2
gdje je $\delta(p,q)=1$ za $p=0$ i $q=0$, dok je, inače, $\delta(p,q)=0$. 

Uključivanjem svih razmatranih specijalnih slučajeva, dobija se da je prosječna vrijednost varijanse (\ref{vg}) konstantna i jednaka:
\begin{align}
\bar{\sigma}_{X^C_0}^{2}&=A^2_1\frac{N_A(MN-N_A)}{M^2N^2(MN-1)^2}\left(MN-\frac{21}{4}\right).
\end{align}
Kako je $MN \gg 1$, prosječna varijansa 2D-DCT koeficijenata koji ne odgovaraju pozicijama komponente signala se može precizno aproksimirati izrazom:
\begin{align}
\sigma^2_N=\bar{\sigma}_{X^C_0}^{2} \approx A^2_1\frac{N_A(MN-N_A)}{M^2N^2(MN-1)}.\label{approx}
\end{align}

\begin{figure}[tbp]
	\centering{\includegraphics[scale=0.9]{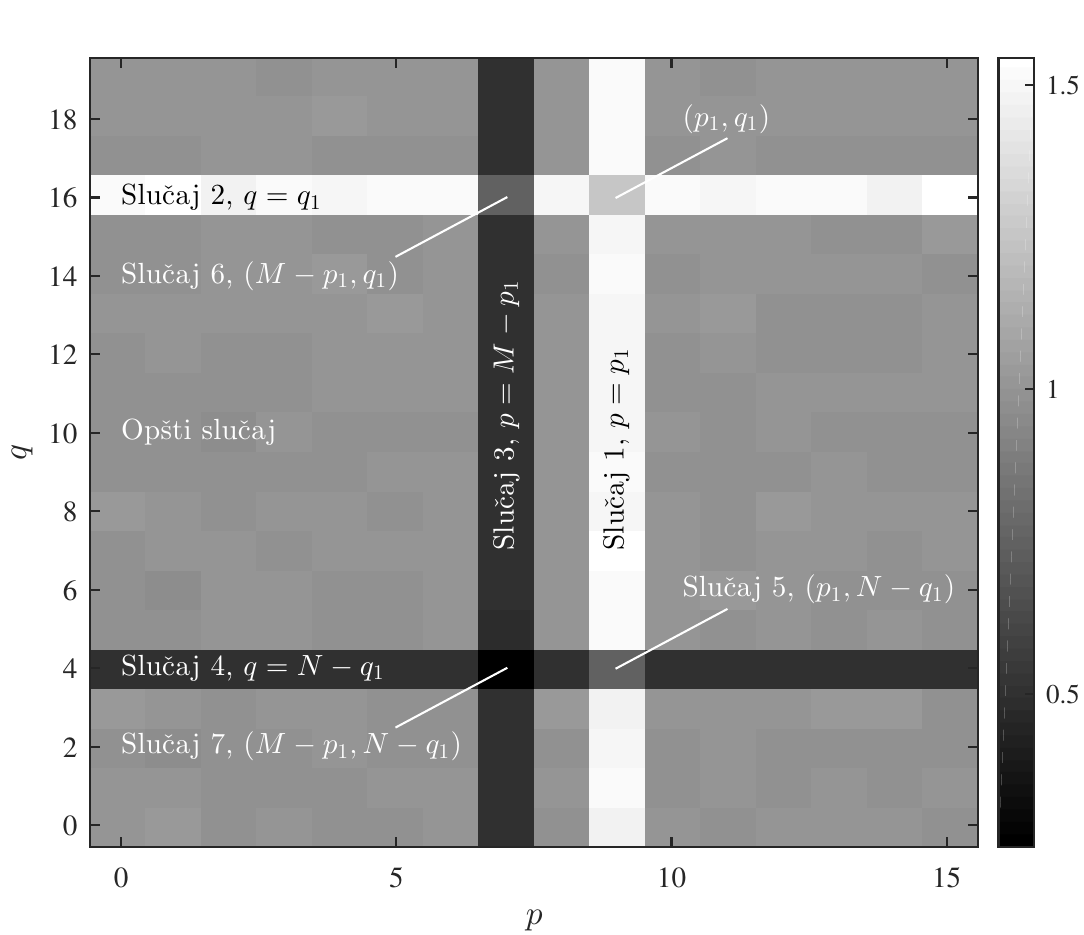}}
	\caption[Varijansa inicijalne estimacije 2D-DCT koeficijenata]{Varijansa inicijalne estimacije 2D-DCT koeficijenata. Dobijena je numerički, na bazi 20000 nezavisnih realizacija monokomponentnog signala dimenzija $M\times N=16\times 20$, rijetkog u 2D-DCT domenu, sa $N_A=128$ slučajno pozicioniranih dostupnih odbiraka.}
	\label{var_image}
\end{figure}

\begin{primjer}
Posmatra se monokomponentni signal rijedak u 2D-DCT domenu, koji je definisan izrazom:
\begin{align}
x(m,n)=A_{1}\varphi_{p_1}(m,M)\varphi_{q_1}(n,N) , \label{ex1}
\end{align}
gdje je $M=16$, $N=20$, $A_1=1$, $p_1=9$ and $q_1=16$. Dostupno je samo $N_A=128$ slučajno pozicioniranih odbiraka signala, i posmatra se 20000 statistički nezavisnih realizacija signala. Na osnovu inicijalnih 2D-DCT estimacija (\ref{ff}), varijansa 2D-DCT koeficijenata je izračunata numerički, usrednjavanjem inicijalnih estimacija iz svih realizacija. Rezultati su prikazani na slici \ref{var_image} i skalirani su konstantnim članom (\ref{v1}). Razmatrani specijalni slučajevi su označeni na slici \ref{var_image}.
\end{primjer}
\paragraph{Multikomponentni signali.}
U slučaju multikomponentnih signala, posmatrana slučajna promjenljiva dobija sljedeći oblik:
\begin{align}
\label{slvar}
X^C_0(p,q)=\sum\limits_{(m,n) \in\mathbb{N}_A}\sum\limits_{l=1}^{K}A_{l}\varphi_M(m,{p_l})\varphi_N(n,{q_l})\varphi_M(m,p)\varphi_N(n,q).
%=\sum\limits_{i=1}^{M}\sum\limits_{l=1}^{K}x(k_l,k,n_i),\label{rv}.
\end{align}

I za multikomponentne signale važi da su koeficijenti na pozicijama koje ne odgovaraju komponentama, odnosno $(p,q)\neq (p_l,q_l)$, slučajne promjenljive sa Gausovom distribucijom i srednjom vrijednošću nula, budući da su formirane sumiranjem nezavisnih Gausovih promjenljivih sa srednjim vrijednostima nula, po indeksu $l$. Naime, sada nedostajući odbirci (tj. pikseli slike) iz svake komponente doprinose nastanku CS šuma. Šum koji potiče iz svake komponente je proporcionalan kvadratu amplitude te komponente, prateći zakonitost (\ref{approx}), sa $A_l, ~l=1,2,\dots,K$. 
Dakle, rezultujuća srednja vrijednost 2D-DCT koeficijenata multikomponentnog signala (slike) može biti zapisana u sljedećem obliku:
\begin{equation}
\mu_{X^C(p,q)}=\frac{N_A}{MN}\sum_{l=1}^{K}A_{l}\delta(p-p_{l},q-q_l).
\end{equation}

Prosječna varijansa koeficijenata na pozicijama koje ne odgovaraju komponentama signala je definisana sljedećim izrazom
\begin{align}
\bar{\sigma}_{X^C_0}^{2}&=\sum_{l=1}^{K}A^2_l\frac{N_A(MN-N_A)}{M^2N^2(MN-1)}.\label{v4}
\end{align}

I ovdje je važno napomenuti da komponente slike pomnožene sa baznim funkcijama transformacije mogu izazvati pojavu efekta uparivanja (\textit{coupling effect}) kao i u jednodimenzionom slučaju, ukoliko su iste locirane na pozicijama koje zadovoljavaju određena svojstva. Kao posljedica toga, ovaj efekat može uzrokovati povećanje prethodno izvedene varijanse na tim pozicijama. Ipak, u slučaju pojave ovog efekta, na primjer, na poziciji $(p_1,q_1)$, i povećanja varijanse na toj poziciji, istovremeno će varijansa na poziciji $(M-p_1,N-q_1)$ biti smanjena za istu vrijednost. Dakle, u daljim razmatranjima se ovaj efekat može zanemariti, budući da izraz (\ref{v4}) važi kao srednja vrijednost.

\subsection{Energija greške u rekonstrukciji signala koji nijesu rijetki}
\label{TheoremSec}

Na ovom mjestu će biti ekspicitno formulisan izraz za očekivanu kvadratnu grešku u rekonstrukciji dvodimenzionih signala (digitalnih slika), koji nijesu rijetki, a rekonstruisani su pod pretpostavkom da su rijetki.

\paragraph{Teorema o energiji greške.} Posmatra se slika koja nije rijetka u 2D-DCT domenu, i čije su najveće amplitude $A_{l}$,
$l=1,2,\dots,K$. Prepostavimo da je dostupno samo $N_A$
od ukupno $MN$ odbiraka (piksela), gdje je $1\ll N_A< MN$. Takođe, pretpostavimo da je slika rekonstruisana kao da je rijetka, sa stepenom rijetkosti $K$. Energija greške u $K$ rekonstruisanih koeficijenata $\left\Vert \mathbf{X}^C_{K}
{-}\mathbf{X}^C_{R}\right\Vert _{2}^{2}$ vezana je za energiju nerekonstruisanih komponenti
 $\left\Vert \mathbf{X}^C_{K0}-\mathbf{X}^C\right\Vert _{2}^{2}$ sljedećom relacijom:
\begin{equation}
\left\Vert \mathbf{X}^C_{K}-\mathbf{X}^C_{R}\right\Vert _{2}^{2} =\frac {K(MN-N_{A})}{N_A(MN-1)}\left\Vert \mathbf{X}^C_{K0}-\mathbf{X}^C\right\Vert_{2}^{2},
\end{equation}
gdje je
\begin{gather}
\left\Vert \mathbf{X}^C_{K}-\mathbf{X}^C_{R}\right\Vert _{2}^{2}=\frac{K(MN-N_{A})}{N_{A}(MN-1)} \sum_{l=K+1}^{MN}A_{l}^{2},\label{vvvar_noise}
~~
\left\Vert \mathbf{X}^C_{K0}-\mathbf{X}^C\right\Vert _{2}^{2}=\sum_{l=K+1}^{MN}
A_{l} ^{2}.
\end{gather}

\paragraph{Dokaz.}
Smatra se da je slika (signal) rekonstruisana pod pretpostavkom da je rijetka, sa stepenom rijetkosti $K$, i da je zadovoljen uslov jedinstvenosti rekonstrukcije u CS teoriji. Broj rekonstruisanih komponenti je $K$. Prema (\ref{v4}), svaka nerekonstruisana komponenta se ponaša kao aditivni Gausov šum srednje vrijednosti nula i varijanse 
$
\sigma_{N}^{2}=A_{l}^{2}\frac{N_A(MN-N_A)}{M^2N^2(MN-1)}. \label{var_noise}
$
Stoga, $MN-K$ nerekonstruisanih komponenti se ponašaju kao šum sa varijansom
\begin{equation}
\sigma_{T}^{2}=\sum_{l=K+1}^{MN}A_{l}^{2}\frac{N_{A}(MN-N_{A})}{M^{2}N^{2}(MN-1)}.
\label{var_noise2}%
\end{equation}

Nakon rekonstrukcije, ukupna energija šuma koji potiče od nerekonstruisanih komponenti 
\begin{align}
\left\Vert \mathbf{X}^C_{K}-\mathbf{X}^C_{R}\right\Vert _{2}^{2} =K\frac{M^2N^2}{N_A^2} \sigma^{2}_{T}=\frac{K(MN-N_{A})}{N_{A}(MN-1)} \sum_{l=K+1}^{MN}A_{l}^{2}\label{vvvar_noise2},
\end{align}
sadržana je u ukupno je $K$ rekonstruisanih komponenti.
Šum u nerekonstruisanim komponentama se može lako povezati sa energijom nerekonstruisanih komponenti:
\begin{align}
\left\Vert \mathbf{X}^C_{K}-\mathbf{X}^C\right\Vert _{2}^{2}=\sum_{l=K+1}^{MN}
A_{l} ^{2}.
\end{align}
Zaključujemo da je ukupna greška u rekonstruisanim komponentama data relacijom
\begin{equation}
\left\Vert \mathbf{X}^C_{K}-\mathbf{X}^C_{R}\right\Vert _{2}^{2} =\frac{K(MN-N_A)}{N_{A}(MN-1)}\left\Vert \mathbf{X}^C_{K}-\mathbf{X}^C\right\Vert_{2}^{2}. \label{TTEE}
\end{equation}
 
%The result can easily be generalized to the noisy
%signal case. For the case when the input signal has some input (additive) noise with values bellow the level of the
%reconstructed component values in the transformation domain.
%\begin{equation}
%\left\Vert \mathbf{C}_{K}\mathbf{-C}_{R}\right\Vert _{2}^{2} =\frac
%{K(MN-N_A)}{N_A(MN-1)}\left\Vert \mathbf{C}_{K}\mathbf{-C}\right\Vert
%_{2}^{2}+\frac{K}{N_{A}} \sigma^{2}_{\varepsilon}MN.
%\label{nonsparse}
%\end{equation}

\subsection{Numerički rezultati}
\label{ResSec}

Teorijski izraz (\ref{TTEE}) će biti numerički evaluiran, korišćenjem odgovarajućeg seta testnih slika. Na osnovu njih, kvadratne greške se numerički računaju za različite stepene rijetkosti $K$, po blokovima. Pretpostavljena veličina bloka je $B\times B$. Kvadratna greška po bloku se računa po formuli:
\begin{equation}
E_{stat}=10\log{\Big(||\mathbf{X}^C_K-\mathbf{X}^C_R||_2^2\Big)},
\label{e_stat}
\end{equation}
za dobijanje numeričkih rezultata, dok se teorijska greška dobija primjenom desne strane izraza (\ref{TTEE}), odnosno, po formuli:
\begin{equation}
E_{teor}=10\log{\Bigg(K\frac{B^2 - N_A}{N_A(B^2-1)}||\mathbf{X}^C_K-\mathbf{X}^C||_2^2\Bigg)}.
\label{e_theor}
\end{equation}

Za svaki pojedinačni blok posmatrane slike, greške se numerički računaju primjenom ovih izraza, a zatim se rezultati usrednjavaju po blokovima.
  Još jedna mjera kvaliteta rekonstrukcije je i tzv. \textit{peak signal-to-noise ratio} (PSNR). U posmatranom eksperimentu, ova veličina se numerički računa po formuli
\begin{equation}
PSNR_{stat}=10\log{\Bigg(\frac{255^2}{||\mathbf{X}^C_K-\mathbf{X}^C_R||_2^2}\Bigg)},
\label{psnr_stat}
\end{equation}
dok je njena teorijska vrijednost definisana izrazom
\begin{equation}
PSNR_{teor}=10\log{\Bigg(\frac{255^2}{K\frac{B^2 - N_A}{N_A(B^2-1)}||\mathbf{X}^C_K-\mathbf{X}^C||_2^2}\Bigg)}.
\label{psnr_teor}
\end{equation}

\begin{figure}[tbp]
	\centering{\includegraphics[]{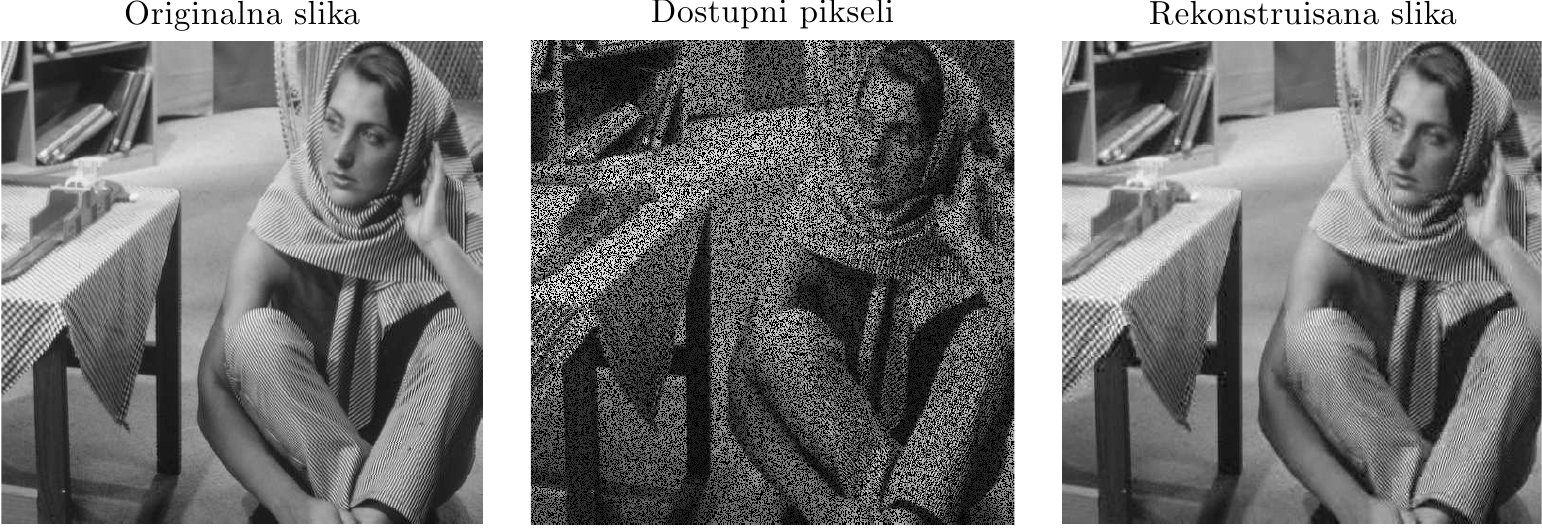}}
	\caption[Rekonstrukcija slike ,,Barbara'' na osnovu $60\%$ dostupnih piksela, i sa pretpostavljenim stepenom rijetkosti $K=16$ po svakom bloku.]{Rekonstrukcija slike ,,Barbara'' na osnovu $60\%$ dostupnih piksela, i sa pretpostavljenim stepenom rijetkosti $K=16$ po svakom bloku dimenzija $16\times16$: Originalna slika (lijevo); dostupni pikseli (u sredini); rekonstruisana slika (desno)}
	\label{imageRecon}
\end{figure}

 Proračun se takođe vrši po svakom bloku, gdje broj $255$ predstavlja maksimalnu vrijednost piksela u slici. Ova veličina će biti korišćena za dodatnu validaciju rezultata. U svim razmatranim slučajevima, rekonstrukcija je obavljena standardnim OMP algoritmom.
 \begin{figure}[tbp]
 	\centering{\includegraphics[]{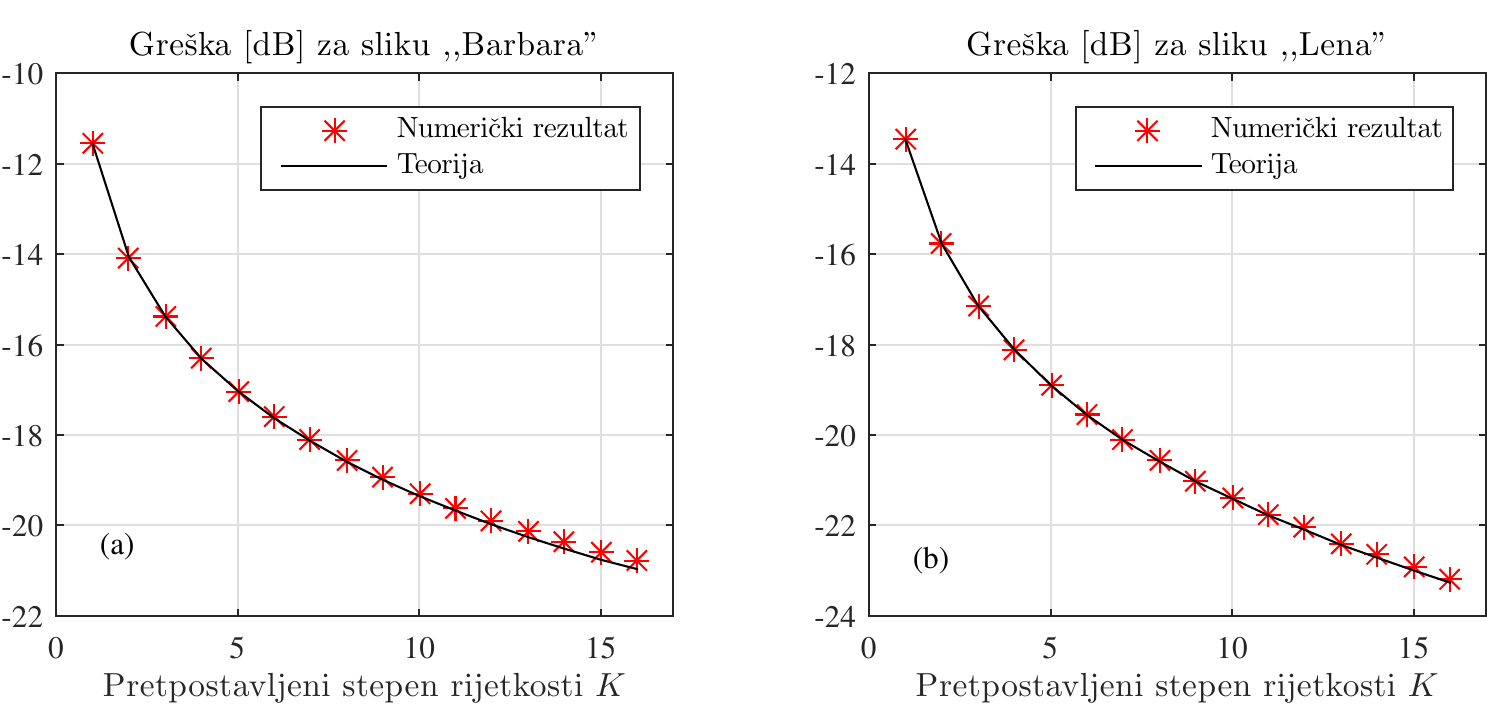}}
 	\caption[Greške nastale usljed nerekonstruisanih komponenti, za različite stepene rijetkosti po bloku u slici]{Greške nastale usljed nerekonstruisanih komponenti, za različite stepene rijetkosti po bloku u slici. Zvjezdicama je označen numerički a punim linijama teorijski rezultat za (a) sliku ,,Barbara'' i (b) ,,Lena''.}
 	\label{errorBarbara}
 \end{figure}
 
\begin{figure}[tbp]
	\centering{\includegraphics[]{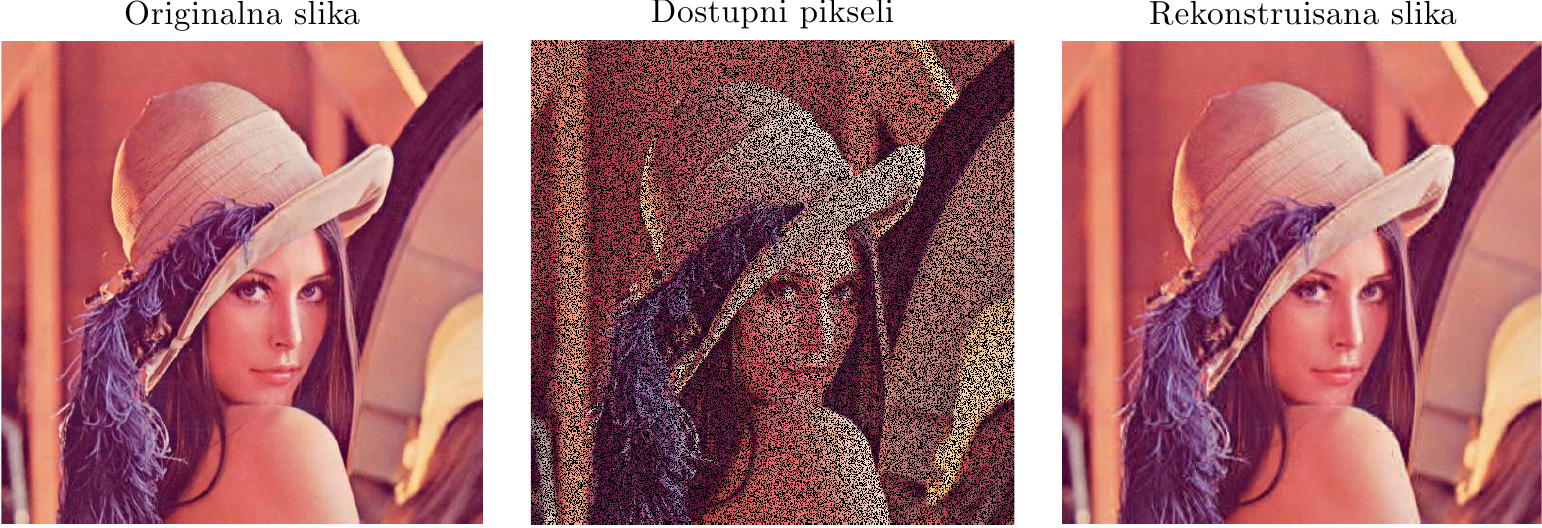}}
	\caption[Rekonstrukcija kolorne slike ,,Lena'' na osnovu $60\%$ dostupnih piksela, i sa pretpostavljenim stepenom rijetkosti $K=16$ po svakom bloku.]{Rekonstrukcija kolorne slike ,,Lena'' na osnovu $60\%$ dostupnih piksela, i sa pretpostavljenim stepenom rijetkosti $K=16$ po svakom bloku dimenzija $16\times16$: Originalna slika (lijevo); dostupni pikseli (u sredini); rekonstruisana slika (desno)}
	\label{imageRecon_Lena}
\end{figure}

\begin{primjer}
Razmatra se sivoskalirana slika ,,Barbara'', veličine $512\times 512$,  standardno dostupna u softverskom paketu MATLAB\textsuperscript{\textregistered}. Slika je prvo podijeljena na blokove dimenzija $B\times B=16\times 16$. Slika je kompresivno odabrana, tako da je dostupno samo $60\%$ njenih piksela. Prilikom rekonstrukcije, pretpostavljeni stepen rijetkosti je $K=16$ po svakom bloku. Originalna ,,Barbara'' je prikazana na slici \ref{imageRecon} (lijevo), dostupni pikseli su prikazani na slici \ref{imageRecon} u sredini (nedostupni pikseli prikazani crnom bojom), dok je rezultat rekonstrukcije na osnovu redukovanog broja piksela i uz pretpostavljeni stepen rijetkosti prikazana na slici  \ref{imageRecon} (desno). 

Numerički dobijena i teorijska energija greške prikazane su na slici \ref{errorBarbara} (a). Prilikom računanja greške, rekonstrukcija je obavljena uz različite pretpostavljene stepene rijetkosti $K$ po bloku, u opsegu od $1$ do $16$. Crvenim zvjezdicama označene su numerički dobijene vrijednosti, dok su teorijski rezultati prikazani crnom linijom.
\end{primjer}

%   \begin{figure}[h]
%   \centering{\includegraphics[]{barbaraGRAY_psnr}}
%  \caption{PSNR with various sparsities per block in image \textquotedblleft Barbara\textquotedblright; stars - statistics, line - theory}
% \label{psnrBarbara}
%\end{figure}

\begin{figure}[tbp]
	\fbox{ \centering{\includegraphics[]{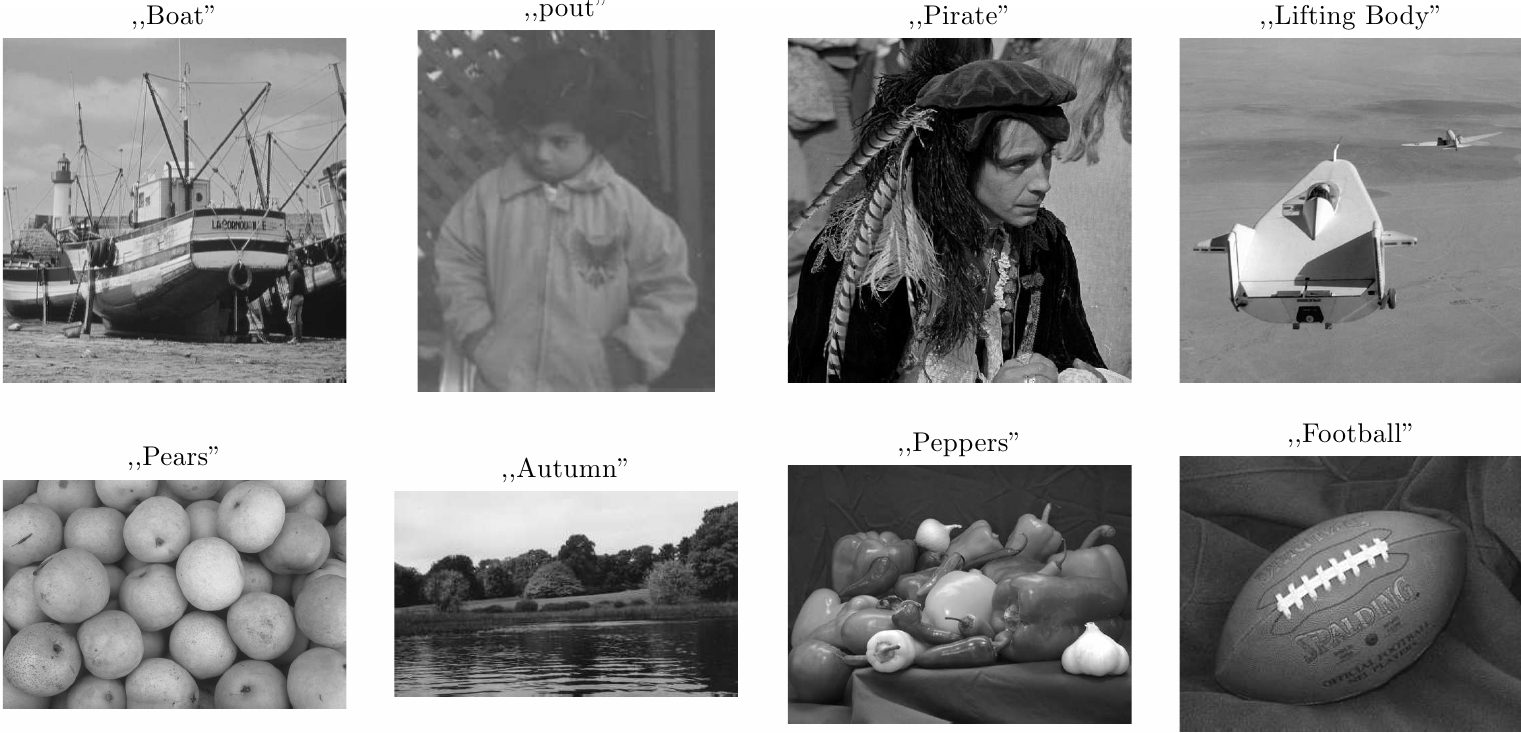}} }
	\caption{Set testnih slika koje se razmatraju u primjeru \ref{slike_ex}.}
	\label{sve_slike}
\end{figure}

\begin{primjer} U ovom primjeru se razmatra standardna kolor (RGB) slika ,,Lena'' dimenzija $512\times 512$. I ova slika je podijeljena na blokove veličine $B\times B=16\times 16$. U eksperimentu se posmatra samo $60\%$ dostupnih piksela slike. Pretpostavljeni stepen rijetkosti je $K=16$ po svakom bloku. Razmatrana ,,Lena''  je prikazana na slici \ref{imageRecon_Lena} (lijevo), dostupni pikseli su prikazani na slici \ref{imageRecon_Lena} u sredini (nedostupni pikseli su obojeni crnom bojom). Rezultat rekonstrukcije na osnovu redukovanog broja piskela i uz pretpostavljeni stepen rijetkosti je prikazan  na slici \ref{imageRecon_Lena}. Može se uočiti da zadati stepen rijetkosti obezbjeđuje visok kvalitet rekonstruisane slike.

Numerički dobijena energija greške, kao i teorijske krive greške, prikazani su na slici \ref{errorBarbara} (b). Prikazane greške su dobijene usrednjavanjem rezultata iz svih blokova i svih kanala. Rekonstrukcije su obavljene za različite pretpostavljene stepene rijetkosti, u opsegu od $1$ do $16$. Crvene zvjezdice prikazuju numerički dobijene rezultate, dok je teorijska kriva greške prikazana punom linijom. I u slučaju   slike u boji, postoji visok stepen poklapanja ovih rezultata.

%       \begin{figure}[h]
%        \centering{\includegraphics[]{psnr_stat_LENACOLOUR_all}}
%        \caption{Error caused by the unreconstructed components with various sparsities per block in image \textquotedblleft Lena\textquotedblright}
%        \label{psnr_Lena}
%\end{figure}

\end{primjer}

\begin{primjer} Validnost teorijskog izraza za energiju greške testirana je i većim setom od 8 standarnih sivoskaliranih slika iz softverskog paketa MATLAB\textsuperscript{\textregistered}, koje su prikazane na slici \ref{sve_slike}. Svaka testna slika je u eksperimentu podijeljena na blokove veličine $B\times B=16\times 16$. Rekonstrukcija je sprovedena na osnovu $60\%$  slučajno raspoređenih dostpunih piksela slika. Prepostavljeni stepen rijetkosti, korišćen u rekonstrukcionom algoritmu je $K=16$. Statistička i teorijska greška su računate prema izrazima (\ref{e_stat}) i (\ref{e_theor}). U eksperimentu je, u cilju potpunije evaluacije, računat i PSNR, na osnovu izraza (\ref{psnr_stat}) i (\ref{psnr_teor}). Rezultati eksperimenta, prikazani u tabeli \ref{Tab:5slika}, potvrđuju visoko poklapanje numeričkih rezultata i teorijskih izraza.
	\label{slike_ex}
\end{primjer}

\begin{table} 
	\small
	\centering
	\caption{Energija greške i PSNR za 8 razmatranih slika u primjeru \ref{slike_ex}.}
	\label{Tab:5slika}
	\begin{tabular}{lcccccc}
		\toprule
		& \multicolumn{2}{c}{Error} & \multicolumn{2}{c}{PSNR} \\
		\cmidrule(lr){2-3}\cmidrule(lr){4-5}
		\textbf{Testna slika}     & \textbf{Statistika} & \textbf{Teorija} & \textbf{Statistika} & \textbf{Teorija}  \\ \midrule
		,,Boat''    &  -19.13   &   -19.20   & 81.97 &  82.13   \\
		,,Pout'' &  -27.32   &   -27.38   & 80.35 &  80.42   \\
		,,Pirate''    &  -10.10   &   -10.23   & 70.97 &  71.10  \\
		,,Lifting Body''   &  -24.78   &   -24.86   & 82.97 &  83.11  \\
		,,Pears''    &  -25.60   &   -25.67   & 78.77 &  78.86  \\
		,,Autumn''     &  -15.80   &   -15.88   & 90.81 &  90.92   \\
		,,Peppers''    &  -22.16   &  -22.22   & 79.16 &  79.23  \\ 
		,,Football''    &  -18.72  & -18.87 &  68.69  & 68.83\\  \bottomrule
	\end{tabular}
\end{table}

\chapter{Vremensko-frekvencijska analiza i rekonstrukcije na bazi mjera koncentracije}

Furijeova transformacija obezbjeđuje jedinstveno mapiranje signala iz vremenskog u frekvencijski domen. Frekvencijski domen pruža informacije o spektralnom sadržaju signala. Budući da se informacije sadržane u faznoj karakteristici ne mogu jednostavno koristiti, Furijeova transformacija gotovo da nije upotrebljiva u analizi i obradi signala sa vremenskim varijacijama spektralnih komponenti \cite{tfsa}. U cilju praćenja i karakterizacije vremenskih distribucija spektralnih komponenti -- uvodi se vremensko-frekvencijska analiza, \cite{tfsa,tfsa_cs1,tfsa_cs2,sire7,sire8,brajovic_tf1,sire10,sire11,sire12,sire13,sire14,sire15,sire16,sire17,sire18,sire19,sire20,sire21,sire22,sire23,sire24,sire25,s-metod,brajovic_tf0,brajovic_tf2,brajovic_tf3,brajovic_tf33,brajovic_tf4,brajovic_tf5,brajovic_tf55,brajovic_tf6,brajovic_tf7,aco1,aco2,brajovic_tf8,brajovic_tf9,brajovic_tf99,brajovic_tf10,brajovic_tf11,brajovic_tf12,brajovic_tf13,brajovic_tf14,brajovic_tf15,glavni,martorella,IT2018,chen1,chen2,ref3,ref4,wang,ref13,csiet1,csiet2,Sucic2,dekompozicija,dekompozicija2,dekompozicijaLFM,dekompozicija3,dekompozicija4,VKLJS1,LJSMDTTB,mandic,TSPMV,bivariate,BB,dekompozicija_radon,dekompozicija_cirplet,iradon1,iradon2,boashmv,XXX}. 

Nakon kratkog osvrta na osnovne koncepte vremensko-frekvencijske analize i definisanja vremensko-frekvencijskih reprezentacija koje su relevantne za dalje izlaganje, poseban akcenat u ovoj glavi je postavljen na problematiku estimacije trenutne frekvencije u uslovima izraženo jakog aditivnog šuma, koja je zasnovana na Wigner-ovoj distribuciji. Predstavljen je originalni algoritam baziran na principima Optimizacije kolonije mrava (engl. \textit{Ant Colony Optimization}, ACO), heurističkog distribuiranog optimizacionog pristupa inspirisanog procesima u prirodi \cite{aco1,aco2}. Koncepti vremensko-frekvencijske analize i kompresivnog odabiranja povezani su kroz razmatranje problematike rekonstrukcije krutog tijela (engl. \textit{rigid body}, RB) dijela ISAR signala radarske mete sa nekompenzovanim ubrzanjem \cite{brajovic_tf3,brajovic_tf6}. Izlaganje se zatim nastavlja sa predstavljanjem originalnog algoritma za dekompoziciju multivarijantnih multikomponentnih signala, koja je zasnovana na mjerama koncentracije vremensko-frekvencijskih reprezentacija \cite{brajovic_tf1,brajovic_tf5,brajovic_tf55}. Pokazano je da multivarijantni signali, koji predstavljaju signale čija se akvizicija vrši pomoću većeg broja senzora, posjeduju karakteristike koje omogućavaju izdvajanje i potpunu rekonstrukciju pojedinačnih komponenti multikomponentnog signala, uprkos činjenici da se one preklapaju u vremensko-frekvencijskoj ravni. Navedeni problem u slučaju univarijantnih (jednovarijantnih) signala nije bilo moguće riješiti u opštem slučaju, odnosno za signale čije su komponente nestacionarne tako da njihova trenutna frekvencija posjeduje zavisnost od vremena koja je proizvoljnog oblika. 

\section{Kratak osvrt na osnovne reprezentacije i koncepte}
Tokom višedecenijskog razvoja vremensko-frekvencijske analize, predložene su mnogobrojne reprezentacije: linearne, kvadratne (distribucije), kao i reprezentacije višeg reda. Međutim, nijedna reprezentacija nije idealna, niti je univerzalno primjenljiva: ograničene su na određene klase signala, a primjenljivost najčešće otvara pitanje pravljenja kompromisa između odgovarajućih prednosti i nedostataka. Mnoge reprezentacije su uvedene u cilju rješavanja konkretnih problema pod konkretnim uslovima, a vrlo često su vezane za karakterizaciju signala kroz estimaciju trenutne frekvencije \cite{tfsa}. 
\subsection{Kratkotrajna Furijeova transformacija i spektrogram}
Jedna od osnovnih vremensko-frekvencijskih reprezentacija je kratkotrajna Furijeova transformacija (engl. \textit{short-time Fourier transform} - STFT). Ona se, za signal $x(t)$, definiše sljedećim izrazom \cite{dos,multimedia,tfsa,brajovic_tf0}:
\begin{equation}
STFT(t,\Omega)=\int_{-\infty}^{\infty}{x(t+\tau)w(\tau){e^{-j\Omega
			\tau}}d\tau}, \label{stft21}%
\end{equation}
gdje je $w(\tau)$ funkcija prozora, koja je najčešće parna funkcija sa maksimumom u $\tau=0$. Za svaki posmatrani trenutak $t$, za dio signala koji je lokalizovan prozorom, centriranim oko tog trenutka, računa se odgovarajuća Furijeova transformacija. Ponavljanjem postupka za svako $t$ dobija se vremensko-frekvencijska reprezentacija signala. STFT je linearna, što znači da za linearnu kombinaciju
\begin{align}
x(t)=\sum\limits_{p=1}^{P}{{a_{p}}{x_{p}}(t)}
\end{align}
važi da je rezultujuća kratkotrajna Furijeova transformacija linearna kombinacija odgovarajućih kratkotrajnih Furijeovih transformacija pojedinačnih komponenti signala:
\begin{align}
STF{{T}_{x}}(t,\Omega)  &  =\sum\limits_{p=1}^{P}STFT{x_p}(t,\Omega).
\end{align}

Originalni signal $x(t)$ se može rekonstruisati na osnovu STFT korišćenjem sljedeće inverzne formule:
\begin{equation}
x(t+\tau)=\frac{1}{2\pi w(\tau)}\int_{-\infty}^{\infty}STFT(t,\Omega)e^{j\Omega \tau}d\Omega.
\end{equation}
\subsubsection{Prozori i problem rezolucije}
Za posmatrani trenutak $t$, funkcija prozora određuje veličinu dijela signala koji je relevantan pri proračunu STFT za taj trenutak. Bolja vremenska rezolucija se postiže odabirom prozora kraćeg trajanja. Ako se STFT redefiniše u obliku:
\begin{equation}
STFT(t,\Omega)=\frac{e^{-j\Omega t}}{2\pi}\int_{-\infty}^{\infty
}X(\Theta)W^{\ast}(\Theta-\Omega)e^{j\Theta t}d\Theta,
\end{equation}
tada ona, mimo člana $e^{-j\Omega t}$, predstavlja inverznu Furijeovu transformaciju od $X(\Theta)W^{\ast}(\Theta-\Omega)$, koji se može tumačiti kao
spektar signala $x(t)$ koji je ograničen prolaskom kroz filtar propusnik
opsega učestanosti $W^{\ast}(\Theta-\Omega)$, sa centralnom učestanošću
$\Omega$, pri čemu $W(\Omega)$ označava FT funkcije prozora $w(\tau).$ Bolja frekvencijska rezolucija se postiže prozorom veće širine (trajanja) u vremenu, kojoj je obrnuto proporcionalna odgovarajuća širina u frekvencijskom domenu. Jedna od standardnih prozorskih funkcija je pravougaoni prozor:
\begin{equation}
w(\tau)=\left\{
\begin{matrix}
1,{\text{ za }}\left\vert \tau\right\vert <T\hfill\\
0,{\text{ ostalo }}\tau.
\end{matrix}
\right.
\end{equation}
FT ove prozorske funkcije je data sljedećim izrazom:
\begin{align}
{W_{R}}(\Omega) =\frac
	{{2\sin(\Omega T)}}{\Omega}.\label{ftwin}
\end{align}

Može se uočiti da ova prozorska funkcija, koja je ograničena na vremenskom intervalu $(-T,T)$, ima beskonačan frekvencijski opseg. Ako se posmatra samo glavna latica u (\ref{ftwin}), jasno je da manje $T$ znači veću širinu glavne latice. U vremensko-frekvencijskoj analizi se koriste i druge prozorske funkcije. Takav je, na primjer, Hanov prozor \cite{tfsa,brajovic_tf0}:%
\begin{equation}
w(\tau)=\left\{
\begin{array}{ll}
\frac{1}{2}{{(1 + \cos(}}\tau\pi{{/T))}},&\left\vert \tau
\right\vert <T\hfill\\
0,&{\text{ostalo }}\tau
\end{array}
\right.
\end{equation}
čija je FT data izrazom:
\begin{align}
W_H(\Omega) =\frac{1}{2}{{W}_{R}}(\Omega)+\frac{1}{4}{{W}_{R}}(\Omega-\pi/T)+\frac
{1}{4}{{W}_{R}}(\Omega+\pi/T), \label{han}%
\end{align}
koji jasno otkriva vezu sa FT pravougaonog prozora. Navedena veza je bitna u kontekstu realizacija STFT, budući da se STFT sa Hanovim prozorom može realizovati pomoću odgovarajućih STFT sa pravougaonim prozorom, korišćenjem relacije:
\begin{align*}
STFT(t,\Omega)={0.5}STF{T_{R}}(t,\Omega)+{0.25}STF{T_{R}}(t,\Omega
-\pi/T)+{0.25}STF{T_{R}}(t,\Omega+\pi/T),
\end{align*}
a slična veza se može uspostaviti i u slučaju drugih vrsta prozorskih funkcija.

\subsubsection{Efektivno trajanje i princip neodređenosti}
Sposobnost vremensko-frekvencijskih reprezentacija da razdvajaju komponente signala, odnosno, rezolucija u vremensko-frekvencijskoj ravni, zavisi od širine prozorske funkcije. Proizvod vremenskog i frekvencijskog trajanja (širine) je konstantan, što povlači činjenicu da bolja vremenska rezolucija znači goru frekvencijsku, i obratno. U opštem slučaju, mogu se uvesti mjere efektivnog trajanja u vremenu i frekvenciji \cite{dos,tfsa,brajovic_tf0}:
\begin{equation}
\sigma_{t}^{2}=\frac{{\int_{-\infty}^{\infty}{{\tau^{2}}{{\left\vert
					{w(\tau)}\right\vert }^{2}}d\tau}}}{{\int_{-\infty}^{\infty
		}{{{\left\vert {w(\tau)}\right\vert }^{2}}d\tau}}},~\sigma_{\Omega}^{2}=\frac
{{\int_{-\infty}^{\infty}{{\Omega^{2}}{{\left\vert {W(\Omega
						)}\right\vert }^{2}}d\Omega}}}{{\int_{-\infty}^{\infty}{{{\left\vert
					{W(\Omega)}\right\vert }^{2}}d\Omega}}}%
\end{equation}
Za bilo koji signal $w(\tau)$ za koji važi da $w(\tau)\sqrt{|\tau|
}\rightarrow0$ za $\tau\rightarrow\pm\infty$, može se pokazati da proizvod mjera efektivnog trajanja zadovoljava princip neodređenosti u obradi signala, koji glasi:
\begin{equation}
\sigma_{T}{\sigma}_{W}\geqslant1/2.
\end{equation}

\subsubsection{Spektrogram}
Energetska verzija STFT, definisana kao kvadrat modula ove transformacije, poznata je pod nazivom spektrogram \cite{tfsa,brajovic_tf0}:
\begin{equation}
SPEC(t,\Omega)={\left\vert {STFT(t,\Omega)}\right\vert ^{2}}.%
\end{equation}
U opštem slučaju, spektrogram nije linearan. Za multikomponentni signal oblika
\begin{equation}
x(t)=\sum\limits_{p=1}^{P}{{x_{p}}}(t),
\end{equation}
spektrogram je dat u obliku:
\begin{align}
SPE{{C}_{x}}(t,\Omega)  =\sum\limits_{p=1}^{P}{SPE{{C}_{{{x}_{p}}}}(t,\Omega)+}\sum\limits_{p=1}%
^{P}{\sum\limits_{\genfrac{}{}{0pt}{}{\scriptstyle k=1\hfill}{\scriptstyle p\neq q\hfill}%
		%EndExpansion
	}^{P}{STF{{T}_{{{x}_{p}}}}(t,\Omega)STFT_{{{x}_{q}}}^{\ast}(t,\Omega),}}
\label{kros}%
\end{align}
gdje desni član predstavlja nepoželjne kros-komponente, nastale usljed interakcija pojedinačnih komponenti, koje su izazvane nelinearnom formom reprezentacije (distribucije).
\subsubsection{Diskretna forma STFT}
Numeričko računanje STFT na računaru podrazumijeva upotrebu diskretne forme ove reprezentacije. Označimo sa $x_a(t)$ analogni signal čija se diskretizacija po vremenu i kašnjenju obavlja sa korakom $\Delta t$, tako da je $t=n\Delta t$ i $\tau=m\Delta t$. U tom slučaju će važiti:
\begin{equation}
STFT(t,\Omega)=\int_{-\infty}^{\infty}{{x_{a}}(t+\tau)w(\tau
	){e^{-j\Omega\tau}}d\tau}\approx\sum\limits_{m=-\infty}^{\infty}{{x_{a}%
	}(n\Delta t+m\Delta t)w(m\Delta t){e^{-j\Omega m\Delta t}}\Delta t}%
\notag\end{equation}
Uvođenjem oznake $
{x_{a}}((n+m)\Delta t)\Delta t={x}(n+m)
$ i normalizacijom ugaone frekvencije  $\omega=\Omega\Delta t$, dobija se STFT diskretnog signala,  koja je periodična po frekvenciji, sa
osnovnom periodom $2\pi$:%
\begin{equation}
STFT(n,\omega)=\sum\limits_{m=-\infty}^{\infty}{{x_{a}}(n+m)w(m){e^{-\omega
			m}.}}%
\end{equation}
Da bi se spriječilo preklapanje STFT perioda, shodno teoremi o odabiranju,
treba da bude zadovoljeno da je perioda odabiranja: $\Delta t=\pi/{\Omega_{0}%
}\leqslant\pi/{\Omega}_{m}$, gdje je ${\Omega_{m}}$ maksimalna frekvencija u
spektru signala ${x_{a}}(n+m)w(m)$. I pored činjenice da funkcija prozora ima beskonačan
spektar, može se pretpostaviti da se iznad neke granične frekvencije
${\Omega_{m}}$ mogu zanemariti spektralne komponente signala ${x_{a}%
}(n+m)w(m)$. Zbog upotrebe prozora konačnog trajanja, diskretni signal je konačne dužine $N=T/\Delta t$. Odgovarajuća DFT u $N$ tačaka za posmatrani indeks $n$ i uz $\omega=\frac{2\pi}{N}k$ vodi do diskretne forme STFT:
\begin{equation}
STFT(n,k)=\sum\limits_{m=-N/2}^{N/2-1}{x(n+m)w(m){e^{-j\frac{{2\pi}}{N}mk},}}
\label{stftd}%
\end{equation}
gdje se, u numeričkoj realizaciji, mogu koristiti postojeće rutine za računanje DFT (FFT). Izraz (\ref{stftd}) je moguće računati i rekurzivnim putem, u cilju povećanja efikasnosti numeričkog proračuna (tzv. numeričke efikasnosti) ove transformacije \cite{brajovic_tf0}.
\subsubsection{Uticaj viših izvoda faze na koncentraciju STFT}
Posmatra se opšti slučaj analitičkog signala:
\begin{equation}
x(t)=A{e^{j\phi(t)},}%
\end{equation}
gdje je $\phi(t)$ diferencijabilna funkcija. STFT ovog signala je \cite{tfsa}:
\begin{align}
STFT(t,\Omega)&=\int_{-\infty}^{\infty}{A{e^{j\phi(t+\tau)}}%
	w(\tau){e^{-j\Omega\tau}}d\tau}=\int_{-\infty}^{\infty}{A{e^{j(\phi(t)+\phi^{\prime
			}(t)\tau+\phi^{\prime\prime}(t)\frac{{{\tau^{2}}}}{{2!}}+\dots)}}w(\tau
	){e^{-j\Omega\tau}}d\tau}\notag\\
&  =A{e^{j\phi(t)}}FT\left\{  {{e^{j\phi^{\prime}(t)\tau}}}\right\}
{\ast_{\Omega}}FT\left\{  {{e^{j\sum_{k=2}^{\infty}\phi^{(k)}(t)\frac{{{\tau^{k}}}}%
			{{k!}}}}}\right\}  {\ast_{\Omega}}FT\left\{  {w(\tau)}\right\}  \notag \\
&  =2\pi A{e^{j\phi(t)}}\delta(\Omega-\phi^{\prime}(t)){%
} {\ast_{\Omega}}W(\Omega)\ast_{\Omega}e^{-j\Omega^2/(2\phi^{\prime\prime})}\sqrt{\frac{2\pi j}{\phi^{\prime\prime}(t)}},
\end{align}
gdje je faza $\phi(t+\tau)$ razvijena u Tejlorov red, u okolini tačke $t$:%
\begin{equation}
\phi(t+\tau)=\phi(t)+\phi^{\prime}(t)\tau+\phi^{\prime\prime}(t)\tfrac
{{{\tau^{2}}}}{2}+\dots+\phi^{(k)}(t)\tfrac{\tau^{k}}{k!}+\dots,
\end{equation}
zatim ${\ast_{\Omega}}$ označava konvoluciju u frekvencijskom domenu, dok je $FT\{\cdot\}$ oznaka za operator Furijeove transformacije.
Uticaj prozora se, dakle, manifestuje kroz rasipanje spektra oko idealno koncentrisanog dijela $\delta(\Omega-\phi^{\prime}(t))$. Viši izvodi faze izazivaju pojavu dodatnog rasipanja. To je rasipanje spektra usljed nestacionarnosti signala. U krajnjem obrascu, uzet je u obzir samo drugi izvod faze, koji najviše i utiče na koncentraciju.

\subsection{Metod stacionarne faze i pojam trenutne frekvencije}
Neka se posmatra opšta forma analitičkog signala:%
\begin{equation}
x(t)=A(t){e^{j\phi(t)}}.
\end{equation}

Kada FT nije moguće jednostavno odrediti analitički, tada se, pod određenim uslovima, može koristiti metod stacionarne
faze u cilju računanja njene aproksimacije. Neka je prozorska funkcija širine $2T$. Ako možemo pretpostaviti da su u okviru prozora $w(\tau)$ varijacije amplitude male a varijacije faze gotovo linearne, odnosno, $A(t)\simeq A(t)$, $\phi(t+\tau)\simeq \phi(t)+ \phi'(t) $, tada je 
\begin{equation}
x(t+\tau)\simeq A(t){e^{j\phi(t)}}{e^{j\phi'(t)}}.
\end{equation}

Signal se u datom trenutku ponaša kao sinusoida sa amplitudom $A(t)$, fazom $\phi(t)$ i frekvencijom $\phi'(t)$. Dakle, prvi izvod faze, $\phi'(t)$, u datom trenutku $t$, unutar posmatranog intervala u njegovoj okolini, ima ulogu frekvencije. Metod stacionarne faze povezuje signal u vremenskom domenu sa spektralnim sadržajem na frekvenciji $\Omega$ relacijom $\phi^{\prime}(t)=\Omega$. 
Signal u vremenskom domenu koji zadovoljava uslove metoda stacionarne faze, u posmatranom trenutku $t$ daje doprinos Furijeovoj transformaciji na odgovarajućoj frekvenciji:
\begin{equation}
\phi^{\prime}(t)=\Omega(t).
\end{equation}
Tada se uspostavlja Furijeov transformacioni par oblika:
\begin{equation}
\int_{-\infty}^{\infty}{A(t){e^{j\phi(t)}}{e^{-j\Omega t}}dt\cong
	A({t_{0}})}{e^{j\phi({t_{0}})}}{e^{-j\Omega{t_{0}}}}\sqrt{\frac{{2\pi j}%
	}{{\phi^{\prime\prime}({t_{0}})}}} \label{msf}%
\end{equation}
gdje$~{t_{0}~}$predstavlja rješenje jednačine: $\phi^{\prime}({t_{0}})=\Omega
$. Ako se posmatra integral sa lijeve strane jednačine (\ref{msf}), može se
uočiti da u slučaju kada je podintegralna funkcija stacionarna, tada $\phi(t)-\Omega t$ ima najveći uticaj
na vrijednost integrala. 
Funkcija je stacionarna ukoliko joj je prvi izvod jednak nuli. Ovo važi zato što se
vrijednosti integrala na intervalima u kojima je faza nestacionarna
usrednjavaju, pa integral nestacionarnog dijela, gledano u cjelini, teži nuli. 

Ukoliko
važi da je faza $\phi(t)-\Omega t$ stacionarna, to znači da se u datom
trenutku $t={t_{0}}$ faza $\phi(t)$ ponaša na isti način kao i $\Omega t$.
Drugim riječima, brzina promjene faze $\phi(t)$ jednaka je frekvenciji
$\Omega$ u tom trenutku: 
$
\frac{{d(\phi(t)-\Omega t)}}{{dt}}{|_{t={t_{0}}}}=0,
$
odnosno, $
\phi^{\prime}(t)=\Omega$.
Formalno govoreći, izraz 
\begin{equation}
\phi^{\prime}(t)=\Omega(t)\label{trenutnafr}%
\end{equation}
predstavlja definiciju trenutne frekvencije \cite{tfsa,brajovic_tf0}.  Estimacija trenutne frekvencije jedna je od najznačajnijih tema u vremensko-frekvencijskoj analizi. U slučaju multikomponentnih signala, razmatraju se trenutne frekvencije pojedinačnih komponenti.

\subsection{Wigner-ova distribucija}

Kvadratne vremensko-frekvencijske reprezentacije su poznate i pod nazivom distribucije. Generalno, one zadovoljavaju niz interesantnih svojstava, među kojima ćemo spomenuti energetski uslov:%
\begin{equation}
\frac{1}{{2\pi}}\int_{-\infty}^{\infty}{\int_{-\infty}^{\infty
	}{P(t,\Omega)dtd\Omega}}=\int_{-\infty}^{\infty}{{{\left\vert
			{x(t)}\right\vert }^{2}}}dt={E_{x},}%
\end{equation}
vremenski marginalni uslov:%
\begin{equation}
\frac{1}{{2\pi}}\int_{-\infty}^{\infty}{P(t,\Omega)d\Omega={{\left\vert
			{x(t)}\right\vert }^{2},}}%
\end{equation}
i frekvencijski marginalni uslov:%
\begin{equation}
\int_{-\infty}^{\infty}{P(t,\Omega)dt={{\left\vert {X(\Omega
				)}\right\vert }^{2},}}%
\end{equation}
gdje $P(t,\Omega)$ označava distribuciju. Među najznačajnijim distribucijama koje zadovoljavaju ova svojstva spada i Wigner-ova distribucija (WD). Konceptualno je preuzeta iz kvantne mehanike. Za signal $x(t)$ ona se definiše izrazom \cite{dos,tfsa}:
\begin{equation}
WD(t,\Omega)=\int_{-\infty}^{\infty}{x(t+\frac{\tau}{2}){x^{\ast}%
	}(t-\frac{\tau}{2}){e^{-j\Omega\tau}}d}\tau. \label{wd}%
\end{equation}
Zbog Hermitske simetrije $R(t,\tau)={R^{\ast}}%
(t,\tau)$, ona je čisto realna vremensko-frekvencijska reprezentacija. Alternativno, data je sljedećim izrazom:
\begin{equation}
WD(t,\Omega)=\frac{1}{2\pi}\int_{-\infty}^{\infty}{X(\Omega
	+\Theta/2){{X}^{\ast}}(\Omega-\Theta/2){{e}^{j\Theta t}}}d\Theta.
\end{equation}

U slučaju linearno frekvencijski modulisanih (LFM) signala, WD produkuje idealnu koncentraciju. Na primjer, za signal $x(t)=A{e^{ja{t^{2}}/2}}$ se dobija:
\begin{equation}
WD(t,\Omega)=\int_{-\infty}^{\infty}{{A^{2}}{e^{jat\tau}}%
	{e^{-j\Omega\tau}}d}\tau=\int_{-\infty}^{\infty}{{A^{2}}{e^{-j(\Omega
			-at)\tau}}d}\tau=2\pi{A^{2}}\delta(\Omega-at)
\end{equation}
što predstavlja idealnu koncentraciju energije  signala duž trenutne
frekvencije $\Omega=\phi^{\prime}(t)$. Wigner-ova distribucija posjeduje niz svojstava po kojima se značajno razlikuje od kratkotrajne Furijeove transformacije.

\subsubsection{Wigner-ova distribucija multikomponentnih signala}
 U opštem
slučaju multikomponentnog signala, koji predstavlja linearnu kombinaciju
\begin{equation}
x(t)=\sum\limits_{p=1}^{P}{{x_{i}}(t)}
\end{equation}
 Wigner-ova distribucija je data sljedećim izrazom:
\begin{align}
WD(t,\Omega) 
  =\sum\limits_{p=1}^{P}{\int\limits_{-\infty}^{\infty}{{{x}_{p}}%
		(t+\frac{\tau}{2})x_{p}^{\ast}(t-\frac{\tau}{2}){{e}^{-j\Omega\tau}}d}}%
\tau
  +\sum\limits_{p=1}^{P}{\sum\limits_{%
		\genfrac{}{}{0pt}{}{\scriptstyle q=1\hfill}{\scriptstyle p\neq q\hfill}%
		%EndExpansion
	}^{P}{\int\limits_{-\infty}^{\infty}{{{x}_{p}}(t+\frac{\tau}{2})x_{q}^{\ast
			}(t-\frac{\tau}{2}){{e}^{-j\Omega\tau}}d}\tau,}}\notag
\end{align}
gdje se jasno izdvajaju korisne komponente -- autočlanovi, dati relacijom:
\begin{equation}
W{D_{AT}}(t,\Omega)=\sum\limits_{p=1}^{P}{\int_{-\infty}^{\infty
	}{{x_{p}}(t+\frac{\tau}{2})x_{p}^{\ast}(t-\frac{\tau}{2}){e^{-j\Omega\tau}}d}%
}\tau.
\end{equation}
i kros-članovi, koji su neželjene komponente nastale usljed nelinearne prirode ove distribucije:
\begin{equation}
W{D_{CT}}(t,\Omega)=\sum\limits_{%
	%TCIMACRO{\QATOP{p=1\hfill}{\hfill}}%
	%BeginExpansion
	\genfrac{}{}{0pt}{}{p=1\hfill}{\hfill}%
	%EndExpansion
}^{P}\sum\limits_{%
	%TCIMACRO{\QATOP{\scriptstyle q=1\hfill}{\scriptstyle k\neq j\hfill}}%
	%BeginExpansion
	\genfrac{}{}{0pt}{}{\scriptstyle q=1\hfill}{\scriptstyle p\neq q\hfill}%
	%EndExpansion
}^{P}{{\int_{-\infty}^{\infty}{{x_{p}}(t+\frac{\tau}{2})x_{q}^{\ast
			}(t-\frac{\tau}{2}){e^{-j\Omega\tau}}d}\tau.}}%
\end{equation}
Pojava kros-komponenti je jedna od glavnih mana Wigner-ove distribucije.
\subsubsection{Uticaj viših izvoda faze na koncentraciju}
Neka se posmatra opšti slučaj analitičkog signala (nelinearnog FM signala):
\begin{equation}
x(t)=A{e^{j\phi(t)}},
\end{equation}
WD ovog signala se može zapisati u formi
\begin{equation}
WD(t,\Omega)=\int_{-\infty}^{\infty
}{{A^{2}}{e^{j\phi(t+\frac{\tau}{2})}}{e^{-j\phi(t-\frac{\tau}{2})}%
	}{e^{-j\Omega\tau}}d\tau}%
\end{equation}
Anuliranjem članova sa parnim izvodima $\phi(t)$, $\phi^{\prime\prime}(t)\frac
{{{\tau^{2}}}}{{4\cdot2!}}\dots$, dalje se dobija \cite{tfsa,brajovic_tf0}:
\begin{align}
WD(t,\Omega)  &  ={{A}^{2}}\int_{-\infty}^{\infty}{{{e}}}^{j\left(
	\phi^{\prime}(t)\tau+2{{\sum\limits_{k=1}^{\infty}{\frac{{{\phi}^{(2k+1)}}%
					(t)}{(2k+1)!}}}}\left(  {{{\frac{\tau}{2}}}}\right)  ^{\left(  2k+1\right)
	}\right)  }{{e}^{-j\Omega\tau}}d\tau\nonumber\\
&  =2\pi{{A}^{2}}\delta(\Omega-\phi^{\prime}(t)){{\ast}_{\Omega}}FT\left\{\exp\left({
2j\sum\limits_{k=1}^{\infty}{\frac{{{\phi}^{(2k+1)}}}{(2k+1)!}{{\left(\frac{\tau}%
			{2}\right)}^{2k+1}}}}\right)\right\},
\end{align}
odakle se zaključuje da rasipanje energije signala oko trenutne frekvencije izazivaju samo neparni izvodi faze, što znači da se postiže bolja koncentracija signala u odnosu na STFT.

\subsubsection{Pseudo-Wigner-ova distribucija i diskretizacija}
U praktičnim realizacijama, WD se redefiniše uključivanjem funkcije prozora u obliku:
\begin{equation}
PWD(t,\Omega)=\int_{-\infty}^{\infty}{w\left(\frac{\tau}{2}\right){w^{\ast}%
	}\left(-\frac{\tau}{2}\right)x\left(t+\frac{\tau}{2}\right){x^{\ast}}\left(t-\frac{\tau}{2}%
	\right){e^{-j\Omega\tau}}d}\tau.
\end{equation}
Ova forma je poznata pod nazivom pseudo-Wigner-ova distribucija (PWD) i ona uzima u obzir konačno trajanje signala i definisana je kroz konačna kašnjenja. U praksi se najčešće koristi diskretna forma ovako definisane WD. Postoji više pristupa diskretizaciji. U daljem izlaganju biće podrazumijevana ova forma Wigner-ove distribucije. Jedan dio pristupa diskretizaciji zahtijeva duplo veću frekvenciju odabiranja od minimalne koja je definisana teoremom o odabiranju, odnosno, preodabiranje signala sa faktorom 2. Za signal odabran sa korakom $\Delta t$, uz odgovarajuću diskretizaciju frekvencije, uzimajući u obzir pri tome adekvatnu diskretizaciju vremenskog kašnjenja u autokorelacionoj funkciji, lako se dobija jedna od poznatih formi diskretne PWD:
\begin{equation}
WD_x(n,k)=\sum\limits_{m=-M/2}^{M/2-1}{x(n+m){x^{\ast}%
	}(n-m){e^{-j\frac{{4\pi}}{M}km},}} \label{PWD}%
\end{equation}
gdje su eventualno izostavljene  konstante koje nastaju tokom diskretizacije, i podrazumijevan je jedinični prozor $w(m)=1,~-N/2\leq m \leq N/2-1$ i $w(m)=0$ za ostalo $m$, dok je $-M/2 \leq k \leq M/2-1  $.
\subsection{Estimacija trenutne frekvencije na osnovu  Wigner-ove distribucije}
Posmatra se diskretna WD oblika (\ref{PWD}) zašumljenog FM signala
\begin{equation}
x(n)=s(n)+\varepsilon(n)=Ae^{j\phi(n)}+\varepsilon(n),
\end{equation}
gdje je $\varepsilon(n)$ bijeli kompleksni Gausov šum sa varijansom $\sigma^2_{\varepsilon}=2\sigma^2$, pri čemu su realni i imaginarni dio šuma sa statistički nezavisnim Gausovim distribucijama, jednakih varijansi $\sigma^2$ i srednje vrijednosti nula. Za posmatrani trenutak $n$, trenutna frekvencija ($\omega(n)$ ili $\Omega(n\Delta t)=\omega(n)/\Delta t$), gdje je $\Delta t$  perioda odabiranja, estimira se na osnovu pozicije maksimuma Wigner-ove distribucije:
\begin{equation}
	\hat{k}=\arg\{\max_k WD_x(n,k)\}.
\end{equation}
Ovaj estimator je u praksi često korišćen matematički alat. Greške u estimaciji mogu nastati zbog više razloga: \textit{bias}, greške usljed varijacija unutar auto-članova signala, greške usljed diskretizacije frekvencije, i greške usljed prisustva šuma. U uslovima jakog šuma, dolazi do grešaka u estimaciji koje su impulsne prirode, i koje nadmašuju sve druge pobrojane oblike grešaka. Estimacija trenutne frekvencije u uslovima jakog šuma je problem koji je privukao dosta pažnje u vremensko-frekvencijskoj analizi. 

\subsubsection{Estimacija trenutne frekvencije u uslovima jakog šuma}
Ovdje će biti predstavljen jedan od pristupa za IF estimaciju u uslovima jakog aditivnog šuma. Optimizacija kolonije mrava (engl. \textit{Ant Colony Optimization}) spada u klasu distribuiranih optimizacionih pristupa i to je paradigma inspirisana biološkim procesima u prirodi. Slijedi kratak opis algoritma koji je predložen u radovima \cite{aco1,aco2}. Vremensko-frekvencijska ravan Wigner-ove distribucije se posmatra kao pravougana mreža dimenzija $N\times M$, sa koordinatama  $(n,k)\in [0,N)\times [-M/2,M/2).$ Agenti (mravi) se inicijalno postavljaju na slučajnim pozicijama, tako da imaju slučajne orijentacije. Tranzicija agenta sa pozicije $(n',k')\in [0,N)\times [-M/2,M/2)$ na novu poziciju $(n,k)$ dešava se u svakoj iteraciji $I$. Jedna iteracija algoritma je završena kada se svi agenti pomjere sa starih pozicija $(n',k')$ na nove pozicije $(n,k)$, ukoliko su im takve tranzicije dozvoljene u datoj iteraciji. U svaku ćeliju (tačku vremensko-frekvencijske ravni) koja je posjećena od strane agenta, dodaje se određena količina feromona. Nivo feromona koji se deponuje na poziciji $(n,k)$ označen je sa $\Phi(n,k)$. U svrhu čuvanja informacija o nivoima feromona u svim tačkama, formira se feromonska matrica (mapa) $\mathbf{\Phi}$. U razmatranom kontekstu, feromonska mapa će biti interpretirana kao nova vremensko-frekvencijska reprezentacija na bazi koje će se vršiti estimacija trenutne frekvencije. Na ovom mjestu važno je istaći da će ovaj pristup obezbijediti eliminaciju nepoželjnih tačaka vremensko-frekvencijske ravni, dok se proces estimacije može interpretirati kao svojesvrstan oblik \textit{rekonstrukcije} trenutne frekvencije, na osnovu preostalih vremensko-frekvencijskih tačaka u feromonskoj mapi. Opšte pravilo tranzicije agenta sa posmatrane pozicije $(n',k')$ na novu poziciju $(n,k)$ definisano je određenom funkcijom mjere, koja će dalje biti označena sa $P_{(n',k')}^{(I)}(n,k)$. U nastavku će biti opisane osnovne veličine i mehanizam formiranja feromonske mape, na bazi koje se vrši estimacija trenutne frekvencije.  Algoritam \ref{aco} sadrži rezime opisa koji slijedi.

\paragraph{Inicijalizacija.} Na početku ACO algoritma, u $\gamma$ procenata tačaka vremensko-frekvencijske ravni se slučajno pozicioniraju inteligentni agenti. Inicijalne pozicije dobijaju se generatorom slučajnih brojeva sa uniformnom distribucijom. Inicijalne pozicije su definisane vrijednostima $P(n,k)$, koje se čuvaju u pomoćnoj matrici $\mathbf{P}$, čiji su elementi
\begin{equation}
P(n,k)=\left\{ \begin{matrix}
r, & (n,k)=({{n}_{0}},{{k}_{0}})  \\
0, & (n,k)\ne ({{n}_{0}},{{k}_{0}})  \\
\end{matrix} \right.
\label{acofor1}
\end{equation}
gdje je $r$ slučajni broj sa uniformnom raspodjelom iz opsega $\{1,2,\dots,8\}$. Svaka vrijednost $r$ odgovara nekoj od mogućih orijentacija agenata (1 odgovara orijentaciju \textit{gore}, 2 označava orijentaciju \textit{lijevo-gore} itd.) U svakoj sljedećoj iteraciji, orijentacija agenta u tački $(n,k)$ je određena pomjeranjem sa stare ćelije $(n',k')$. Ako se agent, na primjer, pomjerio sa pozicije $(n',k')=(n+1,k)$ na poziciju $(n,k)$, tada je orijentisan na \textit{gore}, tj. $P(n,k)=1$. Kada mrav napusti poziciju $(n,k)$ odgovarajuća vrijednost u matrici $\mathbf{P}$ se postavlja na nulu.

Agenti komuniciraju kroz koncept depozicije i evaporacije (isparavanja) feromona.  Inicijalno, za sve pozicije na kojima se nalaze agenti, postavlja se mala konstantna vrijednost feromona,
\begin{equation}
 {{\Phi }^{(0)}}(n,k)=  \begin{cases}
{{\Phi }_{0}}, & (n,k)=({{n}_{0}},{{k}_{0}})  \\
0, & (n,k)\ne ({{n}_{0}},{{k}_{0}}),  \\
\end{cases}
\label{acof2}
\end{equation}
gdje je $0\leq{\Phi }_{0}<1$ inicijalni nivo feromona na početnim pozicijama agenata. Koncept depozicije i evaporacije feromona je krucijalan u kontroli masovnog ponašanja agenata.

\begin{figure}[ptb]%
	\centering
	\includegraphics[scale=1.15
	]%
	{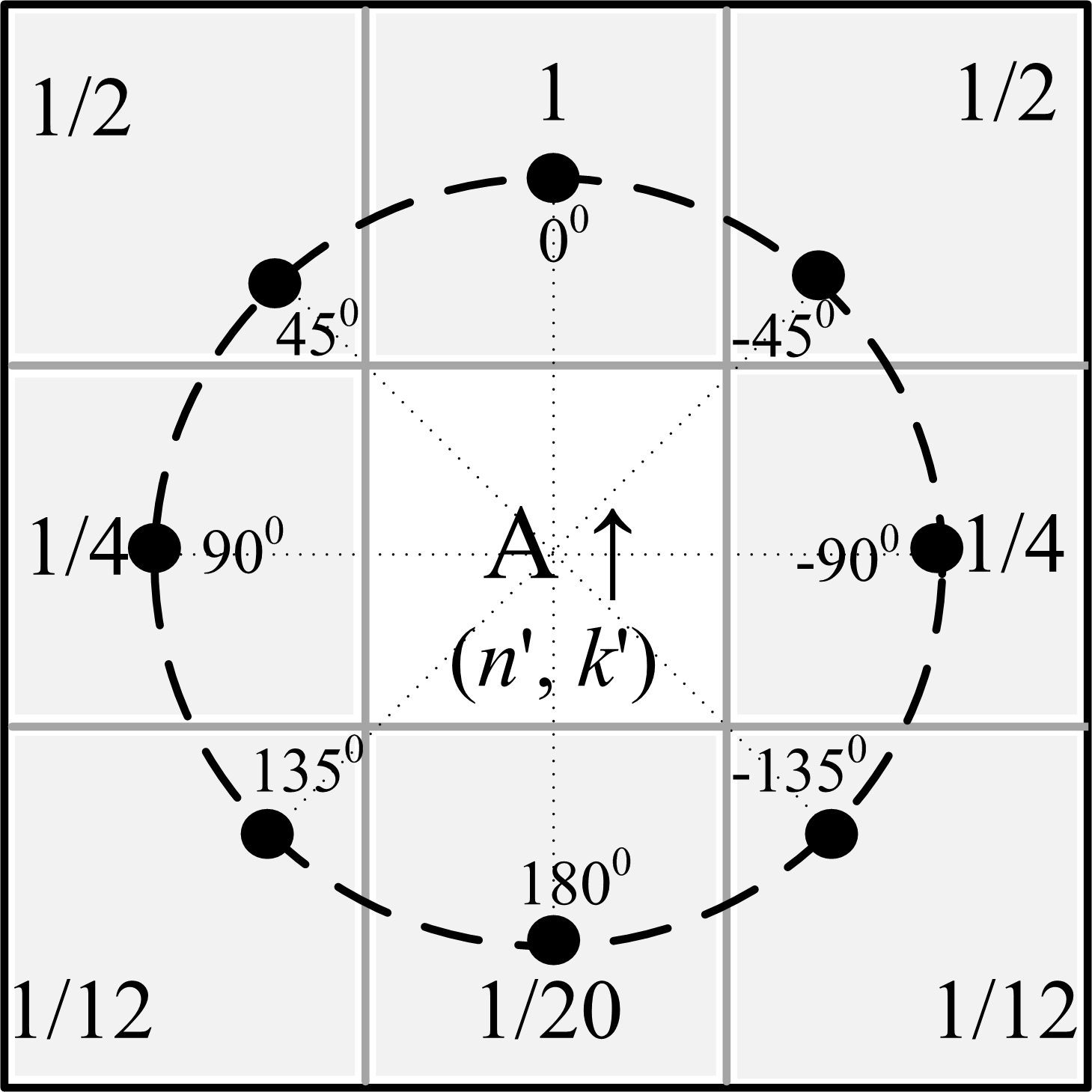}%
	\caption[Ilustracija rotacije agenta.]{Ilustracija orijentacije agenta. Agent je orijentisan sjeverno uz $P(n',k')=1$ uz naznaku vrijednosti $d\left( {{\Theta }_{(n',k')}}(n,k) \right)$ za moguće diskretne pravce. Osjenčena površina označava dozvoljeni domet pomjeranja agenta, $\mathbf{Q}(n ',k')$.}%
	\label{acof1}%
\end{figure}

\paragraph{Pravilo tranzicije agenata.} Nakon inicijalizacije, u svakoj iteraciji algoritma agenti se pomjeraju, prateći strogo određena pravila tranzicije. Dozvoljeno je pomjeranje sa pozicije $(n',k')$ na jednu od tačaka iz susjednog $3\times 3$ okruženja $(n,k)\in \mathbf{Q}(n',k').$ U objašnjenju će biti zanemareni ivični efekti. Prilikom pomjeranja agenta, vrši se depozicija određenog nivoa feromona. U jednoj ćeliji se može nalaziti samo jedan agent, a pomjeranje je zabranjeno ukoliko su sve okolne ćelije već zauzete agentima. Iteracija ACO algoritma se završava kada se svi agenti, kojima je to dozvoljeno, pomjere za jednu poziciju. 	Odluka o pomjeranju se donosi na osnovu nivoa feromona u susjednim ćelijama, i orijentacije agenta \cite{aco1,aco2}. Agent lociran u tački $(n',k')$ može da se pomjeri na jednu od susjednih 8 lokacija iz skupa $\mathbf{Q}(n',k')$, koje su  osjenčene na slici \ref{acof1}. Odgovarajući uglovi su ${{\Theta }_{1}}(n,k)\in \left\{ {{0}^{\circ }},\pm {{45}^{\circ }},\pm {{90}^{\circ }},\pm {{135}^{\circ }},{{180}^{\circ }} \right\}$, (za orijentaciju na \text{gore}), dok se za ostale orijentacije računaju kao  ${{\Theta }_{(n',k')}}(n,k)={{\Theta }_{1}}(n,k)-\left[ P(n',k')-1 \right]\cdot {{45}^{\circ }}$ za $(n,k)\in \mathbf{Q}(n  ',k')$, odnosno $3\times 3$ okruženje posmatrane tačke $(n  ',k')$. Ugao utiče na funkciju koja, pored nivoa feromona, dalje utiče na mjeru koja definiše tranziciju agenta. U radovima \cite{aco1,aco2} se koristi   sljedeća funkcija:
\begin{equation}d\left( {{\Theta }_{(n',k')}}(n,k) \right)= \begin{cases}
1, & {{\Theta }_{(n',k')}}(n,k)={{0}^{\circ }}  \\
1/2, & {{\Theta }_{(n',k')}}(n,k)=\pm {{45}^{\circ }} \\
1/4, & {{\Theta }_{(n',k')}}(n,k)=\pm {{90}^{\circ }} \\
1/12, & {{\Theta }_{(n',k')}}(n,k)\,=\pm {{135}^{\circ }}  \\
1/20, & {{\Theta }_{(n',k')}}(n,k)={{180}^{\circ }} \\
\end{cases}
\label{acof4}
\end{equation}
za  $(n,k)\in \mathbf{Q}(n',k').$ Ova funkcija predstavlja svojevrsnu inerciju orijentacije agenta.
Sljedeći parametar koji definiše vjerovatnoću tranzicije jeste nivo feromona $\Phi(n,k)$.  Uticaj je definisan funkcijom:
\begin{equation}
	g\left( \Phi (n,k) \right)={{\left( 1+\frac{\Phi (n,k)}{1+\delta \Phi (n,k)} \right)}^{\beta }}.
	\label{acof6}
\end{equation}
Velika vrijednost parametra $\beta$ definiše visok nivo privlačućeg uticaja feromona na agente. Parametar $\delta$ definiše osjetljivost agenta na koncentraciju feromona.

Konačno, mjera koja definiše vjerovatnoću tranzicije agenata definisana je na sljedeći način:
 \begin{equation}
  P_{(n',k')}^{(I)}(n,k)=\left\{ \begin{matrix}
 \frac{g\left( {{\Phi }^{(I)}}(n,k) \right)d\left( {{\Theta }_{(n',k')}}(n,k) \right)}{\sum\nolimits_{(n,k)\in \mathbf{Q}\left( n',k' \right)}{g\left( {{\Phi }^{(I)}}(n,k) \right)d\left( {{\Theta }_{(n',k')}}(n,k) \right)}}, & (n,k)\in \mathbf{Q}\left( n',k' \right)  \\
 0, & \text{za ostale tačke}\text{.}  \\
 \end{matrix} \right. 
 \label{acof5}
 \end{equation}

\paragraph{Ažuriranje feromona.} Nakon tranzicije agenta na poziciju $(n,k)\in \mathbf{Q}(n',k')$, feromonski nivo se ažurira po pravilu:
\begin{equation}
{{\Phi }^{(I+1)}}(n,k)={{\Phi }^{(I)}}(n,k)+\mu \nabla (n,k)+\xi.
\label{acofer}
\end{equation}
gdje je $\xi$ mali konstantni nivo feromona, $\nabla (n,k)$ je feromonski gradijent, dok je $\mu$ mali pozitivni korak. Za problematiku estimacije trenutne frekvencije, upravo je podešavanje varijabilnog dijela $\nabla (n,k)$ ključno za rješavanje problema. Na kraju svake iteracije, konstantni nivo $\xi$ feromona isparava iz cijele feromonske mape, odnosno, za svaku tačku mape se obavlja ažuriranje u skladu sa pravilom:
\begin{equation}
	 {{\Phi }^{(I+1)}}(n,k)= \begin{cases}
	{{\Phi }^{(I)}}(n,k)-\xi , & {{\Phi }^{(I)}}(n,k)\ge \xi   \\
	0, & {{\Phi }^{(I)}}(n,k)<\xi .  \\
	\end{cases},
	\label{acofer2}	
\end{equation}
U slučaju kada je novi nivo feromona u feromonskoj mapi na poziciji $(n,k)$ nakon ažuriranja negativan, postavlja se na nulu.

\paragraph{Gradijent depozicije u estimaciji trenutne frekvencije.} U definiciji gradijenta depozicije feromona, u obzir treba uzeti kontekst razmatranog problema, odnosno, estimaciju trenutne frekvencije. U tu svrhu, biće uzeto u obzir nekoliko nepobitnih činjenica \cite{tfsa}. Maksimumi od WD zašumljenog signala su sa velikom vjerovatnoćom dislocirani sa pozicije trenutne frekvencije. Za posmatrani trenutak $n$, međutim, jedna od najvećih vrijednosti WD će ipak biti na trenutnoj frekvenciji.

 Dislokacija je posljedica jakih šumnih impulsa koji svojim intenzitetom mogu prevazići vrijednosti auto-članova WD. Sa druge strane, varijacije trenutne frekvencije u dva susjedna trenutka nijesu previše velike, što je karakteristično u većini praktičnih primjena, \cite{tfsa,aco2}. Uzimajući u obzir ove činjenice, gradijent feromonskog ažuriranja se može definisati na sljedeći način:
\begin{equation}
 \nabla (n,k)=\Psi (n,k)\Xi (n,k)\Lambda (n,k),
 \label{aconabla}
\end{equation}
gdje su:
\begin{equation}
\Psi (n,k)=\frac{1}{27}\prod\limits_{i=-1}^{1}{\sum\limits_{j=-1}^{1}{WD(n+j,k+i)}},
\end{equation}
zatim
\begin{equation}
\Xi (n,k)=\frac{1}{9}\sum\limits_{i=-1}^{1}{\sum\limits_{j=-1}^{1}{WD(n+j,k+i)}},
\end{equation}
dok je posljednji član proizvoda definisan u obliku:
\begin{align}
 \Lambda (n,k)&=\max \left( \left[ \begin{matrix}
\prod\limits_{i=-1}^{1}{WD(n+i,k+i)} & \prod\limits_{i=-1}^{1}{WD(n+i,k-i)} \end{matrix} \right.\right. \notag \\
  &\left. \left. 
 \begin{matrix}
\prod\limits_{i=-1}^{1}{WD(n+i,k-1)} & \prod\limits_{i=-1}^{1}{WD(n+i,k)} & \prod\limits_{i=-1}^{1}{WD(n+i,k+1)}  \\
\end{matrix} \right] \right).
\end{align}

Detaljna diskusija ovih funkcija, odnosno, objašnjenje njihove uloge u procesu estimacije trenutne frekvencije, biće izostavljeno na ovom mjestu, pa se čitalac upućuje na detaljnu diskusiju prezentovanu u  \cite{aco2}.

\begin{algorithm}[!hb]
	\floatname{algorithm}{Algoritam}
	\caption{Formiranje ACO feromonske mape za estimaciju trenutne frekvencije}
	\label{aco}
	\begin{algorithmic}[1]
		\Input
		\Statex
		\begin{itemize}
			\item Matrica diskretne WD, dimenzija $N\times M$
		\end{itemize}
		\Statex
		\State Postaviti agente na slučajnim pozicijama, $(n_0,k_0)$, formiranjem pomoćne matrice $\mathbf{P}$ po pravilu (\ref{acofor1}) i inicijalizovati feromonsku mapu $\mathbf{\Phi}$ i matricu energije $\mathbf{E}$, prema pravilima (\ref{acof2}) i (\ref{acof3}), respektivno. Formirati pomoćnu matricu $\mathbf{d}$ prema (\ref{acof4}).
		\bigskip
		\While{ $I \leq I_{\mathrm{max}}$ i $\sum_n\sum_k |P(n,k)|> \chi N$ } \Comment $0.8\leq\chi \leq 1$, $I_{\mathrm{max}}$ je maks. br. iteracija
		\bigskip	
		\State  \hfill\begin{minipage}{\dimexpr\textwidth-1.5cm} {Za svaki nenulti element u matrici $\mathbf{P}$ izračunati mjeru vjerovatnoće tranzicije $ P_{(n',k')}^{(I)}(n,k)$ po (\ref{acof5}), pomoću  (\ref{acof4}) i (\ref{acof6}), i izvršiti tranziciju agenta na susjednu ćeliju sa najvećom vrijednošću mjere $ P_{(n',k')}^{(I)}(n,k)$, koja nije zauzeta drugim agentima.}\end{minipage} \label{acokorak}
		\bigskip
		\State \hfill\begin{minipage}{\dimexpr\textwidth-1.5cm} {Za svaku tačku vremensko-frekvencijske mreže koju je posjetio agent u koraku \ref{acokorak}, izračunati gradijent $\nabla(n,k)$ prema (\ref{aconabla}) i ažurirati feromonsku mapu $\mathbf{\Phi}$ prema (\ref{acofer}) i matricu energije, pomoću (\ref{acoe1}) i (\ref{acoe2}).}\end{minipage}
		\bigskip
		\State \hfill\begin{minipage}{\dimexpr\textwidth-1.5cm} {Ažurirati feromonsku mapu $\mathbf{\Phi}$ prema (\ref{acofer2}). Ažurirati matricu energije prema (\ref{acoe3}). Ažurirati matricu pozicija $\mathbf{P}$ prema (\ref{acop4}).}\end{minipage}
		\bigskip
		\EndWhile
		\Statex
		\Output
		\Statex
		\begin{itemize}
			
			\item Feromonska mapa $\mathbf{\Phi}$
			
		\end{itemize}
	\end{algorithmic}
\end{algorithm}
\paragraph{Koncept varijabilne populacije agenata.} Varijacija populacije omogućena je kroz koncepte reprodukcije, starenja i umiranja agenata \cite{aco2}. U razmatranom kontekstu, cilj je eliminisati one agente koji se kreću kroz nepoželjne lokacije, odnosno, lokacije kojima odgovaraju niske vrijednosti feromona tokom većeg broja iteracija. Kontrola populacije agenata se može uspostaviti na osnovu feromonske matrice $\mathbf{ \Phi}$. Na početku razmatranog ACO algoritma, svakom agentu se pridružuje određeni fiksni nivo energije $1+\alpha$. Oni se čuvaju u odgovarajućoj pomoćnoj matrici $\mathbf{E}$, sa elementima:
\begin{equation}
	{{E}^{(0)}}(n,k)= \begin{cases}
1+\alpha , & (n,k)=({{n}_{0}},{{k}_{0}})  \\
0, & (n,k)\ne ({{n}_{0}},{{k}_{0}}), \\
\end{cases}  
\label{acof3}
\end{equation}
gdje je $0\leq \alpha <1$.

Tokom tranzicije agenta iz ćelije $(n',k')$ u susjednu ćeliju $(n,k)\in \mathbf{Q}(n',k')$, vrši se ažuriranje energije po pravilu:
\begin{equation}
{{E}^{(I+1)}}(n,k)={{E}^{(I)}}(n',k')+\alpha \mu \nabla (n,k),
\label{acoe1}
\end{equation}
gdje se koristi gradijent $\nabla (n,k)$ koji je već izračunat za potrebe ažuriranja feromonske mape. Nakon toga, 	matrica energije se ažurira tako da se napuštenoj ćeliji $(n',k')$ pridružuje vrijednost nula:
\begin{equation}
{{E}^{(I+1)}}(n',k')=0.
\label{acoe2}
\end{equation}

Na kraju svake iteracije, svim pozicijama $(n,k)\in [0,N)\times [-M/2,M/2)$ se energija globalno umanjuje za konstantnu vrijednost $\alpha$:
\begin{equation}
{{E}^{(I+1)}}(n,k)={{E}^{(I)}}(n,k)-\alpha.
\label{acoe3}
\end{equation}
Ovom prilikom se svi agenti za koje važi	$E(n,k)\leq 0$ uklanjaju, brisanjem odgovarajućih elemenata matrice $\mathbf{P}$, za svako  $(n,k)\in [0,N)\times [-M/2,M/2)$: 
\begin{equation}
{{P}^{(I+1)}}(n,k)=0,\text{ ako je }{{E}^{(I+1)}}(n,k)<0.
\label{acop4}
\end{equation}

Uslov prekidanja može da bude minimalno dozvoljeni broj agenata, npr. $80\%$--$100\%$ od broja vremenskih tačaka $N$. Opisana procedura je rezimirana u Algoritmu \ref{aco}. Detaljna diskusija svih parametara, i rezultati numeričke evaluacije njihovih vrijednosti, mogu se naći u radu \cite{aco2}.

\paragraph{Estimacija trenutne frekvencije iz feromonske mape.} Feromonska mapa $\mathbf{\Phi}$ se interpretira kao nova vremensko-frekvencijska reprezentacija, robustna na uticaj šuma. Problem estimacije trenutne frekvencije sada postaje:
\begin{equation}
\hat{k}=\arg \underset{k}{\mathop{\max }}\,\Phi (n,k).
\end{equation}
Na dobijeni estimat $\hat{k}$ se može primijeniti median filter, kao i odgovarajuća kubična interpolacija \cite{aco2}. U primjeru koji slijedi, algoritam se poredi sa estimacijom zasnovanom na maksimumima WD, kao i sa sofisticiranim Viterbi algoritmom za estimaciju \cite{tfsa,aco2}.
    \begin{figure}[]%
	\centering
	\includegraphics[
	]%
	{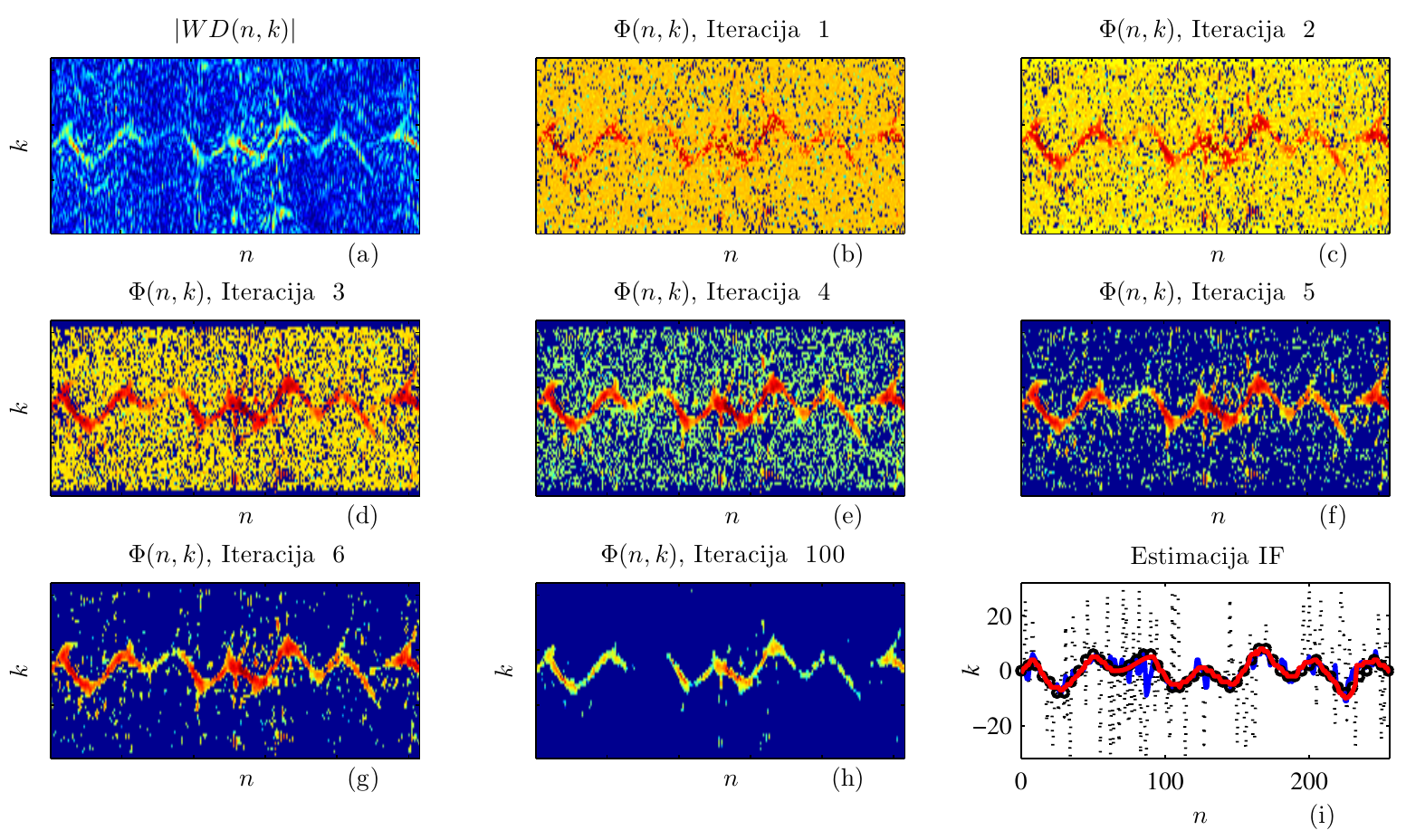}%
	\caption[Primjer estimacije trenutne frekvencije pomoću Wigner-ove distribucije i ACO algoritma, u uslovima šuma velikog intenziteta.]{Primjer estimacije trenutne frekvencije pomoću Wigner-ove distribucije i ACO algoritma, u uslovima šuma velikog intenziteta: (a) Wigner-ova distribucija zašumljenog signala; (b) -- (h) izgled feromonske mape u specifičnim iteracijama; (i) poređenje rezultata dobijenih ACO pristupom (crveno), kompetitivnim Viterbi algoritmom \cite{aco2}, (plavo), maksimumi WD (isprekidana linija) i originalna trenutna frekvencija (kružići). }%
	\label{E33novo}%
\end{figure}
\begin{primjer}
Razmatra se složenoharmonijski FM signal
\begin{equation}
x(n)=\exp \left( j2\sin \left( \frac{13\pi (n+32)}{256} \right)+j3\sin \left( \frac{5\pi (n+32)}{256} \right) \right)+\varepsilon (n)
\end{equation}
dužine $N_s=320$ odbiraka, definisan u trenucima $n\in [-N/2,3N/2)$, pri čemu se WD računa za srednjih $N=256$ tačaka. Signal je zašumljen jakim aditivnim bijelim kompleksnim Gausovim šumom $\varepsilon (n)$. Dužina primijenjenog prozora je $M=64$, a odnos signal-šum je $SNR=-2$ dB. Ilustracija WD zašumljenog signala, izgled feromonske mape u karakterističnim iteracijama i rezultati estimacije trenutne frekvencije prikazani su na slici \ref{E33novo}. ACO pristup obezbjeđuje estimaciju uprkos relativno velikim varijacijama trenutne frekvencije i izraženom uticaju šuma.
\end{primjer}

\subsection{S-metod}
\label{smsmsm}
Kratkotrajna Furijeova transformacija, odnosno spektrogram, u opštem slučaju se karakterišu  slabo koncentrisanom reprezentacijom signala. Sa druge strane, STFT je linearna transformacija, pa je ne karakteriše pojava neželjenih kros-članova u slučaju multikomponentnih signala. WD posjeduje sposobnost znatno boljeg koncentrisanja komponenti signala, a u slučaju LFM signala daje idealnu koncentraciju. Međutim, i pored velikog broja poželjnih osobina, karakteriše je pojava neželjenih kros-članova u slučaju multikomponentnih signala. U cilju očuvanja dobrih i eliminacije loših strana ovih reprezentacija, uveden je S-metod. PWD se može izraziti u funkciji od STFT na sljedeći način:
\begin{equation}
PWD(t,\Omega)=\frac{1}{\pi}\int_{-\infty}^{\infty}STFT(t,\Omega
+\Theta)STFT^{\ast}(t,\Omega-\Theta)d\Theta.
\end{equation}

Prethodna relacija direktno vodi do definicije nove vremensko-frekvencijske reprezentacije, poznate pod nazivom S-metod (SM):
\begin{equation}
SM(t,\Omega)=\frac{1}{\pi}\int_{-\infty}^{\infty}P(\Theta
)STFT(t,\Omega+\Theta)STFT^{\ast}(t,\Omega-\Theta)d\Theta,
\end{equation}
gdje je $P(\Theta)$  frekvencijski prozor ograničene širine. U slučaju pravougaonog prozora sa $P(\Theta)=0,~\left\vert \Theta\right\vert
>L_{p}$, S-metod postaje:%
\begin{equation}
SM(t,\Omega)=\frac{1}{\pi}\int_{-L_{p}}^{L_{p}}P(\Theta)STFT(t,\Omega
+\Theta)STFT^{\ast}(t,\Omega-\Theta)d\Theta. \label{SMc}%
\end{equation}

Ukoliko je $P(\Theta)=\pi\delta(\Theta)$ S-metod se svodi na spektrogram, dok se za
$P(\Theta)=1$ dobija WD. U slučaju ove vremensko-frekvencijske reprezentacije marginalni uslovi ne moraju biti zadovoljeni.

Kada su u pitanju multikomponentni signali, S-metod omogućava dobijanje
reprezentacije signala kod koje je za svaki auto-član postignuta koncentracija
pseudo-Wigner-ove distribucije, dok su kros članovi ili potpuno eliminisani, ili su značajno redukovani. Parametar $L_{p}$ se bira tako da prozor $P(\Theta)$ omogućava potpunu integraciju
(\ref{SMc}) duž autočlanova, ali da je pritom uži od rastojanja između auto-članova,
kako bi se izbjegla pojava kros-komponenti. Dakle, integracijom u izrazu (\ref{SMc}) poboljšava se koncentracija auto-članova sve dok $\Theta$ ne postane
dovoljno veliko tako da $STFT(t,\Omega+\Theta)$ pripada jednoj komponenti, a
$STFT^{\ast}(t,\Omega-\Theta)$ drugoj. Tada bi integracijom njihovog proizvoda
došlo do pojave kros-člana. U slučaju preklopljenih komponenti, u posmatranom trenutku do kros-članova dolazi samo između njih, kao što je to bio slučaj sa spektrogramom.

Za multikomponentni signal oblika
\begin{equation}
x(t)=\sum_{p=1}^{P}x_{p}(t),
\end{equation}
uz pretpostavku da svakoj pojedinačnoj komponenti $x_{p}(t)$ u vremensko-frekvencijskoj ravni pripada odgovarajući region $D_{p}(t,\Omega)$, $p=1,\dots,P$, i uz dodatnu pretpostavku o nepreklapanju ovih regiona, važi sljedeće svojstvo S-metoda:
\begin{equation}
SM_{x}(t,\Omega)=\sum_{p=1}^{P}PWD_{x_{p}}(t,\Omega),
\label{smwd}
\end{equation}
odnosno, S-metod je jednak sumi PWD pojedinačnih komponenti signala, ako je širina pravougaonog prozora $P(\Theta)$ za tačku $(t,\Omega)$
\begin{equation}
L_{p}(t,\Omega)=\left\{
\begin{array}
{ll}%
B_{p}(t)-\left\vert \Omega-\Omega_{0p}(t)\right\vert,&~(t,\Omega)\in D_{m}(t)\\
0,&~\text{ostalo }(t,\Omega),
\end{array}
\right.
\end{equation}
gdje je dužina $p$-tog regiona po frekvenciji $\Omega$ za dato $t$ označena sa
$2B(t)$, a centralna frekvencija sa $\Omega_{0p}(t)$. 
\subsubsection{Diskretna realizacija S-metoda}
U diskretnoj formi, S-metod je definisan sljedećim relacijama:%
\begin{align}
SM(n,k)  &  =\sum_{i=-L_{d}}^{L_{d}}P(i)STFT(n,k+i)STFT^{\ast}%
(n,k-i)\label{SMd}\\
SM(n,k)  &  =\left\vert STFT(n,k)\right\vert ^{2}+2\operatorname{Re}\left[
\sum_{i=1}^{L_{d}}STFT(n,k+i)STFT^{\ast}(n,k-i)\right]  , \label{SMd2}%
\end{align}
gdje je pretpostavljen pravougaoni prozor $P(i)=1$, $-L_{d}\leq i\leq L_{d}.$
Suma u prethodnoj relaciji omogućava poboljšanje koncentracije spektrograma,
do postizanja koncentracije pseudo-Wigner-ove distribucije.

Dok za spektrogram važi perioda odabiranja signala u skladu sa teoremom o
odabiranju, iz (\ref{PWD}) je jasno da je za PWD neophodan duplo manji period
odabiranja. S-metod zahtijeva isti period odabiranja kao i spektrogram, što je
još jedna značajna prednost u odnosu na Wigner-ovu distribuciju.

\section{Rekonstrukcija LFM komponenti ISAR signala  nakon uklanjanja mikro-Doplera}
%\label{section}
Osobina vremensko-frekvencijskih reprezentacija da koncentrišu energiju signala u okolini trenutne frekvencije, sugeriše da ih je moguće razmatrati kao domene  rijetkosti u kontekstu kompresivnog odabiranja i rekonstrukcije rijetkih signala. U ovoj sekciji biće ilustrovan pristup rekonstrukciji nedostajućih tačaka u vremensko-frekvencijskoj ravni, originalno prezentovan u radu \cite{brajovic_tf3}.

Formiranje radarskih slika je klasičan primjer primjene vremensko-frekvencijske analize, \cite{glavni, martorella, IT2018, chen1, chen2,ref3,ref4,wang,csiet1}. Prisustvo rotirajućih ili vibrirajućih djelova radarske mete izaziva pojavu mikro-Dopler (engl. \textit{micro-Doppler}, m-D) efekta, \cite{glavni,martorella, IT2018, chen1, chen2,ref3,ref4,wang,csiet1}. U cilju fokusiranja radarskih slika i poboljšanja njihove čitljivosti, već dugo je aktuelna problematika razdvajanja krutog tijela i m-D dijela signala koji odgovara radarskoj meti. Kruto tijelo (\textit{rigid body}, RB) predstavlja dio radarskog signala koji sadrži informacije o čvrstim djelovima posmatrane mete, koji su reflektovali poslati radarski signal. U slučaju kada je radarski signal obrađen tako da je kompenzovano ubrzanje mete -- RB čine stacionarne komponente. U suprotnom, on je sastavljen od LFM komponenti u posmatranom \textit{range}-u, \cite{glavni}. Više informacija o fizici ISAR radarskih sistema (engl. \textit{Inverse synthetic-aperture radar}) se može naći u literaturi \cite{glavni}. Mikro-Dopler djelovi signala nastaju kao posljedica kretanja djelova radarskih meta i karakteriše ih pretežno visok nivo nestacionarnosti. U slučaju rotirajućih i vibrirajućih reflektora, m-D se uobičajeno modeluje sinusoidalno frekvencijski modulisanim komponentama. 
Razdvajanje RB od m-D, zasnovano na STFT i L-statistici, proučavano je u \cite{glavni}, gdje je predložena tehnika koja daje visoko koncentrisani RB nakon izdvajanja m-D-a.  

Rekonstrukcija RB u slučaju nekompenzovanog ubrzanja mete je razmatrana u \cite{glavni}, gdje je lokal-polinomijalna Furijeova transformacija (engl. \textit{Local Polynomial Fourier Transform}, LPFT) služila kao polazna reprezentacija. Takav pristup zahtijeva poznavanje parametra linearne modulacije, ili tzv. \textit{chirp rate}, kojim je neophodno obaviti kompenzaciju kretanja meta. Ovaj parametar u opštem slučaju nije unaprijed poznat, \cite{glavni} i ne može biti estimiran na osnovu polaznog signala. U ovoj sekciji razmotrićemo mogućnosti njegove estimacije primjenom mjera koncentracije i upotrebe postupka u: (a) aproksimaciji RB komponenti primjenom L-statistike i (b) njegove primjene u kontekstu egzaktne rekonstrukcije RB komponenti zasnovane na konceptima kompresivnog odabiranja.

\subsection{Model signala}
Razmatra se radar sa kontinualnim talasima (engl. \textit{continuous wave - CW radar}) koji šalje signale u formi $N$ koherentnih čirpova (LFM komponenti). 
Ukoliko je udaljenost mete označena sa $d(t)$, a $c$ predstavlja brzinu svjetlosti, tada signal koji se reflektuje od mete kasni za $t_d=2d(t)/c$ u odnosu na poslati signal. U posmatranom modelu podrazumijevane su standardne operacije pretprocesiranja signala (kao na primjer -- demodulacija signala na osnovni opseg), \cite{glavni}. 

Kao što je to standardno podrazumijevano u radarskoj literaturi \cite{glavni}, u cilju analize nestacionarnosti  kros-\textit{range}-a radarske slike, razmatra se samo Doplerov dio primljenog signala tačkaste mete, u kontinualnom vremenu u kojem je meta u dometu radara (engl. \textit{dwell time}):
\begin{equation}
s(t)=\sigma e^\frac{j2d(t)\omega_0}{c}, \label{eq1}
\end{equation}
gdje $\sigma$ označava koeficijent refleksije mete a $\omega_0$ označava frekvenciju na kojoj radar funkcioniše. Pretpostavlja se da je vrijeme ponavljanja impulsa $T_r$, sa $N_c$ odbiraka u svakom čirpu (engl. \textit{chirp}, linearno frekvencijski modulisani signal), i da je vrijeme koherentne integracije (engl. \textit{coherence integration time -- CIT}) dato izrazom $T_c=NT_r$. 

Doplerov dio primljenog signala, koji odgovara RB-u, može se modelovati kompleksnim sinusoidama \cite{glavni}. Međutim, nekompenzovano ubrzanje izaziva pojavu LFM signala u reprezentaciji RB komponenti. Pokretni djelovi mete, pretežno rotirajući i vibrirajući, dovode do pojave mikro-Doplera, koji se manifestuje u vidu dodatnih nestacionarnih komponenti primljenog signala. Rotirajući i vibrirajući djelovi modeluju se sinusoidalnim FM signalima, dok je model kompleksniji u slučaju drugih oblika kretanja. U slučaju RB od $K$ tačaka, i m-D kojeg je izazvalo $D$ reflektora, opšti model signala je dat izrazom \cite{glavni}:
\begin{equation}
s(n)=\sum_{i=1}^{K}\sigma_{B_i}e^{jy_{B_i}n}+\sum_{i=1}^{D}\sigma_{R_i}e^{jA_{R_i}\sin(\omega_{R_{i}}n+\Theta_i)},
\label{model}
\end{equation}
gdje je $n=0,1,\dots,N-1$,  $\sigma_{B_i}$, $\sigma_{R_i}$ su koeficijenti refleksije  RB-a i rotirajućih reflektora, respektivno, $y_{B_i}$ odgovara poziciji RB reflektora, $A_{R_i}$ je proporcionalno rastojanju od rotirajućeg reflektora do centra rotacije. Ugaone frekvencije $\omega_{R_{i}}$ proporcionalne su učestanosti rotacije $i$-tog m-D reflektora. Više informacija o ovom modelu može se naći u literaturi \cite{glavni}.

\subsection{Razdvajanje stacionarnih komponenti od mikro-Doplera}
Pretpostavimo da je primljeni signal $s(t)$ adekvatno odabran, i da se dalja obrada vrši nad diskretnim odbircima signala $s(n)$. Iako su RB komponente u slučaju kompenzovanog ubrzanja mete stacionarne, usljed varijabilnog frekvencijskog sadržaja mikro-Doplera, Furijeovu transformaciju je teško primjenjivati u analizi i obradi ovakvih signala. Stoga se koriste pristupi zasnovani na vremensko-frekvencijskoj analizi. Kao što je to bio slučaj kod DFT, DCT i Hermitske transformacije, sposobnost vremensko-frekvencijske reprezentacije da koncentriše signal se  kvantifikuje mjerama koncentracije:
\begin{equation}
	\mathcal{M}_p^p=\left(\sum_n\sum_p\left|TFR(n,k)^{\frac{1}{p}}\right|\right)^p
\end{equation}
Kod linearnih reprezentacija se može  koristiti $\ell_1$-norma, koja se dobija za  $p=1$, kao u slučaju kompresivnog odabiranja.
Posmatrajmo drugu formu kratkotrajne Furijeove transformacije (STFT) analiziranog signala:
\begin{equation}
STFT(n,k)=\sum_{m=0}^{N-1}s(m)w(m-n)e^{-j2\pi mk/N}, \label{stft}
\end{equation}
gdje se prozorska funkcija $w(m)$ koristi za lokalizaciju frekvencijskog sadržaja. U ovoj formi STFT za prozor važi: $w(n)\neq0$ za $-M/2 \leq m\leq M/2-1$ i on je dopunjen nulama do dužine signala $N$. Originalna koncentracija FT može se dobiti iz (\ref{stft}) pomoću:
\begin{align}
S(k)&=\sum_{n=M/2}^{N-M/2}STFT(n,k)\notag\\       
&=\sum_{m=0}^{N-1}s(m)\bigg[\sum_{n=M/2}^{N-M/2}w(m-n)\bigg]e^{-j2\pi mk/N}.
\label{sumasuma}
\end{align}

{Budući da važi $\sum_{n=M/2}^{N-M/2}w(m-n)\approx{const}$, rezultujući prozor je veoma sličan pravougaonom, s obzirom da je, mimo prelaznih djelova od po $M/2$ tačaka na krajevima, konstantan u posmatranom intervalu. Stoga  se posmatrani izraz (\ref{sumasuma}) može smatrati Furijeovom transformacijom analiziranog signala, sa koncentracijom koja je bliska koncentraciji FT računate za pravougaoni prozor. Tada se m-D komponente iz $STFT(n,k)$ mogu izdvojiti sortiranjem STFT vrijednosti po vremenskom indeksu, i uklanjanjem određenog procenta najvećih vrijednosti. Sumiranjem preostalih tačaka po frekvencijskom indeksu, dobija se aproksimacija Furijeove transformacije krutog tijela.
\subsubsection{Algoritam za razdvajanje mikro-Doplera od stacionarnih komponenti}\label{removal}
U slučaju stacionarnog RB-a, razdvajanje od m-D može biti obavljeno na sljedeći način. Za svaki frekvencijski indeks $k$, označimo odgovarajući set STFT tačaka sa:
\begin{equation} 
\mathbf{S}_k=\{STFT(n,k),~n=M/2,\dots,N-M/2\}.
\end{equation}

Sortiranjem elemenata ovog skupa po vremenskom indeksu, dobija se skup  $\mathbf{\Psi}_k$ čiji su elementi
${\Psi}_k(n_i)\in   \mathbf{S}_k$, $n_i \in \{M/2,\dots,N-M/2\}$. Ovi elementi, za svako $k$ zadovoljavaju
\begin{equation} 
|{\Psi}_k(n_1)| \leq |{\Psi}_k(n_2)|  \leq \dots \leq |{\Psi}_k(n_{N-M})|.
\end{equation}

Razdvajanje RB i m-D pomoću L-statistike podrazumijeva da se iz $\mathbf{\Psi_k}$, za svako $k$, odbacuje $N_U$ najvećih i $N_D$ najmanjih elemenata. Vektor-kolona $\mathbf{\Psi_k}$ je dimenzija $M\times 1$. Ako je $U$ procenat eliminisanih elemenata sa najvećim vrijednostima, a $D$ procenat elminisanih elemenata sa najmanjim vrijednostima, tada se odbacuje ukupno $N_U=\text{int}[(N-M)(1-U)/100]$ najvećih i $N_D=\text{int}[(N-M)(1-D)/100]$ najmanjih elemenata, odnosno, ukupno $Q=D+U$ procenata STFT tačaka.  Za dato $k$, skup dostupnih pozicija je $\mathbb{L}_k$ i on predstavlja podskup od $\{n_1,n_2,\dots,n_{N-M}\}$. Skupovi $\mathbb{L}_k$ za svako $k=0,1,\dots,M-1$ formiraju skup $\mathbb{N}_A$ koji sadrži indekse $(n_i,k_i)$ dostupnih (odnosno zadržanih) vremensko-frekvencijskih tačaka.
Budući da važi
\begin{align}
S_{\Psi}(k)=\sum_{n=M/2}^{N-M/2}STFT(n,k)=\sum_{i=1}^{N-M}{\Psi}_k(n_i),
\label{suma2}
\end{align}
tada se na bazi dobijenih podskupova $\mathbb{L}_k$ skupa $\{n_1,n_2,\dots,n_{N-M}\}$,   računa L-estimacija
\begin{align}
S_L(k)=\sum_{n\in \mathbb{L}_k}STFT(n,k).
\label{suma3}
\end{align}
 
 Ovo je jednostavan način da se izvrši aproksimacija RB komponenti. Stacionarne komponente su, za zadatu frekvenciju, prisutne za sve vremenske indekse. Kako su m-D komponente vremenski promjenljive, sabiranjem STFT tačaka po vremenskom indeksu, na pozicijama RB komponenti dobijaju se koncentrisani pikovi (impulsi), usljed sabiranja koje je u fazi, čak i nakon uklanjanja m-D djelova. Loše koncentrisane m-D komponente koje ostaju nakon uklanjanja $Q\%$ STFT tačaka se za svako $k$ sabiraju sa različitim -- slučajnim fazama, pa se stoga usrednjavaju.
 
\subsection{Izdvajanje mikro-Doplera u slučaju nekompenzovanog ubrzanja radarske mete}\label{estimacija}
Kretanje radarske mete sa ubrzanjem dovodi do pojave da RB komponente uzimaju formu LFM signala. U tom slučaju, u razmatranom modelu (\ref{model}), stacionarne RB komponete se zamjenjuju LFM signalima koji imaju nepoznati \textit{chirp rate} parametar $a$. Rezultujući RB je nestacionaran, pa prethodno opisani postupak uklanja i njegove značajne djelove. U cilju eliminisanja ovog oblika nestacionarnosti, može se koristiti LPFT oblika:
\begin{align*}
LPFT_{\alpha}(n,k)
=\!\!\!\!\!\!\sum_{m=-M/2}^{M/2-1}&s(n+m)w(m)e^{-j2\pi[\frac{m}{M}k+  \alpha (\frac{m}{M})^2]}, \label{lpft}
\end{align*}
kako bi se odredio demodulišući optimalni parametar $\alpha_{opt}=a$. Ovaj parametar ne može biti estimiran na osnovu polaznog signala, već L-statistika mora biti involvirana u postupak pretrage. Jedan pristup je direktna pretraga u zadatom prostoru mogućih vrijednosti parametra, i pronalaženje one vrijednosti koja odgovara najmanjoj mjeri koncentracije RB komponenti. Numerički efikasniji pristup je korišćenje iterativnog algoritma koji je zasnovan na mjerama koncentracije, a koji je inspirisan gradijentnim algoritmima koji su prezentovani u prethodnim glavama ove disertacije \cite{IT2018}:

\noindent\textbf{Korak 0:} Inicijalizovati $\nabla=N/2$ i $\alpha^{(0)}=0$. \\ Zatim, ponavljati korake 1-4, sve dok ne bude ispunjen odgovarajući kriterijum za zaustavljanje.

\noindent\textbf{Korak 1}: Izračunati:
\begin{align}
LPFT_{\alpha^+}(n,k)
=\sum_{m=-M/2}^{M/2-1}&s(n+m)w(m)e^{-j2\pi[\frac{m}{M}k+  (\alpha+\nabla) (\frac{m}{M})^2]}, \notag\\
LPFT_{\alpha^-}(n,k)
=\sum_{m=-M/2}^{M/2-1}&s(n+m)w(m)e^{-j2\pi[\frac{m}{M}k+  (\alpha-\nabla) (\frac{m}{M})^2]}, \notag
\end{align}

\noindent\textbf{Korak 2}: Primijeniti L-statistiku na obije reprezentacije $LPFT_{\alpha^+}(n,k)$ i $LPFT_{\alpha^-}(n,k)$. Polazeći od zadatih LPFT tačaka
\begin{equation} 
\mathbf{L}^{\pm}_k(n)=\{LPFT_{\alpha^{\pm}}(n,k),~n=M/2,\dots,N-M/2\} \notag
\end{equation}
sortirati vrijednosti ovih skupova po indeksu  
$n$ u cilju dobijanja novih, sortiranih skupova, ${\Psi}^{+}_k(n_i)\in      \mathbf{L}^{+}_k(n)$, and ${\Psi}^{-}_k(n_j)\in        \mathbf{L}^{-}_k(n)$, $n_i, n_j \in \{M/2,\dots,N-M/2\}$ koji za dato $k$ zadovoljavaju: 
$
|{\Psi}^+_k(n_1)| \leq |{\Psi}^+_k(n_2)|  \leq \dots \leq |{\Psi}^+_k(n_{N-M})|
$ i $
|{\Psi}^-_k(n_1)| \leq |{\Psi}^-_k(n_2)|  \leq \dots \leq |{\Psi}^-_k(n_{N-M})|.
$

Odbaciti najvećih $N_Q$ vrijednosti iz $\mathbf{\Psi}^+_k(n_i)$ i $N_Q$ vrijednosti iz $\mathbf{\Psi}^-_k(n_j)$, gdje je $N_Q=\text{int}[(N-M)(1-Q)/100]$,  dok je Q procenat odbačenih vrijednosti.
Na osnovu dobijenih podskupova $L^+_k$ i $L^-_k$ of $\{n_1, n_1,\dots,n_{N-M}\}$,  izračunati
\begin{gather}
S^+_L(k)=\sum_{n\in L_k}LPFT_{\alpha^{+}}(n,k),\\
S^-_L(k)=\sum_{n\in L_k}LPFT_{\alpha^{-}}(n,k).
\end{gather}

\noindent\textbf{Korak 3}: Aproksimirati gradijent mjere koncentracije na osnovu razlike
\begin{align}
\nabla=\sum_{k=0}^{M-1}|S_L^{+}(k)|-\sum_{k=0}^{M-1}|S_L^{-}(k)|.
\end{align}

\noindent\textbf{Korak 4}: Ažurirati parametar $\alpha$ korišćenjem pristupa iz metoda najbržeg spuštanja:
\begin{align}
\alpha^{(l+1)}=\alpha^{(l)}-\mu\nabla.
\end{align}

Rezultujući parametar $\alpha$ se dalje koristi za demodulaciju signala i uklanjanje m-D djelova, kao i za rekonstrukciju krutog tijela primjenom algoritama za kompresivno odabiranje. U numeričkom primjeru iz ove sekcije, korišćen je korak $\mu=\frac{M}{N_Q}$. Indeks iteracija je označen sa  $l$. 

%IMA LI SMISLA CRTATI (C) I (D)? NACRTANO DA POPUNI PROSTOR
%\begin{figure}[h]
%       \centering
%       \includegraphics[scale=1]{fig2}
%       \caption{Rigid body and m-D separation in the case of uncompensated rigid body acceleration - noisy signal case. (a) The STFT of the original signal. (b) Sorted STFT values. (c) FT of the original signal. (d) FT obtained summing lowest 40\% of the STFT over time. (e) STFT of the signal dechirped with optimal $\alpha$. (f) Sorted dechirped signal STFT values. (g) FT of the dechirped signal. (h) FT obtained summing lowest 40\% of absolute values of the dechirped signal STFT over time. (i) Parameter $\alpha$ during the optimization algorithm iterations. (j) FT obtained summing lowest 40\% of the dechirped signal STFT over time according to (\ref{suma3}).}
%       \label{fig2}
%\end{figure}
\subsection{Rekonstrukcija krutog tijela}
Neka je sa
$s_{\alpha}(n)=s(n)e^{-j2\pi \alpha (n/N)^2}$
označen signal koji se demoduliše optimalnim parametrom $\alpha$. Odgovarajući vektor signala je $\mathbf{s}_{\alpha}=[s_{\alpha}(0),s_{\alpha}(1),\dots,s_{\alpha}(N-1)]$. Ako je $\mathbf{W}_M$ transformaciona matrica dimenzija $M\times M$ diskretne Furijeove transformacije, sa elementima $\exp(-j2\pi mk/M)$,  tada se  STFT demodulisanog signala 
\begin{equation}
STFT_{\alpha}(n,k)=\sum_{m=0}^{N-1}s(n+m)w(m)e^{-j2\pi mk/N} \label{stft2}
\end{equation}
u matričnom obliku može zapisati kao:
\begin{gather}
\mathbf{STFT}_{\alpha}=\left[
\begin{tabular}
[c]{cccc}%
$\mathbf{W}_{M}$ & $\mathbf{0}$ & $\mathbf{\cdots}$ & $\mathbf{0}$\\
$\mathbf{0}$ & $\mathbf{W}_{M}$ & $\mathbf{\cdots}$ & $\mathbf{0}$\\
$\vdots$ & $\vdots$ & $\mathbf{\ddots}$ & $\vdots$\\
$\mathbf{0}$ & $\mathbf{0}$ & $\mathbf{\cdots}$ & $\mathbf{W}_{M}$%
\end{tabular}
\right]  \mathbf{s}_{\alpha}\nonumber\\
\mathbf{STFT}_{\alpha}\mathbf{={W}s_{\alpha}={W}W}_{N}^{-1}\mathbf{S}_{\alpha}, \label{FMF}%
\end{gather}
gdje je $\mathbf{S}_{\alpha}$ vektor DFT koeficijenata računatih za signal originalne dužine i jedinični pravougaoni prozor. 

Pretpostavljeno je da se  $STFT_{\alpha}(n,k)$ računa u tačkama $0,~M,~2M,\dots,N-M$. Sve STFT vrijednosti se kombinuju u jedan vektor:
\begin{equation}
\mathbf{STFT}_{\alpha}=[\mathbf{STFT}_M(0)^T, \dots,\mathbf{STFT}_M(N-M)^T]^T.
\end{equation}
Za svaki posmatrani trenutak, STFT vektori su
\begin{equation}
\mathbf{STFT}_M(n)=[STFT(n,0),\dots, STFT(n,M-1)],
\end{equation}
i oni se računaju kao 
\begin{equation}
\mathbf{STFT}_M(n)=\mathbf{W}_M\mathbf{s}_{\alpha}(n),
\end{equation}
gdje je $\mathbf{s}_{\alpha}(n)=[s_{\alpha}(n),\dots,s_{\alpha}(n+M-1)]^T$. Razmatrana notacija se lako može generalizovati na slučaj STFT sa preklopljenim prozorima. 
%When other window forms $w(n)$ are used, based on diagonal matrices $\mathbf{H}_M$ containing window samples on the main diagonal, matrix
%\begin{equation}
%       \mathbf{H}=\left[\begin{tabular}
%       [c]{cccc}%
%       $\mathbf{H}_{M}$ & $\mathbf{0}$ & $\mathbf{\cdots}$ & $\mathbf{0}$\\
%       $\mathbf{0}$ & $\mathbf{H}_{M}$ & $\mathbf{\cdots}$ & $\mathbf{0}$\\
%       $\vdots$ & $\vdots$ & $\mathbf{\ddots}$ & $\vdots$\\
%       $\mathbf{0}$ & $\mathbf{0}$ & $\mathbf{\cdots}$ & $\mathbf{H}_{M}$%
%       \end{tabular} \right]
%\end{equation}
%is formed, and the non-overlapped STFT is calculated as
%\begin{equation}
%       \mathbf{STFT}_{\alpha}\mathbf{={W}s_{\alpha}={W}W}_{M}^{-1}\mathbf{S}_{\alpha}.
%\end{equation}

Uočimo da važi:
\begin{equation}
\mathbf{STFT}_{\alpha}\mathbf{={W}s_{\alpha}={W}W}_{N}^{-1}\mathbf{S}_{\alpha}.
\end{equation}
 Ukoliko je $\alpha$ adekvatno određen, DFT vektor $\mathbf{S}_{\alpha}$ je rijedak. Ovaj vektor se može izraziti u obliku:
\begin{equation}
\mathbf{S}_{\alpha}=\mathbf{W}_{N}\mathbf{W}^{-1}\mathbf{STFT}_{\alpha}.
\end{equation}

Uvedimo notaciju $\mathbf{A}=\mathbf{W W}_{N}^{-1}$. Primjenjujući tehniku za uklanjanje m-D djelova, ostaje samo podskup vremensko-frekvencijskih tačaka koje formiraju vektor dostupnih STFT vrijednosti, u oznaci $\mathbf{STFT}_{CS}$. Elementi ovog vektora su $STFT\alpha(i)=STFT(n_i,k_i)$, gdje je $(n_i,k_i)\in \mathbb{N}_A$, skup vremensko-frekvencijskih tačaka koje su zadržane nakon primjene L-statistike (dostupne vrijednosti). One zadovoljavaju:
\begin{equation}
\mathbf{STFT}_{CS}=\mathbf{A}_{CS}\mathbf{S}_{\alpha},
\end{equation} 
gdje je mjerna matrica $\mathbf{A}_{CS}$ formirana na osnovu matrice $\mathbf{A}$ odbacivanjem vrsta koje odgovaraju eliminisanim vremensko-frekvencijskim tačkama.

 Vektor  $\mathbf{STFT}_{CS}$ može biti interpretiran kao vektor dostupnih odbiraka (mjerenja) u kontekstu kompresivnog odabiranja. Tada RB komponente mogu biti rekonstruisane rješavanjem minimizacionog problema  \cite{csiet1}:
\begin{equation}
\textrm{ min} \left\Vert \mathbf{S}_{\alpha} \right\Vert _1 \textrm{  subject to  } \mathbf{STFT}_{CS}=\mathbf{A}_{CS}\, \mathbf{S}_{\alpha}.\label{problem}
\end{equation}

Problem može biti riješen korišćenjem neke od standardnih tehnika za rekonstrukciju, na primjer, OMP Algoritmom \ref{Norm0Alg}.

\subsection{Numerički rezultati}
\begin{primjer}
Razmatra se signal
\begin{align}\label{sigex}
s(n)=\sum_{i=1}^{K}\sigma_{B_i}e^{j2\pi [a(\frac{n}{N})^2+b_i\frac{n}{N}]+j\varphi_i}
+\sum_{i=1}^{D}\sigma_{R_i}e^{jA_{R_i}\sin(\omega_{R_{i}}n+\Theta_i)+j2\pi c_i \frac{n}{N}+j2\pi d_i (\frac{n}{N})^2}.
\end{align}
Ovaj signal odgovara jednom \textit{range} binu radarske slike. Prva suma odgovara modelu RB reflektora sa nekompenzovanim ubrzanjem, gdje figuriše nepoznati \textit{chirp rate} $a$. Druga suma modeluje m-D. Dužina signala je $N=1024$.

Razmatra se RB sa $K=4$ komponente. Parametri komponenti su: $\sigma_{B_i}=[1, 0.5,1.5,1]$, $b_i=[125, -125, 245, -255]$ i $\varphi_i=[0,0, {\pi}/{4},-\pi/{3}]$ for $i=1,2,3,4$, respektivno. Nepoznati \textit{chirp rate} je $a=360$, dok se m-D sastoji od dvije komponente, $D=2$, a njegovi parametri su $\sigma_{R_i}=[7, 5]$, $\Theta_i=[0,\pi/2]$,  $A_{R_i}=[90,160]$, $\omega_{R_{i}}=[2.5,1.95]$, $c_i=[0,0]$ i $d_i=[0,0]$, za $i=1,2$, respektivno. Prvo primjenjujemo L-statistiku za razdvajanje RB i m-D djelova, korišćenjem procedure iz odjeljka \ref{removal}. 
Parametar $\alpha=280$ za demodulisanje (tzv. dečirpovanje) se dobija korišćenjem algoritma prezentovanog u sekciji \ref{estimacija}. Rezultati su predstavljeni na slici \ref{fig1rb}. 

 Inicijalna STFT, računata pomoću prozora širine $M=128$ je predstavljena na slici \ref{fig1rb} (a), dok je odgovarajuća reprezentacija sa sortianim vrijednostima data na slici \ref{fig1rb} (c). STFT signala demodulisanog pomoću optimalnog $\alpha$ predstavljena je na slici \ref{fig1rb} (b), dok je odgovarajuća reprezentacija sa sortiranim STFT vrijednostima data na slici \ref{fig1rb} (d). Furijeova transformacija (DFT) originalnog signala prezentovana je na slici \ref{fig1rb} (e). Odvajanje m-D i RB djelova na bazi STFT ne daje zadovoljavajuće rezultate, pošto su uklonjeni i veći djelovi RB dijela. Furijeova transformacija dobijena sabiranjem  $Q=40\%$ najmanjih apsolutnih vrijednosti sortirane STFT sa slike  Fig. \ref{fig1rb} (d) data je na slici \ref{fig1rb} (g). Furijeova transformacija dobijena sabiranjem  STFT vrijednosti prema  (\ref{suma3}) prikazana je na slici \ref{fig1rb} (h). 
 
 Ovaj primjer ilustruje činjenicu da RB rekonstrukcija korišćenjem  izraza (\ref{suma3}) daje visoko koncentrisane impulse. Međutim, rezultati se mogu poboljšati ukoliko se problem razmatra u kontekstu kompresivnog odabiranja.  STFT signala $s_{\alpha}(n)$ demodulisanog optimalnim $\alpha$, računata sa nepreklopljenim prozorima, predstavljena je na slici \ref{fig2RBB} (a). Ovdje je korišćen prozor širine $M=32$. Procedura za L-statistiku, prezentovana u odjeljku \ref{removal}, primijenjena je na ovu STFT, pri čemu je uklonjeno $U=40\%$ vremensko-frekvencijskih tačaka sa najvećim vrijednostima, i $D=20\%$ tačaka sa najmanjim vrijednostima sortirane STFT, slika \ref{fig2RBB} (b).  
 Nakon uklanjanja m-D dijela signala, problem (\ref{problem}) je riješem OMP algoritmom (odnosno, Algoritam \ref{Norm0Alg}).
 
 Rezultati rekonstrukcije prikazani su na slici \ref{fig2RBB} (e). Kako bi naglasili tačnost rekonstrukcije, na slici \ref{fig2RBB} (c) prikazan je originalni RB definisan izrazom (\ref{sigex}) in Fig. 2c. On se poredi sa nekompenzovanim RB dijelom koji je dobijen računanjem STFT nad inverznom DFT od koeficijenata koji su prikazani na slici \ref{fig2RBB} (e), modulisanih optimalnim parametrom $\alpha$.
 
 \begin{figure}[!htb]
 	\centering
 	\includegraphics[scale=1]{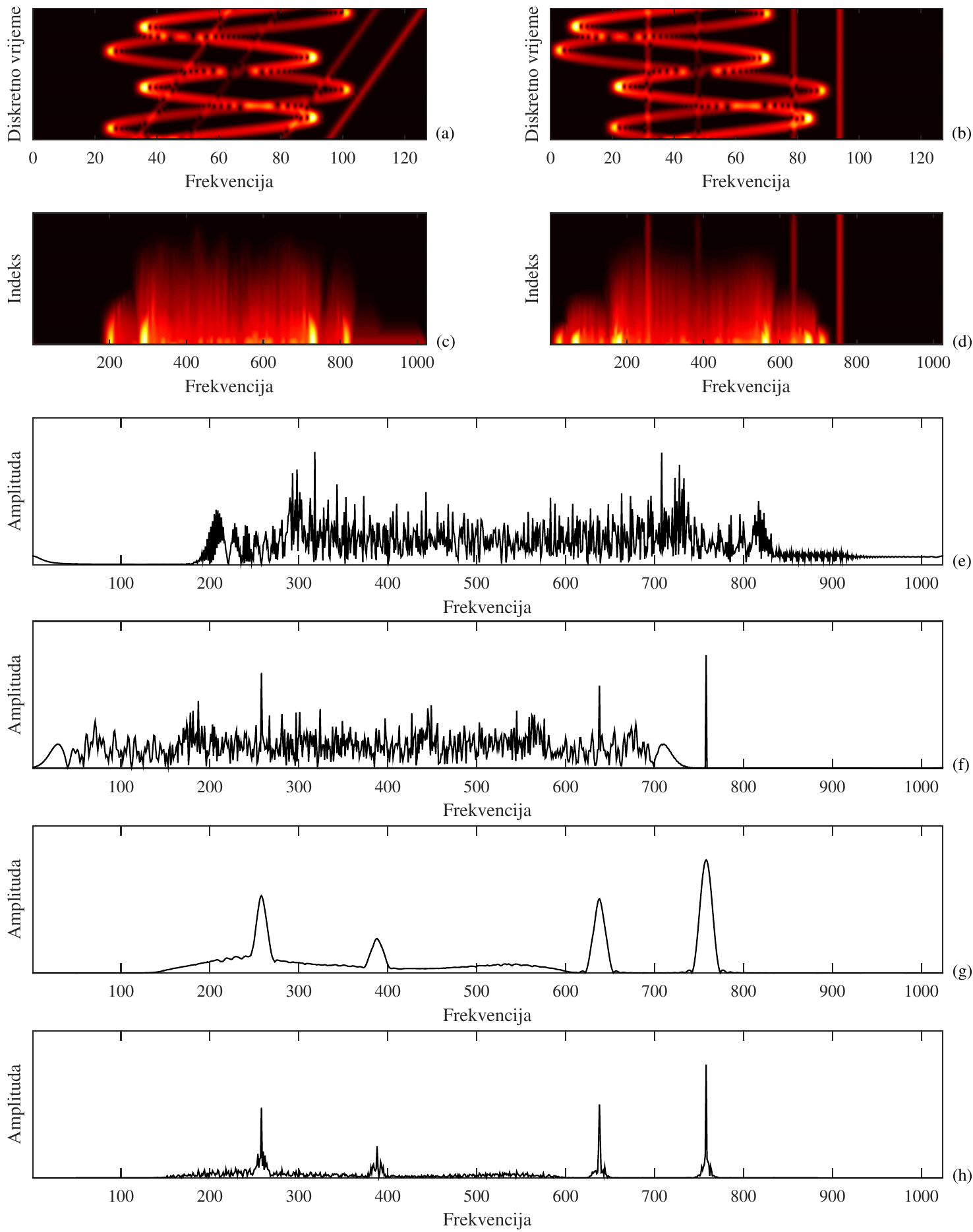} %zove se fig2.m
 	\caption[Razdvajanje RB i m-D djelova u slučaju kada nije kompenzovano ubrzanje radarske mete]{Razdvajanje RB i m-D djelova u slučaju kada nije kompenzovano ubrzanje \text{radarske mete}: (a) STFT originalnog signala, koja je izračunata korišćenjem izraza (\ref{stft}); (b) STFT signala demodulisanog sa pogodno odabranim $\alpha$, korišćenjem prezentovanog algoritma; (c) sortirane vrijednosti originalnog signala; (d) sortirane vrijednosti STFT prikazane na slici (b); (e) Furijeova transformacija originalnog signala; (f) Furijeova transformacija demodulisanog signala; (g) Furijeova transformacija dobijena sabiranjem 40\% najmanjih apsolutnih vrijednosti STFT demodulisanog signala po vremenu; (h) Furijeova transformacija dobijena sabiranjem najmanjih 40\% apsolutnih vrijednosti STFT demodulisanog signala po vremenskom indeksu, dobijene korišćenjem izraza (\ref{suma3}).}
 	\label{fig1rb}
 \end{figure}

 \begin{figure}[tb]
 	\centering
 	\includegraphics[scale=1]{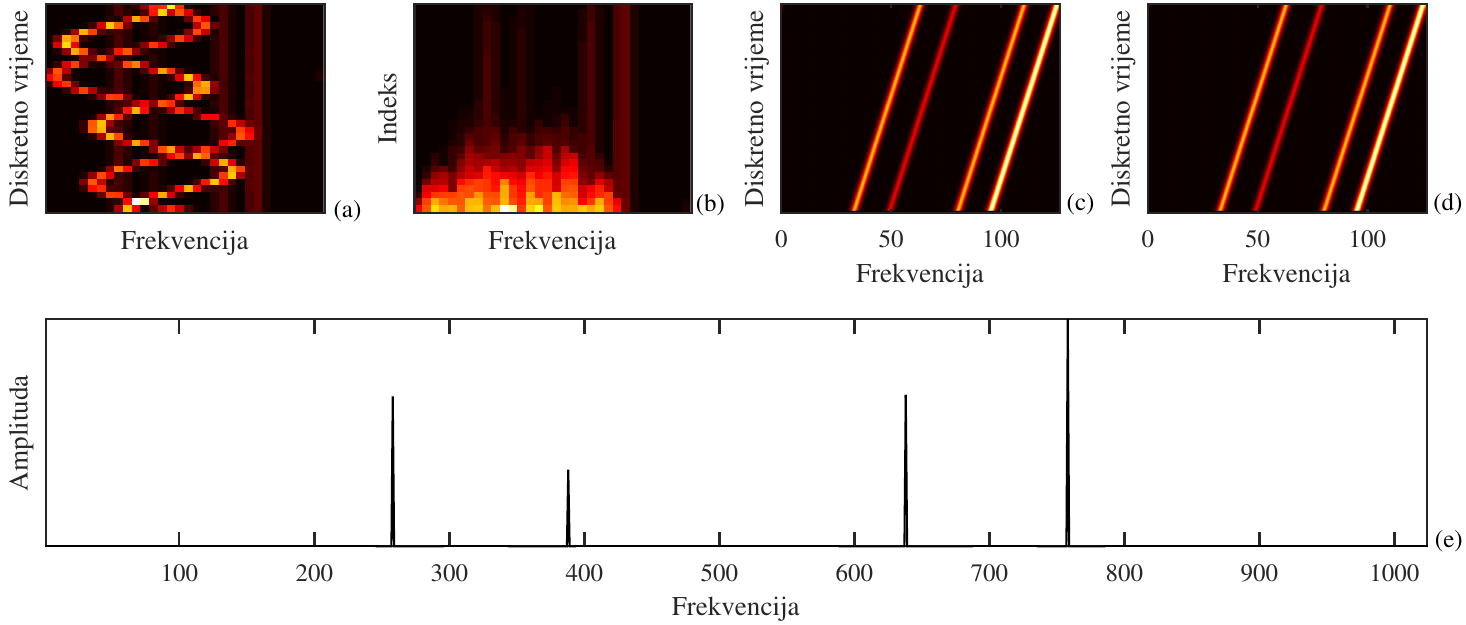}
 	\caption[STFT sa nepreklopljenim prozorom kao baza za CS rekonstrukciju RB dijela signala.]{STFT sa nepreklopljenim prozorom kao baza za CS-rekonstrukciju RB dijela signala: (a) STFT adekvatno demodulisanog signala (pomoću parametra $\alpha$ koji je dobijen prezentovanim algoritmom za pretragu), računata po izrazu (\ref{stft2});  (b) STFT  nakon sortiranja po vremenskom indeksu; (c) STFT originalnog RB dijela definisanog izrazom (\ref{sigex});  (d) STFT rekonstruisanog RB dijela, koja je računata po izrazu (\ref{stft}) za DFT koeficijente prikazane na slici (e) -- ovaj signal je modulisan parametrom $\alpha$ tako da formira originalni RB; (e) DFT koeficijenti koji odgovaraju stacionarnim komponentama demodulisanog signala, koji je rekonstruisan korišćenjem prezentovanog CS pristupa.}
 	\label{fig2RBB}
 \end{figure}
\end{primjer}

\section{Dekompozicija multivarijantnih signala}

Kao što je rečeno, signali sa vremenskim varijacijama spektralnog sadržaja ne mogu se jednostavno okarakterisati standardnom Furijeovom analizom. Oni se uobičajeno proučavaju u okviru vremensko-frekvencijske analize \cite{dekompozicija,dekompozicija2,dekompozicijaLFM,dekompozicija3,dekompozicija4,tfsa,VKLJS1,LJSMDTTB}. Istraživanja u ovoj oblasti dovela su do razvoja brojnih reprezentacija i algoritama koji su namijenjeni procesiranju univarijantnih signala. Oni se uobičajeno opisuju kroz tzv. amplitudske i frekvencijski modulisane oscilacije, \cite{tfsa}, \cite{mandic}.

Skorašnji progres u razvoju senzorske tehnologije izazvao je povećano istraživačko interesovanje za vremensko-frekvencijsku analizu multikanalnih (multivarijantnih ili multidimenzionih) podataka. Razvoj senzorske tehnologije doveo je do pojave koncepta multivarijantnih podataka. Novouvedeni koncepti bivarijantnih i trivarijantnih oscilacija (npr. 3D inercijalni senzori, 3D anemometri \cite{mandic}), ali i generalizacija ovog koncepta na prozivoljan broj kanala, otvorili su mogućnost korišćenja međukanalnih zavisnosti signala, kroz alate vremensko-frekvencijske analize \cite{TSPMV,boashmv,bivariate}. 

Koncept multivarijantnih modulisanih oscilacija predložen je u \cite{TSPMV},  uz pretpostavku da jedna zajednička oscilacija najbolje odgovara individualnim oscilacijama iz svih kanala. Drugim riječima, trenutna frekvencija može okarakterisati multikanalne podatke u smislu objedinjene frekvencije za sve pojedinačne kanale. Ona se definiše kao prosječna trenutna frekvencija svih pojedinačnih kanala, sa pridruženim odgovarajućim težinskim parametrima. U cilju estimacije objedinjene trenutne frekvencije multikanalnih signala, tzv \textit{syncrosqeezed} transformacija, koja spada u kategoriju visoko koncentrisanih vremensko-frekvencijskih reprezentacija, nedavno je proširena na multivarijantni model \cite{mandic}. U cilju ekstrakcije lokalne dinamike oscilacija multivarijatnih signala tzv. \textit{wavelet ridge} algoritam je takođe prilagođen u ovom kontekstu \cite{TSPMV}. Drugi, veoma popularan koncept, tzv.  \textit{empirical mode decomposition -- EMD}, je takođe proučavan za multivarijantne podatke, \cite{emd1,emd2,emd3,emd4,emd5}. Međutim, uspješna dekompozicija multikomponentnih signala zasnovana na EMD pristupu je moguća  samo za signale koji se ne preklapaju u vremensko-frekvencijskoj ravni.

Zbog mogućnosti visoko-koncentrovane reprezentacije signala i drugih poželjnih osobina, Wigner-ova distribucija se uobičajeno koristi u brojnim estimatorima trenutne frekvencije koji su razvijeni u okviru vremensko-frekvencijske analize \cite{tfsa, VKLJS1,LJSMDTTB}. Međutim, u slučaju multikomponentnih signala, kao što je već rečeno, pojavljuju se nepoželjni kros-članovi, nekada u potpunosti maskirajući prisustvo auto-članova. Usljed toga, razvijene su brojne druge reprezentacije, sa namjerom da se očuva koncentracija Wigner-ove distribucije, a istovremeno suzbiju kros-članovi, u koje spada i S-metod \cite{tfsa}, predstavljen u sekciji \ref{smsmsm}, koji je korišćen kao baza za dekompoziciju multikomponentnih signala, \cite{dekompozicija}. Takav oblik dekompozicije omogućava nezavisnu analizu i karakterizaciju komponenti signala, omogućavajući estimaciju trenutne frekvencije svake nezavisne komponente \cite{dekompozicija,dekompozicija2,dekompozicija3,dekompozicija4,dekompozicijaLFM}.

U ovoj sekciji, izučavaćemo Wigner-ovu distribuciju primijenjenu na dekompoziciju multivarijantnih multikomponentnih signala. Biće pokazano da jaka međuzavisnost modulacija pojedinačnih komponenti iz različitih kanala u objedinjenoj vremensko-frekvencijskoj analizi dovodi do redukcije neželjenih kros-članova. 
Matrica inverzne multivarijantne Wigner-ove distribucije će biti dekomponovana na sopstvene vektore, koji sadrže komponente signala u vidu njihove linearne kombinacije. Koeficijenti ove linearne kombinacije će biti proglašeni za optimizacione varijable minimizacije mjere koncentracije primjenom algoritma zasnovanog na metodu najbržeg spuštanja. Ovaj algoritam traži one koeficijente linearne kombinacije koji daju najbolju moguću koncentraciju pojedinačnih komponenti. Upravo zahvaljujući činjenici da koristi međuzavisnosti signala iz pojedinačnih kanala, ova dekompozicija će biti primjenjiva i u slučaju signala koji se preklapaju u vremensko-frekvencijskoj ravni, i to na takav način da integritet izdvojenih komponenti bude očuvan.

Konvencionalna vremensko-frekvencijska analiza se ne može koristiti za razdvajanje komponenti koje se sijeku u vremensko-frekvencijskoj ravni, a pri tome imaju proizvoljne forme (nestacionarnosti). Preklopljene komponente se mogu pojaviti u mnogim primjenama obrade signala. Na primjer, u obradi radarskih signala, ukoliko se potpis radarske mete preklapa sa klaterom (engl. \textit{clutter}). Algoritam koji će biti predstavljen u ovoj sekciji (publikovan u radu \cite{brajovic_tf1}) podrazumijeva da su dostupni signali sa međusobno nezavisnim fazama. Oni se mogu dobiti polarizacijom, ili pomoću sistema sa više antena \cite{R1}. Algoritmom se mogu tretirati i signali sa malim varijacijama frekvencije, kada su promjene amplitude istog reda kao promjene faze. Takvi su, na primjer, EKG signali. Multivarijantne fome ovih signala se dobijaju pomoću više senzora koji su pozicionirani na različitim lokacijama.

\subsection{Multivarijantni signali i Wigner-ova distribucija}
Neka se razmatra multivarijantni signal
\begin{equation}
\mathbf{x}(t)=
\begin{bmatrix}
a_1(t)e^{j\phi_1(t)}\\
a_2(t)e^{j\phi_2(t)}\\
\vdots \\
a_N(t)e^{j\phi_{N_S}(t)}
\end{bmatrix}
\label{MVSIG}
\end{equation}
koji je dobijen snimanjem signala $x(t)$ čije su vrijednosti kompleksne, pomoću $N_S$ senzora, pri čemu svaki senzor mijenja amplitudu i fazu originalnog signala, tako da važi $a_i(t)\exp(j\phi_i(t))=\alpha_ix(t)\exp(j\varphi_i)$.
U slučaju kada je mjereni signal realan, uobičajeno je da se razmatra njegova analitički produžena forma
$x(t)=x_R(t)+j\mathbb{H}\{x_R(t)\},$ gdje je sa $x_R(t)$ označen realni mjereni signal, dok $\mathbb{H}\{x_R(t)\}$ označava njegovu Hilbertovu transformaciju. Analitički signal sadrži samo komponente na nenegativnim frekvencijama, a odgovarajuća realna forma signala može biti jednostavno rekonstruisana. Ovakva forma signala je od velike važnosti za interpretaciju trenutne frekvencije u kontekstu vremensko-frekvencijskih momenata.

Budući da se sve vremensko-frekvencijske reprezentacije mogu posmatrati kao poravnate odnosno "glatke"  (engl. \textit{smoothed}) verzije Wigner-ove distribucije, polazna tačka koncepta analize multivarijantnih signala može biti upravo ova distribucija. Wigner-ova distribucija multivarijantnog signala $\mathbf{x}(t)$ je definisana na sljedeći način:
\begin{align}
\label{mvwd}
WD(\Omega,t)  =\int_{-\infty}^{\infty}\mathbf{x}^{H}(t-\frac{\tau}%
{2})\mathbf{x}(t+\frac{\tau}{2})e^{-j\Omega\tau}d\tau,
\end{align}
gdje $\mathbf{x}^{H}(t)$ označava Hermitsko transponovanje vektora $\mathbf{x}(t)$. 

Inverzna Wigner-ova distribucija je data izrazom:
\begin{align}
\mathbf{x}^{H}(t-\frac{\tau}{2})\mathbf{x}(t+\frac{\tau}{2})  &  =\frac
{1}{2\pi}\int_{-\infty}^{\infty}WD(\Omega,t)e^{j\Omega\tau}d\Omega.
\end{align}

Centar mase na frekvencijskoj osi Wigner-ove distribucije multivarijantnog signala 
$\mathbf{x}(t)$, definisanog sa (\ref{MVSIG}), je
dat izrazom \begin{align}
\left\langle \Omega(t)\right\rangle   =\frac{\int_{-\infty}^{\infty}\Omega
	WD(\Omega,t)d\Omega}{\int_{-\infty}^{\infty}WD(\Omega,t)d\Omega},
\end{align}
odnosno,
\begin{align}
\left\langle \Omega(t)\right\rangle 
=\frac{\frac{d}{jd\tau}\left[  \mathbf{x}^{H}(t-\frac{\tau}{2})\mathbf{x}%
	(t+\frac{\tau}{2})\right]  _{\left\vert \tau=0\right.  }}{\mathbf{x}%
	^{H}(t-\frac{\tau}{2})\mathbf{x}(t+\frac{\tau}{2})_{\left\vert \tau=0\right.
}}  =\frac{1}{2j}\frac{[\mathbf{x}^{H}(t)\mathbf{x}^{\prime}(t)-\mathbf{x}%
	^{\prime H}(t)\mathbf{x}(t)]}{\mathbf{x}^{H}(t)\mathbf{x}(t)},
\label{mulivarpp}
\end{align}
gdje $\mathbf{x}^{\prime}(t)=d\mathbf{x}(t)/dt$ označava izvod vektora signala po vremenu.

Iz izraza (\ref{mulivarpp}) direktno slijedi izraz za trenutnu frekvenciju multivarijantnog signala, dat u sljedećoj formi:
\begin{equation}
\label{mv_if}
\left\langle \Omega(t)\right\rangle =\frac{\sum_{n=1}^{N_S}\phi_{n}^{\prime
	}(t)a_{n}^{2}(t)}{\sum_{n=1}^{N_S}a_{n}^{2}(t)}.
\end{equation}

Ukoliko je multivarijantni signal dobijen tako što je senzorom sniman monokomponentni signal $x(t)$ zadat u formi  $a_i(t)\exp(j\phi_i(t))=\alpha_ix(t)\exp(j\varphi_i)$ gdje je $x(t)=A(t)\exp(j\psi(t))$ i $|dA(t)/dt| \ll |d\psi(t)/dt|$,  tada važi $\left\langle \Omega(t)\right\rangle =d\psi(t)/ dt$, budući da je $d\phi_i(t)/ dt=d\psi(t)/ dt$.

Uslov za varijacije amplituda i faza realnih monokomponentnih signala $a_i(t)\cos(\phi_i(t))$  može se definisati tzv. Bedrosian-ovom produktnom teoremom, \cite{BB}.  Po navedenoj teoremi, kompleksni analitički signal $a_i(t)\exp(j\phi_i(t))=a_i(t)\cos(\phi_i(t))+j\text{H}\{a_i(t)\cos(\phi_i(t))\}$
predstavlja validnu reprezentaciju realnog signala sa amplitudskim i faznim varijacijama $a_i(t)\cos(\phi_i(t))$ ukoliko je spektar dijela $a_i(t)$ nenulti samo unutar frekvencijskog opsega  $|\Omega|<B
$ a spektar od $\cos(\phi_i(t)) $ zauzima nepreklapajući opseg na višim frekvencijama. Signal se može smatrati monokomponentnim ukoliko je spektar od $a_i(t)$ niskopropusnog tipa.

Predstavljena analiza se može lako generalizovati na druge vremensko-frekvencijske i \textit{time-scale} reprezentacije.

Devijacija spektralnog sadržaja signala od trenutne frekvencije opisuje se lokalnim momentima drugog reda (trenutne širine frekvencijskog opsega). Izraz za trenutnu širinu frekvencijskog opsega se dobija na osnovu:
\begin{align}
\sigma_{\Omega}^{2}(t)  &  =\frac{1}{2\pi\mathbf{x}^{H}(t)\mathbf{x}(t)}%
\int_{-\infty}^{\infty}\Omega^{2}WD(t,\Omega)d\Omega-\left\langle
\Omega(t)\right\rangle ^{2}\notag \\
&  =\frac{-\frac{d^{2}}{d\tau^{2}}\left.  \left[  \mathbf{x}^{H}\left(
	t-\frac{\tau}{2}\right)  \mathbf{x}\left(  t+\frac{\tau}{2}\right)  \right]
	\right\vert _{\tau=0}}{\mathbf{x}^{H}(t)\mathbf{x}(t)}-\left\langle
\Omega(t)\right\rangle ^{2}.%
\end{align}

Za signal (\ref{MVSIG}) ova veličina dobija sljedeću formu:
\begin{align}
\sigma_{\Omega}^{2}(t)=\frac{\sum_{n=1}^{N_S}{(a^{\prime}_n(t))^2}-\sum_{n=1}^{N_S}{a_n(t) a_n''(t)}}{2\sum_{n=1}^{N_S}{a_n^2(t)}} .%
\end{align}

Uopšteno govoreći, u slučaju multikomponentnih signala, komponente su lokalizovane oko više od jedne trenutne frekvencije.

\subsection{Multikomponentni signali}
Razmatra se multikomponentni signal
\begin{align}
x(t)= \sum_{p=1}^P x_p(t)
\end{align}
čije komponente imaju formu
$
x_p(t)=A_p(t) e^{j\psi_p(t)},
$
gdje $A_p(t)$ označava amplitude komponenti, koje imaju sporovarirajuću dinamiku u poređenju sa varijacijama faza $\psi_p(t)$, odnosno, $|dA_p(t)/dt| \ll |d\psi_p(t)/dt|$. U posmatranom slučaju, odgovarajući multivarijantni signal je dat sljedećim izrazom:
\begin{equation}
\mathbf{x}(t)=
\sum_{p=1}^P
\begin{bmatrix}
\alpha_{p1}x_p(t)e^{j\varphi_{p1}}\\
\alpha_{p2}x_p(t)e^{j\varphi_{p2}}\\
\vdots \\
\alpha_{p{N_S}}x_p(t)e^{j\varphi_{p{N_S}}}
\end{bmatrix}.
\label{MVmcSIG}
\end{equation}

Pojedinačne komponente $x_1(t),x_2(t),\dots,x_P(t)$,  mjerene različitim senzorima, razlikuju se u amplitudama i fazama, ali dijele zajedničku trenutnu frekvenciju $\Omega_{p}(t)=d\psi_p(t)/dt$ koja odgovara $\left\langle \Omega_p(t)\right\rangle$ u izrazu (\ref{mv_if}), gdje je $p$ indeks komponente.

Wigner-ova distribucija posmatranog multivarijantnog signala je
\begin{equation*}
WD(\Omega,t)=
\sum_{p=1}^P\sum_{q=1}^P \sum_{i=1}^{N_S} \int_{-\infty}^{\infty}
\alpha_{pi}\alpha_{qi}
x_{p}(t+\tfrac{\tau}{2})x^*_{q}(t-\tfrac{\tau}{2})
e^{j(\varphi_{pi}-\varphi_{qi})}
e^{-j\Omega\tau}d\tau,
\end{equation*}
gdje je $i$ indeks senzora. Ona može biti zapisana u vidu sume kros-komponenti i auto-komponenti, odnosno
\begin{align}
%\label{auto_cross}
WD(\Omega,t)  = &
\sum_{p=1}^P  \sum_{i=1}^{N_S} \alpha_{pi}^2 \int_{-\infty}^{\infty}
x_{p}(t+\tfrac{\tau}{2})x^*_{p}(t-\tfrac{\tau}{2})
e^{-j\Omega\tau}d\tau  \label{MVMC}\\
& +\sum_{p=1}^P\sum_{\substack{q=1\\ q\ne p}}^P \sum_{i=1}^{N_S} \alpha_{pi}\alpha_{qi} \int_{-\infty}^{\infty}
x_{p}(t+\tfrac{\tau}{2})x^*_{q}(t-\tfrac{\tau}{2})
e^{j(\varphi_{pi}-\varphi_{qi})}
e^{-j\Omega\tau}d\tau \nonumber\\
&=WD_{AT}(\Omega,t)+WD_{CT}(\Omega,t). \nonumber
\end{align}
Fazni pomjeraji komponenti multivarijantnog signala u izrazu (\ref{MVMC}) se poništavaju u auto-članovima $WD_{AT}(\Omega,t)$. Ovo važno svojstvo zapravo implicira da se auto-članovi, dobijeni iz svakog kanala multivarijantnog signala, sabiraju u fazi, nezavisno od različitih inicijalnih faza u pojedinačnim komponentama signala. U kros-članovima, fazni pomjeraji se ne poništavaju u rezultujućem izrazu $WD_{CT}(\Omega,t)$, a vodi sabiranju koje nije u fazi. Kros-članovi u multivarijantnom slučaju predstavljaju sumu od $N_S$ signala sa proizvoljnim (slučajnim) fazama. Posljedično, kros-članovi će biti redukovani u poređenju sa slučajem Wigner-ove distribucije univarijantnog signala. Dakle, za veliko $N_S$ može se očekivati da auto-članovi budu izraženi, a da kros-članovi teže malim vrijednostima u poređenju sa auto-članovima. 

Očekuje se, drugim riječima, da se kros-članovi, za veliki broj senzora $N_S$, ponašaju kao Gausove slučajne varijable srednje vrijednosti nula, čija varijansa zavisi od vrijednosti kros-članova, $\operatorname{var} \{WD(\Omega,t)\}=\sigma^2(WD_{CT}(\Omega,t))$. Za dati signal, auto-članovi su deterministički, budući da oni ne zavise od slučajnih faza, kao što se može vidjeti u odgovarajućem dijelu izraza posmatrane Wigner-ove distribucije $WD_{AT}(\Omega,t)$. Navedeno znači da za veliko $N_S$ važi:
\begin{align} 
WD(\Omega,t) \sim \mathcal{N} (WD_{AT}(\Omega,t),\sigma^2(WD_{CT}(\Omega,t))). 
\end{align}
\subsection{Inverzija i dekompozicija signala}
Inverzna Wigner-ova distribucija multivarijantnog signala u analognom domenu data je sljedećim izrazom:
\begin{equation}
\mathbf{x}^{H}(t_2)\mathbf{x}(t_1) =\frac{1}{2\pi}\int_{-\infty}^{\infty}WD\left(
\frac{t_{1}+t_{2}}{2},\Omega\right)  e^{j\Omega(t_{1}-t_{2})}d\Omega.
\end{equation}
Diskretizacijom ugaone frekvencije, $\Omega=k\Delta\Omega,$ i vremena, $t_{1}%
=n_{1}\Delta t$, $t_{2}=n_{2}\Delta t$, uz adekvatnu definiciju diskretnih vrijednosti, lako se dobija:
\begin{equation}
\mathbf{x}^H(n_{2})\mathbf{x}(n_{1})=\tfrac{1}{K+1}\sum_{k=-K/2}^{K/2}WD\left(  \frac
{n_{1}+n_{2}}{2},k\right)  e^{j\frac{\pi}{K+1}k(n_{1}-n_{2})}. \label{iwd1}%
\end{equation}
Nakon uvođenja sljedeće notacije:
\begin{equation}
R(n_{1},n_{2})=\tfrac{1}{K+1}\sum_{k=-K/2}^{K/2}WD\left(  \frac{n_{1}+n_{2}%
}{2},k\right)  e^{j\frac{\pi}{K+1}k(n_{1}-n_{2})}, \label{rr}%
\end{equation}
važi da je:
\begin{equation}
R(n_{1},n_{2})=\mathbf{x}^H(n_{2})\mathbf{x}(n_1). \label{eis}%
\end{equation}

Dakle, u slučaju multikomponentnih multivarijantnih signala, inverzija produkuje matricu sa elementima zadatim izrazom:
\begin{gather}
R(n_{1},n_{2})=\sum_{i=1}^{N_S}\sum_{p=1}^P\sum_{q=1}^P\alpha_{pi}\alpha_{qi}x_{p}(n_1)x_{q}^{*}(n_2)e^{j(\varphi_{pi}-\varphi_{qi})}.
\end{gather}
Ukoliko se sada iskoristi pretpostavka da se kros-članovi u WD multivarijantnog signala mogu zanemariti u poređenju sa auto-članovima koji su sumirani u fazi, dobija se:
\begin{gather}
R(n_{1},n_{2})=\sum_{i=1}^{N_S}\sum_{p=1}^Pa_{pi}^{2}x_p(n_1)x_p^{*}(n_2)
=\sum_{p=1}^PB_px_p(n_1)x_p^{*}(n_2)
\end{gather}
gdje je $B_p=\sum_{i=1}^{N_S}\alpha_{pi}^{2}$.
Kao u slučaju bilo koje kvadratne matrice, dekompozicija na sopstvene vrijednosti matrice $\mathbf{R}$ dimenzija $K\times K$ daje
\begin{equation}
\mathbf{R=Q}\mathbf{\Lambda}\mathbf{Q}^{T}=\sum_{p=1}^{K}\lambda_{p}\mathbf{q}%
_{p}(n)\mathbf{q}_{p}^{\ast}(n), \label{eig}%
\end{equation}
gdje $\lambda_{p}$ predstavlja sopstvene vrijednosti, dok $\mathbf{q}_{p}(n)$ označava odgovarajuće sopstvene vektore matrice $\mathbf{R}$. Treba primijetiti da su sopstveni vektori $\mathbf{q}_{p}(n)$ po definiciji ortonormalni.  

Za $P$-komponentni signal, u odsustvu šuma, elementi posmatrane matrice su
\begin{gather}
R(n_{1},n_{2})=\sum_{p=1}^P\lambda_p{q}_p(n_1){q}_p^{*}(n_2).
\end{gather}

Može se razmatrati nekoliko specijalnih slučajeva:
\begin{enumerate}
	
\item U slučaju Wigner-ove distribucije univarijantnog signala, sam signal je jednak sopstvenom vektoru  $\mathbf{q}_{1}(n)$ matrice $\mathbf{R}$, sa faktorom proporcionalnosti koji je kompleksna konstanta \cite{dekompozicija}, gdje su odgovarajuće sopstvene vrijednosti  $\lambda_{1}=E_{x}$, $\lambda_{2}=0,\dots,\lambda_{K}=0$. Činjenica da inverzija Wigner-ove distribucije produkuje samo jednu nenultu sopstvenu vrijednost, koristi se u provjeri da li je data posmatrana dvodimenziona funkcija validna Wigner-ova distribucija.

\item Ukoliko se komponente multikomponentnog univarijantnog signala ne preklapaju u vremensko-frekvencijskoj ravni, tada je moguće iskoristiti svojstvo (\ref{smwd}) koje S-metod definiše kao sumu pseudo-Wigner-ovih distribucija pojedinačnih komponenti signala \cite{dekompozicija}:
\begin{equation}
SM(n,k)=\sum_{p=1}^{P}WD_p(n,k).
\end{equation}

Pošto su nepreklapajuće komponente ortogonalne, dekompozicija na sopstvene vrijednosti i sopstvene vektore u slučaju univarijantnih (i multivarijantnih) multikomponentnih signala će dati:
\begin{equation}
B_p\mathbf{x}_{p}(n)=\lambda_p\mathbf{q}_{p}(n),~ p=1,2,\dots,P,
\end{equation}
gdje je $B_p$ konstanta. Treba primijetiti da je, po definiciji, energija odgovarajućeg sopstvenog vektora jednaka $1$, odnosno,
\begin{equation}
\left\Vert \mathbf{q}_{p}(n)\right\Vert ^{2}=1.
\end{equation}
Može se, dakle, zaključiti da je
\begin{equation}
B_{p}\mathbf{x}_p(n)\mathbf{x}^*_p(n)=\left(  \sqrt{\lambda_{p}}\mathbf{q}%
_{p}(n)\right)  \left(  \sqrt{\lambda_{p}}\mathbf{q}_{p}(n)\right)  ^{\ast}%
\end{equation}
odnosno
\begin{equation}
\lambda_{p}=\left\Vert \sqrt{\lambda_{p}}\mathbf{q}_{p}(n)\right\Vert
^{2}=\left\Vert B_{p}\mathbf{x}_{p}(n)\right\Vert ^{2}=\sum_{n=-K/2}^{K/2}%
B_px_{p}^{2}(n)=B_pE_{{x}_p},\emph{\ }%
\end{equation}
gdje je $E_{x_{p}}$ energija $p$-te komponente signala. Sopstveni vektor $\mathbf{q}_{p}(n)$ je jednak vektoru sa vrijednostima komponente signala $\mathbf{x}_{p}(n)$, uz konstantu proporicionalnosti koja definiše neodređenost u amplitudi i fazi.

\item Ukoliko se komponente signala  $\mathbf{x}_p(n)$ preklapaju  u vremensko-frekvencijskoj ravni, tada dekompozicija signala na pojedinačne komponente nije moguća korišćenjem poznatih tehnika, osim za neke vrlo specifične forme signala, kao što su, na primjer, linearni frekvencijski modulisani signali, korišćenjem \textit{chirplet} transformacije, Radonove transformacije ili sličnih tehnika \cite{dekompozicija_radon}, \cite{dekompozicija_cirplet}, ili sinusoidalno-modulisani signali, korišćenjem inverzne Radonove transformacije \cite{iradon1}, \cite{iradon2}). 

\noindent U opštem slučaju, ove vrste signala ne mogu biti razdvojene (dekomponovane) na pojedinačne komponente u univarijantnom slučaju. Međutim, multivarijantna forma signala mijenja (redukuje) kros-članove u Wigner-ovoj distribuciji, otvarajući na taj način, indirektno, mogućnost dekompozicije signala na komponente čak i u veoma izazovnom kontekstu komponenti preklopljenih u vremensko-frekvencijskoj ravni.   
\end{enumerate}

\subsection{Algoritam za dekompoziciju (razdvajanje i rekonstrukciju)}

Razmatra se multikomponentni signal zadat izrazom (\ref{MVmcSIG}), čije su komponente $\mathbf{x}_{p}$, za $p=1,2,\dots,P$. Podskupovi vremensko-frekvencijskih domena komponenti za koje su one različite od nule, u oznaci $\mathbb D_p$, mogu se djelimično preklapati u vremensko-frekvencijskoj ravni. 

Uvedimo i realnu pretpostavku da ne postoje komponente signala čiji se djelovi vremensko-frekvencijskih domena $\mathbb{D}_p$ potpuno preklapaju sa odgovarajućim djelovima domena drugih komponenti, kao i da važi $D_1 \le D_2 \le \dots \le D_P$, gdje je $D_p$ površina dijela (podskupa) vremensko-frekvencijskog domena $\mathbb D_p$ u kojem je $p$-ta komponenta nenulta. 

Prva komponenta signala može se izraziti kao linearna kombinacija sopstvenih vektora 
$\mathbf{q}_p$ sa koeficijentima $\eta_{1p}$, tako da se dobija:
\begin{equation}
\mathbf{x}_{1}=\eta_{11}\mathbf{q}_{1}+\eta_{21}\mathbf{q}_{2}+\dots+\eta_{P1}\mathbf{q}_{P.}
\end{equation} 

Pošto je realno pretpostaviti da su komponente dobro koncentrisane u vremensko-frekvencijskoj ravni, nepoznati koeficijenti $\eta_{p1}$ se mogu odrediti na osnovu mjera koncentracije. U tu svrhu, formira se linearna kombinacija baznih vektora  $\mathbf{q}_{p}$, sa težinskim koeficijentima $\beta_{p}$,
$p=1,2,\ldots,P$, koja ima sljedeću formu:
\begin{equation}
\label{bete1}
\mathbf{y}=\beta_{1}\mathbf{q}_{1}+\beta_{2}\mathbf{q}_{2}+\dots+\beta_{P}\mathbf{q}_{P,}
\end{equation}
i računa se mjera koncentracije $\mathcal{M}\left\{ \operatorname{TFR}(n,k) \right\}$ vremensko-frekvencijske reprezentacije $\operatorname{TFR}(n,k)$
normalizovanog signala $\mathbf{y}/\Vert \mathbf{y}\Vert_2$. Izbor vremensko-frekvencijske reprezentacije $TFR(n,k)$ ovdje nije krucijalan. Može se koristiti spektrogram kao najjednostavnija vremensko-frekvencijska reprezentacija. Rješavanjem problema minimizacije mjere koncentracije, dobija se globalni minimum koji odgovara najbolje koncentrisanoj komponenti signala. 

Najjednostavniji način za rješavanje ovog problema bilo bi korišćenje ,,$\ell_0$-norme'' kao mjere koncentracije  $\operatorname{TFR}(n,k)$ i direktno pretraživanje po mogućim vrijednostima koeficijenata $\beta_{p}$,
$p=1,2,\ldots,P$. U tom slučaju, koeficijenti $\eta_{p1}$ predstavljaju rješenje minimizacionog problema:
\begin{equation}
[\eta_{11},\eta_{21},\dots,\eta_{P1}]=\arg \min_{\beta_{1},\dots,\beta_{P}}\Vert \operatorname{TFR}(n,k)
\Vert_0.
\end{equation}
Za ove vrijednosti koeficijenata,  $\Vert \operatorname{TFR}(n,k)
\Vert_0$ je jednaka površini $D_1$ dijela domena u najbolje koncentrisane komponente. Ako su bilo koje dvije najmanje površine jednake, i dalje ćemo naći jednu od njih.

Treba uočiti da razmatrani minimizacioni problem ima više lokalnih minimuma, budući da će koeficijenti $\beta_{p}$ u $\mathbf{y}=\beta_{1}\mathbf{q}_{1}+\beta_{2}\mathbf{q}_{2}+\dots+\beta_{P}\mathbf{q}_{P}$
koji odgovaraju bilo kojoj komponenti signala  $\mathbf{x}_{p}$ takođe izazvati pojavu lokalnog minimuma mjere koncentracije, koji je jednak površini dijela vremensko-frekvencijskog domena u kojem je data komponenta različita od nule. Dodatno, i bilo koja linearna kombinacija od  $K<P$ komponenti signala $\mathbf{x}_p$ takođe će produkovati lokalni minimum koji je jednak uniji odgovarajućih dijelova vremensko-frekvencijskih domena u kojima su te komponente različite od nule.

Nakon što je najbolje koncentrisana komponenta detektovana, odgovarajući sopstveni vektor $\mathbf{q}_1$ zamjenjuje se sa izdvojenom komponentom. Izdvojena komponenta se tada uklanja iz preostalih sopstvenih vektora $\mathbf{q}_k$, oduzimanjem projekcije izdvojene komponente na vektore $\mathbf{q}_p$,
$p=2,3,\ldots,P$ (procedura deflacije signala, \cite{SDP}). Zatim se procedura ponavlja sa novim setom vektora 
$\mathbf{q}_p$
formiranjem signala $\mathbf{y}=\beta_{2}\mathbf{q}_{2}+\dots+\beta_{P}\mathbf{q}_{P}$. Variranjem koeficijenata $\beta_{p}$ nalazi se novi globalni minimum mjere koncentracije, koji odgovara drugoj komponenti signala. Procedura se iterativno ponavlja $P$ puta.

 Međutim,  u praktičnim aplikacijama ne može da se koristi ni direktno pretraživanje, ni ,,$\ell_0$-norma'', što je diskutovano u prethodnim poglavljima ove disertacije. U literaturi je razvijeno više pristupa za rješavanje optimizacionih problema sa više lokalnih minimuma. Uopšteno govoreći, mogu se razlikovati tri velike klase ovakvih pristupa: deterministički \cite{Deter}, stohastički \cite{SPST,SPST1} i heuristički (npr. optimizacija mravlje kolonije \cite{ANT}, genetički algoritmi,  simulirano kaljenje \cite{SA}, pretraga planinarenjem \cite{HC}, \textit{particle swarm optimization}, neuralne mreže \cite{neurel_neretva}, itd.). U ovoj disertaciji, predstavićemo adaptiranu formu gradijentnog pristupa rješavanju posmatranog minimizacionog problema. Kao u slučaju kompresivnog odabiranja i rekonstrukcije rijetkih signala, $\ell_0$-norma će biti zamijenjena njenim najbližim ekvivalentom -- $\ell_1$-normom.  U nastavku slijedi detaljan opis pristupa dekompoziciji multikomponentnih multivarijantnih signala. Procedura za dekompoziciju je predstavljena Algoritmom \ref{Alg-MV-Decomp}, a odgovarajuća minimizacija sprovodi se Algoritmom  \ref{Alg-MV-Min}.

\begin{algorithm}
		\floatname{algorithm}{Algoritam}
	\caption{Dekompozicija multivarijantnih signala (rekonstrukcija komponenti signala)}
	\label{Alg-MV-Decomp}
	\begin{algorithmic}[1]
		\Input
		\Statex
		\begin{itemize}
			\item Multivarijantni signal $\mathbf{x}(n)$
		\end{itemize}
		\Statex         
		\State Izračunati S-metod $SM(n,k)$ multivarijantnog signala  $\mathbf{x}(n)$ i matrice  $\mathbf{R}$ sa elementima
		$$
		R(n_1,n_2)=\frac{1}{K+1}\sum_{k=-K/2}^{K/2}
		SM\left(  \frac{n_1+n_2}{2},k\right)  
		e^{j\frac{2\pi}{K+1}k(n_1-n_2)},
		$$
		kao u \cite{dekompozicija}. Ukoliko se koristi Wigner-ova distribucija, tada $SM(n,k)$ treba zamijeniti sa $WD(n,k)$, ili se elementi matrice $\mathbf{R}$ računaju kao $R(n_{1},n_{2})=\mathbf{x}^H(n_{2})\mathbf{x}(n_1)$.
		\State Naći sopstvene vektore $\mathbf{q}_i$ i sopstvene vrijednosti $\lambda_i$ matrice $\mathbf{R}$.
		\State $P \gets$  broj nenultih (odnosno, u slučaju zašumljenog signala - značajnih) sopstvenih vrijednosti
		\Repeat
		\State $N_{U}\gets 0$ \Comment Broj ažuriranja
		\For{ $i=1,2,\ldots,P$}
		\State \parbox[t]{\dimexpr\linewidth-\algorithmicindent-\algorithmicindent}{
			Riješiti minimizacioni problem
			$$
			\min_{\beta_1, \ldots, \beta_P} \mathcal{M}\left\{ \operatorname{TFR}\left\{\frac{1}{C} \sum_{p=1}^{P} \beta_p \mathbf{q}_p \right\} \right\}
			\qquad
			\text{subject to }
			\beta_i=1
			$$
			gdje je $\mathcal{M}\{\cdot\}$ mjera koncentracije, $\operatorname{TFR}\{\cdot\}$ je vremensko-frekvencijska reprezentacija signala koji je proslijeđen kao argument, dok se 
			$$\textstyle C=\sqrt{\left\Vert\sum_{p=1}^{P} \beta_p \mathbf{q}_p \right\Vert_2}$$
			koristi za normalizaciju energije kombinovanog signala na $1$. 
			Koeficijenti $\beta_1,\beta_2,\ldots,\beta_P$ se dobijaju kao rezultat minimizacije.
		}\vspace*{1ex}
		
		\If{bilo koje $\beta_p \ne 0$, $p\ne i$}
		\State
		$ \displaystyle
		\mathbf{q}_i \gets \frac{1}{C} \sum_{p=1}^{P} \beta_p \mathbf{q}_p
		$
		\For{$k=i+1,i+2,\ldots,P$}
		\State $s\gets \mathbf{q}_i^H \mathbf{q}_k$
		\State $\mathbf{q}_k \gets \frac{1}{\sqrt{1-|s|^2}} (\mathbf{q}_k - s\mathbf{q}_i)$
		\EndFor
		\State $N_{U} \gets N_{U}+1$
		\EndIf
		
		\EndFor
		\Until{$N_{U}=0$}
		
		\Statex
		\Output
		\Statex
		\begin{itemize}
			\item Broj komponenti signala $P$
			\item Rekonstruisane komponente signala $\mathbf{q}_1, \mathbf{q}_2, \ldots, \mathbf{q}_P$
		\end{itemize}
	\end{algorithmic}
\end{algorithm}

\begin{algorithm}
	\floatname{algorithm}{Algoritam}
	\caption{Procedura za minimizaciju mjere koncentracije pri dekompoziciji signala}
	\label{Alg-MV-Min}
	\begin{algorithmic}[1]
		\Input
		\Statex
		\begin{itemize}
			\item Vektori $\mathbf{q}_1, \mathbf{q}_2, \ldots, \mathbf{q}_P$
			\item Indeks $i$ gdje odgovarajući vektor $\mathbf{q}_i$ treba da ima fiksiran jedinični koeficijent $\beta_i=1$
			\item Zahtijevana tačnost $\varepsilon$
		\end{itemize}
		\Statex         
		\State 
		$
		\beta_p=
		\begin{cases}
		1 & \text{ for } p=i\\
		0 & \text{ for } p\ne i
		\end{cases}
		, \qquad \text{for } p=1,2,\ldots,P
		$
		\State $M_{old} \gets \infty$
		\State $\Delta=0.1$
		\Repeat
		\State $\displaystyle \mathbf{y}\gets \sum_{p=1}^{P} \beta_p \mathbf{q}_p$
		\State $\displaystyle M_{new} \gets \mathcal{M}\left\{ \operatorname{TFR}
		\left\{\frac{\mathbf{y}}{\Vert\mathbf{y}\Vert_2}\right\} \right\}$\vspace*{1ex}
		\If{$M_{new}>M_{old}$}
		\State $\Delta \gets \Delta/2$
		\State $\beta_p \gets \beta_p + \gamma_p, \qquad \text{for }p=1,2,\ldots,P$
		\Comment{Poništiti posljednje ažuriranje koeficijenata}
		\State $\displaystyle \mathbf{y}\gets \sum_{p=1}^{P} \beta_p \mathbf{q}_p$
		
		\Else
		\State $M_{old}\gets M_{new}$
		\EndIf
		\For{ $p=1,2,\ldots,P$}
		\If{$p\ne i$}
		\State $\displaystyle M_r^+ \gets \mathcal{M}\left\{ \operatorname{TFR}
		\left\{\frac{\mathbf{y} + \Delta \mathbf{q}_p}{\left\Vert \mathbf{y} + \Delta \mathbf{q}_p\right\Vert_2}\right\} \right\}$\vspace*{1ex}
		\State $\displaystyle M_r^- \gets \mathcal{M}\left\{ \operatorname{TFR}
		\left\{\frac{\mathbf{y} - \Delta \mathbf{q}_p}{\left\Vert \mathbf{y} - \Delta \mathbf{q}_p\right\Vert_2}\right\} \right\}$\vspace*{1ex}
		\State $\displaystyle M_i^+ \gets \mathcal{M}\left\{ \operatorname{TFR}
		\left\{\frac{\mathbf{y} + j\Delta \mathbf{q}_p}{\left\Vert \mathbf{y} + j\Delta \mathbf{q}_p\right\Vert_2}\right\} \right\}$\vspace*{1ex}
		\State $\displaystyle M_i^- \gets \mathcal{M}\left\{ \operatorname{TFR}
		\left\{\frac{\mathbf{y} - j\Delta \mathbf{q}_p}{\left\Vert \mathbf{y} - j\Delta \mathbf{q}_p\right\Vert_2}\right\} \right\}$\vspace*{1ex}
		\State $\displaystyle \gamma_p \gets  8\Delta \frac{M_r^+ - M_r^-}{M_{new}} +j8\Delta \frac{M_i^+ - M_i^-}{M_{new}}$\vspace*{1ex}
		\Else
		\State $\gamma_p\gets 0$
		\EndIf
		\EndFor
		\State $\beta_p \gets \beta_p - \gamma_p, \qquad \text{za }p=1,2,\ldots,P$
		\Comment{Ažuriranje koeficijenata}\vspace*{1ex}
		\Until{$ \sum_{p=1}^{P}|\gamma_{p}|^2$ je ispod zadate tačnosti $\varepsilon$}
		
		\Statex
		\Output
		\Statex
		\begin{itemize}
			\item Koeficijenti $\beta_1, \beta_2, \ldots, \beta_P$
		\end{itemize}
	\end{algorithmic}
\end{algorithm}

- U prvom koraku, računa se autokorelaciona matrica $\mathbf{R}$ multivarijantnog signala  $\mathbf{x}(n)$ prema (\ref{rr}) odnosno (\ref{eis}). Broj komponenti signala $P$ jednak je broju nenultih sopstvenih vrijednosti matrice $\mathbf{R}$. U slučaju zašumljenih signala, postoje dva pristupa za određivanje broja komponenti:
\begin{itemize}
\item Broj komponenti je pretpostavljen. Dok god je taj broj veći ili jednak od pravog broja komponenti $P$, algoritam će ispravno funkcionisati, gdje će šum biti izdvojen u vidu dodatnih komponenti.
\item Može se uvesti prag koji će odvojiti sopstvene vrijednosti koje odgovaraju komponentama signala od onih koje odgovaraju šumu. Ovaj prag određuje broj komponenti u dekompoziciji.
\end{itemize}

- Kao vremensko-frekvencijska reprezentacija signala može se koristiti spektrogram, S-metod sa užim frekvencijskim prozorom (na primjer, $L_d=1$), ili bilo koja druga adekvatna reprezentacija. Budući da su ove vremensko-frekvencijske reprezentacije kvadratne prirode, kao mjera koncentracije se koristi  ekvivalent $\ell_1$-norme, definisan sa \cite{XXX}:
\begin{equation}
\mathcal{M}\left\{ \operatorname{TFR}(n,k) \right\}
=\sum_{n}\sum_{k} |\operatorname{TFR}(n,k)|^{1/2},
\end{equation}
gdje se sumiranje vrši po svim dostupnim vremenskim i frekvencijskim indeksima $n$ i $k$.

- Minimizacija mjere koncentracije je implementirana korišćenjem metoda najbržeg spuštanja, koji je prezentovan u Algoritmu \ref{Alg-MV-Min}. Ovdje se fiksiraju koeficijenti $\beta_{p}=1$ i variraju realni i imaginarni djelovi preostalih koeficijenata za $\pm \Delta$. Zatim se računa gradijent normalizovane mjere, $\gamma_p$, i dalje koristi pri ažuriranju koeficijenata. Inicijalna vrijednost parametra $\Delta$ je $0.1$ i ona se redukuje kad god ažuriranje koeficijenata ne vodi smanjenju mjere koncentracije.

- Kada se izdvoji $p$-ta komponenta signala, odgovarajući sopstveni vektor $\mathbf{q}_p$ se zamjenjuje tom izdvojenom komponentom. Ekstraktovana komponenta se zatim uklanja iz preostalih sopstvenih vektora $\mathbf{q}_k$ oduzimanjem projekcije izdvojene komponente na vektore $\mathbf{q}_k$, $k=p+1,p+2,\ldots,P$. Na ovaj način, osiguravamo se da $p$-ta komponenta signala neće biti ponovo detektovana.

- Opisana procedura se ponavlja sve dok više nema ažuriranja vektor\^{a} $\mathbf{q}_k$.

U slučaju dvokomponentnog multivarijantnog signala, razmatrani minimizacioni problem je konveksan, sa jednim, globalnim, minimumom. Za trokomponentni signal postoje lokalni minimumi za sve signale koji se mogu dobiti kao suma bilo koje dvije komponente. Ovo je razlog zbog kojeg se procedura za dekompoziciju ponavlja nakon što je pronađen minimum mjere koncentracije. Naime, u sljedećoj iteraciji, par komponenti koji odgovara lokalnom minimumu se razdvaja kao u dvokomponentnom slučaju. Za veći broj komponenti  signala, broj lokalnih minimuma se povećava. Tada je neophodan veći broj ponavljanja procedure, kako bi se komponente razdvojile iterativnim putem. Treba uočiti da gradijenti algoritam može pronaći bilo koji lokalni minimum, gdje svaki od njih odgovara kombinaciji $K<P$ komponenti signala. Ovo znači da svaki lokalni minimum redukuje složenost dekompozicije vektor\^{a} $\mathbf{q}_p$, vodeći do potpune dekompozicije signala iterativnim putem. Detalji su dati u Algoritmu \ref{Alg-MV-Min}. 

	\begin{figure}[!htb]
	\centering
	\includegraphics[trim={0 0 0 2cm},clip]{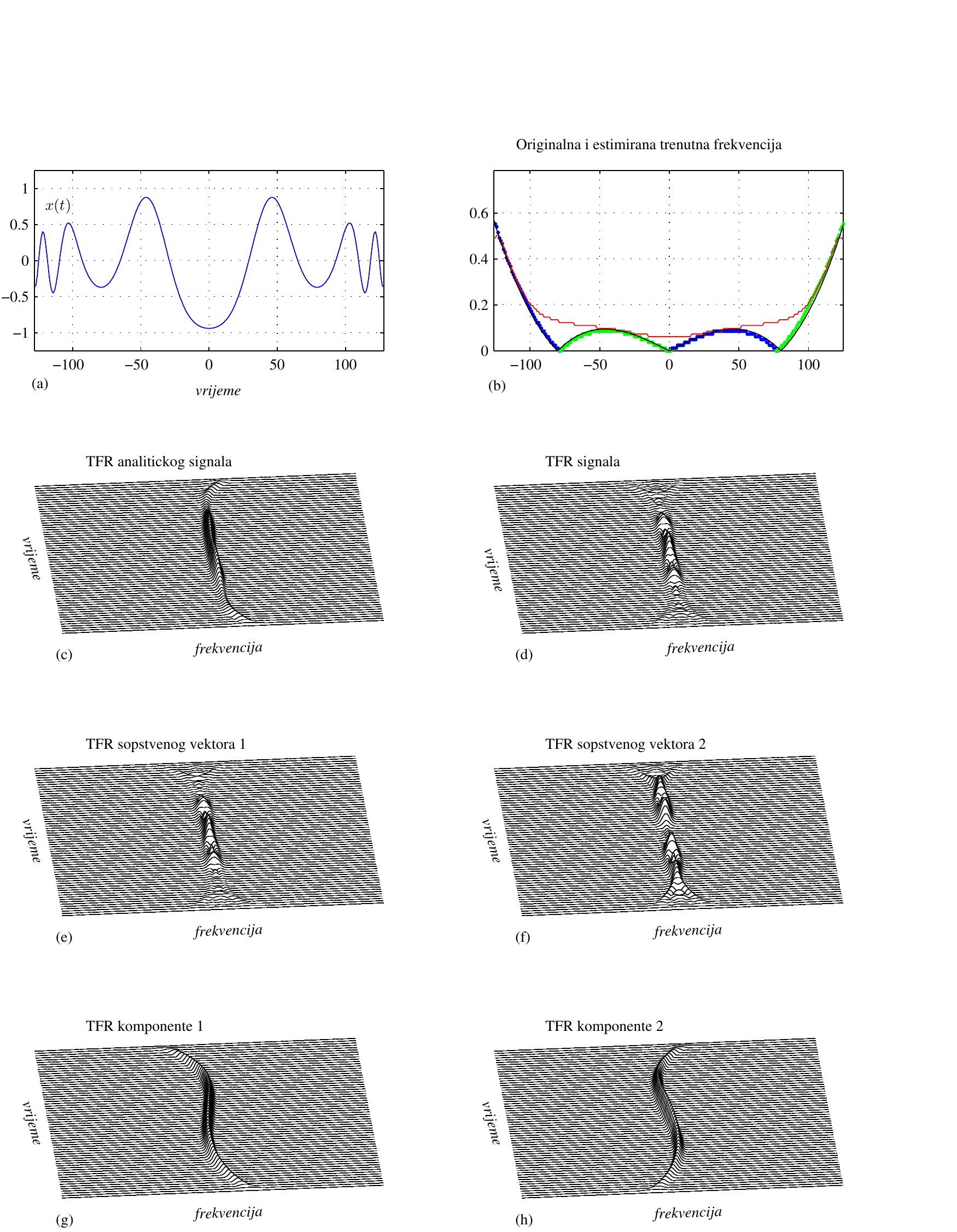}
	\caption[Realni bivarijantni signal analiziran u primjeru \ref{multivare1}.]{Realni bivarijantni signal analiziran u primjeru \ref{multivare1}: (a) signal prikazan u vremenskom domenu; (b) estimacija trenutne frekvencije: crno -- tačna IF, crveno - estimacija IF korišćenjem analitičkog signala, zeleno i plavo --  estimacija IF zasnovana na komponentama koje su izdvojene opisanim pristupom; (c) PWD analitičkog signala; (d) PWD originalnog signala; (e) i (f) PWD sopstvenih vektora; (g) i (h) PWD komponenti koje su izdvojene pristupom opisanim u ovoj sekciji.}
	\label{dec0}
\end{figure}
\subsection{Numerički rezultati}
\begin{primjer}
	\label{multivare1}

\begin{figure}[tb]
	\centering
	\includegraphics{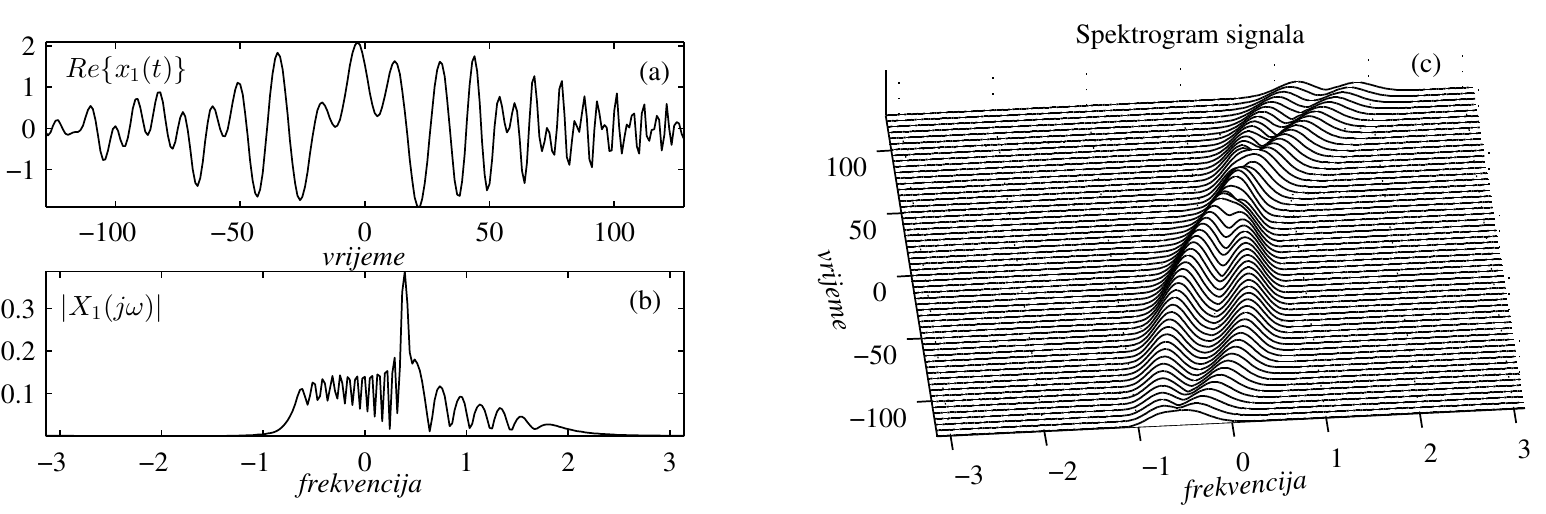}
	\caption[Dvokomponentni signal predstavljen u različitim domenima.] {Dvokomponentni signal predstavljen u: (a) vremenskom domenu; (b) frekvencijskom domenu; (c) vremensko-frekvencijskom domenu (spektrogram).}
	\label{fig2}
\end{figure}
Razmatra se realni bivarijantni signal $\mathbf x(t)=[x_1(t),~x_2(t)]^T$, u $i$-tom kanalu definisan na sljedeći način:
\begin{align}
\label{closesig}
x_{i}(t)&=e^{-(t/128)^2}\cos \left((t/16)^4/128-8\pi (t/16)^2/64+\varphi_{i}\right)\\
&=0.5e^{-(t/128)^2}\left[e^{j((t/16)^4/128-8\pi (t/16)^2/64+\varphi_{i})}+e^{-j((t/16)^4/128-8\pi (t/16)^2/64+\varphi_{i})}\right] \nonumber\\
&=x_{1i}(t)+x_{2i}(t),~i=1,~2,\nonumber
\end{align}
za $-128\leq t\leq128$. Signal iz prvog kanala je prikazan na slici \ref{dec0} (a). Faze $\varphi_{1}\ne \varphi_{2}$ predstavljaju slučajne brojeve iz intervala $[0,~2\pi]$, sa uniformnom raspodjelom. Budući da je posmatrani signal realan, u njegovoj Furijeovoj transformaciji, kao i u vremensko-frekvencijskim domenima, postoje dvije simetrične komponente $x_{1i}(t)$ i $x_{2i}(t)$. Međutim, ove komponente se djelimično preklapaju, pa su stoga nerazdvojive poznatim tehnikama koje su zasnovane na ovim reprezentacijama.

Razmotrimo uobičajeni problem estimacije trenutne frekvencije signala (IF). U tom cilju, za realne signale se uobičajeno koristi njihova analitička forma dobijena na bazi Hilbertove transformacije. Tačna trenutna frekvencija je prikazana na slici \ref{dec0} (b), crnom linijom. Vremensko-frekvencijska reprezentacija (TFR) ovog analitičkog signala prikazana je na slici \ref{dec0} (c). Međutim, rezultat estimacije trenutne frekvencije zasnovan na analitičkom signalu, prikazan na slici \ref{dec0} (b) crvenom linijom, značajno se razlikuje od tačne trenutne frekvencije. Naime, estimacija trenutne frekvencije, zasnovana na maksimumima vremensko-frekvencijske reprezentacije, ne prati varijacije trenutne frekvencije na odgovarajući način, budući da su one izgubljene u vremensko-frekvencijskoj ravni usljed značajnog preklapanja komponenti i činjenice da su varijacije amplitude i faze signala istog reda. Zapravo, uslovi Bedrosianove produktne teoreme za amplitudu i fazu signala nijesu zadovoljeni u ovom slučaju.

Ukoliko se, sa druge strane, izračuna vremensko-frekvencijska reprezentacija originalnog signala (\ref{closesig}), dvije komponente $x_{1i}(t) $ i $x_{2i}(t)$ se preklapaju u vremensko-frekvencijskoj ravni, što je prikazano na slici \ref{dec0} (d). Ove komponente su i nelinearne, te stoga nijedna poznata tehnika ne može biti primijenjena za njihovo razdvajanje u cilju estimacije trenutne frekvencije. Pošto se komponente značajno preklapaju, one nijesu ortogonalne, pa zato dekompozicija zasnovana na S-metodu \cite{dekompozicija} ne može biti direktno primijenjena. 

Sa druge strane, ključno je uočiti da su značajno izmijenjeni kros-članovi u Wigner-ovoj distribuciji, kao i da se pojavljuju samo dvije sopstvene vrijednosti različite od nule. Dva odgovarajuća sopstvena vektora, čije su pseudo-Wigner-ove distribucije prikazane na slikama \ref{dec0} (e) i (f), sadrže obije komponente, koje se pojavljuju u vidu linearne kombinacije. Korišćenjem Algoritma \ref{Alg-MV-Decomp}, odnosno procedure iz Algoritma \ref{Alg-MV-Min}, izračunati su koeficijenti $\beta_{1}$ i  $\beta_{2}$, formirajući linearnu kombinaciju (\ref{bete1}) sopstvenih vektora. Minimum mjere koncentracije, odnosno rijetkosti, odgovara dvijema razdvojenim komponentama, kao što je prikazano na slikama \ref{dec0} (g) i (h). Može se uočiti da je estimacija trenutne frekvencije na osnovu maksimuma vremensko-frekvencijskih reprezentacija ovih komponenti (korišćenjem pozitivnih djelova trenutne frekvencije), tačna do očekivanog \textit{biasa} usljed nelinearnosti trenutne frekvencije, koji može biti dalje redukovan korišćenjem poznatih tehnika za estimaciju trenutne frekvencije \cite{tfsa}. Rezultat je prikazan na slici \ref{dec0} (b), zelenim i plavim linijama.
\end{primjer}
\begin{figure}[!h]
	\centering
	\includegraphics{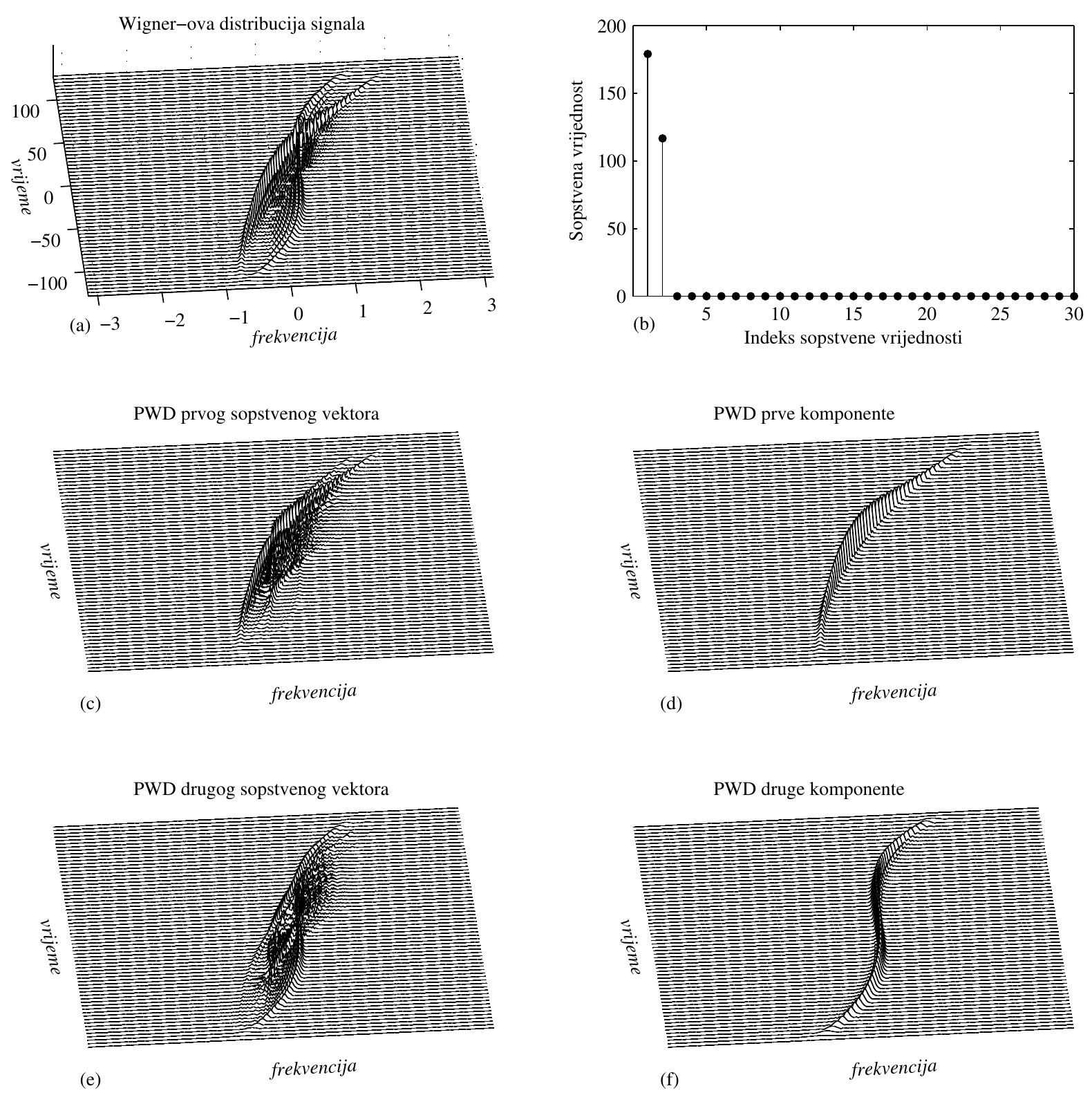}
	\caption[Dekompozicija bivarijantnog dvokomponentnog signala iz primjera \ref{multivarex2}.]{Dekompozicija bivarijantnog dvokomponentnog signala iz primjera \ref{multivarex2}: (a) WD analiziranog signala; (b) Sopstvene vrijednosti autokorelacione matrice $\mathbf R$; (c) i (e) PWD prvog i drugog sopstvenog vektora; (d) i (f) PWD izdvojenih komponenti signala.}
	\label{fig3}
\end{figure}

\begin{primjer}
 \label{multivarex2}
Razmatra se bivarijantni dvokomponentni signal $\mathbf x(t)$, pod pretpostavkom da svaki senzor mjeri: 
\begin{equation}
x_i(t)=x_{1i}(t)+x_{2i}(t),~i=1,~2
\end{equation}
gdje su komponente date izrazima:  
\begin{align}
x_{1i}(t)=1.2e^{-(t/96)^2}e^{-j12\pi (t/16)^2/25+jt^3/{256^2}+\varphi_{1i}},\label{comp1}\\
x_{2i}=0.9e^{-(t/128)^2}e^{-j\pi t/8+j(t/16)^4/{100}+\varphi_{2i}},\label{comp2}
\end{align}
sa fazama $\varphi_{1i},~\varphi_{2i},~i=1,~2$ koje se simuliraju kao slučajni brojevi sa uniformnom distribucijom iz intervala $[0,~2\pi]$. Realni dio signala iz prvog kanala, kao i odgovarajuća FT, prikazani su na slikama Fig. \ref{fig2} (a) i (b), dok je multivarijantni spektrogram prikazan na slici \ref{fig2} (c). Može se uočiti da komponente signala ne mogu biti razdvojene primjenom spektrograma, a da pri tom postupku ne dođe do značajne degradacije autočlanova. Treba uočiti da dvije komponente signala imaju nelinearnu frekvencijsku modulaciju, te su stoga nerazdvojive uobičajenim algoritmima za dekompoziciju komponenti.

Prilikom primjene prezentovanog algoritma za dekompoziciju multikomponentnih signala, u skladu sa prezentovanom teorijom, WD signala se koristi kao inicijalna vremensko-frekvencijska reprezentacija za dekompoziciju na sopstvene vrijednosti i sopstvene vektore. Wigner-ova distribucija analiziranog signala predstavljena je na slici \ref{fig3} (a), dok su sopstvene vrijednosti autokorelacione matrice $\mathbf R$ prikazane na slici \ref{fig3} (b). Uočava se da postoje dvije nenulte sopstvene vrijednosti. Njima odgovaraju dva sopstvena vektora koji sadrže linearnu kombinaciju komponenti signala. 

Za posmatrani signal se računa vremensko-frekvencijska reprezentacija, a zatim primijenjuje prezentovana procedura za minimizaciju, u cilju nalaženja koeficijenata koji prave linearnu kombinaciju sopstvenih vektora (\ref{bete1}) sa najboljom koncentracijom komponenti. Numerički eksperimenti su pokazali da se veoma slične performanse minimizacionog postupka ostvaruju primjenom Algoritma \ref{Alg-MV-Min} ukoliko se Wigner-ova distribucija, spektrogram i S-metod koriste kao vremensko-frekvencijske reprezentacije posmatranih sopstvenih vektora. Na slici  \ref{fig3}, prikazani su rezultati dobijeni primjenom WD. U cilju bolje vizuelne prezentacije rezultata, na slikama  \ref{fig3} (c) i (e) je prikazana PWD računata sa Hanovim prozorom dužine 256 odbiraka, za svaki od razmatranih sopstvenih vektora, iako se u minimizacionoj proceduri koristila WD. Slični rezultati dobili bi se primjenom bilo koje druge vremensko-frekvencijske reprezentacije u minimizacionom koraku. Pseudo-WD svake razdvojene komponente je prikazana na slikama \ref{fig3} (d) i (f), za signale $x_{1i}(t)$ i $x_{2i}(t)$, respektivno.

\begin{figure}[tb]
	\centering
	\includegraphics{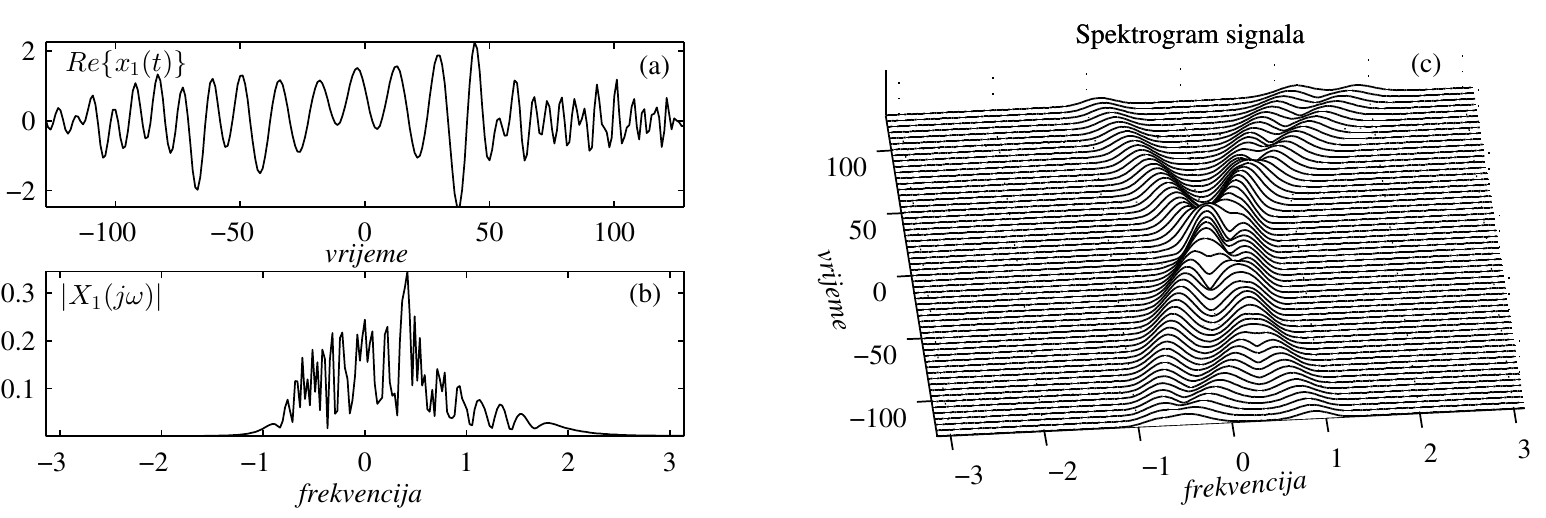}
	\caption [Multivarijantni signal sa tri komponente, razmatran u primjeru \ref{multivarex3}, prikazan u različitim domenima.]{Multivarijantni signal sa tri komponente, razmatran u primjeru \ref{multivarex3}, sa ${N_S}=4$, prikazan u: (a) vremenskom domenu; (b) frekvencijskom domenu; (c) vremensko-frekvencijskom domenu (spektrogram).}
	\label{defs3}
\end{figure}
\end{primjer}
\begin{figure}[!htb]
	\centering
	\includegraphics[]{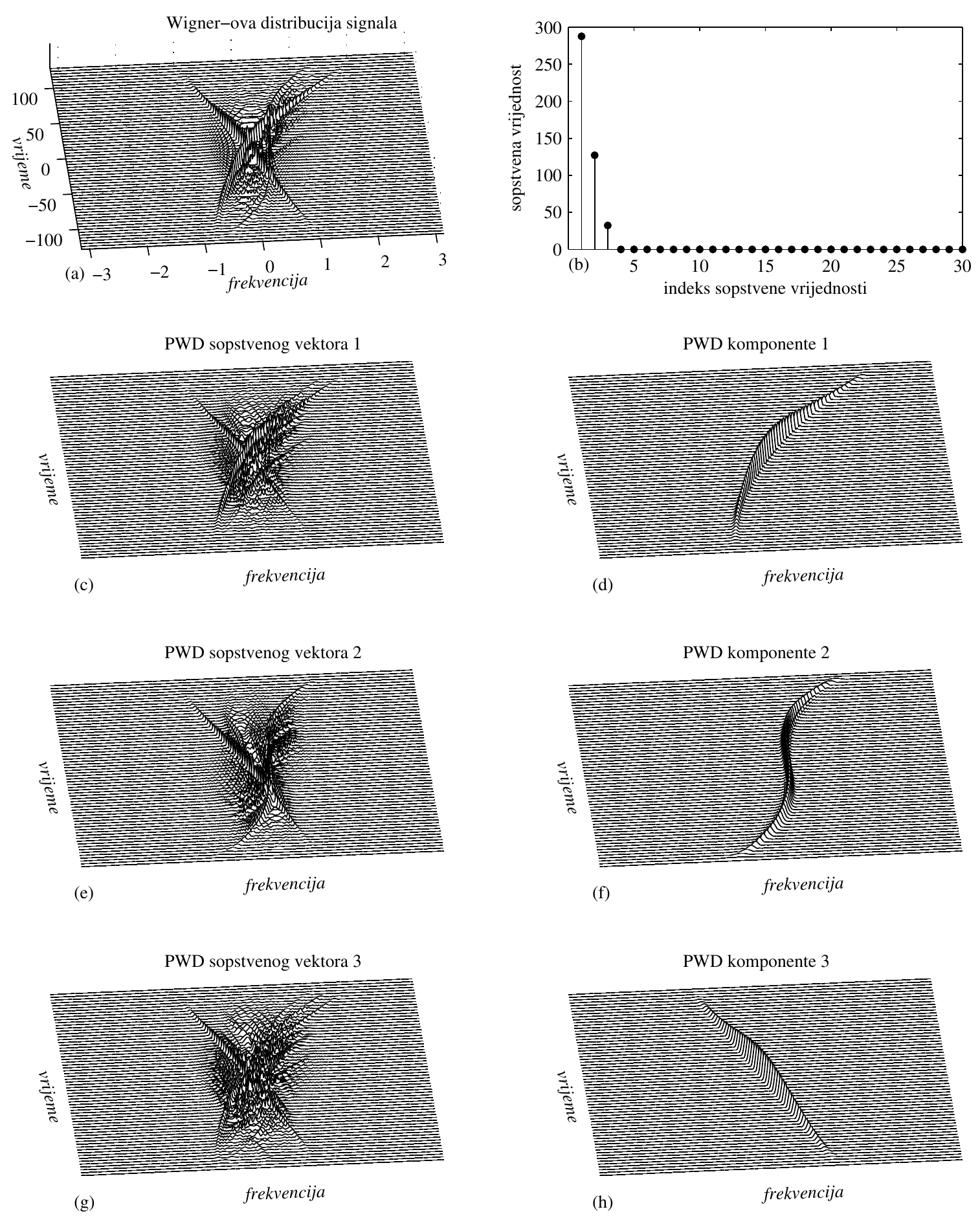}
	\caption [Dekompozicija multivarijantnog signala iz primjera \ref{multivarex3}, koji ima tri komponente (autočlana) i 4 kanala.]{Dekompozicija multivarijantnog signala iz primjera \ref{multivarex3}, koji ima tri komponente (autočlana) i ${N_S}=4$ kanala: (a) Wigner-ova distribucija signala; (b) sopstvene vrijednosti autokorelacione matrice $\mathbf R$; (c), (e) i (g): PWD prvog, drugog i trećeg sopstvenog vektora; (d), (f) i (h) PWD svih izdvojenih komponenti.}
	\label{dec3}
\end{figure}
\begin{primjer}
	\label{multivarex3}
Razmatra se multivarijantni trokomponentni signal $\mathbf x(t)$ za brojem kanala ${N_S}=4$, pri čemu je signal iz $i$-tog kanala definisan izrazom:
\begin{equation}
x_i(t)=x_{1i}(t)+x_{2i}(t)+x_{3i}(t),~i=1,\dots,4,
\end{equation}
pri čemu su komponente $x_{1i}(t)$ i $x_{2i}(t)$ date relacijama (\ref{comp1}) i (\ref{comp2}), za $i=1,\dots,~4$, dok je treća komponenta data sljedećim izrazom:
\begin{align}
x_{3i}=0.9e^{-(t/128)^2}e^{-j\pi t/8+j(t/16)^4/{100}+\varphi_{3i}},\label{comp22}
\end{align}
pri čemu i ona ima fazu $\varphi_{3i},~i=1,\dots,4$ koja je simulirana kao slučajni broj sa uniformnom raspodjelom, iz intervala $[0,~2\pi]$. Signal iz prvog kanala, njegova FT i multivarijantni spektrogram prikazani su na slikama \ref{defs3} (a) -- (c), respektivno.

\begin{figure}[htb]
	\centering
	\includegraphics[]{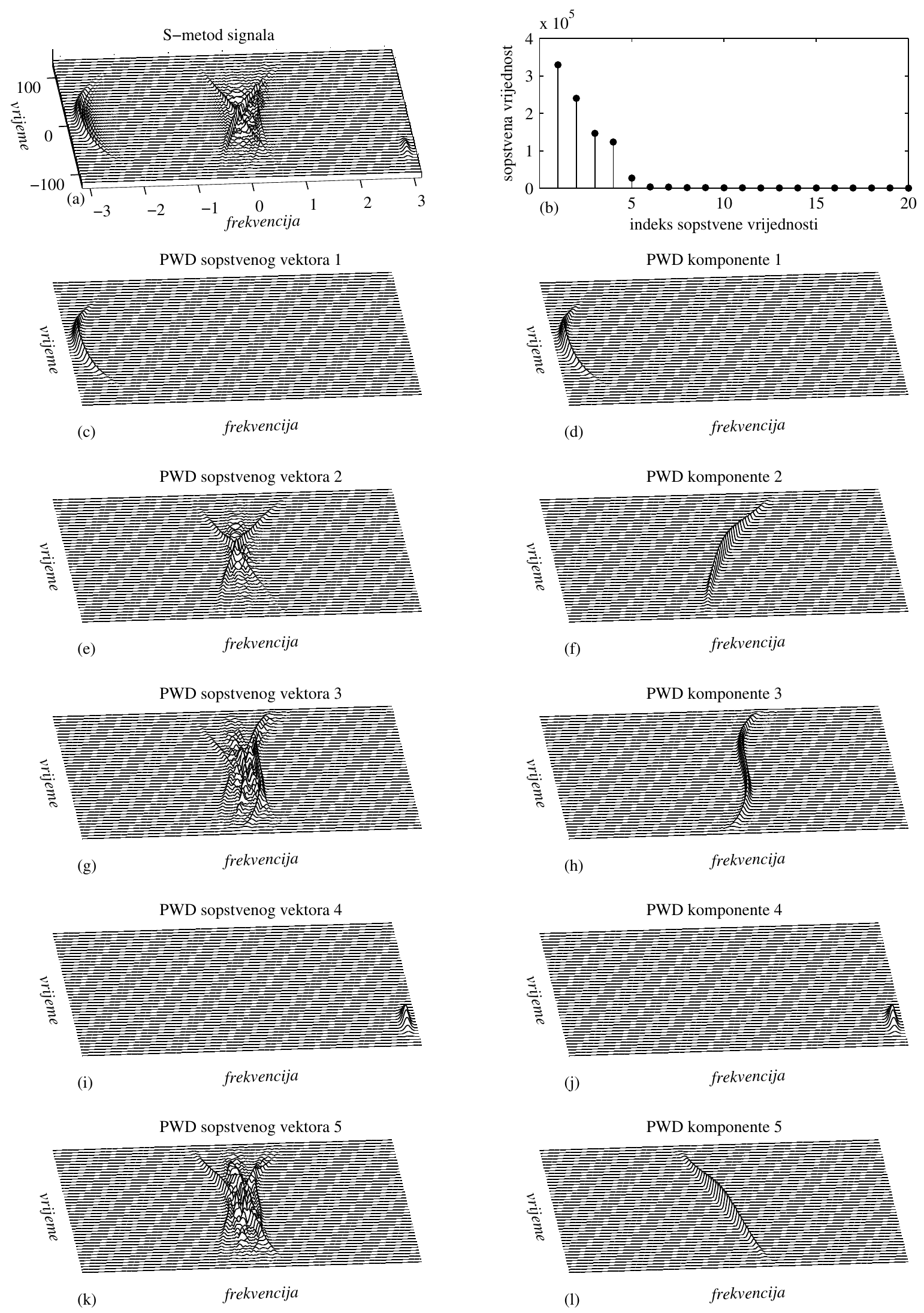}
	\caption[Dekompozicija multivarijantnog signala iz primjera \ref{primjer4mv} sa  $N=3$, zasnovana na S-metodu kao polaznoj transformaciji]{Dekompozicija multivarijantnog signala iz primjera \ref{primjer4mv} sa  $N=3$, zasnovana na S-metodu kao polaznoj transformaciji: (a) S-metod analiziranog signala; (b) sopstvene vrijednosti autokorelacione matrice $\mathbf R$; (c), (e), (g), (i), (k) PWD sopstvenih vektora koji odgovaraju najvećim sopstvenim vrijednostima; (d), (f), (h), (j) i (l) PWD izdvojenih komponenti.}
	\label{dec4}
\end{figure}

\begin{figure}[htb]
	\centering
	\includegraphics[]{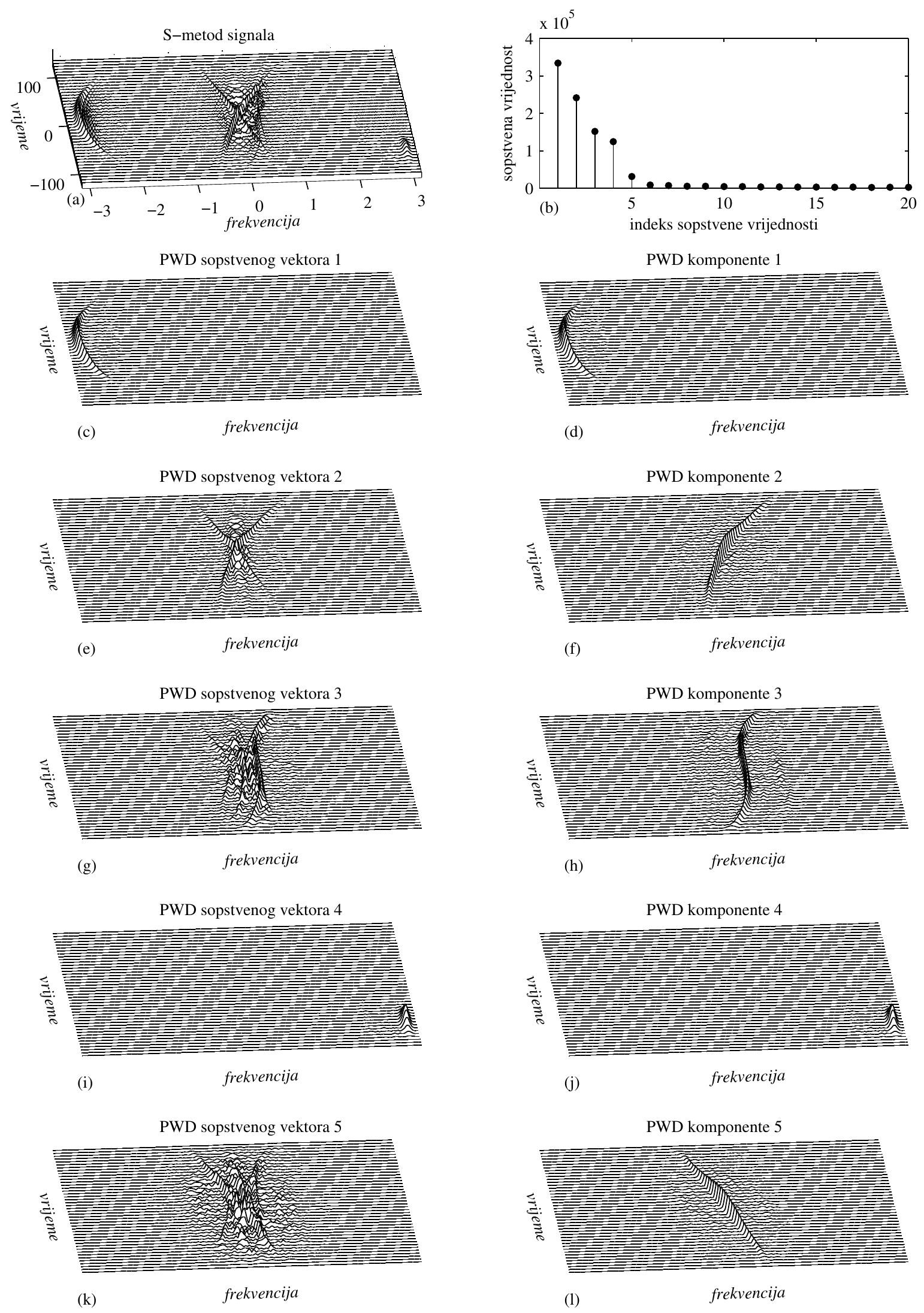}
	\caption[Dekompozicija zašumljenog multivarijantnog petokomponentnog signala iz primjera \ref{primjer4mv} sa ${N_S}=3$, zasnovana na S-metodu kao inicijalnoj reprezentaciji]{Dekompozicija zašumljenog multivarijantnog petokomponentnog signala iz primjera \ref{primjer4mv} sa ${N_S}=3$, zasnovana na S-metodu kao inicijalnoj reprezentaciji: (a) S-metod analiziranog signala; (b) sopstvene vrijednosti autokorelacione matrice $\mathbf R$; (c), (e), (g), (i), (k) PWD sopstvenih vektora koji se odnose na pet najvećih sopstvenih vrijednosti; (d), (f), (h), (j) i (l) PWD izdvojenih komponenti signala.}
	\label{dec41}
\end{figure}

Wigner-ova distribucija analiziranog signala, čija je matrica  $\mathbf R$ predmet dekompozicije na sopstvene vrijednosti i sopstvene vektore,  prikazana je na slici \ref{dec3} (a). Dobijene sopstvene vrijednosti prikazane su na slici \ref{dec3} (b), dok slike \ref{dec3} (c), (e) i (g) prikazuju pseudo-Wigner-ove distribucije sopstvenih vektora koji imaju najveće sopstvene vrijednosti na slici  \ref{dec3} (b),  ilustrujući činjenicu da komponente nijesu razdvojene. Naime, kao i u prethodnom primjeru, komponente koje se sijeku nijesu ortogonalne, pa kao posljedica toga svaki razmatrani sopstveni vektor sadrži linearnu kombinaciju komponenti signala. Na dobijene sopstvene vektore se primjenjuje predložena minimizaciona procedura, u cilju nalaženja seta koeficijenata linearne kombinacije sopstvenih vektora koji daje najbolju koncentraciju komponenti signala. Sve komponente signala su uspješno izdvojene, i one su prikazane na slici \ref{dec3} (d), (f) i (h).
\end{primjer}

\begin{primjer}
	\label{primjer4mv}
Razmatra se multivarijantni signal $\mathbf x(t)$, sastavljen od tri komponente koje se sijeku i dvije komponente koje se ne sijeku u vremensko-frekvencijskoj ravni, dat izrazom:
\begin{equation}
x_{i}(t)=x_{1i}(t)+x_{2i}(t)+x_{3i}(t)+x_{4i}(t)+x_{5i}(t).
\end{equation}
Signali u svakom od ${N_S}=3$ kanala su definisani na sljedeći način:
\begin{align}
x_{1i}(t)&=e^{-(t/96)^2}e^{j(-\pi(t/16)^2/5+\varphi_{1i})}\\
x_{2i}(t)&=1.2e^{-(t/96)^2}e^{j(\pi(t/16)^3/32+3\pi(t/16)^2/10+\varphi_{2i})}\\
x_{3i}(t)&=0.9e^{-(t/128)^2}e^{j(\pi(t/16)^4/200+\pi t/8+\varphi_{3i})}\\
x_{4i}(t)&=e^{-(t/16)^2}e^{j(3\pi t/4+\varphi_{4i})}\\
x_{5i}(t)&=e^{-(t/96)^2}e^{j(-6\pi(t/16)^2/25+\pi t/4+\varphi_{5i})}
\end{align}
gdje $i=1,~2,~3$ označava indeks kanala.
U ovom primjeru, S-metod se koristi kao inicijalna vremensko-frekvencijska reprezentacija, i prikazan je na slici \ref{dec4} (a). Budući da je posmatrani signal sa tri kanala, a pritom ima pet komponenti, primjena S-metoda u ovom kontekstu je od ključnog značaja. S-metod može izdvojiti sve nepreklapajuće komponente u jednoj realizaciji. Zatim, se dostupne realizacije koriste samo za preklapajuće komponente. Dekompozicija na sopstvene vrijednosti i sopstvene vektore inverzne autokorelacione matrice S-metoda, $\mathbf{R}$, daje pet sopstvenih vektora koji odgovaraju najvećim sopstvenim vrijednostima prikazanim na slici \ref{dec4} (b).

Budući da su dvije nepreklopljene komponente međusobno ortogonalne, ali i ortogonalne sa ostatkom presječenih komponenti, na osnovu teorije iz \cite{dekompozicija}, postoje tačno dva sopstvena vektora koji odgovaraju ovim komponentama (po jedan sopstveni vektor za svaku od dvije komponente). Pseudo-Wigner-ove distribucije ovih sopstvenih vektora prikazane su na slici \ref{dec4} (c) i (i). Dakle, ove dvije komponente su jednostavno izdvojene, što je i prikazano na slici \ref{dec4} (d) i (j). Tri preostale komponente se dobijaju na osnovu adekvatne linearne kobinacije preostala tri sopstvena vektora, korišćenjem koeficijenata $\beta_i$  koji se dobijaju minimizacionom procedurom koja je predstavljena u ovoj sekciji. PWD tri preostala sopstvena vektora prikazane su na slici \ref{dec4} (e), (g) i (k), dok su odgovarajuće razdvojene komponente, dobijene na osnovu adekvatne linearne kombinacije sopstvenih vektora, prikazane na slici \ref{dec4} (f), (h) i (l). 

Isti eksperiment je ponovljen za zašumljeni signal $  \hat{\mathbf x} (t)=\mathbf x(t)+\mathbf{\varepsilon}(t)$. Signal iz svakog kanala je oštećen aditivnim, bijelim, kompleksnim Gausovim šumom srednje vrijednosti nula, sa identičnom raspodjelom realnog i imaginarnog dijela, pri čemu oba dijela imaju istu varijansu $\sigma^2=0.15^2$. SNR za jednu (linearnu) komponentu je $7.13$ dB,  što je relativno niska vrijednost. Rezultati dobijeni primjenom predloženog algoritma za dekompoziciju prikazani su na slici \ref{dec41}, koja ilustruje robustnost algoritma na uticaj aditivnog Gausovog šuma.
\end{primjer}

\chapter*{Zaključak\markboth{Zaključak}{}}
Ova disertacija sadrži veći broj originalnih doprinosa u oblasti rekonstrukcije signala koji su rijetki, odnosno, visoko koncentrisani u različitim transformacionim domenima: Furijeovom, Hermitskom, zatim u domenima jednodimenzione i dvodimenzione diskretne kosinusne transformacije kao i u vremensko-frekvencijskim domenima, uključujući Wigner-ovu distribuciju, S-metod i kratkotrajnu Furijeovu transformaciju. Rezultati teze nijesu ograničeni samo na rekonstrukciju u kontekstu kompresivnog odabiranja. Pokazano je da mjere koncentracije, kao fundamentalan koncept vezan za transformacione domene signala, mogu biti primijenjene i u optimizaciji parametara transformacija, uklanjanju šuma ili dekompoziciji nestacionarnih signala korišćenjem vremensko-frekvencijske analize. Navedeno znači da se koncept mjera koncentracije može koristiti u okviru rješavanja različitih problema. Rezultati primjene predloženih algoritama na realne signale -- biomedicinske, audio signale, digitalne slike, telekomunikacione, radarske i druge signale, koji su predstavljeni u tezi, dodatno potkrepljuju ovu činjenicu. Drugi aspekti praktičnih primjena predmet su tekućih i budućih istraživanja autora.

U kontekstu kompresivnog odabiranja, pored sličnosti nekih opštih koncepata, određene specifičnosti transformacija zahtijevaju razvijanje posebne teorije koja će osvijetliti proces rekonstrukcije, uticaj nedostajućih odbiraka u signalima, kao i performanse algoritama za rekonstrukciju. Jedan od fundamentalnih izazova, odnosno istraživačkih pravaca koji je ova teza postavila, jeste i mogućnost generalizacije predstavljenih rezultata, odnosno, njihovog objedinjavanja u konzistentnan i cjelovit teorijski okvir.

I pored velikog broja doprinosa, koji su publikovani kroz veći broj radova u renomiranim časopisima i koji su izloženi na većem broju uglednih međunarodnih konferencija, istraživanja u sklopu ove teze otvorila su brojna nova pitanja i istraživačke izazove. U tom smislu, planira se nastavak istraživanja u svim razmatranim oblastima. Započeto je proširivanje prezentovane teorije na Hermitsku transformaciju drugog tipa, u cilju razvoja algoritama za optimizaciju njenih parametara i analize uticaja nedostajućih odbiraka, gdje su ostvareni inicijalni rezultati. Jedan dio novijih rezultata vezanih za rekonstrukciju signala koji se odabiraju mimo uniformnog vremenskog grida nije uključen u ovu disertaciju. U budućim istraživanjima će biti pokazano da predstavljeni izrazi za grešku u slučaju uniformnog odabiranja zapravo predstavljaju specijalni slučaj izraza koji će biti izvedeni za opštiji slučaj neuniformnog, odnosno, slučajnog odabiranja.

Još jedan aspekat kompresivnog odabiranja koji nije istražen u tezi jeste i uticaj kvantizacije na mogućnost, tok i ishode rekonstrukcije. Planira se ispitivanje uticaja ograničavanja broja bita kojim se reprezentuju mjerenja i sa kojima rade algoritmi koji se koriste za rekonstrukciju. U tom smislu, moguće je izvesti uslove za rekonstrukciju, izraze za vjerovatnoće grešaka u rekonstrukciji, energije grešaka, kao i druge relevantne izraze. U tom slučaju, moguće je odrediti koji faktor (broj nedostajućih odbiraka, pretpostavljeni stepen rijetkosti, broj bita kojima se reprezentuju mjerenja, varijansa aditivnog šuma itd.) dominantno utiče na mogućnost i performanse procesa rekonstrukcije.

Mogućnost da se  komponente multikomponentnih multivarijantnih signala predstave kao linearne kombinacije sopstvenih vektora otvorila je prostor istraživanja primjene različitih optimizacionih pristupa, kao što su genetički algoritmi, u pretrazi prostora koeficijenata te linearne kombinacije, u cilju minimizacije mjere koncentracije vremensko-frekvencijske reprezentacije (što odgovara linearnim kombinacijama sopstvenih vektora koje predstavljaju komponente signala).
Naročito je interesantno ispitivanje mogućnosti primjene prezentovane dekompozicione tehnike na realne signale koji su dobijeni snimanjem pomoću više senzora.

 Uklanjanje impulsnih smetnji u slučaju audio signala, odnosno, mogućnost njihove rekonstrukcije, otvara dodatne istraživačke teme i u ovoj oblasti. To se prevashodno odnosi na mogućnost razvijanja algoritama za detekciju impulsnih smetnji, koji bi bili zasnovani, na primjer, na diferenciranju signala i detekciji impulsa pomoću mjera koncentracije u DCT domenu. 
 Još jedna vrlo insteresantna istraživačka tema jeste i rekonstrukcija signala čiji transformacioni koeficijenti nijesu pozicionirani na frekvencijskom gridu diskretne Furijeove transformacije, što može biti posebno izazovno, budući da je nepoznata odstupanja od grida takođe neophodno procijeniti. 
 Na kraju, treba spomenuti da se jedan od budućih istraživačkih pravaca tiče   obrade signala na grafovima. U tom pogledu, mnogi koncepti izloženi u ovoj tezi mogu biti prenijeti na opštiji slučaj ove vrste signala.
 
\addcontentsline{toc}{chapter}{Zaklju\v{c}ak}

\clearpage
\thispagestyle{empty}
\begin{center}
\textbf{KRATKA BIOGRAFIJA AUTORA}
\end{center}

	Miloš Brajović je rođen 24.05.1988. godine u Podgorici, Republika Crna Gora. Nakon završene osnovne škole ,,Njegoš'' u Spužu, 2003. godine upisao je opšti smjer gimnazije ,,Petar I Petrović Njegoš'' u Danilovgradu. Dobitnik je diploma ,,Luča'', za ostvarene rezultate u osnovnoj i srednjoj školi. Od strane nastavničkog vijeća gimnazije proglašen je za učenika generacije.
	
	Elektrotehnički fakultet u Podgorici, na odsjeku Elektronika, telekomunikacije i računari, upisuje septembra 2007. godine. Specijalističke studije, na smjeru Računari, upisuje nakon završenih osnovnih studija, i završava ih u junu 2011. godine, odbranom specijalističkog rada ,,\textit{Vremensko-frekvencijska analiza i predstavljanje trenutne frekvencije}'', čime je stekao stepen Specijaliste (Spec. Sci) za oblast Elektronike, telekomunikacija i računara.
	
	Od septembra 2011. godine angažovan je kao saradnik u nastavi na Elektrotehničkom fakultetu u Podgorici, gdje iste godine upisuje i magistarske studije, na smjeru Računari, pod mentorstvom prof. dr Miloša Dakovića. Magistarsku tezu sa temom: ,,\textit{Rekurzivno izračunavanje vremensko-frekvencijskih reprezentacija}'' odbranio je u oktobru 2013. godine. Počevši od 2012. godine, bio je, ili je još uvijek, angažovan na preko 10 naučno-istraživačkih i stručnih projekata. 
	
	Miloš Brajović obavlja dužnost recezenta za više renomiranih međunarodnih časopisa sa SCI/SCIE liste, za koje je do sada recenzirao preko 70 naučnih radova. Među ovim časopisma su: \textit{IEEE Transactions on Signal Processing, IEEE Signal Processing Letters, IEEE Transactions on Geoscience and Remote Sensing, Signal Processing, Digital Signal Processing, IET Signal Processing, Electronics Letters, IET Image Processing, IET Radar Sonar \& Navigation, Signal image and video processing, Biomedical Signal Processing and Control, Journal of Visual Communication and Image Representation} i drugi. Obavljao je dužnost recezenta i za domaće časopise, kao i za veći broj međunarodnih i domaćih konferencija. 
	
	Miloš Brajović je do sada publikovao 8 radova u međunarodnim časopisima sa SCI/SCIE liste (sa kumulativnim IMPACT Factor-om od 21.27), 4 rada u ostalim časopisima, 28 radova na međunarodnim konferencijama (indeksiranim u SCOPUS-u i IEEE Xplore bazama) i 11 radova na regionalnim i lokalnim konferencijama. Više radova, na kojima je jedan od autora, nalazi se u postupku recenziranja.  Kompletan spisak referenci dostupan je na sajtu TFSA grupe {\url{http://www.tfsa.ac.me/milosb_papers.html}}. 
	
	Miloš Brajović je drugorangirani dobitnik nagrade u konkurenciji za najuspješnijeg naučnika do 30 godina života, koju mu je dodijelilo Ministarstvo nauke za 2018. godinu. Miloš Brajović je 2018. godine dobio i Plaketu Univerziteta Crne Gore, za doprinos u broju publikovanih radova u renomiranim časopisima sa SCI/SCE liste u oblasti tehničkih nauka.

\clearpage
\thispagestyle{empty}

\clearpage
\thispagestyle{empty}
\begin{center}
	\textbf{IZJAVA O AUTORSTVU}
\end{center}

\vspace{2cm}

\begin{flushleft}
	\begin{minipage}[t]{5.5cm}
	\flushleft
	Potpisani/a:\\
	\vspace{6pt}
	Broj indeksa: \\

	\vspace{4pt}

\end{minipage}
\null\hfill
\begin{minipage}[t]{10cm}
	\flushleft
	{Miloš Brajović} \\
	\vspace{5pt}
	{2/2013}\\
	\vspace{5pt}
\end{minipage}
\end{flushleft}
	\vspace{2cm}
\begin{center}
	\textbf{Izjavljujem}
\end{center}
	\vspace{1cm}
da je doktorska disertacija pod naslovom:

 	\vspace{1cm}
 	
\noindent\textbf{Analiza algoritama za rekonstrukciju signala rijetkih u Hermitskom i Furijeovom transformacionom domenu}

\begin{itemize}
\item[--]	rezultat sopstvenog istraživačkog rada;
\item[--] da predložena disertacija ni u cjelini ni u djelovima nije bila predložena za dobijanje bilo koje diplome prema studijskim programima drugih ustanova visokog obrazovanja;
\item[--]	da su rezultati korektno navedeni, i 
\item[--]	da nijesam povrijedio autorska i druga prava intelektualne svojine koja pripadaju trećim licima.
\end{itemize}

\vfill

\begin{flushleft}
	\begin{minipage}[t]{8.5cm}
\noindent	Podgorica, \\
oktobar 2018. godine\\
\end{minipage}
\null\hfill
\begin{minipage}[t]{5.8cm}
	\flushleft
	Potpis doktoranda:\\
	\vspace{3mm}
	\line(1,0){160} \\
	%	Ime i prezime:\\
	%	Datum i mjesto ro\dj enja: \\
	%	Naziv zavr\v{s}enog postdiplomskog studijskog programa:\\
	%	Godina zavr\v{s}etka>
\end{minipage}
\end{flushleft}

\clearpage
\thispagestyle{empty}

\begin{center}
\textbf{IZJAVA O ISTOVJETNOSTI ŠTAMPANE I 
	ELEKTRONSKE VERZIJE DOKTORSKOG RADA}
\end{center}

\vspace{2cm}

\begin{flushleft}
	\begin{minipage}[t]{5.5cm}
		\flushleft
		Ime i prezime autora:\\
		\vspace{6pt}
		Broj indeksa/upisa:\\
		\vspace{6pt}
		Studijski program:\\
		\vspace{6pt}
		Naslov rada: \\
		\text{ }\\
		Mentor:\\
		Potpisani:\\
		\vspace{6pt}
	
	\end{minipage}
	\null\hfill
	\begin{minipage}[t]{10cm}
		\flushleft
		Miloš Brajović\\
	\vspace{6pt}
	2/2013\\
	\vspace{6pt}
Doktorske studije elektrotehnike\\
	\vspace{6pt}
	\textbf{Analiza algoritama za rekonstrukciju signala rijetkih u Hermitskom i Furijeovom transformacionom domenu}\\
	Prof. dr Miloš Daković\\
	Miloš Brajović\\
	\vspace{6pt}
	\end{minipage}
\end{flushleft}
\vspace{2cm}
\noindent Izjavljujem da je štampana verzija mog doktorskog rada istovjetna elektronskoj verziji koju sam predao za objavljivanje u Digitalni arhiv Univerziteta Crne Gore.

\vspace{3mm}

\noindent Istovremeno izjavljujem da dozvoljavam objavljivanje mojih ličnih podataka u vezi sa dobijanjem akademskog naziva doktora nauka, odnosno zvanja doktora umjetnosti, kao što su ime i prezime, godina i mjesto rođenja, naziv disertacije i datum odbrane rada.

\vfill

\begin{flushleft}
	\begin{minipage}[t]{8.5cm}
		\noindent	Podgorica, \\
		oktobar 2018. godine\\
	\end{minipage}
	\null\hfill
	\begin{minipage}[t]{5.8cm}
		\flushleft
		Potpis doktoranda:\\
		\vspace{3mm}
		\line(1,0){160} \\
		%	Ime i prezime:\\
		%	Datum i mjesto ro\dj enja: \\
		%	Naziv zavr\v{s}enog postdiplomskog studijskog programa:\\
		%	Godina zavr\v{s}etka>
	\end{minipage}
\end{flushleft}

\clearpage
\thispagestyle{empty}

\begin{center}
	\textbf{IZJAVA O KORIŠĆENJU}
\end{center}

\vspace{2cm}

\noindent Ovlašćujem Univerzitetsku biblioteku da u Digitalni arhiv Univerziteta Crne Gore pohrani moju doktorsku disertaciju pod naslovom:

\vspace{1cm}

\noindent\textbf{Analiza algoritama za rekonstrukciju signala rijetkih u Hermitskom i Furijeovom transformacionom domenu}

\vspace{1cm}

\noindent koja je moje autorsko djelo.

\vspace{1cm}

\noindent Disertaciju sa svim prilozima predao sam u elektronskom formatu pogodnom za trajno arhiviranje.

\vspace{3mm}

\noindent Moju doktorsku disertaciju pohranjenu u Digitalni arhiv Univerziteta Crne Gore mogu da koriste svi koji poštuju odredbe sadržane u odabranom tipu licence Kretivne zajednice (Creative Commons) za koju sam se odlučio.

\vfill

\begin{enumerate}
	\item	Autorstvo
	\item	Autorstvo -- nekomercijalno
	\item	\textbf{Autorstvo – nekomercijalno -- bez prerade}
	\item	Autorstvo -- nekomercijalno -- dijeliti pod istim uslovima
	\item	Autorstvo -- bez prerade
	\item	Autorstvo -- dijeliti pod istim uslovima	
\end{enumerate}

\vfill

\begin{flushleft}
	\begin{minipage}[t]{8.5cm}
	
		\noindent	Podgorica, \\
		oktobar 2018. godine\\
	\end{minipage}
	\null\hfill
	\begin{minipage}[t]{5.8cm}
		\flushleft
		Potpis doktoranda:\\
		\vspace{3mm}
		\line(1,0){160} \\
		%	Ime i prezime:\\
		%	Datum i mjesto ro\dj enja: \\
		%	Naziv zavr\v{s}enog postdiplomskog studijskog programa:\\
		%	Godina zavr\v{s}etka>
	\end{minipage}
\end{flushleft}

\clearpage
\thispagestyle{empty}
\begin{enumerate}
\item \textbf{Autorstvo}. Licenca sa najširim obimom prava korišćenja. Dozvoljavaju se prerade, umnožavanje, distribucija i
javno saopštavanje djela, pod uslovom da se navede ime izvornog autora (onako kako je izvorni
autor ili davalac licence odredio).
Djelo se može koristiti i u komercijalne svrhe.

\item  \textbf{Autorstvo -- nekomercijalno}. Dozvoljavaju se prerade, umnožavanje, distribucija i javno saopštavanje djela, pod uslovom da se
navede ime izvornog autora (onako kako je izvorni autor ili davalac licence odredio).
Komercijalna upotreba djela nije dozvoljena.

\item  \textbf{Autorstvo -- nekomercijalno -- bez prerade}. Licenca kojom se u najvećoj mjeri ograničavaju prava korišćenja djela. Dozvoljava se umnožavanje, distribucija i javno saopštavanje djela, pod uslovom da se navede ime izvornog autora (onako kako je izvorni autor ili davalac licence odredio). Djelo se ne može mijenjati, preoblikovati ili koristiti u drugom djelu. Komercijalna upotreba djela nije dozvoljena.

\item  \textbf{Autorstvo -- nekomercijalno -- dijeliti pod istim uslovima}. Dozvoljava se umnožavanje, distribucija, javno saopštavanje i prerada djela, pod uslovom da se
navede ime izvornog autora (onako kako je izvorni autor ili davalac licence odredio). Ukoliko se
djelo mijenja, preoblikuje ili koristi u drugom djelu, prerada se mora distribuirati pod istom ili
sličnom licencom.
Djelo i prerade se ne mogu koristiti u komercijalne svrhe.

\item  \textbf{Autorstvo -- bez prerade}. Dozvoljava se umnožavanje, distribucija i javno saopštavanje djela, pod uslovom da se navede ime
izvornog autora (onako kako je izvorni autor ili davalac licence odredio). Djelo se ne može mijenjati,
preoblikovati ili koristiti u drugom djelu.
Licenca dozvoljava komercijalnu upotrebu djela.

\item \textbf{Autorstvo -- dijeliti pod istim uslovima}. Dozvoljava se umnožavanje, distribucija i javno saopštavanje djela, pod uslovom da se navede ime
izvornog autora (onako kako je izvorni autor ili davalac licence odredio). Ukoliko se djelo mijenja,
preoblikuje ili koristi u drugom djelu, prerade se moraju distribuirati pod istom ili sličnom licencom.
Ova licenca dozvoljava komercijalnu upotrebu djela i prerada. Slična je softverskim licencama,
odnosno licencama otvorenog koda.

\end{enumerate}

\end{document}